\documentclass[a4paper,12pt]{article}

\usepackage{a4}
\usepackage{axodraw}
\usepackage{amsfonts}
\usepackage{amssymb}
\usepackage{graphicx}

\newcommand{\drawsquare}[2]{\hbox{%
\rule{#2pt}{#1pt}\hskip-#2pt
\rule{#1pt}{#2pt}\hskip-#1pt
\rule[#1pt]{#1pt}{#2pt}}\rule[#1pt]{#2pt}{#2pt}\hskip-#2pt
\rule{#2pt}{#1pt}}
\newcommand{\fund}{\raisebox{-.5pt}{\drawsquare{6.5}{0.4}}}
\newcommand{\Ysymm}{\raisebox{-.5pt}{\drawsquare{6.5}{0.4}}\hskip-0.4pt%
        \raisebox{-.5pt}{\drawsquare{6.5}{0.4}}}
\newcommand{\Yasymm}{\raisebox{-3.5pt}{\drawsquare{6.5}{0.4}}\hskip-6.9pt%
        \raisebox{3pt}{\drawsquare{6.5}{0.4}}}
\newcommand{\antifund}{\overline{\fund}}

\usepackage{bm}
\def\p{\partial}

\def\ba{\begin{array}}
\def\ea{\end{array}}

\def\ICC{{\bf C}}

\newcommand{\be}{\begin{equation}}
\newcommand{\zbe}{\begin{equation}}
\newcommand{\ee}{\end{equation}}

\def\fc#1#2{{\frac{#1}{#2}}}
\def\lf{\left}\def\ri{\right}
\def\h{\fc{1}{2}}
\def\req#1{(\ref{#1})}

\def\im{{\rm Im}}
\def\ie{{\it i.e.\ }}\def\eg{{\it e.g.\ }}
\def\vev#1{\langle #1 \rangle}
\def\Kc{{\cal K}}\def\Oc{{\cal O}}\def\Mc{{\cal M}}\def\Ac{{\cal A}}
\def\Dc{{\cal D}}\def\Bc{{\cal B}}\def\Cc{{\cal C}}\def\Vc{{\cal V}}
\def\Tr{{\rm Tr}}

\def\ds{\displaystyle}
\def\cff{{\it c.f.\ }}
\def\ie{{\it i.e.\ }}
\def\eg{{{\it e.g.\ }}}
\def\eqq{{{\it Eq.\ }}}
\def\ra{\rightarrow}
\def\lra{\longrightarrow}

\def\re{{\rm Re}}
\def\im{{\rm Im}}
\def\beqn{\begin{eqnarray}}
\def\eeqn{\end{eqnarray}}
\def\nnn{\nonumber}

\def\eqn#1#2{\zbe\label{#1} #2  \ee}

\def\tb{type IIB\ }\def\ta{type IIA\ }
\def\npp{{h_+^{1,1}({\cal X})}}
\def\mmm{{h_-^{1,1}({\cal X})}}
\def\IZ{\mathbb{Z}}\def\ICC{\mathbb{C}}\def\IR{\mathbb{R}}
\def\Om{\Omega}\def\th{\theta}\def\om{\omega}\def\si{\sigma}\def\La{\Lambda}\def\Si{\Sigma}
\def\al{\alpha}\def\bet{\beta}\def\eps{\epsilon}
\def\ap{\al'}
\def\ninepoint{}

\def\tf{\tilde f}
\def\fao{^{{\cal X}}\hskip-0.15cm f^a}
\def\fo{^{{\cal X}}\hskip-0.15cm f}

\newcommand{\n}{{\bf n}}

\newcommand{\ov}{\overline}
\newcommand{\ca}{{\cal A}}

\newcommand{\cf}{{\cal F}}
\newcommand{\cg}{{\cal G}}

\newcommand{\ch}{{\rm ch}}

\newcommand{\ck}{{\cal K}}
\newcommand{\cl}{{\cal L}}
\newcommand{\cm}{{\cal M}}
\newcommand{\cn}{{\cal N}}
\newcommand{\co}{{\cal O}}
\newcommand{\cp}{{\cal P}}

\newcommand{\cR}{{\cal R}}
\newcommand{\cs}{{\cal S}}
\newcommand{\ct}{{\cal T}}

\newcommand{\cv}{{\cal V}}
\newcommand{\cw}{{\cal W}}
\newcommand{\cx}{{\cal X}}
\newcommand{\cy}{{\cal Y}}
\newcommand{\cz}{{\cal Z}}
\def\IC{\relax\,\hbox{$\inbar\kern-.3em{\rm C}$}}
\def\IR{\relax{\rm I\kern-.18em R}}
\def\IN{\relax{\rm I\kern-.18em N}}
\def\IP{\relax{\rm I\kern-.18em P}}
\def\a{{\alpha}}

\def\d{{\delta}}
\def\e{{\epsilon}}

\def\vf{{\varphi}}
\def\g{{\gamma}}
\def\i{{\iota}}
\def\k{{\kappa}}
\def\l{{\lambda}}
\def\m{{\mu}}
\def\n{{\nu}}
\def\o{{\omega}}
\def\p{{\psi}}
\def\r{{\rho}}
\def\s{{\sigma}}
\def\bs{\bar{\sigma}}
\def\t{{\tau}}
\def\th{{\theta}}

\def\x{{\xi}}

\def\D{{\Delta}}

\def\G{{\Gamma}}

\def\L{{\Lambda}}

\def\O{{\Omega}}
\def\P{{\Psi}}

\def\S{{\Sigma}}
\def\Th{{\Theta}}

\def\bbe{{\bar \beta}}

\def\Im{{\rm Im}}
\def\Re{{\rm Re}}
\newcommand{\tr}{{\rm tr}}

\newcommand{\tht}{\vartheta}

\newcommand{\thba}[2]{\vartheta[\!\!\begin{array}{c}
     {\phantom{}\vspace{-.5mm}\scriptstyle#1}%
                        \\[-1.6mm]{\scriptstyle #2}\end{array}\!\!]}

\newcommand{\zba}[2]{[\!\!\begin{array}{c}{\scriptstyle#1}
                        \\[-1.6mm]{\scriptstyle #2}\end{array}\!\!]}

\newcommand{\non}{\nonumber\\ }

\newcommand{\bea}{\begin{eqnarray}}
\newcommand{\eea}{\end{eqnarray}}
\newcommand{\mbb}{\mathbb}
\newcommand{\reef}[1]{(\ref{#1})}

\newcommand{\ti}{\times}

\def\<{\langle}
\def\>{\rangle}
\newcommand{\pa}{\partial}
\newcommand{\lab}{\label}

\newcommand{\gs}{Green-Schwarz }
\newcommand{\FI}{Fayet-Iliopoulos }

\begin{document}

\begin{flushright} \vspace{-2cm}
{\small
CERN-PH-TH/2006-218\\
LMU-ASC 66/06 \\
MPP-2006-135 }
\end{flushright}
\vspace{0cm}
\begin{center}
{\bf\Large
Four-dimensional String Compactifications with \\[.2cm]
D-Branes, Orientifolds and Fluxes\\[1cm]
}

Ralph Blumenhagen$^1$, Boris K\"ors$^{2}$, Dieter L\"ust$^{1,3}$
and
Stephan Stieberger$^{2,3}$\\[0.5cm]

{\small
\emph{$^1$ Max-Planck-Institut f\"ur Physik, \\
F\"ohringer Ring 6, D-80805 M\"unchen, Germany}\\
\vspace{0.1cm}
\emph{$^2$  Physics Department, Theory Division, CERN\\
CH-1211 Geneve 23, Switzerland}\\
\vspace{0.1cm}
\emph{$^3$ Arnold-Sommerfeld-Center for Theoretical Physics, Department f\"ur Physik,\\
Ludwig-Maximilians-Universit\"at M\"unchen, Theresienstra{\ss}e
37, 80333 M\"unchen, Germany} }

\vspace{1cm}

{\bf Abstract}
\end{center}

\vspace{-.3cm}

\noindent
This review article provides a pedagogical introduction into
various classes of chiral string compactifications to four
dimensions with D-branes and fluxes. The main concern 
is to provide all necessary technical tools to explicitly construct
four-dimensional orientifold vacua, with the final aim to come as close 
as possible to the supersymmetric Standard Model.
Furthermore, we  outline  the available  methods to derive the resulting 
four-dimensional effective action.
Finally,  we summarize  recent attempts
to address the string vacuum problem via the statistical approach
to D-brane models. 

\vskip 0.2cm
\noindent
{\small Pacs numbers: 11.25.Mj, 11.25.-w, 11.25.Wx}
\vskip1.2cm
\vfill
\protect\hspace{0cm}\rule{5.5cm}{0.1mm}  
\vskip -0.3cm
\noindent
{\small e-mail address of the review: revibw@mppmu.mpg.de}
\thispagestyle{empty}
\clearpage

\tableofcontents
\clearpage


\section{INTRODUCTION}

The history of theoretical particle physics has been extremely
successful. Based on the principles of quantum  mechanics and its
relativistic generalization in the form of quantum field theory a
unified framework could be developed over the second half of the
twentieth century allowing the prediction of many experimental
data with amazing precision. In the so-called electro-weak
Standard Model (SM) of particle physics the fundamental particles,
the quarks, leptons and the Higgs scalar, interact via three types
of gauge interactions, namely the strong, the weak and the
electromagnetic interaction. The fermionic matter particles come
in three identical copies which only differ by their mass. The
third family is hierarchically heavier than the first
two\footnote{If this also holds for the neutrino member remains
to be seen.}. All stable particles we observe in our universe
consist only of fermions from the first and lightest family. The
only ingredient of the SM not yet detected experimentally is the
Higgs particle, a scalar boson that triggers spontaneous gauge
symmetry breaking at the electro-weak scale by a vacuum
expectation value and gives masses to the gauge bosons of the weak
interactions as well as to all the matter fields.

Given the fact that the SM is very powerful in explaining a
surprisingly large number of independent experimental data, one
may still feel not quite satisfied with a purely phenomenological
approach. From a more conceptual point of view we do not know the
principles which fix the  numerical parameters
that define the Lagrangian of the SM to the values they
have in our universe. In the SM Lagrangian they appear simply as
free parameters like coupling constants and mixing angles which we
fix a posteriori by observation. Is it possible to actually
calculate their values at some higher scale from a more
fundamental theory?\footnote{One would still need to evolve the
values to the electro-weak scale by renormalization group
running.} Moreover, the choice we make when we single out the
field theory to describe particle interactions involves an even
larger degree of arbitrariness. The $SU(3)\times SU(2)\times
U(1)_Y$ gauge invariant and renormalizable Lagrangian with the
given matter spectrum of quarks and leptons in three generations
is just one specific model out of the infinite class of possible
local quantum field theories. Beyond these issues of arbitrary
choices there is also the question of naturalness. On a technical
level, it refers to the necessity of fine-tuning tree-level
parameters to accommodate for experimentally acceptable values
given the size of the perturbative quantum corrections. This
reasoning has motivated most of the explicit models for extensions
of the SM.

Any truly fundamental theory should somehow be able to incorporate
quantum gravity at very high energies. But despite all the
successes of the SM it does not seem to be a good candidate for a
complete theory of elementary particle physics simply because
gravity does not appear to fit into the framework of perturbative
local quantum field theories in any obvious way. Trying to
quantize Einstein's theory of general relativity perturbatively
around a flat background one encounters infinities in the
resulting Feynman diagrams, that cannot be cured by the usual
renormalization procedure.

Adding up the evidence, the SM together with classical general
relativity appears to be an excellent field theory to describe our
universe up to the electro-weak energy scale of $100$GeV. In a
way, it works much better than we may have expected and until to
date has needed only very minor modifications to explain all the
low energy data\footnote{Such a statement depends on what may
still be counted as part of the SM and what is considered an
extension. An example would be the addition of right-handed
neutrinos and Majorana neutrino masses.}. On the other hand, the SM
is unsatisfactory from the perspective of searching for a
fundamental theory. So we expect new physics, i.e.\ new particles,
new interactions, or other new effects, at energies only little
beyond the $100$GeV threshold, at most two orders of magnitude
above it. The SM is thus expected to be only an effective theory.

The motivation for this expectation does not only follow from the
theoretical reasoning but also from recent cosmological
observations. From the analysis of supernovae at large red shifts
and the recent precision measurement of the temperature
fluctuations in the cosmic microwave background radiation, a
cosmological standard scenario has been derived. It implies that
roughly $70\%$ of the current energy density in the universe
consists in a form of dark energy that basically behaves like a
cosmological constant $\L$, and that $26\%$ come in the form of an
unknown kind of dark matter. The rest is accounted for by ordinary
particles, i.e.\ mostly baryons. The most appealing version of
this model is then known as $\L$CDM, the cosmological constant
$\L$ together with cold dark matter. There is no straightforward
realization of such a scenario within the SM, in particular due to
the lack of a candidate particle species to serve as cold dark
matter\footnote{Neutrinos lead to hot dark matter.}. Beyond that,
there are various other problems which do not find good answers
within the SM, such the issue of baryogenesis which requires a
strong first order phase transition not found in the SM.

Another widely accepted, but still more speculative ingredient of
the standard cosmology is the paradigm of inflation. It says that
in a rather early period of its evolution our universe has
undergone a phase of accelerated expansion. In the simplest
scenarios this could have been triggered by the vacuum energy of
an unidentified scalar field, the so-called inflaton. The only
scalar that exists in the SM is the Higgs scalar, and it does not
seem to be a reasonable inflaton candidate. All this shows that
there must be physics beyond the SM.

One candidate for the new physics at the TeV scale, mainly
motivated by the naturalness problem of the SM, is a
supersymmetric quantum field theory that includes the SM particles
as a subset. In its simplest form it assumes for each known
elementary particle the existence of a superpartner with the
opposite spin-statistics, i.e.\ the fermionic quarks and leptons
have bosonic scalar partners called squarks and sleptons (s for
scalar), and the gauge bosons have fermionic partners, the
gauginos. The Higgs scalar would come with a fermionic partner as
well, a higgsino\footnote{As is very well known, anomaly
considerations force the doubling of scalar degrees of freedom,
such that the minimal extension of the SM has two complex Higgs
scalars instead of only one, two charged real scalars and two
neutral real scalars.}. This model is called the minimal
supersymmetric extension of the SM, the MSSM. One of its solid
prediction is that the mass of the lightest neutral Higgs scalar
should be around the weak scale, at least not heavier than, say,
$150$GeV\footnote{Further extensions of the Higgs sector of the
MSSM allow one to weaken this bound.}. Supersymmetric field
theories provide a potential candidate particle species to serve
as dark matter, namely the lightest supersymmetric particle (LSP).
Assuming a plausible discrete symmetry called R-parity this
particle is absolutely stable. Within the MSSM a very attractive
concrete possibility for the LSP is the lightest neutral fermion
among the superpartners, the lightest neutralino, which could
produce just the right amount of cold dark matter, given favorable
choices of parameters.

Since we do not observe the superpartners directly, supersymmetry
has to be broken at low energies. In order not to spoil the
original motivation, the breaking has to be ``soft''. One option
to achieve this is to start from an extended model with more
fields in a so-called hidden sector which then undergoes
spontaneous supersymmetry breaking. Integrating out the extra
fields leads to soft breaking in the visible sector, ideally the
MSSM. An example is the minimal supergravity model
(mSUGRA) with spontaneous supersymmetry breaking at an
intermediate scale $10^{11}-10^{13}$GeV which is communicated to
the MSSM by gravitational interactions only, leading to an
effective breaking scale just around a TeV. To parameterize the
Lagrangian of the MSSM including these effects one has to add all
the soft breaking terms to the supersymmetric theory. This entire
procedure introduces many new parameters into the model, partly
due to the supersymmetrization, partly due to the breaking of
supersymmetry. Eventually, they would need to be determined by
experiments, a formidable task beyond any present plans for future
experiments. However, if supersymmetry in form of the MSSM is
realized at all will be tested at the large hadron collider (LHC).

Eventually, it might turn out that a supersymmetric generalization
of the SM, which is still a local quantum field theory, indeed
pushes our understanding of elementary particle physics many
orders of magnitude higher in energy, maybe even not far from to
the Planck scale of about $10^{19}$GeV where quantum effects of
gravity become important. But it does not solve any of the
fundamental problems related to the shortcomings of quantum field
theories, in particular supersymmetric extensions of gravity,
supergravity theories, do not appear to be perturbatively
renormalizable either. This may be related to the extremely small
value of the cosmological constant whose unnaturalness is not
cured by supersymmetry in the broken phase. In order to make
progress in the direction of reconciling quantum field theory with
quantum gravity one has to give up one of the implicit underlying
assumptions. To take into account that space-time itself is
expected to fluctuate and deviate from the classical picture of
smooth geometry at the Planck scale, one can contemplate various
approaches. For example, one may want to change the smoothness of
the space-time itself at small Planckian distance scales and
discretize in some manner. One might hope that for instance
concepts like  non-commutative geometry, which assumes  that the
space-time coordinates do not commute and instead obey an
uncertainty principle, lead to a description of quantum gravity.
Whether such a formulation exist is not clear at the moment,
though interesting first results have been obtained. Similarly,
the approach of loop quantum gravity leads to a quantization of
space-time at very small distances which may or may not lead to a
consistent theory of quantum gravity.

String theory starts from a rather different point of view. It
postulates that the fundamental objects in nature are not
point-particles but one-dimensional strings, at least this is the
perturbative definition of the theory. Space-time itself together
with the fields of general relativity and quantum field theories
are emergent phenomena that arise as effective descriptions of
string dynamics. Fundamental strings are of finite length $\ell_s$
and thus cannot resolve distances smaller than $\ell_s$, the
string scale. Below this scale, there is no meaning to the
geometry of space-time in perturbative string theory. String
perturbation theory in fact means a quantum theory of small
fluctuations of elementary strings around a given background. At
larger distances such a theory is again described by an effective
field theory, a quantum field theory plus general relativity and
potentially with supersymmetry built in. A priori, these are the
ingredients needed for a unified theory of all forces and
particles.

After its discovery thirty years ago, it became clear that this at
first sight quite ad hoc construction leads to many interesting
consequences. By quantizing the fundamental string moving in a
flat Minkowski space the first proponents of string theory
realized that strings live in more than four space-time dimensions
and that the space-time only becomes stable in the presence of
supersymmetry. More specifically, closed strings incorporate
gravity and open strings potentially non-abelian gauge
interactions. For supersymmetric strings the critical space-time
dimension turned out $D=10$. Moreover, the perturbative expansion
of superstring theory was argued to be finite, which was shown
explicitly up to two-loop order\footnote{This means that
individual loop diagrams are free of ultra-violet (UV)
divergences. It does not say anything about the convergence of the
perturbation series and does not claim that perturbative string
theory is complete by itself.}. Nowadays
superstring theory is considered to be a very promising, if not
the most successful, candidate for the fundamental unified theory
underlying particle physics and gravity, the theory of everything.
Indeed, it is very encouraging that such a simple idea in
principle automatically incorporates all the features we know of
that a fundamental theory must have, such as local
reparametrization invariance and non-abelian gauge symmetry, plus
some other ingredients which are maybe not equally essential, but
which we still like, for instance supersymmetry and extra
dimensions.

Around 1995 it was realized that string theory is more complex and
more general than anticipated before. It not only contains strings
as fundamental degrees of freedom, but also higher-dimensional
objects called $p$-branes \cite{JP95} (see \cite{pcj96,JP96} for reviews).
These are not present in the
perturbative spectrum, since their masses scale inversely with the
coupling constant. They are only relevant in the non-perturbative
regime. Moreover, supersymmetry was used to argue for certain
duality relations between different string theories and different
string backgrounds, essentially claiming that these are only
apparently different descriptions of identical physics. Everything
eventually pointed towards a yet unknown theory that unifies all
known string theories, called M-theory, with an eleven-dimensional
effective description motivated by the maximal dimension of
supergravity. The various string theories in ten dimensions are
considered to be only perturbative limits of this M-theory. After
all, this has also raised a number of new questions that need to
be answered in order to make a completely convincing case that
string theory is really the fundamental theory we are longing for.
We still have no conclusive idea about the mathematical framework
in which to formulate the quantum theory in eleven dimensions.

The problem of much more practical urgency on the contrary is to
relate the higher-dimensional theories
to our macroscopic four-dimensional universe, and to the experiments we
are planning to perform in the near future.
This review article is concerned with recent progress
in improving our understanding in this respect, we concentrate on what
is called string phenomenology.

To make contact with four-dimensional physics starting from ten
dimensions, we have to explain what happens to the other six
dimensions. Performing a dimensional reduction in the spirit of
Kaluza-Klein (KK) field theories, one studies string theories on a
compact six-dimensional internal space of very tiny dimensions.
Our visible world would effectively be four-dimensional. Among the
infinite towers of KK states that follow from the expansion of the
higher-dimensional fields one only keeps the states of lowest
mass. All excitations along the internal space have masses which
are parametrically given by the compactification scale $1/R$, $R$
being the average linear scale of the internal space, the radius.
If this is small enough only massless modes will be of direct
phenomenological relevance\footnote{Massive modes may contribute
to quantum corrections by ``running in the loop''. This could be
important for example for precise gauge coupling unification.}.
Such a description in principle allows to determine at least the
classical couplings in the effective four-dimensional theory from
a dimensional reduction. This provides a geometric interpretation
for some of the parameters and other characteristics of the SM.
For instance, the spectrum of massless chiral four-dimensional
fermion fields is determined by topological invariants of the
compact space. Then also the number of generations of matter
particles gets a geometric explanation.

The next question to address then is to find the allowed (and
interesting) compactifications. In the semi-classical regime one
has to solve the string equations of motion and then test whether
the solutions are able to describe our universe to the measured
accuracy.
The, sort of, conservative viewpoint regarding supersymmetry in
this process goes as follows: The fundamental string scale is
assumed to be rather close to the Planck scale. The background
that is used in the compactification is required to preserve
(minimal local) supersymmetry, such that the four-dimensional
theory is supersymmetric at the compactification scale.
Supersymmetry is eventually broken in one way or another such that
the visible sector with the MSSM or a moderate extension thereof
receives soft breaking corrections with an effective scale close
to the electro-weak scale. Spontaneous breaking in a hidden sector
like the moduli sector of string compactifications, followed by
gravitational mediation, would be an attractive possibility, but
not the only one, and not without drawbacks. In any case, we will
mostly stick to the paradigm that string theory has to be
compactified on a supersymmetric background to start
with\footnote{There are alternative proposals in the framework of
effective field theory models based on the assumption that the
string scale could be much smaller than the Planck scale or even
close to the TeV scale. In this  case one can contemplate to start
right away with a non-supersymmetric string background, i.e.\
break supersymmetry at the compactification scale. However, when
one tries to find explicit string theoretic realizations such
models usually have serious stability problems.}. A major challenge
always remains, namely to explain how the dynamics of models
relevant at the string scale looks at low energies. The hardest of
these riddles is probably to understand how supersymmetry can be
broken without generating an unacceptably large cosmological
constant. All this is part of what we call the string vacuum
problem. If all the constraints imposed by low energy physics
could be solved for one concrete string compactification, this
would be a great advance towards the understanding of our
universe, it would among others involve solutions to the
cosmological constant problem and the hierarchy problem.

There now exist two main classes of string compactifications with
serious hope to find realistic four-dimensional physics, which
have so far received the largest amount of attention. The first
one has been pursued since the mid of the eighties already. Its
starting point is the discovery of the  cancellation of
gravitational and gauge anomalies
in ten-dimensional ${\cal N}=2$ type II supergravity theories as well as
in ten-dimensional ${\cal N}=1$ supersymmetric Yang-Mills gauge theories
with gauge groups $SO(32)$ and
$E_8\times E_8$ \cite{Alvarez-Gaume:1983ig,Green:1984sg}.
This was followed by
the subsequent construction of
the heterotic string in ten dimensions with
gauge group $E_8\times E_8$ \cite{GrossDD}.
It is compactified on a so-called
Calabi-Yau manifold, which is the unique supersymmetric background
where only the internal metric is non-trivial, all other fields
vanish \cite{Candelas:1985en}.
Of course, there are many six-dimensional Calabi-Yau
manifolds. In addition, they have to get endowed with a
non-trivial profile for the gauge fields of $E_8\ti E_8$, a vector
bundle, that breaks the gauge symmetry down to subgroups. A
central piece of motivation for this model comes from the $E_6$
grand unification scenario which can be embedded here.
Subsequently, exact heterotic string solutions on six-dimensional
orbifold spaces were constructed
\cite{Dixon:1985jw,Dixon:1986jc,Ibanez:1986tp}, which can be regarded  as being singular
limits of specific smooth Calabi-Yau spaces.
Finally, it was shown that one can construct a very large number
of four-dimensional heterotic string vacua directly in four dimensions
by using for the internal degrees of freedom either free fermions
\cite{Kawai:1986va,Antoniadis:1986rn}
or bosons on a covariant lattice
\cite{Lerche:1986cx} \footnote{This reference estimates
the number of four-dimensional heterotic strings to be of order
$10^{1500}$.}.

The other
possibility is that of orientifold compactifications respectively
open string descendants (see e.g. 
\cite{Sagnotti:1987tw,Polchinski:1987tu,Pradisi:1988xd,Horava:1989vt,Bianchi:1990yu,Bianchi:1991eu,as02}), as they were called before
the advent of D-branes in the mid of the
nineties (see e.g. \cite{Gimon:1996rq,Gimon:1996ay,Berkooz:1996dw,Angelantonj:1996uy,Kakushadze:1997wx,Aldazabal:1998mr} and the review \cite{as02} for
a detailed discussion of the history of this field). It can be interpreted as a generalization of
compactifications of the type I string with gauge group $SO(32)$
in ten dimensions.
The type I string  is the unique string theory in ten
dimensions which contains unoriented closed string and open
strings. Open strings in general have their ends on certain kinds
of $p$-branes, called D$p$-branes, and their lowest excitation
modes gives rise to massless gauge fields (and their fermionic
superpartners).
This makes them promising candidates for realistic string compactifications.
Indeed it was realized  that on the intersection of two such D-branes,
chiral fermions appear \cite{bdl96} which are another main feature of the SM.
All these aspects have been applied to concrete globally consistent string
compactifications in many examples starting with the early work
of \cite{CB95,bgk99,GP99,bgk99a,bgkl00a,aads00,fhs00,bkl00,afiru00,afiru00a,imr01,bklo01,csu01,csu01a}.

In general, any string model contains more than just the SM
sector, and other sectors usually have dramatic physical effects.
In particular, there is a rather large number of unobserved light
neutral scalar particles, called moduli fields. Geometrically,
their expectation values  parameterize the size and shape of the
compactification manifold or positions of D-branes, and similar
data. These fields would mediate long range forces and, due to
their very weak couplings, would be dominating the energy density
of the universe to an unacceptable degree (``They would overclose
the universe.''). Not only are the moduli phenomenologically
unacceptable, but their expectation values also determine
parameters like gauge couplings and masses of the effective
four-dimensional theory. Without uniquely determining these
expectation values by means of minimising an effective potential,
which would then also induce mass terms for the moduli, string
models are not predictive. This is the moduli problem of
Calabi-Yau compactifications.

Another big advance during the last five years has been the
discovery  of a controllable mechanism which generates a potential
for moduli fields. It requires to go beyond the purely geometric
Calabi-Yau compactifications where only the metric (and possibly
the gauge connection in the gauge bundle) are non-trivial. The
spectrum in ten dimensions also contains the various
anti-symmetric tensor fields, so called $p$-form fields $C_p$. One
has to allow the corresponding field strengths, schematically
$F_{p+1}=dC_p$, to take non-trivial expectation values along the
internal space, avoiding the breaking of four-dimensional Lorentz
invariance (heterotic flux compactifications were already
discussed in the mid eighties by A. Strominger \cite{Strominger:1986uh}).
These fluxes induce the interesting effects and modify
the metric via the Einstein equations in a way that can be
interpreted as a scalar potential for the moduli fields in the
effective four-dimensional theory \cite{drs99,gvw99,tv99,Peter,gss00,cklt00,gkp01,kst02}. The most general potential
known for such a flux compactification can have both stable
supersymmetric as well as meta-stable non-supersymmetric minima.
This is an important step towards realistic string
compactifications with fixed moduli, such that at least in
principle predictions about the low energy Lagrangian in a given
flux compactification can be made. However, flux compactifications
are in various aspects not understood as good as Calabi-Yau
compactifications, e.g.\ quantum corrections to the background and
back-reaction among fluxes and D-branes could be severe and are
much harder to obtain. So a lot of work remains to be done on the
subject of flux vacua.

When flux vacua were taken more and more seriously a number of
string theorists changed  their view towards the string vacuum
problem. Most attempts before had concentrated on the study of
classes of Calabi-Yau string vacua. Simply assuming some
unspecified low energy effect to take care of the moduli
stabilization, one can estimate the total number of vacua in that
case. A reasonable approximation for the degeneracy of Calabi-Yau
vacua seems to be of the order of around $10^{10}$. On the
contrary, the scalar potentials generated by fluxes have of the
order of $10^{500}$ different supersymmetric minima\footnote{This
is based on classical supergravity formulas for the potential.
Assuming randomly distributed quantum corrections it appears very
unlikely that they can reduce the number of minima substantially.}.
The search for realistic flux vacua has thus led string theory to
face an enormous vacuum degeneracy. In a sense, this is the other
face of the vacuum problem. Various still heavily debated
proposals to address this situation were made, ranging from a
statistical treatment of the properties of these vacua to the use
of the anthropic principle to eliminate undesirable solutions. It
is not our aim with this article to enter into this sometimes
rather philosophical debate. Instead, we wish to review the string
theoretic foundations of the recent developments in model building
with D-branes and fluxes, in order to provide the reader with a
comprehensive compendium of techniques, methods, some examples,
and an overview of achievements and shortcomings of various
approaches.

In this review we will to large extent focus on string
compactifications based on orientifolds with D-branes. Since there
have been interesting parallel developments recently, we have also
included a short section on heterotic string model building.
Our intention is that this article should provide a broad overview
and a deep introduction into the subject starting from elementary
concepts. It should be equally suitable for students and advanced
researches. We hope the reader may be able to use it to either
enter this active field of research or only get an idea about what
theoretical notions the debate about the string vacuum is based
on, according to his taste.

Of course, it is impossible to cover this extremely vast topic
from first principles in all its variety. It was mandatory to
leave out various aspects, and the selection of topics clearly
reflects our personal approach to the field. There are various
other aspects of string model building, most notably the study of
local models with D-branes at singularities, compactifications of
M-theory in the framework of heterotic M-theory or on manifolds of
$G_2$ holonomy, local non-compact as well as compact models of the
heterotic string, or even F-theory compactifications. Some of
these can be related via dualities to models we discuss, but we
will not try to unravel this structure in any detail. Other
interesting developments in string phenomenology are not covered
at all.
We neither deal with most of the more phenomenologically motivated
work on D-brane models (which was partially reviewed in \cite{bcls05}),
but stick to the conceptual issues.
Instead of going into detail about the physical interpretation of
the low energy Lagrangian, we provide techniques for deriving it.
Most of the physics in the end depends crucially on the concrete
model and the way supersymmetry is broken eventually, a question
no-one can answer conclusively to date. It is very interesting
though to contemplate string theoretic features that are common in
all string models, or at least common in an entire class of
compactification. These could lead to a ``smoking gun'' of string
theory. Recent attempts to find such generic signatures involve a
possible stringy correction to the proton decay rate
\cite{kw03,Cvetic:2006iz} or the presence of many scalars with
behavior similar to the standard axion. They could possibly serve
as dark matter or quintessence candidates
\cite{Conlon:2006tq,Svrcek:2006yi,Svrcek:2006hf,Conlon:2006ur}.

Let us provide a brief guide through  this long article. To begin
with, we are assuming that the reader is familiar with the basic
notions of string theory as for instance provided by the textbooks
\cite{gsw87,gsw87a,Lust:1989tj,JP98,JP98a,CJ03,BZ04}. Most of our conventions
are taken from the books by Polchinski \cite{JP98,JP98a}. We are
trying to be comprehensive and pedagogic in the exposition of the
topic, which makes some overlap with existing reviews unavoidable,
notably with review articles on D-branes \cite{pcj96,JP96}, on
orientifolds \cite{AD98,as02}, on D-brane model building
\cite{Quevedo,AU03,EK03,DL04,RB04,bcls05} (see also the PhD theses \cite{FM03,OTT03,LG04}) , or on fluxes \cite{grana,Douglas:2006es} (see also the PhD theses \cite{Frey:2003tf,Susanne}). On the other
hand, it is impossible to cover all the topics we are dealing with in
an exhaustive way, so we have to refer to other reviews like the
above, or the original literature in a number of places, where we
would rather like to go into more detail ourselves.
In such a long article, sometimes some repetitions are not only inevitable
but are intended to keep the average readers on track.

In section 2 we introduce the basic concepts relevant to the class of
models we are dealing with in the later sections. We start off with
D-branes from first principles, their description via
boundary states, as well as the way they appear in effective field
theory. Next we discuss the general concept of orientifolds of type II
string theories, using either simple examples from conformal
field theory or the effective approximation within
supergravity. Finally, we generalize the previous two subsections into
the subject of intersecting and magnetized D-branes that can exist in
orientifolds. Essential pieces needed later for the construction of
models include the conditions for supersymmetry in these models, and
the basic field theoretic formulation of the four-dimensional \gs mechanism.

Section 3 treats a number of approaches to string model building
with type II orientifolds and D-branes. The first discussed class
of models are supersymmetric intersecting D6-brane models in
type IIA orientifolds
on general Calabi-Yau manifolds (in the suergravity regime).
In the first part
we are describing such models in  most general terms without
specifying any concrete background.
We derive the tadpole cancellation conditions, give the general
rules for the determination of the chiral massless spectrum,
work out in detail the Green-Schwarz mechanism for the cancellation
of field theory abelian anomalies. We discuss
the supersymmetry conditions and employing supersymmetry
derive the tree level gauge couplings.
Eventually, we discuss D-term potentials resulting from the
anomalous $U(1)$s and in addition provide an outlook on F-term
potentials generated by world-sheet and space-time instantons.

As concrete examples we briefly present some aspects of
intersecting D6-brane on toroidal orbifold  backgrounds, which is
the class mostly studied so far in the literature, but clearly
constitutes only a very tiny fraction of all imaginable
intersecting D-brane models on generic Calabi-Yau spaces. As a
prototype model serves the $\mbb T^6/\mbb Z_2\times \mbb Z_2$
orientifold, which we discuss for the two possible choices of
discrete torsion giving rise to different kinds of D-branes.

Next we  describe in general terms the mirror symmetric
compactifications, which are given by type IIB orientifolds with
either O9/O5 or O7/O3-planes. The new issue is that the D-branes
are now wrapping even dimensional cycles of the Calabi-Yau and
also carry non-trivial vector bundles on their world-volume. We
give the general description for  the case with O9/O5 planes, as
here the D-branes can easily be described in terms of vector
bundles on Calabi-Yau spaces \footnote{To our knowledge, the case
with O7/O3-planes has not been worked out
  in full generality yet and we will only cover certain aspects of this
   type of orientifolds later in the article.}.
We systematically provide the same information as for the type IIA
case and point out the appearing differences and analogues.

So far the discussion was based on supergravity and therefore
is valid and trustable in the large radius regime.
For certain Calabi-Yau space, which are not toroidal orbifolds,
the exact conformal field theory is known at special points
in the moduli space. These are the so-called Gepner models.
We provide some of the technical details of the construction
of  orientifolds of these Gepner models (in the formal approach which
is closest to our expositions for orientifold constructions described
in the second section.)

At the very end of the string model building section we also
briefly mention recent advances in geometric heterotic
string constructions. First, using sophisticated vector bundle constructions
a number of concrete models have been found
which are quite close to the MSSM.  Second,
 by extending the set of vector bundles to those with $U(n)$ structure
groups, these heterotic models by S-duality were argued  to
 have very  similar  features than the type IIB orientifolds.

In section 4 we elaborate on the technical methods
to extract more information about the low energy effective action which
cannot be seen by dimensional reduction of the tree level
$D=10$ supergravity action and the one-loop Green-Schwarz counter terms.
While it is rather straightforward to construct the $D=4$ low--energy effective action
by a dimensional reduction of the supergravity action of the underlying string theory in
$D=10$, some truly stingy effects cannot be captured by this method.
A dimensional reduction  is always limited by
the fact, that already the effective action in $D=10$ is only known up to a certain order
in $\ap$. Moreover, this procedure does not take into account
in an appropriate way truly stringy effects such as string--loops or effects from
the string world--sheet.
In Section 4 we shall especially be interested in such effects and obtain non--trivial
coupling functions capturing stringy effects for the matter field metrics, Yukawa couplings
and one--loop gauge threshold corrections.

In section 5 we provide  a rough introduction into
flux compactifications. Since there exists a very good review
article on general flux compactifications \cite{grana}, here we mostly
stick to the best understood case of three-form fluxes in type IIB
orientifolds with O7/O3-planes. Only at the very end
we briefly summarize advances towards the understanding of
type IIA and heterotic flux vacua.
We discuss how the presence of fluxes modifies  the model building
rules outlined  in section 3. This includes new contributions
to the tadpole cancellation conditions and additional supersymmetry
constraints on the fluxes. In addition one encounters the generation
of a tree level flux induced superpotential giving rise to a moduli dependent
scalar potential and new consistency conditions for the
presence of both fluxes and D-branes. We also review 
moduli stabilization in Type IIB orientifolds, and how 
supersymmetry breaking fluxes can induce soft supersymmetry
breaking terms on the world-volume of D3 and D7-branes.

Finally, in section 6 we summarize the main technical arguments underlying
 one of the most controversial conclusions drawn
from the immense proliferation of the number of flux vacua, namely
that it is very unlikely that we will ever find {\it the} realistic
string model, but instead can only try to find  statistical
arguments for their existence.  According to
the topic of this review article, we put less emphasis on the
closed string sector (this has been reviewed in \cite{Kumar:2006tn,Douglas:2006es}), but
briefly discuss statistical methods developed to estimate
the distributions of physical parameters for the open
string sector in intersecting D6-brane models.

Section 7 contains our conclusions, which, of course, can only
reflect the contemporary state of the art in our approach
to the string vacuum problem.

\clearpage

\setcounter{equation}{0}


\section{BASIC CONCEPTS}
\lab{secbasic}

We start by introducing D-branes, orientifolds, and a number of
other concepts and notions of string theory. Clearly, the
presentation cannot be exhaustive and cover all aspects of these
topics. Nevertheless we try to be self-contained in what is really
essential to the specific models discussed later. Necessarily, we
have to leave much interesting extra material on the various
subjects these models are based on to the more specialized
literature. Some background material is collected in appendices.

The crucial property of D-branes for string phenomenology is the
fact that their world volume zero-modes form a potentially
supersymmetric and non-abelian gauge theory. How this arises and
how various aspects like the gauge symmetry, the matter spectrum,
conditions for supersymmetry and anomaly freedom are determined,
is the subject of this section.


\subsection{D-branes}

There are various different aspects to the nature of D-branes in
string theory. They can be interpreted in a microscopic way as
boundaries of the world sheet of fundamental strings. This
provides a definition in terms of the conformal field theory (CFT)
on the respective world sheet, which is perturbative in nature.
D-branes are also solitonic solutions to macroscopic equations of
motion for the supergravity theory defined on the target space.
This ``geometric'' description is effective and only involves the
degrees of freedom visible at low energies. In this domain
D-branes are related to objects like black holes, cosmic strings,
monopoles, instantons or domain walls.

For our purposes the definition of D-branes in a CFT and in the
effective geometric language are equally useful. We start with the
former point of view and introduce the notion of boundaries in the
world sheet of closed strings, and deduce from there the other
relevant properties of D-branes.


\subsubsection{Closed and open string world-sheets}

In string theories with open and closed strings Riemann surfaces
with and without boundaries are both included in the perturbative
definition of the string theoretic path integral. In generality,
Riemann surfaces are topologically classified by the number of
handles $g$, boundaries $b$ and cross caps $c$. The presence of
cross-caps makes a surface non-orientable. The order at which a
given topology appears in string perturbation theory is given by
the Euler characteristic
\beqn\lab{euler}
\chi(\Sigma) = 2-2g-b-c\ .
\eeqn
Any string diagram is weighted by a factor $g_s^{-\chi}$. The string
coupling $g_s$ is related to the vacuum expectation value of the
dilaton scalar field $\Phi$ by $g_s=\< e^\Phi\>$.

The leading terms of the non-linear world sheet sigma-model action of the
bosonic string on a general background and with a
potential boundary is given by
\cite{Fradkin:1985qd,Callan:1985ia,Abouelsaood:1986gd,Dorn:1986jf,Leigh:1989jq}\footnote{We
are working in conventions of \cite{JP98,JP98a} in most respects.}
\beqn \label{sigmaaction}
&& \hspace{-.6cm}
\cs = \frac{1}{4\pi\alpha'} \int_\Sigma d\sigma_1 d\sigma_2\, \sqrt{h}
 \Big[ \Big( h^{\a\beta} g_{MN}(X) + \epsilon^{\a\beta} B_{MN}(X) \Big)
         \partial_\a X^M \partial_\beta X^M  +
  \alpha' R(h) \Phi(X) \Big] \non
&&\hspace{1cm}
 + \int_{\partial \Sigma} d\sigma\, A_M(X)\partial_\sigma X^M \ .
\eeqn
The parameter $\a'$ is related to the string length scale $\ell_s$ via
\beqn \lab{strscale}
\ell_s = 2\pi \sqrt{\a'} \ .
\eeqn
The metric on the world sheet is denoted
$h_{\a\beta}(\sigma_1,\sigma_2)$. The target space background fields
are the closed string metric $g_{MN}$ and antisymmetric two-form
tensor $B_{MN}$, and the open string (abelian) vector field $A_M$
on the boundary. It has a field strength $F_{MN} = 2
\partial_{[M} A_{N]} = \partial_{M} A_{N} - \partial_{N}
A_{M}$. We use conventions where the fields $g_{MN}$, $B_{MN}$ and
$2\pi\a' F_{MN}$ are dimensionless. The equations of motion for
these fields are derived from the conditions that the
beta-functions of the sigma-model vanish.

A major part of this chapter will only deal with non-compact empty
Minkowski space-time, with $g_{MN} = \eta_{MN}$, $\eta_{MN}$
denoting the flat Minkowski metric (in ``mostly plus''
conventions), $B_{MN}=0$ and constant $\Phi$. As compact spaces we
will consider tori and toroidal orbifold backgrounds with constant
metric and $B$-field, or otherwise the effective low energy
supergravity limit where all fields only vary very slowly over the
internal space. Later, also backgrounds defined by Gepner models
will be discussed.

A closed superstring propagating in ten-dimensional Minkowski
space or a torus is described by the free CFT of the ten world
sheet coordinates $X^M(\s_1,\s_2)$ plus their fermionic counter
parts under world sheet supersymmetry $\p^M(\s_1,\s_2)$ and
$\tilde\p^M(\s_1,\s_2)$, plus the reparametrization ghost. Often
the world sheet coordinates $\s_1$ and $\s_2$ are complexified
into $z,\, \bar z$ as given in \reef{cplws}. By use of the
equations of motion the world sheet fields can be split into
chiral and anti-chiral fields $X^M = X^M_L
(z) + X^M_R(\bar z)$ and $\psi^M = \psi^M(z)$,
$\tilde\psi^M = \tilde\psi^M(\bar z)$. We have included some basic
definitions in appendix \ref{cftbasics} to settle our conventions,
following
\cite{JP98,JP98a}.

The closed string world sheet $\Sigma$, a Riemann surface, is
(locally) parameterized by coordinates $(\sigma_1,\sigma_2)$ with
\beqn
0 \leq \sigma_1 \leq 2\pi \ , \quad
\s_1 \equiv \s_1 + 2\pi \ , \quad
-\infty < \s_2 < \infty \ .
\eeqn
Such a patch forms an infinite cylinder. For a surface with a
boundary $\pa\S$ we pick (local) coordinates $(\s,\t)$ such that
$\partial\Sigma$ is at $\tau=0$, with
\beqn
0 \leq \sigma \leq 2\pi \ , \quad
\s \equiv \s + 2\pi \ , \quad
0 \leq \tau < \infty \ .
\eeqn
Now $\tau$ is a time variable for the evolution of the string,
describing a closed string emitted from the boundary, as depicted
in figure \ref{figclstr}.

\begin{figure}[h]
\vspace{.5cm}
\begin{picture}(180,180)(0,0)

\Oval(100,100)(50,20)(1)
\Oval(350,100)(50,20)(1)
\Oval(348,100)(50,20)(1)
\Line(100,50)(350,50)
\Line(100,150)(350,150)

\Line(60,30)(60,180)
\Line(132,45)(132,155)
\Line(60,30)(132,45)
\Line(60,180)(132,155)

\Vertex(180,80)2
\LongArrow(180,80)(210,80)
\LongArrow(180,80)(180,110)

\Text(182,115)[1]{$\s$}
\Text(184,75)[1]{$\t$}
\Text(90,170)[1]{$\pa\S = \{ \t=0\}$}

\end{picture}
\vspace{-1cm}
\caption{Closed string emitted from a D-brane: semi-infinite cylinder
\lab{figclstr}}
\end{figure}
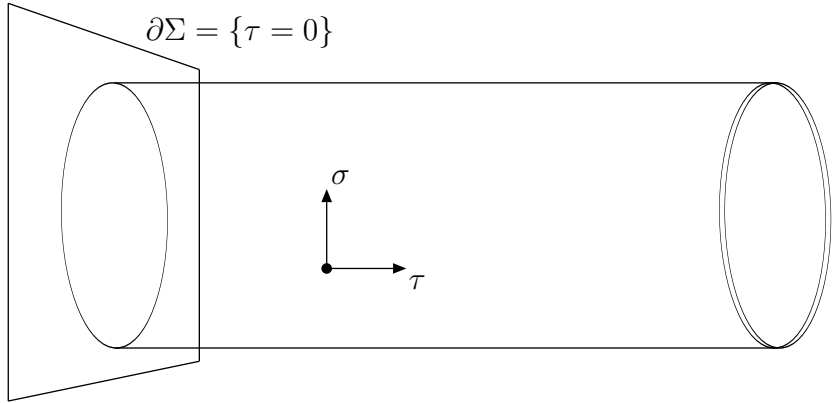

Each component of the boundary can couple to a different gauge
field $A_M$ but we refrain from introducing an extra label to
distinguish the various boundary components at this point. The action has the
abelian gauge invariance of the vector potential at the boundary
(independently at each component of the boundary)
\beqn
\delta A_M = \partial_M \lambda\ ,
\eeqn
and the combined two-form gauge invariance of the antisymmetric
tensor $B_{MN}$, which also involves a boundary term,
\beqn
\delta B_{MN} = \partial_M \zeta_N - \partial_N \zeta_M \
, \quad
\delta A_M = -\frac{1}{2\pi\alpha'} \zeta_M \ .
\eeqn
Therefore, the gauge invariant field strength is \cite{Witten:1995im}
\beqn \label{calF}
2\pi\alpha' \cf_{MN} = B_{MN} + 2\pi\alpha' F_{MN}\ .
\eeqn
The boundary condition that follow from the variations of
the world sheet action are \cite{Abouelsaood:1986gd}
\beqn \label{mixedbc}
g_{MN} \partial_\tau X^M + 2\pi\a' \cf_{MN} \partial_\sigma X^M =
0
\bigg|_{\partial\Sigma} \ .
\eeqn
They interpolate between Neumann (N) and Dirichlet (D) boundary
conditions, namely
\beqn \label{NDbc}
{\rm N:} \quad \partial_\tau X^M  = 0 \Big|_{\partial\Sigma}
\ , \quad
{\rm D:} \quad \partial_\sigma X^M = 0 \Big|_{\partial\Sigma}
\ .
\eeqn
To evaluate (\ref{NDbc}) for the momenta $p^M$ and winding modes
$w^M$ of the closed string one may use the mode expansion
\reef{modes} to get
\beqn\label{bcpw}
{\rm N:} ~~ \quad p^M=\frac12 \Big( p^M_L+p^M_R  \Big) =0 \ , \quad
{\rm D:} ~~ \quad w^M=\frac12 \Big( p^M_L-p^M_R \Big) =0 \ .
\eeqn
Neumann conditions allow no energy-momentum
transfer at the boundary, while the closed string can move freely
in Dirichlet directions.

One can now construct states made out of closed string oscillator
and zero-mode excitations that automatically satisfy the above
boundary conditions
\cite{Polchinski:1987tu,Callan:1988wz,Cardy:1989ir,Li:1995pq,Green:1996um}.
They take the form of products of coherent states. We define a
state $|B p\>$ that satisfies
\beqn
\partial_\tau X^M |B p\> = 0 \quad {\rm for}\ M=0,\, ...\, , p\ ,
\quad
\partial_\sigma X^M |Bp\> = 0 \quad {\rm for}\ M=p+1, \, ...\, ,9\ ,
\eeqn
to describe a D$p$-brane. The piece $|B p \>_{\rm osc}$ of $|B
p\>$ involving the bosonic string oscillators modes can be written
\beqn \lab{bosbst}
|Bp\>_{\rm osc} = \exp\Big( S_{MN} \sum_{n>0} \frac1n \a^M_{-n} \tilde\a^N_{-n}
\Big) |0\> \ ,
\eeqn
using \reef{modes} and \reef{wsalgebra}. The matrix $S_{MN}$
encodes the boundary conditions and takes the simple form
\beqn\lab{Smn}
S = {\rm diag}( -1,...,-1,1,...,1)\ ,
\eeqn
with eigenvalue $-1$ for Neumann and $+1$ for Dirichlet directions.
Furthermore, the state \reef{bosbst} needs to be multiplied by
delta-functions in momentum or coordinate space to impose
\reef{bcpw}. The proper normalization of the state is fixed by
comparing the tree-channel transition function that defines a
cylinder diagram to a loop calculation, see e.g.\
\cite{Gaberdiel:2000jr} for more details on these issues.


\subsubsection{Open bosonic strings ending on D-branes}

We now reconsider the description of boundaries in the string
world sheet in a dual language by switching to the open string
picture. The open string world sheet on an infinite strip we
parameterize by coordinates $(\t,\s)$ with
\beqn
-\infty < \t < \infty\ ,\quad
0 \leq \s \leq \pi \ .
\eeqn
The boundary has two components $\pa\S_1$ and $\pa\S_2$ at
$\s=0,\pi$.

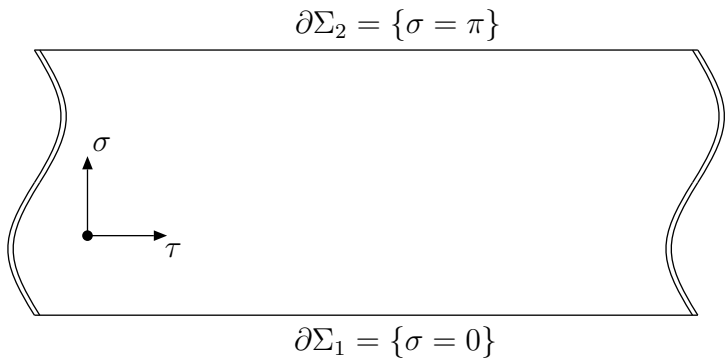
\begin{figure}[ht]
\begin{picture}(180,180)(0,0)

\Line(100,50)(350,50)
\Line(100,150)(350,150)
\Photon(100,50)(100,150){10}{1}
\Photon(102,50)(102,150){10}{1}
\Photon(350,50)(350,150){10}{1}
\Photon(348,50)(348,150){10}{1}

\Vertex(120,80)2
\LongArrow(120,80)(150,80)
\LongArrow(120,80)(120,110)

\Text(122,115)[1]{$\s$}
\Text(122,75)[1]{$\t$}
\Text(180,160)[1]{$\pa\S_2 = \{ \s=\pi\}$}
\Text(152,40)[1]{$\pa\S_1 = \{ \s=0\}$}

\end{picture}
\vspace{-1cm}
\caption{Open string: infinite strip with two boundaries}
\end{figure}

The world sheet sigma-model contains two boundary terms with
potentially different gauge potentials,
\beqn\lab{openbt}
- \int_{\pa\S_1} d\t\ A^a_M(X) \pa_\t X^M + \int_{\pa\S_2} d\t\
A_M^b(X) \pa_\t X^M \ .
\eeqn
The relative sign that appears in \reef{openbt} reflects the
orientation of the open string or the relative charge of the two end
points.

The boundary conditions
(\ref{mixedbc}) can now be applied independently at both ends of
the open string,
\beqn
\s=0 \ : && \quad
g_{MN} \partial_\s X^M + 2\pi\a' \cf^a_{MN} \partial_\t
X^M = 0 \bigg|_{\partial\Sigma_1} \ , \non
\s=\pi \ : && \quad
g_{MN} \partial_\s X^M + 2\pi\a' \cf^b_{MN} \partial_\t
X^M = 0 \bigg|_{\partial\Sigma_2} \ .
\eeqn
Since they involve the field strength from (\ref{calF}) one can
distinguish open strings that stretch between one and the same
type of boundary when $a=b$, or two different types with different
gauge fields $\cf_{MN}^a\neq \cf^b_{MN}$.

For Neumann or Dirichlet conditions at both ends the open string
mode expansion
\reef{NDopenmodes} leads to opposite results regarding open string
momenta $p^M$ and winding $w^M$, as compared to
(\ref{bcpw}),
\beqn\label{bcNDopen}
{\rm N:} ~~ \quad w^M=0\ , \quad
{\rm D:} ~~ \quad p^M=0 \ .
\eeqn
Thus, open strings have non-vanishing momenta and can move along
the Neumann directions, whereas Dirichlet condition freeze the
motion of the open string and fix the coordinate to a constant
value, only allowing winding.

This implies that the ends of open strings are not always free to
move throughout the full space-time, but may be confined to
certain regions if Dirichlet boundary conditions apply. These
regions are the D-branes, more specifically D$p$-branes
\cite{Dai:1989ua,JP95,JP96}. At the classical level, they are
geometric submanifolds of the total space-time, and their
dimensionality is given by the number $p+1$ of directions with
Neumann boundary conditions at a given position in space-time.
Even if there are no directions with Dirichlet conditions one uses
the term D-brane, in that case D9-brane (for a total space-time
dimension 10). There are extra names for some special cases: $p=2$
is a membrane, $p=1$ is a D-string, $p=0$ is a D-particle, and
$p=-1$ is a D-instanton. For the purpose of models with
intersecting D-branes in type IIA string theory the D6-branes will
be most important, in type IIB D-branes of dimensions $p=3,5,7,9$
will be considered in the models we discuss.

The Dirichlet boundary conditions are an unavoidable consequence
if one wants to make open string theory compactified on a circle
(or more generally a torus) invariant under T-duality
\cite{Dai:1989ua}.
A T-duality on a circle is simply the inversion of its radius in string units.
On a circle momentum and winding states are labelled by integers
$(m,n)$. Left- and right-moving momenta are defined
\beqn\lab{KKcirc}
p_L = \frac mR + n \frac R{\a'}\ ,
\quad
p_R = \frac mR  - n \frac R{\a'} \ ,
\eeqn
such that $p = m/R$ and $w = n R/\a'$. T-duality is the flip of momentum and
winding states of the string mode expansion,
\beqn
R \mapsto \frac{\alpha'}{R}
\quad\quad\Longleftrightarrow\quad\quad (p_L,p_R) \mapsto
(p_L,-p_R)\ .
\eeqn
Regarding (\ref{bcpw}) this just exchanges Dirichlet and
Neumann boundary conditions on the momentum and winding modes,
i.e.\ it maps a D$p$-brane to a D$(p+1)$-brane or a
D$(p-1)$-brane.

In general one can define T-duality on a circle along $X^M$ as a
reflection of the right-moving chiral world sheet fields,
\beqn\lab{Tdual}
R \mapsto \frac{\alpha'}{R}
\quad\quad\Longleftrightarrow\quad\quad
\left\{
\begin{array}{c}
(X_L^M,X_R^M)\mapsto (X_L^M,-X_R^M) \\
(\psi^M,\tilde \psi^M) \mapsto (\psi^M ,-\tilde\psi^M)
\end{array}\right.\ .
\eeqn
The mode expansions of the fields on Minkowski space
are defined in (\ref{modes}). Since $\partial_\tau X_R^M =
- \partial_\sigma X_R^M$ and $\partial_\tau X_L^M =
\partial_\sigma X_L^M$ this flips the Neumann and Dirichlet boundary
conditions. This operation can also directly be applied to the
boundary state \reef{bosbst} where it flips the signs of $S_{MN}$
as required.


\subsubsection{Superstrings with boundaries}

We will be dealing with D-branes in supersymmetric type II string
theories later, so let us briefly discuss the extension of bosonic
strings to superstrings. The world sheet sigma-model for the
fermions can be obtained from the bosonic one in
(\ref{sigmaaction}) by supersymmetrization in the RNS (Ramond and
Neveu-Schwarz) formalism. Some relevant material for flat target
space backgrounds has been collected in appendix \ref{cftbasics}.
The chiral closed string world sheet coordinates in that case are
$X_L^M(z)$ and $X_R^M(\bar z)$, accompanied by fermionic partners
$\psi^M(z)$ and $\tilde\p^M(\bar z)$, see \reef{modes}.

The boundary conditions for the world sheet fermions are the
analogue of (\ref{mixedbc}), but there is a sign ambiguity
referring to the possibility to have periodic or anti-periodic
world sheet fermion modes at the boundary. It is reflected in
$\eta=\pm 1$ labelling the so-called spin structure
\cite{Seiberg:1986by}. The Neumann or Dirichlet boundary
conditions on flat space-time are
\beqn\label{mixedfermbc}
{\rm N}: ~~\quad \p^M + i\eta \tilde\p^M  =0\Big|_{\pa\S}  \ ,\quad
{\rm D}:~~\quad \p^M - i\eta \tilde\p^M =0\Big|_{\pa\S}\ .
\eeqn
The boundary state that satisfies the bosonic and fermionic boundary
conditions of \reef{NDbc} and \reef{mixedfermbc} can be written
\beqn \lab{oscbst}
|Bp,\eta\>_{\rm osc} = \exp\Big( S_{MN} \sum_{n>0} \frac1n \a^M_{-n} \tilde\a^N_{-n}
+i\eta S_{MN} \sum_{r>0} \p^M_{-r} \tilde\p^N_{-r}
\Big) |Bp,0,\eta\> \ ,
\eeqn
generalizing \reef{bosbst}. The ground state $|B,0,\eta\>$ is a
tensor product of the NSNS and RR ground states, as dictated by the
GSO projection. This is the place where type IIA and IIB differ by a
sign,
\beqn \label{gso}
\cp^{\rm cl}_{{\rm GSO}} = \frac12 {[1+ (-1)^F]}
\ti \frac12
\left\{
\begin{array}{l}
[1+ (-1)^{\tilde F}] \quad {\rm for\ NSNS} \\
\big[ 1\mp (-1)^{\tilde F}\big] \quad {\rm for\ RR}
\end{array}
\right.
 \ ,
\eeqn
with minus sign for IIA and plus for IIB in the right-moving R
sector. The world sheet fermion number operators $F$ and $\tilde
F$ are given in (\ref{numop}). They act on the NSNS and RR
components of the boundary states by (see e.g.\
\cite{Gaberdiel:2000jr})
\beqn \lab{GSOeta}
(-1)^F | Bp,0,\eta\>_{\rm NSNS} &=&
(-1)^{\tilde F} | Bp,0,\eta\>_{\rm NSNS} ~=~ - | Bp,0, -\eta\>_{\rm
NSNS} \ , \non
(-1)^F | Bp,0,\eta\>_{\rm RR} &=& | Bp,0, -\eta\>_{\rm RR} \ ,  \non
(-1)^{\tilde F} | Bp,0,\eta\>_{\rm RR} &=&
(-1)^{p+1} | Bp,0, -\eta\>_{\rm RR}\ .
\eeqn
One has to form linear combinations of states with $\eta=\pm 1$ to
get invariant states. It follows that in the RR component this can
only be achieved for even $p$ in IIA and odd $p$ in IIB,
restricting the dimensionalities of supersymmetric D$p$-branes to
the well known values.

The open superstrings are similarly obtained from open bosonic
strings via supersymmetrization with a single holomorphic world
sheet fermion. Some details on open strings in flat backgrounds
are collected in appendix \ref{openCFT}. The spectrum of physical
states is generated by the zero-modes and the mode operators
$\a^M_{-n}$ and $\p^M_{-r}$ acting on the groundstate, subject to
the open string GSO projection onto states of even world sheet
fermion number. The projector reads
\beqn
\cp_{\rm GSO}^{\rm op} = \frac12 [ 1+(-1)^F ] \ .
\eeqn
The simplest case is an open string with both ends on the same
D-brane with $\cf_{MN}=0$. The mode expansion has only integer
moded bosons and integer and half-integer moded fermions in the NS
and R sectors. The mass of an open string excitation is
\beqn \lab{massopen}
\a' M^2 = \sum_{n>0} \a^M_{-n} \a_{Mn} + \sum_{r>0}
\p^M_{-r} \p_{Mr} -
\left\{
\begin{array}{l}
\frac12 \quad {\rm for\ NS}\\
0 \quad {\rm for\ R}
\end{array}
\right.
\ .
\eeqn
The Lorentz index $M$ runs only over transverse excitations in
light-cone gauge.\footnote{One way to fix the local world sheet
reparametrization gauge invariance is to use light-cone gauge.
This means that we eliminate the light-cone direction $M=0,1$ from
the physical polarization of the world sheet fields $X^M$ and
$\psi^M,\, \tilde\p^M$ but do not consider any ghost fields
explicitly.} For most phenomenological considerations only the
massless fields are of any relevance.

All excitations with bosonic oscillators are massive. Only the
lowest excitations of the NS sector and the degenerate groundstate
of the R sector produce massless particles. The NS states can be
written
\beqn\label{nsgrst}
{\bf 8}_V : \quad \psi^M_{-1/2} | 0 \rangle_{\rm NS}\ , \quad
M=2,\, ...\, ,9\ .
\eeqn
It carries a ten-dimensional space-time vector index and is
eight-fold degenerate. The fermionic zero modes of the R sector of
an open superstring satisfy a Clifford algebra
\beqn
\{ \psi^M_0, \psi^N_0\} = \eta^{MN}\ .
\eeqn
One can define raising and lowering generators in a standard
fashion by $\psi^I_\pm = \psi_0^{2I} \pm i \psi_0^{2I-1},$ $I=1,\,
...\, ,4$, which form an oscillator algebra. Its vacuum, the R
groundstate, is annihilated by one half of the operators,
$\psi^I_{-} | 0\rangle_{\rm R} =0$. The massless states of the R
ground state are written
\beqn\lab{Rvac}
\prod_{I=1}^4 (\psi_+^I)^{s_I+1/2} |0\>_{\rm R} \ ,
\quad s_I = \pm\frac12\ ,
\eeqn
which produces $2^4=16$ states. These form a
spinor representation of the $SO(8)$ little group, and
thus behaves like a space-time fermion.

The GSO projection acts on the R vacuum as the chirality
projector, and restricts the number of raising operators
$\psi_+^I$ acting on the lowest weight state to be even. The R
vacuum of the open superstring in ten dimensions is thus $2^3=8$
fold degenerate, i.e.\ transforms as an irreducible
ten-dimensional Majorana-Weyl spinor {\bf 8}$_s$. We denote it by
$|\alpha \rangle_{\rm R}$, where $\alpha$ is the relevant spinor
index, or equivalently by the weights $s_I$ of the spinor
representation,
\beqn \lab{Rweights}
{\bf 8}_s: \quad  |\alpha \rangle_{\rm R} =
|s_1,s_2,s_3,s_4\rangle_{\rm R}\ .
\eeqn
The massless spectrum of an open string with identical boundary
conditions on both ends can finally  be written in light-cone
gauge
\beqn \label{openspec}
({\bf 8}_V \oplus {\bf 8}_s) : \quad
A_M \psi_{-1/2}^M | 0\rangle_{\rm NS}
~~\oplus~~
\chi_\a | \a\rangle_{\rm R} \ .
\eeqn
The fields $(A_M,\chi_\a)$ together form a vector or spin one
supermultiplet under ten-dimensional $\cn=1$ supersymmetry.

For a D$p$-brane $9-p$ of the polarizations of the vector field
are actually transverse components from the point of view of the
D-brane world volume, i.e.\ scalars under the $SO(p,1)$ Lorentz
group on the brane. Geometrically, they parameterize the location
of the branes in the transverse space. For non-abelian gauge
symmetries, these scalars can play the role of adjoint Higgs
fields. Their vacuum expectation values break the gauge symmetry
spontaneously.

In any string compactification, the target space is split into
$3+1$ large dimensions and 6 internal compact directions. This
gives a split $SO(9,1)$ into $SO(3,1)\times SO(6)$. In the
situations we consider, the $3+1$ macroscopic dimensions are part
of the $p+1$ world volume dimensions of any D$p$-branes present,
in particular $p\geq 3$. Together, $SO(9,1)$ reduces to
$SO(3,1)\times SO(p-3)\times SO(9-p)$. From the four-dimensional
point of view, the ten-dimensional vector $A_M$ transforms as a
four-dimensional vector plus six scalars, $A_\m$ and $A_i$. The
ten-dimensional spinor ${\bf 16}$ decomposes into ${\bf
(2,4)\oplus (\bar 2,\bar 4)}$. One is left with two
representations each of either four-dimensional spin, a total of
four four-dimensional Weyl fermions from a single ten-dimensional
Majorana-Weyl spinor. This is a non-chiral spectrum. Together with
the bosons they fill out an $\cn=4$ supersymmetric vector
multiplet, forming a $\cn=4$, $d=4$ supersymmetric gauge theory on
a D3-brane.

This reduction is part of the ``chirality problem'' of
Kaluza-Klein (KK) reduction \cite{Witten:1981me}. It means the
challenge to produce a fermion spectrum that involves states of
different representations under the gauge symmetry for the
particles of left- respectively right-handed four-dimensional
chirality, as in the Standard Model. The decomposition of the
ten-dimensional multiplets showed that a trivial dimensional
reduction of the ten-dimensional theory on a torus cannot provide
this, and one needs more complicated internal structures for the
six-dimensional compactification space.


\subsubsection{Chan-Paton labels: non-abelian gauge symmetries}

Whenever a number of identical D-branes is located in the same
position in space-time (and all have an identical world volume
field configuration) the individual branes are indistinguishable.
We say they form a stack. The abelian $U(1)$ gauge symmetry of a
single D-branes then gets promoted to a non-abelian gauge group
\cite{Witten:1995im}.

An open string can have either one of its ends on any individual
D-brane in a stack. To distinguish the open strings that connect
the various D-branes one introduces the concept of Chan-Paton (CP)
labels and assigns a formal label $\lambda^A$ to every open string
\cite{Paton:1969je}. These CP labels can be represented by
matrices that satisfy a Lie algebra as a symmetry group of open
string interactions, i.e.\ the $\l^A$ can be chosen as hermitian
matrix generators and $A$ is the adjoint index. The symmetries of
open string scattering amplitudes turn out to be compatible with
symmetry algebras $U(N)$, $SO(N)$ or $Sp(N)$
\cite{Schwarz:1982md,Marcus:1982fr}.
The symmetry is a global symmetry from the point of view of the
world sheet sigma-model, but local in the ten-dimensional target
space-time. Thus, the theory of open strings with ends on a stack
of D-branes has a non-abelian gauge symmetry by means of the CP
labels.

The degeneracy of open string states results from distinguishing
strings that run from a brane $i$ to another brane $j$ by their
label $|ij\>$. Using the representation matrices $\l_{ij}^A$ one
can choose a basis of states
\beqn
| A\> = \l_{ij}^A | ij \>\ ,
\eeqn
where all other quantum numbers have been suppressed. The massless
open string states from \reef{openspec} with this extra degeneracy
are
\beqn \label{CPspec}
A^A_M \psi_{-1/2}^M \lambda_{ij}^A | 0,ij\rangle_{\rm NS}
~~\oplus~~
\chi^A_\alpha \lambda_{ij}^A | \alpha,ij \rangle_{\rm R} \ ,
\eeqn
The CP label $\l^A$ determine the representation of the respective
multiplets $(A^A_M, \chi^A_\a)$ under the gauge symmetry.

The total number of oriented open strings in a background with $N$
D-branes is thus $N^2$. One can deduce the dimensions of the
representations carried by the various strings from counting
degeneracies.\footnote{For a more complete discussion see e.g.\
\cite{as02}.} For an open string with one end attached to a stack
of $N_a$ D-branes the degeneracy is $N_a$, the dimension of the
fundamental irreducible representation (denoted $\fund_a$) of the
gauge group. The orientations of the strings that go into or go
out off a given stack are related by a CPT transformation. This
implies that the representation of the opposite orientation is the
conjugate representation, the anti-fundamental $\bar{\fund}_a$.

The representations for general open string states are then the
tensor products of fundamental and anti-fundamental
representations from the two endpoints. If the open string has
both ends on the same set of branes and is $N_a^2$ degenerate it
carries a representation of $(\fund_a,\bar{\fund}_a)$. It has the
dimension of the adjoint for a $SU(N_a)$ gauge group plus a
singlet. In this case \reef{CPspec} provides the vector multiplet
of an $SU(N_a)\times U(1)_a=U(N_a)$ gauge group. The explicit
$U(1)_a$ factor is always chosen as the diagonal proportional to
the $N_a\ti N_a$ unit matrix. The charges are normalized such that
a representation $\fund_a$ has charge $+1$ and $\bar{\fund}_a$ has
$-1$. In the presence of charged matter fields the $U(1)_a$ are
very often anomalous, and the anomaly-free symmetry group can
reduce to the simple group $SU(N_a)$. This will be discussed in
section \ref{secgs}. For open strings with ends on two different
branes, the CP factor will carry the representation
$(\fund_a,\bar{\fund}_b)$ or $(\fund_a,{\fund}_b)$, a
bifundamental representation.\footnote{Note that the spectrum of
the Standard Model or the MSSM is assembled entirely out of
bifundamental representations. This is, however, not the case for
some of the most attractive grand unified models, such as the
$SO(10)$ which cannot be realized in D-brane models because it
involves a spinor representation ${\bf 16}$.}

In theories of unoriented strings the components $\l^A_{ij}$ and
$\l_{ji}^A$ can get identified up to a sign. The $N\times N$
matrix is then projected to either a symmetric or anti-symmetric
matrix. This breaks the adjoint representation of $U(N)$ down to
the adjoint of the $SO(N)$ or $Sp(N)$ (in conventions, where
$Sp(N)$ has rank $N/2$) gauge symmetry. In this way, the most
general gauge group that can appear in any open string theory is
of the form
\beqn \lab{gengauge}
\cg = \prod_a U(N_a) \ti \prod_b SO(N_b) \ti \prod_c Sp(N_c) \ .
\eeqn
The only representations that can appear are adjoint
representations, symmetric representations $\Ysymm_a$,
anti-symmetric representations $\Yasymm_a$, or finally
bifundamental representations $(\fund_a,\fund_b)$ or
$(\fund_a,\bar{\fund}_b)$. These are the ingredients to start any
model building with D-branes.

A rather efficient tool to summarize the data of the massless open
string sector including gauge group and spectrum are quiver
diagrams \cite{Douglas:1996sw}. In figure \ref{randomquiver} we
have depicted a random example for a piece of such a diagram.

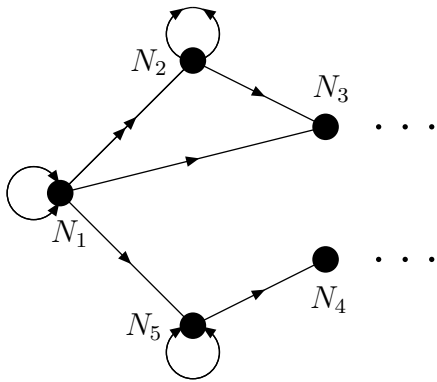
\begin{figure}[ht]
\vspace{0cm}
\hspace{2cm}
\begin{picture}(180,180)(0,0)

\Vertex(150,150)5
\Vertex(100,100)5
\Vertex(150,50)5
\Vertex(200,125)5
\Vertex(200,75)5

\Vertex(220,125)1
\Vertex(230,125)1
\Vertex(240,125)1
\Vertex(220,75)1
\Vertex(230,75)1
\Vertex(240,75)1

\ArrowLine(98,98)(148,148)
\ArrowLine(102,102)(152,152)
\ArrowLine(100,100)(150,50)
\ArrowLine(150,50)(200,75)
\ArrowLine(100,100)(200,125)
\ArrowLine(150,150)(200,125)

\LongArrowArc(90,100)(10,0,340)
\LongArrowArcn(90,100)(10,360,20)
\LongArrowArc(150,40)(10,90,70)
\LongArrowArcn(150,40)(10,90,110)
\LongArrowArc(150,160)(10,90,70)
\LongArrowArcn(150,160)(10,90,110)

\Text(100,85)[1]{$N_1$}
\Text(103,150)[1]{$N_2$}
\Text(145,140)[1]{$N_3$}
\Text(117,60)[1]{$N_4$}
\Text(20,50)[1]{$N_5$}

\end{picture}
\vspace{-1cm}
\caption{Open string spectrum represented by quiver diagram\lab{randomquiver}}
\end{figure}

The nodes stand for the factors in the gauge group $\cg$  with the
adjoint vector multiplets implicit. Each line represents a matter
field chiral multiplet connecting the two factors of $\cg$ it
transforms non-trivially under. Arrows indicate the orientation
and can distinguish $(\bar{\fund}_a,\fund_b)$ from
$(\fund_a,\bar{\fund}_b)$ and $\Ysymm_a$ from $\Yasymm_a$ in an
obvious manner. For the use of quiver diagrams in orientifold
models see e.g.\ the appendix of \cite{Klein:2000hf}.

A configuration of D-branes in this way has an alternative
interpretation apart from cutting holes in closed string world
sheets. At low energies and small string coupling, only the
massless gauge and matter fields on the world volume are visible.
In this regime, a D-brane can be characterized by its location in
space-time and the gauge field configuration on its world volume.
In other words, a stack of D-branes is given by specifying a
submanifold of the sigma-model target space together with a gauge
bundle with support on this submanifold, the so-called Chan-Paton
bundle. In more complicated configuration with several stacks this
type of data is needed for every stack. To be physically
acceptable the submanifold has to satisfy additional criteria such
as being a spin manifold, i.e.\ admitting spinors. An obstruction
to this is a non-vanishing second Stiefel-Whitney class. We will
later see how this condition emerges from anomaly cancellation
conditions in the effective theory on the D-brane world volume.
There are in fact more refined versions of such a definition of a
D-brane ``at large volume'' which involve more sophisticated
mathematics such as K-theory, sheaves or even derived categories.


\subsubsection{The effective DBI and CS action}

The dynamics of the massless open string modes, the
ten-dimensional gauge field multiplet or its lower-dimensional
descendants, is described by the Dirac-Born-Infeld (DBI)
\cite{Fradkin:1985qd,Leigh:1989jq} plus the Chern-Simons (CS) action
\cite{Douglas:1995bn,Green:1996dd,Cheung:1997az,Morales:1998ux,Stefanski:1998yx,
Scrucca:1999uz,Scrucca:1999jq}.
Together they form the relevant Lagrangian at leading classical
order in the string coupling (disk level) and at leading order in
derivatives. The known expressions do in fact contain terms with
more than two derivatives, but further corrections at the same
higher orders are expected to exist. The two pieces involve
different background fields which the open string modes couple to,
\beqn
\cs_{\rm eff} = \cs_{\rm DBI}[g,\Phi,B] + \cs_{\rm CS}[C_p]\ .
\eeqn
The DBI action contains the coupling of the open string degrees of
freedom to the bulk NSNS fields, the dilaton, metric and two-form,
while the CS action involves the RR $p$-forms $C_p$. In
particular,  the DBI action is only well understood for a single
brane with only abelian gauge symmetry.

Let us start with the DBI action, a generalization of Maxwell
theory with higher derivative couplings. The bosonic part in
string frame is given by
\beqn \label{dbi}
\cs_{\rm DBI} = - \mu_p \int_{\cw} d^{p+1}\x\,
e^{-\Phi(X)} \sqrt{-{\rm det}( g_{ab}(X) +
2\pi\a'\cf_{ab}(X))} \ ,
\eeqn
where we split the ten-dimensional indices $\{M,N,...\}$ for the
space-time directions into the brane world volume with labels
$\{a,b,...\}$, running from 0 to $p$ and the transverse space
$\{i,j,...\}$, running from $p+1$ to 9. The prefactor of the
dilaton identifies \reef{dbi} as the open string tree-level
effective action, i.e.\ resulting from disk diagrams only. In the
non-abelian case it contains a single trace over gauge group
indices or CP labels. The fields are defined
\beqn
g_{ab} = \partial_a X^M \partial_b X^N g_{MN} \ ,
\quad
B_{ab} = \partial_a X^M \partial_b X^N B_{MN} \
,
\eeqn
and the overall (dimensionfull) parameter
\beqn\lab{mup}
\m_p = 2\pi
\ell_s^{-p-1} \times
\Bigg\{ {~~1~\quad {\rm for\ type\ II} \atop
           \frac{1}{\sqrt 2} \quad {\rm for\ type\ I} } \ .
\eeqn
We use conventions where the integration measure $\int d^{p+1}x$
produces a factor $\ell_s^{p+1}$. The functions $X^M(\x)$ are the
coordinates of the $(p+1)$-dimensional D$p$-brane world volume
$\cw$, parameterized by the coordinates $\x^a$, in the
ten-dimensional target space $\cy$. Denoting the embedding by $f$
this means, $f : \cw \mapsto
\cy$, $\x^a \mapsto X^M(\x^a)$. The metric $g_{ab}$ is the
pull-back of the ten-dimensional metric $g_{MN}$ under $f$, etc.
\footnote{See \cite{Myers:1999ps} as a useful reference.}

The bosonic degrees of freedom of the D-brane, the massless open
string fields, are the $(p+1)$-dimensional gauge field $A_a(\x)$
and the fluctuations of the transverse coordinates $X^i(\x)$. The
latter are describing the motion and deformation of the brane.
Linearized in transverse fluctuations $\Phi^i$ one can expand
\beqn
X^a = \d^a_b \xi^b\ ,
\quad X^i = x^i + 2\pi\a' \Phi^i(\xi)+\,\cdots
\eeqn
with constant $x^i$. There is one scalar field $\Phi^i$ for each
of the $9-p$ transverse direction of the
D$p$-brane.\footnote{These scalars should not be confused with the
ten-dimensional dilaton field $\Phi$.} Together the $A_a$ and the
$\Phi^i$ comprise the eight bosonic degrees of freedom found in
the open string spectrum in (\ref{openspec}).

In order to extract the two-derivative leading order Lagrangian
out of (\ref{dbi}) one can perform an expansion in powers of the
field strength by use of
\beqn
{\rm det}(1+M) = 1 + {\rm tr}(M) - \frac12 {\rm tr}(M^2) + \, \cdots\
.
\eeqn
If four-dimensional components are involved one needs to take care of
the sign in the ``mostly plus'' metric.
The simplest case is a D9-brane as appears in type I string
theory. It has only gauge fields, no transverse scalars, and all
pull-backs are trivial. On a flat background with flat metric and
vanishing vacuum expectation value for $\cf_{MN}$ the expansion
(in the abelian case) just produces
\beqn\lab{dbiexp}
\cs_{\rm DBI} = - \mu_9 \int d^{10}x\,
e^{-\Phi} \sqrt{-g} \Big[1 + \frac14 (2\pi\a')^2 \cf_{MN}\cf^{MN}
+ \, \cdots
 \Big] \ .
\eeqn
These are the kinetic terms for the gauge fields and a term
proportional to the volume of the brane, in this case the total
infinite space-time volume. The latter is a contribution to the
vacuum energy of the theory, the tension of the D-brane. One can
read off the ten-dimensional gauge coupling of type I string
theory as
\beqn
g_{\rm YM}^{-2} = \m_9 (2\pi\a')^2 e^{-\Phi} \ .
\eeqn
Formally, the formula (\ref{dbi}) is also the form of the
(disk-level) action expected for the non-abelian case, however, it is not
fully known how to define the trace over gauge group indices in
that case. Various different approaches have been attempted
\cite{Tseytlin:1997cs,Hashimoto:1997gm,Myers:1999ps,Denef:2000rj,Bergshoeff:2000ik,Stieberger:2002fh,Stieberger:2002wk,Coletti:2003ai}.
In any case, one expects corrections to the DBI action from higher
order string diagrams and from higher derivative interactions.

The second piece of the open string effective action is the CS
action of a D-brane, sometimes also called Wess-Zumino action. It
is essential in obtaining a supersymmetric theory and furthermore
plays an important role in the process of anomaly cancellation via
generalizations of the Green-Schwarz mechanism. This CS action is
given by
\cite{Douglas:1995bn,Green:1996dd,Cheung:1997az,Morales:1998ux,Stefanski:1998yx}
\beqn \label{cs}
\cs_{\rm CS} = - \mu_p \int_{\cw} {\rm ch}\left( 2\pi \a' {\cf}\right) \wedge
\sqrt{\frac{\hat A({\cal R}_T)}{\hat A({\cal R}_N)}} \wedge
  \bigoplus_{q} C_q \ ,
\eeqn
where we are using standard differential form calculus. The
curvature two-forms $2\pi\a' \cf$ and ${\cal R}={\ell_s^2} R$
appearing in the Chern character A-roof genus are
made dimensionless, the $p$-form potentials $C_p$ are dimensionless as
well. The indices $N,\, T$ on $\cR$ stand for the
curvature form of the tangent or normal bundle of $\cw$.
We will not need explicit formulas for these.
The Chern character and the A-roof genus are defined in \reef{Chernch} and
\reef{Aroof}. The sum over the RR $q$-forms is over all the potentials
that appear in either type IIA or IIB theory. The exotic and
non-dynamical forms of high degree, the ten-form and eight-form of
IIB, and the nine-form and seven-form of IIA, are meant to be
included. The CS action does not involve the metric and is thus of
topological nature. It measures the charge of a D-brane.

The supersymmetrization of the DBI action including the world volume
fermions was originally derived in a superfield formulation
\cite{Cederwall:1996pv,Aganagic:1996pe,Cederwall:1996ri,Bergshoeff:1996tu,Aganagic:1996nn}
much alike (\ref{dbi}) and (\ref{cs}) with bosonic fields replaced
by superfields. Its expansion in terms of component fermionic
fields has, for instance, been partly performed in
\cite{Grana:2002tu}.


\subsubsection{D-branes as charged BPS states}
\lab{secBPS}

What makes D-branes so extremely useful tools in the study of
string dualities is the fact that they carry conserved charges of
topological nature \cite{JP95}. In the context of extended supersymmetry
the charge can be related to the central charge of the superalgebra
\cite{deAzcarraga:1989gm}.
This implies that D-branes satisfy BPS conditions, they can be
grouped into short multiplets and one can benefit from
non-renormalization theorems for such states. For the construction
of D-brane models, string compactifications with D-branes in the
vacuum, an important consequence of this is the so-called
``no-force law'', the statement that two BPS states do not exert
any force towards each other. Thus, BPS D-branes are static and
can be superposed without creating an instability.

Polchinski's discovery of the charged nature of D-branes
\cite{JP95,pcj96,JP96} was a
major step in the development of modern string theory since 1995.
The RR charge of D-branes emerged from an analysis of the one-loop
annulus diagram of strings stretching between two stacks of
parallel D-branes, such as two space-time filling D9-branes. This
amplitude vanishes by the cancellation of an attractive
gravitational and a repulsive electromagnetic force. The latter is
due to the RR fields of the theory which behave like generalized
electromagnetic fields, which  couple to D-branes as generalized
charged particles.

At low energies D-branes are described by solitonic supersymmetric
solutions to the effective supergravity equations of motion that carry RR
charge. The field content of these theories contains the metric and
dilaton plus various anti-symmetric tensor fields.
The classical equations of motion can be written (see e.g.\
\cite{Dabholkar:1990yf,Horowitz:1991cd,Callan:1991at,Stelle:1998xg})
\beqn
&& R_{MN} ~=~ \frac12 \pa_M \Phi \pa_N \Phi
+ \frac1{2(p+1)!}
 e^{A\Phi} \Big( F_{MN_2...N_{p+2}} F_N{}^{N_2...N_{p+2}} \\
&& \hspace{8cm}
-  \frac{p+1}{8(p+2)} g_{MN} F_{N_1...N_{p+2}}F^{N_1...N_{p+2}}\Big) \ ,
\non
&& \nabla_M \big( e^{A\Phi} F^{MN_2...N_{p+2}} \big) ~=~ 0 \ ,
\quad \nonumber
\nabla^M \pa_M \Phi ~=~ \frac A{2(p+2)!} e^{A\Phi}
F_{N_1...N_{p+2}}F^{N_1...N_{p+2}}\ .
\eeqn
where $A=(3-p)/2$ for a D$p$-brane. To obtain this form, one has
to use the ten-dimensional Einstein-frame, $\Phi$ is the dilaton,
and $F_{M_1...M_{p+2}}$ the RR (or NSNS) $(p+2)$-form field
strengths. These are the Einstein equation, and the equations of
motion for $\Phi$ and $F_{p+2}$. More details of the underlying
type II supergravity theories from which this set of equations
derives will be discussed in section \ref{secIaII}.\footnote{The
string frame Lagrangian is given in (\ref{IIaction}) to which the
Weyl rescaling \reef{weyl} has to be applied.}

To write a solution, one splits indices into the world volume
directions $a,b=0,...,p$ and the transverse space
$i,j=p+1,...,9$. The metric, dilaton and $p$-form are then
determined by a single harmonic function $H(x^i)$,
\beqn \lab{Hfu}
H(x^i) = 1 + Q r^{p-7}\ , \quad r = (\d_{ij}x^ix^j)^{1/2}\ , \quad Q>0\ , \ p\leq 7\ ,
\eeqn
$Q$ denoting the charge unit of the brane. The (electric) solution then reads
\beqn \lab{Dpsol}
ds^2 &=& H^{\frac{p-7}{2\D}}\, \eta_{ab} dx^adx^b +
H^{\frac{p+1}{2\D}}\, \d_{ij} dx^idx^j \ ,
\non
e^\Phi &=& H^{-\frac{2A}\D} \ ,\quad
F_{i a_1...a_{p+1}} ~=~ - \e_{a_1...a_{p+1} i} \pa^i H^{-1} \ .
\eeqn
Here $\D = A^2 + (p+1)(p-7)/4$. The function of the internal
radial distance in front of the four-dimensional piece
$\eta_{ab}dx^adx^b$ is called the warp factor. Together, D-branes
warp the geometry of the transverse space, for $p\neq3$ they lead
to a non-trivial dilaton profile, and a $(p+1)-$dimensional
D$p$-brane induces a non-trivial $(p+2)$-form $F_{p+2}$. The
charge can be computed by integrating the RR flux through an
$(8-p)$-dimensional sphere at transverse infinity,
\beqn \lab{RRcharge}
\int_{\mbb S^{8-p}} * F_{p+2} = Q \ .
\eeqn
Comparing to \reef{cs} allows to identify the charge $Q$ of the
soliton with the coupling strength $\m_{6-p}$ in the world volume
action.

The no-force law for two D-branes that preserve mutual
supersymmetry now follows from the DBI and CS actions (\ref{dbi})
and (\ref{cs}). One can use one D-brane as a probe of a background
created through the presence of other branes if the backreaction
with respect to the probe can be neglected. The effective
potential energy of such a configuration is obtained by inserting
the solution for the background fields into the DBI plus CS action
of the probe. The simplest case is a D3-brane in a background of a
number of parallel D3-branes at some distance in the transverse
space. One finds a flat potential for the scalar field that
parameterizes the distance due to a cancellation between the
attractive DBI and repulsive CS pieces.

In this section we have used the solitonic $p$-brane solution to
illustrate the type of backreaction that appears unavoidably
whenever D-branes are present, non-trivial warp factors, dilaton
profiles and anti-symmetric tensor fields. For four-dimensional
models, the internal space has to be replaced with a compact space
where the branes wrap submanifolds. In such a situation explicit
solution are not known except from special examples and the
backreaction cannot be computed in the same way, not even at the
classical level. Nevertheless, similar effects on the metric, the
dilaton and other fields like in flat space are expected.


\subsection{Orientifolds}

The importance of orientifolds\footnote{There are a number of
excellent other review
  articles available which are entirely devoted to the subject of
  orientifolds. For more details on the subject we like to refer to
  \cite{Dabholkar:1997zd,as02}.}
lies in the fact that type II compactifications with D-branes are
often inconsistent or at least unstable. Both of these deficits
can be repaired by performing an orientifold projection. It
introduces a background charge and background energy density which
together allow to stabilize D-brane configurations. The same is
true for compactifications with background fluxes which we come to
later. The charge and tension that allows to balance the D-branes
or fluxes are carried by the so-called orientifold planes or
O-planes.


\subsubsection{Orientifolds of type IIA and type IIB}

An orientifold is by definition obtained from one of the two type
II superstring theories by performing a projection that involves
the world sheet parity operator $\Omega$ that just swaps the left-
and right-moving sectors of a closed string and flips the two ends
of an open string via
\beqn\lab{Omaction}
{\rm Closed:}\quad &&
\Omega: (\s_1,\s_2) ~\mapsto~ (2\pi- \s_1,\s_2) \
, \non
{\rm Open:}\quad &&
\Omega: (\t,\s) ~\mapsto~ (\t,\pi - \s) \ .
\eeqn
The orientifolds we will discuss will all be built on type II
string theories and their supersymmetric compactifications to
four-dimensional Minkowski space-time. The standard choices of
such compactifications without background fluxes or other
modifications are schematically
\beqn\lab{IIcomp}
{\rm Type\ II\ on}~~~~
\cy = \mbb R^{1,3} \times \cx \ ,\quad
\cx = \left\{
\begin{array}{ll}
{\rm Calabi-Yau~\ for}& \cn=2\\
{\rm K3} \times \mbb T^2{\rm ~\ for}& \cn=4\\
\mbb T^6~\ {\rm for}& \cn=8
\end{array}
\right. \ .
\eeqn
We are not going to provide an extensive introduction into the
compactification of type II strings here, but some more material
will be added in later sections. In a type II compactification one
half of the gravitinos in the spectrum always comes from the
left-moving the other from the right-moving world sheet sector.
Thus, identifying fields under the world sheet parity always
reduces the number of gravitinos, and thus the number of
supercharges to one half of the values given in \reef{IIcomp}.

In general, $\O$ can be combined with any other discrete symmetry
of the background to form the orientifold projection. These
operations may be defined geometrically by using isometries of the
metric on $\cx$, or other symmetries of the background field
configuration if fields beside the metric are turned on. Or, in
case the compactification is given in form of an abstract world
sheet CFT like a Gepner model, it can be specified by describing
its action on the fields of the CFT. Both cases will be considered
later on.

Only rather specific choices of orientifold projections will be
used, and only such which preserve supersymmetry in the
construction. This does not necessarily mean that all the models
have to be automatically supersymmetric, since the D-brane
content, the open string sector, can break supersymmetry. But all
the orientifolds will at least possess a supersymmetric ground
state.

We will use the notation $\O\s$ or $\O\bs$ for the combined
operation in IIB or IIA respectively. When acting on the
background geometry $\O$ and $\s,\, \bar\s$ are both of order two
and commute, i.e.\ $\O^2=\s^2=\bs^2=1,$ $\O\s\O^{-1}\s^{-1} =
\O\bs\O^{-1}\bs^{-1} = 1$.
When they act on fermions these relations may hold only up to phase
factors. The prototype example of an orientifold is of course the type I string theory in ten
dimensions, which is obtained from type IIB with $\s$ the identity.

One can also include other identifications which do not include
$\O$ when performing an orientifold of some type II
compactification. Let these form the group $G$. We then denote the
full orientifold group of, say, a IIB orientifold by\footnote{See
e.g.\
\cite{Bianchi:1990yu,Gimon:1996rq,Gimon:1996ay,Angelantonj:1996uy,Zwart:1997aj,Aldazabal:1998mr,Klein:2000qw}
for a number of examples of II orientifolds.}
\beqn \label{origroup}
G_\Omega = G \cup \Omega \sigma G \ .
\eeqn
Of course, one can think of the orientifold by $G_\O$ as first
performing an orbifold of type II on $\cx$ by $G$ and afterwards
its orientifold by $\O\s$, but it can be useful to treat this as a
one step procedure.

The simplest situation is a toroidal orientifold compactification,
where the type II theory is compactified on a six-dimensional
torus $\mbb T^6$, $G$ is a cyclic group $\mbb Z_N =
\langle\Theta\rangle=\{\Theta, \Theta^2,...,\Theta^N=1\}$
or a product of two $\mbb Z_N\times \mbb Z_M =\langle \Theta_1,
\Theta_2\rangle$ \cite{Dixon:1985jw,Dixon:1986jc}, and $\s$ another
isometry. If it is the identity, the model can sometimes be
interpreted as type I string theory on an orbifold, but the
concept of an orientifold is more general.

We will explicitly use two qualitatively different world sheet
parity operators, referring to IIA or IIB orientifolds. In the
context of toroidal orientifolds both can be derived from $\O$ as
a symmetry of IIB by T-dualities. Since a T-duality is a
right-moving reflection of the free world sheet fields as in
\reef{Tdual} the T-dual of $\O$ is just equal to $\O$ times a
reflection along the dualized circles. When acting on the R ground
states such a reflection generates phase factors which have to be
properly included. A way to make the microscopic definition of the
T-duality consistent with the effective description is to combine
any reflection along a complex coordinate with a phase factor
$(-1)^{F_L}$, $F_L$ the left-moving space-time fermion number. In
the effective description a RR $p$-form $C_{M_1\cdots M_p}$ maps
to forms $C_{M_1\cdots M_p N_1 N_2},$ $C_{M_1\cdots M_{p-1} N_1},$
$C_{M_1\cdots M_{p-2}}$ of degrees $p+2,\, p,\, p-2$ upon two
T-dualities. If the original $p$-form was even under $\O$, and the
$(p\pm2)$-forms odd, the signs resulting from the reflection along
the dualized circles and from $(-1)^{F_L}$ just work out to keep
all three T-dual components in the spectrum, as desired. However,
we will often suppress the extra phase factor and just write
$\O\s$ or $\O\bs$ for the two operations we are using.

Introducing a complex structure on the $\mbb T^6$ we can employ
complex coordinates $Z^I$, $I=1,2,3$, for the bosonic fields of
the world sheet sigma-model. The action of $\s$ and $\bs$ on these
fields can be chosen
\beqn \lab{genOm}
{\rm IIA}: &&\quad \bs Z^I \bs^{-1} = \pm \bar Z^I \quad{\rm for}\
I=1,2,3\ ,
\non
{\rm IIB}: &&\quad \s Z^I \s^{-1} = \left\{
\begin{array}{l}
\pm Z^I \quad {\rm with\ even\ number\ of\ minus\ signs}:~~~ {\rm O5/O9}\\
\pm Z^I \quad {\rm with\ odd\ number\ of\ minus\ signs}:~~~ {\rm O3/O7}
\end{array}
\right.
\ .
\eeqn
It extends to the fermionic fields by supersymmetry. This explains
the use of $\s$ or $\bs$, the IIA operation is anti-holomorphic
while the IIB operation is holomorphic. In IIB there is the
distinction referring to the number of complex directions that are
reflected being even or odd, leading to orientifold models which
either permit compactifications with D5- and D9-branes or D3- and
D7-branes. This definition of the ``dressed'' world sheet parity
on a toroidal orientifold will be generalized to type II
Calabi-Yau compactifications later.


\subsubsection{Type I superstrings as a type IIB orientifold}

The world sheet parity $\Omega$ is defined to act by exchanging
the left- and right-moving world sheet fields of a closed string, such
as
\beqn\lab{OmLRflip}
\Omega  X^M_L \Omega^{-1} =  X^M_R \ ,
\quad
\Omega \psi^M \Omega^{-1} =  \tilde\psi^M \ .
\eeqn
In the free CFT of closed strings in a flat background, one can
easily solve for the invariant degrees of freedom. States that are
invariant under this operation are those with
\beqn \label{omegasym}
\alpha^M_n = \tilde\alpha^M_n \ , \quad
\psi_r^M = \tilde\psi_r^M \ , \quad
p_L^M = p_R^M \ .
\eeqn
Let us introduce some relevant elements of type II string theories to
explain, why this operation is actually a symmetry of
the type IIB superstring but not of type IIA.
From the algebra of the fermionic world sheet oscillators in
(\ref{wsalgebra}) it follows that zero mode oscillators
$\psi^M_0$ and $\tilde\psi^M_0$ in the two R sectors satisfy two
independent Clifford algebras.
After imposing light-cone gauge one can combine the transverse
polarizations of the zero-modes into
raising and lowering operators
\beqn \label{ladop}
\psi^I_\pm = \psi^{2I}_0 \pm i \psi^{2I-1}_0 \ , \quad
\tilde\psi^I_\pm = \tilde\psi^{2I}_0 \pm i \tilde\psi^{2I-1}_0 \ ,
\quad I=1,\, ...\, ,4 \ .
\eeqn
The ground states of the left- and right-moving R sectors are then
defined as for the open string, and carry a eight-fold degeneracy
after imposing the GSO projection \reef{gso}. The GSO projection
again acts as the chirality projection on the spinor
representations of the R ground states, projecting onto identical
chiralities in the left- and right-moving R vacuum in IIB and
opposite in IIA. The exchange of the two sectors is a symmetry of
type IIB, but not of type IIA. As representations of $SO(8)$ the R
vacua transform as ${\bf 8}_c$ or ${\bf 8}_s$, the indices $s$ and
$c$ distinguishing the two chiralities. More generally, since
$\bs$ flips the chirality by exchanging raising and lowering
operators in \reef{ladop}, but not so $\s$, $\O\bs$ is a symmetry
of IIA and $\O\s$ of IIB.

D-branes in type I string theory can be defined by boundary states
in the closed string theory just as in type II. They take the same
form as in \reef{oscbst}. The states then also have to be
compatible with the world sheet parity projection. This singles
out the D1-, D5- and D9-branes as the only supersymmetric
D$p$-branes of type I string theory.\footnote{There are however
boundary states for D-branes of different dimensions which are not
supersymmetric \cite{Frau:1999qs,Lerda:1999um}. This matches with
classification of D-branes by K-theory (see section
\reef{ktheorya}).} The D$p$-branes which come into play by the use
of the world sheet parity \reef{genOm} can be deduced from these
by the T-duality that maps $\O$ to $\O\s$ or $\O\bs$ by reflecting
two, three, four or all six circles of the $\mbb T^6$.

As an alternative definition of the orientifold projection
performed on type IIB one can view type I as type IIB with extra
orientifold planes. A path on the string world sheet from $\s_1$
to $2\pi -\s_1$ already forms a closed loop in the orientifolded
theory. One can thus restrict the closed string world sheet to
positive values of $\s_1$ which is equivalent to inserting a
cross-cap into the world sheet at $\s_1=0$. In type I the image of
such a path in the target space-time is free to extend throughout
the entire space-time, but in general its endpoints are restricted
to the fixed locus of $\s$ in order to form a closed loop. This
fixed locus (or each disconnected component) is called an
orientifold plane, an O$p$-plane of dimension $p+1$. In type I
where $\s$ is trivial this is the entire target space-time, an
O9-plane. With an O9-plane the fields of type IIB are projected
down to the fields of type I invariant under $\O$ at every point
in space-time, whereas in theories with lower-dimensional
O$p$-planes also type II fields odd under $\O\s$ can be present in
the bulk away from the O-planes.

The O-planes can be described by cross-cap states $|Cp\>$, similar
to the boundary states $|Bp\>$ introduced earlier. The periodicity
condition along the cross-cap lead to relations for the oscillator
modes like Neumann or Dirichlet boundary conditions but with extra
phase factors, namely \cite{Polchinski:1987tu}
\beqn\lab{periodcc}
\a^M_n \pm  e^{i\pi n}\tilde\a_{-n}^M = \p^M_{r} \pm i\eta e^{i\pi r}\tilde\p^M_{-r}
= 0 \ .
\eeqn
A state that satisfies these conditions as operator equations is the
cross-cap state whose oscillator contribution can be written
\beqn \lab{crcap}
|Cp,\eta\>_{\rm osc} = \exp\Big( S_{MN} \sum_{n>0} \frac1n e^{i\pi n} \a^M_{-n} \tilde\a_{-n}^N
+i\eta S_{MN} \sum_{r>0} e^{i\pi r} \p^M_{-r} \tilde\p^N_{-r}
\Big) |Cp,0,\eta\> \ .
\eeqn
The NSNS and RR components are again built out of linear combinations with
$\eta=\pm1$ to get GSO invariant states, and $S_{MN}$ is as in
\reef{Smn}. The above oscillator part of the state needs to be complemented with
delta-functions for momenta or coordinates and with proper
normalization factors as well.


\subsubsection{Effective action for type I and II closed superstrings}
\lab{secIaII}

The properties of the closed string part of an orientifold
are rather directly obtained from the underlying type II theory,
in particular the massless spectrum and the form of the effective
action.

In generality, the spectrum of any closed string theory in ten
dimensions is just obtained by tensoring the left- and
right-moving massless states and applying the GSO projection. In
the NSNS sector of type II theories this leads to the universal
result
\beqn
{\bf 8}_V \otimes {\bf 8}_V ~~~\longrightarrow~~~
\{ g_{MN} , \ B_{MN} , \ \Phi\} \ .
\eeqn
The massless spectrum consists of the ten-dimensional metric, the
antisymmetric NSNS two-form, sometimes called just $B$-field, and
the dilaton. These fields appeared in the DBI action
\reef{dbi}. The massless states in the RR sector are
given by tensoring the two R ground states, as discussed above,
\beqn\lab{IIspec}
{\bf 8}_c \otimes
\Bigg\{ { {\bf 8}_s~~~{\rm for\ IIA} \atop {\bf 8}_c~~~{\rm for\ IIB}}
~~~\longrightarrow~~~
\Bigg\{
\begin{array}{l}
\{ (C_1)_M , \ (C_3)_{MNR}\}~~~{\rm for\ IIA} \\
\{C_0 , \ (C_2)_{MN} , \ (C_4)_{MNRS}\}~~~{\rm for\ IIB}
\end{array}
\ .
\eeqn
It contains the antisymmetric RR $p$-form potentials. The
four-form of IIB carries a self-duality constraint, only half of
the degrees of freedom survive. These are the RR forms that couple
to the various D$p$-branes via the CS action
\reef{cs}. The fermionic spectrum in the RNS and NSR sectors just
consists of the two gravitinos of the $\cn=2$ supergravities,
which are of equal or opposite chirality respectively.

The effective Lagrangian (at the two-derivative level, and at
tree-level in the string coupling) is basically dictated by
$\cn=2$, $d=10$ (local) supersymmetry. The full action is the sum
of the three pieces, the kinetic terms of the NSNS and RR fields
and the CS action,
\beqn\lab{IIaction}
\cs = \frac1{2\k_{10}^2} \int d^{10}x\, \sqrt{-g} \Big[ \cl_{\rm NSNS} +\cl_{\rm RR} +\cl_{\rm CS}
\Big] \ ,
\eeqn
for either IIA or IIB. The gravitational coupling is
\beqn \lab{kappa10}
\k_{10}^2 = \frac1{4\pi}(4\pi^2\a')^4 = \frac{\ell^8_s}{4\pi} \ .
\eeqn
For later reference we also note the four-dimensional
gravitational coupling
\beqn\lab{kappa4}
\k_4^2 = \ell_s^{-6} \k_{10}^2 = M_{\rm Pl}^{-2} = (8\pi G)^{-1} \
,
\eeqn
where $G$ is Newton's constant and $M_{\rm Pl} \simeq 2\ti
10^{18}\, {\rm GeV}$. The generic NSNS part of the Lagrangian in
string frame is
\beqn\lab{IINSNS}
\cl_{\rm NSNS} = e^{-2\Phi} \Big[ R + 4 \pa_M \Phi
\pa^M \Phi - \frac12 \frac 1{3!} H_{MNR}H^{MNR}\Big] \ ,
\eeqn
where the field strength of the NSNS two-form is defined
\beqn
H_3 = dB_2\ , \quad H_{MNR} = 3 \pa_{[M} B_{NR]}\ .
\eeqn
The RR $p$-forms appear in the
action via their field strengths in the kinetic terms
\beqn\lab{IIRR}
\cl_{\rm RR} = - \frac12 \sum_p \frac 1{p!}
F_{M_1...M_p}F^{M_1...M_p} = - \frac12 \sum_p |F_p|^2 \ ,
\eeqn
the sum running over $p=2,4$ in IIA or $p=1,3,5$ in IIB. This
action also appears at the sphere-level, but for the proper loop
counting with RR forms one has to absorb a factor $g_s$ in $F_p$.
The field strengths are actually defined in a way that $F_p$ also
involves the RR potentials $C_q$ of degree $q=p-(2n+1)$, and the
NSNS two-form $B_2$. But setting $B_2$ to zero they collapse to
$F_p=dC_{p-1}$. For later use, we only record the definitions of
the IIB field strengths $F_1$, $F_3$ and $F_5$,
\beqn\lab{IIBRR}
F_1 = dC_0\ , \quad
F_3 = dC_2 - C_0 dB_2 \ , \quad
F_5 = dC_4 -\frac12 C_2\wedge dB_2 + \frac12 B_2 \wedge dC_2\ .
\eeqn
The RR forms also appear in the Chern-Simons
action, which is
\beqn\lab{IICS}
{\rm IIA}: \quad \cl_{\rm CS} &=& -\frac12 \frac1{2!4!4!} \e^{M_0...M_9} B_{M_0M_1} F_{M_2...M_5}
F_{M_6...M_9}\ , \non
{\rm IIB}: \quad \cl_{\rm CS} &=& -\frac12 \frac1{4!3!3!} \e^{M_0...M_9} C_{M_0...M_3} H_{M_4M_5M_6}
F_{M_7M_8M_9}\ .
\eeqn
The only subtlety that remains, is the
self-duality constraint of $F_5$ in type IIB,
\beqn
F_5 = * F_5\ ,\quad
F_{M_0...M_4} = \frac1{5!} \epsilon_{M_0...M_4}{}^{M_5...M_9} F_{M_5...M_9} \ ,
\eeqn
which has to imposed after deriving the equations of motion from
the action. For some purposes it can also be implemented by an
extra factor $\frac12$ in the kinetic term.

To put the actions into ten-dimensional Einstein frame one has to perform the
rescaling of the metric by
\beqn\lab{weyl}
g_{MN}^{\rm E} = e^{-\Phi/2} g_{MN} \ .
\eeqn
Once this is done, the type IIB action can be written in the form
that makes its $SL(2,\mbb R)$ symmetry manifest. To do so, one
defines the complex scalar and three-form
\beqn \lab{dilG3}
\t = C_0 + ie^{-\Phi} \ , \quad
G_3 = F_3 - \t H_3 \ .
\eeqn
Then the IIB action in Einstein frame can be formulated\footnote{For a
complex $p$-form we define $|F_p|^2 = \frac1{p!} F_{M_1...M_p}\bar F^{M_1...M_p}$.}
\beqn \lab{IIBSL2}
\cs_{\rm IIB} &=& \frac1{2\k_{10}^2} \int d^{10}x\ \sqrt{-g^{\rm E}} \Big[
R^{\rm E} - \frac{\pa_M\t\pa^M\bar\t}{2({\rm Im}\, \t)^2} - \frac12
\frac{|G_3|^3}{{\rm Im}\, \t} - \frac12 |F_5|^2 \Big]
\non
&&
+ \frac1{8i\k_{10}^2} \int \frac1{{\rm Im}\, \t} C_4 \wedge G_3 \wedge
\bar G_3 \ .
\eeqn
Again, the self-duality of $F_5$ still needs to be imposed. This form
of the action is the best starting point for studying IIB flux
compactification.

There is another way of encoding the same set of type II equations of
motion in a pseudo-Lagrangian, i.e.\ an action plus a set of
duality constraints \cite{Fukuma:1999jt,Bergshoeff:2001pv}, which is better adapted to deal with D-branes
(see e.g.\ \cite{Hassan:2000zk,Berg:2003ri,Jockers:2005pn}).
The reason being that the D-brane CS-action (\ref{cs}) is
written with RR forms of all degrees ranging from $0$ to $9$. The
kinetic terms of the RR forms in string frame can also be written in such a way.
One simply replaces (\ref{IIRR}) by
\beqn\lab{IIRRpr}
\cl'_{\rm RR} = - \frac12 \sum_p |F_p|^2 \
,\quad {\rm IIA:}\ p=0,2,...,10\ , \quad
{\rm IIB:}\ p=1,...,9\ ,
\eeqn
with $p$ taking all even or odd values, even if the $C_p$ do not
represent propagating degrees of freedom. After deriving equations of
motion, one imposes the constraints
\beqn\lab{RRHodge}
&&{\rm IIA:}~~~~~ F_2 = * F_8 \ , \quad F_4 = -*F_6 \ , \non
&&{\rm IIB:}~~~~~ F_1 = * F_9 \ , \quad F_3 = -*F_7\ , \quad F_5 = *F_5 \
.
\eeqn
This leads to identical classical dynamics, exchanging Bianchi
identities and equations of motion. Now the action contains a
kinetic term for each RR $p$-form potential, which then couples to
a D$p$-brane via (\ref{cs}). Note that in this ``democratic''
version of the RR Lagrangian there is no CS term in the action.
Its effects are reproduced by integrating out the RR forms of
degree higher then five \cite{Fukuma:1999jt,Bergshoeff:2001pv}.

The action for the bulk fields of the type I orientifold theory is
given by projecting out all degrees of freedom of the type IIB
theory that are odd under the world sheet parity. These are the
non-vanishing independent states of the type II spectrum with
(\ref{omegasym}) imposed. The bosonic spectrum is given by
\beqn \label{typIspec}
{\rm Closed\ string\ type\ I\ spectrum:}~~~~~ \{g_{MN}, \ \Phi, \ (C_2)_{MN}\}\
,
\eeqn
plus a single gravitino in the fermionic sector.

The Lagrangian is identical to the relevant pieces of the type IIB
theory, i.e.\ in string frame
\beqn\lab{typeILag}
\cl_{\rm I} = e^{-2\Phi} \Big[ R + 4 \pa_M \Phi
\pa^M \Phi \Big] - \frac12 |F_3|^2 \ .
\eeqn
The type IIB RR three-form field strength is defined $F_3=dC_2
- C_0 dB_2$, but since $C_0$ and $B_2$ are projected out, the type
I three-form in (\ref{typeILag}) is just $dC_2$ (in the absence of
open string degrees of freedom). Note that there is no CS term in
type I. One can also write the RR part of the type I action in a
democratic version, where then in addition to the two-form $C_2$ a
six-form $C_6$ appears, and the constraint is $dC_2 = - *dC_6$.
The two-form and six-form then couple to D-strings and D5-branes.
Beyond that, there is also a non-dynamical ten-form $C_{10}$
coupling to D9-branes.

Note that (\ref{typeILag}) is the action for the dynamical degrees
of freedom that survive the projection in a trivial background.
Even though it does not contain for example a second two-form
tensor $B_{MN}$ from the NSNS sector one can still introduce a
non-trivial discrete background and define orientifolds on such a
background. In this way, orientifold compactifications can contain
background fields of type II which are no longer dynamical. In
more general orientifolds with $\O\s$ or $\O\bs$ projection one
can introduce background profiles for fields of type II that are
odd under the projection and are projected out of the spectrum.
This includes background fluxes which we will study later.


\subsubsection{Cancellation of charge and tension with O-planes}
\lab{sectad}

The most important consistency condition that has to be regarded
in any orientifold compactification is the condition of RR charge
cancellation. In terms of the effective field theory of the
massless modes, it arises as a consequence of the Gauss law.

The simplest field theoretic
analogue of this charge cancellation criterion is a scalar
field $\phi$ on a compact space with
a source term $\phi J$. The equation of motion, schematically
$\Box\phi = -J$, leads to the condition that $J$ integrates to zero.
In the same sense a non-vanishing tadpole for a RR form is an
inconsistency of the vacuum at the classical level. For more details
see \cite{Rabadan:2002wy,Dudas:2004nd}.

For simplicity, let us look at a single RR form field with field
strength $F_{p+2}=dC_{{p+1}}$.
The potentials couple to sources as in the CS action of a
D$p$-brane in \reef{cs}. The generic form of the relevant action is
\beqn
\cs = - \frac{1}{4\kappa_{10}^2} \int d^{10}x\, \sqrt{-g} | F_{p+2}
|^2 - \m_{q} \sum_a \int C_{p+1} \wedge \pi^a_{9-p}  \ ,
\eeqn
where $\pi_{9-p}^a$ is a closed $(9-p)$-form, $d\pi^a_{9-p}=0$.
The explicit expression follows from \reef{cs} in a given
background.\footnote{In a flux compactification the CS action
\reef{IICS} can also function as a source for RR forms.}
The label $a$ stands for different sources, different stacks of
D-branes. This leads to an equation of motion
\beqn\lab{eomCp}
d*d C_{p+1} = 2\m_{q} \k_{10}^2 \sum_a \pi^a_{9-p} \ .
\eeqn
One can integrate this equation over the compactification manifold
or any $(9-p)$-dimensional compact submanifold of it. As long as $\pi_{9-p}^a$ is closed, the
integral will only depend on the topological class in homology. Denoting the
cohomological class of $\pi_{9-p}^a$ as $\Pi_a=[\pi_{9-p}^a]$ it follows
\beqn\label{Loctad}
\sum_a \Pi_a = 0 \ .
\eeqn
One gets a set of conditions on the topological data that define
the gauge bundle and the geometry of the cycle wrapped by the
brane, by integrating the equation \reef{eomCp} over a basis of
cycles of proper dimension. These are the RR tadpole cancellation
conditions. For example, a D6-brane in type IIA can wrap any
three-cycle and the $\pi_3^a$ are just the Poincar\'e-dual
three-forms of these three-cycles. Expanding in a basis of the
third cohomology one finds $b_3$ charge cancellation conditions,
$b_3$ the third Betti number of the compactification manifold.

The condition (\ref{Loctad}) is the main restriction why there are
no supersymmetric compactifications of type II string theories
with D-branes that fill out the four-dimensional space-time. The
integral in (\ref{eomCp}) only vanishes if the brane charges
effectively add up to zero, which is in conflict with
supersymmetry. Branes of negative charge, anti-branes, preserve
different supersymmetries than branes of positive charge, and
hence models with both types of branes tend to be unstable.

This inconsistency is repaired in the process of orientifolding
type IIB to obtain type I by adding orientifold planes, O9-planes.
They do not carry any new dynamical degrees of freedom, and there
is no field theory of massless modes associated to their world
volume. In this sense O-planes should not directly be interpreted
as physical objects, but as auxiliary formal constructions that
cover essential aspects of the vacuum of orientifolds. In
particular, O-planes carry charge and tension, and thus couple to
the bulk closed string fields including the RR forms. The action
which describes this coupling is formally identical to the DBI and
CS action after setting world volume fields to zero \cite{Scrucca:1999uz,Scrucca:1999jq}
\beqn \label{Oplact}
\cs^{\rm Opl}_{\rm DBI} &=& -T_p \mu_p \int_{\cw} d^{p+1}\xi\,
e^{-\Phi} \sqrt{-{\rm det}(g_{ab})} \
, \non
\cs^{\rm Opl}_{\rm CS} &=& - Q_p \mu_p \int_{\cw}
\sqrt{\frac{L({\cal R}_T/4)}{L({\cal R}_N/4)}} \wedge \bigoplus_{q} C_q
\ ,
\eeqn
with the Hirzebruch $L$-polynomial
in \reef{Hirzebruch}. The world volume $\cw$ of an O$p$-plane
is defined as the fixed locus of an element of the
orientifold group $\O\s G$ in \reef{origroup}. For every element that is
not freely acting there is an O-plane of proper dimension.
For example, for the dressed world sheet parity $\O\s$ one has
an O-plane located at the fixed locus of $\s$, denoted $\cw_{\O\s}$, or formally
\beqn \label{orifix}
\cw_{\O\s} = {\rm Fix}(\s) \ .
\eeqn
Since \reef{orifix} uniquely fixes $\cw_{\O\s}$ the O-planes
cannot fluctuate and their coordinates are not dynamical.

For the world sheet parity operations that we use in toroidal
models as given in \reef{genOm} there are the following types of
O-planes: In type IIA $\bs$ is the complex conjugation up to a
possible sign, and its fixed locus on a $\mbb T^6$ is
three-dimensional, an O6-plane along the real or imaginary axes.
In IIB $\s$ is a reflection along zero, one, two, or all three
complex directions inside $\mbb T^6$, leading to a O9-, O7-, O5-,
or O3-planes. Whenever the orbifold group $\mbb Z_N$ that can
appear in a supersymmetric orientifold on $\mbb T^6$ contains an
element of order two, $\Th^2=1$, then $\Th$ is a reflection along
two complex directions of the $\mbb T^6$. In IIA $\bs\Th$ also
only leads to O6-planes, while $\s\Th$ in IIB adds O5-planes to
models with O9-planes, and O3-planes to models with O7-planes, and
vice versa. This reasoning now also explains the use of the two
different world sheet projections for type IIB in \reef{genOm}.

The coefficients $T_p$ and $Q_p$ measure the charge and tension of
an O$p$-plane relative to that of a D$p$-brane. In supersymmetric
models they have opposite sign compared to D-brane charge and
tension. They may thus balance the tadpole conditions
(\ref{loctad}).\footnote{If one wants to keep track of all
possibilities one sometimes distinguishes O-planes by the sign of
their charge and tension, $\{
++,+-,-+,--\}$.}
In a model with a non-vanishing source for the RR form $C_{p+1}$
from O-planes, the action \reef{Oplact} contains a term
\beqn
- Q_{q} \m_{q} \int C_{p+1} \wedge \pi^{{\rm O}p}_{9-p}  \ ,
\eeqn
where again $d\pi^{{\rm O}p}_{9-p}=0$, and $\Pi_{{\rm O}p} = [\pi^{{\rm O}p}_{9-p}]$.
The charge neutrality condition then becomes
\beqn\lab{loctad}
\sum_a \Pi_a + Q_{q} \Pi_{{\rm O}p} = 0 \ .
\eeqn
For supersymmetric compactifications, all D-branes have to wrap
cycles with the same orientation as the O-planes, such that
solutions will only exist for negative O-plane charge $Q_{q} \leq
0$. The actual value of the charge of an O$p$-plane turns out
\beqn\lab{Oplcharge}
Q_p = - \frac{32}{2^{9-p}} \ .
\eeqn
Note that the set of orientifold planes in an orientifold model is
uniquely specified when the type II background and the world sheet
parity operation are given. Thus, the closed string sector is
determined. However, the open string sector or the content of
D-branes in the model may differ. The condition of charge
neutrality only puts constraints on the overall RR charge of all
branes. Even with  the requirement of supersymmetry there can be
many different solutions, as well shall see later. In this way one
can engineer a large amount  of string models with a single
orientifold by varying the D-brane content.

In type I string theory the fixed locus of $\O$ is the entire
ten-dimensional space-time, an O9-plane, with $T_9=Q_9=-32$. By
the presence of 32 D9-branes its charge and tension is balanced,
the right-hand-side of
(\ref{loctad}) becomes proportional to $\m_9(N_{\rm D9} - 32)$.
The values for $T_9,\, Q_9$ are determined by an explicit
calculation of one-loop string diagrams, performed in section
\ref{sec1l}. The relative charge is determined as the relative
sign of the exchange of massless RR states in the one-loop
diagrams with two boundaries (annulus) or one boundary and one
cross-cap (M\"obius strip).

Related to the tension of the branes and planes, there can also be
tadpoles for the NSNS fields, i.e.\ the metric, dilaton and
two-form. An NSNS tadpole signals that one has not solved the
equation of motion for the corresponding ten-dimensional fields
respectively that the model does not sit in a stationary point of
the effective four-dimensional potential. Therefore the true
vacuum (if it exists) deviates from a flat four-dimensional
Minkowski space times and internal (conformal) Calabi-Yau with
constant two-form and dilaton.

However, this instability may at least in principle be cured by
including other stabilizing effects, and does not necessarily
render the model fatally inconsistent, such as a RR tadpole does.
The simplest NSNS tadpole that can appear is a tadpole for the
dilaton field. The dilaton appears as a prefactor in the DBI
action (\ref{dbi}) whose leading term in an expansion in
derivatives describes the tension of the D$p$-brane. Let us look at
type I with $N_{\rm D9}$ D9-branes as a simple example to illustrate the problem. The
action for $\Phi$ in Einstein frame is schematically of the form
\beqn
\cs = - \frac{1}{2\kappa_{10}^2}
\int d^{10}x\, \sqrt{-g^{\rm E}} \Big[ \frac12 \partial_M \Phi
\partial^M\Phi + \m_9 e^{3\Phi/2} (N_{\rm D9} +T_9) +\, \cdots\Big]\ .
\eeqn
We have just kept the kinetic term for $\Phi$ from \reef{IINSNS}
and the tension terms from \reef{dbiexp} and \reef{Oplact},
transformed to Einstein frame via \reef{weyl}. In the
supersymmetric case $N_{\rm D9}=-T_9=32$, but otherwise a
non-vanishing tadpole remains for the dilaton. In that case a
constant dilaton is no solution anymore.

In four-dimensional language the tension plays the role of a
scalar potential. For constant dilaton there is a non-vanishing
potential energy in the vacuum except if the tensions cancel out.
This is not in accord with a compactification to a
four-dimensional Minkowski space-time with vanishing cosmological
constant. It could for instance be countered by other terms in the
action which depend on $\Phi$, or by a space-dependent profile for
the background of $\Phi$
\cite{Fischler:1986tb,Fischler:1986ci,Dudas:2000ff,Blumenhagen:2000dc}.\footnote{In
that situation care has to be exercised to use an effective action
derived by assuming the background to be trivial to describe other
vacua.}


\subsubsection{Compactification ansatz in orientifolds}

A Calabi-Yau compactification starts from an ansatz for the
ten-dimensional metric that splits into a six-dimensional
Ricci-flat internal piece and a four-dimensional Minkowski space
which is a solution to the vacuum equations of motion of
supergravity. This ansatz will however no longer be a solution
when the D-branes are added due to a backreaction of the geometry
towards their charge and energy density. The same is true for
fluxes. In the absence of orientifold-planes there even exist
no-go theorems for branes and fluxes in type II string theories on
a compact internal space
\cite{deWit:1986xg,Maldacena:2000mw,Ivanov:2000fg}.
Starting from a warped compactification ansatz for the
ten-dimensional metric with four-dimensional Minkowski space to
preserve four-dimensional Lorentz symmetry, such as
\beqn\lab{metricansatz}
ds^2 = g_{MN} dx^Mdx^N = \D^{-1}(x^k) \eta_{\m\n} dx^\m dx^\n +
\D(x^k) \hat g_{ij} dx^i dx^j \ ,
\eeqn
one can write the ``trace reversed'' Einstein equations
\beqn \lab{Estequ}
R_{\m\n} &=& \frac12 g_{\m\n} g^{ij} \hat\nabla_i \pa_j \ln(\D)
 ~=~ \k_{10}^2 ( \ct_{\m\n} - \frac18 \eta_{\m\n} \ct_M{}^M) \ ,
\non
R_{ij} &=& \k_{10}^2 ( \ct_{ij} - \frac18 g_{ij} \ct_M{}^M)\ .
\eeqn
Hence, $g^{\m\n}R_{\m\n}$ is a total derivative on the internal space
$\cx$. If this is compact the combination $\ct_{\m}{}^\m - \frac12 \ct_M{}^M =
\frac12 ( \ct_\m{}^\m - \ct_i{}^i)$ has to integrate to zero,
\beqn \lab{constr}
0 = \int_{\cx} d^6x \sqrt{\hat g}\, ( \ct_\m{}^\m - \ct_i{}^i) \ .
\eeqn
In a standard Calabi-Yau compactification\footnote{The proper
definition of a Calabi-Yau three-fold is that of a complex
K\"ahler manifold with vanishing first Chern class, or
equivalently with holonomy group $SU(3)$. The additional
information is required to ensure supersymmetry, here we only look
at equations of motions.} the solution is
\beqn\lab{Rflat}
\hat R_{ij} = 0 \, , \quad \D = {\rm const}\ .
\eeqn
No fields other than the metric are turned on,
\beqn \lab{noflux}
e^\Phi = g_s = {\rm const} \ , \quad
F_p = H_3 = 0\ .
\eeqn
This is not an arbitrary choice, but a direct consequence of the
condition \reef{constr}. Evaluating this expression for the DBI
action (\ref{dbi}) with $\cf_{MN}=0$ one finds
\beqn
g^{\m\n} \ct_{\m\n}^{\rm DBI} - \frac12 g^{MN} \ct_{MN}^{\rm DBI} =
\frac{p-7}2 \m_p e^{(p-3)\Phi/4} \frac1{\sqrt{g_\perp}} \d^{(9-p)}(x)\ .
\eeqn
With only D-branes of positive tension $\m_p>0$ (and $p<7$) it is
impossible to satisfy (\ref{constr}). The same is true for NSNS
and RR field strengths. The contribution of the kinetic action $-
\frac12 \sqrt{-g} |F_p|^2$, with internal components of $F_p$
potentially non-vanishing, to the right-hand-side of the trace of
the first equation in
\reef{Estequ} is
\beqn
g^{\m\n} \ct_{\m\n}^{\rm flux} - \frac12 g^{MN} \ct_{MN}^{\rm
flux} = - \frac{p-1}2 |F_p|^2 \ ,
\eeqn
This cannot be compensated for, which leads to the no-go theorem
that a warped Calabi-Yau compactification of type II with D-branes
or fluxes (with $p>1$) and a four-dimensional Minkowski space does
not exist. The problem can be circumvented by a negative
four-dimensional cosmological constant so that anti-de Sitter
compactifications are possible.\footnote{There are other caveats,
mainly the possibility that higher derivative terms in the
Lagrangian can lead to important modifications.}

In orientifolds there are orientifold planes which can have negative
tension. Their contribution to the effective action
(\ref{Oplact}) can balance the tension of a D-brane if $T_p<0$.
For example, in type I the right-hand-side of \reef{constr}
becomes proportional to $\m_9(N_{\rm D9}+T_9)$ again. In a very
similar way, internal fluxes become possible when O-planes with
negative tension are present. The example that is best understood
involves vacuum expectation values for the three-forms $H_3$ and
the $F_p$ of type IIB \cite{gkp01}. We will come to study this in
more detail later.

If (\ref{constr}) is not satisfied the compactification is
unstable, but not necessarily inconsistent. In the effective
four-dimensional action this is a situation where some scalar
fields deviate from minima and generate a non-vanishing vacuum
energy, as described in section \reef{sectad}. If a stable
configuration can be reached, the model can find a stable minimum,
if not the potential will show a run-away behavior.

In most interesting orientifold models the D-branes and O-planes
are not lying on top of each other, such that the right-hand-side
of (\ref{Estequ}) is practically never vanishing locally, even if
it integrates to zero. This implies that the background metric is
always warped, $\D(x^i)\neq{\rm const}$. Furthermore, the equation
of motion of the dilaton will also not allow for constant
solutions with the exception of models with only D3-branes. Thus,
in orientifolds in which charge and tension do not cancel locally
the backreaction of the bulk fields with respect to the presence
of the D-branes and O-planes leads to a warped background metric,
a non-trivial dilaton profile, and non-trivial RR forms as
discussed in section \ref{secBPS}. In most instances of model
building all these elements of backreaction are ignored and a
``perturbative'' approach is taken in that we assume the
deviations of the solution from a Calabi-Yau metric with constant
warp factor and from a constant dilaton to be small. This
approximation is expected to be accurate at large radius and small
coupling, when all fields and the warp factor vary slowly over the
internal space and interactions are weak \cite{gkp01}.


\subsubsection{Elements of effective actions and Calabi-Yau compactifications}
\lab{sugrapot}

In this subsection we now provide some basic material on the structure
of the four-dimensional effective theory that describes the light
modes of a compactification on Calabi-Yau spaces. This is basically
for later use when we discuss the effective action of orientifolds in
more detail.

As displayed in \reef{IIcomp} type II string theories lead to
$\cn=2$ supersymmetry in four dimensions which is reduced to
$\cn=1$ by the orientifold projection. We therefore start by
introducing the relevant notation and terms of the $\cn=1$
supergravity Lagrangian following the standard textbooks
\cite{JP98,JP98a}.
Furthermore, it will also be useful to add some information about
the moduli spaces of type II compactifications on Calabi-Yau
spaces, since the general structure already implies stringent
restrictions on the potential and in particular its quantum
corrections that can appear in either theory.

First of all, the general Lagrangian contains in its bosonic part
at the two-derivative level
the kinetic action of the gauge fields and scalars plus a potential,
\bea\label{eff4}
\cl_{\rm SG} &=&
\frac1{2\kappa_4^{2}} R
- G_{\a\bbe}(\phi,\bar\phi)  D_\mu\phi^\a D^\mu \phi^{\bbe} -
\cv_{\rm SG} (\phi,\bar\phi) \non
&&
- \frac18 \Re\, f_{ab}(\phi) F^a_{\mu\nu}F^{b\mu\nu}
- \frac18 \Im\, f_{ab}(\phi) \e^{\m\n\r\s} F^a_{\m\n} F^b_{\r\s} + \ \cdots
\eea
These are the types of terms we are mostly going to concentrate
on. The scalar fields $\phi^\a$ are complex coordinates of the
sigma-model target space with metric $G_{\a\bbe}$ which is given
as second derivative of the K\"ahler potential $K(\phi,\bar\phi)$,
\beqn \lab{metric}
G_{\a\bbe} = \frac{\pa^2 K(\phi,\bar\phi)}{\pa \phi^\a
\pa\phi^\bbe}\ .
\eeqn
The gauge fields are in general labelled by the adjoint
index $a$. The matrix of gauge couplings and theta-angles $f_{ab}$,
the gauge kinetic functions, has
only off-diagonal elements for abelian factors in the gauge
group, otherwise we write $f_{ab}=\d_{ab} f_a$.
The functions $f_{ab}$ are holomorphic in the $\phi^\a$.
Note that the terms multiplied by the imaginary parts is odd
under CP.

The general form of the scalar potential has two
pieces which are referred to as F-terms and D-terms,
\beqn\lab{SGpot}
\cv_{\rm SG}(\phi,\bar\phi) = \cv_{\rm F} + \cv_{\rm D} \ .
\eeqn
The two pieces are written in terms of the K\"ahler potential
$K(\phi,\bar\phi)$ and the superpotential $W(\phi)$. The latter is
also holomorphic. The F-term potential is\footnote{The dimensions
work out as follows: The fields $\phi^\a$ have dimension one, the
K\"ahler potential dimension two, and the superpotential dimension
three. The factors of $\k_4=1/M_{\rm Pl}$ are put in to give the
potential dimension four. Later on we will often use dimensionless
scalar fields such that $D_\a W$ has the same dimension as $W$.}
\beqn\lab{Fterm}
\cv_{\rm F} = e^{\k_4^2 K} \Big( G^{\a\bbe} D_\a W D_\bbe \bar
W - 3 \k_4^2 |W|^2 \Big) \ .
\eeqn
The covariant derivative
\beqn\lab{Kaecov}
D_\a W = \pa_\a W + \k_4^2 K_\a W = F_\a
\eeqn
indicates that $W$ is actually not a function but a section of a
holomorphic line bundle over the sigma-model target space. The
$F_\a$ are the auxiliary complex scalar fields in the chiral
multiplets, thus the name F-terms, and a non-vanishing value
indicates (spontaneously) broken supersymmetry.

The D-term potential is written in terms of the auxiliary
$D^a$-fields in the vector multiplets as
\beqn\lab{Dterms}
\cv_{\rm D} = \frac12 \big(\Re(f)^{-1}\big)_{ab}\, D^a D^b\ .
\eeqn
The $D^a(\phi,\bar\phi)$ are the auxiliary D-fields. Denoting the
holomorphic Killing vectors by $X^{a\a}$, the $D^a$ satisfy
$\pa_\a D^a =- i K_{\a\bbe} X^{a\bbe}$. For linearly transforming
scalars and diagonal gauge kinetic matrix the D-term potential
becomes
\beqn
\cv_{\rm D} = \frac18 (\Re\, f_{a})^{-1} ( K_\a T^a \phi^\a + \,
{\rm h.c.}\, )^2 \ .
\eeqn
The $T^a$ are the constant representation matrices of the gauge
symmetry. For an abelian gauge symmetry the D-term can always be shifted
by the \FI parameter $D^a \rightarrow D^a + \xi^a$. A non-vanishing value
of $D^a$ means supersymmetry is broken in the vacuum. Together the
conditions for unbroken supersymmetry are
\beqn
F_\a = 0 \ , \quad D^a=0\ .
\eeqn
In a supersymmetric Minkowski vacuum the vacuum energy proportional to $\k_4^2
|W|^2$ also has to vanish which implies $W=0$.

By general non-renormalization theorems there are no perturbative quantum corrections
to the superpotential at all, and no corrections to the gauge kinetic
functions beyond one-loop. The K\"ahler potential can in principle
have corrections at any loop order. Furthermore, all three quantities
can have corrections from non-perturbative quantum effects such as
instantons. This set of statements is summarized in table
\ref{tabSGcorr}.

\begin{table}
\centering
\begin{tabular}{|c|c|}
\hline
& \\[-.45cm]
Quantity  & Quantum correction \\
\hline
& \\[-.45cm]
K\"ahler potential $K(\phi,\bar\phi)$ & Any perturbative and
non-perturbative
\\[.05cm]
Gauge kinetic function $f_a(\phi)$ & One-loop and non-perturbative \\[.05cm]
Superpotential $W(\phi)$ & Non-perturbative \\[.05cm]
\hline
\end{tabular}
\caption{Quantum corrections in general $\cn=1$ supergravity\lab{tabSGcorr}} 
\end{table}

A case that is important throughout the various classes of models
we discuss is the gauging of a shift symmetry in the context of the
Green-Schwarz mechanism. It is described by
a constant Killing vector
\beqn\lab{constKV}
X^{a\a} = i Q^{a\a}\ ,
\eeqn
for some real constant $Q^{a\a}$.
A simple case where this
happens is a (dimensionless) scalar $S$ with K\"ahler potential
\beqn\lab{dilKpot}
\k_4^2 K(S,\bar S) = -\ln(S+\bar S)
\eeqn
such that $S$ parameterizes the space $SU(1,1)/U(1)$. One finds
immediately
\beqn\lab{dilDterm}
i K_S X^{aS} = \k_4^{-2} \frac{Q^{aS}}{S+\bar S}  \ .
\eeqn
This is the famous \FI parameter of the heterotic string
\cite{Dine:1987xk}. Strictly speaking it is just a D-term but after
stabilizing $S$ it may be viewed as a constant \FI term. Similar
\FI terms are always induced along with the
\gs mechanism. Another useful way to derive them is to replace the
K\"ahler potential \reef{dilKpot} by the gauge invariant
expression
\beqn\lab{invK}
\k_4^2 \tilde K ( S,\bar S,V) = -\ln(S+\bar S - Q^{aS} V^a)\ ,
\eeqn
where $V^a$ are the abelian vector fields that gauge the shift
isometry of $S$. In superfield notation the gauge transformations
are $\d S = Q^{aS} \L$ and $\d V^a = \L + \bar\L$ for a chiral
multiplet $\L$. One can then use $\tilde K$ instead of $K$ as a
superspace density to derive the Lagrangian, schematically in the
form $\int d^2\th d^2\bar\th\, \tilde K$. The D-term contribution
is
\beqn\lab{FIfromK}
\frac{\xi_a}{g_a^2} = \frac{\pa \tilde K}{\pa V^a} \Big|_{V^a=0}\ .
\eeqn
Due to the fact that these \FI terms are related to anomalies in the
\gs mechanism they are also protected from perturbative quantum
corrections beyond one-loop
\cite{Witten:1981nf,Fischler:1981zk,Dine:1987xk}.

We now turn to the more specific case of type II string theories
on Calabi-Yau manifolds. In the first place they lead to $\cn=2$
supersymmetry. It has a more restrictive structure for the scalar
fields contained in vector multiplets (VM) and hypermultiplets
(HM). The total moduli space is a direct product of two factors,
one for each type of moduli. The two manifolds are of particular
geometric nature with either special K\"ahler geometry (VM) or
quaternion K\"ahler (or shorter quaternionic) geometry (HM). The
Lagrangian in particular for the VM is given through a single
holomorphic function, the so-called prepotential which serves as a
chiral superspace density and benefits from non-renormalization
theorems similar to the superpotential of $\cn=1$.

As mentioned earlier, in IIA the complex structure moduli of the
Calabi-Yau enter the HM, the K\"ahler deformations go into VM, and
vice versa in IIB. This structure is summarized in table
\ref{tabIICY}.
In both cases the ten-dimensional dilaton, i.e.\ the string
coupling constant, is part of the universal HM. This implies that
quantum corrections in the string coupling, either perturbative or
non-perturbative, do not depend on the VM and only affect the HM
moduli space. On the other hand, corrections in the derivative
expansion of the effective action, quantum corrections in the
sigma-model expansion parameter $\a'$ only depend on K\"ahler
moduli. This leads to the pattern of possible corrections also
displayed in table \ref{tabIICY}.

\begin{table}[h]
\centering
\begin{tabular}{|c|c|c|}
\hline
&& \\[-.45cm]
IIA & IIB & Multiplets \\
\hline
&& \\[-.45cm]
K\"ahler: $\a'$ & Complex structure: None & VM:
Special K\"ahler
\\[.05cm]
Complex structure: $g_s$ & K\"ahler: $g_s$ and $\a'$ & HM: Quaternionic \\[.05cm]
\hline
\end{tabular}
\caption{Moduli spaces and quantum corrections
of Calabi-Yau compactifications\lab{tabIICY}} 
\end{table}

Furthermore, the complexified K\"ahler moduli always include
internal components of anti-symmetric tensor fields, either of the
NSNS two-form or of the RR forms. They enjoy additional abelian
gauge symmetries in ten dimensions which descend to shift
symmetries of the effective theory that are exact in perturbation
theory. Because the superpotential is holomorphic it can not
depend on the K\"ahler moduli perturbatively in order to be
invariant. Non-perturbatively a superpotential can be induced by
world sheet instanton corrections or space-time non-perturbative
effects, as we will discuss in some detail later on.

For later reference we write the
K\"ahler potentials and some more geometrical data
for type II Calabi-Yau compactifications
written in terms of integrals of the K\"ahler two-form $J_2$ and the
holomorphic three-form $\O_3$ that are characteristic of a Calabi-Yau.
In our conventions these form have dimension two and three
respectively, and their periods are integer after pulling out factors
of $\ell_s$. The complex structure K\"ahler potential is
\beqn\lab{KpotOm}
\k_4^2 K_{\rm CS} = -\ln \Big[ -i \frac1{\ell_s^6} \int_\cx \O_3\wedge\bar \O_3\Big]\ ,
\eeqn
and the K\"ahler moduli space potential given by
\beqn\lab{KpotK}
\k_4^2 K_{\rm K} = - \ln \Big[ \frac1{3!} \frac1{\ell_s^6}
 \int_\cx J_2\wedge J_2\wedge J_2 \Big]\
.
\eeqn
The potential for the complex dilaton of IIB is
\beqn\lab{Kpotdil}
\k_4^2 K_{\rm Dil} = -\ln [-i (\t-\bar\t)]\ .
\eeqn
In IIA the universal HM originally arises in the form of a tensor
multiplet that needs to be dualized. The full K\"ahler potential
is simply the sum of three pieces, and the metric therefore
block-diagonal.
Both $J_2$ and $\O_3$ are closed and can be expanded in terms of
(dimensionless) harmonic two-forms $\o_A$ or three-forms $(\a_\L,\bet^\S)$ as
\beqn \lab{expJOm}
\frac1{\ell_s^2} J_2 &=& v^A \o_{A} \ , \quad A=1,..., h^{1,1}\ , \non
\frac1{\ell_s^3} \O_3 &=& X^\L(z) \a_\L - \cf_\S(z) \bet^\S \ , \quad \L,\S=0,...,h^{2,1} \ .
\eeqn
The $v^A$ are the (dimensionless) real K\"ahler moduli scalars
which get complexified by axionic scalars into K\"ahler moduli
scalars $t^A$. The $X^\L(z)$ are projective coordinates on the
complex structure moduli space with holomorphic dependence on the
(dimensionless) moduli $z^K$, $K=1,...,h^{2,1}$, defined as
periods of $\O_3$ along a dual basis of three-cycles $(A^\L,B_\S)$
via
\beqn \lab{perOm}
X^\L = \frac1{\ell_s^3} \int_{A^\L} \Omega_3 \ , \quad
\cf_\S  = \frac1{\ell_s^3} \int_{B_\S} \Omega_3 \ .
\eeqn
One can choose conventions such that the forms used in the expansion
satisfy the relations
\beqn \lab{3int}
\int_\cx \a_\L \wedge \bet^\S = \d^\S_\L \ , \quad
\int_\cx \a_\L \wedge \a^\S = 0 = \int_\cx \bet_\L \wedge \bet^\S \ ,
\eeqn
and
\beqn \lab{2int}
&&
\hspace{-1cm}
\ck_{ABC} = \int_\cx \o_A \wedge \o_B \wedge \o_C \ , \quad
\ck_{AB} = \frac1{\ell_s^2} \int_\cx \o_A \wedge \o_B \wedge J_2 = \ck_{ABC} v^C\ ,
\\
&&
\hspace{-1cm}
\ck_{A} = \frac1{\ell_s^4} \int_\cx \o_A \wedge J_2 \wedge J_2 =
\ck_{ABC} v^B v^C \ , \quad
\ck = \frac1{\ell_s^6} \int_\cx J_2 \wedge J_2 \wedge J_2 = \ck_{ABC} v^A v^B
v^C \ .
\nonumber
\eeqn
Analogously, the intersection pairing of the three-cycles satisfies $A^\Lambda\circ A^\Sigma=0, \,
B_\Lambda\circ B_\Sigma=0$ and $A^\Lambda\circ B_\Sigma=\delta^\Lambda_\Sigma$.

Thus, the argument of the logarithm in \reef{KpotK} is given by the
tripel intersections numbers $\ck_{ABC}$ of harmonic two-forms, and
the intersections of three-forms are normalized to a
delta-function. The metric is in in either case computed by using
\reef{metric}.

Part of this structure survives in orientifold models with $\cn=1$
supersymmetry as long as additional open string (or bundle) moduli
are neglected. In that case, the closed string sector is only a
truncation of the type II compactification and one expects its
properties to survive, even though in general $\cn=1$
supersymmetry is not as restrictive and all scalars come in chiral
multiplets (We do not consider the possibility of linear
multipets.). When open string scalars also enter, the direct
product structure of the moduli space into complex structure and
K\"ahler deformations is broken and one has to deal with a single
manifold for the sigma-model target space. In section
\ref{secZ22} and more generally in \ref{OPEN}
we will give an example of this phenomenon and show how the open
string moduli lift the factorization of the corresponding type II
moduli space.


\subsubsection{Closed string spectra in Calabi-Yau orientifolds}
\lab{oriclspec}

We now discuss how to determine the spectrum of closed string
modes in an orientifold that starts from a supersymmetric type II
compactification on a Calabi-Yau or a space that permits even more
supersymmetries, K3$\times \mbb T^2$ or a $\mbb T^6$.

The type II compactification is orientifolded by identifying
states under $G_\O$ from (\ref{origroup}). The number of
supercharges is cut in half by the orientifold projection in any
case. One may view this as a quotient of the type II
compactification on $\cx/G$ by $\O\s$ or $\O\bar\s$.

The spectrum of the closed string sector consists first of all of
the parent type II spectrum projected onto states invariant under
the orientifold group. If the orbifold group $G$ is non-trivial
and its elements have fixed loci, there also exist twisted
sectors. Twisted states satisfy periodicity conditions on the
covering space only up to elements of $G$. On a torus one can
write explicitly for the world sheet bosons
\beqn\lab{twper}
X^M(\sigma_1,\sigma_2) = \Theta X^M(\sigma_1+2\pi,\sigma_2)
\Theta^{-1} \ ,
\eeqn
for any $\Th\in G$. More generally, such twisted periodicity
condition would have to be applied to any field in the theory. We
will first discuss the case where the spectrum of type II on
$\cx/G$ has already been determined (or where $G$ is trivial) and
only the projection onto states invariant under the world sheet
parity needs to be performed.

Let us start with a simple illustrative case where everything is
very explicit, the toroidal compactification of type IIA with
$\cn=4$ supersymmetry. The naked world sheet parity $\Omega$ acts
together with the complex conjugation $\bar\s$ introduced in
\reef{genOm}. For a toroidal background space, it is related to
$\O$ in type IIB by three T-dualities. Denote the three direction
of the $\mbb T^6$ that are reflected by $\bs$ as
$i,j\in\{4,6,8\}$, the three other internal directions
$a,b\in\{5,7,9\}$, while the four-dimensional space-time is
$\mu,\nu\in\{0,1,2,3\}$. The closed string spectrum that descends
from the IIA fields of \reef{IIspec} is then determined by
decomposing the fields into internal and four-dimensional
components and keep the even fields. It comes out as
\beqn\lab{IIAonT6}
{\rm NSNS:} && \{G_{\mu\nu} ,\  G_{ij} , \ G_{ab}, \
B_{\mu a}, \ B_{ia}, \ \Phi\} \ , \\
{\rm RR:} && \{(C_1)_a, \ (C_3)_{\mu ab}, \ (C_3)_{iab}, \ (C_3)_{\mu ij} , \
(C_3)_{\mu\nu i}, \ (C_3)_{ijk}\} \ ,
\nonumber
\eeqn
which comprises 35 scalars, 12 vectors, plus 3 antisymmetric
tensors, which in four dimensions can be dualized into scalars,
and finally the metric. This spectrum consists of a spin two
$\cn=4$ multiplet and six abelian spin one multiplets.\footnote{See also
\cite{Kachru:2002he,Berg:2003ri}
for a similar IIB orientifold on $\mbb T^6$ studied in the frame
work of flux compactifications.}

This procedure can be generalized to specify the spectrum of IIA
and IIB orientifolds on Calabi-Yau manifolds \cite{gl04,Grimm:2004ua}. Morally speaking,
the projection will be performed in a very similar fashion.
We will make use of the standard results for the
dimensional reduction of type II theories on Calabi-Yau spaces
\cite{Candelas:1985en}. The spectrum of closed string modes is stated in terms of
supermultiplets of the $\cn=2$ supersymmetry of the type II model
which is then reduced to $\cn=1$ by the projection. These are the
gravity multiplet, vector multiplets and
hypermultiplets.\footnote{We do not distinguish here between
hyper- and tensormultiplets assuming that all anti-symmetric
tensors have been dualized into axionic scalars via
four-dimensional Hodge duality of their field strengths. Similarly
we do not distinguish between chiral and linear multiplets in $\cn
=1$ spectra.} The vectormultiplets contain a four-dimensional
vector field plus a complex scalar, the hypermultiplets two
complex scalars as bosonic components.

For type IIA the dimensional reduction on a Calabi-Yau leads to
$h^{1,1}$ vector multiplets ($h^{p,q}$ being the Hodge numbers of
the Calabi-Yau) whose vectors come from the reduced RR three-form
$C_3$ and whose scalars come from the NSNS two-form $B_2$ and the
K\"ahler deformations. Further, there are $h^{2,1}+1$
hypermultiplets which include the internal components of $C_3$ and
the complex structure moduli (the one extra hypermultiplet is the
universal one that includes the dilaton). The $\cn=2$
supersymmetry requires that the moduli space is of a direct
product form, one for each type of multiplet. The vector
multiplets enjoy so-called special K\"ahler geometry while the
hypermultiplet parameterize a quaternionic manifold.

\begin{table}[h]
\begin{center}
\begin{tabular}{|c|c|c|}
\hline & & \\[-.4cm]
Multiplicity & Multiplet & Moduli \\[.1cm]
\hline
\hline
$1$ & {\rm gravity\ multiplet} & \\
$h^{1,1}$ & {\rm vector\ multiplets} & K\"ahler \\
$h^{2,1}+1$ & {\rm hyper\ multiplets} & Complex Structure\\
\hline
\end{tabular}
\end{center}
\caption{$\cn=2$ spectrum of IIA Calabi-Yau compactification\lab{specIIACY}}
\end{table}

\noindent
The operation $\bs$ acts on the
cohomology groups by
\beqn\lab{sigA}
{\rm IIA}:\quad \bs : H^{p,q} \mapsto H^{q,p}\ .
\eeqn
This implies that $(1,1)$-forms are mapped onto themselves, but
$(3,0)$- and $(2,1)$-forms are swapped with the complex conjugate.
Thus, $H^{1,1}$ can be split into subspaces of eigenvalues $\pm 1$
with dimensions $h_\pm^{1,1}$. For $\cn=2$ vectormultiplets that
come from a reduction with internal component in $H^{1,1}_+$ the
vector field survives as bosonic component the projection with
$\O\s$, whereas for $H_-^{1,1}$ the scalar survives. Among the
hypermultiplets one can form linear combinations such that
precisely one half of the scalars is even under $\O\bs$.

Putting the pieces together, the $\cn=1$ supergravity theory of
the massless closed string modes of a type IIA Calabi-Yau
orientifold with orientifold group generated by $\O\bs$ alone is
summarized in table \ref{specIIA}.

\begin{table}[h]
\begin{center}
\begin{tabular}{|c|c|}
\hline & \\[-.4cm]
Multiplicity & Multiplet \\[.1cm]
\hline
\hline
$1$ & {\rm gravity\ multiplet}\\
$h_+^{1,1}$ & {\rm vector\ multiplets}\\
$h_-^{1,1}+h^{2,1}+1$ & {\rm chiral\ multiplets}\\
\hline
\end{tabular}
\end{center}
\caption{Spectrum of IIA Calabi-Yau orientifolds\lab{specIIA}}
\end{table}

\noindent
The classical moduli space of the closed string fields is still of
a direct product form, involving $h_-^{1,1}$ K\"ahler moduli and
$h^{2,1}$ complex structure moduli that can be viewed as the real
or imaginary parts of the $2h^{2,1}$ deformations present in the
IIA compactification
\cite{Grimm:2004ua}.

In type IIB orientifolds one starts with the spectrum of table
\ref{specIIBCY} identical to the IIA case after flipping the two Hodge
numbers.

\begin{table}[h]
\begin{center}
\begin{tabular}{|c|c|c|}
\hline & & \\[-.4cm]
Multiplicity & Multiplet & Moduli \\[.1cm]
\hline
\hline
$1$ & {\rm gravity\ multiplet} & \\
$h^{2,1}$ & {\rm vector\ multiplets} & Complex Structure \\
$h^{1,1}+1$ & {\rm hyper\ multiplets} & K\"ahler \\
\hline
\end{tabular}
\end{center}
\caption{$\cn=2$ spectrum of IIB Calabi-Yau compactification\lab{specIIBCY}}
\end{table}

\noindent
Here the situation is slightly different
because $\s$ contains an even number of reflections.
It is holomorphic and maps
\beqn\lab{sigB}
{\rm IIB}:\quad \s : H^{p,q} \mapsto H^{p,q}\ .
\eeqn
Now, all the cohomology groups split into even and odd subspaces
under $\s$. There is a further ambiguity that referring to the two
versions mentioned in \reef{genOm}, the O5/O9 or the O3/O7
orientifold projection. One then gets by similar reasoning as
above the spectrum of an $\cn=1$ IIB orientifold as displayed in
table \ref{specIIB} \cite{gl04}. The upper sign refers to the
O3/O7 version the lower to O9/O5.

\begin{table}[h]
\begin{center}
\begin{tabular}{|c|c|}
\hline & \\[-.4cm]
Multiplicity & Multiplet \\[.1cm]
\hline
\hline
$1$ & {\rm gravity\ multiplet}\\
$h_\pm^{2,1}$ & {\rm vector\ multiplets}\\
$h_\mp^{2,1}+h^{1,1}+1$ & {\rm chiral\ multiplets}\\
\hline
\end{tabular}
\end{center}
\caption{Spectrum of IIB Calabi-Yau orientifolds\lab{specIIB}}
\end{table}

The effective Lagrangian that captures the classical dynamics of
these fields has been obtained by explicit dimensional reduction
from the ten-dimensional type II parent theories \cite{gl04,Grimm:2004ua}.

The situation where $G$ is non-trivial and has elements that are
not free we only treat briefly by an example, the case of a
toroidal orientifold model based on an orbifold $\mbb T^6/\mbb
Z_N$. Fixed points correspond to the twisted sectors of this
orbifold. In the geometric large volume regime one can deal with
these without explicit CFT calculations needed.\footnote{The
supersymmetric CFT solutions for such an orientifold, however,
does not have to be ``in the geometric phase''. This means, it can
lead to different Hodge numbers than predicted by the geometrical
method, see e.g.\
\cite{Klein:2000qw}.} An isolated fixed point of the orbifold
group is resolved by replacing it with a $\mbb {CP}^2$ which
contributes $h^{1,1}=1$ to the relevant Hodge numbers. Thus, in
IIA there is an extra vector multiplet, in IIB a hypermultiplet.

The spectrum of the orientifold in IIA now depends on if a fixed
point of some element $\Th$ of $G$ is fixed under $\bs$ as well.
If not, the fixed point and its image form a pair and one half of
the states survive the projection, i.e.\ the two $\cn=2$ vector
multiplets decompose into a chiral and vector multiplet under
$\cn=1$, the pair carries $h_+^{1,1}=h_-^{1,1}=1$. An invariant
fixed point simply supports a single chiral multiplet. If a fixed
point of $\Th$ is invariant under $\bs$ or not, in turn depends on
the complex structure of the underlying torus. To give a concrete
example, the orientifold $\mbb T^6/\mbb Z_3$ in
\cite{bgk99a,GP99,bgk00} has $3^3=27$ isolated fixed points,
leading to $h^{1,1}=27$. But there are four distinct choices for
the complex structure of the underlying torus compatible with the
orientifold projection $\O\bs$ which produce 13, 12, 9 or zero
$\cn=1$ vector multiplets. Many more such examples can be found in
the literature, for instance in
\cite{bcs04}. In IIB the distinction is less important since only
chiral multiplets are present.


\subsubsection{Open strings in orientifolds}
\lab{secopenori}

One may formally interpret closed strings that are twisted under
the world sheet parity $\O$ as open strings. A closed string
twisted by $\O$ satisfies the periodicity condition
\beqn
X^M(\sigma_1,\sigma_2) = \Omega X^M(\sigma_1+2\pi,\sigma_2)
\Omega^{-1} = X^M(-\sigma_1,\sigma_2)\ .
\eeqn
The mode expansion then becomes identical to that of an open string with
Neumann boundary conditions along all space-time directions, i.e.
like an open string with ends on a D9-brane \cite{Horava:1989vt}.

This is another way to argue that an orientifold automatically
introduces open strings or D-branes into type II
compactifications. When D-branes are added the closed string
spectrum
(\ref{typIspec}) is extended by massless open string modes such as
(\ref{openspec}). The open strings introduce non-abelian gauge
symmetry and charged matter fields, while the closed string modes
have at most abelian gauge symmetries.

Since $\O$ inverts the orientation of the closed string  world
sheet, one has to include all oriented and unoriented Riemann
surfaces in the perturbative loop expansion of string scattering
amplitudes in orientifolds, weighted by their Euler number
\reef{euler}. The closed string tree-level is given by the sphere
diagram with $g=b=c=0$ or $\chi=2$. The open string tree-level is
the disk with $g=c=0,\, b=1$ and cross-cap $g=b=0,\, c=1$ both
with $\chi=1$. The sphere leads to the supergravity Lagrangian of
the closed string or bulk fields, such as (\ref{typeILag}) for the
type I theory. It is blind to open strings. The disk and cross-cap
produce the DBI plus CS action of the D-branes in (\ref{dbi}) and
(\ref{cs}) and of the O-planes in \reef{Oplact}. The former
involve open string fields, i.e. the gauge fields and scalars plus
fermions on D-branes, and their couplings to the bulk fields. The
CS interactions between closed and open string modes are actually
required by ten-dimensional $\cn=1$ supersymmetry
\cite{Bergshoeff:1981um} and play an important part in the
Green-Schwarz anomaly cancellation mechanism \cite{Green:1984sg}.

Let us again treat the case of type I string theory in ten
dimensions first. Here, there are 32 D9-branes whose multiplicity
will be determined by a calculation of the one-loop divergences in
section \ref{sec1l}. The massless states on the D9-branes are the
analogue of \reef{CPspec} subject to the orientifold projection.
It inverts the orientation of the world sheet fields
\beqn\lab{Omopen}
\Omega X^M(z) \Omega^{-1} = X^M(\pi-\bar z)\ ,\quad
\Omega \psi^M(z) \Omega^{-1} = \tilde\psi^M(\pi-\bar z)
= \psi^M(z-\pi)\ ,
\eeqn
which leads to
\beqn \lab{Omosc}
\O \a^M_n \O^{-1} &=& \left\{
\begin{array}{ll}
e^{i\pi n} \a^M_n & {\rm for\ NN\ boundary\ conditions} \\
e^{i\pi} e^{i\pi n} \a^M_n & {\rm for\ DD\ boundary\ conditions}
\end{array}
\right. \ , \non
\O \p^M_r \O^{-1} &=& \left\{
\begin{array}{ll}
e^{i\pi r} \p^M_r & {\rm for\ NN\ boundary\ conditions} \\
e^{i\pi} e^{i\pi r} \p^M_r & {\rm for\ DD\ boundary\ conditions}
\end{array}
\right. \ ,
\eeqn
for the oscillators.\footnote{The last equality in \reef{Omopen}
is due to the doubling trick, see section \ref{openCFT}.} Of
course, on a D9-brane all directions have NN boundary conditions.
Furthermore, there is an extra phase factor $-i$ on the NS vacuum,
such that the massless vector $\p^M_{-1/2} |0\>_{\rm NS}$ is odd,
and a factor $-1$ on the R vacuum. An unambiguous way to compute
these phase factors is to read off the relative signs of diagrams
with boundaries or cross-caps in the one-loop calculation of
section \ref{sec1l}. The CP matrices are a priori $32\times 32$
unconstrained matrices that form the adjoint of $U(32)$ in type
IIB. On these $\O$ acts by transposition
\beqn
\O \l^A_{ij} | ij\> = \l^A_{ij} |ji\> = \l^A_{ji}|ij\>\ .
\eeqn
It follows that invariant states in the massless spectrum satisfy
\beqn\lab{CPproj}
\lambda^A_{ij} = - \lambda^A_{ji}\ .
\eeqn
Only the $32(32-1)/2$ antisymmetric generators among the $32^2$ of
$U(32)$ survive the projection. The gauge symmetry is thus broken
to $SO(32)$ and the vector gauge field and gaugino as given in
(\ref{CPspec}) transform in its adjoint
representation.\footnote{By a different choice of O9-planes,
namely with negative RR charge but positive tension, supersymmetry
is broken but the RR charge neutrality constraint \reef{loctad}
still maintained. This non-supersymmetric ten-dimensional
orientifold of type IIB has a vector gauge boson in the adjoint of
the gauge group $Sp(32)$ \cite{Sugimoto:1999tx}.} In type I these
are all open string modes.

The CS interactions with the bulk modes are described by modifying
the three-form field strength of the RR two-form $C_2$ by
\beqn\lab{mod3form}
\tilde F_3 = dC_2 - \frac{\k_{10}^2}{g_{10}^2} \Big(
\o_3^{\rm YM} - \o_3^{\rm L}\Big) \ ,
\eeqn
where $\o^{\rm YM}_3$ and $\o_3^{\rm L}$ are the Yang-Mills and
Lorentz CS three-forms. The value of the coefficient is
\beqn
\frac{\k_{10}^2}{g_{10}^2}  = \frac{\a'}4 \ .
\eeqn
The CS three-forms are defined by their properties
\beqn
d \o^{\rm YM}_3 = {\rm tr}\, F \wedge F\ ,\quad
d \o^{\rm L}_3 = {\rm tr}\, R \wedge R\ .
\eeqn
Here $F$ is the Yang-Mills two-form field strength $F=\frac12
F_{MN} dx^M\wedge dx^N$ and $R$ the curvature two-form not to be
confused with the curvature scalar that appears in the
Einstein-Hilbert action. This modifies the Bianchi identity of
$\tilde F_3$ to become
\beqn\lab{F3BI}
d\tilde F_3 = \frac{\k_{10}^2}{g_{10}^2}  \Big( {\rm tr}\, F \wedge F\ -{\rm tr}\, R \wedge R\
\Big) \ .
\eeqn
It implies a coupling of $C_2$ to (two or more) gauge bosons and gravitons.
The gauge (and Lorentz) transformations are defined in such a way that
$\tilde F_3$ is automatically gauge invariant, formally
\beqn
\d \o_3^{\rm YM} = d\o_2^{\rm YM} \ , \quad
\d C_2 = \frac{\k_{10}^2}{g_{10}^2}  \o_2^{\rm YM}\ ,
\eeqn
and similar for $\o_3^{\rm L}$.

The full bosonic Lagrangian of type I string theory in string
frame, to leading order in the string coupling (at sphere plus
disk level) and up to two derivatives can now be collected. Since the tension terms from
D9-branes and O9-planes just cancel, the only terms that survive at
the two-derivative level from the open string sector are the gauge
kinetic terms from \reef{dbiexp}. Together with \reef{typeILag} and
\reef{F3BI} one has
\beqn\lab{tILag}
\hspace{-.3cm}
\cs_{\rm I}= \frac{1}{2\k_{10}^2} \int d^{10}x \sqrt{-g} \Bigg[ e^{-2\Phi} \Big[ R + 4 \pa_M \Phi
\pa^M \Phi \Big] - \frac12 |\tilde F_3|^2
- \frac{2\k_{10}^2}{4g_{\rm YM}^2} {\rm tr}\, F_{MN}F^{MN} \Bigg]
\ .
\eeqn
This is the form dictated by local $\cn=1$ supersymmetry in ten
dimensions \cite{Bergshoeff:1981um}.

A very important derivative correction to \reef{tILag} is the
CS term $C_2 \wedge F^4$ that is included in \reef{cs}. It completes the
couplings necessary for a contribution to the six-point function
(hexagon diagram) of gauge bosons via tree-level exchange of $C_2$, which is the
essence of the Green-Schwarz anomaly cancellation \cite{Green:1984sg}.

The type I action can now be rewritten in the democratic version
analogous to \reef{IIRRpr} with the RR six-form $C_6$. This allows
one to leave out the correction \reef{F3BI} of the three-form
field strength and use the full expression of the equally
democratic CS action \reef{cs}. The six-form appears in \ref{cs}
with a term $C_6\wedge F^2$. Its equation of motion $d*dC_6 +
(\k_{10}^2/g_{10}^2) F^2 = 0$ then reproduces the Bianchi identity
\reef{F3BI} after replacing $*dC_6 = -F_3$ (leaving out the gravitational correction).
At the same time, the equation of
motion of $C_2$ is the Bianchi identity of $C_6$, including the $C_2
\wedge F^4$ coupling. In this way, the
CS correction of the three-form field strength is automatically
induced by the CS action of the six-form, see \cite{jl04} for the
relevant analysis in case of D7-branes. This point of view lends
itself much better to generalizations for other orientifolds with
D$p$-branes of lower dimensions since one can start from the CS
action \reef{cs} and the democratic RR kinetic action
\reef{IIRRpr} with all RR forms in a given orientifold, and does
not need to know explicitly how the Bianchi identities of the
field strengths get modified by open string fields.

Let us now only briefly turn to other orientifold compactifications
and their open string spectra. We will be more explicit later on.
Since our central subject is the construction of models of
particle physics from string compactification, we are mainly
interested in the gauge group and the spectrum of charged
particles of any given model. To determine the open string
spectrum and gauge group for models with a CFT description one has
to analyze the spectrum of open string states supported by the
boundary states of the theory. In geometric models one needs to
find gauge bundles on the internal compactification space and
their topological properties.

The simplest example is again provided by toroidal orientifold
models with orientifold group (in IIB for definiteness) $G_\O =
\O\s\cup \O\s \mbb Z_N$, $\mbb Z_N=\<\Th\>$. An element of the
orientifold group acts on an open string state by acting on the
coordinates themselves, i.e.\ on the representation under the
Lorentz group, and on the CP label, the representation under the
gauge group. To describe the latter there is an elaborate
formalism which denotes the operation of any element on the CP
labels by a gamma-matrix $\g_{\O\s}$, $\g_{\O\s\Th}$, $\g_{\Th}$,
etc. More precisely, the gamma-matrices form a projective
representation of the orientifold group, which means they satisfy
the algebra of $G_\O$ only up to phase factors. There is a unitary
$N\times N$ $\g$-matrix, $N$ the number of branes that are added
to the parent type II theory, one for each element of $G_\O$,
\beqn\lab{gammas}
\Omega\s\Theta \mapsto \gamma_{\Omega\s\Theta} \ , \quad
\Theta \mapsto \gamma_{\Theta} \ .
\eeqn
Algebraic consistency implies, for instance, $\g_{\O\s}^2 =\pm 1$,
$\g_{\Th}^N=\pm 1$, $\g_{\O\s\Th} = \g_{\O\s} \g_\Th$. In type I
string theory there is only a single element $\O$ and $\g_\O$ can
be chosen the identity matrix of dimension $32$. In generality,
states invariant under the action of the elements of $G_\O$ have
to satisfy
\beqn
G\ : \quad \psi^M_{-r} \lambda_{ij}^A | 0, ij \rangle
&=&
[\Theta \psi^M_{-r} \Theta^{-1}] [(\gamma_{\Theta})_{ik}
\lambda_{kl}^A (\gamma_{\Theta}^{-1})_{lj}] | \Th,ij \rangle
\ , \\
\O\s G\ : \quad
\psi^M_{-r} \lambda_{ij}^A | 0,ij \rangle
&=&
[(\Omega\s\Theta) \psi^M_{-r} (\Omega\s\Theta)^{-1}]
[(\gamma_{\Omega\s\Theta})_{ik} \lambda^A_{lk}
(\gamma_{\Omega\s\Theta}^{-1})_{lj}] | \O\s\Th,ij \rangle\ .
\nonumber
\eeqn
Determination of the gauge group and spectrum then
requires an analysis of all the elements of the orientifold
group and its representations. Different representations, i.e.\
choices of $\g$-matrices, can lead to different gauge symmetries
and spectra, referring to different gauge bundles used in the
compactification.

However, we will not use this formalism. Instead, we will stick to
a more geometric approach. It is a little easier to explain for
type IIA orientifolds. Starting from a type IIA compactification
on $\cx/G$ with D6-branes wrapping three-cycles characterized by
topological classes $\Pi_a$ a lot about the spectrum can be
determined without much effort. The stacks in IIA carry $U(N_a)$
gauge symmetries with adjoint gauge multiplets, plus potential
non-chiral matter which we ignore for the moment. Under $\bs$ the
stacks may or may not be invariant. Denoting the image of a stack
$a$ by $a'$ the operation of $\O\bs$ on the various open strings
can be summarized schematically
\beqn \lab{Ommatrix}
\O\bs : \left(
\begin{array}{cccc}
aa & a'a & ba & b'a \\
aa' & a'a' & ba' & b'a' \\
\end{array}
\right) \mapsto
\left(
\begin{array}{cccc}
a'a' & a'a & a'b' & a'b \\
aa' & aa & ab' & ab \\
\end{array}
\right)\ .
\eeqn
If a stack $a$ is not invariant, one half of the states among the
$aa$ strings and their images are invariant, no projection
applies. Thus one just inherits the diagonal $U(N_a)$ gauge
symmetry from the pair of stacks in IIA. In a similar way, the
spectrum of strings in any $ab$ sector is mapped to the $b'a'$
sector, and $a'b$ to $b'a$, again no projection applies to these
fields. One half of the IIA states survives the projection.

The only sectors that require extra care are $aa'$ strings which
mapped among themselves under $\bs$. The case of invariant stacks
where $a=a'$ is a special case of this situation. Here, the CP
matrices satisfy $\l_{ij}^A = \pm
\l_{ji}^A$ the sign deciding over either symmetric representations
$\Ysymm_a$ or anti-symmetric ones $\Yasymm_a$. For stacks of
invariant branes this distinguishes an $SO(N_a)$ gauge group like
in type I from an $Sp(N_a)$ gauge group. As explained for the case
of type I, the sign is determined in a CFT from a one-loop
calculation by comparing the relative phase of these states in the
annulus and M\"obius strip diagrams in the open string
loop-channel. If the stacks and O-planes are defined by boundary
and cross-cap states, one needs to compute their overlap in the
tree-channel and perform the modular transformation to the
loop-channel to read off the sign. In this way, the use of
gamma-matrices can completely be avoided.


\subsubsection{Cancellation of one-loop divergencies}
\lab{sec1l}

In the previous section we have discussed the presence of tadpoles
on the level of the classical tree-level effective action, where
they signal a potential violation of the bulk equations of motion
induced by adding D-branes and O-planes to a Calabi-Yau
compactification of type II theories. Whenever charge and tension
cancel it is permissable to add them to the type II Calabi-Yau
compactification. Otherwise a net charge results in an
inconsistency of the model and with a net energy density from
left-over tension the background has to be modified.

The conventional and most rigorous way to detect the presence of a
tadpole in a compactification for which a description in terms of
a CFT is given consists of calculating not the one point functions
of the massless fields at tree-level, but the one-loop vacuum
amplitude, the partition function. In a supersymmetric vacuum it
vanishes by a cancellation of the NSNS and RR contribution. But
this does not imply the cancellation of the tadpoles. A non-zero
tadpole for a massless field manifests itself through a divergence
in the one-loop amplitude. The cancellation of the divergences
insures the absence of such tadpoles. In the language of
four-dimensional field theory the divergences can be interpreted
as quadratic UV divergences for a massless field at one-loop,
resulting from the integrated propagator $\int d^4k/k^2
\sim \Lambda_{\rm UV}^2$.

At the one-loop level in string perturbation theory, with
$\chi=0$, there are four diagrams that interfere, the torus with
$g=1$, the Klein bottle with $c=2$, the annulus or cylinder with
$b=2$, and finally the M\"obius strip with $b=c=1$. The first two
are closed string diagrams, while the latter have
boundaries (see e.g.\
\cite{Burgess:1986ah,Burgess:1986wt,Horava:1989vt,as02}).
The one-loop partition function then is
\beqn
\cz_{\rm 1-loop} = \ct + \ck + \ca + \cm \ .
\eeqn

Let us go through the calculation for type I string theory in ten
dimensions and on a toroidal compactification on $\mbb T^6$,
which are the simplest cases to consider. In four-dimensional terms
these models have $\cn=4$ supersymmetry.
To be more precise, since type I has only D9-branes and O9-planes, the
annulus and M\"obius strip diagrams are over $99$ open strings and
should be denoted $\ca_{99}$ and $\cm_9$ but we will leave out the
indices. To start with, the torus diagram is defined
\beqn
\ct &=& \int_{\cf_\ct} \frac{d^2\tau}{4\tau_2}
 {\rm Tr}_{\rm cl} \Big[ \frac12 \cp^{\rm cl}_{\rm GSO}
 e^{2\pi i\tau L_0} e^{-2\pi i\bar\tau\bar L_0}
 \Big] \ .
\eeqn
Since the fundamental domain $\cf_\ct$ of conformally inequivalent
tori does not include $\t=\t_1+i\t_2=0$ the integration does not
lead to UV divergences for small $\t_2$ and the torus diagram can
be ignored. The other three diagrams are defined through
\beqn \label{loopchannel}
\ck &=& \int_0^\infty \frac{d\tau_2}{2\tau_2}
 {\rm Tr}_{\rm cl} \Big[ \frac{\O}{2} \cp^{\rm cl}_{\rm GSO}
 e^{2\pi i\tau L_0} e^{-2\pi i\bar\tau\bar L_0}
 \Big] \ , \non\vspace{.3cm}
\ca &=& \int_0^\infty \frac{d\tau_2}{2\tau_2}
 {\rm Tr}_{\rm op} \Big[ \frac12 \cp^{\rm op}_{\rm GSO}
 e^{-2\pi \tau_2 L_0} \Big] \ , \non\vspace{.3cm}
\cm &=& \int_0^\infty \frac{d\tau_2}{2\tau_2}
 {\rm Tr}_{\rm op} \Big[ \frac{\O}{2} \cp^{\rm op}_{\rm GSO}
 e^{-2\pi \tau_2 L_0} \Big] \ .
\eeqn
A number of explanations are in order: The traces are over the
Hilbert space before applying the GSO projections, for the closed
strings in the Klein bottle, and the open strings in the two
diagrams with boundaries. It includes an integral over space-time
momenta, sums over Kaluza-Klein states, and traces over all open
string D-brane sectors and the CP labels. The GSO projection is
then performed explicitly. It is further understood that
space-time fermions from the open string R sector enter with
opposite sign.

The equivalent formulation of the amplitude in terms
of tree-channel overlap of boundary and cross-cap states made of
closed strings is
\beqn \label{treechannel}
\hspace{-.5cm}
\ck + \ca + \cm =
\Bigg[ \sum_{{\rm O}p} \langle C | +
  \sum_{{\rm D}p} \langle B | \Bigg] \int_0^\infty dl\, e^{-2\pi
    l(L_0+\bar L_0)}   \Bigg[ \sum_{{\rm O}p} | C \>  + \sum_{{\rm D}p} | B \> \Bigg] \ .
\eeqn
Pictorially this is shown in figure \ref{fig1l}.

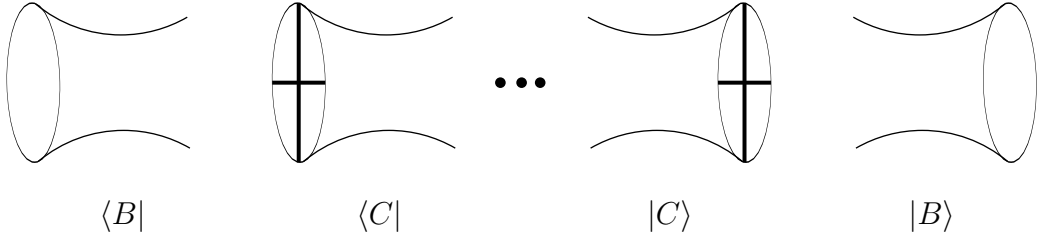
\begin{figure}[ht]
\vspace{-1cm}
\hspace{-.5cm}
\begin{picture}(180,180)(0,0)

\Oval(50,100)(30,10)(1)
\CArc(83,168)(50,230,300)
\CArc(84,32)(50,60,130)

\Oval(150,100)(30,10)(1)
\CArc(183,168)(50,230,300)
\CArc(184,32)(50,60,130)
\def\axowidth{1.5 }
\Line(140,100)(160,100)
\Line(150,70)(150,130)
\def\axowidth{0.5 }

\Oval(418,100)(30,10)(1)
\CArc(384,168)(50,240,310)
\CArc(385,32)(50,50,120)

\Oval(318,100)(30,10)(1)
\CArc(284,168)(50,240,310)
\CArc(285,32)(50,50,120)
\def\axowidth{1.5 }
\Line(308,100)(328,100)
\Line(318,70)(318,130)
\def\axowidth{0.5 }

\Vertex(226,100)2
\Vertex(234,100)2
\Vertex(241,100)2

\Text(80,50)[1]{$\< B|$}
\Text(150,50)[1]{$\< C |$}
\Text(233,50)[1]{$| C \>$}
\Text(305,50)[1]{$| B\>$}

\end{picture}
\vspace{-1cm}
\caption{One-loop partition function: tree-channel interpretation\lab{fig1l}}
\end{figure}

\noindent
There is one boundary state $| B\rangle$, one for each set of
D$p$-branes in the model, and a cross-cap state $| C \rangle$ for
each O-plane, in type I of course only the D9-branes and the
O9-plane. Formulas for their oscillator pieces were given in
\reef{oscbst} and \reef{crcap}. In this formulation, the tree- or
closed string channel, the one-loop amplitude appears as a
transition function of boundary and cross-cap states, and it forms
a perfect square. The interference of two boundary states is the
annulus, two cross-caps form the Klein bottle, and one of each
gives the M\"obius strip.

Some properties of orientifold planes find an explanation in the
one-loop diagrams: The Klein bottle and M\"obius strip have
contributions only from states that are invariant under the world
sheet parity and are localized at its fixed locus on the O-plane.
It is also evident from (\ref{treechannel}) that a cancellation of
contributions from various fields in the amplitude needs opposite
couplings to boundary and cross-cap states, to O-planes and
D-branes, which is why their RR charges are required to be
opposite.

Due to normalization issues in defining the boundary and cross-cap
states, the actual calculation of the partition function better starts from
the loop-channel. We first treat the case of ten-dimensional Minkowski
space and compactify afterwards. To evaluate (\ref{loopchannel}) we use
\beqn \lab{traces}
{\rm Tr}\, e^{-2\pi \t_2 \a' p^2} ~\equiv~ \int \frac{d^Dp}{(2\pi)^D}
e^{-2\pi \t_2\a' p^2} &=& \frac{V_D}{(8\pi^2\a' \t_2)^{D/2}} \ ,
\non
{\rm Tr}\, \exp\Big( 2\pi i\tau \Big[ \sum_{n\neq 0} \a^M_{-n} \a_{M n}
- \frac{D-2}{24} \Big] \Big) &=& \frac1{\eta(\t)^{D-2}} \ ,
\non
{\rm Tr}\zba{\a}{\beta}\, \exp\Big( 2\pi i\tau \Big[ \sum_{r\neq 0} r\psi^M_{-r} \psi_{M r}
- \frac{D-2}{48} \Big] \Big) &=&
\eta_{\a\beta} \frac{\thba{\a}{\beta}(0,\t)^{D/2}}{\eta(\t)^{D/2}}
\ .
\eeqn
The dimension is $D=10$ and $V_D$ the regularized $D$-dimensional
volume. The labels $\a,\beta$ take values $0,\frac12$ independently
and run over the four spin structures. The interpretation is
\beqn
\a = \left\{ \begin{array}{ll}
0 & {\rm trace\ over\ NS\ sector}\\
\frac12 & {\rm trace\ over\ R\ sector}
\end{array}
\right.
\ , \quad
\beta = \left\{ \begin{array}{ll}
0 & {\rm trace\ without}\ (-1)^F\\
\frac12 & {\rm trace\ with}\ (-1)^F
\end{array}
\right. \ .
\eeqn
We have further defined
\beqn\lab{etasigns}
\eta_{\a\beta} = (-1)^{2\a+2\beta+4\a\beta}\ .
\eeqn
In the tree-channel the four contribution come from the
interference of the various terms with different spin structure,
i.e.\ $\eta=\pm 1$, as follows from
\reef{GSOeta}. The same value of $\eta$ in the in- and out-state maps to the
loop-channel trace over NS fields, opposite values to the trace of
R fields, the NSNS component maps to the trace without insertion
of $(-1)^F$ the RR component to the terms with $(-1)^F$ insertion
(see e.g.\ \cite{Polchinski:1987tu,Gaberdiel:2000jr}). The
relative sign between charge and tension corresponds to the sign
of the GSO projection in the loop channel.

For the annulus the argument $\t$ in \reef{traces} is to be
replaced with $i\t_2$. To treat traces in the Klein bottle, note
that the invariant closed string states are of the type
\beqn
\O \a_n^\m \tilde\a_n^\m \O^{-1} = \a_n^\m \tilde\a_n^\m \ , \quad
\O \p_r^\m \tilde\p_r^\m \O^{-1} = \p_r^\m \tilde\p_r^\m \ ,
\eeqn
which means that one can replace $\O e^{2\pi i\t L_0}e^{-2\pi i \bar\t
\bar L_0}$ by $e^{-4\pi\t_2 L_0}$ in the Klein bottle, and only trace
over left-movers, such that $\t$ in \reef{traces} gets replaced by
$2i\t_2$ (except in the first line). For the M\"obius strip, one needs \reef{Omosc}.
The argument $\tau$ of the M\"obius amplitude then turns out to be
$\frac12 +i\t_2$ on the right-hand-side of
(\ref{traces}). We refrain from using the machinery of gamma-matrices
for the CP labels, and just introduce the multiplicities by hand.
The D9-brane boundary states simply come with an extra CP
degeneracy factor $N_{\rm D9}$.
Putting pieces together, one has for (\ref{loopchannel}) the explicit
result
\beqn \lab{1loop}
\ck &=& \frac{V_{10}}{4(4\pi^2\a')^5}\int_0^\infty \frac{dt}{t^6}
\frac1{\eta(2it)^{12}} \sum_{\a,\beta} \eta_{\a\beta}\thba{\a}{\beta}(0,2it)^4 \ ,
\\
\ca &=& \frac{V_{10}}{4(8\pi^2\a')^5} N_{\rm D9}^2 \int_0^\infty \frac{dt}{t^6}
\frac1{\eta(it)^{12}} \sum_{\a,\beta} \eta_{\a\beta}\thba{\a}{\beta}(0,it)^4 \ ,
\non
\cm &=& \pm \frac{V_{10}}{4(8\pi^2\a')^5} N_{\rm D9}
\int_0^\infty \frac{dt}{t^6}
\frac1{\eta(\frac12+it)^{12}} \sum_{\a,\beta}
\eta_{\a\beta}\thba{\a}{\beta}(0,{1/2} + it)^4 \  .
\nonumber
\eeqn
Each of these amplitudes vanishes by virtue of (\ref{jacobi}) as a
consequence of space-time supersymmetry. In order to extract the UV
divergences one has to look at the contributions from small proper
time parameters $t\rightarrow 0$ in the integrals. This is done by
performing the modular transformation that translates the loop-channel
formulation (\ref{loopchannel}) into the tree-channel
(\ref{treechannel}). It consists of replacing
\beqn \lab{modtrafo}
t \mapsto
\left\{
\begin{array}{ll}
l^{-1} & {\rm for}\ \ca \\[.1cm]
(2l)^{-1} & {\rm for}\ \ck \\[.1cm]
(4l)^{-1} & {\rm for}\ \cm
\end{array}
\right.
\ .
\eeqn
One then utilizes (\ref{Seta}), \reef{S} and \reef{SMoeb} to
recast the eta- and theta-functions as functions of $l$, leading
to\footnote{For more details see again
\cite{as02}, in particular regarding the phase factors in the M\"obius
diagrams.}
\beqn \lab{1looptree}
\tilde\ck &=& \frac{V_{10}}{4(4\pi^2\a')^5} 2^5 \int_0^\infty dl
\frac1{\eta(il)^{12}}\sum_{\a,\beta} \eta_{\a\beta}\thba{\beta}{\a}(0,il)^4 \ ,
\\
\tilde\ca &=& \frac{V_{10}}{4(8\pi^2\a')^5} N_{\rm D9}^2 \int_0^\infty dl
\frac1{\eta(il)^{12}}\sum_{\a,\beta} \eta_{\a\beta}\thba{\beta}{\a}(0,il)^4 \ ,
\non
\tilde\cm &=& \pm \frac{V_{10}}{4(8\pi^2\a')^5} 2^6 N_{\rm D9}
\int_0^\infty dl
\frac1{\eta(\frac12+il)^{12}}\sum_{\a,\beta} \eta_{\a\beta}\thba{\a}{\beta}(0,1/2+il)^4 \ ,
\nonumber
\eeqn
According to the definition of the theta-functions in
\reef{thetaprod} the upper characteristic distinguishes world
sheet fermions with either half-integer or integer mode expansion
from the NSNS or RR sectors. This identifies the integer-moded
states as the exchange of the RR sector. Expanding the integrand
for large $l$ the constant terms refer to the propagation of
massless modes which lead to divergences. If one cuts off the
integral at the upper bound by replacing $\int^\infty dl
\rightarrow \int^{\Lambda_{\rm UV}^2}dl $, these are quadratic in
the cut-off.

Adding the three amplitudes, one finds the famous tadpole cancellation
for the type I string
\beqn
\cz_{\rm 1-loop} = \frac{V_{10}}{4(8\pi^2\a')^{5}}
( 1_{\rm NSNS} - 1_{\rm RR} ) \int_0^\infty dl\, \Big[ 16 (N_{\rm D9} \pm 32)^2 + \, \cdots \Big] \ ,
\eeqn
The solution is $N_{\rm D9}=32$ and the sign in the M\"obius strip
$-1$. This sign is the phase factor that appeared in
(\ref{CPproj}) for the action of $\O$ on the massless fields in
the open string spectrum. Starting from $32$ D9-branes with a
primordial gauge symmetry $U(32)$ the sign in the open string NS
sector distinguishes between the $SO(32)$ and $Sp(32)$ subgroups
thereof, where $-1$ chooses $SO(32)$. One may think of the
condition $N_{\rm D9} -32 = N_{\rm D9} + Q_9 = 0$ as a degenerate
version of the general topological charge cancellation condition
\reef{loctad}, expressed for the case of space-time filling
charges.

It is straightforward to generalize the above calculation done for
ten-dimensional Minkowski space-time to a compactification on an
internal six-dimensional torus $\mbb T^6$ (leaving out Wilson
lines). To be general, and because this will be of relevance
later, we will also include the NSNS $B$-field. Even though it is
not in the spectrum, i.e.\ its fluctuations are frozen, it can be
non-vanishing.

In computing the diagrams in a toroidal compactifications, the
only modification of the above calculation in ten-dimensional
Minkowski space needed is to replace the first equation of
(\ref{traces}) with
\beqn \lab{KKsum}
\int \frac{d^4p}{(2\pi)^4}
e^{-2\pi \t_2 \a' p^2} {\rm Tr}\, e^{-2\pi \t_2 \a' M_{\rm KK}^2} =
\frac{V_4}{(8\pi^2 \a' \t_2)^2}
{\rm Tr}\, e^{-2\pi \t_2 \a' M_{\rm KK}^2}
\ ,
\eeqn
where $M_{\rm KK}^2$ is the mass operator for the KK modes that
propagate in the loop-channel. The
tree-channel KK mass operator we always denote $\tilde M_{\rm
KK}^2$. One needs to be careful that the KK spectra can be different for
different diagrams, and that there is an extra factor $\frac12$ in the
closed string Hamiltonian in front of $\a' p^2$.

Just for simplicity, we assume that the torus factorizes into
two-dimensional tori, $\mbb T^6=\bigotimes_{I=1}^3 \mbb T^2_I$, by
which we mean that the metric $g_{ij}$ is block-diagonal, as well
as the $B$-field.
The closed string momenta on a torus are\footnote{Note that the
momenta in \reef{KKcirc} were written in terms of an ein-bein on the
circle, i.e.\ the momenta there are $p_a = p_i {\bf e}_a^i$ with ${\bf
e}_a^i = \sqrt{\a'}/R$, etc. Also remember that $g_{ij}$ and $B_{ij}$
are dimensionless, hence no factors of $\a'$.}
\beqn
\sqrt{\a'} (p_L)_i = m_i + ( g_{ij} + B_{ij} ) n^j \ , \quad
\sqrt{\a'} (p_R)_i = m_i - ( g_{ij} - B_{ij} ) n^j \ .
\eeqn
The $(m_i,n^j)$ are a priori arbitrary integers.
In order that the theory is symmetric under $\O$ the spectra of
left- and right-moving momenta have to match. This leads to the
requirement that
\beqn
2 B_{ij}n^j \in \mbb Z \ .
\eeqn
The entries of $B$ can thus be half-integer or integer
\cite{Bianchi:1991eu,Kakushadze:1998bw,Angelantonj:1999jh,Kakushadze:1999if,ab99,Kakushadze:2000hm,bkl00}.

Invariant
states in the Klein bottle have to satisfy (\ref{omegasym}) or
$n^j = 0$. Therefore, the Klein bottle amplitude does not see the
$B$-field at all.
The mass spectrum in the Klein bottle loop-channel can then be written
\beqn \lab{KKspecKB}
\ck: \quad \a' M^2_{\rm KK} = \a' p_i g^{ij} p_j = m_i g^{ij} m_j \ .
\eeqn
The Neumann boundary conditions (\ref{bcpw}) on the contrary
require that left- and right-movers are opposite equal, $m_i =
B_{ij} n^j$. For $B_{ij}=0$ this just imposes $m_i=0$. For
$B_{ij}$ half-integer the condition leads to $n^j \in
2\mbb Z$. The closed string winding modes on the D9-brane
boundaries, $\sqrt{\a'}w_i = g_{ij} n^j$ are doubled if $B_{ij}\neq 0$.
The KK mass spectrum in the tree-channel then reads
\beqn \lab{KKann}
\tilde\ca: \quad \a' \tilde M_{\rm KK}^2 = \a' w_i g^{ij} w_j = n^i g_{ij} n^j \
,
\quad n^j \in \left\{ \begin{array}{cl}
\mbb Z & {\rm for}\ B_{ij} =0 \\
2\mbb Z & {\rm for}\ B_{ij} \neq 0 \\
\end{array}\right. \ .
\eeqn
In the open string loop-channel the mass spectrum becomes
identical to that of the Klein bottle amplitude in \reef{KKspecKB}
but up to a possible factor $\frac14$ for half-integer momentum
states. Via Poisson resummation using
(\ref{poisson}) this produces a total prefactor $2^{{\rm rk}(B)} = ({\rm rk}(B))^2$ in
the tree-channel annulus diagram. For the boundary state from
\reef{oscbst} that describes the D9-brane in this background a factor ${\rm
rk}(B)$ has to be incorporated into its overall normalization,
schematically
\beqn \lab{Bbst}
| B9\> ~\propto~  {\rm rk}(B) | B9\>_{\rm osc}\ .
\eeqn
Since the Klein bottle is invariant the cross-cap state $|C9\>$
does not carry such a factor. In the M\"obius strip this leads to
a relative factor ${\rm rk}(B)$ in the tree-channel.\footnote{See
for instance \cite{Angelantonj:1999jh} for an alternative
derivation of this effect.}

Making the replacement in (\ref{1loop}) it is
straightforward to follow through the modular transformation
(\ref{modtrafo}) and derive the analogue of (\ref{1looptree}).
The tadpole cancellation condition is only modified by the relative
factors in the M\"obius and annulus diagrams. It reads
\beqn
\cz_{\rm 1-loop} = \frac{V_{10}}{4(8\pi^2\a')^{2}}
( 1_{\rm NSNS} - 1_{\rm RR} ) \int_0^\infty dl\, \Big[ 16 ( N_{{\rm
D}9}\,  {\rm rk}(B) \pm 32)^2 + \, \cdots \Big] \ .
\eeqn
The solution is $N_{\rm D9}=32/{\rm rk}(B)$ with a gauge group
$\cg=SO(32/{\rm rk}(B))$ in the supersymmetric case. In this
manner, the discrete NSNS $B$-field can reduce the rank of the gauge
group \cite{Bianchi:1991eu}.

Let us also briefly discuss the additional elements that appear when
D5-branes and O5-planes are included. Of course, D5-branes follow from
D9-branes by T-duality along a $\mbb T^4$ inside the $\mbb T^6$. Such
a T-duality transforms $\O$ into $\O\Th$ where $\Th$ is the reflection
along the four directions of the $\mbb T^4$. Such an operation appears
in every orientifold compactification on a $\mbb T^6/\mbb Z_{2N}$
where $\Th$ is the $N$-th power of any generator of $\mbb Z_{2N}$. Any
such model has $16=2^4$ O5-planes along the fixed $\mbb T^2$ of $\O\Th$ and
D5-branes along the same torus and at arbitrary points on $\mbb T^4$.

Every amplitude now comes with an insertion of the orbifold
projector onto invariant states in the loop-channel, which at
least involves an insertion of $\Th$. Let us ignore the other
insertions for the moment. The Klein bottle amplitude with the
insertion of $\Th$, call it $\ck_\Th$, is, of course, T-dual to
the Klein bottle with the identity. The states invariant under
$\O$ are also invariant under $\Th$, except for KK modes. Here the
condition $w_i=0$ is replaced by $p_i=0$ for directions along the
$\mbb T^4$, and thus the relevant piece of the KK sum in $\ck_\Th$
has winding instead of momentum modes. In the presence of a
$B$-field it behaves like the KK sum
\reef{KKann} in $\tilde\ca_{99}$. It will give rise to an extra
factor $2^{-{\rm rk}(B|_{\rm \mbb T^4})}$ in the tree-channel
diagram, $B|_{\mbb T^4}$ the $B$-field restricted to the ${\mbb
T}^4$. This factor has to be included into the normalization of
the cross-cap state of the O5-plane,
\beqn
|C5\> ~\propto~  \frac1{{\rm rk}(B|_{\mbb T^4})} | C5\>_{\rm osc}\ ,
\eeqn
whose charge and tension is thus diminished in the presence of a
$B$-field. Analogously, one finds that the 55 annulus diagram
$\tilde \ca_{55}$ has KK modes with $n^j=0$ along Dirichlet
directions and is blind to the $B$-field there. Therefore, the
extra normalization factor in $|B5\>$ only depends on the
$B$-field along the Neumann directions,
\beqn
|B5\> ~\propto~  {\rm rk}(B|_{\mbb T^2}) | B5\>_{\rm osc}\ ,
\eeqn
as opposed to \reef{Bbst}. Following this reasoning, the tadpole cancellation
condition for the RR charge of O5-planes and D5-branes reads
\beqn \lab{tadD5}
N_{\rm D5} \pm \frac{32}{{\rm rk}(B)} = 0 \ .
\eeqn
Of course, one could also add D5-branes to a given model without
O5-planes, in which case the tadpole cancellation is $N_{\rm
D5}=0$. Branes and anti-branes are needed in equal numbers and
supersymmetry is broken. The sign $-1$ for the M\"obius strip
contribution turns out to imply an opposite projection for the CP
label of 55 open strings compared to 99 strings, namely $\l_{ij}^A
= \l_{ji}^A$ \cite{Gimon:1996rq}. The gauge group on a stack of
D5-branes is then $Sp(N_{\rm D5})$.

We have so far ignored all other contributions to the partition
function present on $\mbb T^6/\mbb Z_{2N}$ orientifolds, other
insertions of elements of the orbifold group in the Klein bottle,
and the other terms of the open string diagrams. They turn out to
produce contributions to the exchange of twisted closed string
fields in the tree-channel. The tadpole cancellation conditions
for these twisted tadpoles are often expressed in terms of
conditions on the gamma-matrices that are introduced according to
\reef{gammas}. We will not go through this, and instead later
interpret these conditions as coming from the so-called
exceptional divisors of the orbifold background.


\subsubsection{The toroidal orientifolds on $\mbb T^6/(\mbb Z_2\times
\mbb Z_2)$}
\lab{secZ22}

In the course of this review we will frequently make use of a specific
simple set of Calabi-Yau orientifolds, namely the type II
orientifolds built on the toroidal orbifold background
\beqn
\cx = \frac{\mbb T^6}{\mbb Z_2 \times \mbb Z_2}\ .
\eeqn
It is defined by the two generators of the $\mbb Z_2$ denoted $\Th_1$
and $\Th_2$, such that $G= \mbb Z_2 \times \mbb Z_2 = \{
1,\Th_1,\Th_2,\Th_1\Th_2\}$. Their geometric operation on the bosonic world
sheet coordinates $Z^I = x^{2I-1}+u^I y^{2I},\, I=1,2,3,$ (the
so-called point group) is given by
\be\lab{reflZ22}
\Theta_1 :\cases{ Z^1\mapsto -Z^1 \cr
                 Z^2\mapsto -Z^2 \cr
                  Z^3\mapsto +Z^3 \cr }\ , \quad\quad\quad\quad
\Theta_2:\cases{ Z^1\mapsto +Z^1 \cr
                 Z^2\mapsto -Z^2 \cr
                  Z^3\mapsto -Z^3 \cr }\ .
\ee
States in the spectrum of the type II theory on the $\mbb T^6$ are
then projected onto invariant states. In addition, the orbifold
theory contains twisted sectors satisfying periodicity conditions
\reef{twper} for some element of the orbifold group $G$. Since each
generator of $G$ has $2^4=16$ fixed two-dimensional tori there are
$48$ such twisted sectors with fields localized along the fixed loci.

The invariant untwisted fields are just the parameters in the
diagonal $2\times 2$ blocks of the six-dimensional metric, plus
diagonal components of the NSNS two-forms and the invariant
components of the various RR forms. We will be more specific on
how to parameterize the two-dimensional torus metric in section
\ref{secinttor}. Here we only note that there are three complex
structure moduli $U^I$ and three (complexified) K\"ahler moduli
$T^I$ in addition to the scalar $S$ that involves the
dilaton.\footnote{We use capital $T^I$ and $U^I$ to denote the
properly defined and normalized scalars in the chiral multiplets
and $t^I$ and $u^I$ for the naked geometric quantities of the
tori.} The explicit definitions of the fields in terms of
geometric quantities depend on the concrete model. This spectrum
also follows directly from the number of invariant harmonic two-
and three-forms to be $h^{1,1}_{\rm untw}=h^{2,1}_{\rm untw}=3$.
The K\"ahler potential for the fields $S,\, U^I,\, T^I$ can be
deduced from explicit dimensional reduction \cite{seealso}
\beqn\lab{KpotZ22}
\k_4^2 K = -\ln(S+\bar
S)-\sum_{I=1}^3\ln(T^I+\bar T^I)-\sum_{I=1}^3\ln(U^I+\bar U^I)\ .
\eeqn
This indicates that the seven untwisted scalars parameterize the
space $(SU(1,1)/U(1))^7$. It arises as remnant of the moduli space
of the $\mbb T^6$ (with dilaton) which is $SU(1,1)/U(1)\times
SO(6,6)/(SO(6)\times SO(6))$.

To construct orientifolds from this type II orbifold
compactifications let us briefly discuss the issue of discrete
torsion \cite{Vafa:1994rv,Klein:2000qw}. In the CFT definition of
the type II orbifold discrete torsion refers to the freedom to add
extra phase factors to the sum over insertions of elements of $G$
and over twisted sectors in the total partition function.
Effectively, there appears a freedom of a single sign choice
$\e=\pm1$ in this orbifold, $\e=1$ meaning no discrete torsion.
The choice of sign leads to relative signs in the projection onto
invariant twisted sector contributions. Consistency implies that
still all twisted sectors are treated with the same relative sign,
such that either the states of a vector multiplet or a
hypermultiplet survive as invariant. In other words, depending on
the presence or absence of discrete torsion the Hodge numbers of
the compactification receive a contribution
$(h^{1,1},h^{2,1})=(1,0)$ or $(h^{1,1},h^{2,1})=(0,1)$ from each
twisted sector.

In defining the orientifold one has the further freedom of
choosing the orientifold projection $\O\bs$ in IIA or $\O\s$ in
IIB. The fixed loci of the elements of $\s G$ define the location
of O9- and O5-planes, that of $\bs G$ O6-planes only. Their
charges are cancelled by adding the D-branes of appropriate
dimensions in the standard approach. Tadpole cancellation now
relates the sign $\e=\pm1$ to minus the product of the charges
(normalized to $\pm1$) of the three O-planes located at $\s\Th_1$,
$\s\Th_2$ and $\s\Th_1\Th_2$, and similar in the case of IIA.
Thus, without discrete torsion one is forced to add D5-branes and
anti-D5-branes at the same time and supersymmetry is broken in the
spectrum. Only discrete torsion allows to construct a
supersymmetric type II orientifold on the $\mbb Z_2\times \mbb
Z_2$ orbifold. We summarize the situation in table
\ref{tabZ22models}. For more details of this analysis see
\cite{Klein:2000qw}.

\begin{table}[h]
\centering
\begin{tabular}{|c|c|c|}
\hline
&& \\[-.45cm]
Type IIA models & Type IIB models & Supersymmetry \\
\hline
&& \\[-.45cm]
$(h^{1,1},h^{2,1})=(51,3)\, ,~~~~\e=+1$ &
$(h^{1,1},h^{2,1})=(3,51)\, ,~~~~\e=-1$ & $\cn=1$ \\[.05cm]
$(h^{1,1},h^{2,1})=(3,51)\, ,~~~~\e=-1$ &
$(h^{1,1},h^{2,1})=(51,3)\, ,~~~~\e=+1$ & $\cn=0$ \\[.05cm]
\hline
\end{tabular}
\caption{Type II orientifolds on $\mbb T^6/(\mbb Z_2\times \mbb Z_2)$
\lab{tabZ22models}}
\end{table}

The maximal gauge group in the supersymmetric models is $Sp(16)^4$
each factor supported on a stack of D9- or D5-branes in IIB and
D6-branes in IIA. When general intersecting D6-branes or magnetic
background fluxes are admitted, as we will discuss in the next
section, the constraints of table \ref{tabZ22models} can be lifted in
that there are supersymmetric solutions to the tadpole cancellation
conditions for either choice of sign $\e=\pm1$. This allows to work
with either set of Hodge numbers in this orbifold model.

The orientifold $\mbb T^6/(\mbb Z_2\times \mbb Z_2)$ is also a
good example to illustrate how the factorization of the type II
moduli space with $\cn=2$ supersymmetry is broken when open string
moduli are added, as discussed in section \ref{sugrapot}. In the
standard orientifold of IIB with $\s$ the identity the orientifold
involves D9-branes and D5-branes. Concentrating on the Wilson line
moduli $A^I$ of the D9-branes\footnote{Since the background space
is a Calabi-Yau there are no actual non-contractible one-cycles
and thus no Wilson lines. Nevertheless, the open string moduli are
often referred to as Wilson lines. More accurately, we could call
them bundle moduli.} and suppressing a label for the individual
D-brane stacks, it is also known from dimensional reduction how
the K\"ahler potential gets modified. The direct product structure
of the type II compactification from \reef{KpotZ22} is lost and
the potential is modified to the form given in \reef{finalD9}.
The modification is due to a redefinition \reef{redT} of the
proper K\"ahler coordinates and
also lifts the no-scale structure of the scalar
potential in type II.


\subsection{Intersecting and magnetized D-branes in orientifolds}\label{sec23}

After briefly introducing the foundations of D-branes and
orientifold models we now turn to the subject of intersecting and
magnetized D-branes. D-branes at angles or magnetized D-branes are
interesting ingredients to model building within string theory,
because they allow to solve the chirality problem in a simple and
very geometrical way. We will study the simplest prototypes here
first, D-branes in flat Minkowski space and on (factorized) tori.


\subsubsection{General boundary conditions and chirality}
\lab{secbcangles}

As a starting point to set up the notion of intersecting and
magnetized D-branes \cite{bdl96} we like to consider the simplest
configuration, which is a D-brane in a flat two-dimensional plane
(or on a two-dimensional torus). Either the brane is extended over
the two-dimensional plane or it is a one-dimensional submanifold
of the plane. The situation when it is point-like will arise as a
limiting case of the first option.

Starting with a D-brane that covers the plane, this is described by
the boundary conditions (\ref{NDbc}) which one can also write
\beqn
( g_{ij} + 2\pi \a' \cf_{ij} ) \pa_\t X_L^j +
( g_{ij} - 2\pi \a' \cf_{ij} ) \pa_\t X_R^j = 0 \Big|_{\pa\S}
\ .
\eeqn
By introducing a (constant) world volume zwei-bein $e_a^j$, $a,b=1,2$,
$i,j=1,2$, and world volume
coordinates $X^a = e^a_j X^j$ one can rewrite this
\beqn
( \eta_{ab} + 2\pi\a' \cf_{ab} ) \pa_\t X_L^b +
( \eta_{ab} - 2\pi\a' \cf_{ab} ) \pa_\t X_R^b = 0 \Big|_{\pa\S}
\ ,
\eeqn
with $\eta_{ab} = e_a^i g_{ij} e^j_b$ the flat metric and
$\cf_{ab} = e_a^i \cf_{ij} e^j_b$. One can divide by the
determinant and obtain a rotation matrix $\Th_{ab}$,
\beqn \lab{rotmatrix}
\Th_{ab} = \frac{\eta_{ab}+2\pi\a' \cf_{ab}}{{\rm det}(\eta+2\pi\a' \cf)} \ .
\eeqn
The boundary conditions read
\beqn \lab{Thbc}
\pa_\t ( \Th_{ab} X_L^b + \Th^{-1}_{ab} X_R^b) = 0 \Big|_{\pa\S}\ .
\eeqn
We write $\Th_{ab}$ in terms of an angle $\vf$,
\beqn \lab{22rot}
\Th_{ab} = \left( \begin{array}{cc}
\cos(\vf) & \sin(\vf) \\
-\sin(\vf) & \cos(\vf) \end{array}
\right)_{ab} \ ,
\eeqn
where
\beqn \lab{Fphi}
2\pi\a' \cf_{ab} = \tan(\vf) \e_{ab}\ , \quad \e_{12}=-\e_{21}=1\ .
\eeqn
This shows that the presence of the world volume gauge field
background $\cf_{ab}$ can be interpreted as an asymmetric rotation
of the world sheet coordinates, where left- and right-moving modes
are treated with opposite rotation angles \cite{bgkl00}. The two limiting
cases of pure Neumann or pure Dirichlet boundary conditions are
\beqn
{\rm N:} \quad \cf_{12} = 0 ~\Leftrightarrow~ \vf=0 \ , \quad
{\rm D:} \quad \cf_{12} = \infty ~\Leftrightarrow~ \vf=\frac\pi2 \
.
\eeqn
Of course, the angle $\vf$ is only defined modulo $\pi$ here,
$-\frac\pi2 <\vf \leq \frac\pi2$.

The other situation of a one-dimensionally extended brane, a
straight line, can be obtained by T-duality from (\ref{Thbc}). A
T-duality along, say, $X^1$ reflects $X_R^1$ which can be
equivalently described by replacing (\ref{Thbc}) by
\beqn \lab{Thbc2}
\pa_\s \Th_{1b} X^b = \pa_\t \Th_{2b} X^b = 0 \Big|_{\pa \S}\ ,
\eeqn
which is just a rotation of ND boundary conditions along $X^1$ and
$X^2$ through an angle $\vf$.

Now consider the intersection of two intersecting D6-branes in
flat space-time, filling out a common 3+1 subspace. For simplicity
assume that the rotation matrices that describe their location in
the six-dimensional transverse space are both block-diagonal at
the same time. Each D6-brane is then described by three rotation
angles $\vf^I_a$, where $a=1,2$ distinguishes the two individual
branes, and $I=1,2,3$. As is explained in appendix
\ref{appinters}, the quantization of the open string modes that
live on strings stretching between the two branes differs from
that of open strings with both ends on either brane in important
ways \cite{Abouelsaood:1986gd,bdl96}. The mode expansion of the
world sheet fields along the plane labelled by $I$ gets shifted by
\beqn \lab{delAB}
\d_{ab}^I = \frac{\vf_{ab}^I}{\pi} =
\frac{\vf_{a}^I-\vf_{b}^I}{\pi}\ .
\eeqn
In particular, there are no longer any fermionic zero-modes in the
R sector along the internal space directions. The Clifford algebra
generated by the R zero-modes is thus restricted to the 3+1
directions along the the common world volume. Taking the
GSO-projection into account removes one four-dimensional
chirality, leaving two degrees of freedom, denoted $|\a\>_{\rm R}
= |s_0,s_1\>_{\rm R}$ with $s_0=\pm s_1$ with a fixed relative
sign depending on the chirality. The R groundstate is thus a
single chiral, say left-handed, Weyl fermion,
\beqn\lab{RRchiral}
{\rm R\ ground\ state:} \quad \chi^A_\a \l^A_{ij}| \a,ij\>_{\rm R}\ , \quad
\a = 1,2\ ,
\eeqn
where we have reinstalled the CP labels for intersecting stacks of
multiple D6-branes. Since it connects two different stacks of
D-branes the state transforms in the bifundamental representation
$(\fund_a,\bar{\fund}_b)$ or $(\fund_a,\fund_b)$ of the gauge
group factors on the two stacks.\footnote{In more general
situations this can also be a self-intersection of a single
stack.} This shows how intersecting D-branes can lead to chiral
spectra in non-abelian gauge groups. The same holds for branes
with constant magnetic world volume flux via T-duality. This
observation is the starting point for model building with magnetic
background fields \cite{CB95}.


\subsubsection{Supersymmetry}
\lab{kappa}

The concept of supersymmetry in orientifolds has two aspects, the
supersymmetry of the background bulk theory, i.e.\ the type II
theory supergravity subject to the orientifold projection, and the
supersymmetry of the world volume theory on the D-branes. We will
only be dealing with compactifications on Calabi-Yau spaces which
guarantees $\cn=1$ supersymmetry in the bulk.\footnote{As remarked
earlier, we will ignore the effects of backreaction for most of
what we have to say.}

Regarding the supersymmetry of the background type II theory, let
us only briefly repeat some well known statements about Calabi-Yau
compactifications. The starting point is the ten-dimensional type
II theory with $D=10$, $\cn=2$ supersymmetry. In the absence of
expectation values for the NSNS and RR field strengths and for
constant dilaton, the supervariations of the two gravitinos
(both denoted collectively by $\psi_M$) simply read
\beqn
\d \psi_M = \nabla_M\e\ ,
\eeqn
where $\nabla_M$ is the covariant derivative of the Levi-Civita
connection. The full ten-dimensional metric is the direct product
metric from \reef{metricansatz} with $\D(x^i)=1$. The internal
metric is the Ricci-flat Calabi-Yau metric. The orientifold
projection in any case flips the NSR and RNS sectors of the closed
string spectrum, and thus identifies the two gravitinos projecting
to $D=10$, $\cn=1$ supersymmetry. The existence of the
ten-dimensional constant spinor $\e$ follows from $\d\psi_M =0$.
It guarantees an internal six-dimensional constant spinor
$\eta_\pm$,
\beqn \lab{Killspin}
\nabla_i \eta_\pm = 0 \ .
\eeqn
We will be more explicit about the notation $\eta_\pm$ in section
\ref{subsubsecsusy}. Using this spinor, one can construct the K\"ahler form $J_2$,
\beqn \lab{J2}
(J_2)_{ij} ~\propto~ \eta_\pm^\dag \G_{ij} \eta_\pm \ ,
\eeqn
and the holomorphic three-form $\O_3$,
\beqn \lab{O3}
(\O_3)_{ijk} ~\propto~ \eta_-^\dag \G_{ijk} \eta_+ \ .
\eeqn
The gamma-matrices are anti-symmetrized products of
six-dimensional gamma-matrices and the factors of proportionality
will be fixed later. Both forms are closed,
\beqn\lab{OmJclosed}
d\O_3 = dJ_2 =0\ ,
\eeqn
as a consequence of (\ref{Killspin}). With respect to the complex
structure of the Calabi-Yau, they are of Hodge type $(3,0)$ and
$(1,1)$. With this property, $J_2$ and $\O_3$ can be used as
defining quantities for the Calabi-Yau as well. The condition
\reef{OmJclosed} makes the internal space a complex K\"ahler
manifold with $SU(3)$ holonomy. So much about the supersymmetry
conditions for the background geometry, let us turn to the
D-branes.

For the world volume theory on a D-brane the existence of unbroken
supersymmetry relies on the existence of another local symmetry,
the so-called $\k$-symmetry
\cite{Cederwall:1996pv,Aganagic:1996pe,Cederwall:1996ri,Bergshoeff:1996tu,Aganagic:1996nn,Bergshoeff:1997kr}.
The simultaneous supersymmetry and
$\k$-symmetry variation of a spinor $\th$ on the brane world
volume reads
\beqn
\d \th = (1+\G) \k + \e \ ,
\eeqn
where $\k$ and $\e$ are the parameters of $\k$-symmetry and
supersymmetry, functions of the world volume and space-time
coordinates. $\G$ satisfies $\G^2=1$ and will be defined in a
moment. From $\d\th=0$ it follows
\beqn \lab{branesusy}
( 1-\G )\e =0 \ ,
\eeqn
which is the defining property of world volume supersymmetry for a
D-brane. Note that the background dependence with respect to the
bulk fields enters via $\e$ which (up to normalization) is the
bulk Killing spinor $\eta_\pm$ from (\ref{Killspin}). The world
volume fields on the brane, in particular the gauge field strength
$\cf_{ab}$ and the pull-back $g_{ab}$ of the background metric,
enter via $\G$. There are a number of ways to write $\G$. The most
suited formulation for us is
\cite{Bergshoeff:1997kr}
\beqn
\G = e^{-\frac12 y} \G_{(0)}' e^{\frac12 y} \ , \quad
y = \left\{ \begin{array}{ll}
- Y_{ab} \g^{ab} \G_{11} & {\rm for\ IIA} \\
  Y_{ab} ( \s_3\otimes \g^{ab} ) & {\rm for\ IIB}
\end{array} \right.
\eeqn
with
\beqn
\G_{(0)}' = \left\{\begin{array}{ll}
(\G_{11})^{(p-2)/2} \G_{(0)} & {\rm for\ IIA} \\
\s_3^{(p-3)/2} \s_2 \otimes \G_{(0)} & {\rm for\ IIB}
\end{array}\right.
\eeqn
and
\beqn
\G_{(0)} = \frac1{(p+1)! \sqrt{g}} \e^{m_1...m_{p+1}}
\g_{m_1...m_{p+1}} \ , \quad
\g_m = E^M_m \G_M \ .
\eeqn
The $\G_M$ are ten-dimensional gamma-matrices, $M=0,...,9$, the
$\g_m$, $m=0,...,9$, are defined by pulling back with the
ten-dimensional viel-bein, and $\g_a$, $a=0,...,p$, are the world
volume gamma-matrices $\g_a = e^m_a\g_m$. The anti-symmetric
two-form $Y_{ab}$ can be put into block-diagonal shape, when it
becomes
\beqn \lab{phidiag}
Y = \bigoplus_{I=1}^3 \left( \begin{array}{cc}
0 & \vf^I \\ - \vf^I & 0
\end{array}\right) \ .
\eeqn
The three angles $\vf_I$ are defined by the magnetic field
strength $\cf_{ab}$ as in (\ref{Fphi}), i.e.\
\beqn \lab{Fdiag}
2\pi\a' \cf = \bigoplus_{I=1}^3 \left( \begin{array}{cc}
0 & 2\pi\a'  \cf_I \\ -2\pi\a'  \cf_I & 0
\end{array}\right)
= \bigoplus_{I=1}^3 \left( \begin{array}{cc}
0 & \tan(\vf^I) \\ -\tan(\vf^I) & 0
\end{array}\right) \ .
\eeqn
This is a heavy machinery, but one can learn a number of things
without doing much.

Consider constant $\cf_{ab}$ and $g_{ab}$. For a single D-brane
one can always remove the field-dependence from $\G$ via
\beqn
\e' = e^{\frac12 y} \e\ ,
\eeqn
and multiply (\ref{branesusy}) by $e^{-\frac12 y}$. Then, $\G$
gets replaced by the constant $\G_{(0)}'$ while (\ref{Killspin})
is not affected. This implies that on a single supersymmetric
D$p$-brane in a type II theory constant background fields
strengths always preserve supersymmetry. Examples are a D-brane on
flat ten-dimensional Minkowski space-time or compactified on a
torus, and with constant magnetic field on its world volume.

Non-trivial conditions arise when two or more (stacks of) D-branes
are present simultaneously. For any set of two branes, labelled by
$a,b$,\footnote{The reader should try not to confuse the labels
for the D-brane stacks and the Lorentz indices for the world
volume coordinates.} the conditions
(\ref{branesusy}) can be put into the form
\beqn
\Big( 1-\G'_{(0)} \Big) \e = \Big( 1- e^{-\frac12 y_{ab}} \G_{(0)}'
e^{\frac12 y_{ab}} \Big) \e = 0 \ ,
\eeqn
where we have redefined $\e$ as above, and $y_{ab}$ is defined
similar to $y$ but with $\vf^I$ replaced by $\vf_{ab}^I =
\vf^I_a-\vf^I_b$ in (\ref{phidiag}). Since $y_{ab}$ and
$\G_{(0)}'$ anti-commute, this just means that
\beqn
e^{y_{ab}} \e = \exp\left( i \sum_{I=1}^3 2s_I \vf^I_{ab}
\right) \e = \e \ .
\eeqn
The state $\e = |s_1,s_2,s_3\>$ is labelled by the weights
$s_I=\pm\frac12$ of the six-dimensional spinor representation, the
eigenvalues of the generators of rotations. Conditions for
unbroken supersymmetry are then
\beqn
\sum_{I=1}^3 s_I \vf^I_{ab} = 0\ {\rm mod}\ \pi\ ,
\eeqn
or
\beqn \lab{anglesusy}
\pm~ \vf^1_a \pm \vf^2_a \pm \vf^3_a = \th \ {\rm mod}\ 2\pi\ ,
\eeqn
for all $a$ with fixed $\th$. To convert this into a condition on
$\cf$ one can use some trigonometry to get (we suppress the label $a$)
\beqn \lab{fluxsusy}
{\sum_{I=1}^3 2\pi\a'  \cf_I - \prod_{I=1}^3 (2\pi\a'  \cf_I)}
= \tan(\th) \Big[ 1 - \sum_{I=1}^3
\d^{IJK} (2\pi\a'  \cf_J)( 2\pi\a'  \cf_K) \Big] \ ,
\eeqn
using
\beqn\lab{dIJK}
\d^{IJK} = \left\{\begin{array}{ll}
1 & {\rm for}\ I\neq J\neq K\neq I \\
0 & {\rm else}
\end{array}\right. \ .
\eeqn
To summarize, any pair of branes conserves mutually supersymmetry
if (\ref{anglesusy}) is satisfied for both branes with the same
$\th$ and some combination of signs. The number of supercharges
preserved is given by the number of sign choices allowed. This
classification of supersymmetric D-brane configurations is equally
valid for D-branes with constant world volume magnetic fields
background or for D-branes intersecting at relative angles (or
D-branes with magnetic field and angles)  via the T-duality that
was discussed in section \ref{secbcangles}. It applies to flat
Minkowski space or to a (factorized) torus.

For non-constant world volume fields non-trivial conditions
necessarily arise already for single D-branes, since
(\ref{Killspin}) forbids a field-dependent rescaling of $\e\propto\eta_\pm$. One
then can make use of the properties (\ref{J2}) and (\ref{O3}) of
the Killing spinor and replace gamma-matrices acting on $\e$ by
the geometric quantities $\O_3$ and $J_2$. The conditions that
follow from (\ref{branesusy}) are in general called calibrations.

Since we will be interested in D6-branes of IIA on
three-cycles\footnote{There is actually another possibility to
have a supersymmetric brane configuration in IIA on a Calabi-Yau,
namely a so-called co-isotropic D8-brane (see \cite{Font:2006na}
in this context). It wraps a trivial five-cycle and is stabilized
by a world volume gauge field.} and in D$p$-branes of IIB with
$p=\,$3, 5, 7, 9 on zero-, two-, four-, and six-cycles in
Calabi-Yau orientifold compactifications, let us quote the results
for these cases from
\cite{Marino:1999af}. For a D6-brane wrapping a three-manifold the
calibration condition is
\beqn \lab{susyA}
f^* \O_3 = e^{i\th} d{\rm vol}|_3 \ , \quad
\cf = 0 \ .
\eeqn
The pull-back of the holomorphic three-form has to be equal to the
volume form $d{\rm vol} = \sqrt{g} d^6x$ restricted to the
three-manifold up to a phase factor, and the gauge bundle flat.
This makes $\O_3$ a calibration form for supersymmetric
three-manifolds which are called special Lagrangian (sLag)
manifolds. For the IIB D$p$-branes on $(2q)$-manifolds one has
\beqn \lab{susyB}
\exp ( f^*J_2 + 2\pi i \a' \cf ) |_{2q} = e^{i\th} \sqrt{{\rm det}(
\eta + 2\pi \a' \cf )} d{\rm vol}|_{2q} \ , \quad
\cf^{(2,0)} = 0 \ .
\eeqn
Here $\eta_{ab} +2\pi\a' \cf_{ab}=e_a^i(g_{ij}+2\pi \a' \cf_{ij})e^j_b$ is
expressed via the world volume viel-bein $e_a^i$, as before. The
calibration form is $\exp(J_2 + 2\pi i \a' \cf)$ which on a supersymmetric
cycle equals the argument of the DBI action (\ref{dbi}) up to its
dilaton prefactor and the phase. In addition, the world volume
gauge field strength is restricted to Hodge type $(1,1)$. Note
also that the volume form of the entire Calabi-Yau is
\beqn
d{\rm vol} = \frac1{3!} J_2^3\ .
\eeqn
Even-dimensional calibrated submanifolds with $\cf=0$ are holomorphic
curves or holomorphic planes (divisors).

Since the calibration forms are always closed (which is part of
their definition) the volume of a calibrated submanifold is
already specified by its homological class. This is why one mostly
talks about wrapping a D-brane on a supersymmetric $q$-cycle
$\S_q\in H_q(\cx)$ (instead of a calibrated submanifold of dimension $q$).

So far, all the calibration conditions leave a $U(1)$ ambiguity
that allows to choose the overall angle $\th$ as a free parameter.
In orientifold models this parameter is fixed by the presence of
orientifold planes. This is another way of saying that the D-brane
configuration have to be symmetric under the action of the dressed
world sheet parity $\O\s$ or $\O\bs$. Whenever there is an
orientifold plane present the D-branes have to satisfy the
calibration conditions with the same phase as the orientifold.
Since the orientifold has no world volume gauge fields, it will
always either wrap sLag or holomorphic submanifolds calibrated
with $\O_3$ or $\exp(J_2)$ for some value of $\th$ depending on
conventions. This then fixes the ambiguity.

In this section we have presented the supersymmetry condition from
a geometrical point of view using the $\k$-symmetry of the DBI
action. This relies on the assumption that classical geometry is
applicable and corrections due to quantum effects or higher
derivative terms are small. Whenever a CFT description of a given
model is available one can also define the supercharges on a
microscopic level by operators of the CFT, or derive the the
supersymmetry conditions in other ways. In the following section
\ref{secangle1loop} we compute the one-loop partition function for
toroidal models and show that it vanishes when the calibration
conditions are satisfied. Since the torus is flat, geometrical and
CFT methods lead to the same conclusions because corrections are
absent.

The calibration conditions derived above are conditions on the
moduli parameters of the background Calabi-Yau and the D-branes
that enter the effective action as scalar fields. The consistency
of the four-dimensional effective action requires that
supersymmetric D-brane configurations in a Calabi-Yau
compactification to four-dimensional Minkowski space correspond to
minima of the effective potential with vanishing vacuum energy.
Therefore, one should be able to derive the above conditions also
from minimizing the potential of the relevant moduli scalars. We
will later see how this works out.


\subsubsection{Intersecting and magnetized D-branes on tori}
\lab{secinttor}

In this section we incorporate relative angles of intersecting
branes or magnetic background fields into the simplest case of
compactified D-branes, namely toroidal compactification.

Let us start with a two-dimensional torus, the easiest model which
captures many features of higher-dimensional tori (but also misses
some of their structure).

\begin{figure}[ht]
\vspace{1cm}
\hspace{1cm}
\begin{picture}(200,200)(0,0)

\DashLine(50,50)(320,50)5
\DashLine(50,100)(320,100)5
\DashLine(50,150)(320,150)5
\DashLine(50,200)(320,200)5

\DashLine(70,30)(134,220)5
\DashLine(150,30)(214,220)5
\DashLine(230,30)(294,220)5

\def\axowidth{1.5 }
\LongArrow(77,50)(157,50)
\LongArrow(77,50)(94,100)
\def\axowidth{0.5 }

\Vertex(77,50)3
\Vertex(287,200)3
\Vertex(253,100)3

\def\axowidth{1.5 }
\Line(77,50)(287,200)
\Line(130,50)(166,75)
\Line(85,75)(121,100)
\Line(104,50)(174,100)

\Line(77,50)(253,100)
\Line(85,75)(174,100)
\def\axowidth{0.5 }

\Text(73,92)[1]{$\vec{\bf e}_2$}
\Text(110,42)[1]{$\vec{\bf e}_1$}

\end{picture}
\vspace{-1cm}
\caption{Two-dimensional torus $\mbb T^2 = \mbb C/\L^2$ with two intersecting D-branes\lab{T2}}
\end{figure}
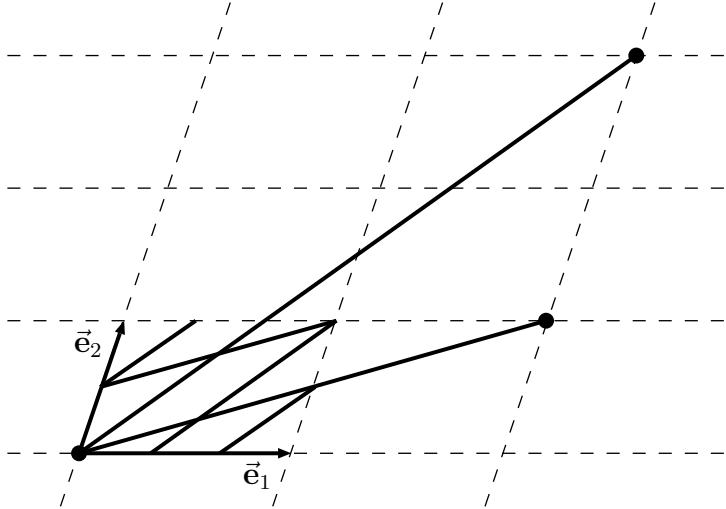

A complex torus is defined by three real parameters, the complex
structure and the volume. The latter is complexified as well by
the component of the NSNS $B$-field. Two lattice vectors $\vec{\bf
e}_i$ form the basis of the two-dimensional lattice $\L^{(2)}$
that defines $\mbb T^2$ via $\mbb T^2 = \mbb C/\L^{(2)}$. A sketch
of this, together with two intersecting D-branes of co-dimension
one is shown in figure \ref{T2}. We write the zwei-bein in
components as
\beqn
\vec{\bf e}_1 = \frac1{\sqrt{\a'}}(0, R_1) \ , \quad
\vec{\bf e}_2 = \frac1{\sqrt{\a'}}(R_2 \sin(\th) , R_2 \cos(\th) ) \ .
\eeqn
They define the parameters\footnote{Note again that $u$ and $t$
are not the K\"ahler coordinates that appear in the effective
action as bosonic components of chiral multiplets. These are
denoted $U$ and $T$ (with indices) and are obtained after
including factors of the string coupling, proper normalization,
and in our conventions, a flip of imaginary and real part.}
\beqn\lab{cplstr}
u = \frac{R_2}{R_1} e^{i\th} \ , \quad
{\rm Im}\,  t = \frac1{\a'} \sin(\th) R_1 R_2 \ ,
\eeqn
and the dimensionless metric
\beqn\lab{tormetric}
g_{ij} = \vec{\bf e}_i\cdot \vec{\bf e}_j = \frac{{\rm Im}\,
t}{{\rm Im}\, u} \left(
\begin{array}{cc}
1 & {\rm Re}\, u \\
{\rm Re}\, u & |u|^2
\end{array} \right) \ .
\eeqn
The complex K\"ahler parameter is
\beqn\lab{Kahmod}
t = b + i \frac1{\a'} \sin(\th) R_1 R_2 \ ,
\eeqn
where $b$ is the component of $B_{ij}$ along the $\mbb T^2$.

D-branes of interest are now either of co-dimension one or wrap
the entire torus. In the latter case they can carry magnetic
background fields. Let us first look at the first type.
Topologically, co-dimension one branes are described simply by the
homology class of the one-cycle they are wrapped on. If we denote
the classes that are generated by the two lattice vectors
$\vec{\bf e}_1$ and $\vec{\bf e}_1$ by $e_1$ and $e_2$, a general
one-cycle $\Pi_a \in H_1(\mbb T^2,\mbb Z)$ is just a linear
combination with integer coefficients
\beqn
\Pi_a = p_a e_1 + q_a e_2 \ .
\eeqn
Whenever we just talk about a single brane we drop the label $a$.
In principle $p$ and $q$ can be any integers. But whenever the two
are not co-prime, say $p = n_1r$ and $q=n_2r$, the single brane
configuration would decay into a stack of $r$ D-branes with winding numbers
$(n_1,n_2)$ and gauge symmetry $U(r)$ instead of $U(1)$.
Therefore, we require the two co-prime from the start,
$(p,q)=(1)$. The number of intersection points (counted with
orientation) of two such branes is
\beqn \lab{int1}
I_{ab} = \Pi_a \circ \Pi_b = p_a q_b - q_a p_b \ .
\eeqn
Of course, this relies on $e_1\circ e_2 = 1$ and the antisymmetry of
the intersection pairing. An example with two D-branes with winding numbers $(2,1)$ and
$(2,3)$ is drawn in figure \ref{T2}.
A co-dimension one brane of minimal energy is a straight line. Let us
also note some of its elementary geometrical data. The coordinate
of its termination point is
\beqn
p \vec{\bf e}_1 + q \vec{\bf e}_2 = \frac1{\sqrt{\a'}}
(q R_2 \sin(\th), p R_1 + q R_2 \cos(\th)) \ .
\eeqn
Thus, its (dimensionless) length $L$ and the distance $D$ to the next
copy in the elementary cell satisfy
\beqn\lab{lendis}
L^2 = \frac{{\rm Im}\, t}{{\rm Im}\, u} | p + q u |^2 \ , \quad
D^2 = \frac{({\rm Im}\, t)^2}{L^2} = \frac{{\rm Im}\, u\,  {\rm
Im}\, t}{| p + q u |^2}\ .
\eeqn
From these formulas one can immediately read off the normalization
of KK momentum modes in the open string spectrum, quantized as
$m/L$ for integer $m$, and winding modes quantized as $nD$. The
tan of the angle of intersection with vertical axis can be
expressed
\beqn\lab{tanphi}
\tan(\vf) = \frac{p}{q} \frac1{{\rm Im}\, u} + \cot(\th) \ .
\eeqn
These data are illustrated in figure \ref{T2angle} for the D-brane
with winding numbers $(2,1)$.

\begin{figure}[ht]
\hspace{2cm}
\begin{picture}(200,200)(0,0)

\DashLine(50,50)(280,50)5
\DashLine(50,100)(280,100)5

\DashLine(70,30)(100,120)5
\DashLine(150,30)(180,120)5
\DashLine(230,30)(260,120)5

\def\axowidth{1.5 }
\LongArrow(77,50)(157,50)
\LongArrow(77,50)(94,100)
\def\axowidth{0.5 }

\Vertex(77,50)3
\Vertex(253,100)3

\def\axowidth{1.5 }
\Line(77,50)(253,100)
\def\axowidth{0.5 }

\Line(77,20)(77,130)

\LongArrowArc(77,50)(35,0,70)
\LongArrowArc(77,50)(45,15,90)

\Text(83,109)[1]{$\vec{\bf e}_2$}
\Text(113,42)[1]{$\vec{\bf e}_1$}
\Text(40,67)[1]{$\th$}
\Text(27,91)[1]{$\vf$}

\end{picture}
\vspace{-1cm}
\caption{Geometry of D-brane with co-dimension one\lab{T2angle}}
\end{figure}
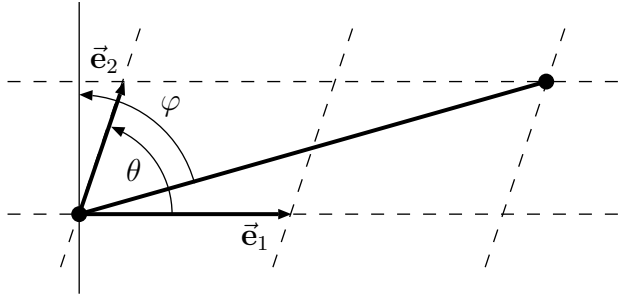

We have seen in section \ref{secbcangles} that magnetized D-branes
are obtained from rotated D-branes of one dimension less by
T-duality. Let us apply this to obtain a D-brane wrapping the full
$\mbb T^2$ and with a magnetic background field. A T-duality along
the axis parallel to $\vec{\bf e}_1$ takes $R_1 \mapsto
R_1'=\a'/R_1$ and thus
\beqn\lab{TonUT}
u ~\mapsto~ t' = \frac1{\a'} R_1 R_2 e^{i\th} \ , \quad t~\mapsto~
u' = {\rm Re}\, u' + i \sin(\th) \frac{R_2}{R_1} \ .
\eeqn
This flip of complex and K\"ahler structure leads to a torus with
${\rm Re}\, u' = b$ and $b' = \frac1{\a'} R_1 R_2 \cos(\th)$. A
tilted torus with $\th \neq \frac\pi2$ becomes a torus with
non-vanishing $b'$, and vice versa. The T-dual of the D-brane
rotated relative to the axis orthogonal to $\vec
{\bf e}_1$ by the angle $\vf$ from \reef{tanphi} now has a background gauge
field given by
\beqn
2\pi\a'  \cf' = \tan(\vf) = \frac1{{\rm Im}\, t'}  \left(
\frac{p}{q} + b'\right) \ .
\eeqn
We can read off
\beqn
2\pi\a'  F' = \frac{p}{q} \frac1{{\rm Im}\, t'} \ , \quad B' =
\frac{b'}{{\rm Im}\, t'}\ ,
\eeqn
for the components of $F'_{ab}$ and $B'_{ab}$ along the D-brane.
The factor of the volume ${\rm Im}\, t'$ in the denominators is
made up for by the pull-back with $\vec{\bf e}_i$ that converts
$\cf'_{ab}$ into $\cf'_{ij}$. The normalization is such that the
integral on the torus has periods
\beqn\lab{DiracF}
\frac{2\pi\a'}{\ell_s^2}  \int_{\mbb T^2} \frac12 \cf_{ij} dx^i\wedge dx^j =
\frac{p + bq}q\ ,
\eeqn
omitting the prime. Note that the brane wraps the torus with
multiplicity $q$ such that the integral of $F$ over the world
volume is always $2\pi$ times an integer.

This has the following interpretation as a gauge bundle on a stack
of D-branes. One starts with $q$ D-branes wrapped on the torus and
their $U(q)$ gauge symmetry. This decomposes into the abelian
$U(1)$ and the $SU(q)$ precisely as $U(q) = (U(1)\times
SU(q))/\mbb Z_{q}$. We often ignore the discrete $\mbb Z_{q}$
identification but here it is crucial. The gauge field background
is classified by the first Chern number $C_1$
(proportional to the integral of $F'$ over the $\mbb T^2$) taking
values in $\mbb Z/q$, in the present case $C_1 = {p}/{q}$. When
the corrected field strength $\cf'$ is used $F'$ is simply
corrected by adding $B$. The intersection number
(\ref{int1}) has an interpretation as a chiral index. For more details
on the dual gauge bundles on tori we refer for example to
\cite{Hashimoto:1997gm,RR01}.

To implement the effects of the orientifold, we need to take care
of the invariance of the tori and D-brane configurations under the
world sheet parity. In case of the IIA version the D-branes on
one-cycles are relevant and $\O\bs$ comes with a reflection along
one of the axes, namely the direction that was T-dualized. In IIB
there is a reflection along either none or both directions of the
$\mbb T^2$. Starting with IIA, $\bs$ acts by reflection along
$\vec{\bf e}_1$,
\beqn \lab{reflcycle}
\bs: (q {\bf e}^1_2 , p {\bf e}^2_1 + q {\bf e}^2_2) ~\mapsto~
(q {\bf e}^1_2 , - p {\bf e}^2_1 - q {\bf e}^2_2) \ .
\eeqn
To make the lattice invariant under $\bs$ the vector $(0, 2{\bf
e}^2_2)$ needs to be a lattice vector, too.

\begin{figure}[ht]
\vspace{.5cm}
\begin{picture}(250,150)(0,0)

\DashLine(50,50)(200,50)5
\DashLine(50,100)(200,100)5

\DashLine(70,30)(70,130)5
\Line(120,10)(120,150)
\DashLine(170,30)(170,130)5

\def\axowidth{1.5 }
\LongArrow(120,50)(120,100)
\LongArrow(120,50)(170,50)
\def\axowidth{0.5 }

\LongArrowArc(120,50)(72,50,130)
\LongArrowArcn(120,50)(72,130,50)

\Vertex(170,100)3
\Vertex(70,100)3

\DashLine(250,50)(400,50)5
\DashLine(250,100)(400,100)5

\DashLine(250,30)(300,130)5
\DashLine(300,30)(350,130)5
\DashLine(350,30)(400,130)5
\Line(310,10)(310,150)

\def\axowidth{1.5 }
\LongArrow(310,50)(360,50)
\LongArrow(310,50)(335,100)
\def\axowidth{0.5 }

\LongArrowArc(310,80)(31,50,130)
\LongArrowArcn(310,80)(31,130,50)

\Vertex(335,100)3
\Vertex(285,100)3

\Text(105,92)[1]{$\vec{\bf e}_2$}
\Text(129,42)[1]{$\vec{\bf e}_1$}
\Text(70,128)[1]{$\bar\s$}

\Text(235,92)[1]{$\vec{\bf e}_2$}
\Text(234,42)[1]{$\vec{\bf e}_1$}
\Text(178,118)[1]{$\bar\s$}

\end{picture}
\vspace{0cm}
\caption{Two choices of complex structure compatible with symmetry under $\bar\s$}
\end{figure}
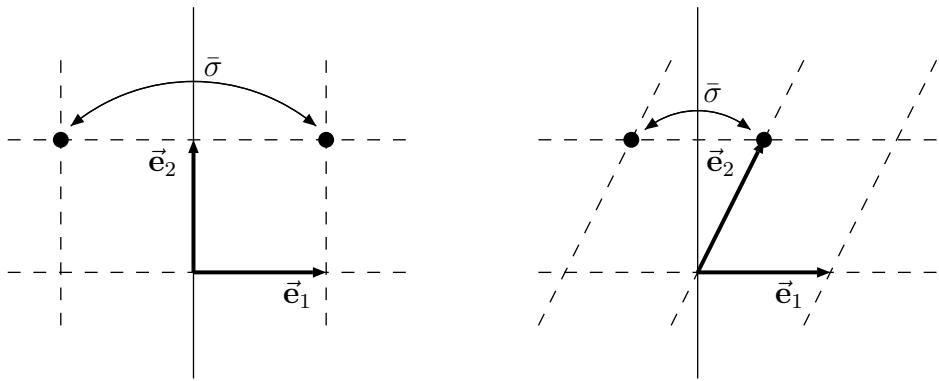

\noindent
This implies that the real part of $u = u_1+iu_2$ satisfies
\beqn
u_1 = \left\{ \begin{array}{l} 0 \\ \frac12 \end{array} \right.
{\rm mod}\ \mbb Z \ .
\eeqn
By the T-duality along the reflected axis, this translates into the
quantization of the NSNS $B$-field in IIB (dropping the prime on
$b'$),
\beqn
b = \left\{ \begin{array}{l}
0 \\ \frac12 \end{array} \right.
{\rm mod}\ \mbb Z \ .
\eeqn
On the elementary cycles, $\bs$ then acts by $e_1 \mapsto -e_1$
and $e_2\mapsto -2u_1e_1 +e_2$, hence, any brane along a cycle
labelled $(p,q)$ is accompanied by another copy wrapped on the
image,
\beqn
\bs:(p,q) \mapsto (-p - 2 u_1 q, q) \ .
\eeqn
It is helpful to define a new winding number
\beqn \lab{ptilde}
\tilde p = p + u_1 q\ ,
\eeqn
on which $\bs$ operates by
\beqn \lab{Omangle}
\bs:(\tilde p,q) \mapsto (-\tilde p, q) \ ,
\eeqn
irrespective of the value of $u_1$. Using $\tilde p$ the formula
for the relative angle \reef{tanphi} can be rewritten
\beqn \lab{tanphi2}
\tan(\vf) = \frac{\tilde p}{q u_2} \ .
\eeqn
We usually label the image of the a stack $a$ by $a'$.

An invariant brane obviously satisfies $\tilde p = p + u_1 q= 0$.
These branes have Dirichlet boundary conditions along the
T-dualized circle only, they correspond to fully wrapped D-branes
with vanishing world volume flux before the T-duality. A brane
with $q=0$ has Dirichlet conditions along the orthogonal direction
and Neumann conditions along the T-dualized circle, it refers to a
fully localized D-brane before the duality. These branes get
reflected onto themselves under $\bs$ which only flips the
orientation.

Let us now look at six-dimensional compactification tori,
concentrating again on direct products of two-dimensional ones,
such that the metric and $B$-field respect the factorization
\beqn\lab{torfact}
\mbb T^6 = \bigotimes_{I=1}^3 \mbb T^2_I\ .
\eeqn
If the brane configurations also have this product structure, one
only needs to keep track of the index $I=1,2,3$ labelling $t^I$
and $u^I$, winding numbers $(p_a^I,q_a^I)$, and all other
quantities above. Note that these are not the most general
configurations of branes that are compatible with supersymmetry on
tori, and they also do not produce a complete set on toroidal
orientifolds with $\cn=1$ supersymmetry. Nevertheless it is useful
to start with these.

First of all, one can easily identify the D-branes known from type I
in this more general framework. By the above reasoning,
D9-branes and D5-branes of IIB orientifolds on \reef{torfact} can be
identified by their dual winding numbers as follows,
\beqn
{\rm D9}: && (\tilde p^I_a, q^I_a) ~=~ ( 0 , 2^{2u^I_1})\quad {\rm for}\
I=1,2,3\ ,
\non
{\rm D5}: && (\tilde p^I_a, q^I_a) ~=~ \left\{
\begin{array}{l}
( 0 , 2^{2u^I_1})\quad {\rm for}\ I=J_1 \\
( 1 , 0)\quad {\rm for}\ I=J_2, J_3
\end{array}
\right. \ .
\eeqn
The above describes a D5-brane wrapped on the torus labelled $J_1$.
For completeness we also denote the intersection number for two
factorizable three-cycles on the six-dimensional torus
\beqn \lab{inter}
I_{ab} = \prod_{I=1}^3 \Big( p_a^I q_b^I - p_b^I q^I_a \Big) \ ,
\eeqn
which immediately follows from \reef{int1}. More formally the
gauge connection on the torus giving rise to a field strength
$\cf$ taking values only in the diagonal $U(1)$ inside a $U(N)$
(or a $U(1)$ along the Cartan subalgebra of $SO(2N)$) defines
a line bundle on the torus, i.e.\ the structure group of the gauge
bundle is $U(1)$. The gauge symmetry that survives in the
lower-dimensional theory in this background is the commutant of
the structure group inside the higher-dimensional gauge group.
Since it is abelian in this case, the rank is not reduced. So
starting from type I in ten dimensions we can split $SO(32)$ into
\beqn\lab{SO}
SO(32) ~\longrightarrow~ SO(32-2N) \times \prod_a U(N_a)\ , \quad
\sum_a N_a = N \leq 16\ .
\eeqn
If we also have O5-planes and D5-branes with maximal symmetry
gauge $Sp(32)$ this can also be decomposed in the same manner as
\beqn\lab{Sp}
Sp(32) ~\longrightarrow~ Sp(32-2N) \times \prod_a U(N_a)\ , \quad
\sum_a N_a = N \leq 16\ .
\eeqn
Together the product of \reef{SO} and \reef{Sp} forms the gauge
symmetry of intersecting or magnetized D-brane models, in full
generality and not just in toroidal models. This is the explicit
version of \reef{gengauge}. In this way, compactifying on a line
bundle (or tensor products thereof) breaks the primordial gauge
symmetry down to unitary groups. The intersection form
\reef{inter} has an interpretation as the index of the Dirac
operator in this background
\beqn
I_{ab} = \int_\cw {\rm ch}(2\pi\a'\cf_a \otimes 2\pi\a'\bar \cf_b) \wedge \hat A(\cR)\ .
\eeqn
The curvature forms are normalized to give integer periods. The
integral runs over the world volume of the D9-branes and
covers the torus multiple times. This geometric point of view is
valid beyond the toroidal CFT models discussed in this
introductory section and will also be useful later for
orientifolds on other Calabi-Yau manifolds.

Let us now discuss the spectrum of the lightest states in the NS
sector of an open string connecting two different stacks of
rotated or magnetized D-branes on a torus of complex dimensions
one to three. These strings transform as bifundamental
representations $(\fund_a,\bar{\fund}_b)$ or $(\fund_a,{\fund}_b)$
of the gauge groups on the two stacks. As explained earlier, the
ground state of the R sector is a massless chiral fermion in four
dimensions. In the presence of supersymmetry it is clear that it
must be accompanied by a complex scalar to form a chiral
multiplet, which is provided by the lightest, in that case
massless, NS state.

In general, the mode expansion of the oscillator modes generating
the open string states has modes shifted by \reef{delAB} as
defined via the relative angle. The complexified world sheet
coordinate fields $Z^I$ and $\P^I$ have mode operators
$\a^I_{n+\d^I_{ab}}$, $\bar\a^I_{n-\d^I_{ab}}$ and
$\psi^I_{r+\d_{ab}^I}$, $\bar\psi^I_{r-\d_{ab}^I}$ respectively.
The shift vector $\d_{ab}^I$ is defined modulo $2\mbb Z$, since
the angles themselves are defined modulo $2\pi\mbb Z$. However,
whenever any $\d_{\a\beta}^I$ becomes greater than 1 two level in the
spectrum cross. It is therefore more convenient to pick
conventions in which the shift vectors are defined modulo $\mbb
Z$. This is done by flipping the orientation of a D-brane whenever
a relative angle does not satisfy
\beqn
-\frac\pi 2 < \vf_{ab}^I \leq \frac\pi2\ .
\eeqn
With this prescription, the relative angles $\vf_{ab}^I$ are
defined modulo $\mbb Z$ even though the absolute angles $\vf_a^I$
are given modulo $2\mbb Z$. This means that one exchanges a brane
for an anti-brane with any such flip and thus has to use the GSO
projection with opposite sign.\footnote{From the boundary state
one can deduce that a rotation by an angle $\pi$, a flip of
orientation, changes the sign of the RR component since it acts by
a phase factor $i$ on the fermionic R groundstate. In an annulus
diagram, this change of sign leads to the opposite GSO projection
in the open string loop-channel, as explained below
\reef{etasigns}.} Depending on the number of such flips being even
or odd either the NS groundstate or the first oscillator level is
the lightest physical state in the NS spectrum. Following this
reasoning the mass of the lightest state is easily found using the
formula
(\ref{vacen}) for the zero point energy to be
\beqn\lab{tachmass}
\a' M^2_0 = \left\{ \begin{array}{ll}
\frac12 \sum_I |\d^I_{ab}| - {\rm max}_I\{ |\d^I_{ab}| \} & {\rm
for\ at\ most\ one}\ |\d^I_{ab}|>\frac12 \\
1+\frac12 ( |\d^I_{ab}| - |\d^J_{ab}| - |\d^K_{ab}| ) & {\rm
for}\ |\d^I_{ab}| \leq\frac12 ,\ |\d^J_{ab}|,\, |\d^K_{ab}| >\frac12 \\
1-\frac12 \sum_I |\d^I_{ab}| & {\rm for\ all\ three}\
|\d^I_{ab}|>\frac12
\end{array}\right. \ .
\eeqn
It follows (for small angles) that with only a single relative
angle non-vanishing, i.e.\ two intersecting branes on a $\mbb
T^2$, the scalar is always of negative mass $-\frac12 |\d_{ab}|$.
On a $\mbb T^4$ the mass is always proportional to the difference
of the two angles $-\frac12 ||\d^1_{ab}|-|\d^2_{ab}|| \leq 0$ and
negative or zero. But this is no longer true for two branes on
$\mbb T^6$, where the mass for small angles is $\frac12 (
|\d^I_{ab}| + |\d^J_{ab}| - |\d^K_{ab}| )$ if $|\vf^K_{ab}|$ is
the largest of the three angles in absolute value. The
supersymmetry conditions
(\ref{anglesusy}) imply the existence of a massless NS groundstate
on $\mbb T^4$ and $\mbb T^6$, but they have no solution on $\mbb
T^2$ except for vanishing relative angles. The degeneracy of the
massless groundstate is four-fold on $\mbb T^4$ (two complex
scalars to form a hyper multiplet) and two-fold on $\mbb T^6$ (one
complex scalar for a chiral multiplet).

It is interesting to compare these results derived from an exact
quantization of open strings in a flat background geometry to mass
spectra derived from effective field theory or even quantum
mechanics. The mass spectrum of open strings connecting two
magnetized branes of different world volume gauge field strengths
has been used to test and better understand the non-abelian DBI
effective action \cite{Hashimoto:1997gm,Denef:2000rj,el03}.
Furthermore, one can identify the lowest
oscillator excitation with a semi-classical quantum mechanical
mass formula (see \cite{CB95})
\beqn
M^2_{\rm QM} = (2n+1) |e F| + 2seF \ ,
\eeqn
which is the standard formula for the Landau levels of a point
particle in a background magnetic field $F$ and with spin $s$ and
charge $e$.

As a consequence of \reef{tachmass} configurations of intersecting
or magnetized D-branes on an internal six-dimensional torus  can
in principle be meta-stable with respect to small deformations of
open string modes even without supersymmetry. However, by their
tension these configurations are also sources for the background
closed string bulk fields which are destabilized by
non-supersymmetric brane configurations in the absence of other
effects to counter the brane tension.

The case of a tachyonic mass for a scalar field at the
intersection has an interpretation as a spontaneous symmetry
breaking scenario with a bifundamental Higgs scalar \cite{cim02a}.
The negative mass signals an instability that leads to a
non-vanishing vacuum expectation value for the respective scalar
field. The recombination breaks the gauge symmetry to the
diagonal, e.g.\ $U(1)_1\times U(1)_2\rightarrow U(1)_{\rm diag}$.
This process is of phenomenological interest since it closely
resembles the Higgs mechanism of the Standard Model or the MSSM
when applied to $U(2)\ti U(1)_Y$.

Geometrically, the two stacks of D-branes are deformed into a
single stack in the same topological class when the tachyonic
scalar takes a non-vanishing value. On a two-dimensional torus
with two intersecting branes this is just the decay of the two
into a single straight line given by adding the winding numbers
$(p_a.q_a)$ of the two stacks.

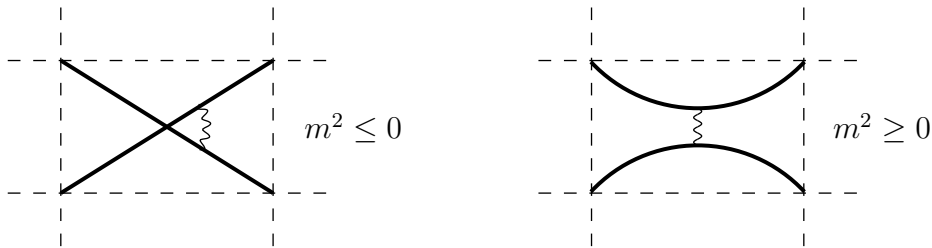
\begin{figure}[ht]
\vspace{-2.5cm}
\begin{picture}(200,200)(0,0)

\DashLine(50,50)(170,50)5
\DashLine(50,100)(170,100)5

\DashLine(70,30)(70,120)5
\DashLine(150,30)(150,120)5

\def\axowidth{1.5 }
\Line(70,50)(150,100)
\Line(70,100)(150,50)
\def\axowidth{0.5 }

\PhotonArc(110,75)(15,325,35){1.5}{4}

\DashLine(250,50)(370,50)5
\DashLine(250,100)(370,100)5

\DashLine(270,30)(270,120)5
\DashLine(350,30)(350,120)5

\def\axowidth{1.5 }
\CArc(310,13)(55,43,137)
\CArc(310,137)(55,223,317)
\def\axowidth{0.5 }

\Photon(310,68)(310,82){1.5}{3}

\Text(177,75)[1]{$m^2 \leq 0$}
\Text(350,75)[1]{$m^2 \geq 0$}

\end{picture}
\vspace{-1cm}
\caption{Recombination of two intersecting D-branes\lab{recomb}}
\end{figure}

In general, starting from the flat factorized sLag three-cycles
described above, the recombination can lead to more general
calibrated sLag cycles which are no longer of the factorized type.
This has been explored in a number of approaches to model building
in order to obtain more general classes of cycles, e.g.\ in
\cite{bgo02,ho04}.


\subsubsection{One-loop divergences with intersecting and magnetized D-branes}
\lab{secangle1loop}

In this section we compute the one-loop partition function for a
type II orientifold with intersecting or magnetized D-branes on an
internal six-dimensional torus. Essential steps were developed in
\cite{bgkl00a,aads00,bkl00}. We impose the factorization
\reef{torfact} for the full background, the NSNS fields and the brane
configurations.

For definiteness, we adopt the language of type IIA with
D6-branes. Each D6-brane wraps a sLag three-cycle, a
one-dimensional straight line on each $\mbb T^2_I$, $I=1,2,3$. The
world sheet parity comes with the complex conjugation $\bs$ from
(\ref{genOm}). O6-planes are located along the axes fixed under $\bs$,
they are T-dual to O9-planes. The stacks of branes are labelled by
$a,b,...$ with brane multiplicity $N_a$. Boundary conditions for
the rotated D6-branes are given by (\ref{Thbc2}) each brane
characterized by three angles $\vf_a^I$. The angles are defined
via the two co-prime integers $(p_a^I, q_a^I)$ through
\reef{tanphi2}, alternatively we use $(\tilde p_a^I,q_a^I)$ which
in case $u_1^I=\frac12$ do not necessarily have to be co-prime and
$\tilde p_a^I$ can then also be half-integer. The three-cycles
wrapped by the O6-planes are denoted $\Pi_{\rm O6}$, they are
given by winding numbers $(0,2^{2u_1^I})$ for $I=1,2,3$. The world
sheet parity $\O\bs$ reflects the angles by \reef{Omangle}, i.e.\
$\O\bs : \vf^I_a \mapsto- \vf^I_a =
\vf^I_{a'}$. It identifies two stacks of branes at opposite
angles.

With these preliminaries, we have to compute the three diagrams, Klein
bottle, annulus and M\"obius strip. The Klein bottle has no
boundaries. It is not at all sensitive to the relative angles of
the D-branes, and thus identical to the Klein bottle of type I
string theory on a six-dimensional torus as computed in section
\ref{sec1l}, only up to the T-duality that exchanges the D9-branes into
D6-branes. This T-duality along the three imaginary parts of the
coordinates $Z_I$ maps $\O$ to $\O\bs$, and D9-branes with
magnetic fluxes as described by (\ref{Thbc}) to D6-branes with
relative angles defined through (\ref{Fphi}).

Thus, the Klein bottle amplitude is still given though
(\ref{loopchannel}). The oscillator traces are given in
\reef{traces} and the trace over zero-modes in \reef{KKsum}.
In the Klein bottle the KK modes are obtained by just taking the
T-dual of \reef{KKspecKB}. In the loop-channel the mass spectrum can be
written
\beqn
\ck: \quad \a' M^2_{\rm KK} = \sum_{I=1}^3 \frac{|n_I + m_I
t^I|^2}{t^I_2} \frac1{u_2^I} \ .
\eeqn
The momentum and winding numbers $(m_I,n_I)$ run over integers,
$t^I$, $u^I$ are the K\"ahler and the complex structure moduli
defined through \reef{tormetric} and \reef{Kahmod}. The modular
transformation of the open string partition function to the
tree-channel via \reef{poisson} produces a moduli-dependent
prefactor
\beqn \lab{KBpre}
\tilde \ck ~\sim~ \prod_{I=1}^3 {u_2^I}
 \ .
\eeqn
All the numerical factors just work out as for the type I string
theory on $\mbb T^6$. This is just the T-dual of the prefactor
obtained in type I which was the overall volume of space-time.
Here we find the ratio of the three-dimensional internal volume
covered by O6-planes dual to the O9-planes, divided by the volume
of the transverse internal space.

In the annulus diagram we have to distinguish two different types
of open strings, those with both ends on the same stack of branes,
and those connecting different stacks, $\ca_{aa}$ or $\ca_{ab}$
diagrams. The $\ca_{aa}$ annuli are identical to the diagrams in
type I in (\ref{1loop}) and (\ref{1looptree}) up the bosonic zero
modes. Here the mass operator of the open string KK modes follows
from \reef{lendis},
\beqn
\ca_{aa}: \quad \a'M_{\rm KK}^2 &=& \sum_{I=1}^3 \Big[
\frac{m_I^2}{(L^I_a)^2} + n_I^2 (D^I_a)^2 \Big] \non
&=&
\sum_{I=1}^3 \frac{|n_I + m_I t^I|^2}{t^I_2}\frac{u_2^I}{|p^I_a + q^I_a
u^I|^2} \ .\label{2192}
\eeqn
The closed
string KK spectrum can be obtained by modular transformation or
directly from the boundary conditions
(\ref{Thbc2}) \cite{bgkl00} as
\beqn
\tilde \ca_{aa}: \quad  \a' \tilde M_{\rm KK}^2 =  \sum_{I=1}^3 \frac{|n_I + m_I
t^I|^2}{t^I_2}\frac{|p^I_a + q^I_a u^I|^2}{u_2^I}
\ .
\eeqn
The modular transformation from the loop-channel to the closed
string tree-channel by means of (\ref{poisson}) produces a
prefactor
\beqn \lab{annpre}
\tilde \ca_{aa} ~\sim~ \prod_{I=1}^3 \frac{V_a^I}{D_a^I} ~=~
\prod_{I=1}^3 \frac{(V_a^I)^2}{t_2^I}
~=~ \prod_{I=1}^3 \frac{|p^I_a + q^I_a u^I|^2}{u_2^I} \ .
\eeqn
For the T-dual of a D9-brane with $(p_a^I + u_1^I q_a^I,q_a^I) =
(0,2^{2u_1^I})$ this collapses to the same moduli dependence
$\prod_I u_2^I$ as in the Klein bottle. In case that a $u_1^I =
\frac12$ there appears the extra factor $(q_a^I)^2=4$ which we
have argued about around equation \reef{Bbst}. For a dual D5-brane
one finds the inverse moduli dependence with respect to the moduli
of the $\mbb T^2_I$ with Dirichlet directions and no extra factor
$4$.

The role of the missing annulus diagrams $\tilde\ca_{ab}$ and of
the M\"obius diagrams $\tilde\cm_a$ is now to complete
(\ref{KBpre}) and (\ref{annpre}) into a sum of perfect squares. The normalization
of the cross-cap states of the O6-planes is proportional to the
square root of \reef{KBpre}, schematically,
\beqn\lab{normcc}
|C 6\> ~\propto~ \prod_{I=1}^3 \sqrt{u_2^I}\, |C 6\>_{\rm osc}\ .
\eeqn
The boundary states for rotated D6-branes has to
carry factors proportional to
\beqn\lab{normbst}
|B 6,a\> ~\propto~ \prod_{I=1}^3 \frac{|p^I_a + q^I_a
u^I|}{\sqrt{u_2^I}} |B 6,a\>_{\rm osc}\ .
\eeqn
Let us still demonstrate the appearance of the complete square by
calculating the loop-channel diagrams. The annuli $\ca_{ab}$ do
not have any bosonic zero modes along the internal space, i.e.\
there is no KK sum to perform. The traces over the oscillator
modes are modified compared to (\ref{traces}) because the modings
are shifted by $\d_{ab}^I$ from (\ref{delAB}). They become
\beqn \lab{angletr}
&& {\rm Tr}\, \exp\Big( 2\pi i\tau \sum_{I=1}^3\Big[ \sum_{n\neq 0}\Big(
\a^I_{-n-\d_{ab}^I} \bar\a^I_{n+\d_{ab}^I}+ \a^I_{-n+\d_{ab}^I}
\bar\a^I_{n-\d_{ab}^I}\Big)
- a(\d_{ab}^I) \Big] \Big)
\non
&&
\hspace{7cm}
~=~ \prod_{I=1}^3 \Bigg[ e^{i\pi(\d_{ab}^I + \frac12)}
\frac{\eta(\t)}{\thba{1/2+\d_{ab}^I}{1/2}(0,\t)}\Bigg]
\ ,
\non
&& {\rm Tr}{\zba\a\beta}\, \exp\Big( 2\pi i\tau \sum_{I=1}^3\Big[
\sum_{r\neq 0} \Big(r\psi^I_{-r-\d_{ab}^I} \bar\psi^I_{r+\d_{ab}^I} +
 r\psi^I_{-r+\d_{ab}^I} \bar\psi^I_{r-\d_{ab}^I}\Big)
- a(\d^I_{ab})\zba\a\beta \Big] \Big) \non
&&
\hspace{7cm}
~=~
\eta_{\a\beta} \prod_{I=1}^3 \Bigg[ e^{2\pi i(\a+\d_{ab}^I)\beta}
\frac{\thba{\a+\d_{ab}^I}{\beta}(0,\t)}{\eta(\t)}
\Bigg] \ .
\nonumber
\eeqn
The mode generators are now complexified, see appendix
\ref{appinters}. Phase factors are basically determined by demanding
that the tree-channel is properly normalized.
The zero-point energy $a(\d_{ab}^I)$ defined in
\reef{vacen} depends on the spin structure. The modular
parameter for the annulus is $\t=i\t_2=it$. With this the annulus diagrams
are
\beqn
\ca_{ab} = \frac{-i V_{4}}{4(8\pi\a')^2}  N_a N_b I_{ab} \int_0^\infty \frac{dt}{t^3}
\sum_{\a,\beta} \eta_{\a\beta}
\frac{\thba\a\beta (0,it)}{\eta(it)^3}\prod_{I=1}^3 e^{i\pi\d_{ab}^I}
\frac{\thba{\a+\d_{ab}^I}{\beta}(0,it)}{\thba{1/2+\d_{ab}^I}{1/2}(0,it)}  \ ,
\eeqn
in the loop channel and
\beqn
\tilde\ca_{ab} = \frac{V_{4}}{4(8\pi\a')^2}  N_aN_b  I_{ab}
\int_0^\infty dl\,
\sum_{\a,\beta} \eta_{\a\beta} \frac{\thba{-\beta}\a(0,il)}{\eta(il)^3} \prod_{I=1}^3
\frac{\thba{-\beta}{\a+\d_{ab}^I}(0,il)}{\thba{-1/2}{1/2+\d_{ab}^I}(0,il)}  \ ,
\eeqn
in the tree channel. The amplitude with opposite orientation,
$\ca_{ba}$, has to be added. The intersection number $I_{ab}$ is defined by
the topological intersection of the two three-cycles, given in
\reef{inter}.

From here the normalization of the boundary states can be recovered as
the pre\-factor of the bosonic part of the partition function,
\beqn
I_{ab} \prod_{I=1}^3
\frac2{\thba{1/2}{1/2+\d_{ab}^I}(0,il)}
&=& I_{ab}\prod_{I=1}^3 \frac1{\sin(\vf_{ab}^I)} + \  \cdots
\non
&=& \prod_{I=1}^3 \frac{| p_a^I + q_a^I u_2^I|| p_b^I + q_b^I u_2^I|}{u_2^I}
+ \  \cdots
\eeqn
in accord with \reef{normbst}. In this way, the intersection
numbers can also be calculated from the proper normalization of
the boundary states. The contribution of the RR sector comes from
the spin structure with upper characteristic $\frac12$ in the
theta-functions. Using
\beqn
I_{ab}\prod_{I=1}^3 \frac1{\tan(\vf_{ab}^I)} 
&=& \prod_{I=1}^3 \frac{\tilde p_a^I \tilde p_b^I + q_a^I q_b^I (u_2^I)^2}{u_2^I}
\eeqn
one can extract the moduli dependence.

When the supersymmetry condition (\ref{anglesusy}) is satisfied the
shifts $\d_{ab}^I$ sum to zero such that
one can use the Jacobi identity \reef{jacobi} to show that the
amplitude vanishes. This signals that the
NSNS and RR sectors precisely cancel. Thus, the exact no-force law
at one-loop among intersecting D-branes arises as a consequence of
the calibration condition \reef{anglesusy}. The loop calculation
on the toroidal background includes all orders of $\a'$ while
the derivation of \reef{anglesusy} was based on classical
geometry. Only in the case of an exactly flat background like a
torus the two can become equivalent in sigma-model
perturbation theory.

Since $\O\bs$ flips the value of the angle of any brane relative
to the real axes of the three tori, the M\"obius strip
loop-channel can only have contribution from $aa'$ open strings
that connect branes with their images. It is quite tedious to
evaluate the M\"obius strip directly in the loop-channel, getting
all phase factors right.\footnote{For example, there are extra
phase factor from the action of $\O\bs$ on fields with mixed
boundary conditions, interpolating between the two cases given in
\reef{Omosc}.} So we prefer to write directly the result for the
tree channel expression,
\beqn
\tilde\cm_a &=& \pm \frac{V_{4}}{4(8\pi\a')^2} 2^6 N_a \prod_{I=1}^3 \tilde
p_a^I
\\
&& \hspace{2cm} \times
\int_0^\infty dl\,
\sum_{\a,\beta} \eta_{\a\beta} \frac{\thba{\a}{\beta}(0,\frac12 +
il)}{\eta(\frac12 + il)^3}
\prod_{I=1}^3
\frac{\thba{\a}{\beta+\vf_{a}^I/\pi}(0,\frac12  + il)}{\thba{1/2}{1/2+\vf_{a}^I/\pi}(0,\frac12  + il)}
\ .
\nonumber
\eeqn
The overall numerical factor has been fixed in order to reproduce
the normalization of the M\"obius strip in type I in
\reef{1looptree} in the limit where only D6-branes are present
that are directly T-dual to D9-branes without any world volume
magnetic fields. Again, the bosonic partition function gives the
moduli-dependent normalizations of the cross-cap and boundary
states,
\beqn
\frac{\tilde p_a^I}{\sin(\vf_a^I)} = \frac{|p_a+q_au_2^I|}{u_2^I}u_2^I\ .
\eeqn
The RR contribution has a moduli dependence
\beqn
\frac{\tilde p_a^I}{\tan(\vf_a^I)} =
q_a^I u_2^I
\ .
\eeqn
Putting all the pieces together, the RR tadpole cancellation
condition comes out as
\beqn
\Bigg[ \tilde\ck + \sum_{a,b} \tilde\ca_{ab} + \sum_a
\tilde\cm_a\Bigg]_{\rm RR}
&=&
\\
&& \hspace{-4cm}
\frac{8 V_{4}}{4(8\pi\a')^{2}} \prod_{I=1}^3 u_2^I
\int_0^\infty dl\, \Bigg[ \Bigg(
 \sum_a N_a \prod_{I=1}^3 {q_I^a} \pm 16 \Bigg)^2 +
\Bigg( \sum_a N_a \prod_{I=1}^3 \frac{\tilde p_a^I}{u_2^I}\Bigg)^2
\non
&& \hspace{-2cm}
+
\sum_{I=1}^3 \Bigg(  \sum_a N_a q_a^I \prod_{J \neq I} \frac{\tilde p_a^J}{u_2^J} \Bigg)^2
+
\sum_{I=1}^3 \Bigg(  \sum_a N_a \frac{\tilde  p_a^I}{u_2^I} \prod_{J \neq I} q_a^J
\Bigg)^2 +
\, \cdots \Bigg] \ .
\nonumber
\eeqn
Terms of massive modes are left out in the integrand. This is a
sum of eight squares, corresponding to the eight components of the
RR seven-form with one leg along each of the three $\mbb T^2_I$,
i.e.\ $8=2^3$. These are the massless modes that propagate in the
tree channel as intermediate states. Topologically the eight terms
correspond to the eight factorizable three-cycles of the torus.
The solution is given by any choice of co-prime $(p_a^I, q_a^I)$
that satisfies the four independent conditions
\beqn\lab{tadtor}
\sum_a N_a \prod_{I=1}^3 q_a^I ~=~ 16 \ , \quad
\sum_a N_a q_a^I \prod_{J \neq I} \tilde p_a^J ~=~ 0 \ .
\eeqn
The other four conditions which are linear or cubic in the $\tilde
p^I_a$ are automatically satisfied due to the symmetry
(\ref{Omangle}) under $\O\bs$. This set of conditions is the an
explicit way of writing \reef{loctad} for D6-branes with flat
world volume CP bundle on factorizable three-cycles of a torus
\cite{afiru00a}.\footnote{Note that the rank reduction
for T-dual D9-branes through non-vanishing $u_1^I$, a dual NSNS
$B$-field background, is built in automatically since in that case
the relevant $q_a^I$ is doubled.} The counting of branes is
adapted to \reef{SO} and \reef{Sp} now, we only count the pairs
$a$ and $a'$ as one brane so that the total number is 16 instead
of 32. The gauge symmetry involves a factor $SO(2N)$ for the
invariant D6-branes that are just T-dual to D9-branes, $Sp(2N)$
for dual D5-branes, and factors $U(N_a)$ for pairs of stacks not
invariant.

Similarly, the massless contribution in the NSNS sector can be written
as a sum of squares corresponding to the propagation of the dilaton
and the moduli scalar fields. Given what we said about adding
O5-planes and D5-branes to a toroidal compactification at the end of section
\ref{sec1l} one can rather
easily see that O5-planes add a background charge to the second
set of equations in \reef{tadtor}. There are three potential O5-planes
which map to three potential types of O6-planes, each providing a
source for one of the three tadpoles. Furthermore, as is evident from
\reef{tadD5}, a non-vanishing $u_1^I$ along a $\mbb T^2_I$ transverse to the
dual O5-plane depletes the charge by a factor $\frac12$. The cross-cap
states $|C6,I\>$, $I=1,2,3$, are
then normalized
\beqn
|C 6,I\> ~\propto~  \frac{\sqrt{u_2^I}}{
\prod_{J\neq I} 2^{2u_1^J} \sqrt{u_2^J}} |C 6,I\>_{\rm osc}\ ,
\eeqn
Topologically, these O6-planes wrap cycles characterized by winding
numbers $(0,1)$ along the $\mbb T^2_J$, $J\neq I$, and D6-branes on
top of these O6-planes will give rise to symplectic gauge groups to
start with. We will study
models of this type in section \ref{IntersectingZ2xZ2}.

The open string spectrum can in principle be read off from the
partition functions calculated above. The theta-functions for the
non-compact directions with upper characteristic $\frac12$ provide
a factor 2 as the degeneracy of the massless open string modes of
the R sector. It stands for the two polarization states of the
four-dimensional Weyl spinor in \reef{RRchiral}. The only other
source of a degeneracy of this state comes from the intersection
numbers in the loop channel. We have already outlined the method
to determine the open string spectrum at the end of section
\ref{secopenori}. Strings with both ends on a given stack $a$ not invariant under
$\O\bs$ form the adjoint of the gauge group $U(N_a)$.
Open strings connecting two different stacks are bifundamental. We
distinguish those that connect $ab$ from the $ab'$ by either using
$(\fund_a,\bar{\fund}_b)$ or $(\fund_a,\fund_b)$. The multiplicity of
chiral Weyl fermions (or chiral multiplets) in these representations
is given by the relevant intersection number \reef{inter}.

Only if a stack is invariant under $\O\bs$ a projection $\l^A_{ij}
= \pm \l^A_{ji}$ applies to the CP label, where the sign is fixed
by comparing the relative signs in the annulus and M\"obius strip
diagrams in the loop-channel. Since a D6-brane with vanishing
relative angle $\vf_a^I$ is T-dual to a D9-brane, the sign is
identical to type I and the gauge group on such a stack is
$SO(2N_a)$ and the gauge boson (or vector multiplet) transforms in
the anti-symmetric representation $\Yasymm_a$. Similarly, massless
states on dual D5-branes transform as $\Ysymm_a$ for $Sp(2N_a)$.
Whenever the supersymmetry condition
\reef{anglesusy} on the relative angles is satisfied, the
bifundamental fermions are completed into chiral multiplets with
massless NS scalars. Open string states at intersections of
branes $a$ with their images $a'$ are also subject to the projection onto
either symmetric or anti-symmetric states. We will be more concrete
later on.


\subsubsection{Anomaly cancellation and \gs mechanism}
\lab{secgs}

String theory starts out as an anomaly free ten-dimensional theory
of gravity and gauge fields. An essential piece in the anomaly
cancellation in ten dimensions is the Green-Schwarz mechanism
\cite{Green:1984sg}. It is known that in type I vacua essentially all
the RR scalars can participate in its generalized lower-dimensional
versions
\cite{Sagnotti:1992qw,Ibanez:1998qp,Ibanez:1999pw,Scrucca:1999zh,Antoniadis:2002cs}.
Here, we will only consider the four-dimensional effective
theories, and describe the anomaly structure of their gauge
currents, as well as the four-dimensional remnant of the
Green-Schwarz mechanism needed to make the theories consistent (see
e.g.\
\cite{Derendinger:1991kr,Louis:1996ya,Klein:1999im,Coriano':2005js,Anastasopoulos:2006cz}).

In the presence of chiral fermions one has to worry about
anomalies in the theory. Anomalies in gauge theories are
inconsistences of the theory which spoil the validity of current
conservation, or Ward identities, at the quantum level.
%
An anomaly is induced by triangle diagrams (and one-loop diagrams
with four and five external legs). For a theory with left-handed
chiral fermions $\psi_L$ the anomaly of the current
$\bar\psi_L\bar\s^\m\psi_L$ can be written\footnote{We use the
form of the so-called covariant anomaly.}
\beqn
(D_\m J^\m)^A = \frac{i}{32\pi^2} \epsilon^{\m\n\r\s} {\rm tr} (
T^A F_{\m\n} F_{\r\s} ) \ .
\eeqn
The field strength is $F_{\m\n} = \pa_\m A_\n - \pa_\n A_\m +
[A_\m,A_\n]$ and the trace is over all fermion representation of
the spectrum.\footnote{Besides the actual gauge anomalies there
can also be sigma-model and K\"ahler anomalies in orientifolds.
Their cancellation can also be important and may require a
separate \gs mechanism.}

We are interested in a theory with gauge group $G \times U(1)^3$
where $G$ is a simple non-abelian group, $SU(N)$, $SO(N)$ or
$Sp(N)$. This covers all anomalies that can arise from the general
gauge group \reef{gengauge} that can appear in orientifold models.
In general each stack $a$ supports a gauge group $U(N_a) =
SU(N_a)\ti U(1)_a$, $SO(2N_a)$ or $Sp(N_a)$. In this section we
will denote abelian vector field of the $U(1)_a$ by $C_\m^a$ and
the non-abelian gauge field $(A_\m)^A (T^a)^A = A_\m T^a$ where
the matrices $T^a$ are traceless hermitian generators. The charge
operator of the $U(1)_a$ is $Q^a$. $A$ is the adjoint index.
For the field strengths we write $F_{\m\n}^A (T^a)^A = F_{\m\n} T^a$ and $C_{\m\n}^a$.

The anomaly of the non-abelian gauge current is not relevant for
the Green-Schwarz mechanism. The only condition that arises is the
standard cancellation condition for the irreducible non-abelian
anomaly of each factor $a$, namely,
\beqn\lab{irranom}
\ca^{aaa} = {\rm tr}((T^a)^A \{(T^a)^B,(T^a)^C\}) = 0\ .
\eeqn
This condition is only non-trivial for $SU(N_a)$ with $N_a>2$ in
the cases we consider. The current conservation laws for the
abelian $U(1)_a$ currents $J^a_\m$ contain various terms, namely
\beqn \lab{gano}
\pa_\m J^{a\m} = \frac{i}{32\pi^2} \epsilon^{\m\n\r\s} {\rm tr} \Big(
Q^a (T^b)^A (T^b)^B F_{\m\n}^A  F_{\r\s}^B + Q^a Q^b Q^c C^b_{\m\n}
C^c_{\r\s}
\Big) \ ,
\eeqn
Sums over repeated stack indices $b,c$ are meant to be implicit.
One method to derive this is the Wess-Zumino descent formalism.
One starts from a gauge invariant six-form, the only candidate for
the mixed $U(1)_a-SU(N_b)^2$ anomaly being $I_6={\rm tr}\, F\wedge
F\wedge C$, where $F$ and $C$ are the field strength two-forms.
This can be written as derivative of a five-form,
\beqn
I_6= {\rm tr}\, F\wedge F\wedge C = a_1 d( C_\m dx^\m \wedge {\rm
tr}\, F
\wedge F ) + a_2 d( C\wedge \o_3^{\rm YM} ) = dI_5 \ ,
\eeqn
with $a_1+a_2=1$. Its gauge variation under abelian gauge
transformations is again the derivative of a four-form which
defines the anomaly\footnote{More precisely, the descent formalism
gives the formula for the consistent version of the anomaly,
whereas we are using the covariant form for simplicity only.}
\beqn
\d I_5 = a_1 d(\L_a {\rm tr}\, F \wedge F) =dI_4 \ .
\eeqn
The freedom to choose the coefficients reflects the possibility to
shuffle anomalies from the abelian to the non-abelian current and
vice versa. Our choice is actually $a_2=0$ and $a_1=1$. The cubic
abelian anomaly follows even simpler from $I_6=C^a\wedge C^b\wedge
C^c$.

Furthermore, the currents are also coupled to gravity. The
triangle diagrams with one gauge current and two gravitons lead to
\beqn \lab{grano}
\pa_\m J^{a\m}|_{\rm grav} &=& \frac{i}{768\pi^2} \epsilon^{\m\n\r\s}
{\rm tr}(Q^a) R_{\m\n\k\t} R_{\r\s}{}^{\k\t} \ .
\eeqn
The various anomalies are divided into the mixed
$U(1)_a-SU(N_b)^2$ anomaly $\ca^{abb} =
{\rm tr}(Q^a T^b T^b)$, mixed abelian $U(1)^3$ anomalies
$\ca^{abc} = {\rm tr}(Q^a Q^b Q^c)$, and the mixed
$U(1)-$gra\-vi\-tational anomaly are proportional to $\ca^{agg} =
{\rm tr}(Q^a)$.

To evaluate the anomalies in practice only the mixed $U(1)-SU(N)^2$
needs some extra techniques, all other anomalies just require summing
over $U(1)$ charges. For a given multiplet of charge $Q^a$ one can
employ \reef{anomtr} to convert all traces into traces over the
fundamental representation of $SU(N)$ and then sum over the
charges of these.
The consistency of the theory now demands that the right-hand-sides of
(\ref{gano}) and (\ref{grano}) either vanish due to cancellations among
the various fermion species, or need to be balanced by the
Green-Schwarz mechanism.

The latter adds a piece $\cl_{\rm GS}$ to the classical Lagrangian
which is not invariant under abelian gauge transformations,
schematically (suppressing identical gauge group indices in the
following)
\beqn \lab{gsvar}
\d_a \cl_{\rm GS} = \L_a \Big( k^{abb} \epsilon^{\m\n\r\s} F^b_{\m\n} F^b_{\r\s}
+ k^{abc} \epsilon^{\m\n\r\s} C^b_{\m\n} C^c_{\r\s}
+ k^{agg} \epsilon^{\m\n\r\s} R_{\m\n\k\t} R_{\r\s}{}^{\k\t} \Big) \ .
\eeqn
Choosing the coefficients appropriately, one can
achieve
\beqn
\L_a \pa_\m J^{a\m} + \L_a \pa_\m J^{a\m}|_{\rm grav} + \d_a \cl_{\rm GS} = 0 \
.
\eeqn
This is the essence of the four-dimensional \gs mechanism. The two
contributions are diagrammatically displayed in figure
\ref{figgs}.

\begin{figure}[ht]

\begin{picture}(200,200)(0,0)

\Photon(120,150)(180,150){2}{8}
\Photon(220,180)(280,180){2}{8}
\Photon(220,120)(280,120){2}{8}
\Vertex(180,150)2
\Vertex(220,180)2
\Vertex(220,120)2
\ArrowLine(180,150)(220,180)
\ArrowLine(220,180)(220,120)
\ArrowLine(220,120)(180,150)

\Photon(120,50)(180,50){2}{8}
\Photon(220,50)(280,80){2}{9}
\Photon(220,50)(280,20){2}{9}
\Vertex(180,50)2
\Vertex(220,50)2
\Line(180,50)(220,50)

\end{picture}
\caption{Interference of Feynman diagrams in Green-Schwarz mechanism\lab{figgs}}
\end{figure}
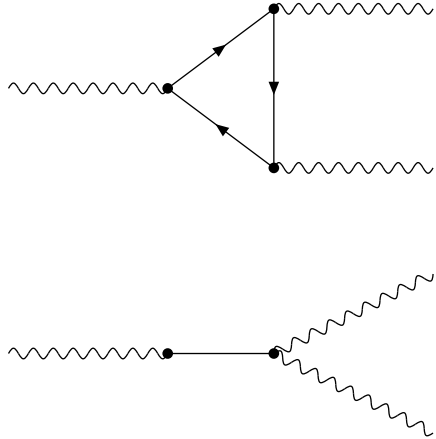

The variation (\ref{gsvar}) can be obtained by gauging the shift
symmetries of RR scalars in the theory if these appear in the
gauge kinetic functions of the gauge fields. Written in terms of
four-dimensional tensor fields $B_{\m\n}^I$ with field strengths
$F_{\m\n\r}^I$ the relevant piece of the Lagrangian looks
\beqn \lab{gsLag}
\cl_{\rm GS} = - \frac1{12} \sum_{I} F^I_{\m\n\r} F^{I\m\n\r} + \frac1{6} \sum_I k^{Ia}
\epsilon^{\m\n\r\s} C_\m^a F^I_{\n\r\s} \ ,
\eeqn
The $F^I_{\m\n\r}$ are gauge invariant and satisfy
\beqn \lab{modbianchi}
\frac23 \epsilon^{\m\n\r\s} \pa_{\m} F^I_{\n\r\s} =
k^{Ibb} \epsilon^{\m\n\r\s} F^b_{\m\n} F^b_{\r\s}
+ k^{Iab} \epsilon^{\m\n\r\s} C^a_{\m\n} C^b_{\r\s}
+ k^I \epsilon^{\m\n\r\s} R_{\m\n\k\t} R_{\r\s}{}^{\k\t} \ .
\eeqn
This is very similar to the Chern-Simons modification of the
ten-dimensional three-form of type I in \reef{mod3form}. We are
allowing complete flexibility for the coefficients of the various
terms. Applying $\d_a C_\m^a = \pa_\m \L_a$, the second term in
(\ref{gsLag}) produces the gauge transformation (after partial
integration)
\beqn
~~~\d_a \cl_{\rm GS} = - \frac1{4} \L_a \sum_I k^{Ia} \Big( k^{Ibb} \epsilon^{\m\n\r\s} F^b_{\m\n} F^b_{\r\s}
+ k^{Ibc} \epsilon^{\m\n\r\s} C^b_{\m\n} C^c_{\r\s}
+ k^I \epsilon^{\m\n\r\s} R_{\m\n\k\t} R_{\r\s}{}^{\k\t} \Big) \ .
\nonumber
\eeqn
Ignoring numerical coefficients, one can read off the anomaly
cancellation conditions for the conservation of the abelian
current to be
\beqn \lab{anomcan}
U(1)_a-SU(N_b)^2: && \ca^{abb} = {\rm tr}( Q^a T^bT^b) ~\propto~ \sum_I k^{Ia} k^{Ibb}\ ,
\non
U(1)_a-U(1)_b-U(1)_c : && \ca^{abc} = {\rm tr}( Q^aQ^bQ^c ) ~\propto~ \sum_I k^{Ia} k^{Ibc}\ ,
\non
U(1)_a-{\rm grav}: && \ca^{agg} = {\rm tr}( Q^a ) ~\propto~ \sum_I k^{Ia} k^{I}\ .
\eeqn
In the orientifold models we are interested in, the trace over the
fermion spectrum and the couplings on the right-hand-sides of
\reef{anomcan} are both expressed in terms of topological data such as
winding numbers of D-branes. In order for a \gs mechanism to work
their dependence on these data have to match, which is what we
will demonstrate in various classes of examples.

An important consequence of the \gs mechanism is the fact that a
mass is generated for the abelian gauge boson whose current would
otherwise not be conserved. Thus, for each $B_{\m\n}^I$ involved
in the anomaly cancellation, one abelian vector $C_\m^a$ will become
massive, the two combining into a massive vector field. This
can easier be demonstrated by replacing the $B_{\m\n}^I$ by scalars
$a^I$, schematically related via $\pa_\m a^I \sim
\epsilon_{\m\n\r\s}\pa^\n B^{I\r\s}$. These are the RR axions.
The Lagrangian (\ref{gsLag}) is
replaced by
\beqn\lab{gsLag2}
\cl_{\rm GS} &=& - \frac12 \sum_I (\pa_\m a^I - k^{Ia} C^a_\m )^2\non
&&
-\frac14 \sum_I a^I \epsilon^{\m\n\r\s} \Big( k^{Ibb}
F^b_{\m\n}F^b_{\r\s}
+ k^{Iab} C^a_{\m\n} C^b_{\r\s}
+ k^I R_{\m\n\k\t} R_{\r\s}{}^{\k\t} \Big) \
.
\eeqn
The gauge transformation of the participating RR scalars $a^I$ are
\beqn\lab{axshift}
\d_a a^I = k^{Ia} \L_a \ ,
\eeqn
leading to the variation (\ref{gsvar}) as before. In this
language, the \gs mechanism consists of gauging the non-linear
shift symmetries of the RR scalars $a^I$ combined with non-minimal
(and non gauge invariant) gauge kinetic functions. The $k^{Ia}$
are the constant Killing vectors that generate the shift
symmetries. One can decouple the scalars and vectors. They absorb
the scalars and become massive. The mass term reads
\beqn\lab{Stmass}
\frac12 \a'M^2_{ab} C^a_\m C^{b\m} = \frac12 \sum_I k^{Ia} k^{Ib}
C^a_\m C^{b\m}\ .
\eeqn
Regarding the couplings $k^{Ia}$ as vectors in the charge space,
they span a subspace of the dimension given by the number of
scalars $a^I$ with linearly independent couplings. The linear
combinations of abelian vectors orthogonal to these remain
massless, while those along the subspace become massive. A very
important observation is here that the masses only depend on the
couplings $k^{Ia}$ while the currents $J^{a\m}$ can be free of
anomalies if the couplings in (\ref{modbianchi})
vanish.\footnote{In that case there is a relation to
six-dimensional anomalies \cite{}.} Therefore, abelian gauge
bosons can become massive and decouple even if their currents are
not anomalous (those with anomalous currents must become massive).
This provides the opportunity to remove all unwanted abelian
factors from the gauge group. On the other hand, one has to take
care off the hypercharge gauge boson in this process
\cite{imr01}.


\clearpage
\setcounter{equation}{0}


\section{MODEL BUILDING}

After having laid down the basic string theoretic notions for the
construction of string compactifications with D-branes, the purpose of
this section is to explain in some technical detail the main rules
for model building. We will try to be as general as possible and
discuss not only type IIA orientifolds with intersecting
D6-branes, but also the general structure of type IIB orientifolds
with magnetized D-branes, both in the large radius regime as well
as at certain points deep inside the K\"ahler moduli space, the
so-called Gepner model orientifolds. For completeness, we will
also have a brief section on recent advances in heterotic string
constructions. Toroidal orbifolds will only appear as specific
examples.

Some of the features in the effective four-dimensional field
theory  of D-brane models are purely topological in nature. These
are the ones which can be determined for fairly general background
spaces. Among such topological aspects are the gauge group, the
chiral matter content, the anomalies with the generalized
Green-Schwarz anomaly cancellation mechanism, and the leading and
next-to-leading order perturbative contributions to gauge kinetic
functions and Fayet-Iliopoulos couplings. As we will discuss, due
to non-renormalization theorems these can already be exact in
sigma-model or string perturbation theory.

For the ``non-holomorphic'' data of the effective action like the
K\"ahler potential more knowledge of the underlying geometry is
necessary which, except in the large radius limit or for some
simple (toroidal) orbifolds is still beyond our contemporary
computational capacities. In the geometric domain at large radius
some important information can be gained by dimensional reduction
of the effective ten-dimensional actions
\cite{gl04,jl04,Grimm:2004ua}. But for quantities without a
ten-dimensional origin, like the matter fields localized at the
intersections of D-branes, such an approach is not applicable.

As one important piece of information about the vacuum structure
one in general expects a superpotential for some of the closed and
open string moduli to be generated. For intersecting D-brane
models on general Calabi-Yau spaces this can lead to a lifting of
moduli scalars, or even to a destabilization of the background.
For concrete compact models essentially nothing is known so far
about the concrete form of this superpotential.

From a phenomenological point of view, the  main question is
whether in the plethora of various string compactifications that
we summarize in the following realistic (and possibly
supersymmetric) ``Standard-like'' models can be found. One
economic way to realize a model with gauge symmetries close to the
Standard Model from D-branes is to start with four stacks of
D-branes with initial gauge symmetry $U(3)\times U(2)\times
U(1)\times U(1)$ or $U(3)\times Sp(2)\times U(1)\times U(1)$. The
Standard Model matter fields can be realized by bifundamental
fermions. Such a configuration can be encoded in a quiver diagram
in an obvious manner, see figure \ref{quiver}. (Alternative 
D-brane realizations of the Standard Model can be found in 
\cite{Antoniadis:2000en,Krause:2000gp,Antoniadis:2002qm}.) In this case there
are no chiral fermions transforming in the symmetric or
antisymmetric representations of the unitary gauge symmetries
(which would be arrows starting and ending on the same node of the
quiver). 
Since the multiplicities of the fields are
given by topological intersection numbers of three-cycles such as
\reef{int1}, this puts constraints on the allowed
three-cycles. There should be three bifundamental representations
of quarks and leptons, and anti-symmetric or symmetric matter
absent.

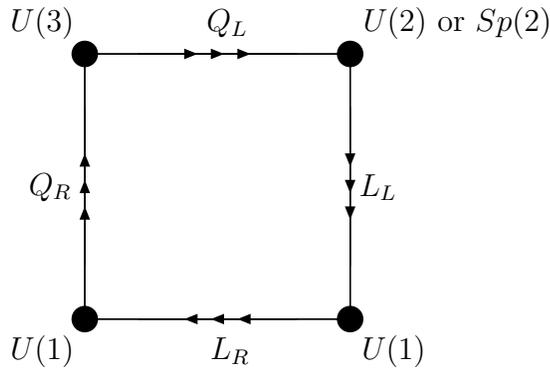
\begin{figure}[ht]
\vspace{-2cm}
\hspace{6cm}
\begin{picture}(180,180)(0,0)

\Vertex(0,100)5
\Vertex(100,100)5
\Vertex(0,0)5
\Vertex(100,0)5

\ArrowLine(0,0)(0,100)
\ArrowLine(0,20)(0,100)
\ArrowLine(0,0)(0,80)

\ArrowLine(0,100)(100,100)
\ArrowLine(20,100)(100,100)
\ArrowLine(0,100)(80,100)

\ArrowLine(100,100)(100,0)
\ArrowLine(100,80)(100,0)
\ArrowLine(100,100)(100,20)

\ArrowLine(100,0)(0,0)
\ArrowLine(100,0)(20,0)
\ArrowLine(80,0)(0,0)

\Text(-20,-12)[1]{$U(1)$}
\Text(86,-12)[1]{$U(1)$}
\Text(-74,112)[1]{$U(3)$}
\Text(56,112)[1]{$U(2)$ or $Sp(2)$}
\Text(-125,50)[1]{$Q_R$}
\Text(-84,-12)[1]{$L_R$}
\Text(-55,50)[1]{$L_L$}
\Text(-139,112)[1]{$Q_L$}

\end{picture}
\vspace{1cm}
\caption{Standard Model quiver with four stacks and three generations\lab{quiver}}
\end{figure}

Since concrete searches for semi-realistic examples and their
phenomenological consequences have been reviewed elsewhere
\cite{Blumenhagen:2005mu,bbkl02a}, here we will mainly focus  on the
technical aspects of these constructions.


\subsection{Type IIA orientifolds with intersecting D6-branes}
\label{sectypeiia}

Type IIA orientifolds with intersecting D6-branes are the
prototype of D-brane model building. They have been studied most
during the last couple of years. The CFT aspects of such models
have been discussed in the previous section in the framework of
toroidal orientifolds. Many of the important features of these
models can be already derived from the classical DBI action
\reef{dbi} on the D6-branes.  Let us demonstrate  how this can be done
quite generically for intersecting D6-branes on arbitrary
Calabi-Yau three-folds. Here we will restrict ourselves 
to compact manifolds but generalizations to non-compact ones
(having compact three-cycles) have been proposed in \cite{AU02,AU03a}
(see also \cite{cgqu03,Verlinde:2005jr,Garcia-Etxebarria:2006rw,Buican:2006sn} for some recent semi-realistic 
local Type IIB orientifold models).

The following presentation is mainly in the spirit of
\cite{bbkl02}. Consider the compactification of type IIA string theory on $\cy = \mbb R^{1,3}\times
\cx$ with $\cx$ a Calabi-Yau three-fold. Now we  perform an
orientifold $\Omega \bar\sigma (-1)^{F_L}$ where $\bs$ is the
complex conjugation as defined in \reef{genOm}. The K\"ahler form
$J_2$ and holomorphic $(3,0)$-form $\Omega_3$ of the Calabi-Yau
from \reef{J2} and \reef{O3} transform as
\bea
                \bar\sigma(J_2)~=~-J_2\ , \quad
                \bar\sigma(\Omega_3)~=~e^{2i\th}\bar\Omega_3\ .
\eea
The resulting O6-plane preserves ${\cal N}=1$ supersymmetry and
wraps the fixed locus of the anti-holomorphic involution
$\bar\sigma$. Its topological class in homology is denoted
$\Pi_{\rm O6}$, the Poincar\'{e} dual 3-form of $\Pi_{\rm O6}$  is
called $\pi_{\rm O6}$ in the following.


\subsubsection{Tadpole cancellation}

As discussed in generality in section \ref{sectad} the orientifold
plane induces a tadpole for the RR 7-form potential $C_{7}$ with four legs on
the flat uncompactified part  ${\mbb R^{1,3}}$.
This can be derived from the  CS-terms \reef{Oplact} on the O6-plane
\bea
\label{cs1a}
  \cs_{\rm CS}^{\rm O6}=  -4\mu_6\int_{\cy}
       C_{7} \wedge \pi_{\rm O6}\ ,
\eea
In order to cancel this tadpole we add stacks of $N_a$ D6-branes
wrapping three-cycles $\Pi_a$ in the internal manifold $\cx$ and
with flat gauge connections on their world volume. For
$\bar\sigma$ to be a symmetry, one has to introduce ``mirror
branes'' wrapping the three-cycles $\Pi'_a$, images under the
action induced by $\bar\sigma$ on the homological classes. The CS
action \reef{cs} on a stack of D6-branes simplifies to
\bea
\label{cs2a}
  \cs_{\rm CS}^{{\rm D6}_a} =    \mu_6\,\int_{\cy}
       C_{7} \wedge \pi_a
\eea
with a similar contribution from the mirror cycles $\pi'_a$.
From these CS terms it is straightforward to derive the tadpole
cancellation condition for $C_{7}$, the analog of \reef{loctad}.
It is
\bea
\label{tad1a}
             \sum_{a} N_a\, (\Pi_a+\Pi'_a) = 4\, \Pi_{\rm O6}\
             .
\eea
The total number of stacks is $2K$, split into equal number of
branes and their images. Whenever a stack is invariant under
$\O\bs$ one has to correct by a factor $\frac12$. The condition
\reef{tad1a} simply means that the total homology class of all
D6-branes and the orientifold plane has to vanish.


\subsubsection{Massless spectrum}

The massless spectrum in the closed string sector was already
described in section \ref{oriclspec} following
\cite{bbkl02,Grimm:2004ua}.
Besides the universal chiral multiplet there are $h^{2,1}$ complex
structure moduli each consisting out of a real complex structure
modulus and a RR scalar. In addition, there are $h_+^{1,1}$ vector
multiplets and the $h_-^{1,1}$ chiral multiplets containing the
complexified K\"ahler moduli.

In the open string sector one gets various non-abelian gauge
fields in addition to chiral charged matter. If a D6-brane wraps a
submanifold that is invariant under $\O\bs$ the gauge symmetry is
$SO(2N_a)$ or $Sp(2N_a)$.\footnote{Note that it is not sufficient
that the cycle $\Pi_a$ is mapped to itself under the induced
action of $\O\bs$ in homology. The submanifold representative of
the cycle has to be invariant.} In general, the cycles are mapped
non-trivially, $\Pi_a'\ne\Pi_a$, and the gauge symmetry is
$U(N_a)$. The chiral massless spectrum is given by the topological
intersection numbers. For a gauge group $\cg=\prod_{a=1}^K U(N_a)$
it is given in table \ref{tcs}.

\begin{table}
\centering
\begin{tabular}{|c|c|c|}
\hline
Non-abelian representation & $U(1)$ charges &    Multiplicity \\
\hline \hline
$\Yasymm_a$
 & $(2_a)$ & ${1\over 2}\left(\Pi'_a\circ \Pi_a+\Pi_{{\rm O}6}
\circ \Pi_a\right)$  \\
$\Ysymm_a$
   & $(2_a)$ &   ${1\over 2}\left(\Pi'_a\circ \Pi_a-\Pi_{{\rm O}6} \circ \Pi_a\right)$   \\
$(\antifund_a,\fund_b)$
 & $(-1_a,1_b)$ & $\Pi_a\circ \Pi_{b}$   \\
 $(\fund_a,\fund_b)$
 & $(1_a,1_b)$ & $\Pi'_a\circ \Pi_{b}$
\\
\hline
\end{tabular}
\caption{Chiral spectrum for intersecting D6-branes} 
\label{tcs}
\end{table}

The common situation is an open string connecting two distinct branes,
identified under $\O\bs$ with another string between the two
images. These open string states are in bifundamental representations
of the two factors in the gauge group.
Open strings stretched between a D-brane and its image under
$\bar\sigma$ are the only ones left invariant under the combined
operation $\Omega\bar\sigma (-1)^{F_L}$. Therefore, they transform
in the antisymmetric or symmetric representation of the gauge
group. More concretely, it turns out that chiral states localized at
 intersection points invariant under
$\Omega\bar\sigma (-1)^{F_L}$ transform in the antisymmetric representation
and intersections points which are anti-invariant give rise to
chiral states in the symmetric representation.

Additional non-chiral matter transforming in the adjoint
representation of $U(N_a)$ arises from open strings stretched
between branes in the same stack, i.e.\ branes lying on top of
each other. Geometrically, these correspond to deformations of the
three-cycle inside the Calabi-Yau, respectively to Wilson lines
along non-trivial one-cycles inside the three-cycles. For
supersymmetric cycles the number of these two kinds of moduli are
equal. They combine into complex scalars in the adjoint whose
multiplicity is then given by the first Betti number $b_1(\Pi_a)$
of the three-cycle.

For the spectrum of charged matter fields in table~\ref{tcs} 
the RR tadpole cancellation condition (\ref{tad1a}) guarantees the absence of non-abelian gauge
anomalies. Thus, the condition \reef{irranom} is satisfied for any
factor $a$ of the gauge group by virtue of \reef{tad1a}. Using
\reef{anomtr} the contribution of the states listed in table
\ref{tcs} to the anomaly \reef{irranom} for the factor $SU(N_a)$ in the gauge group is proportional to
\bea
 {\cal A}^{aaa}
 &\propto& \sum_{b\ne a} N_b \Big[-\Pi_a\circ\Pi_b+\Pi'_a\circ\Pi_b\Big] \nonumber\\
             &&  +~ {{N_a-4\over 2}} \Big[\Pi'_a\circ\Pi_a+\Pi_{\rm O6}\circ \Pi_a\Big]
                 +{{N_a+4\over 2}} \Big[\Pi'_a\circ\Pi_a-\Pi_{\rm O6}\circ \Pi_a\Big] \nonumber\\
        &=& -\Pi_a \circ \Big(\sum_b N_b \Big[\Pi_b+\Pi_b'\Big] -4\,  \Pi_{\rm O6} \Big) ~=~ 0 \ .
\eea
There can be additional non-chiral massless matter fields whose
spectrum cannot  be determined from
topology, instead one really has to compute the number of points
in which two submanifolds intersect geometrically (not just the
intersection number in topology which counts intersections with
orientation) or use conformal field theory methods. In the latter
case the combination of the annulus and the M\"obius strip
amplitude allows one to read off the complete massless spectrum,
as was discussed in simple toroidal models in section
\ref{secangle1loop}.


\subsubsection{K-theory constraints}
\label{ktheorya}

The topological classification of D-branes via cohomology actually
has to be refined by using K-theory instead \cite{Minasian:1997mm,Witten:1998cd}.
This means that besides the usual RR tadpole cancellation
conditions (which are conditions on the homology of the cycles
wrapped by the branes) additional constraints arise due to torsion
factors in the K-groups of the Chan-Paton bundles of the D-branes.
For more background material on K-theory and the relevant
applications to string theory and in particular orientifolds we
refer the reader to \cite{Witten:1998cd}.

For instance, for the type I string there are various K-theory
groups consisting of finite abelian groups (torsion)
\bea
    KO(\mbb S^1)=\mbb{Z}_2\ , \quad  KO(\mbb S^2)=\mbb{Z}_2\ , \quad
KO(\mbb S^9)=\mbb{Z}_2\ , \quad
    KO(\mbb S^{10})=\mbb{Z}_2\ .
\eea
These imply that there exist non-BPS D8-, D7-, D0- and
D$(-1)$-branes carrying a discrete $\mbb{Z}_2$ charge. Due to the
conservation of this charge the branes are stable.

For type IIA orientifolds something similar is expected to happen.
For a compact Calabi-Yau  it is in general quite difficult to
explicitly compute the K-theory groups. However, it has been
pointed out in \cite{Uranga:2000xp} that often the K-theory
constraints guarantee the absence of discrete anomalies on string
theory probe branes. For four-dimensional intersecting D-brane
models this in particular refers to the absence of $Sp(2N)$ global
Witten anomalies \cite{Witten:1982fp} which could exist on probe
branes with symplectic gauge fields. It is not clear in general
whether this captures all K-theory constraints for such models,
but at least it provides a number of additional constraints not
visible to the RR charge cancellation conditions. For the
$\mbb Z_2\times \mbb Z_2$ orientifold it has been shown in \cite{Maiden:2006qe}
that indeed the probe brane argument provides all K-theory constraints.
In practice, one
first classifies all D-branes carrying symplectic gauge symmetries
in a given model, and then requires that the number of fundamental
representations is even.
For the F/M theory dual origin of these K-theory  constraints see \cite{Garcia-Etxebarria:2005qc}.

\subsubsection{Green-Schwarz mechanism}

Given the chiral spectrum of table \ref{tcs}, we have shown that
the non-abelian gauge anomalies of all $SU(N_a)$ factors in the
gauge group vanish. On the other hand, the pure abelian, and the
mixed anomalies with abelian and non-abelian gauge fields and with
abelian gauge fields and gravitons do not cancel among the charged
fields of table \ref{tcs} alone. However, string theory provides
the \gs mechanism to cancel the mixed anomalies, as discussed in
section \ref{secgs}. It works for orientifold models with
intersecting D-brane \cite{afiru00a} (see \cite{Anastasopoulos:2006cz}
for a more general recent study of $U(1)$ anomalies). To demonstrate that we now
discuss in some detail the relevant axionic counter terms that
allow to confirm \reef{anomcan}.

The $U(1)_a-SU(N_b)^2$ anomalies $\ca^{abb}$ (with one abelian and
two non-abelian gauge bosons) result from the second triangle
diagram shown in figure \ref{fgsmc}.

\begin{figure}[ht]
\begin{picture}(200,200)(0,0)

\Photon(120,50)(180,50){2}{8}
\Gluon(220,80)(280,80){2}{8}
\Gluon(220,20)(280,20){2}{8}
\Vertex(180,50)2
\Vertex(220,80)2
\Vertex(220,20)2
\ArrowLine(180,50)(220,80)
\ArrowLine(220,80)(220,20)
\ArrowLine(220,20)(180,50)

\Photon(120,150)(180,150){2}{8}
\DashLine(220,180)(280,180)3
\DashLine(220,120)(280,120)3
\Vertex(180,150)2
\Vertex(220,180)2
\Vertex(220,120)2
\ArrowLine(180,150)(220,180)
\ArrowLine(220,180)(220,120)
\ArrowLine(220,120)(180,150)

\Text(108,50)[1]{$C_\m^a$}
\Text(276,20)[1]{$A^A_\m (T^b)^A$}
\Text(249,80)[1]{$A^A_\m (T^b)^A$}

\Text(026,150)[1]{$C_\m^a$}
\Text(182,120)[1]{$g_{\m\n}$}
\Text(155,180)[1]{$g_{\m\n}$}

\end{picture}
\caption{One-loop diagrams for mixed gravitational and $U(1)_a-SU(N_b)^2$ anomalies \lab{fgsmc}}
\end{figure}
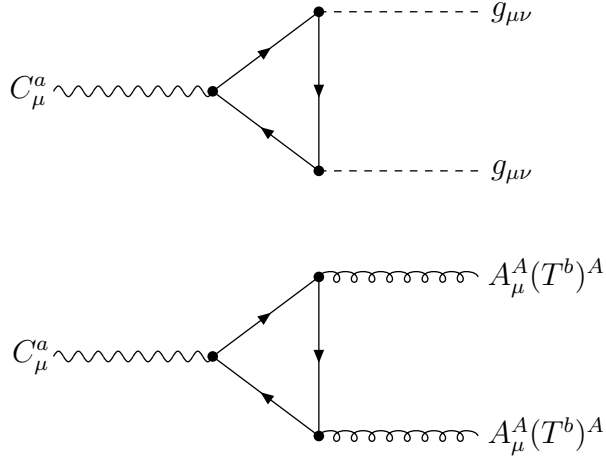

\noindent
The anomaly itself for any two stacks $a$ and $b$ is proportional
to
\bea
 \label{mixedan}
\ca^{abb}~\propto~ {N_a \over 2}\Big[
-\Pi_a+\Pi'_a\Big]\circ \Pi_b\ .
\eea
The mixed $U(1)_a$-gravitational anomaly $\ca^{agg}$ depicted by
the first diagram in figure \ref{fgsmc} is given by
\bea
\label{gravian}
\ca{}^{agg}~\propto~ {N_a }\, \Pi_{\rm O6} \circ \Pi_a\ .
\eea
To derive the Green-Schwarz couplings in the four-dimensional
effective action it is convenient to use the integral basis
$(A^\L,B_\L)$ of the homology $H_3(\cx,\mbb{Z})$ introduced in 
\reef{perOm}. 
In terms of the Poincar\'e dual basis of \reef{3int} 
the $(3,0)$-form $\O_3$ can be expanded as in \reef{expJOm}. 
The tree-level K\"ahler potential on the
complex structure moduli space of type IIA compactified on $\cx$
was also given in \reef{KpotOm}. 
Similarly, the three-forms that define the cycles wrapped by
D6-branes can be expanded in that basis as
\bea
\pi_{a}=\pi_a^\Lambda\,  \alpha_\Lambda - \pi_{a\Lambda}\, \beta^\Lambda\ ,
\quad\quad
\pi'_{a}=\pi'^{\Lambda}_a\,  \alpha_\Lambda - \pi'_{a\Lambda}\, \beta^\Lambda\ ,
\eea
and analogously for the O6-plane with coefficients $\pi_{\rm
O6}^\L$ and $\pi_{{\rm O6}\L}$.

To describe the physical degrees of freedom of the RR $p$-forms
reduced to four dimensions it is useful to employ the democratic
description discussed in section \ref{secIaII}. As follows from
the spectrum of the IIA orientifold on $\mbb T^6$ in
\reef{IIAonT6} the RR form $C_3$, and thus also its partner under
Hodge duality $C_5$, can be reduced along three-cycles. They
decompose into four-dimensional axions $C^{(0)}$ and two-forms
$C^{(2)}$ labelled by $\L$,
\bea
      {1\over \ell_s^3}\,  C_3&=& C^{(0)\Lambda} \alpha_\Lambda -
          C^{(0)}_\Lambda \beta^\Lambda \ , \quad
         {1\over \ell_s^3} C_5~=~ C^{(2)\Lambda}  \alpha_\Lambda -
          C^{(2)}_\Lambda \beta^\Lambda\  . \nonumber
\eea
In four dimensions $(C^{(0)}_\Lambda, -C^{(2)\Lambda})$ and
$(C^{(0)\Lambda}, C^{(2)}_\Lambda)$ are related by Hodge duality
to each other, by inheriting the duality \reef{RRHodge} of $F_4$
and $F_6$ from ten dimensions. Only half of these fields are
independent degrees of freedom.

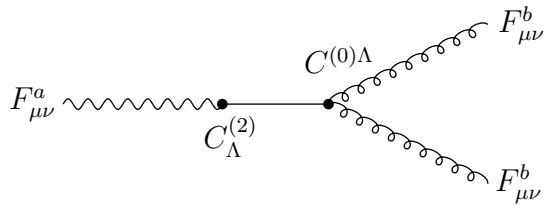
\begin{figure}[ht]
\hspace{1cm}
\begin{picture}(200,100)(0,0)

\Photon(120,50)(180,50){2}{8}
\Gluon(220,50)(280,80){2}{9}
\Gluon(220,50)(280,20){2}{9}
\Vertex(180,50)2
\Vertex(220,50)2
\Line(180,50)(220,50)

\Text(105,50)[1]{$F_{\m\n}^a$}
\Text(194,67)[1]{$C^{(0) \L}$}
\Text(126,38)[1]{$C^{(2)}_\L$}
\Text(208,20)[1]{$F_{\m\n}^b$}
\Text(182,80)[1]{$F_{\m\n}^b$}

\end{picture}
\caption{Tree-level anomalous contribution of RR axions \label{fgsmb}}
\end{figure}

The D6-brane couplings  \reef{cs} can now be dimensionally reduced
to four dimensions. In contrast to the derivation of the tadpole
cancellation condition, now also the gauge fields and curvature
two-forms with legs along the four-dimensional Minkowski space
have to be taken into account. Eventually, one obtains axionic
couplings of the form
\bea
\label{vertexak}
 \cs_{\rm ax}^{\rm D6} &=&\\
 &&\hspace{-.5cm}
 {1 \over   2\,(2\pi)} \sum_a
      \int_{\mbb R^{1,3}}  \Big[ {\rm tr}_{N_a} F^2_a -
            {N_a\over 48} {\rm tr} R^2 \Big] \Big( C_\Lambda^{(0)} \bigg( \pi^\Lambda_a + \pi'^\Lambda_a \bigg )
            + C^{(0)\Lambda} \bigg( \pi_{a\Lambda} + \pi'_{a\Lambda} \bigg) \Big) \ .
\nonumber
\eea
The orientifold action gives rise to an additional piece
\bea
\label{vertexako6}
 \cs^{\rm O6}_{\rm ax} &=&
  {1 \over  \, 48\,(2\pi)} \sum_a
      \int_{\mbb R^{1,3}} {\rm tr} R^2  \Big(
        C_\Lambda^{(0)} \bigg(\pi^\Lambda_a + \pi'^\Lambda_a\bigg)  +  C^{(0)\Lambda} \bigg(\pi_{a\Lambda} + \pi'_{a\Lambda}\bigg)\Big)\ .
\eea
Denoting the field strength two-form of the diagonal $U(1)_a$ in
$U(N_a)$ as $f_a$, the axion-gauge boson mixing terms
(Stueckelberg mass terms for the gauge bosons) can be written as
\bea
\label{masst}
\cs_{\rm mass}^{\rm D6} = {1\over \ell_s^2}\, \sum_a N_a
      \int_{\mbb R^{1,3}}  \Big(
        C_\Lambda^{(2)}\wedge  f_a \bigg(\pi^\Lambda_a - \pi'^\Lambda_a\bigg)+
       C^{(2)\Lambda} \wedge f_a \bigg(\pi_{a\Lambda} - \pi'_{a\Lambda}\bigg)\Big)\ .
\eea
The axionic couplings in \reef{vertexak} and \reef{vertexako6} are
not invariant under the gauge symmetries and contribute to the
mixed gauge anomalies at tree-level.  The contribution to the
mixed gauge anomaly is diagrammatically depicted in figure
\ref{fgsmb}.

\noindent
Adding up all the terms and taking the RR tadpole condition into
account, one can show that the result has precisely the form
(\ref{mixedan}) and (\ref{gravian}) and can cancel these field
theoretic anomalies. This provides an explicit check of the
relations \reef{anomcan} up to numerical factors. Said
differently, the mixed gauge anomalies are canceled by gauging
the shift symmetries $C^{(0)}\to C^{(0)}+{Q\over 2}\L_a$ of the
axions through the $U(1)_a$ gauge bosons $C^a_\mu \to
C^a_\mu+\partial_\mu\L_a$ as in \reef{axshift}. The charges $Q$
take the role of the Killing vectors of the gauging.\footnote{More
precisely, the Killing vectors are imaginary constants
proportional to $iQ$.}

As in \reef{Stmass} some linear combinations of the RR axions
provide the longitudinal modes for the gauge bosons of the
anomalous $U(1)_a$ which become massive via the couplings
\reef{masst}. As explained earlier, not only anomalous gauge bosons
can get massive through the couplings to the axions. Therefore, to
determine the low energy spectrum, one has to take the  couplings
(\ref{masst}) into account. Before the Green-Schwarz mechanism
is taken into account all perturbative string correlation
functions obey the selection rules for the $U(1)$ gauge
symmetries. After these have become massive via the axionic
couplings these selection rules still apply, since there is no
breaking of the symmetries in the \gs Lagrangian. Thus the massive
$U(1)$ symmetries still give rise to perturbatively exact global
symmetries of the low energy effective action. The continuous
shift symmetry of the RR axions is in fact known to be broken to a
discrete shift symmetry by instantons. Therefore, also the
continuous global $U(1)$ gauge symmetry under which the axions
transform is broken to a discrete subgroup. In the case of
intersecting D6-brane in type IIA string theory, the relevant
instantons are Euclidean D2-branes wrapped on three-cycles of the
underlying Calabi-Yau manifold, to which the three-form $C_3$
couples.


\subsubsection{Supersymmetry}

We have discussed the conditions under which D-branes preserve
supersymmetry in generality in section \ref{kappa}. Let us here
specialize the formalism to type IIA with calibrated D6-branes.

As explained earlier, type IIA string theory compactified on a
Calabi-Yau three-fold preserves ${\cal N}=2$ supersymmetry in four
dimensions which is broken by the orientifold projection to ${\cal
N}=1$ supersymmetry. One way to see this is to notice that the
three-cycle invariant under an anti-holomorphic involution has the
special Lagrangian property, which is known to be
the supersymmetry condition for a general three-cycle to
preserve supersymmetry \cite{bbs95,Kachru:1999vj}. A three-dimensional submanifold
$\S_a$ is called Lagrangian if the restriction of the K\"ahler
form vanishes,
\zbe
J_2 \vert_{\S_a} =0  \ .
\ee
If the submanifold in addition minimizes the volume among such
Lagrangian manifolds it is called special Lagrangian. This can be
formulated as the condition
\bea
\label{calia}
\Im(e^{-i\th_a}\, \Omega_3)\vert_{\S_a} =0  \ .
\eea
The constant parameter $\th_a$ parameterizes a $U(1)$ that
determines which ${\cal N}=1\subset {\cal N}=2$ supersymmetry is
preserved by the brane. Since the calibration forms are closed we
usually only refer to $\S_a$ as a three-cycle denoted $\Pi_a =
[\S_a]$.

Thus, different three-cycles  with different values for $\th_a$
preserve different ${\cal N}=1$ supersymmetries. The condition
(\ref{calia}) implies that the volume form on the three-cycle is given
by
\zbe\label{calib} d{\rm vol}|_{\S_a} =
\Re(e^{-i\th_a} \Omega_3)|_{\S_a}  \ ,
\ee
the three-cycle is calibrated with respect to the three-form
$\Re(e^{-i\th_a} \Omega_3)$. In order to preserve an overall
${\cal N}=1$ supersymmetry, all stacks of D6-branes have to wrap
special Lagrangian three-cycles with the angle defined by the
orientifold planes, $\th_a=\th$. Note that these
conditions are first of all valid in the geometric regime at large
radius and for a single D6-brane, i.e.\ in the abelian limit.

In the case supersymmetry is preserved also the NSNS tadpole
vanishes along with the RR tadpole, as shown in section
\ref{secangle1loop} for toroidal orientifolds. This can be
demonstrated
by using the above calibration conditions for curved background
spaces as well. The calibration condition \reef{calib} together
with the flatness of the Chan-Paton world volume gauge bundle of
sLag D6-branes implies that the DBI action \reef{dbi} evaluated on
the world volume is proportional to the right-hand-side of
\reef{calib}. In particular, it only depends on the topological
class of the submanifold wrapped by the brane as well. In that
case, it follows from the RR charge cancellation constraint
\reef{tad1a} that the sum of the D6-brane and O6-plane tension
vanishes.

It has been pointed out that there is another possibility for
supersymmetric D-branes in IIA Calabi-Yau orientifolds besides
D6-branes with flat gauge connections. These are the so-called
co-isotropic D-branes which, for instance, are D8-branes
stabilized by a non-trivial gauge bundle, i.e.\ endowed with
magnetic flux ${\cal F}$ (see also \cite{cu04} for a general
discussion on generalized calibrated submanifolds). Even though on a genuine Calabi-Yau
manifold there are no non-trivial five-cycles and thus no source
for the nine-form $C_9$ can be induced, these D8-branes do not
decay through the non-trivial D6-brane charge induced by the
magnetic flux. The supersymmetry conditions for these branes read
\bea
     d{\rm vol}\vert_{\Sigma_5} &=&2\pi\alpha' {\cal F}\wedge \Omega_3\vert_{\Sigma_5} \ ,  \\
     (2\pi\alpha' {\cal F} +i J)^2\vert_{\Sigma_5}&=&0\ .
\eea
Concrete D-brane model building using these kinds of branes has been performed
in \cite{Font:2006na}.


\subsubsection{Gauge couplings}

Each factor of the gauge group is supported on a stack of
D6-branes and comes with its own four-dimensional gauge coupling
$g_a$. The classical tree-level (disc diagram) expression can be
deduced from the DBI action.
In any supersymmetric gauge theory the
gauge coupling can be combined with an axionic theta-angle into
the holomorphic gauge kinetic function. In principle, the gauge
kinetic functions are not diagonal among different abelian factors
of the gauge group and are collected into a matrix $f_{ab}$. The
kinetic action of the gauge fields is given by the general form 
of the $\cn=1$ supergravity Lagrangian in \reef{eff4}. 
The matrix of gauge kinetic function actually turns out to be
diagonal at tree-level, $f_{ab} = \d_{ab} f_a$. Performing the
dimensional reduction of the DBI action the resulting classical
gauge coupling
is given by \cite{cim02,bbkl02,bls03}
\bea
\lab{kinYM}
{\rm Re}(f_a) ={1 \over 2\pi \ell_s^3}\,  e^{-\Phi}
                        \left\vert   \int_{\Pi_a}   {\Omega}_3 \right\vert\ ,
\eea
where $e^\Phi = g_s$. Thus, the gauge coupling depends on the
volume of the sLag three-cycles. Using (\ref{vertexak}) for the
axionic couplings of the RR fields, the gauge kinetic function for
a D6-brane wrapping a calibrated three-cycle is 
simply given by
\bea\lab{gcpl}
f_a = {1 \over 2\pi \ell_s^3}
\left[ e^{-\Phi} \int_{\Pi_a}   \Re( e^{-i\th_a}\, {\Omega}_3) - i\,
                           \int_{\Pi_a}  C_3 \right] \ .
\eea
The real part depends on the $h^{2,1}$ real complex structure
moduli of the Calabi-Yau manifold. They combine with the
components of $C_3$ along the three-cycles into complex scalars.
Moreover, it is known that in supersymmetric theories the gauge
kinetic function receives quantum corrections only at the one-loop
level.
Using the normalization ${\rm Tr}(T^A\, T^B)={1\over 2}\, \delta^{AB}$
for the generators of the non-abelian gauge groups the physical gauge couplings
become
\bea
    {4\pi\over g_a^2}={1 \over g_s\, \ell_s^3} \int_{\Pi_a}   \Re( e^{-i\th_a}\, {\Omega}_3).
\eea



\subsubsection{D-terms and Fayet-Iliopoulos couplings}

The tension of wrapped D6-branes induces a vacuum energy from the
effective four-dimensional point of view. It is a contribution to
the effective potential. We will see in a moment that in the case
of calibrated D6-branes this vacuum energy only depends on the
complex structure of the Calabi-Yau three-fold. From the arguments
in \cite{Brunner:1999jq,Kachru:2000ih} one therefore expects that
in the effective ${\cal N}=1$ field theory this vacuum energy does
not arise from a superpotential but instead from a D-term for the
$U(1)$ gauge groups localized on the D6-branes.
Since it is independent of K\"ahler moduli there are no $\alpha'$
corrections to this Fayet-Iliopoulos term, neither perturbative
nor non-perturbative (world sheet instantons).
For globally supersymmetric field theories with an anomalous $U(1)$ gauge
group a \FI term is dynamically generated only at the one-loop level \cite{Witten:1981nf,Fischler:1981zk}.
This statement has been generalized to \FI terms  
in string perturbation theory \cite{Dine:1987xk}.

The D-term scalar potential for scalars $\Phi_\a$ that transform
linearly and with charges $q_a^\a$ under the gauge transformations
of some $U(1)_a$ has the following general form
\bea
\label{dterm}
{\cal V}_{\rm D} =
\sum_a {1\over 2\, g_a^2}\biggl(\sum_\a q^\a_a|\Phi_\a|^2+{\textstyle {\xi_a}}\biggr)^2\ .
\eea
For later purposes we are working in the convention where in the
Lagrangian the kinetic terms of both the gauge fields and
the charged matter fields are multiplied by ${1/g_a^2}$.
In the minimum of
the D-term potential the $\Phi_\a$ can obtain a positive or
negative mass term for non-vanishing $\xi_a$, depending on the
sign of their $U(1)_a$ charges $q^\a_a$ and the sign of $\xi_a$. Supersymmetry
will only be unbroken, if the potential vanishes in the
groundstate. On the other hand, any charged condensate breaks the
gauge symmetry.

As we will discuss below, an uncanceled tension of the wrapped D6-brane and
O6-plane configuration  can
be interpreted as a non-vanishing D-term  potential energy. When
the vacuum expectation values of the charged fields vanish, i.e.\ with unbroken gauge symmetry, it
corresponds to the term ${1\over g_a^2} \xi_a^2$ in
(\ref{dterm}).
The scalars $\Phi_\a$ are
open strings fields which become massive, massless or tachyonic,
depending on their values in the minimum.

Let us determine the Fayet-Iliopoulos term for a brane
configuration which slightly breaks supersymmetry by violating the
calibration condition \reef{calia} by only a bit, i.e.\ we are
performing an expansion in (the integrals over)
$\Im(e^{-i\th}\Omega_3)$, where $\th$ is fixed by the
calibration of the orientifold plane and we choose it to be zero.
The disk-level scalar potential for
D6-branes wrapping sLag cycles is just the DBI action integrated
over the three-cycle. Summing over all D6-branes and the O6-planes
it can be written as \cite{bbkl02,Villadoro:2006ia,Diaconescu:2006nk}
\bea
\label{fayetil}
{\cal V}_{\rm DBI} &=& {1\over 2}
\m_6\, e^{-\Phi}\, \left[
\sum_a {N_a   \left| \int_{\Pi_a}  \Omega_3 \right| } +
 \sum_a{N_a   \left| \int_{\Pi'_a}   \Omega_3 \right| } -
4\, \int_{\Pi_{{\rm O}6}} \Re( \Omega_3) \right] \nonumber \\
&=&
 \m_6\, e^{-\Phi}\,
\sum_a N_a \left(  \left|  \int_{\Pi_a} \Omega_3 \right|-
                           \int_{\Pi_a} \Re(\Omega_3)
\right)  \ .
\eea
Using
\bea
     \left| \int_{\Pi_a}  \Omega_3 \right|=\sqrt{ \left( \int_{\Pi_a} \Re( \Omega_3) \right)^2 +
                        \left( \int_{\Pi_a} \Im( \Omega_3) \right)^2 }
\eea
and taking only the leading order term in
$\Im(\Omega_3)$ one obtains
\bea
{\xi^2_a \over 2 g_a^2} ~\simeq~ {2\pi  \over 2\, \ell_s^7 }\, e^{-\Phi}\,
 { \left( \int_{\Pi_a} \Im(\Omega_3) \right)^2 \over
\int_{\Pi_a} \Re(\Omega_3) }
+ \ \cdots \ .
\eea
Using the expression \reef{gcpl} for the gauge coupling, again to
leading order, one realizes that
\bea
\xi_a ~=~ {1  \over 2\pi \alpha' } \, {\int_{\Pi_a} \Im(
\Omega_3)\over  \int_{\Pi_a} \Re( \Omega_3)}+ \ \cdots \ ,
\eea
so that the Fayet-Iliopoulos term indeed vanishes if the cycle
$\Pi_a$ is special Lagrangian. Thus, the calibration condition on
the three-cycle can be rederived to leading order in the effective field theory
from minimization of the scalar potential.
This can be written in a compact way. Defining a complex valued central charge as
\bea
      Z_a = {1\over 2\pi\, \ell_s^3\, g_s} \int_{\Pi_a} \Omega_3\ ,
\eea
the $U(1)_a$ gauge coupling and the Fayet-Iliopoulos are given by
\bea
     {1\over g_a^2}=\left\vert Z_a \right\vert\ , 
\quad\quad 2\pi\alpha'\, \xi_a=\arg\left(Z_a \right)\ .
\eea
In this way it becomes clear that a choice of the angle variable $\th$ in 
the calibration condition fixes the \FI parameter.


\subsubsection{F-terms}

For compactifications with ${\cal N}=1$ supersymmetry in four
dimensions possible chiral massless fields $\Phi_i$  
can appear in the superpotential
$W(\Phi_i)$. It appears in the superspace action as a chiral
density
\bea
\cs=\int d^4x\, d^2\theta \  W(\Phi_i)+\ {\rm h.c.}\ .
\eea
A superpotential generates a scalar potential \reef{Fterm}
which, in case $\cv_{\rm F-term}$ shows a runaway behavior, can
also destabilize the vacuum \cite{Dine:1986zy,Dine:1987bq}. Therefore, for any potential ${\cal
N}=1$ vacuum it is a very important whether such an F-term
potential is generated, and if so, what the resulting surviving
moduli space is. For general type IIA orientifolds this is a
largely unsolved problem. 

In addition, in an instanton sector
charged matter couplings might be generated
which are absent perturbatively. In particular, we have seen
that anomalous $U(1)$ gauge symmetries due to the Green-Schwarz
mechanism survive as perturbative global symmetries, often
prohibiting certain charged matter couplings in the superpotential.
These global symmetries involve the axionic shift symmetries, which
are broken by the corresponding instantons.

\begin{figure}[ht]
\begin{picture}(200,170)(0,0)

\Line(50,100)(350,100)
\Line(50,100)(100,130)
\Line(100,130)(173,130)
\Line(277,130)(400,130)
\Line(400,130)(350,100)
\DashLine(277,130)(173,130)5

\Oval(225,70)(80,80)(0)
\Oval(225,115)(10,65)(0)
\ArrowLine(224,105)(226,105)

\Text(225,115)[1]{$D_\a$}
\Text(233,106)[1]{$\g_\a$}
\Text(280,93)[1]{D6-brane on $\Pi_\a$}

\end{picture}
\vspace{.5cm}
\caption{World sheet disk instantons \label{fgsmd}}
\end{figure}
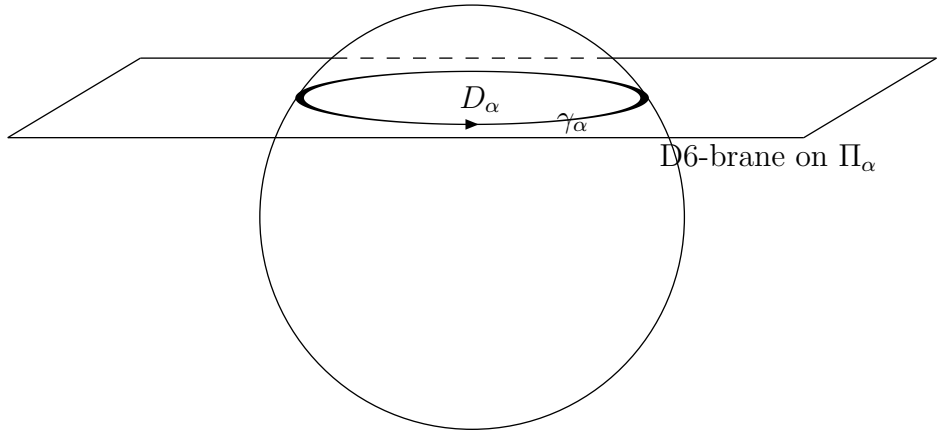

Let us discuss in some more detail  such superpotentials for
type IIA orientifolds with intersecting D6-branes. In this case,
besides the closed string moduli of complex structure and K\"ahler
deformations, one has open string moduli which describe the
infinitesimal deformations of the special Lagrangian three-cycles,
which are counted by $b_1(\Pi_a)$. Following the discussion in
\cite{Kachru:2000ih,Kachru:2000an}, and as shown in figure
\ref{fgsmd}, let us denote by $\{\gamma_\a\}$ a basis of one-cycles
inside $\Pi_a$. Since $b_1(\cx)=0$,
 one can find a volume minimizing disk $D_\a$ inside $\cx$,
which satisfies $\partial D_\a=\gamma_\a$. Then the volume of this
disc is given by $\omega_\a=\int_{D_\a} J_2$. This area defines a
natural candidate for an open string modulus. It is complexified
by the Wilson line $a_\a=\int_{\gamma_\a} A$ of the gauge
connection $A$ on the D-brane around the one-cycle $\gamma_\a$.
Very similar to the closed string case, the complex variable
\bea
   \Delta_\a=\omega_\a + i\, a_\a
\eea
is the bosonic part of a chiral superfield and the imaginary part
has a Peccei-Quinn shift symmetry. The latter implies that the
superpotential can only be a function of $\exp(-\Delta_\a/\a')$, so
that there can only be world-sheet instanton corrections to the
superpotential, no perturbative $\a'$-corrections.
To summarize, $W$ cannot be corrected by perturbative world-sheet
or string loop corrections.

In addition to these world sheet instantons, there can also
be space-time instantons, which are non-perturbative
in $g_s$. These are given by Euclidean D2-branes, called E2-branes,  wrapping
special Lagrangian three-cycles $\Xi$ of the Calabi-Yau manifold.
In order for such an instanton to contribute to the
superpotential, it must be 1/2-BPS. 
Any correlation function in this single-instanton background
contains the classical factor
\bea
\label{expfac}
             e^{-\cs_{\rm E2}}=\exp\left[ -{2\pi\over \ell_s^3}
           \left( {1\over g_s}\int_{\Xi} \Re(\Omega_3) - i \int_{\Xi} C_{3}
         \right) \right] \ ,
\eea
which depends exponentially on the complex structure moduli.
One can show that due to the GS mechanism, 
this instanton factor transforms under a  $U(1)_a$ gauge symmetry with 
infinitesimal parameter $\L_a$ as
\bea
       e^{-\cs_{\rm E2}}\to  e^{i\, Q_a({\rm E2})\,\Lambda_a}  \,\, e^{-\cs_{\rm E2}}
\eea
with
\bea
\label{chargee}
          Q_a({\rm E2})= {\cal N}_a\,\, \Xi\circ (\Pi_a - \Pi'_a)\ .
\eea
Therefore, the $U(1)$ charge of this term is given by the topological
intersection number of the three-cycle wrapped by the E2-instanton  and
the three-cycles wrapped by the D6-branes.  
If any of the charges $Q_a({\rm E2})$ is non-vanishing, it is clear
that a purely exponential superpotential of the form 
$W=c\cdot \exp ( -\cs_{\rm E2})$ cannot be generated due to
gauge invariance. Only if the exponential factor gets multiplied
by a suitable combination of matter superfields $\Phi_{ab}$ also charged under $U(1)_a$ can
a  superpotential
\bea
\label{superproduct}
                 W=  \prod_{ab} \Phi_{ab}\   e^{-\cs_{\rm E2}}
\eea
be gauge invariant.
The microscopic origin of the $U(1)_a$ charges
of the instanton factor is the appearance of extra
fermionic zero modes $\lambda_a$ living on the intersection of the instanton
with the D6-branes. This was worked out in detail in \cite{Blumenhagen:2006xt}
(see also \cite{Florea:2006si}).
The total number of such charged fermionic zero modes is displayed  in table
\ref{tablezero}, where we have split the 
topological intersection number as
\bea
             \Xi\circ \Pi_a =[\Xi\cap\Pi_a]^+ - [\Xi\cap \Pi_a]^-\ .
\eea
Note that,  if the instanton wraps a cycle which is also wrapped by a D6-brane,
additional bosonic and fermionic zero modes arise. To keep the presentation simple,
we neglect this possibility in the following.

\begin{table}[ht]
\centering
\begin{tabular}{|c|c|c|}
\hline
Zero modes&  Representation & Multiplicity   \\
\hline \hline
$\lambda_{a,I}$ &     $(-1_E,\fund_a)$     & $I=1,\dots, [\Xi\cap \Pi_a]^+$    \\
$\ov{\lambda}_{a,I}$ &  $(1_E,\antifund_a)$   & $I=1,\dots, [\Xi\cap \Pi_a]^-$    \\
\hline
$\lambda_{a',I}$ & $(-1_E,\antifund_a)$   & $I=1,\dots, [\Xi\cap \Pi'_a]^+$    \\
$\ov{\lambda}_{a',I}$    & $(1_E,\fund_a)$    & $I=1,\dots,[\Xi\cap \Pi'_a]^-$    \\
\hline
\end{tabular}
\caption{Zero modes on E2-D6 intersections
\label{tablezero} } 
\end{table}

\begin{figure}[h!]
  \begin{center}
    \includegraphics[width=50mm]{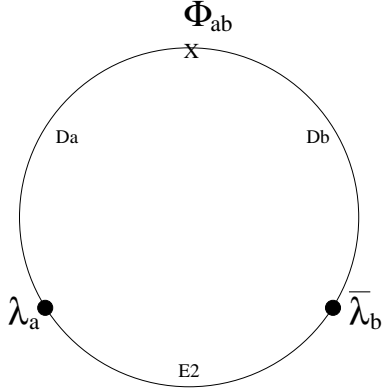}
    \caption{Standard disk tadpole.}
    \label{figga}
  \end{center}
\end{figure}

The single rigid E2-instanton contribution
to the charged matter superpotential was determined in \cite{Blumenhagen:2006xt}
in terms of open string CFT correlators.
In the simplest case, the E2 has to be placed in an $\ov\sigma$ invariant position
with gauge group $O(1)$. When each matter field soaks up
precisely two fermonic zero modes, the relevant instanton amplitude
 is given by  the following zero mode integral over disk and one-loop
open string CFT amplitudes
\bea
\label{instam}
     && \langle \Phi_{a_1,b_1}\cdot\ldots\cdot   \Phi_{a_M,b_M}
\rangle_{E2-{\rm inst}} = \nonumber \\
&& 
 = \int d^4 x\, d^2\theta \,\,
       \sum_{\rm conf.}\,\,  {\textstyle
  \prod_{a} \bigl(\prod_{i=1}^{ [\Xi\cap
             \Pi_a]^+}  d\lambda_a^i\bigr)\,
               \bigl( \prod_{i=1}^{ [\Xi\cap
             \Pi_a]^-}  d\overline{\lambda}_a^i\bigr) } \ \,\, \exp ({-S_{E2}}) \,
         \times \, \exp \left({Z'_0}\right) \,  \nonumber \\
&&\phantom{aaa}
\times \langle \Phi_{a_1,b_1}   \rangle_{\lambda_{a_1},\overline{\lambda}_{b_1}}\cdot
            \ldots \cdot  \langle \Phi_{a_M,b_M}
          \rangle_{\lambda_{a_M},\overline{\lambda}_{b_M}} \,
\eea
with the one-loop Pfaffian
\bea
\exp \left({Z'_0}\right)=\exp \left( {\textstyle \sum_a \left[ {Z'}^A ({\rm E2},{\rm D6}_a) \right] +  {Z'}^M({\rm E2},{\rm O6})} \right)\ .
\eea
Here in an annulus partition function ${Z'}^A ({\rm E2},{\rm D6}_a)$ for open strings between
the E2-instanton and a D6-brane, the zero modes have to be removed as
the integral over them is carried out explicitly. 
The formula (\ref{instam}) contains from the disk the
exponential instanton action and
the combinatorics of disk tadpole diagrams with two charged
fermionic zero modes attached to each disk (see figure \ref{figga}).
At the  one-loop level, the annulus and M\"obius strip
amplitudes provide the non-vanishing  exponential vacuum contribution.
This is nothing else than the one-loop Pfaffian/determinant over the fluctuations
around the instanton background.

The open string moduli dependence can be uncovered by additional $\Delta_{\alpha}$
insertions along the D6-boundaries, both in the disk and in the annulus diagrams.
The schematic  form of the superpotential, making the moduli dependence explicit, is
\bea
W=\sum_{\rm E2} \sum_{\rm discs}  e^{-\cs_{\rm E2}(U) }\, f\left(\exp\left( - {\textstyle {T\over \alpha'}}\right),
 {\rm tr}\left[ \exp\left( - {\textstyle { \Delta_\a\over \alpha'}}\right)\right], \Phi_{ab} \right)\ ,
\eea
where the exponential dependences are enforced by the fact that the imaginary
parts of the complex scalars are either axions or Wilson lines having a
shift symmetry.

The string tree-level superpotential, which is solely
generated by world-sheet instantons, has been 
discussed in \cite{Ooguri:1999bv,Aganagic:2000gs,Aganagic:2001nx,Lerche:2001cw}.
Using open string mirror symmetry,  in \cite{Aganagic:2000gs,Aganagic:2001nx} 
for a  non-compact example and  a specific D6-brane, the discs
instantons sum could really be extracted.

Let us sumarize some of the main effects of E2-instantons.
Instantons wrapping rigid supersymmetric three cycles can generate charged
matter couplings in the superpotential, which can
potentially destabilise the vacuum or lead to  new effects
in the four-dimensional action which are absent in perturbation theory. 
In this case their contribution,
though exponentially suppressed, yields the leading order terms.
As an example, E2-instantons  with appropriate intersections
with the two D-branes supporting the right-handed
neutrinos in an MSSM like intersecting D-brane model can give rise to
Majorana mass-terms \cite{Blumenhagen:2006xt,Ibanez:2006da}. The scale of these masses
is
\bea
           M_M \propto M_s\, e^{-{2\pi\over \ell_s^3\, g_s} {\rm Vol}_{\rm E2} }
\eea
with a constant of proportionality that is expected to be of order $\co(1)$.
These  can easily generate a hierarchy between the string and
the Majorana mass scale.

In general, summing up all multiple E2- and world-sheet instanton corrections,
one expects a
complicated F-term potential over the combined open string, K\"ahler and complex structure
moduli space which is largely unexplored.
To compute this instanton expansion directly is still a horrendous task,
as one does not understand the background well enough to
determine all non-trivial holomorphic discs and special Lagrangian three-cycles. 


\subsection{Toroidal models}

So far we have presented the general formalism for the
construction of intersecting D6-brane models on type IIA
orientifolds. As we have seen, the effective DBI action allows one
to derive some of the phenomenologically important aspects of the
models quite easily, at least at the classical level. Many
concrete intersecting D-brane models have been discussed in detail
in the literature. For most of them the underlying Calabi-Yau
manifold is a toroidal orbifold. The main reason for this choice
is of technical nature, as for toroidal models the formulae from
the last section can be evaluated quite explicitly, since all
background fields are constant. As the simplest example and since
it underlies most of the constructions existing in the literature,
let us go through the formalism on the six-dimensional torus as 
presented in \cite{bgkl00a}.


\subsubsection{Intersecting brane models on the torus}
\label{sssibt}

For simplicity we assume that the six-dimensional torus factorizes
as before in \reef{torfact}, by which we mean that the background
metric, NSNS $B$-field are block-diagonal and the brane
configurations also respect the factorization.
As in section  \ref{secinttor}, for each $\mbb T_I^2$ we choose the lattice vectors
\beqn
\vec{\bf e}^{\, I}_1 = \frac1{\sqrt{\a'}}(0, R^I_1) \ , \quad
\vec{\bf e}^{\, I}_2 = \frac1{\sqrt{\a'}}(R^I_2 \sin(\th) , R^I_2 \cos(\th) ) \ .
\eeqn
defining a torus with complex structure
\beqn\lab{cplstrI}
u^I = u_1^I + i u_2^I = \frac{R^I_2}{R^I_1} e^{i\th^I}.
\eeqn
Invariance of the tori under the orientifold projection $\vec
{\bf e}^{\, I}_1 \mapsto -\vec {\bf e}^{\, I}_1$ and $\vec {\bf e}^{\, I}_2\mapsto
-2u_1^I \vec {\bf e}^{\, I}_1 +\vec {\bf e}^{\, I}_2$ requires
$u_1^I\in\{0,\frac12\}$.
Factorizable D6-branes are specified by co-prime integer wrapping
numbers $(p_a^I,q_a^I)$ along the fundamental one-cycles $ e^{\,
I}_1$ and $e^{\, I}_2$.
It is useful to express also the branes for the tilted tori in
terms of the untilted one-cycles.
Then a three-cycle is specified by a triplett
of wrapping numbers
\zbe \label{torcycle}
     (\tilde p_a^I, q_a^I)\ ,
     \quad I=1,2,3
\ee
along the two fundamental one-cycles on each $\mbb T_I^2$.
with
\beqn
\tilde p^I_a = \left\{
\begin{array}{ll}
p^I_a & {\rm for}\ \Re(u^I)=0\\
p^I_a+{1\over 2} q^I_a & {\rm for}\ \Re(u^I)=\frac12
\end{array}\right. \ .
\eeqn
Note that $\tilde p_a^I$ can be half-integer whenever
$u_1^I=\frac12$.
The intersection number between two three-cycles can be computed
as
\zbe
\label{inti}
    \Pi_a\circ \Pi_b =\prod_{I=1}^3 (\tilde p^I_a\,  q^I_b-  q^I_a\, \tilde p^I_b)
           =\prod_{I=1}^3 (p^I_a\, q^I_b- q^I_a\, p^I_b)\ .
\ee
To work out the tadpole cancellation conditions one has to
determine the three-cycle of the O6-plane and the action of the
anti-holomorphic involution on the  D6-branes.
Independent of the
tilt on each $\mbb T_I^2$, the O6-plane stretches along the real axis.

The action of $\bar\sigma$ on a general three-cycle is simply
$(\tilde p^I, q^I)\to
(-\tilde p^I, q^I)$. Expanding the general tadpole cancellation
condition for the homological RR charges (\ref{tad1a}) one obtains
the four independent equations
\bea
\label{tad}
&& \sum_{a}   N_a \prod_I q^I_a = 16\ , \nonumber \\
&& \sum_{a}   N_a\, \tilde p^I_a  \tilde p^J_a q^K_a=0\quad {\rm for}\ J\neq I\neq K\neq J\ .
\eea
These conditions were also derived in section \ref{secangle1loop}
from the requirement that the divergences of the RR contribution
to the one-loop partition function cancel.

As discussed in subsection \ref{ktheorya}, the tadpole conditions
are supplemented with the following  K-theory constraints
\cite{FM03}. Employing the probe brane argument, one first has to
determine the branes yielding $Sp(N)$ gauge groups.
Requiring that the resulting
total number of fundamental representations with respect to the
symplectic gauge groups is even leads to the following three
constraints
\bea
\label{ktheory}
 \sum_{a}   N_a\tilde
p^I_a\,  q^J_a q^K_a =0\ {\rm mod}\ 2 \quad {\rm with}\ I\ne J\ne
K\ne I
\ .
\eea
Concerning supersymmetry, The Lagrangian condition
$J_2\vert_{\Pi_a}=0$ is always satisfied for flat factorizable
branes. Using the metric in \reef{tormetric} the calibration
condition $\Im(\Omega_3)\vert_{\Pi_a}=0$ can be brought to the
following form
\zbe
\label{susya}
\prod_{I=1}^3 \tilde p^I_a - \sum_{I\ne J\ne K \ne I} \tilde p^I_a q^J_a q^K_a\,
   u_2^J u_2^K =0\ .
\ee
A further constraint arises from the condition $\Re(\Omega_3)|_{\pi_a}>0$,
which takes the form
\zbe
\label{susyb}
\prod_{I=1}^3  q^I_a - \sum_{I\ne J\ne K \ne I} q^I_a\, {\tilde p^J_a\tilde p^K_a
   \over u_2^J u_2^K } >0\ .
\ee
We conclude that for a  given D-brane with definite wrapping
numbers the supersymmetry condition (\ref{susya}) puts a
constraint on the complex structure moduli. This constraint is
required to make the vacuum energy induced by the brane tension
vanish, as discussed earlier.


\subsubsection{Non-supersymmetric intersecting D-brane constructions}
\label{sssg}

Toroidal intersecting D-brane models were the first models that
have been studied systematically \cite{bgkl00a,aads00}. It was realized
that it is not possible to get any chiral supersymmetric model on
this background, mainly for the reason that there is only one
O6-plane along the  real axes. Nevertheless, by ignoring the
instabilities induced by the uncanceled NSNS tadpoles
\cite{bklo01} (see also \cite{ekn02,Axenides:2003hs}), there have been several attempts to construct
semi-realistic non-supersymmetric intersecting D-brane models on
tori. Since these more phenomenological aspects have been covered
in the review \cite{Blumenhagen:2005mu}, let us here only briefly
summarize some of the main ideas.

The minimal realization of the Standard Model gauge group started
from four stacks of D-branes as in figure \ref{quiver}. The
straightforward generalization of the four stack realization of
the Standard Model is to use more than four stacks of D6-branes
\cite{CK02a,CK02b}. Similarly, one can try to find models with
characteristics and gauge groups of Grand-Unified-Theories (GUT)
in this toroidal set-up \cite{CK02,CK02d}. As one is giving up
supersymmetry there is a priori no need to introduce orientifold
planes in the first place, since the RR charge can be canceled
among branes and anti-branes. One can simply start with
intersecting D6-branes in type IIA \cite{afiru00,afiru00a}.

Another approach is not to work with D6-branes but instead with
D4-branes in IIA, respectively D5-branes in IIB. In order to
achieve chirality one has to perform an additional orbifold
projection in the transverse space \cite{afiru00,afiru00a}.
Therefore, the models constructed in
\cite{afiru00,afiru00a,fhs01,bkl01,GH01,GH02,cim02b,CK02c,bkl02a,bkl02b,DB02,bkl02c}
can be regarded as a hybrid of these two methods to obtain chiral
fermions, namely as intersecting branes at singularities. Without
supersymmetry one can also study orientifolds of type 0 string
theory \cite{bkl02}.

A peculiarity about the nature of supersymmetry breaking in
intersecting D-brane models has been pointed out in
\cite{cim02,cim02a,cim02c}, namely that one can build models in which at each
intersection between two D6-branes some ${\cal N}=1$ of the
initial $\cn=2$ supersymmetry is preserved, but not all
intersections preserve the same $\cn=1$ supersymmetry. In such
models the absence of one-loop corrections to the Higgs mass
weakens the gauge hierarchy problem and allows one to enhance the
string scale up to ${\rm 10\, TeV}$. These models were called
quasi-supersymmetric. 

An alternative way to break supersymmetry is via the Scherk-Schwarz
mechanism. This has been generalized to the intersecting
D-brane context in \cite{Angelantonj:2005hs}. 
Magnetized D-branes have also been argued to be a stringy
realization of the split-supersymmetry scenario 
\cite{Arkani-Hamed:2004fb,Antoniadis:2004dt,Antoniadis:2006eb}.

Another interesting idea has been presented in \cite{Ott:2005bf},
where it was proposed that a non-supersymmetric intersecting
D-brane model on $\mbb T^9$ might explain the emergence of four
large and six compact dimensions via its scalar potential.


\subsubsection{Intersection numbers on orbifolds}

In order to introduce additional orientifold planes one can
perform a toroidal orbifold. As explained in section
\ref{secangle1loop}, for $\Theta$ generating a $\mbb{Z}_N$
action on $\mbb T^6$ there are new orientifold planes located at
the fixed point loci of $\Omega\bar\sigma\Theta^k$ with
$k=0,\ldots,\,  N-1$. In order to construct the orientifold one
can either use conformal field theory techniques or, for just the
chiral spectrum and RR tadpole constraints, one can utilize the
homological formalism introduced in section \ref{sectypeiia}.

The spectrum in table \ref{tcs} can be applied to orbifold models
as well. In doing so one has to use the intersection numbers on
the resolved orbifold space and not on the ambient torus. Some
three-cycles $\Pi_a$ on the orbifold space are inherited from the
torus. Under a $\mbb{Z}_N$ orbifold group, three-cycles $\Pi^{\mbb
T}_a$ of the torus are arranged in orbits of length $N$ \cite{bbkl02}, i.e.
\be
\Pi^{\rm O}_a = \sum_{j=0}^{N-1}
\Theta^j\,\Pi^{\mbb T}_a\ .
\ee
Such an orbit can then be considered as a three-cycle of the
orbifold, the intersection numbers given by
\zbe
\Pi^{\rm O}_a\circ\Pi^{\rm O}_b={1\over N}
\left(\sum_{j=0}^{N-1} \Theta^j\, \Pi^{\mbb T}_a
  \right) \circ \left(\sum_{k=0}^{N-1} \Theta^k\, \Pi^{\mbb T}_b \right) .
\ee
Beside these untwisted three-cycles, certain twisted sectors of
the orbifold action can give rise to additional twisted
three-cycles. They correspond to massless fields in the twisted
sectors of the orbifold and will be discussed in some more detail
in section \ref{secrigid}.


\subsubsection{Intersecting D6-branes on the $\mbb{Z}_2\times \mbb{Z}_2$ orbifold}
\label{IntersectingZ2xZ2}

As pointed out at the end of the last section, to obtain
non-trivial supersymmetric models one needs more orientifold
planes altogether extending along the real and imaginary axes. The
easiest way to obtain   these is by considering not just tori but
toroidal orbifolds, of which the $\mbb{Z}_2\times \mbb{Z}_2$
orbifold is the simplest one \cite{fhs00,csu01,csu01a}. The general
features of this model have been introduced already in section
\ref{secZ22}. The
orbifold action of the two $\mbb{Z}_2$ symmetries is defined in
\reef{reflZ22}. The two generators we here denote as $\Th$ and $\Th'$
for convenience. 
Together with the world sheet parity $\O\bs$ these generate the
orientifold group $G_\O$ as in \reef{origroup}.
First we consider the model in which the $(1,1)$-forms in the twisted
sectors are invariant and the $(2,1)$-forms removed from the
spectrum. According to table \ref{tabZ22models} this model with Hodge
numbers $(h^{2,1},h^{1,1})=(3,51)$ has no
discrete torsion \cite{Vafa:1994rv}.
There are precisely eight three-cycles in the untwisted sector.

In order to deal with three-cycles on the orbifold space we have
to carefully distinguish between three-cycles on the covering
space and three-cycles on the actual orbifold. In the particular
case at hand, under the action of $\mbb{Z}_2\times \mbb{Z}_2$ a
factorizable three on $\mbb T^6$ has three images, all of them
with the same wrapping numbers as the initial three-cycle.
Therefore, a three-cycle $\Pi_a^{\rm B}$ in the  bulk of the
orbifold space can be identified with $\Pi_a^{\rm B} = 4\,
\Pi_a^{\mbb T}$. Computing the intersection number we get
\zbe \label{prod}
\Pi_a^{\rm B} \circ \Pi_b^{\rm B} =
4\prod_{I=1}^3 \bigg(p^I_a\, q^I_b
- q^I_a\, p^I_b\bigg)\ ,
\ee
Therefore, the cycles $\Pi_a^{\rm B}$ do not span an  integral
homology lattice,
which suggests that there exist smaller three-cycles in the
orbifold space. This is indeed the case. By choosing the
three-cycles to run through fixed points of either $\mbb{Z}_2$, we
obtain three-cycles which are given by $\Pi^{\rm O}_a={1\over
2}\Pi^{\rm B}_a$, which have intersections on the orbifold
$\Pi^{\rm O}_a\circ \Pi^{\rm O}_b =\Pi_a^{\mbb T}\circ \Pi_b^{\mbb
T}$. 

Working out the fixed point locus of the four non-trivial elements of the
orientifold group $G_\O$, $\Omega\bar\sigma(-1)^{F_L}$, $\Omega \bar\sigma\Theta (-1)^{F_L}$,
$\Omega \bar\sigma\Theta'(-1)^{F_L}$, $\Omega
\bar\sigma\Theta\Theta'(-1)^{F_L}$, and expressing everything
in terms of three-cycles $\Pi^{\rm O}_a$, we obtain
the four tadpole cancellation
conditions for the homological RR charges
\bea
\label{tadorb}
&& \sum_{a}   N_a \prod_{I=1}^3 q^I_a = 8\ , \\
&& \sum_{a}   N_a\, q^I_a\,  \tilde p^J_a\,
\tilde p^K_a = -2^{3-2u_1^J-2 u_1^K}\ ,\quad      {\rm with}\ I\ne
J\ne K \ne I\ .
\nonumber
\eea
D6-branes on top of one of the four orientifold planes always
yield an $Sp(2N_a)$ gauge symmetry \cite{Berkooz:1996dw}. One can
also compute the four K-theory constraints for these factors in
the gauge group. The supersymmetry conditions are the same as for
$\mbb T^6$ shown in (\ref{susya},\ref{susyb}).
The minus sign in the second line of \reef{tadorb} refers to the
convention to choose the orientation of the O6-plane such
that relative rotation angles add up to zero. 

One of the phenomenological
disadvantages of this $\mbb{Z}_2\times \mbb{Z}_2$ orientifold is
that none of the three-cycles in rigid, which means that one
always gets one chiral multiplet in the adjoint representation of
the $U(N_a)$ gauge groups. As worked out in
\cite{Blumenhagen:2005tn}, the other $\mbb{Z}_2\times \mbb{Z}_2$
orientifold, where one keeps instead of the twisted sector
two-cycles the twisted three-cycles, allows for rigid three-cycles.
Let us demonstrate this in more detail, as it also shows how the
twisted sector three-cycle can be dealt with in the topological
formalism.


\subsubsection{Rigid three-cycles}
\label{secrigid}

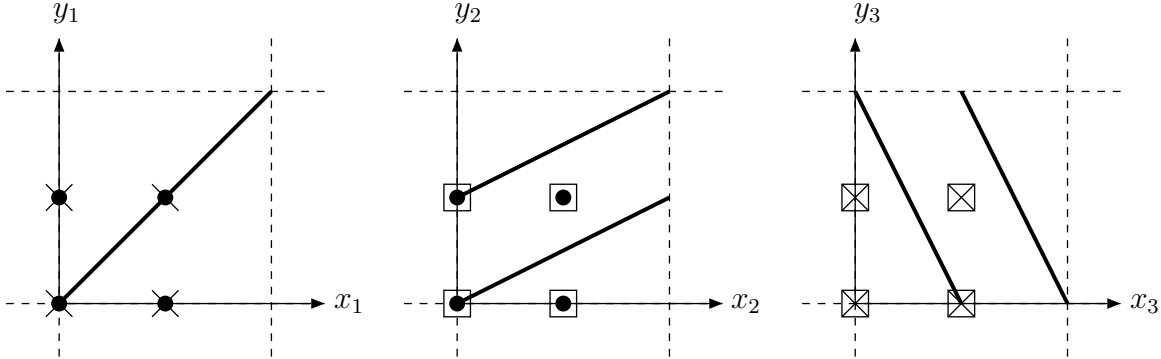
\begin{figure}[ht]
\hspace{.5cm}
\begin{picture}(200,150)(0,0)

\LongArrow(0,20)(100,20)
\LongArrow(150,20)(250,20)
\LongArrow(300,20)(400,20)
\LongArrow(0,20)(0,120)
\LongArrow(150,20)(150,120)
\LongArrow(300,20)(300,120)

\DashLine(-20,20)(100,20)3
\DashLine(-20,100)(100,100)3
\DashLine(0,0)(0,120)3
\DashLine(80,0)(80,120)3

\DashLine(130,20)(250,20)3
\DashLine(130,100)(250,100)3
\DashLine(150,0)(150,120)3
\DashLine(230,0)(230,120)3

\DashLine(280,20)(400,20)3
\DashLine(280,100)(400,100)3
\DashLine(300,0)(300,120)3
\DashLine(380,0)(380,120)3

\Vertex(0,20)3
\Vertex(40,20)3
\Vertex(0,60)3
\Vertex(40,60)3

\Vertex(150,20)3
\Vertex(190,20)3
\Vertex(150,60)3
\Vertex(190,60)3

\EBox(145,15)(155,25)
\EBox(185,15)(195,25)
\EBox(185,55)(195,65)
\EBox(145,55)(155,65)

\EBox(295,15)(305,25)
\EBox(335,15)(345,25)
\EBox(335,55)(345,65)
\EBox(295,55)(305,65)

\Line(-5,15)(5,25)\Line(-5,25)(5,15)
\Line(35,55)(45,65)\Line(35,65)(45,55)
\Line(35,15)(45,25)\Line(35,25)(45,15)
\Line(-5,55)(5,65)\Line(-5,65)(5,55)

\Line(295,15)(305,25)\Line(295,25)(305,15)
\Line(335,55)(345,65)\Line(335,65)(345,55)
\Line(335,15)(345,25)\Line(335,25)(345,15)
\Line(295,55)(305,65)\Line(295,65)(305,55)

\def\axowidth{1.5 }
\Line(0,20)(80,100)
\Line(150,20)(230,60)
\Line(150,60)(230,100)
\Line(300,100)(340,20)
\Line(340,100)(380,20)
\def\axowidth{0.5 }

\Text(-5,130)[1]{$y_1$}
\Text(75,20)[1]{$x_1$}
\Text(198,20)[1]{$x_2$}
\Text(66,130)[1]{$y_2$}
\Text(295,20)[1]{$x_3$}
\Text(163,130)[1]{$y_3$}

\end{picture}
\caption{Fractional brane passing through four fixed points of each
element of the orbifold group. Fixed points of $\Theta$ are
denoted by dots, those of $\Theta'$ by squares, those of
$\Theta\Theta'$ by crosses. \label{fixedpoints}}
\end{figure}

Now consider the  type IIA $\mbb{Z}_2\times \mbb{Z}_2$ orientifold
with discrete torsion, i.e.\ now there are also three-cycles in the
three twisted sectors. More concretely, in addition to the
untwisted cycles we have 32 independent collapsed three-cycles for
each of the three twisted sectors, $\Theta$, $\Theta'$, and
$\Theta\Theta'$. Let us first consider the $\Theta$-twisted
sector. We denote the 16 fixed points on $(\mbb T^2_1 \times
\mbb T^2_2)/\mbb{Z}_2$ by $[e^\Theta_{ij}]$, with $i,j\in\{1,2,3,4\}$
(see figure \ref{fixedpoints}). After blowing up the orbifold
singularities, these become two-cycles with the topology of $\mbb S^2$.
Each such four-dimensional $\mbb T^4/\mbb Z_2$ is an orbifold of K3
before taking the other elements of the orientifold group into
account.

With our choice of discrete torsion, these two-cycles are
combined with a one-cycle  $(p^3,q^3)$ of $\mbb T^2_3$,
in order to form a three-cycle in the
$\Theta$-twisted sector. Let us denote a basis of such twisted
three-cycles as
\bea
[\alpha^\Theta_{ij,\,p}] = 2\, [e^\Theta_{ij}]\otimes  [\vec
{\bf e}_1^{\, 3}]\ , \quad\quad
[\alpha^\Theta_{ij,\,q}] = 2\, [e^\Theta_{ij}]\otimes [\vec {\bf
e}^{\, 3}_2]\ ,
\eea
where the factor of two is due to the action of the second  $\mbb{Z}_2$.
Analogously, we define the basic twisted three-cycles in the $\Theta'$ and $\Theta\Theta'$
twisted sectors as
\bea
\begin{array}{lcl} \vspace*{.2cm}
\quad [\alpha^{\Theta'}_{ij,\,p}] = 2\, [e^{\Theta'}_{ij}]\otimes
[\vec {\bf e}^{\, 1}_1]\ , & &
[\alpha^{\Theta'}_{ij,\,q}] = 2\, [e^{\Theta'}_{ij}]\otimes [\vec {\bf e}^{\, 1}_2]\ , \\
\quad [\alpha^{\Theta\Theta'}_{ij,\,p}] = 2\,
[e^{\Theta\Theta'}_{ij}]\otimes [\vec {\bf e}^{\, 2}_1]\ ,
& &
[\alpha^{\Theta\Theta'}_{ij,\,q}] = 2\, [e^{\Theta\Theta'}_{ij}]
\otimes [\vec {\bf e}^{\, 2}_2]\ .
\end{array}
\eea
The intersection number between a pair of such cycles is easy to
compute knowing that the collapsed two-cycles of the K3-orbifold have
self-intersection number $[e_{ij}] \circ [e_{kl}] = -2 \d_{ik}
\d_{jl}$ in each twisted sector, and that two-cycles of different twisted sectors do not
intersect. For the three-cycles $[\Pi^g_{ij,\,a}]= p_a^{I_g}
[\alpha_{ij,\,p}] +  q_a^{I_g} [\alpha_{ij,\,q}]$ and
$[\Pi^h_{kl,\,b}]= p_b^{I_h}[\alpha_{kl,\,p}] +  q_b^{I_h}
[\alpha_{kl,\,q}]$, with $g,h = \Th, \Th', \Th\Th'$, we find
\bea
\label{inttwist}
[\Pi^g_{ij,\,a}] \circ [\Pi^h_{kl,\,b}]\,
&=&
=\, 4\,  \d_{ik} \d_{jl} \d^{gh} \,
(p_a^{I_g}\, q_b^{I_g} - q_a^{I_g}\, p_b^{I_g})\ .
\eea
In this notation, for the sectors twisted by $g = \Theta, \Theta',\Theta\Theta'$ one has
$I_g = 3,1,2$, respectively.

Equipped with the above description of the untwisted and twisted
sector three-cycles, one can  build rigid D6-branes. Namely, one
considers fractional D6-branes charged under all three
twisted sectors of the orbifold. In order to construct such
D-branes, let us start with a factorizable three-cycle, described
by three pairs of wrapping numbers  $(p_a^I, q_a^I)$. A fractional
D6-brane should be invariant under the orbifold action, and hence
it must run through four fixed points for each twisted sector, as
illustrated in fig. \ref{fixedpoints}.

Let us denote such a set of four fixed points of the element $g$
(each labelled by a pair $(i,j)$) as $S_g^a$. Then the entire
three-cycle that such a fractional D-brane is wrapping is of the
form
\bea
\label{rigid}
\Pi_a\, &=&\\
&& \hspace{-.5cm}
{1\over 4}\, \Pi^{\rm B}_a +
{1\over 4} \sum_{(i,j) \in S_\Theta^a} \epsilon^\Theta_{a,ij}\,
\Pi^\Theta_{ij,\,a} +
 {1\over 4}  \sum_{(j,k)\in S_{\Theta'}^a} \epsilon^{\Theta'}_{a,jk}\,
\Pi^{\Theta'}_{jk,\,a} +
 {1\over 4} \sum_{(i,k)\in S_{\Theta\Theta'}^a}
\epsilon^{\Theta\Theta'}_{a,ik}\, \Pi^{\Theta\Theta'}_{ik,\,a}\ ,
\nonumber
\eea
where the signs
$\epsilon^\Theta_{a,ij},\,\epsilon^{\Theta'}_{a,jk},\,\epsilon^{\Theta\Theta'}_{a,ik}\,=\,\pm
1$ define the charge of the fractional brane $a$ with respect to
the massless fields living at the various fixed points.
Geometrically, these numbers indicate the two possible
orientations with which the brane can wrap around the blown-up
$\mbb S^2$. Clearly, only those fixed points appear in
(\ref{rigid}), which the D6-brane is passing through. Since the brane
is stuck at the orbifold fixed points in all three $\mbb T_I^2$,
there are no adjoint scalars appearing in the massless spectrum.
This can also be confirmed  by a direct CFT computation of the
corresponding boundary states. There appear additional constraints
which the interested reader can find in
\cite{Blumenhagen:2005tn}. Here we just wanted to explain how
twisted three-cycles and rigid three-cycles can be constructed in
a prototype model. In \cite{Blumenhagen:2005tn} all the other
details of this model have been worked out and supersymmetric
Pati-Salam type models were constructed. In \cite{Dudas:2005jx,Blumenhagen:2006ab} 
it was shown that  $\mbb Z_2\times \mbb Z_2$ shift  orbifolds \cite{Antoniadis:1999ux,GP02,LG03}
do also admit rigid cycles.


\subsubsection{Supersymmetric intersecting D-brane models on toroidal orbifolds}
\label{sssto}

As first shown in \cite{csu01,csu01a}, the first discussed
$\mbb{Z}_2\times \mbb{Z}_2$ orientifold without  discrete torsion
admits chiral supersymmetric intersecting D-brane models.
Given its relative simplicity, it is the background studied most
for systematic searches for Standard Model- or MSSM-like
\cite{cps02,Cvetic:2002qa,Cvetic:2002wh,cp03a,cp03a,clll04} and GUT-like
\cite{cps02,cll04,Chen:2005ab,Chen:2005mm} intersecting
D-brane configurations. For the phenomenological details of these
models we refer  the reader to the original literature or to the
review article \cite{Blumenhagen:2005mu}.

One way to generalize the $\mbb{Z}_2\times \mbb{Z}_2$ orientifolds
is to include additional shift symmetries in the
$\mbb{Z}_2$ actions \cite{GP02,GP03,LG03,Dudas:2005jx,Blumenhagen:2006ab}. These have the effect
of eliminating some of the orientifold planes present in the
original models, which makes it much harder to find interesting
supersymmetric models. On the other hand, it also gives rise to
twisted sector three-cycles, which allows for more general
fractional D6-branes.

Employing the topological methods introduced in section \ref{sectypeiia},
chiral supersymmetric intersecting D-brane models have been
studied on the $\mbb{Z}_4$ \cite{bgo02}, $\mbb{Z}_4\times \mbb{Z}_2$
\cite{GH03,GH03a,GH04}, $\mbb{Z}_6$ \cite{ho04} and $\mbb{Z}'_6$ \cite{Bailin:2006zf} 
toroidal orbifolds. In the
first two cases, semi-realistic MSSM-like models can
only be achieved after certain D-brane ``recombination
processes'' were taken into account (see the original papers for
more details). For the $\mbb{Z}_6$ model \cite{ho04}
an exhaustive search for MSSM-like models was performed and
a class of interesting D-brane configurations found. It gives
rise to the MSSM spectrum without the complication of brane
recombinations. In this example the K-theory constraints remain to
be checked explicitly.

Moreover, there are both four- and six-dimensional toroidal
backgrounds, where so far only non-chiral models with D6-branes parallel to
the orientifold planes have been constructed \cite{bgk99a,bgk00,fhs00,bgkl00,bcs04}.


\subsection{Type IIB orientifolds with magnetized D-branes}
\label{sectypeiib}

So far we have mainly discussed type IIA string model building
with intersecting D6-branes. The advantage of this is a
very clear geometric picture, where D-branes wrap
geometric three-cycles and carry only flat gauge bundles. The chiral
fermions are then given by the topological intersection number
between pairs of such cycles. However, supersymmetry requires that
these three-cycles are special Lagrangians. For general compact
Calabi-Yau manifolds finding sLag submanifolds is a very hard task and
no systematic classification exists. Mathematically, one
of the reasons is that three-cycles cannot be described by complex geometry with
all its powerful tools.

For model building on non-orbifold Calabi-Yau spaces it
appears to be  more appropriate to study the mirror symmetric
orientifold compactifications of type IIB.\footnote{Actually, the first models
of this type on a toroidal background have been discussed in the beautiful early work \cite{CB95}.
More recent Type IIB orientifold models with magnetized D-branes include
\cite{Antoniadis:2004pp,Bianchi:2005yz} .}
According to \reef{genOm} the world-sheet
parity transformation can be combined with a holomorphic
involution $\sigma$ of the Calabi-Yau three-fold. The action of
$\sigma$ on the K\"ahler form and the holomorphic three-form are
\bea
\sigma(J_2)&=&J_2\ , \quad~~~
\sigma(\Omega_3)~=~
\left\{\begin{array}{ll}
+ \Omega_3 & {\rm for\ O9/O5} \\
- \Omega_3 & {\rm for\ O3/O7}
\end{array}\right.
\ .
\eea
Note  that in the case with O3- and O7-planes the complete
orientifold projection is actually $\Omega\sigma (-1)^{F_L}$, for
O9/O5 models it is only $\O\s$. These two different classes of IIB
orientifolds both preserve ${\cal N}=1$ supersymmetry in four
dimensions.

The tadpoles are canceled by introducing stacks of so-called B-type
D$p$-branes, $p=3,5,7,9$, which are wrapping even-dimensional holomorphic cycles
of the underlying Calabi-Yau. They are endowed with
(stable) holomorphic vector bundles with a general structure group $G\subset U(N)$.
From the mathematical point of view, a vector bundle on a
D$p$-brane can be described as a coherent sheaf on a D9-brane
whose fiber is non-zero only on a codimension $9-p$ submanifold of
the Calabi-Yau manifold. Thus, coherent sheaves allow for a
unified description of all B-type branes on the Calabi-Yau
manifold\footnote{What we have merely described here is the
supergravity, i.e. large radius description of topological B-type
branes. It has been pointed out that the precise definition of
such branes, consistent with open string mirror symmetry, is
provided by the mathematical notion of derived categories of
coherent sheaves. The objects (D-branes) in this category are
complexes of coherent sheaves ${\cal E}^\bullet$ and the morphisms
(open strings) are given by the  the global Ext$^n({\cal
E}^\bullet, {\cal F}^\bullet)$ groups. For more information on
this issue please consult the review article
\cite{Aspinwall:2004jr} and the references therein.}.

The advantage of IIB orientifolds clearly is that one is
dealing with complex geometry. However, there is always a prize to pay,
namely that besides the holomorphic cycles one also has to
deal with the Chan-Paton gauge bundles on their world volume.
Recall that for the type IIA models the
string tree level F-terms depend on the K\"ahler moduli and are therefore subject
to world-sheet instanton corrections, whereas the D-terms
depend on the complex structure moduli
and are therefore exact at string sigma-model tree level
\cite{Ooguri:1999bv,Kachru:2000ih}. However, there can be
stringy non-perturbative corrections from E2-instantons.
For type IIB models at string tree level, the F-terms (the superpotential) depend on complex structure
moduli and are exact at string tree-level. They are  only corrected 
non-perturbatively by E1-, E5-instantons depending on the K\"ahler moduli.
The string tree level D-terms now
depend on the K\"ahler moduli and are therefore corrected by
higher loops and world-sheet instanton corrections. This implies
that the stability of D-branes depends on K\"ahler moduli. The
relevant mathematical notion of stability for the gauge bundles
(or for coherent sheaves) in the large radius regime is the so-called
Mumford- or $\mu$-stability. It gets corrected to what is called
$\Pi$-stability \cite{Douglas:2000ah} due to the
$\alpha'$-corrections (see section
\ref{subsecbsus}).\footnote{\lab{mstab}While the notion of $\Pi$-stability
will be explained in the course of this section, let us spell out
what $\mu$-stability means. The slope of a coherent sheaf ${\cal
F}$ with respect to a K\"ahler form $J_2$ on a manifold ${\cx}$ is
defined as
\bea
      \mu({\cal F})={1\over {\rm rk}({\cal F})} \int_{\cx}
          J_2\wedge J_2\wedge c_1({\cal F})\  .
\eea
A vector bundle $V$  is called $\mu$-stable if for each coherent
subsheaf ${\cal F}$ of $V$ with $0<{\rm rk}({\cal F})<{\rm
rk}({V})$ one has
\bea
\mu ({\cal F})<\mu({V})\ .
\eea
We denote the sheaf itself and the curvature form by the same
symbol.}.
Beyond these corrections, one also expects stringy non-perturbative
corrections from E1- and E5-instantons.

The ultimate goal in a systematic analysis of supersymmetric type
IIB orientifold compactifications on Calabi-Yau three-folds is to
determine for each point in K\"ahler moduli space the set of
stable B-type D-branes
(i.e. the $\Pi$-stable coherent sheaves in the $g_s<<1$ regime) and classify all
solutions to the tadpole and K-theory consistency conditions. At
the moment we are still far away from such a goal. However, so far
we do have the means to study the large radius supergravity regime
as well as some special points inside the K\"ahler moduli space.

Following \cite{Blumenhagen:2005zh,Blumenhagen:2005zg} let us now
review rules for model building in type IIB orientifolds, where
$\s$ is just the identity $(\O\s = \O)$, but on general smooth
Calabi-Yau spaces in the large radius regime. 
For the 
orientifold projection with O3/O7-planes, the analogous formulas
have not yet been worked out in full generality. In section
\ref{secgepner} we discuss IIB orientifold models at very small
radius by use of CFT methods.


\subsubsection{Tadpole cancellation}

In this section, we consider compactifications of the type I
strings to four space-time dimensions on a  Calabi-Yau manifold
$\cx$. Therefore, the  total ten-dimensional space is ${\cal Y}=\mbb R^{1,3}\times \cx$.
We start with the ambient model, which is
the type IIB string divided by the world-sheet parity
transformation $\Omega$. As is well known, this induces a tadpole for the RR 10-form $C_{10}$,
and, since  the Calabi-Yau is generically curved, a
tadpole for the 6-form $C_{6}$. Quantitatively, these tadpoles are
given by the CS action from \reef{Oplact} on the O9-plane
\bea
\label{cs1}
  \cs_{\rm CS}^{\rm O9}=  -32 \mu_9 \int_{{\cal Y}} \bigoplus_{p=0}^2
       C_{4p+2} \wedge \sqrt{ {L}
      \bigg({{\cal R}_{T({\cal Y})}  \over 4}\bigg) }\ ,
\eea
with $\mu_9$ given in \reef{mup} and ${\cal R}=2\pi\alpha' R$.
The Hirzebruch $L$-polynomial is defined in (\ref{Hirzebruch}).

In order to cancel these tadpoles, one introduces D9-branes
endowed with  holomorphic vector bundles on their world-volume. In
the following we now use labels $i,\, j,...$ for D9-branes and
$a,\, b,...$ for D5-branes. Concretely, we take stacks of
$M_i=N_i\,n_i$ branes and allow a non-trivial holomorphic vector
bundle $V_i$ along the diagonal $U(n_i)$. This breaks the gauge
group to $\prod_i U(N_i)$. If the gauge field strength on such a
stack is $F_i$, then under the action of $\Omega$ this stack is
mapped to a different stack with field strength $-F_i$. Therefore,
we have to introduce these stacks in pairs with vector bundles
$V_i$ and $V_i^*$ supported on their world-volume. The
Chern-Simons action on the D9-branes reads
\bea
\label{cs2}
\cs_{\rm CS}^{\rm D9}=  2\mu_9\,\int_{\cal Y} \bigoplus_{p=0}^2
       C_{4p+2} \wedge \ch(2\pi \a' {\cal F}_i)\wedge \sqrt{\hat{\cal A}
      ({{\cal R}_{T({\cal Y})} }) }.
\eea
The Chern character and the A-roof
genus are defined in (\ref{Chernch}) and \reef{Aroof}.

In addition, we allow stacks of $2N_a$ D5-branes wrapping
holomorphic two-cycles $\Gamma_a$ inside $\cx$. The total six-dimensional
world-volume of the D5-branes is ${\cal Z}_a=\mbb R^{1,3}\times \Gamma_a$
and the corresponding CS action  reads
\bea
\label{cs3}
\cs_{\rm CS}^{{\rm D5}}&=&
-\mu_5\,\int_{{\cal Z}_a}
\bigoplus_{p=0,1} C_{4p+2} \wedge  \left(2\, N_a+
     {\ell_s^{4}\over 2\, (2\pi)^2}{\rm tr}_{Sp}\, F_a^2\,
  \right)\wedge \sqrt{\hat{\cal A}
      \left( {\cal R}_{T({\cal Z}_a)} \right) \over
       \hat{\cal A}\left(  {\cal R}_{N({\cal Z}_a)} \right)} \ .
\eea
Here $T({\cal Z}_a)$ denotes the tangent bundle and $N({\cal
Z}_a)$  the  normal bundle of the D5-brane in $\mbb R^{1,3}\times
\cx$. The gauge group on a stack of D5-branes is $Sp(2N_a)$.
Similarly, $2M_i$ D9-branes with flat gauge bundle support
$SO(2M_i)$ gauge symmetries. For brevity, we do not explicitly
include these cases in our formulas, but this is easily
accomplished. Note that the overall minus sign in (\ref{cs3})
relative to the gauge bundles on the Calabi-Yau reflects the fact
that a D5-brane wrapping a holomorphic two-cycle $\Gamma_a$ can be
considered as a small instanton ${F}_a$ with Chern class
$c_2({F}_a/2\pi)=\gamma_a$, where $\gamma_a$ denotes the
Poincar\'e dual four-form of the two-cycle $\Gamma_a$.

From the CS terms it is straightforward to derive the tadpole cancellation
condition for $C_{10}$ and $C_6$
\bea
\label{tad1}
             \sum_{i=1}^K N_i\, n_i &=& 16\ , \\
         \sum_{i=1}^K N_i\, \ch_2(V_i)-\sum_{a=1}^L
               N_a\, \gamma_a  &=& -c_2(\cx)\ , \nonumber
\eea
the concrete form of \reef{loctad} in this class of models.
Here we have used the notation $\ch_2(V)$ for the Chern character $\ch_2(F_V/2\pi)$  of the
curvature $F_V$ of the vector bundle $V$.


\subsubsection{Massless spectrum}

The spectrum of massless closed string modes that survive the
orientifold projection in IIB orientifolds was described in section
\ref{oriclspec}. There are $h^{1,1}$ chiral superfields whose
scalars involve the K\"ahler moduli and the universal dilaton superfield. For the case at hand,
orientifolds with O9- and O5-planes, one gets $h_-^{2,1}$ vector
superfields and $h_+^{2,1}$ chiral superfields with complex
structure deformations (vice versa in the case of O7- and
O3-planes).
In addition to these closed string fields there will
be massless chiral multiplets for the open string moduli of the
D9- and D5-branes.

The  chiral massless spectrum resulting from open strings stretched between
stacks of D9-branes carrying gauge bundles is determined by the respective Euler characteristics
\bea
            \chi(\cx,W)=\sum_{r=0}^3 (-1)^r {\rm dim} \, H^r(\cx,W)=
\int_\cx \left( \ch_3(W)+{1\over 12}\, c_1(W)\wedge c_2(\cx ) \right)\ .
\eea
where the appearing vector bundles $W$ can be read off from the
decomposition of the adjoint representation of the ten-dimensional
gauge symmetry $SO(32)$. For  the case of interest these are
displayed in table \ref{Tchiral1}. The complete massless spectrum
is given by the cohomology groups $H^r(\cx,W)$, which is the
cohomology of holomorphic $p$-forms over $\cx$ with values in the
holomorphic vector bundle $W$.

\begin{table}
\centering
\begin{tabular}{|c|c|c|}
\hline
Non-abelian representation & $U(1)$ charges & Vector bundle $W$ \\
\hline \hline
&& \\[-.45cm]
$\Yasymm_a$ & $(2_i)$ & $\bigotimes^2_{\rm sym}  V_i$  \\[.05cm]
$\Ysymm_a$  & $(2_i)$ &   $\bigwedge^2 V_i$   \\[.05cm]
$(\antifund_i,\fund_j)$ & $(1_i,1_j)$ & $V_i \otimes V_j$   \\[.05cm]
$(\fund_a,\fund_b)$     & $(1_i,-1_j)$ & $V_i \otimes V_j^{\ast}$
\\[.05cm]
\hline
&& \\[-.45cm]
$(\fund_i,\fund_a)$  & $(1_i)$ & $V_i\otimes {\cal F}_a$ \\[.05cm]
\hline
\end{tabular}
\caption{Chiral massless spectrum of type IIB orientifold. $\bigotimes^2_{\rm sym}  V_i$ denotes
the symmetric tensor product bundle.\label{Tchiral1}} 
\end{table}

For computing the massless D5-brane matter we use the description
of  the D5-brane wrapping the two-cycle $\Gamma_a$ by the sheaf
${\cal F}_a$ mentioned above. The relevant Euler characteristic is
\bea
\chi(\cx, V_i \otimes {\cal F}_a)=-\int_{\Gamma_a} c_1(V_i)\ .
\eea
The full chiral spectrum is listed in table \ref{Tchiral1} in terms of representations under
the gauge symmetry
\beqn
\cg = \prod_{i=1}^K \bigg[ SU(N_i)\times U(1)_i \bigg] \times \prod_{a=1}^L Sp(2N_a)\ .
\eeqn
Again, one can show that for the chiral matter in table \ref{Tchiral1}
the non-abelian gauge anomalies in four dimensions precisely
cancel if the tadpole cancellation conditions (\ref{tad1}) are
satisfied.

The non-chiral massless spectrum can be determined from the
respective cohomology groups $H^*(\cx,W)$. In addition, there
exists non-chiral adjoint matter given  by the bundle deformations
$H^1(\cx,V_i\otimes V_i^*)$ for the D9-branes. For D5-branes one
gets antisymmetric matter counted by
\bea
H^1(\Gamma_a,{\cal O})+H^0(\Gamma_a,{\rm N}\Gamma_a)\ .
\eea
These two types of moduli correspond to Wilson lines along the two-cycle
and deformations of the holomorphic curve inside the Calabi-Yau.


\subsubsection{K-theory constraints}

Recall that K-theory constraints follow from the absence of
$Sp(2N)$ global Witten anomalies \cite{Witten:1982fp}. In our case
symplectic gauge symmetries arise from D5-branes wrapping two-cycles of the Calabi-Yau
$\cx$. Therefore, the cancellation of the Witten anomaly leads to
\bea
    \sum_{i=1}^K N_i \, \chi(\cx,V_i\otimes {\cal F}_a)
        =0\ {\rm mod}\ 2
\eea
for every two-cycle $\Gamma_a$, again represented by $\cf_a$. This
condition is the criterion for the entire vector bundle
$W=\bigoplus_{i=1}^K N_i\, V_i$ to admit spinors,\footnote{Note
that for complex vector bundles the second Stiefel-Whitney class
is given by the first Chern class modulo two.}
\bea
c_1(W)=\sum_{i=1}^K N_i\, c_1(V_i) =0\ {\rm mod}\ 2\ .
\eea
Let us comment that for the heterotic string this condition was
derived from the absence of anomalies in  the two-dimensional
non-linear world sheet sigma-model
\cite{Witten:1985mj,Freed:1986zx}.


\subsubsection{Green-Schwarz mechanism}

Since all these string models naturally  contain abelian gauge
groups, one also has mixed abelian-non-abelian, mixed
abelian-gravitational and cubic abelian anomalies. These anomalies
only cancel after axionic couplings are taken into account
\cite{Witten:1984dg}. The mixed $U(1)_i-SU(N_j)^2$ anomaly $\ca^{ijj}$ for
$i\ne j$ is given by
\bea
\label{mixeda}
\ca^{ijj}
&\propto&
      2N_i\, \int_\cx \left[n_j\ch_3(V_i) + c_1(V_i)\wedge \ch_2(V_j) +
     {n_j\over 12}\, c_1(V_i)\wedge c_2(T)\right]\ .
\eea
The last expression also holds for the case $i=j$, where
the contribution from the symmetric and antisymmetric matter fields
have to be taken into account. The mixed $U(1)_i-Sp(N_a)^2$ anomaly is
\bea
\label{mixedd}
\ca^{iaa}
&\propto &
    -N_i \int_\cx c_1(V_i)\wedge \gamma_a\ .
\eea
For the mixed $U(1)_i-$gravitational anomaly one finds
\bea
\label{mixedb}
\ca^{igg}&\propto&
N_i\, \int_\cx \left[24\, \ch_3(V_i) +
     {1\over 2}\, c_1(V_i)\wedge c_2(T)\right]\ .
\eea
These anomalies have to be canceled by axionic Green-Schwarz
couplings arising from the dimensional reduction of the three
kinds of CS terms in (\ref{cs1},\ref{cs2},\ref{cs3}). We expand the
relevant forms $C_2$ and $C_6$ as
\bea
   C_2=C_0^{(2)} + \ell_s^2 \sum_{k=1}^{h^{1,1}} C_k^{(0)}\, \omega_k\ ,\quad\quad
   C_6=\ell_s^6\, C_0^{(0)}\, d{\rm vol}_\cx + \ell_s^4
       \sum_{k=1}^{h^{1,1}} C_k^{(2)}\, \hat\omega_k\ ,
\eea
where $d{\rm vol}_\cx$ denotes the normalized volume from in the Calabi-Yau manifold.
Note that for the pure $\Omega$ orientifold $h^{1,1}_-=0$. 
Moreover,  $\omega_k$ and $\hat\omega_k$ form a normalized
basis of harmonic two- and four-forms on $\cx$ with
\mbox{$\int
\omega_k\wedge  \hat\omega_l=\delta_{kl}$}. The four-dimensional
two-forms  $(C_0^{(2)},C_k^{(2)})$ are Hodge dual to the
four-dimensional scalars  $(C_0^{(0)},C_k^{(0)})$.

By dimensional reduction we obtain the following Stueckelberg mass
terms in four-dimensions\footnote{Here we are using the type IIB
variables. If we were consistently using the Type I parameters,
the equations
(\ref{mass}) and (\ref{vertexk}) would have an extra factor of two.}
\bea
\label{mass}
\cs_{\rm mass} ={1\over 6\, (2\pi)^5\alpha'} \sum_i
   N_i \int_{\mbb R^{1,3}}  C_0^{(2)}\wedge f_i \ \int_X \left[
         {\rm tr}_{n_i} \bar F^3_i -{1\over 16}\, {\rm tr}_{n_i} \bar F_i\wedge
           {\rm tr} \bar R^2\right]\ ,
\eea
for $C_0^{(2)}$ and
    \bea
\cs_{\rm mass} = {1\over (2\pi)^2\, \alpha'} \sum_{i,k}
   N_i \int_{\mbb R^{1,3}}  C_k^{(2)}\wedge f_i \
         \left[{\rm tr}_{n_i} \bar F_i\right]_k\ ,
\eea
for $C_k^{(2)}$.
where $f_i$ denotes the field strength of the $U(1)_i$ factor in the
gauge group and $\bar F_i$ the internal field strength. The traces are over the fundamental
representation of the structure group $U(n_i)$ or the internal Lorentz
group. We have also expanded
\bea
{\rm tr} \bar F_i=(2\pi)\sum_k \big[{\rm tr} \bar F_i\big]_k\,
          \omega_k\ .
\eea
Similarly one obtains the axionic couplings of  $C_0^{(0)}$ as
\bea
\label{vertex0}
\cs_{\rm ax} ={1 \over  2\, (2\pi)} \sum_i
   n_i \int_{\mbb R^{1,3}}
        C_0^{(0)}\wedge {\rm tr}_{N_i}  F^2_i -
   {1\over 4\,  (2\pi)}  \int_{\mbb R^{1,3}}
        C_0^{(0)}\wedge  {\rm tr}   R^2
\eea
and for $C_k^{(0)}$
\bea
\label{vertexk}
\cs_{\rm ax}&=&{1 \over  4\, (2\pi)} \sum_{i,k}
      \int_{\mbb R^{1,3}}
        C_k^{(0)} \wedge  {\rm tr}_{N_i}  F^2_i \
         \left[ {\rm tr}_{n_i} \bar F^2_i-{n_i\over 48}
         {\rm tr} \bar R^2 \right]_k \\
&&
-\frac1{2\pi}
      \int_{\mbb R^{1,3}}
        C_k^{(0)} \wedge \Big( \frac14 \sum_a {\rm tr}_{Sp(2N_a)}  F^2_a \
         \left[ \gamma_a \right]_k
   +{1 \over 768}  {\rm tr}  R^2 \
         \left[  {\rm tr} \bar R^2 \right]_k \Big) \ . \nonumber
\eea
Here, we have expanded
\bea
{\rm tr} \bar F^2_i=(2\pi)^2\sum_k \big[{\rm tr} \bar F^2_i\big]_k\,
          \hat\omega_k\ ,\quad\quad
\gamma_a= \sum_k [\gamma_a]_k\, \hat\omega_k\ ,
\eea
and similarly for the internal curvature. Note that in the axionic
couplings also the D5-branes give a
contribution. The tadpole cancellation conditions have
be used to bring the expression to its final form (\ref{vertexk}).
Combining the Stueckelberg mass terms and the axionic couplings
provides counter terms for the triangle anomalies. Adding
up all contributions yields precisely the right form to cancel the
mixed anomalies (\ref{mixeda},\ref{mixedd},\ref{mixedb}).

As usual the gauge fields of anomalous gauge symmetries become
massive via the Green-Schwarz couplings (\ref{mass}) by absorbing
the axions as longitudinal modes and give rise to perturbative
global $U(1)$ symmetries. Similar to the type IIA case, 
these are broken non-perturbatively by wrapped Euclidean E1- and E5-branes.


\subsubsection{Gauge kinetic functions}

Let us now give the expressions for  the gauge kinetic functions
for the non-abelian $SU(N)$ gauge groups.
The holomorphic gauge kinetic function ${f}_i$ appears in the
four dimensional effective field theory as in \reef{kinYM}.
The complexified dilaton and  K\"ahler moduli appearing in the
${\cal N}=1$ four-dimensional superfields are\footnote{In the presence
of open string moduli these K\"ahler coordinates will get
corrected. See the discussion at the end of section \ref{secZ22}.}
\bea
 S={1\over 2\pi}\Big[ e^{-\Phi} {{\rm vol}({\cx}) \over
             \ell_s^6 } + {i}\, C^{(0)}_0 \Big]\ ,
\qquad
 T_k={1\over 2\pi}\Big[ - e^{-\Phi} \alpha_k + {i} C^{(0)}_k \Big]\
,
\eea
with $\alpha_k=\ell_s^{-2}\,\int_\cx  J_2\wedge \widehat\omega_k$ and
${\rm vol}(\cx) = \frac1{3!}\int_\cx J_2\wedge J_2\wedge J_2$.
Then the gauge kinetic functions can be deduced from the imaginary parts in
the axionic couplings (\ref{vertex0},\ref{vertexk}). One finds
\bea
\label{ggg}
{f}_{i} &=& n_i\, S +{1\over 2} \sum_{k=1}^{h^{1,1}} T_k\,
\left[{\rm tr}_{n_i} \bar F_i^2-\frac{n_i}{48} {\rm tr} \bar
R^2\right]_k \ .
\eea
The real part of the holomorphic gauge kinetic function ${f}_i$
can be cast into the form
\bea
\label{gal}
{\rm Re}\, {f}_{i} = {1\over 2\pi\ell_s^6 g_s}\left[ {n_i\over 3!}\,
 \int_{\cx} J_2 \wedge J_2 \wedge J_2 -
{(2\pi\alpha')^2\over 2} \int_{\cx} J_2 \wedge
       \left( {\rm tr}_{n_i} \bar {F}_i^2  - {n_i \over 48}  {\rm tr}{\bar R}^2 \right)
\right]
\eea
and further be written as\footnote{Working with Type I parameters
amounts to replacing $\alpha'$ by $\sqrt{2}\alpha'$ in
(\ref{pigauge}).}
\bea
\label{pigauge}
  {\rm Re}\, {f}_{i} ={1\over 2\pi\ell_s^6 g_s}
    {\rm Re} \int_{\cx} {\rm tr}_{n_i}
\left[   e^{J_2 + 2\pi i \alpha' F_i }\,
         \sqrt{\hat A(\cx)} \right].
\eea
For the D5-branes the gauge couplings are
given by
\bea
   {\rm Re}\, {f}_{a} = {1\over 2\pi\ell_s^2 g_s} \int_{\Gamma_a} J_2\ ,
\eea
which consistently is the small instanton limit of the second term
in (\ref{gal}).


\subsubsection{D-terms and Fayet-Iliopoulos couplings}
\label{seciibfi}

In deriving the Fayet-Iliopopoulos terms we follow the discussion
of \cite{Dine:1987xk}.
Let us just briefly recall the basic \gs mechanism as explained in
much more detail in section \ref{secgs}. 
The gauge invariance of the quadratic axion-gauge boson mixing terms
\reef{gsLag2} forces the axion to transform under an otherwise
anomalous $U(1)_m$ gauge symmetry by a shift \reef{axshift}. 
As in \reef{invK} this phenomenon can be described by a gauge
invariant K\"ahler function where the chiral and vector superfields
appear in the combination $S+\bar S-Q^S_m V_m$ etc. 

The tree-level K\"ahler potential for the ten-dimensional dilaton $\t$
and the K\"ahler moduli of a type II Calabi-Yau compactification was given in \reef{KpotK} and
\reef{Kpotdil}. The K\"ahler potential of $S$ is identical to that of
$\t$ and the potential for the $T_k$ is just the truncation of
\reef{KpotK} onto modes that survive the orientifold projection.  
Due to the non-trivial gauge transformations (\ref{axshift}) of the
axions inside $S$ and $T_k$ it is, however, not any longer gauge
invariant. 

Without any explicit computation, one can write an invariant K\"ahler
potential of the form 
\bea
\kappa_4^2\, {K} &=&
-  \ln\Big(S+\bar S-\sum_i Q^i_0\, V_i\Big)
-  \ln\Bigg[-\sum_{k,l,m=1}^{h^{1,1}}
{{\cal K}_{klm}\over 6}
\bigg(  T_k+\bar T_k-\sum_i Q^i_k\, V_i\bigg) \nonumber \\
&&
\hspace{3cm}
\times
\bigg(  T_l+\bar T_l-\sum_i Q^i_l\, V_i\bigg)
\bigg(  T_m+\bar T_m-\sum_i Q^i_m\, V_i\bigg) \Bigg] \ .
\eea
The charges $Q^i_k$ (up to proportionality 
the components of the constant Killing vectors) are defined via
\bea
\cs_{\rm mass}=\sum_{i}  \sum_{k=0}^{h_{11}}
            {Q^i_k\over 2\pi\alpha'}
                \int_{\mbb R^{1,3}} f_i \wedge C^{(2)}_k
\eea
and can be read off from the mass terms (\ref{mass}).
For such a K\"ahler potential the coefficients $\xi_i$ of the \FI terms can be derived from the relation
\reef{FIfromK}. This results in the \FI terms
\bea
\label{fiterms}
   {\xi_i\over g_i^2}&=&{1\over g_s\, (2\pi)^8\, (\alpha')^4}\Biggl[ {(2\pi\alpha')\over 2} \int_X J\wedge J\wedge {\rm tr}_{n_i}
          \bar F_i - \nonumber  \\
    && \phantom{aaaaaaaaaaaaa} {(2\pi\alpha')^3\over 3!}\, \int_\cx
   \biggl[{\rm tr}_{n_i} \bar F^3_i -{1\over 16}\, {\rm tr}_{n_i} \bar F_i\wedge
           {\rm tr} \bar R^2\biggr] \Biggr]\ ,
\eea
which have a tree-level and a one-loop term in sigma model perturbation
theory. According to \cite{Dine:1987xk} there are no higher loop contributions.
However, one expects these expressions to be corrected by world-sheet
and space-time instanton contributions. 

Supersymmetry of the four-dimensional effective theory
implies that the D-terms have to vanish, which for zero expectation values for
charged matter fields means that all FI-terms have to vanish.
Similarly to the type IIA case, the $U(1)_a$ gauge couplings and
FI-terms can be expressed in terms of a complex valued  central charge
\bea
\label{fiterms2}
   Z_i={1\over 2\pi g_s\, \ell_s^6} 
      \int_{\cx} {\rm tr}_{n_i}
\left[  e^{J_2 + 2\pi i\alpha' F_i }\,
         \sqrt{\hat A(\cx)} \right]
\eea
as
\bea
     {1\over g_i^2}=\left\vert Z_i\right\vert\ , \quad\quad 
2\pi\alpha' \xi_i=\arg\left(Z_i \right)\ .
\eea
This is precisely the perturbative part of
the expression appearing in the $\Pi$-stability condition of
\cite{Douglas:2000ah} (which also has no higher loop contributions).


\subsubsection{F-terms}
\label{sectypeiibfterm}

For type IIA we have discussed the possibility of  both a
world-sheet and a space-time instanton generated superpotential.
For the mirror dual type IIB orientifolds the story simplifies in
that the sigma-model superpotential is believed to only depend on
the complex structure moduli \cite{Kachru:2000ih} and has the
simple form
\bea
\label{superb}
   \kappa^2_{10}\, W_{\rm tree}=\int_{\cx} \tilde F_3\wedge \Omega_{3}\ ,
\eea
where $\tilde F_3$ is the RR three-form field strength
\reef{mod3form} of type I. Therefore, the superpotential is
exact at sigma-model tree-level and can only have additional
contributions from space-time instantons.
The E2-instantons for Type IIA orientifolds map to E1- and E5-instanton
for Type I models, where the E5-instanton can carry in addition
a non-trivial vector bundle. These instanton break the global $U(1)$
symmetries, which have become massive due to 
St\"uckelberg mass couplings to the
K\"ahler and universal axions. 
Therefore, here the general form of the superpotential is
\bea
\label{superpotiib}
W=\sum_{\rm E1} e^{ -S_{\rm E1}(T) }\,\, f_1(U,B,\Phi_{ab} ) +  \sum_{\rm
E5} e^{-S_{\rm E5}(S,T) }\, f_2(U,B,\Phi_{ab} )\ ,
\eea
where $B$ denote the vector bundle moduli.

In order to further evaluate the tree level superpotential (\ref{superb}), note that   
the Bianchi identity for
$\tilde F_3$ (the equation of motion for the Hodge dual six-form
potential $C_6$) reads
\bea
\label{hhh}
{1\over \ell_s^2}\, d\tilde F_3 = {1\over 2\, (2\pi)^2 }\, \Big[
{\rm tr}_{SU(3)}  R^2 - \sum_i N_i\, {\rm tr}_{n_i}  \bar F_i^2  \Big]
+ \sum_a N_a \,\gamma_a \ .
\eea
Locally one can write the right-hand-side in terms of CS 3-forms,
${\rm tr}_{n_i} \bar F_i^2=d\omega^{\rm YM}_{3i}$ and ${\rm
tr}_{SU(3)}  R^2=d\omega_3^{\rm L}$, so that the superpotential is
given by
\bea
\label{superc}
  \kappa^2_{10}\, W_{\rm tree}={\ell_s^2\over 2\, (2\pi)^2 }\, \int_{\cx}
\Big[ \omega^{\rm L}_3 -  \sum_i N_i \, \omega^{\rm YM}_{3i}   \Big]
\wedge \Omega_3+\ell_s^2
\sum_a \int_{C_a} \Omega_{3}\ .
\eea
In the last term the integrals are over three-cycles $C_a$ satisfying
$\pa C_a=\Gamma_a$. Note that
(\ref{superc}) is gauge invariant, as the variation of the CS 3-forms
is exact and $\Omega_3$ is closed.

This superpotential reflects two possible sources for non-trivial
relations between open string and closed string complex structure
moduli. Say one has, for instance, satisfied the tree-level
supersymmetry conditions for a particular choice of complex
structure and vector bundle moduli. This in particular means that
the vector bundle $V_i$ is holomorphic, its curvature is of type
$(1,1)$. Infinitesimally, such a bundle has
dim\,$H^1(\cx,V_i\otimes V_i^*)$ vector bundle moduli which by
definition preserve the holomorphy of the bundle.

However, it is not necessarily true that all these deformations
can be integrated. In fact, it can happen that at a some order $n$
in a power series expansion in a deformation parameter $\e$ the
curvature ceases to be purely of type $(1,1)$, instead
\bea
\label{deforma}
           F_i=F_i^{1,1}+\epsilon^n\, F_i^{0,2}\ .
\eea
At order $n$ there is an obstruction to this deformation of the
bundle, and one expects a term in the superpotential of the form
$\Phi^n$ with $\Phi$ denoting the vector bundle modulus in
question. One can easily verify that the deformation
(\ref{deforma}) leads to a non-vanishing superpotential once
inserted into (\ref{superc}). Similarly, the D5-brane term in
(\ref{superc}) reflects the fact that the local deformations of a
holomorphic curve counted by $H^0(\Gamma_a, N\Gamma_a)$ cannot
necessarily  be integrated to finite deformations. Very similarly
to this argument for bundle moduli, by deforming the complex
structure of the Calabi-Yau, a $(1,1)-$form does not necessarily
stay of this type and can pick up a component of type $(0,2)$.

Thus, we have seen that on general grounds the tree-level
superpotential is expected to be non-vanishing. Mirror symmetry
relates the tree-level superpotential of IIB to the world-sheet
instanton corrected superpotential of the corresponding type IIA
orientifolds. Though exact at sigma-model tree-level, the type IIB
superpotential can receive corrections from space-time
E1- and E5-instantons introducing an exponential dependence on the
K\"ahler moduli.

\subsubsection{Supersymmetry}
\label{subsecbsus}

For the mostly studied case of just $U(1)$ bundles on D9-branes the
supersymmetry condition is simply the vanishing of the \FI terms
(\ref{fiterms}). This is an effective criterion obtained after integrating over
the internal space.

However, for non-abelian bundles the supersymmetry condition is more involved
and very similar in spirit to the situation for the heterotic string
(see section \ref{sechet}).
In analogy one proposes \cite{Blumenhagen:2005pm}  that the local
supersymmetry condition is the integrand of (\ref{fiterms2}),
so that the resulting matrix equation reads
\bea
\label{SUSYloc}
{\rm Im}\Bigg(  e^{-i\th} \, e^{2\pi\alpha' F_i^{ab} -iJ_2}\,
         \sqrt{\hat A(\cx)} \Bigg)\Big|_{\rm top} =0\ .
\eea
Note that this is nothing else than the
non-abelian generalization of the MMMS equation
\cite{Marino:1999af} also including curvature terms.

Since this equation is related to a D-term in the effective low
energy action, one does not expect higher sigma-model
loop-corrections, whereas non-perturbative world-sheet instanton
corrections will in general appear.

In analogy to the $\mu$-stability condition explained in footnote
\ref{mstab} one defines the $\Pi$-slope of a sheaf ${\cal V}$ with respect to a
K\"ahler form $J_2$ as
\bea
\label{pistable}
\xi({{\cal V}}) = \arg\left[
    \int_{\cx} {\rm tr}_{n_{\cal V}} \left[ e^{2\pi\alpha' F_{\cal V} -iJ_2}\,
         \sqrt{\hat A(\cx)}\right] + {\co}\big(e^{-{1/\alpha'}},e^{-{1/g_s}} \big) \right] .
\eea
In analogy to the theorem by Donaldson-Uhlenbeck and Yau, one
expects that, if the integrated supersymmetry condition is
satisfied for a bundle $V$, i.e. $\xi(V)=0$, and the bundle is
$\Pi$-stable, then there exist a unique solution to the local
supersymmetry condition for the curvature $F_V$. Here a bundle $V$
is called $\Pi$-stable, if for any subsheaf ${\cal V}\subset V$
with ${\rm rk}({\cal V})<{\rm rk}({V})$ the $\Pi$-slope is
smaller, i.e. $\xi({{\cal V}})<\xi({V})$.

This is the type I generalization of the  $\Pi$-stability
condition for B-type D-branes  as introduced in
\cite{Douglas:2000ah}. The difference lies in the fact that for
the type I string, the orientifold plane already breaks the
supersymmetry down to ${\cal N}=1$, so that in order for the
D-branes to preserve the same supersymmetry the slope of the
vector bundle $V$ is fixed at $\xi(V)=0$.

These stability conditions  are to be understood  as  restrictions
first  on the vector bundles and  second on the K\"ahler moduli. A
vector bundles is in general only stable on a subset of the entire
K\"ahler moduli space. Since the study of $\Pi$-stable bundles for
concrete Calabi-Yau spaces is just beginning to evolve, most
concrete model building attempts are solely using line bundles
$L_i$ which are trivially stable and $\xi(L_i)=0$ only imposes a
condition on the K\"ahler parameters.

From the model building point of view it would be very useful to
have control over these models not only in the large radius
regime. Though we do not understand the stringy regime in
generality, at least there exists a class of conformal field
theories, the so-called Gepner models, which correspond to exactly
solvable points in the moduli space of Calabi-Yau
compactifications. Using conformal field theory techniques, these
models are discussed in the next section.


\subsection{Gepner model orientifolds}
\label{secgepner}

In the last section we have outlined the general model building
rules for type IIB Calabi-Yau orientifolds at large radius. As we have
mentioned, both the supersymmetry conditions as well as the gauge
couplings will receive world-sheet instanton corrections.

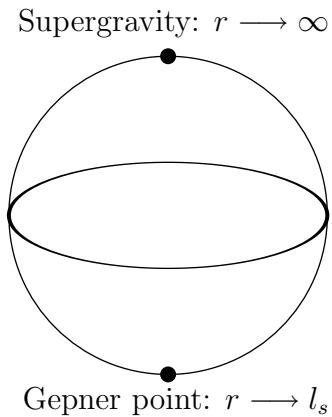
\begin{figure}[ht]
\begin{picture}(200,150)(0,0)

\Oval(225,70)(60,60)(0)
\Oval(225,70)(20,60)(0)

\Vertex(225,10)3
\Vertex(225,130)3

\Text(225,0)[1]{Gepner point: $r \longrightarrow l_s$}
\Text(197,143)[1]{Supergravity: $r \longrightarrow \infty$}

\end{picture}
\vspace{.5cm}
\caption{One-dimensional K\"ahler moduli space \label{figgepner}}
\end{figure}

How can we  learn something about type IIB orientifolds on
Calabi-Yau manifolds at small radii? It is known that at special
points with enhanced symmetry in the K\"ahler moduli space of
certain Calabi-Yau manifolds the non-linear sigma-model is exactly
solvable. It is given by an ${\cal N}=(2,2)$ supersymmetric
rational conformal field theories. These conformal field theories
have been discovered by Gepner \cite{Gepner:1987qi,Gepner:1987vz}
and are called Gepner models. In figure \ref{figgepner} a sketch
of a (compactified) complex one-dimensional K\"ahler moduli space
is given, where the large radius limit and the Gepner model are
the two points under fairly good control.

\noindent
This situation occurs literally in the quintic Calabi-Yau which
has precisely one K\"ahler modulus. As demonstrated in
\cite{Brunner:1999jq}, there are so-called lines of marginal
stability in the K\"ahler moduli space. Crossing these lines by
changing the K\"ahler modulus formerly stable D-brane
configurations become unstable and decay into other, now stable,
states. Thus, the notion of stability depends on the
moduli.\footnote{The notion of lines of marginal stability relies
on the presence of $\cn=2$ supersymmetry.}

The Gepner models and their D-branes have been under intense
investigation during the last years. All this work would easily
fill a review article of its own. Therefore, we restrict ourselves
to a brief review of the main aspects of Gepner models and in
particular concentrate on the orientifolds that can be constructed
from these.

After some earlier studies \cite{abpss96,bw98} which proceeded model
by model, various equivalent frameworks have been developed for the
systematic construction of Gepner model orientifolds
\cite{Govindarajan:2003vp,aaln03,RB03a,bhhw04,bw04,dhs04,aaj04,bw04a,dhs04a,Dijkstra:2004ym,Gato-Rivera:2005qd,Aldazabal:2006nz,Anastasopoulos:2006da}.
Some of them are mathematically more abstract and general than others.
In order to avoid too much of advanced conformal
field theory techniques we stick to a very direct method
which follows straightforwardly the usual steps for the
construction of conformal field theory orientifolds as described
in section \ref{secbasic}.

Starting from the Klein bottle partition function one defines
cross-cap states, and analyzing consistent boundary conditions one
obtains boundary states that describe the D-branes. From computing
their overlap analogous to \reef{treechannel} conditions for RR
charge cancellation are derived. The entire procedure is in the
same spirit as the steps performed in sections \ref{sec1l} and
\ref{secangle1loop} to define orientifolds of toroidal (orbifold) type II
compactifications.

Moreover, here we just stick to the usual Gepner models, i.e.\
without explicitly presenting the analogous results for orbifolds
thereof or simple current extensions. Those can be found in the
existing literature \cite{Fuchs:2000cm,bhhw04,bw04,dhs04a}. The
simple current extensions actually give rise to the plethora of
different Gepner model orientifolds computed in
\cite{dhs04,dhs04a}.


\subsubsection{Review of Gepner models}

Let us  briefly review some aspects of Gepner models needed in the
subsequent sections. Here we  assume that the reader is familiar
with the basic notions of both ordinary and boundary conformal
field theory. If some readers find this section rather technical
and advanced they can directly jump to section \ref{sechet}.

In light-cone gauge, for a compactification to four space-time
dimensions there remain two dynamical  non-compact directions.
From the world-sheet point of view, these correspond to two free
bosons $X^\mu$ and two free fermions $\psi^\mu$ with $\mu=2,3$.
The two fermions form an $SO(2)_1=U(1)_2$ current algebra with the
current given by the normal ordered product $j(z)=\,:\!\!\psi^2\psi^3\! :\!(z)$.
The internal six dimensions are compactified on a
Calabi-Yau three-fold, which on the world-sheet is described by
an abstract ${\cal N}=2$  supersymmetric conformal field theory
(SCFT) with central charge $c=9$. The minimal symmetry algebra of
such an ${\cal N}=2$ SCFT is the ${\cal N}=2$ extension of the
Virasoro algebra, generated by the  components $L_m$ of the
energy momentum tensor, a $U(1)$ current $J_m$ and two fermionic
primary fields $G^\pm_r$ of conformal dimension $h=\frac32$. The
algebra reads
\bea
 \left[ L_m,L_n\right]&=&(m-n)\, L_{m+n} +{c\over 12}\, m(m^2-1)\,
\delta_{m+n,0}\ ,
 \nonumber \\
\left[ J_m,J_n\right]&=& {c\over 3}\, m\, \delta_{m+n,0}\ , \quad
[L_m,J_n]~=~  -n\, J_{m+n}\ ,
\quad \{G^\pm_r,G^\pm_s\}~=~ 0\ , \nonumber \\
   \left[ L_m,G^\pm_r \right] &=&  -\left( r-{m\over 2} \right)\,
G^\pm_{m+r}\ ,\quad
  \left[ J_m,G^\pm_r\right]~=~  \pm G^\pm_{m+r}\ , \nonumber \\
  \{G^+_r,G^-_s\}&=& 2\, L_{r+s} + (r-s)\, J_{r+s} +
       {c\over 3} \left(r^2-{1\over 4}\right)\, \delta_{r+s,0}\ .
 \eea
Space-time supersymmetry is realized on the world-sheet by the
existence of a spectral flow for the ${\cal N}=2$  Virasoro
algebra
\bea
     L'_m=L_m+\eta\,  J_m + {c\over 6}\eta^2\, \delta_{m,0}\ , \ \
     G'^\pm_{r}=G^\pm_{r\pm\eta}\ , \ \ J'_m=J_m+ {c\over 3}\eta\,
\delta_{m,0}\ ,
\eea
which for $\eta=\frac12$ provides an isomorphism between the NS- and
R-sector of the ${\cal N}=2$  SCFT. This implies a one-to-one map between space-time bosons and
space-time fermions.

A comparably simple class of SCFTs is given by tensor
products of the rational models of the two-dimensional ${\cal
N}=2$  super Virasoro algebra with total central charge $c=9$
\cite{Gepner:1987qi,Gepner:1987vz}. To summarize, a Gepner model
has the conformal field theory building blocks shown in table
\ref{tabgepnera}.
\begin{table}[htb]
\renewcommand{\arraystretch}{1.5}
\begin{center}
\begin{tabular}{|c|c|}
\hline
 Central charge & CFT    \\
\hline
 $c= 2$ &   $X^{\mu},\  \mu=2,3$   \\
 $c= 1$ &   $\psi^{\mu},\  \mu=2,3\  \simeq   \ U(1)_2$  \\
 $c= 9$ &   $(2,2)$ SCFT: $\bigotimes_{j=1}^5 (k_j)$ \\
\hline
\end{tabular}
\caption{SCFT for ${\cal N}=2$ type II Gepner models}
\label{tabgepnera}
\end{center}
\end{table}

Now, let us discuss in some detail the unitary models of the
${\cal N}=2$ super Virasoro algebra. The minimal models are
parameterized by the level $k=1,2,\ldots$ and have central charge
\bea
\label{num}
c={3k\over k+2}\ .
\eea
Such a model is denoted by $(k)$.
Since $c<3$, one achieves the required value $c=9$ by using
tensor products of such minimal models $\bigotimes_{j=1}^r
(k_j)$. The  irreducible representations of the ${\cal N}=2$
Virasoro algebra of each unitary model are labelled by the three
integers $(l,m,s)$ in the range
\bea
\label{range}
l=0,\ldots k\ , \quad m=-k-1,-k,\ldots k+2\ , \quad
              s=-1,0,1,2\ ,
\eea
with $l+m+s=0$ mod $2$. Actually, the identification between
$(l,m,s)$ and $(k-l,m+k+2,s+2)$ reveals that the range
(\ref{range}) is a double covering of the allowed
representations. The conformal dimension and charge of the highest
weight state with label $(l,m,s)$ are given by
\bea
\label{quannum}
\Delta^l_{m,s}=\Big[{l(l+2)-m^2\over 4(k+2)} + {s^2\over 8}\Big]\ {\rm mod}\ 1 , \quad
 q^l_{m,s}=\Big[{m\over k+2}-{s\over 2}\Big]\ {\rm mod}\ 2\ .
\eea
To obtain the precise conformal dimension $h$ and
charge  from (\ref{quannum}) one  first shifts  the labels into
the standard range $|m-s|\le l$ by using the shift symmetries
$m\to m+2k+4$, $s\to s+4$ and the reflection symmetry. The
NS-sector consists of those representations with even $s$, while
those with odd $s$ make up to the R-sector.

In addition to the internal ${\cal N}=2$ sector, one has the
contributions with $c=3$ from the two uncompactified directions.
The two world-sheet fermions $\psi^\m$ generate $U(1)_2$ whose
four irreducible representations are labelled by $s_0=-1,0,1,2$.
The highest weight and charge are, respectively,
\bea
\label{fermi}
\Delta_{s_0}={s_0^2\over 8} \ {\rm mod}\ 1 \ ,
             \quad\quad q_{s_0}=-{s_0\over 2}\ {\rm mod}\ 2\ .
\eea
The GSO projection in Gepner models projects onto
states with odd overall $U(1)$ charge
$Q_{\rm tot}=q_{s_0}+\sum_{j=1}^r q^{l_j}_{m_j,s_j}$. To
have a good space-time interpretation one has to ensure that in
the tensor product only states from the NS respectively the R
sectors couple among themselves.

These projections are described most conveniently in the following notation.
First one defines some multi-labels
\bea
\label{mlabels}
\lambda=(l_1,\ldots,l_r)\ ,\quad
\mu=(s_0;m_1,\ldots m_r;
                s_1,\ldots,s_r)\ ,
\eea
and the respective characters
\bea
\label{mchar}
\chi^\lambda_\mu(q)=\chi_{s_0}(q)\, \chi^{l_1}_{m_1,s_1}(q)
                        \ldots \chi^{l_r}_{m_r,s_r}(q)\  .
\eea
In terms of the vectors
\bea
\label{mvec}
\beta_0=(1;1,\ldots,1;1,\ldots,1)\ , \quad
            \beta_j=(2;0,\ldots,0;0,\ldots,0,\smash{\underbrace{2}_{j^{\rm
th}}},0,\ldots,0)\ ,
\eea
and the following product
\bea
\label{mprod}
Q_{\rm tot}=2\beta_0\bullet \mu =-{s_0\over 2}-\sum_{j=1}^r{s_j\over 2}
              +\sum_{j=1}^r {m_j\over k_j+2}\ ,\quad
\beta_j\bullet \mu =-{s_0\over 2}-{s_j\over 2}\ ,
\eea
the projections are
$Q_{\rm tot}=2\beta_0\bullet \mu\in 2 \mbb{Z}+1$ and $\beta_j\bullet
\mu \in  \mbb{Z}$ for all $j=1,\ldots, r$. The first is the GSO
projection and the second guarantees that only NS and R
sectors from the different tensor products are coupled among each
other, respectively. Gepner has shown that the following
partition function
\bea
\label{parti}
 \cz_C(\tau,\o{\tau} )&=&
\\
&&
\hspace{-.5cm}{1\over 2^r}
{ ({\rm Im}\, \tau)^{-2} \over |\eta(q)|^4 }
     \sum_{b_0=0}^{K-1} \sum_{b_1,\ldots,b_r=0,1} {\sum_{\lambda,\mu}}^\beta
    (-1)^{s_0} \ \chi^\lambda_\mu (q)\, \chi^\lambda_{-\mu+b_0\beta_0
              +b_1 \beta_1 +\cdots +b_r\beta_r} (\o q)
\nonumber
\eea
is indeed modular invariant and vanishes due to space-time
supersymmetry. Here $K={\rm lcm}(4,2k_j+4)$, and ${\sum}^\beta$
means that the sum is restricted to those $\lambda$ and $\mu$ in
the range (\ref{range}) satisfying $2\beta_0\bullet \mu\in 2
\mbb{Z}+1$ and $\beta_j\bullet \mu \in  \mbb{Z}$. The factor
$1/2^r$ due to the field identifications guarantees the correct
normalization of the amplitude. In the partition function
(\ref{parti}) states with odd charge are arranged in orbits under
the action of the $\beta$-vectors. Therefore, although the
partition function is non-diagonal in the original characters, it
can be written as a diagonal partition function for all odd levels
in terms of the GSO-orbits under the $\beta$-vectors (\ref{mvec})
which in this case have all equal length $2^r K$. Geometrically,
these models correspond to Calabi-Yau three-folds defined by
Fermat type hypersurfaces
\bea
\sum_i z_i^{k_i+2} =0
\eea
in the appropriate weighted projective space $\mbb
P_{w_1,\ldots,w_5}[d]$,
with $d=\sum_i w_i$.

The rules for applying the modular transformation $S:\tau\to -1/\tau$
to the characters involved in (\ref{parti}) are as follows. For the
$SU(2)_k$ Kac-Moody algebra the $S$-matrix  is given by
\bea
\label{ssmatrix}
  S_{l,l'}=\sqrt{2\over k+2}\, \sin (l,l')_k\ ,
\quad
(l,l')_k={\pi(l+1)(l'+1)\over
k+2}\ .
\eea
For the ${\cal N}=2$ minimal model with level $k$,  the
modular $S$-matrix reads
\bea
\label{smatrix}
S^{U(1)_2}_{s_0,s_0'}&=&{1\over 2} e^{-i\pi{s_0 s'_0
        \over 2}}\ , \nonumber \\
        {S}_{(l,m,s),(l',m',s')}&=&{1\over 2\sqrt{2k+4}}\, S_{l,l'}\,
        e^{i\pi{m\, m' \over k+2}}\, e^{-i\pi{s\, s' \over 2}} \ .
\eea
The loop-channel and tree-channel M\"obius amplitudes are related by the $P$-matrix
$P=T^{1\over 2}S\,T^2\,S\,T^{1\over 2}$, which for
the $SU(2)_k$ Kac-Moody algebra is given by
\bea
\label{ppmatrix}
 P_{l,l'}={2\over \sqrt{k+2}}\,
     \sin\left[ {1\over 2}(l,l')_k \right]\, \delta_{l+l'+k,0}^{(2)}\
,
\eea
and for the ${\cal N}=2$ unitary models reads
\bea
\label{pmatrix}
 P^{U(1)_2}_{s_0,s'_0}&=&{1\over \sqrt{2}}\, \sigma_{s_0}\sigma_{s'_0}
  e^{-i\pi{s_0 s'_0 \over 4}}\, \delta_{s_0+s_0',0}^{(2)}\ , \\ \nonumber
        { P}_{(l,m,s),(l',m',s')}&=&{1\over 2\sqrt{2k+4}}\, \sigma_{(l,m,s)}\,
   \sigma'_{(l',m',s')} \, e^{{i\pi\over 2}{m\, m' \over k+2}}\,
    e^{-i\pi{s\, s' \over 4}}\, \delta_{s+s',0}^{(2)} \\
    &&~~~~~ \times \left[
    P_{l,l'}\, \delta_{m+m'+k+2,0}^{(2)}+ (-1)^{l'+m'+s'\over 2}\,
     e^{i\pi{m+s\over 2}}\, P_{l,k-l'}\, \delta_{m+m',0}^{(2)}
   \right]\  .\nonumber
\eea
The extra sign factors in (\ref{pmatrix}),
\bea
\label{sino}
 \sigma_{s_0}=(-1)^{h_{s_0}-\Delta_{s_0}}\ , \quad
    \sigma_{(l,m,s)}=(-1)^{h^l_{m,s}-\Delta^l_{m,s}}\ ,
\eea
stem from the roots of the modular $T$-matrix in the definition
of $P$.


\subsubsection{Orientifolds of Gepner models}

Now we are in the position to describe orientifolds of the 168
different Gepner models with $c=9$. Let us start with some general
comments on Gepner model orientifolds.

The partition function (\ref{parti}) is the so-called  charge
conjugate invariant in the sense that it combines left- and
right-moving states with opposite $U(1)$ charge. Geometrically,
the resulting model describes the type II string compactified on a
Calabi-Yau space $\cx$. Besides the charge conjugate  partition
function $\cz_C$, there also exists the diagonal partition
function $\cz_D$ where one combines left- and right-moving states
with the same  $U(1)$ charges in each tensor factor. As is well
known this describes the type II string on the mirror manifold
$\cw$. Moreover, under mirror symmetry type IIB with the diagonal
invariant is mapped to type IIA with the charge conjugate
invariant and vice versa.

These relations continue to hold if we perform an orientifold
projection which breaks the space-time supersymmetry down to
${\cal N}=1$. As usual, in an orientifold one  takes the quotient
by the world-sheet parity transformation $\Omega$\footnote{As we
will see in the following, also for Gepner models there exist more
general orientifold models where one combines $\Omega$ with some
other $\mbb{Z}_2$ symmetry of the internal SCFT.}.  The
orientifold projection of type IIB with $\cz_{D}$ by $\O$ is exchanged with the
orientifold projection of type IIA with $\cz_{C}$ by $\O\bs$, and vice
versa, where
$\bar\sigma$ denotes the charge conjugation of the $U(1)$ charge in each tensor
factor. For reasons which will become clear below  we call the
type IIB orientifold with $\cz_D$ the B-type model and the
orientifold with $\cz_C$ the A-type model. All these relations are
summarized in table \ref{gepnerori}.

\begin{table}[htb]
\renewcommand{\arraystretch}{1.5}
\begin{center}
\begin{tabular}{|c|c|c|c|c|}
\hline
 & IIB A-type & IIB B-type & IIA A-type & IIA B-type \\
\hline \hline
 Projection &  $\Omega$ &$\Omega$ &  $\Omega\bar\sigma$ &  $\Omega\bar\sigma$ \\
\hline
Partition function & $\cz_C$   &  $\cz_D$   &  $\cz_D$ & $\cz_C$  \\
\hline
Calabi-Yau &  $\cw$    &  $\cx$  &  $\cx$  &  $\cw$  \\
\hline
\end{tabular}
\caption{\small Gepner model orientifolds}
\label{gepnerori}
\end{center}
\end{table}

Due to this relation via mirror symmetry we can restrict ourselves
to the discussion of type IIB orientifold models. Since in type
IIB one has even-dimensional orientifold planes and D-branes on
the Calabi-Yau, the number of tadpole conditions
is related to the number of even cycles in the Calabi-Yau, the Hodge
number $h^{1,1}$. For Gepner
models with  $\cz_D$ this number is generally rather small, whereas
for $\cz_C$ it is rather large. Therefore, the A-type
orientifold models are much more restrictive than B-type
orientifolds. It is known that by successive orbifolding one can
reduce $h^{1,1}$ while increasing $h^{2,1}$. For
instance, the first four-dimensional concrete examples appearing in
the literature \cite{bw98} were constructed for the $(3)^5$ Gepner
model. There,  for the diagonal partition function $\cz_D$ one gets
$h^{1,1}=1$. The only two resulting tadpole conditions allowed
a non-chiral gauge group as big as $SO(20)\times SO(12)$.
However, for the charge conjugate partition function $\cz_C$ with
$h^{1,1}=101$ the only solution was found to have gauge group $SO(4)$.

After these generalities, let us now compute the Klein bottle
amplitude for the A-type orientifold models in detail. A Gepner
model $\bigotimes_{j=1}^r (k_j)$, $r=5,9$, with only four tensor
factors can  be treated as if having five tensor factors, the
fifth being given by $k_5=0$. Our starting point  is the charge
conjugate modular invariant
(\ref{parti}). As is evident from (\ref{parti}) the states surviving the
$\Omega$-projection have to satisfy
\bea
\label{kaa}
\mu \cong  -\mu + b_0 \beta_0 + b_1 \beta_1 + \cdots + b_r \beta_r\ ,
\eea
i.e. they have to be equal up to reflections and shifts.
This means
\bea
\label{kab}
m_j &=& b   + \frac{1}{2}\eta_j (k_j +2) \ {\rm mod} \ (k_j +2)  \quad
{\rm for\ all}\  j\ ,  \non
   s_0 &=& b + \sum_i b_i   \ {\rm mod} \ 2\ ,  \non
   s_j &=& b + b_j + \eta_j\ {\rm mod} \, 2 \quad {\rm for\ all}\  j\ ,
\eea
for some $b$ in the range $\{0, \ldots, \frac{K}{2} -1 \}$, $b_j =
0,1$. The parameters $\eta_j$ take values $\eta_j =0,1$ in every
tensor factor where $l_j =\frac1{2}k_j$ and are zero  otherwise.
Therefore, they  are  only  present for even $K' = {\rm lcm}(k_j
+2)$. The origin of $\eta_j$ is due to the fact that for even
levels the value $l_j =\frac1{2}k_j$ is invariant under the
reflection symmetry $(l_j,m_j,s_j)\to (k_j-l_j,m_j+k_j+2,s_j+2)$,
thus leading to the existence of shorter simple current orbits.
The constraints on $s_j$ and $s_0$ imply
\bea
\sum_j \eta_j = 0 \ {\rm mod} \, 2\ .
\eea
However, the orientifold projection is by no means unique in the
sense that one is always free to dress the characters which
survive the projection with additional signs consistent with the
fusion rules \cite{Sagnotti:1996eb}.

In view of the free parameters in (\ref{kaa}) and the various relations
(\ref{kab})
between them,  we define the orientifold projection
$\Omega_{{\Delta}_j, \omega,
\omega_\alpha}$ by including the sign factors
\bea
 (-1)^{ {\omega} \, (b +s_0) + \sum_j  {\Delta}_j \eta_j }
\eea
for $\Delta_i,\, \omega= 0,1$. Note  that the $\Delta_j$ only have a
non-trivial effect if $k_j$ is even. Moreover, the combination $(b
+s_0)$ is just right for the dressing with $\omega$ to preserve
supersymmetry of the resulting Klein-bottle amplitude. It is only
well-defined for $K'$ even. The resulting overall A-type Klein
bottle can be written as
\bea
\ck^{\rm A}({\Delta}_j, \omega) &=& 4 \int_0^{\infty} \frac{dt}{t^3} \frac{1}{2^{r+1}}
  \frac{1}{\eta( 2it)^2}  {\sum_{\lambda,\mu}}^{\beta}\,
\sum_{\eta_1,\ldots,\eta_r=0,1}
\sum_{b=0}^{\frac{K}{2}-1}  \,\,
(-1)^{s_0}\, (-1)^{\omega  (b +s_0)} (-1)^{ \sum_j  {\Delta}_j \eta_j} \nonumber \\
&&
\hspace{-2cm}
\times
   \delta^{(2)}_{\sum_j \eta_j, 0} \,\, \left( \prod_{k<l} (-1)^{\eta_k \eta_l} \right)
  \left( \prod_j \delta_{l_j \eta_j, \frac{k_j}{2} \eta_j} \right)
 \left( \prod_{j=1}^r \, \delta^{(k_j +2)}_{m_j, b + \eta_j \frac{1}{2}(k_j+2)}\right)  \,
 \chi^{\lambda}_{\mu} ( 2 i t )\ ,
\eea
where generally $\delta^{(n)}$ denotes the delta-function modulo
$n$. The  tree-channel amplitude can be obtained as usual by
applying a modular $S$-transformation
\bea
\label{kak}
\tilde{\ck}^{\rm A}({\Delta}_j, \omega)&=&
 \frac{2^4 }{ 2^{\frac{3r}{2}}
 \prod_j \sqrt{ k_j + 2}}
\int_0^{\infty} dl \frac{1}{\eta(2il)^2}
{\sum_{\lambda',\mu'}}^{\rm ev}\,
\sum_{\eta_1, \ldots, \eta_r=0,1} \,
\sum_{\nu_0 = 0}^{K-1} \,
\sum_{\nu_1, \ldots, \nu_r=0,1} \,
\sum_{\epsilon_1, \ldots, \epsilon_r=0,1}  \nonumber \\
&&
\times \prod_{k<l} (-1)^{\eta_k \eta_l}
\delta_{\sum_j \eta_j, 0}^{(2)}\,\,
\delta^{(4)}_{s'_0 + \nu_0 + 2 \sum \nu_j +2, 2\omega }\,\,
\delta^{(2)}_{ \sum_j \frac{1}{k_j +2} (m'_j  + (1-
\epsilon_j)(k_j + 2)), \omega}  \nonumber \\
&&
\times
\prod_{j=1}^r  \Bigg[ \frac{ P_{l_j',\epsilon_j k_j} P_{l_j', (\epsilon_j +
\eta_j) k_j}}{S_{l_j',0}} \,\, \delta^{(2)}_{\eta_j k_j,0} \,\,
(-1)^{\eta_j \big( \frac{m_j'}{2} + \nu_0 + \Delta_j +(1-\epsilon_j) \big)}   \\
&&
\hspace{3.5cm} \times
\delta^{(2)}_{m'_j + (1- \epsilon_j)(k_j + 2),0 } \,\,
\delta^{(4)}_{s'_j+  \nu_0 +2\nu_j + 2(1-\epsilon_j),0} \Bigg]\,\,
\chi^{\lambda'}_{\mu'}(2il)\ .
\nonumber
\eea
From the tree-channel Klein bottle, we can read off the cross-cap
state up to an overall phases factor. One can for instance follow the
method presented in \cite{RB03a} which was shown to work in the
NSNS sector and is therefore sufficient for supersymmetric models.
The cross-cap state takes the form
\bea
\label{kam}
\big| C; {\Delta}_j,\omega  \big>_{\rm NSNS} &=& \frac{1}{\kappa^{\rm A}_c}
{\sum_{\lambda',\mu'}}^{\rm ev}\,
\sum_{\nu_0 = 0}^{{K\over 2}-1} \,
\sum_{\nu_1, \ldots, \nu_r=0,1} \,
\sum_{\epsilon_1, \ldots, \epsilon_r=0,1}\
(-1)^{\nu_0} (-1)^{\sum_j \nu_j} \prod_{k<l} (-1)^{\nu_k \nu_l} \,
\nonumber \\
&& \hspace{-3.5cm}
\times (-1)^{\omega \frac{s'_0}{2}}\,
e^{i \pi \sum_j \frac{\Delta_j}{k_j +2} (m_j' + (1-\epsilon_j)(k_j
 +2))} \, \delta^{(4)}_{s'_0 + 2 \nu_0 + 2 \sum \nu_j +2, 2 \omega }
\delta^{(2)}_{ \sum_j \frac{1}{k_j +2} (m'_j  + (1-
\epsilon_j)(k_j + 2)), \omega } \\
&&\hspace{-3.5cm}
\times
\prod_{j=1}^r  \Biggl( \sigma(l_j', m_j', s_j')
\frac{P_{l'_j, \epsilon_j \, k_j}}{\sqrt{S_{l'_j,0}}} \,
(-1)^{\epsilon_j \frac{m_j' + s_j'}{2}}
\nonumber
\delta^{(2)}_{m'_j + (1- \epsilon_j)(k_j + 2),0 } \,
 \delta^{(4)}_{s'_j+ 2\nu_0 +2\nu_j + 2(1-\epsilon_j),0} \Biggr)\,\,
\big|{\lambda'},{\mu'}\big>\big>_c\ ,
\eea
where $\big|{\lambda'},{\mu'}\big>\big>_c$ denote the
Cardy type cross-cap states in  the tensor product CFT and
\bea
\label{kah}
\left({\kappa_c^{\rm A}}\right)^{-2} =
\frac{ 2^5  }{ 2^{\frac{3r}{2}}K
 \prod_j \sqrt{k_j + 2} }\ .
\eea


\subsubsection{Open string one-loop amplitudes}

As usual, in order to cancel the charge and tension of the
orientifold planes one introduces appropriate D-branes. In the
CFT D-branes are described by boundary states.
Starting with the so-called Cardy boundary states
for the individual tensor factors, the A-type boundary states for
Gepner models were constructed in \cite{Recknagel:1997sb}. They
are essentially given by the modular $S$-matrix as
\bea
\label{ana}
 \big|B,a\big> &=& \Big|S_0; \prod_{j=1}^r (L_j, M_j, S_j)\Big> \non
&=&
\frac{1}{\kappa_{a}}
{\sum_{\lambda',\mu'}}^{\beta}\,
(-1)^{\frac{s'^2_0}{2}} e^{ -i\pi \frac{s'_0 S_0}{2}}
\prod_{j=1}^r \Bigg[ \frac{S_{l'_j, L_j}}{\sqrt{S_{l'_j,0}}} \,\,
e^{i \pi \frac{m'_j M_j}{k_j +2} } \, e^{-i \pi \frac{s'_j S_j}{2} }\Bigg]
\big|{\lambda',\mu'}\big>\big>  \nonumber
\eea
with normalization
\bea
\label{and}
\left(\kappa_{a}\right)^{-2 } =
\frac {K }{2^{\frac{r}{2} +1} \prod_j \sqrt{k_j +2} }\ .
\eea
In order to finally read off the massless spectrum, we have to
transform the annulus diagram to the loop-channel
\bea
\label{anc}
\ca_{\tilde{a}\, a} &=& N_{a} N_{\tilde{a}}\, {1\over 2^{r+1}}\,
\int_0^{\infty} \frac{dt}{t^3} \frac{1}{\eta(it)^2}
 {\sum_{\lambda,\mu}}^{\rm ev}\,
\sum_{\nu_0 = 0}^{K-1}
\sum_{\nu_1, \ldots, \nu_r = 0,1}
\sum_{\epsilon_1, \ldots,\epsilon_r=0,1} \,
( -1)^{\nu_0}  \\
&&\hspace{-1.6cm}
\nonumber
\times
\delta^{(4)}_{s_0, 2+ \tilde{S}_0 - S_0 - \nu_0 - 2 \sum_j \nu_j}
\prod_{j=1}^r \bigg[ N^{|\epsilon_j k_j - l_j|}_{L_j, \tilde{L}_j}
\delta ^{(2k_j +4)}_{m_j + M_j - \tilde{M}_j + \nu_0 + \epsilon_j
(k_j+2), 0}
\delta^{(4)}_{s_j, \tilde{S}_j - S_j - \nu_0 - 2 \nu_j + 2 \epsilon_j} \bigg]
\chi_{\mu}^{\lambda} (it)\ .
\eea
There is one subtlety here which appears if levels are
even. In this case  some of the boundary states (\ref{ana}) are
not fundamental. This can for instance be deduced from
degeneration of the vacuum state in the annulus amplitude
$\ca_{aa}$. These boundary states have to be further
resolved which was discussed in
\cite{Brunner:2000nk,Fuchs:2000fd}. Keeping in mind this subtlety,
we still work with the unresolved
boundary states (\ref{ana}) of Recknagel and Schomerus
\cite{Recknagel:1997sb,Brunner:1999jq} to keep the presentation simple.

Let us now address the issue  of the action of
$\Omega_{{\Delta}_j, \omega}$ on a boundary state. For this
purpose, we compute the overlap of a boundary state with the
cross-cap state (\ref{kam}).\footnote{This remains the same for
the resolved boundary states, as the cross-cap state only contains
untwisted contributions.} After transforming into loop channel we
obtain
\bea
\label{moe}
\cm_a^{\rm NS}( {\Delta}_j, \omega)   &=& (-1)^s N_{a} \, \,
\frac{1}{2^{r+1}}
\int_0^{\infty} \frac{dt}{t^3} \frac{1}{\hat\eta(it+\frac{1}{2})^2}
{\sum_{\lambda,\mu}}^{\rm ev}\,  \sum_{\nu_0 = 0}^{\frac{K}{2}-1}
\sum_{\epsilon_1, \ldots, \epsilon_r=0,1} \, (-1)^{\omega (\nu_0+ \frac{s_0}{2})}     \nonumber \\
&&
\hspace{-2.5cm}
\times
\delta^{(2)}_{\sum_j \rho_j, 0}\, \,
\delta^{(2)}_{s_0,0}\,
\prod_{k<l} (-1)^{\rho_k \rho_l}
\prod_{j=1}^r  \Biggl( \sigma_{(l_j, m_j, s_j)}\, Y^{l_j}_{L_j,
\epsilon_j k_j}
\delta^{(2)}_{s_j,0}  \,\,
\delta^{(2k_j +4)}_{2 (M_j- \Delta_j)+ m_j + 2 \nu_0 +
\epsilon_j (k_j+2), 0} \nonumber  \\
&&
\hspace{-1cm}
\times
(-1)^{\frac{\epsilon_j}{2} [2S_j  -s_j- 2\epsilon_j]}
(-1)^{\frac{(1-\epsilon_j)}{2} [2M_j  -m_j+ \epsilon_j (k_j+2)]} \Biggr)\,\
\hat{\chi}^{\lambda}_{\mu} (it+{1}/{2})\ ,
\eea
where
\bea
\label{rhoo}
  r = 4 s +1\ , \quad
  \rho_j={s_0+s_j\over 2} + \omega+ \epsilon_j-1\ .
\eea
The real hatted characters $\hat\chi$ are defined as
\bea
\hat{\chi} (it+{1}/{2})=e^{-i\pi \left( h-{c\over 24}\right) }\, {\chi} (it+{1}/{2})
\eea
and similarly for $\hat\eta$.
The $Y$-tensor is defined as 
\bea
 Y_{l_1,l_2}^{l_3}=\sum_{l=0}^k  {S_{l_1,l}\, P_{l_2,l} \, P_{l_3,l}\over
                               S_{0,l} }\ .
\eea
Requiring that the M\"obius amplitude (\ref{moe}) is consistent
with the annulus amplitude (\ref{anc}) for a D-brane and its
image under $\Omega_{{\Delta}_j, \omega}$, one can derive
the action of $\Omega_{{\Delta}_j, \omega}$ on a boundary state.
First note that $\Omega$ reverses the sign of the labels
$S_0, M_j, S_j$. The phase dressings shift the $M_j$ to
$M_j+2\Delta_j$ and  the $\omega$ dressing changes the GSO
projection in (\ref{moe}). It therefore maps a brane to an
anti-brane, which can also be described by the shift $S_0\to
S_0+2$. To summarize, the entire action of $\Omega_{{\Delta}_j,
\omega}$ on a boundary state is given by
\bea
\label{acti}
\hspace{-.5cm}
\Omega_{{\Delta}_j,
\omega} : \Big| S_0, \prod_j (L_j, M_j, S_j)\Big> ~\mapsto~ \Big| -S_0 +2\omega,
 \prod_j (L_j,- M_j +2\Delta_j,-S_j)\Big> \ .
\eea
In particular, the invariant branes of the
Gepner model are now classified by
\bea
\label{bb}
\Big| S_0, \prod_{j=1}^{r'}\Big(\frac{k_j}{2}, \Delta_j + \frac{k_j
 +2}{2}, S_j\Big), \prod_{j=r'+1}^r (L_j, \Delta_j, S_j) \Big>
\eea
for $(r'-\omega)$ even and the $M_j$ chosen modulo $(k_j+2)$.
The massless spectra for the various open strings  stretched between
pairs of boundary states can be computed from the annulus (\ref{ana}) and
M\"obius strip amplitude (\ref{moe}). Note that they  contain both
information about the chiral and the non-chiral spectrum.

In order for the entire background to be supersymmetric, the
boundary states have to preserve the same supersymmetry as the
cross-cap state, which boils down to the simple condition
\bea
  {S_0-\omega\over 2} -\sum_j \Big[ {M_j-\Delta_j \over k_j+2} +
      {S_j \over 2}\Big]  =0\ {\rm mod}\ 2 \ .
\eea
From this  expression it is clear that the phase dressings can be
thought of as a rotation in the $M_j$ planes, whereas the $\omega$
quantum  dressing similarly can be considered as a rotation in the
$S_0$ plane. Therefore, from the CFT point of
view the phase shifts and the quantum dressing are completely
analogous.


\subsubsection{Tadpole cancellation conditions}

The tadpole cancellation conditions contain both the contribution
from the D-branes and from the orientifold planes. They take the
general schematic form
\bea
{\rm Tad}_{\rm D}(\lambda,\mu)-4\, {\rm Tad}_{\rm O}(\lambda,\mu)=0
\eea
for all the massless fields $(2)(0,0,0)^5$ and $(0)\prod_j(l_j,
l_j, 0)$ with $\sum_j \frac{l_j}{k_j +2} = 1$. The NSNS tadpoles
of the orientifold plane read
\bea
\label{kal}
{\rm Tad}_{\rm O}(\lambda,\mu) &=& (-1)^{(1+ \frac{s_0}{2})(1+\omega)} \sum_{\epsilon_1, \ldots,
\epsilon_r = 0,1} \, \,
e^{i \pi \sum_j \frac{\Delta_j}{k_j +2} (1-\epsilon_j)(k_j
 +2)}
\delta^{(2)}_{\sum_j \epsilon_j, \omega + \frac{s_0}{2}}
\prod_{k<l} (-1)^{\epsilon_k \epsilon_l}
\non
&&
\hspace{-1.5cm}
\times
\prod_j \Bigg(  {\rm sin}\Big[\frac{1}{2}(l_j, \epsilon_j k_j)_{k_j} \Big]\, \,
\delta^{(2)}_{l_j+(1-\epsilon_j) k_j,0} \,
\delta^{(2)}_{m_j + (1-\epsilon_j)(k_j+2), 0 } \,\,(-1)^{\epsilon_j
 \frac{m_j}{2} } \Bigg)\ .
\eea
Note that for even $k_j$ only those massless states with even $m_j$
have a non-vanishing tadpole on the orientifold plane.
Collecting all terms from the boundary states and their
images under $\Omega_{{\Delta}_j,\omega}$ the massless tadpoles read
\bea
\label{bc}
{\rm Tad}_{\rm D}(\lambda,\mu)&=&
 \sum_{a} 2N_{a}\, {\rm cos}\Big[\pi
     \sum_j \frac{m_j (M_j^a-\Delta_j)}{k_j+2} \Big] \,
\prod_j {\rm sin}(l_j,L_j^a)_{k_j}\ .
\eea
As has been checked in many examples, even though the tadpole equations
seem to contain non-integer coefficients,
one can  bring them in a form with only integer coefficients by
forming appropriate linear
combinations.

As we
mentioned earlier, besides the tadpole cancellation conditions,
there are also K-theory constraints, which for Gepner
model orientifolds so far can only be derived in a case by case
analysis utilizing the probe brane argument and the vanishing of
the global Witten anomaly \cite{Gato-Rivera:2005qd}. For Gepner models
themselves, i.e.\ without the orientifold projection, the
algebraic K-theory groups have been computed in \cite{Braun:2005eg}.


\subsubsection{Examples}

Let us assume that all the levels of the Gepner model are odd,
i.e.\ we are choosing a model from table \ref{tabgepner}.
\begin{table}[htb]
\renewcommand{\arraystretch}{1.5}
\begin{center}
\begin{tabular}{|c|c|c|}
\hline
 Levels & $(h^{2,1},h^{1,1})$ & Calabi-Yau \\
\hline
\hline
 $(1,1,1,1,1,1,1,1,1)$ &  $(84,0)$ & $-$ \\
 $(1,1,3,7,43)$ &  $(67,19)$ &  $\mbb P_{1,5,9,15,15}[45]$ \\
 $(1,1,3,13,13)$ &  $(103,7)$ &  $\mbb P_{1,1,3,5,5}[15]$ \\
 $(1,1,5,5,19)$ &  $(65,17)$ &  $\mbb P_{1,3,3,7,7}[21]$ \\
 $(1,1,7,7,7)$ &  $(112,4)$ &  $\mbb P_{1,1,1,3,3}[9]$ \\
 $(1,3,3,3,13)$ &  $(75,3)$ &  $\mbb P_{1,3,3,3,5}[15]$ \\
 $(3,3,3,3,3)$ &  $(101,1)$ &  $\mbb P_{1,1,1,1,1}[5]$ \\
 \hline
\end{tabular}
\caption{\small Gepner models with only odd levels. }
\label{tabgepner}
\end{center}
\end{table}
For the massless states $(0)\prod_j(l_j,l_j, 0)$ an odd number of
the $l_j$ are odd and an even number
even. One can then show that there  exists a generic solution to the
tadpole cancellation condition. Just introducing four
D-branes invariant under the world sheet parity operation and with
\bea
    L_j={k_j\mp1\over 2}\ ,\quad  M_j=S_j=S_0=0
\eea
for all $k_j=4n_j\pm 1$. This solution corresponds geometrically to
placing the D-branes on top of the orientifold
plane. One finds an $SO(4)$ gauge group for five tensor factors
and $Sp(4)$ for the single model with nine tensor factors.


\subsubsection{MSSM-like Gepner model orientifolds}
\label{sssgmo}

Clearly, the supersymmetric Gepner model orientifolds constitute
a very large class of consistent string compactifications. The
abstract CFT description gives some insights into D-branes deep inside
the truly stringy regime, at least at special points in moduli space. This by itself is
quite remarkable.

However, one can do more and search
systematically for models with characteristics close to the MSSM. As a
starting point we require the realization of the Standard Model gauge
symmetry and of its quiver from figure \ref{quiver} in the chiral
matter spectrum. Then we try to
satisfy the tadpole cancellation condition by introducing
additional stacks of hidden branes \cite{dhs04a}, with the
restriction of only vector-like matter charged under the Standard
Model gauge group.

An exhaustive search for Gepner model orientifolds including
simple current extensions using grid computing technology has been
carried out in \cite{dhs04a} (see also
\cite{Anastasopoulos:2006da}). The large number of $180.000$
MSSM-like models was found. The statistical distributions of
certain gauge theoretic quantities for this ensemble feature very
similar patterns as the intersecting D-brane models on the
$\mbb{Z}_2\times \mbb{Z}_2$ orientifold, all described in section
\ref{secstatistic}.

In all the models found in \cite{dhs04a} extra vector-like matter
appeared, which is probably related to the fact that Gepner models
lie at very specific, symmetric points in the Calabi-Yau moduli
space. It would be interesting to explore whether any of the many
models found can accommodate a realistic pattern of Yukawa
couplings. One would need to to develop the necessary tools
to compute the couplings in the CFT \cite{Brunner:2000wx}.


\subsection{Heterotic string compactifications}
\label{sechet}

Though it is a bit off the main track of this review, for
completeness and to show that many aspects are analogous to the
type II orientifolds, we  include a brief section about certain
aspects of heterotic string model building. Before the advent of
D-branes,  heterotic string theory, in particular the $E_8\times
E_8$ heterotic string, was considered to be the essentially the
only promising  candidate for realizing realistic four-dimensional
compactifications with GUT gauge groups like $SO(10)$ or $SU(5)$.
Various types of compactifications have been discussed in the
literature, which include toroidal orbifolds, Gepner models, or
hypersurfaces in toric varieties, just to name a few prominent
ones. It is not our purpose to review all these different
constructions, but just focus on some aspects which are
reminiscent to orientifold models we have discussed so far.


\subsubsection{String model building  constraints}

The rules for model building in the geometric large radius domain are very similar to type IIB
orientifolds as discussed in section \ref{sectypeiib}. In fact,
the $Spin(32)/\mbb Z_2$ heterotic string is related to the type I string by S-duality
\cite{Polchinski:1995df}, such that similarity among their
compactifications is to be expected. Let us summarize these rules briefly.

To preserve ${\cal N}=1$ supersymmetry in four dimensions, one
compactifies the $E_8\times E_8$ or $Spin(32)/\mbb Z_2$ heterotic string on a
Calabi-Yau three-fold $\cx$. Due to right moving supersymmetry the
right moving fermions couple to the tangent bundle of the
Calabi-Yau space. However, to completely define the supersymmetric
non-linear world
sheet sigma-model one also has to specify to which bundle the left
moving world-sheet fermions couple. The heterotic non-linear
sigma-model takes the following general form
\bea
\cs&=&{i\over 2\pi} \int_\S d^2z \biggl[  {1\over 2}\,
G_{i\bar \jmath}\,
         (\partial X^i\, \bar{\partial} X^{\bar \jmath} +
         \partial X^{\bar \jmath}\, \bar{\partial} X^{i})
           - {1\over 2}\, B_{i\bar \jmath}\,
         (\partial X^i\, \bar{\partial} X^{\bar \jmath} -
         \partial X^{\bar \jmath}\, \bar{\partial} X^{i}) \nonumber \\
       &&\phantom{aaaaaaaaaa} + i \left( \lambda_a\, \bar{D}\lambda^a + \psi_{\bar \imath}
           D \psi^{\bar \imath} \right)+
       F^{a {\bar \imath}}_{b{\bar \jmath}} (X)\, \lambda_a\, \lambda^b\,
     \psi_{\bar \imath}\, \psi^{\bar \jmath} \biggr]
\eea
where the coordinates $X^i(z,\bar z)$ define the embedding from
the two-dimensional world-sheet $\S$ into the Calabi-Yau manifold.
The right-moving fermions $\psi^{\bar\imath}$ couple to the
pull-back of the Levi-Civita connection
\bea
D\psi^{\bar \imath}=\partial \psi^{\bar \imath} + \partial X^{\bar \jmath}\,
      \Gamma^{\bar \imath}_{\bar \jmath \bar k}(X)\, \psi^{\bar k}
\eea
and the left-moving fermions to a holomorphic connection
\bea
\bar D\lambda^{a}=\bar\partial \lambda^{a} + \partial X^{i}\,
      A^{a}_{bi}(X)\, \lambda^{b}
\eea
with curvature $F^{a}_{bi{\bar \jmath}}(X)$. If, like in type II string
theory, this bundle is identified with the tangent bundle, one gets
$(2,2)$ supersymmetry on the world-sheet and a gauge
symmetry $E_6\times E_8$ or $SO(26)\times U(1)$, respectively
\cite{Candelas:1985en}. However, this is only a very specific
subclass of heterotic string models and in general the vector
bundle $W$ can be different from the tangent bundle $T\cx$. In most cases
studied $W$ is of the form
\bea
\label{gbundle}
W = \bigoplus_{i=1}^K  V_{n_i}\ ,
\eea
where the $V_{n_i}$ are $SU(n_i)$ and $U(n_i)$ bundles
respectively. In addition, there can also be stacks of $N_a$
heterotic five-branes wrapping holomorphic
two-cycles $\Gamma_a$ of the Calabi-Yau. On the six-dimensional
world volume of these five branes there are $N_a$ massless
tensormultiplets plus hypermultiplets in the case of $E_8\times E_8$
and a vectormultiplet plus a hypermultiplet in
the anti-symmetric representation of $Sp(2N_a)$ in the case of
$Spin(32)/\mbb Z_2$. To leading
order in $\alpha'$ the string equations of motion, respectively the
supersymmetry conditions, provide several constraints on the vector
bundle $W$.

First, the structure group of $W$ has to be embedded into $SO(32)$
or $E_8\times E_8$, respectively. The gauge group $H$ in
four dimensions is the commutant of the structure group inside
$Spin(32)/\mbb Z_2$ or $E_8\times E_8$. By embedding single $SU(n)$ factors into $E_8\times E_8$
a set of gauge groups appears familiar from GUTs,
\bea
     SU(n)\times E_{9-n} \subset E_8\ ,
\eea
where we defined $E_5=SO(10)$, $E_4=SU(5)$ and $E_3=SU(3)\times
SU(2)$. This can be modified by further splitting
$SU(n)=U(n-1)\times U(1)$ and considering $U(n-1)\times U(1)$ bundles
\bea
\label{embed0}
V_{n-1}\oplus L\ \ {\rm with}\ \ c_1(V_{n-1})+c_1(L)=0.
\eea
Then the structure group is
$SU(n-1)\times U(1)$ and the commutant in $E_8$ becomes
$E_{9-n}\times U(1)$.

For $SO(32)$ there also exists a very natural class of
embeddings of $U(N)$ bundles which underlies also the type IIB
orientifolds. Decomposing the group $SO(32)$  as
\bea
\label{embed1}
SO(2M) \times \prod_{i=1}^{K} U(M_i)
\eea
with \mbox{$M+\sum_{i=1}^{K} M_i=16$} and embedding  a vector bundle with structure group
\mbox{$\prod_{i=1}^{K} U(n_i)$}
with $M_i=N_i\, n_i$ results in the non-abelian gauge group
\bea
\label{embed2}
SO(2M) \times \prod_{i=1}^K U(N_i)\ .
\eea
For both heterotic theories, anomaly freedom of the two
dimensional sigma-model implies that
\bea
\label{kthe}
       c_1(W)\in H^2(\cx,2\mbb{Z})\ ,
\eea
which means that the bundle has to admit spinors. For  $Spin(32)/\mbb Z_2$
the same constraint can be derived from the vanishing of the
global Witten anomaly for probe heterotic five-branes
\cite{Blumenhagen:2005zg}.

At string tree-level, the connection of the vector bundle has to satisfy
the hermitian Yang-Mills equations
\bea
\label{HYM}
F_{ij}=F_{\bar \imath\bar \jmath}=0\ ,\quad     g^{i\bar\jmath}\, F_{i\bar\jmath}=0\ .
\eea
The first equation implies that each term in (\ref{gbundle}) has
to be a holomorphic vector bundle. One can  identify this
constraint as an F-term in the effective ${\cal N}=1$ supergravity
description, which does not receive any perturbative
corrections in $\alpha'$ or in the string-loop expansion. For the
second equation in (\ref{HYM}) to hold, the so-called
Donaldson-Uhlenbeck-Yau (DUY) condition,
\bea
\label{DUY}   \int_{\cx}  J_2\wedge J_2 \wedge c_1(V_{n_i}) = 0
\eea
necessarily has to be satisfied. If so, a theorem by
Uhlenbeck and Yau guarantees a unique solution to (\ref{HYM})
provided each term is a $\mu$-stable vector bundle.
Completely analogous to the type I string discussed in section
\ref{sectypeiib}, this second constraint arises  from a D-term in
the effective supergravity description. Perturbatively it might
therefore be subject to at most one-loop corrections for anomalous
$U(1)$ gauge factors.

As has been pointed out in
\cite{Blumenhagen:2005ga,Blumenhagen:2005pm}, for $U(n)$ bundles
there indeed exists a one-loop correction to the DUY equation giving
rise to a loop correction to the $\mu$-slope.
For the specific embedding of $U(n)$ bundles into $SO(32)$ (\ref{embed1},\ref{embed2}) the
one-loop corrected central charge can be defined as
\bea
\label{fiterms4}
Z^{Spin(32)/\mbb Z_2}_{i}={1\over 2\pi g_s\, \ell_s^6} 
      \int_{\cx} {\rm tr}_{n_i}
\left[  e^{J_2 + 2\pi i\alpha' g_s F_i }\,
         \sqrt{\hat A(\cx)} \right] + {\co}\big(e^{-{1/\alpha'}},e^{-{1/g^2_s}} \big)
\eea
so that the gauge couplings and FI-terms of the $U(1)_i\subset U(N_i)$ subgroups are given by
\bea
     {1\over g_i^2}=\left\vert Z_i\right\vert\ , 
\quad\quad 2\pi\alpha' \xi_i=\arg\left(Z_i \right)\ .
\eea
Note, that for $SU(n)$ bundles with non-vanishing third Chern character the
additional constraint $\xi(V)=0$ cannot be satisfied. This one loop corrected heterotic slope  
can be viewed as the
S-dual of the type I $\Pi$-slope.

One gets a very similar result for the $U(n-1)\oplus U(1)$ bundles (\ref{embed0}) embedded
in for instance the first of the two  $E_8$  factors \cite{Blumenhagen:2005ga}
\bea
\label{fiterms5}
Z^{E_8}&=&{1\over 2\pi g_s\, \ell_s^6} 
      \int_{\cx}   e^{J_2}\, (1 + \bar f)
      \left({\textstyle 1-{1\over 8} \left[{\rm tr}_{E_8} \bar F^2 
                    - {1\over 2} {\rm tr}_{SO(1,9)} \bar{R}^2
\right]}\right) \\
&& \hspace{8cm}+ 
 {\co}\big(e^{-{1/\alpha'}},e^{-{1/g^2_s}} \big) 
\nonumber  
\eea
with the rescaled curvatures $\bar{f}=2\pi i\alpha' g_s f_{U(1)}$, 
$\bar{F}=2\pi i\alpha' g_s F_{U(n-1)\oplus U(1)}$ and $\bar{R}=2\pi i\alpha' g_s R$
containing in particular a factor of $g_s$. 
In contrast to (\ref{fiterms4}), the one-loop correction
contains a sum over all vector bundles embedded into $E_8$. Note,
that for $SU(n)$ bundles both the tree-level and the one-loop
contribution  vanish. The five-brane contributions to 
(\ref{fiterms5}) have been computed in  \cite{Blumenhagen:2006ux}
and the generalization to K3 manifolds was worked out  in \cite{Honecker:2006dt}.

The Bianchi identity for the NS three-form field strength
$H_3$
in the  $E_8\times E_8$ heterotic string reads
\bea
\label{TCC}
\frac1{l_s^2} dH_3&=& {1\over4  (2\pi)^2}\left[
{\rm tr} R^2 -{\rm tr} F_1^2 -{\rm tr} F_2^2 \right]+ \sum_a N_a
\,\gamma_a\ ,
\eea
For $SO(32)$ there is only one field strength and the term ${\rm
tr}(F_2^2)$ is absent. Here we have also introduced stacks of
heterotic (or better M-theory) 5-branes wrapping holomorphic, effective
curves in the Calabi-Yau manifold with Poincar\'e-dual four-forms $\gamma_a$. It
imposes the so-called tadpole condition for the background
bundles. For direct sums of $SU(n)$ bundles the resulting tadpole
cancellation condition takes the familiar form
\bea
\sum_{i=1}^K c_2(V_{n_i})+\sum_{a=1}^L N_a \gamma_a =c_2(T)\ .
\eea
The rules for computing the chiral massless spectrum are  very
similar to the type IIB orientifolds and for the $SO(32)$
heterotic string the table \ref{Tchiral1} still applies.

For $U(n)$ bundles in both heterotic string theories one finds
abelian anomalies, which are canceled by a generalized
Green-Schwarz mechanism involving the universal axion and the
internal axions that complexify the K\"ahler moduli. Analogously to
the type II orientifolds one can compute the
gauge kinetic functions which receive one-loop threshold corrections
as well. See \cite{Witten:1984dg,Andreas:2004ja,Andreas:2006dm,Blumenhagen:2006ux,Tatar:2006dc} for more details on
four-dimensional heterotic string models with
$U(N)$ bundles.

As for the type II orientifolds, it is a very hard problem to have
control over instanton corrections to the superpotential which
might lift some (or even all) of the classical moduli. The moduli
include the complex structure,  the K\"ahler, and the
bundle moduli. For heterotic compactifications,
described by $(0,2)$ supersymmetric world sheet linear sigma-models no superpotential
is generated for neutral scalars by world-sheet instantons
\cite{Silverstein:1995re,Basu:2003bq,Beasley:2003fx}. This result
is supported  by the construction of exactly solvable
superconformal field theories, $(0,2)$ Gepner models, for some
examples in this class \cite{Blumenhagen:1995tt,Blumenhagen:1995ew}. As mentioned, a
similar strong result for type II orientifolds is still lacking.
In addition one expects heterotic five-brane instanton corrections, which
are beyond analytic  control so far.


\subsubsection{Progress in semi-realistic heterotic model building}

During the last years there has been some progress in constructing
heterotic compactifications with realistic gauge groups and
particle spectra. In order to proceed from the general model
building rules summarized in the last section, one needs to have
some mathematical control over Calabi-Yau manifolds and their
vector bundles.

A class of Calabi-Yau three-folds for which constructive methods
to obtain stable vector bundles exist is given by elliptic
fibrations. In this case,  Freedman, Morgan, and Witten
\cite{Friedman:1997yq} have defined stable vector bundles by the
so-called spectral cover  construction. It is beyond the scope of
this review article to  cover this mathematical method, but let us
stress that it  plays an essential role in almost all recent
developments about compactifications  of the heterotic string.
Therefore, for more details we refer the interested reader to the
literature, where these methods have been refined  for the search
of realistic compactification of the $E_8\times E_8$ heterotic
string (see
\cite{Curio:1998vu,Ovrut:1999xu,Donagi:1999fa,Donagi:2000zf,Donagi:2000zs,Ovrut:2002hi,Donagi:2004qk,
Donagi:2004ia,Donagi:2004ub}
for a representative selection).
Alternative interesting realistic heterotic string models based on the
$\IZ_6$ orbifold have been constructed in 
\cite{Kobayashi:2004ya,Buchmuller:2005jr,Buchmuller:2006ik}.
Large classes of $SO(32)$ heterotic string vacua have been explored
in \cite{Nilles:2006np}.

If one uses $SU(5)$ or $SU(4)$ bundles to break the observable
gauge group down to $SU(5)$ or $SO(10)$, respectively, the problem
arises that there does not exist a Higgs field in the massless
spectrum, which can break the GUT model down to the MSSM.
Therefore, in these compactifications, one has to implement an
alternative way to break the GUT gauge symmetry. This is done by
turning on discrete Wilson lines, which however only exist if the
Calabi-Yau contains non-contractable one-cycles, i.e. if the first
fundamental group is non-trivial
(It can only be finite group, as the first Betti number vanishes.)

Concretely, if the Calabi-Yau manifold admits a
$\mbb Z_2$ \cite{Bouchard:2005ag} or $\mbb Z_3\times \mbb Z_3$
\cite{Braun:2005ux,Braun:2005bw,Braun:2005zv} Wilson line
it allows to break the $SU(5)$ or $SO(10)$ GUT gauge symmetry down
to the Standard Model gauge symmetry. Such manifolds can be
constructed by taking free quotients of simply connected
Calabi-Yau manifolds. It has been demonstrated in a series of
papers that such manifolds can be defined. Using sophisticated
mathematical tools for constructing both $SU(5)$
\cite{Bouchard:2005ag,Bouchard:2006dn} and $SO(10)$
\cite{Braun:2005ux,Braun:2005bw,Braun:2005zv} bundles,
string models can be defined whose particle content comes
remarkably close to the MSSM. Indeed, the charged massless
spectrum in the observable supersymmetric sector consists
precisely of three generations of Standard Model matter even
without additional vector-like matter. These are very interesting
advances and it remains to be seen how realistic the further
details are going to be.

All  the compactifications mentioned in the last paragraph start
with $SU(5)$ or $SU(4)$ bundles. An alternative way to get GUT
string models has been proposed in \cite{Blumenhagen:2006ux},
where also bundles with $U(N)$ structure groups have been allowed.
Without going into the details let us mention that it is possible
to get, for instance, so-called ``flipped'' $SU(5)$ GUT models at
the string scale whose gauge symmetry can be spontaneously broken
to the Standard Model by a vacuum expectation value for a Higgs
field in the anti-symmetric representation of $SU(5)$. For more
details on all these heterotic string model building attempts we
refer the reader to the still growing literature.

\clearpage
\setcounter{equation}{0}

\section{LOW--ENERGY EFFECTIVE ACTION}

In the previous sections we have developed the techniques for constructing
consistent string vacua with D--branes and orientifolds. 
In trying to use these theories to describe the observed physics
or the measurements in future accelerator experiments
like the LHC, it is of fundamental importance to obtain the
low--energy field theory of each given class of $D=4$ string vacuum.

For a given string vacuum the physics below the Planck scale may be described 
by a low--energy effective action, which is valid below a certain cut--off scale.
The latter may be the string scale $M_s\sim \ap^{-1/2}$. 
The effective action then describes the dynamics of all fields with masses
below this scale. 

Hence in defining a field theory originating from string theory one first has to 
find all string states with masses below this scale. These states  include all the massless
closed and open string states of a given string vacuum. In addition, there may be 
states, which become light or even massless in a certain region of the string theory moduli space.
Moreover, heavy string states like string oscillators with masses close to or above the 
string scale may have a considerable impact on the low--energy 
couplings\footnote{The low--energy (bare) effective field--theory couplings
are  generically related to the string coupling $g_s$.}
through one--loop threshold effects.  

While finding all massless or light string states of a given string vacuum is rather 
straightforward even for non-toroidal backgrounds,  
finding their interactions turns out to be a non--trivial program.
There are two efficient ways to construct the effective interaction terms.
One way to proceed is to start with the effective action of the underlying
$D=10$ string theory
and perform a dimensional reduction on all interaction terms. This provides 
the effective low--energy action up to a certain accuracy. The latter is limited by
the fact, that already the effective action in $D=10$ is only known up to a certain order
in $\ap$. Moreover, this procedure does not take into account 
in an appropriate way truly stringy effects coming from string--loops
or effects from the string world--sheet.
Nevertheless, qualitative important results may be obtained this way.
The second method to construct the $D=4$ effective action uses the string $S$--matrix 
approach, \ie computing string scattering amplitudes
involving massless string states as external states. A string $S$--matrix 
represents a perturbative expansion in $\ap$ and the string coupling constant $g_s$.
From this expansion one may extract for a given order in $\ap$ and $g_s$
the relevant interaction terms of the low--energy effective action. This way
the low--energy effective action becomes a perturbative series in  $\ap$ and $g_s$.
The $S$--matrix approach may be carried out to an arbitrary order in $\ap$ and 
$g_s$, but it requires the knowledge of the vertex operators and their interactions 
within the underlying conformal field theory.
About ten years ago the $S$--matrix approach was pursued with great detail
and also some success, see e.g  \cite{LauerKH,Lauer:1987kh,DKL,IbanezHC,HPN} 
for $D=4$ heterotic string vacua \cite{GrossDD}.

In this section by using the above described two methods we shall construct 
the effective low--energy action for type II orientifolds with D--branes.
In particular we shall present non--trivial 
coupling functions capturing stringy effects for the matter field metrics, 
Yukawa couplings and one--loop gauge threshold corrections.

\subsection{Low--energy effective field theory}\label{LOW}

It is well known \cite{CremmerEN} that any $\cn=1$ supergravity action
in four space--time dimensions
is encoded by three functions, namely the K\"ahler potential $K$,
the superpotential $W$, and the (matrix of) gauge kinetic function $f$.
We have collected some basic elements of the action that is derived
from these in section \ref{sugrapot} which we refer to in the
following. 

When such an effective action arises from a higher dimensional string theory 
these three functions usually depend non--trivially
on moduli fields $\phi^\a,\, \ov \phi^{\bar\a}$ describing 
the background of the present string model.
It is the aim of this section to present the moduli 
dependence of the K\"ahler potential, superpotential, and 
gauge kinetic function $f$ at string tree--level and one--loop. 
The K\"ahler potential will always be chosen as a gauge invariant function of the chiral
superfields. In addition to the metric also the sigma model couplings of the
chiral superfields are derived from it (\cff subsection \ref{SCATT}). 
The terms of the second line of \req{eff4} represent the gauge kinetic terms
for the world volume gauge fields of D-branes with their individual gauge symmetry. 
Gauge couplings are encoded in the holomorphic gauge kinetic functions $f_a(\phi)$. 
Finally, the function $\cv_{\rm SG}(\phi,\bar\phi)$ is the scalar potential accounting for F- and 
D-terms given in \reef{Fterm} and \reef{Dterms}. 
Due to $\cn=1$ supersymmetry the fermionic terms that we leave out 
are related to \req{eff4} by supersymmetry.

It is useful to split the set of scalar fields $\phi$ into a set of neutral
moduli fields $\Mc$ and into charged matter fields $\Cc$.
While the set of fields $\Mc$ refers to the dilaton
field and the geometric moduli fields of the compactification manifold ${\cal X}$,
the set of fields $\Cc$  accounts for all kinds of charged chiral matter fields
whose vacuum expectation values would change the gauge symmetry.
If the gauge symmetry is unbroken, the vevs of the matter fields $\Cc$ vanish. 
We therefore
may expand the K\"ahler potential and the superpotential with respect to  small $\Cc$.

The most general renormalizable superpotential involving the chiral superfields $\Cc$~is:
\beqn \lab{mostW}
W(\Mc,\Cc)&=&\\
&& 
\hspace{-2cm}
\sum_\al a_\al(\Mc)\ \Cc^\al+\h\ \sum_{\al,\bet} \mu_{\al\bet}(\Mc)\ \Cc^\al\ \Cc^\bet+
\fc{1}{3}\ \sum_{\al,\bet,\gamma} W_{\al\bet\gamma}(\Mc)\ \Cc^\al\ \Cc^\bet\
\Cc^\gamma+\, \cdots\ .
\nonumber
\eeqn
Similarly we expand the K\"ahler potential in terms of the matter
fields as: 
\beqn\lab{mostK}
K(\Mc,\bar \Mc,\Cc,\bar \Cc)&=&\\
&&
\hspace{-3.3cm}
K_0(\Mc,\bar \Mc)+\sum_{\al,\bet} G_{\Cc^\al
  \Cc^\bbe}(\Mc,\bar \Mc)\ \Cc^\al\ \Cc^\bbe + \lf(\h\ H_{\al\bet}(\Mc,\bar \Mc)\
\Cc^\al\ \Cc^\bet+{\rm h.c.} \ri)+\, \cdots\ .
\nonumber
\eeqn
Explicitly, we are interested to compute the coefficient functions of
these expansions, at least their dependence on the moduli scalars. 
Higher powers (denoted by the dots) in the matter fields $\Cc$ both in the superpotential 
\req{mostW} and in the K\"ahler potential \req{mostK} may come from higher order string 
corrections. 
The form \req{mostW} and \req{mostK} makes sure, that supersymmetry is unbroken
explicitly. In \req{mostW} the second and third terms give rise to 
supersymmetric mass terms and Yukawa couplings in the scalar potential \req{SGpot}.
On the other hand the first term of \req{mostW} generates non--vanishing F-terms $F_\Cc$ 
for the matter fields $\Cc$ and supersymmetry may be broken dynamically.
In subsection \ref{SOFT} the superpotential is extended by further 
potential supersymmetry breaking terms generating also F-terms for the closed string 
moduli fields $F_\Mc$.
Furthermore, the elements of the action are subject to
the non-renormalization constraints mentioned in section
\ref{sugrapot} and table \ref{tabSGcorr}.


\subsection{Closed string moduli space}\label{GEOMETRY}

Starting from the parent type II compactification
on the Calabi-Yau ${\cal X}$ in this subsection we shall perform the truncation to the
orientifold theory. Most of the non-trivial information is in the
proper definition of the K\"ahler variables in the $\cn=1$
Lagrangian. 

\subsubsection{Calabi--Yau compactification of type {IIB} strings}
\label{IIBCYM}

We start with a \tb compactification on a Calabi--Yau (CY) manifold ${\cal X}$.
This leads to ${\cal N}=2$  supersymmetry in $D=4$ dimensions. The geometry
of the manifold ${\cal X}$ is described by $h^{1,1}({\cal X})$ K\"ahler moduli $T^I$
and $h^{2,1}({\cal X})$ complex structure moduli $U^\Si$. These moduli fields 
represent scalar components of ${\cal N}=2$ hyper-- and vector multiplets, respectively. 
Together  with the universal hypermultiplet we have 
$h^{1,1}({\cal X})+1$ hypermultiplets and $h^{2,1}({\cal X})$ vector
multiplets.
The full moduli space $\Mc$ is a direct product
\eqn{FULLM}{\Mc=\Mc^{h^{2,1}({\cal X})}_{CS} \otimes \Mc^{h^{1,1}({\cal X})+1}_{KM}}
of a special K\"ahler manifold $\Mc_{CS}$ of (complex) dimension $h^{2,1}({\cal X})$ and
a quaternionic manifold $\Mc_{KM}$ of (quaternionic) dimension $h^{1,1}({\cal X})+1$.

The holomorphic $3$--form $\Om_3$ may be expanded with respect to  a real
symplectic basis $(\alpha_\La,\beta^\La)$ \cite{Cand}
\eqn{Omega}{
\Om_3=\sum_{\La=0}^{h^{2,1}({\cal X})}X^\La\ \alpha_\La-F_\La\ \beta^\La\ ,}
with the sections $(X^\La,F_\La)$ depending holomorphically on the complex
structure deformations  $u^\Si$. Here $F_\La$ is the first derivative of
the holomorphic prepotential $F(X)$ with respect to  $X^\La$. 
A set of special coordinates, with $X^\La=(1,u^\La)$, may be introduced.
The metric $g_{u^\La \ov u^{\ov \Si}}(u,\ov u)$ on the space of complex structure deformations 
$u^\La$ is derived from the K\"ahler potential
\eqn{}{
\kappa_4^2\ K_{CS}=-\ln\lf(-i\int_{{\cal X}} \Om_3\wedge\ov\Om_3\ri)\ ,}
with 
\eqn{}{
g_{\La\ov\Si}=\fc{\partial^2}{\partial_{u^\La}\partial_{\ov u^{\ov \Si}}}\  K_{CS}(u,\ov u)\ .}

On the other hand, the
metric $g_{I\ov J}\equiv g_{t^I t^J}(t,\ov t)$ on the space of K\"ahler deformations
is derived from the K\"ahler potential \cite{Cand}
\eqn{KMKM}{
\kappa_4^2\ K_{KM}(t)=-\ln\left( \fc{1}{6}\int_{{\cal X}} J_2\wedge J_2\wedge J_2\right)\, ,}
with the K\"ahler form
\eqn{KF}{
J_2=\sum_{I=1}^{h^{1,1}({\cal X})} \im(t^I)\ \omega_I\ ,} 
the complexified K\"ahler coordinates $t^I$ and the $h^{1,1}({\cal X})$ harmonic 
$(1,1)$--forms $\omega_I$.
Let us introduce the triple intersection numbers
$\Kc_{IJK}$ and the intersection form $\Kc_{IJ}$
\eqn{intersection}{
\Kc_{IJK}=\int_{{\cal X}}\omega_I\wedge\omega_J\wedge\omega_K\ \ \ ,\ \ \ 
\Kc_{IJ}=\int_{{\cal X}}\omega_I\wedge\omega_J\wedge J_2\ ,}
and 
\eqn{intersectionI}{
\Kc_{I}=\int_{{\cal X}}\omega_I\wedge J_2 \wedge J_2\ \ \ ,\ \ \ 
\Kc=\int_{{\cal X}} J_2\wedge J_2\wedge J_2\ .}
Finally, with this information we may write  the metric for the K\"ahler moduli\footnote{
As we shall see in the next subsections, in orientifold compactifications
new K\"ahler moduli $T$ have to be introduced, since the K\"ahler moduli $t$ do not 
represent proper scalars of ${\cal N}=1$ chiral multiplets. In the following only 
the K\"ahler deformations $\im(t)$ will 
be relevant. By abuse of notation we shall simply write $t$ for the K\"ahler deformations $\im(t)$
of the CY manifold  ${\cal X}$.} $t^I$:
\eqn{}{
g_{I\ov J}=-\fc{\partial^2}{\partial_{t^I}\partial_{\ov t^{\ov J}}}\  K_{KM}(t,\ov t)=-\fc{3}{2}\ \kappa_4^{-2}\ 
\lf(\ \fc{\Kc_{IJ}}{\Kc}-\fc{3}{2}\ \fc{\Kc_I\ \Kc_J}{\Kc^2}\ \ri)\ .}

\subsubsection{Calabi--Yau orientifolds of type IIB superstring theory}\label{IIORI}

To arrive at ${\cal N}=1$ supersymmetry in $D=4$ we introduce an orientifold projection $\Oc$.
The orientifold projection $\Oc$ acting on the closed \tb string states is given 
by a combination of world--sheet parity transformation $\Om$ and a reflection $\si$
in the internal CY space.
Here, $\Om$ describes a reversal of the 
orientation of the closed string world--sheet and $\si$ is an internal symmetry
of the manifold ${\cal X}$. More precisely, consistency requires $\sigma$ to act as 
an isometric and holomorphic involution on ${\cal X}$.
We shall also label the CY geometry ${\cal X}$ modded out by the
additional involution $\sigma$ by ${\cal X}$.
The transformation $\si$ leaves the K\"ahler form invariant, but may act non--trivially 
on the holomorphic three--form $\Om_3$. 
In a local coordinate patch of ${\cal X}$, the transformation $\sigma$ 
may be represented as a reflection $\sigma\equiv I_n$ of $n$ internal
coordinates $x^i,\ i=1,\ldots,6$, 
supplement with some additional transformations in the extra coordinates 
parameterizing the local patch.
For $\Omega I_n$ to represent a symmetry of the original theory, $n$
has to be an even integer in \tb. Moreover, in order that $\Om I_n$ becomes also a $\IZ_2$--action
in the fermionic sector, the action $\Om I_n$ has to be supplemented by the operator
$[(-1)^{F_L}]^{\lf[\fc{n}{2}\ri]}$. 
Here, $\lf[\fc{n}{2}\ri]$ represents the integer part of $n/2$.
The operator $(-1)^{F_L}$ assigns a $(+1)$-eigenvalue to states from the NSNS--sector and a 
$(-1)$ to states from the RR--sector.
With these details, subject to the transformation 
behavior of $\Om_3$, two choices for $\si$ are possible   \cite{Acharya:2002ag,Brunner:2003zm}: 
\bea\label{Action}
(i)&&\Oc=(-1)^{F_L}\ \Om\ \sigma^\ast\ \ \ ,\ \ \ \sigma^\ast\ \Om_3=-\Om_3\ \ \ \Longrightarrow 
\sigma=I_6,I_2\ ,\\[3mm]
(ii)&&\Oc=\Om\ \sigma^\ast\ \ \ ,\ \ \ \sigma^\ast\ \Om_3=\Om_3\ \ \ \Longrightarrow 
\sigma=I_0,I_4\ . 
\label{Action2}
\eea
Here, $\sigma^\ast$ is the action of $\sigma$ on the cohomology $H^{p,q}({\cal
  X})$ of the CY manifold ${\cal X}$
(pullback of $\sigma$).
Generically, the projection $\Oc$ produces orientifold fixed planes [O(9-n)--planes],
placed at the orbifold fixpoints of the double cover ${\cal X}/I_n$.
Case $(i)$ leads to a set of orientifold 
O3/O7--planes, while case $(ii)$ yields O5/O9--planes.
The orientifold planes have negative D--brane charge, which 
has to be balanced by introducing positive tension objects.
Candidates for the latter may be collections of D(9-n)--branes and/or non--vanishing
three--form fluxes $H_3$ and $C_3$.
In order to obtain a consistent low--energy supergravity description, the above objects
are subject to the supergravity equations of motion. Eventually, this puts restrictions on the 
possible choices of fluxes, to be discussed later.

Due to the holomorphic action of $\si^\ast$, the latter splits the cohomology groups
$H^{p,q}({\cal X})$ into a direct sum of an even eigenspace $H^{p,q}_+({\cal X})$
and an odd eigenspace $H^{p,q}_-({\cal X})$ \cite{Brunner:2003zm}.
Hence, the action $\si$ splits the $h^{1,1}({\cal X})$ harmonic $(1,1)$ 
forms $\omega_I$ of ${\cal X}$
into a set of $h^{1,1}_{+}({\cal X})$ even forms $\omega_i$ and into a set of 
$h^{1,1}_{-}({\cal X})$ 
odd forms $\omega_a$.
Since the K\"ahler form $J_2$ is invariant under the orientifold action, 
it is expanded with respect to  
a basis of $H^{1,1}_+({\cal X})$
\eqn{expandJ}{
J_2=\sum\limits_{i=1}^\npp t^i\ \omega_i\ ,}
\ie only the even harmonic forms $\omega_i$ survive in the expansion \req{KF}.
On the other hand, the action $\si^\ast$ splits the $h^{2,1}({\cal X})$
harmonic $(2,1)$ 
forms $d_\La$ of ${\cal X}$ into a set of $h^{2,1}_{+}({\cal X})$ even forms $d_\alpha$ and
into a set of $h^{2,1}_{-}({\cal X})$ odd forms $d_\lambda$.
Hence, in case $(i)$ the three--form $\Om_3$ is expanded with respect to  a basis $\{d_\lambda\}$
of $H^{3}_-({\cal X})$, while 
in case $(ii)$ it is expanded with respect to  a basis $\{d_\alpha\}$ of $H^{3}_+({\cal X})$.
So, the orientifold action splits the forms into the subsets shown in Table \ref{tab1a}.

\begin{table}[h]
\centering
\begin{tabular}{|ccc|}
\hline
$H^{1,1}({\cal X})$ & $H_+^{1,1}({\cal X})$ & $H_-^{1,1}({\cal X})$ \\ [2pt]
\hline
$\om_I$ & $\om_i$ & $\om_a$ \\ [2pt]
\hline
\end{tabular}
\begin{tabular}{|ccc|}
\hline
$H^{2,1}({\cal X})$ & $H_+^{2,1}({\cal X})$ & $H_-^{2,1}({\cal X})$ \\ [2pt]
\hline
$d_\La$ & $d_\al$ & $d_\lambda$ \\ [2pt]
\hline
\end{tabular}
\caption{Splitting of $H^{1,1}({\cal X})$ and $H^{2,1}({\cal X})$ under the orientifold action.\label{tab1a}}
\end{table}
The same pattern holds for the K\"ahler $t^I$ and complex structure moduli $u^\Lambda$, as shown
in the next Table \ref{tab1b}.

\begin{table}[h]
\centering
\begin{tabular}{|ccc|}
\hline
$h^{1,1}({\cal X})$ & $h_+^{1,1}({\cal X})$ & $h_-^{1,1}({\cal X})$ \\ [2pt]
\hline
$t^I$ & $t^i$ & $t^a$ \\ [2pt]
\hline
\end{tabular}
\begin{tabular}{|ccc|}
\hline
$h^{2,1}({\cal X})$ & $h_+^{2,1}({\cal X})$ & $h_-^{2,1}({\cal X})$ \\ [2pt]
\hline
$u^\Lambda$ & $u^\alpha$ & $u^\lambda$ \\ [2pt]
\hline
\end{tabular}
\caption{Splitting of geometric moduli under the orientifold action.\label{tab1b}}
\end{table}

Finally let us remark, that the situation of $h^{1,1}_{-}({\cal X})\neq
0$ occurs quite generically for orientifolds of 
resolved orbifolds (see also subsection \ref{ORBIFOLDS} and \cite{blowup}).

\subsubsection{Calabi--Yau orientifolds of type IIB with D3/D7--branes}\label{IIB37}

In this subsection we shall discuss the low--energy effective action of the closed string moduli
space $\Mc$ after applying the orientifold projection $(i)$, defined in \eqq \req{Action}.
To cancel tadpoles and to construct models of phenomenological
interest we also add stacks of D3-- and D7--branes. 

The scalars $\phi,C_0$, the CY metric $g$ and the four--form $C_4$ are even under
the action $(-1)^{F_L}\ \Om$. On the other hand, the two--forms $B_2$ and $C_2$
are odd under $(-1)^{F_L}\ \Om$.

Only $h^{2,1}_{-}({\cal X})$ complex structure deformations $U^\lambda$
survive the orientifold projection  \req{Action}.
Because of \req{Action}, the holomorphic three--form \req{Omega} may be 
expanded with respect to   a real symplectic basis $(\alpha_\lambda,\beta^\lambda)$ of 
$H^{3}_-({\cal X})$ , \ie
\eqn{expandO}{
\Om_3=\sum_{\lambda=0}^{h^{2,1}_{-}({\cal X})} X^\lambda \alpha_\lambda-
F_\lambda\beta^\lambda\ ,}
with $(X^\lambda,F_\lambda)$ the periods of the original CY manifold  ${\cal X}$.
Furthermore, in 
\tb orientifolds with D3-- and D7--branes, under the orientifold action \req{Action}
the NSNS two--form $B_2$ and the RR two--form
$C_2$ are odd under the projection $\Om(-1)^{F_L}$. Hence, under $\sigma^\ast$  
these forms  must transform with a minus sign, \ie they are expanded with respect to  
a basis  of the cohomology $H^{1,1}_-({\cal X})$:
\eqn{expandB}{
B_2=\sum\limits_{a=1}^\mmm b^a\ \omega_a\ \ \ ,\ \ \ C_2=\sum\limits_{a=1}^\mmm c^a\ \omega_a\ .}
Similarly, a two--form flux is expanded with respect to  a basis
of the odd cohomology $H^{1,1}_-({\cal X})$:
\eqn{2flux}{
F_2=\sum\limits_{a=1}^\mmm f^a\ \omega_a\ .}
Clearly, D3--  and D7--branes may only be wrapped around
four--cycles whose Poincar\'e dual two--forms are elements of $H^2_{+}({\cal X})$. 
On a stack of D7--branes,
there are two kinds of world--volume $U(1)$ two--form fluxes $\fo$ and $\tilde f$. 
The flux $\fo$ is inherited from the ambient CY space ${\cal X}$, while the flux $\tilde f$
is a harmonic two--form of  the four--cycle $C_k$, which means homologically that
its Poincare dual two cycle is not the intersection of $C_k$ with
another four-cycle in ${\cal X}$.  We refer the reader to  
\cite{Jockers:2005zy,Jockers:2005pn}  
for a description of the orthogonal  splitting of a general two--form flux $f$ into
$f=\fo+\tilde f$, with $\fo\in im(\iota^\star),\ \tilde f\in coker(\iota^\star)$,
with the map $\iota^\star: H^{2}_-({\cal X})\ra  H^{2}_-(C_k)$.
Therefore, in the above expansion \req{2flux} we assume the orthogonal
decomposition $f^a=\fao+\tilde f^a$. Let us also introduce the instanton number
\eqn{inst}{Q_{\tilde f,k}=(2\pi\ap)^2\ \int_{C_k} \tilde f\wedge \tilde f\ ,}
and the combination
\eqn{COMB}{\Bc^a=b^a-2\pi\ap\ \fao\ ,}
which will become relevant later

In \tb orientifolds the fields $b^a,\ c^a$ and the flux $\fo^a$ give rise to $h^{1,1}_-({\cal X})$ 
complex scalars \cite{Jockers:2005zy,Jockers:2005pn} 
\eqn{combined}{
G^a=i\ c^a-S\ \Bc^a\ \ \ ,\ \ \ a=1,\ldots,h^{1,1}_-({\cal X})}
of ${\cal N}=1$ chiral multiplets \cite{gl04,GL}.
The moduli space of the dilaton field\footnote{We reserve the
letter $\t$ for the ten--dimensional complex dilaton scalar of type IIB
as defined in \req{dilG3}, while $S$ denotes the four-dimensional
K\"ahler coordinate and often the corresponding chiral
superfield, that measures the string coupling.} 
\eqn{DILA}{
S=e^{-\Phi}+i\ C_0\equiv i\ \ov\tau}
is locally spanned by the K\"ahler coset:
\eqn{Scoset}{
\Mc_S=\fc{SU(1,1)}{U(1)}\ .}
To summerize so far, for the \tb compactifications with the orientifold projection
\req{Action} the spectrum consists of the following moduli fields: the dilaton $S$, 
$h^{2,1}_{-}({\cal X})$ complex structure 
deformations $u^\lambda$, $\npp$ K\"ahler moduli $t^i$ and $h^{1,1}_-({\cal X})$ K\"ahler moduli $b^a$ 
and $c^a$.
With the dilaton field $S$ a CY orientifold compactification ${\cal X}$ has $\npp$ K\"ahler moduli $t^k$, 
$h^{1,1}_-({\cal X})$ scalars $G^a$ and $h^{2,1}_{-}({\cal X})$ complex structure moduli
$u^\lambda$.
As shown in Table \ref{tab2}, under the orientifold action \req{Action} 
the original set of $h^{1,1}({\cal X})+1$ ${\cal N}=2$ hypermultiplets  and $h^{2,1}({\cal X})$ ${\cal N}=2$
vectormultiplets is split into a set of ${\cal N}=1$ chiral and
vectormultiplets. 
\begin{table}[h]
\centering
\vbox{\ninepoint{$$
\vbox{\offinterlineskip\tabskip=0pt
\halign{\strut\vrule#
&~$#$~\hfil 
&\vrule$#$ 
&~$#$~\hfil 
&\vrule$#$ 
&~$#$~\hfil 
&\vrule$#$
&~$#$~\hfil 
&\vrule$#$\cr
\noalign{\hrule}
&\ 1\      &&\ {\rm dilaton} && S   &&  {\rm chiral\ multiplet}&\cr
&\ &&\ &&\ &&\ &\cr 
&\ h_{-}^{2,1}({\cal X})\ &&\ {\rm CS\ moduli}\ && 
\ u^\lambda\ &&{\rm  chiral\ multiplets}&\cr
&\ &&\ &&\ &&\ &\cr 
&\ h_{+}^{1,1}({\cal X})\   &&\ {\rm K\ddot{a}hler\ moduli}\ T^k\ 
&&\ t^k,\rho^k\ &&{\rm chiral/linear\ multiplets}&\cr
&\ &&\ &&\ &&\ &\cr  
&\ h_{-}^{1,1}({\cal X})\ &&\ {\rm add.\ moduli}\ G^a\ &&\ b^a,c^a\ 
&&{\rm chiral\ multiplets}&\cr
\noalign{\hrule}
&\ h_{+}^{2,1}({\cal X})\ &&\ {\rm add.\ vectors}\ 
&& V_\mu^{\tilde j}&&{\rm vector\ multiplets}&\cr
\noalign{\hrule}}}$$}}
\vskip-6pt
\caption{Moduli of Calabi--Yau orientifold ${\cal X}$ with O3/O7--planes.\label{tab2}}
\end{table}
The additional $h_{+}^{2,1}({\cal X})$ vectors (and their magnetic duals)
arise from the Ramond $4$--form $C_4$ reduced with respect to  the cohomology $H^{3}_+({\cal X})$.

The intersection properties \req{intersection} and \req{intersectionI}
generically change under the orientifold action.
First of all, only $\Kc_{ijk}$ and $\Kc_{iab}$ may  be
non--vanishing, while $\Kc_{abc}=0$ and $\Kc_{aij}=0$.
Second, the non--vanishing triple intersections $\Kc_{ijk}$ and
$\Kc_{iab}$
may change their values after the orientifold projection.  Concrete examples,
where this happens, represent the resolved orbifolds \cite{blowup}.

In the K\"ahler potential for the moduli fields
\eqn{FULL}{
\kappa_4^2\ 
K=-\ln(S+\ov S)-2\ln \left( \fc{1}{6}\int_{{\cal X}} J_2\wedge J_2\wedge J_2\right)-
\ln\lf(-i\int_{{\cal X}} \Om_3 \wedge \ov \Om_3\ri),}
besides the dilaton field $S$ 
only the $\npp$ invariant K\"ahler moduli  $t^k$ and the $h^{2,1}_{-}({\cal X})$ invariant
complex structure moduli enter explicitly.
However, the string theoretical K\"ahler moduli $t^j$ are not yet scalars 
of an ${\cal N}=1$ chiral multiplet.
After defining the proper holomorphic moduli fields $T^j$ (in the string frame\footnote{
In the Einstein frame the K\"ahler moduli $t^k$ are multiplied with $e^{-\h\Phi}$.
In the Einstein frame the CY volume reads  ${\rm Vol}({\cal X})=\fc{1}{6}e^{-\fc{3}{2}\Phi}\ 
\Kc_{ijk} t^it^jt^k$.}) \cite{gl04,GL} (\cff also \cite{lmrs04,lrs04} 
for the case with $h_{-}^{1,1}({\cal X})=0$)
\eqn{defmod}{
T^j=\fc{3}{4}\ \ \Kc_{jkl}\ t^k\ t^l-\fc{3}{8}\ e^{\Phi}\ 
\Kc_{jbc}\ \ov G^b\ (G+\ov G)^c+\fc{3}{2}\ i\ \lf(\ \rho^j-\h\
\Kc_{jbc}\ c^b\ \Bc^c\ \ri)-\fc{3}{4}\ \ov S\ \delta^j_k\ Q_{\tilde f,k}\ ,}
the second term $\kappa_4^2\ K_{KM}=-2\ln {\rm Vol}({\cal X})=-2\ln\fc{1}{6}\Kc_{ijk} t^it^jt^k$ 
in \req{FULL} may be expressed in terms of the ${\cal N}=1$ fields $T^j$ by eliminating the moduli
$t^j$. This way, 
in the low--energy effective action of \tb CY orientifolds, the fields $G^a$ do enter the
K\"ahler potential 
for the K\"ahler moduli $t^k$ 
through eliminating the moduli $t^k$ via the definition \req{defmod}.
By that the K\"ahler potential $K_{KM}$ for the $\npp$ K\"ahler moduli $T^j$  
becomes a complicated function $K_{KM}(S,T^j,G^a)$ depending on the dilaton $S$, 
the $\npp$ moduli 
$T^j$ and the $h^{1,1}_-({\cal X})$ moduli $G^a$ \cite{gl04,GL}. 
In \req{defmod} the axion $\rho^j$ originates from integrating the RR four-form 
along the four--cycle $C_j$. Eventually, the full K\"ahler potential 
\eqn{FULLK}{
\kappa_4^2\ K=-\ln(S+\ov S)-2\ln {\rm Vol}({\cal X})+K_{CS}}
for the dilaton $S$, $\npp$ K\"ahler moduli $T^j$, 
$h^{1,1}_-({\cal X})$ scalars $G^a$ and $h^{2,1}_{-}({\cal X})$ complex structure moduli
takes the form \cite{gl04,GL}:
\eqn{fullK}{
\kappa_4^2\ K=-\ln(S+\ov S)+K_{KM}(S,T^j,G^a)+K_{CS}\ .}
Finally, from the form of this K\"ahler potential we deduce, that the
full 
closed string moduli space $\Mc$ has the form \cite{gl04,GL}:
\eqn{closedMOD}{
\Mc=\Mc_{CS}^{h^{2,1}_{-}({\cal X})}\otimes \Mc_{KM}^{h^{1,1}({\cal X})+1}\ .}
Each factor is a K\"ahler manifold.

The structure of the terms in the definition of the K\"ahler modulus 
\req{defmod} may be easily anticipated from studying the holomorphic gauge
couplings on a D7--brane. On a D7--brane, which is wrapped around the
four--cycle $C_j$ the world--volume Chern--Simons couplings $\int_{\mbb R^{1,3}\times C_j}
C_p\wedge e^{-(B-2\pi\ap F)}$
give rise to the following series of CP--odd gauge couplings
\zbe
\int_{\mbb R^{1,3}} F^j\wedge F^j \  \int_{C_j}
\lf(\ C_4- C_2\wedge \Bc_2+\h\ C_0\ \Bc_2\wedge\Bc_2+\fc{(2\pi\ap)^2}{2}\ C_0\ \tilde f\wedge\tilde f\ \ri)\ ,
\ee
with $\Bc=B-2\pi \ap F$ and $F^j$ the space--time gauge field strength
on the D7--brane wrapping the four--cycle $C_j$.
The terms in the bracket may be identified
with the imaginary part $\im T^j$ of the K\"ahler modulus \req{defmod}:
\zbe
\im(T^j)\equiv \fc{3}{2}\ \int_{C_j}
\lf( C_4- C_2\wedge \Bc_2+\h\ C_0\ \Bc_2\wedge\Bc_2+\fc{(2\pi\ap)^2}{2}\ C_0\ \tilde f\wedge\tilde f\ \ri)\ .
\ee
Since $\im(T^j)$ couples to the CP--odd coupling $F^j\wedge F^j$, according to ${\cal N}=1$
supersymmetry in $D=4$, $\re(T^j)$ is related  to the CP even gauge
coupling. This may be seen by studying the calibration condition for 
the D7--brane wrapped around the four--cycle $C_j$.
More precisely, from the Born--Infeld action 
$$-\mu_7\ 
e^{-\Phi}\ \int_{\mbb R^{1,3}\times C_j}\ d^8\xi\ [-\det(G+B-2\pi\ap F)]^{1/2}$$
we extract the CP--even gauge--coupling $\re(T^j) F^j_{\mu\nu}F^{j\mu\nu}$.
In order for the D7--brane to respect $1/2$ of the supersymmetry
of the bulk theory, which is ${\cal N}=2$ in $D=4$, the internal four--cycle $C_j$ the D7 brane
is wrapped on has to fulfill the calibration conditions \cite{Marino:1999af,Jockers:2005zy,Jockers:2005pn}:
\eqn{calibration}{
e^{-\Phi}\ \int_{C_j} d^4\xi\  \det(G+B-2\pi\ap F)^{1/2}=\h\ \int_{C_j}\
J_2\wedge J_2-\fc{e^{-\Phi}}{2}\ \int_{C_j} \lf(\Bc_2\wedge \Bc_2+(2\pi\ap)^2\tilde f\wedge\tilde f\ri)\ ,}
and $\Bc_2\wedge J_2=0$.
Hence, the real part of the correct holomorphic modulus $T^j$, describing the 
CP even gauge coupling, is given by
\eqn{calibrationi}{
\re(T^j)\equiv \fc{3}{4}\ 
\int_{C_j}\ J_2\wedge J_2-\fc{3}{4}\ e^{-\Phi}\ \int_{C_j} 
\lf(\ \Bc_2\wedge \Bc_2+(2\pi\ap)^2\tilde f\wedge\tilde f\ \ri)\ ,}
in agreement with the definition \req{defmod}.
With $\omega_j$ the Poincar\'e dual
two--form of the four--cycle $C_j$, we have:
$\int\limits_{C_j}\ J_2\wedge J_2=\int\limits_{{\cal X}} \omega_j \wedge J_2\wedge J_2$. 
With \req{expandJ} we may write
\eqn{NICE}{
{\rm Vol}(C_j)=\fc{3}{4}\ 
\int_{C_j}\ J_2\wedge J_2=
\fc{3}{4}\ \Kc_{jkl}\ t^k\ t^l=\fc{1}{4}\ \fc{\partial}{\partial t^i}\ \int_{{\cal X}}\ J_2\wedge J_2\wedge
J_2=\fc{3}{2}\ 
\fc{\partial}{\partial t^j}\ {\rm Vol}({\cal X})\ ,}
which gives the volume ${\rm Vol}(C_j)$ of the four--cycle $C_j$ in string units.

To conclude, the holomorphic gauge kinetic function $f_j$ for a gauge group
on a D7--brane, which is wrapped on the four--cycle $C^j$ is:
\eqn{gaugef}{f_{D7,j}=T^j\ .}
On the other hand, for a space--time filling D3--brane the gauge kinetic function
is given by the dilaton field $S$:
\eqn{gauge3}{f_{D3}=S\ .}

Finally, the presence of the background two--form fluxes \req{expandB} and \req{2flux}
gives rise to the D--term potential 
\eqn{Dterm}{
{\cal V}_{\rm D}\sim \int J_2\wedge  \Bc_2}
on the D7--world volume (for more details see subsection \ref{OPEN}).
In \cite{LRSSii} this potential has been used to stabilize some of the $h^{1,1}_-({\cal X})$ K\"ahler moduli
$G^a$.

\subsubsection{Calabi--Yau orientifolds of type IIB with D5/D9--branes}\label{IIB59}

In this subsection we shall discuss the low--energy effective action of the closed string moduli
space $\Mc$ after applying the orientifold projection $(ii)$, defined in \eqq \req{Action2}.
To cancel tadpoles and to construct models of phenomenological
interest we also add D9--branes and stacks of D5--branes. 
The simplest example is the one with just the undressed $\Omega$ projection
discussed in length in section \ref{sectypeiib}.

Only $h^{2,1}_{+}({\cal X})$ complex structure deformations $u^\gamma$
survive the orientifold projection  \req{Action}.
Because of \req{Action}, the holomorphic three--form \req{Omega} may be 
expanded with respect to   a real symplectic basis $(\alpha_\gamma,\beta^\gamma)$ of 
$H^{3}_+({\cal X})$ , \ie
\eqn{expandOO}{
\Om_3=\sum_{\gamma=0}^{h^{2,1}_{+}({\cal X})} X^\gamma \alpha_\gamma-
F_\gamma\beta^\gamma\ ,}
with $(X^\lambda,F_\lambda)$ the periods of the original CY manifold ${\cal X}$.
Furthermore, in 
\tb orientifolds with D9-- and D5--branes, under the orientifold action \req{Action2}
the NSNS two--form $B_2$ is odd under $\Om$, while the RR two--form
$C_2$ is even under $\Om$. Hence, under $\sigma^\ast$  
these forms  must transform with a minus and plus sign, respectively.
Therefore these forms are expanded with respect to  
a basis  of the cohomology $H^{1,1}_-({\cal X})$ and $H^{1,1}_+({\cal X})$, respectively, \ie
\eqn{expandBB}{
B_2=\sum\limits_{a=1}^{h^{1,1}_-({\cal X})} b^a\ \omega_a\ \ \ ,\ \ \ C_2=\sum\limits_{i=1}^\npp c^i\ \omega_i\ .}
Similarly, a two--form flux from the ambient space ${\cal X}$ is expanded with respect to  a basis
of the odd cohomology $H^{1,1}_-({\cal X})$, \ie it assumes the expansion \req{2flux}.
To summarize so far, for the \tb compactifications with the orientifold projection
\req{Action2} the spectrum consists of the following moduli fields: the dilaton $S$, 
$h^{2,1}_{+}({\cal X})$ complex structure 
deformations $u^\lambda$, $\npp$ K\"ahler moduli $t^i$ and $c^j$ and $h^{1,1}_-({\cal X})$ K\"ahler moduli
$b^a$.
As shown in Table \ref{tab3}, under the orientifold action \req{Action2} the original set of 
$h^{1,1}({\cal X})+1$ ${\cal N}=2$ hypermultiplets  and $h^{2,1}({\cal X})$ ${\cal N}=2$
vectormultiplets is split into a set of ${\cal N}=1$ chiral and
vectormultiplets.  (\cff Table \ref{tab3}).
\begin{table}[h]
\centering
\vbox{\ninepoint{$$
\vbox{\offinterlineskip\tabskip=0pt
\halign{\strut\vrule#
&~$#$~\hfil 
&\vrule$#$ 
&~$#$~\hfil 
&\vrule$#$ 
&~$#$~\hfil 
&\vrule$#$
&~$#$~\hfil 
&\vrule$#$\cr
\noalign{\hrule}
&\ 1\      &&\ {\rm dilaton} && S   &&  {\rm chiral/linear\ multiplet}&\cr
&\ &&\ &&\ &&\ &\cr 
&\ h_{+}^{2,1}({\cal X})\ &&\ {\rm CS\ moduli}\ && 
\ u^\lambda\ &&{\rm  chiral\ multiplets}&\cr
&\ &&\ &&\ &&\ &\cr 
&\ h_{+}^{1,1}({\cal X})\   &&\ {\rm K\ddot{a}hler\ moduli}\ T^k\   
&&\ t^k,c^k\ &&{\rm chiral\ multiplets}&\cr
&\ &&\ &&\ &&\ &\cr  
&\ h_{-}^{1,1}({\cal X})\ &&\ {\rm add.\ moduli}\ G_a&&\ b^a,\rho^a\ 
&&{\rm chiral/linear\ multiplets}&\cr
\noalign{\hrule}
&\ h_{-}^{2,1}({\cal X})\ &&\ {\rm add.\ vectors}\ 
&& V_\mu^{\tilde j}&&{\rm vector\ multiplets}&\cr
\noalign{\hrule}}}$$}}
\vskip-6pt
\caption{Moduli of Calabi--Yau orientifold ${\cal X}$ with O5/O9--planes.\label{tab3}}
\end{table}
\ \\
As in the case with O3/O7--planes the complex structure moduli $u^\lambda$ are already
good K\"ahler coordinates, while for the other fields new K\"ahler coordinates have to be
defined (in the Einstein frame) \cite{gl04,GL} (\cff also \cite{cim02,lmrs04} for the case 
with $h_{-}^{1,1}({\cal X})=0$):
\bea\label{defmodd}
T^j&=&e^{-\Phi}\ t^j-i\ c^j\ \ \ ,\ \ \ 
G_a=e^{-\Phi}\ \Kc_{ab}\ b^b+i\ \lf(\rho_a-\Kc_{abj}\ b^b\ c^j\ri)\ ,\\[4mm]
S&=&e^{-\Phi}\ {\rm Vol}({\cal X})-\h\ e^{-\Phi}\ \Kc_{ab}\ b^a\ b^b+
i\ \lf(c_6-\rho_a\ b^a+\h\ \Kc_{abj}\ b^a\ b^b\ c^j\ri)\ .\nnn
\eea
Here, $c_6$ is the integrated $6$--form $c_6=\int_{{\cal X}} C_6$, while $c^j$ are the
integrated two--forms $c^j=\int_{C_2^j} C_2$ over the two--cycle $C_2^j$.
In this coordinates the  K\"ahler potential for the dilaton and K\"ahler moduli
becomes a complicated function $K(S,T^j,G^a)$. 
Starting from the K\"ahler potential (\cff \req{FULL} or \req{FULLK})
\zbe
\kappa_4^2\ K=-\ln\lf(2\ e^{-\Phi}\ri)-2\ \ln e^{-{3\over 2}\Phi}\ {\rm Vol}({\cal X})-
\ln\lf(-i\int_{{\cal X}}\Omega_3\wedge\ov\Omega_3\ri)\ ,
\ee
the full K\"ahler potential for the dilaton $S$, the $\npp$ K\"ahler moduli $T^j$, 
$h^{1,1}_-({\cal X})$ scalars $G_a$ and $h^{2,1}_{+}({\cal X})$ complex structure moduli becomes:
\bea
\kappa_4^2\ K&=&-\ln\lf[\ S+\ov S+\fc{e^{\Phi}}{4}\ (G+\ov G)_a\ \Kc^{ab}\ (G+\ov G)_b\ \ri]\\
&-&\ln\lf[\ \fc{1}{48}\ \Kc_{ijk}\ (T^i+\ov T^i)\ (T^j+\ov T^j)\ (T^k+\ov T^k)\ \ri]
-\ln\lf(-i\int_{{\cal X}}\Omega_3\wedge\ov\Omega_3\ri)\ .\nnn
\eea
Furthermore,  after introducing the $D=4$ dilaton field
\zbe\label{4dil}
\Phi_4=\Phi-\h\ \ln {\rm Vol}({\cal X})
\ee
we may also write \cite{lmrs04}
\zbe
\kappa_4^2\ K=-\ln\lf(2\ e^{-4\Phi_4}\ri)-\ln\lf(-i\int_{{\cal X}}\Omega_3\wedge\ov\Omega_3\ri)\ ,
\ee
which makes the duality to the \ta and heterotic string manifest.
Finally, from the form of this K\"ahler potential we deduce, that the
full  closed string moduli space $\Mc$ has the factorized form \cite{gl04,GL}
\eqn{closedMODD}{
\Mc=\Mc_{CS}^{h^{2,1}_{+}({\cal X})}\otimes \Mc_{KM}^{h^{1,1}({\cal X})+1}\ ,}
with each factor a K\"ahler manifold.

Let us now discuss the gauge kinetic function.
Generically, we have stacks of D5--branes 
and one stack of D9--branes wrapped around the full CY ${\cal X}$.
The gauge kinetic function for the stack of D9--brane is given by:
\zbe
f_{D9}=S\ ,
\ee
while for the D5--branes wrapped around the two--cycle $C_2^j$ we have:
\zbe
f_{D5,j}=T^j\ .
\ee
Again, the specific forms  of the holomorphic moduli \req{defmodd} may be understood 
from studying the CP--even and CP--odd gauge couplings.
{\it E.g.} the imaginary part of the dilaton $S$ follows from the CS--couplings 
$\int_{\mbb R^{1,3}\times {\cal X}}C_p\wedge e^{-(B-2\pi\ap F)}$ on the D9--brane. The latter
give rise to the following series of CP--odd gauge couplings
\zbe
\int_{\mbb R^{1,3}} F\wedge F \  \int_{{\cal X}}
\lf(\ C_6- C_4\wedge \Bc_2+\h\ C_2\wedge \Bc_2\wedge\Bc_2\ \ri)\ ,
\ee
with $F$ the space--time gauge field strength on the D9--brane wrapping the 
full CY manifold ${\cal X}$.
The terms in the bracket may be identified
with the imaginary part $\im S$ of the dilaton \req{defmodd}:
\zbe
\im(S)\equiv \ \int_{C_j} 
\lf( C_6- C_4\wedge \Bc_2+\h\ C_2\wedge \Bc_2\wedge\Bc_2\ \ri)\ .
\ee
Similarly, the CS--couplings 
$\int_{\mbb R^{1,3}\times C_2^j }C_p\wedge e^{-(B-2\pi\ap F)}$ on the D5--brane, which 
is wrapped around the two--cycle $C_2^j$, give rise to the CP--odd gauge coupling
\zbe
\int_{\mbb R^{1,3}} F\wedge F \  \int_{C_2^j} C_2\ ,
\ee
The term in the bracket may be identified
with the imaginary part $\im T^j$ of the K\"ahler modulus  \req{defmodd}:
\zbe
\im(T^j)\equiv  -\int_{C_j} C_2\ .
\ee
The real part of the dilaton follows from the calibration condition for the D9--brane.

\subsubsection{Type IIB orientifolds of toroidal orbifolds}\label{ORBIFOLDS}

In this section we shall discuss \tb orientifolds of toroidal orbifolds with O3/O7 or
O9/O5--planes.
Specifically, we start with the \tb superstring compactified on the six--dimensional
orbifolds 
\eqn{orbi}{
{\cal X}=\fc{\mbb T^6}{\IZ_N}\ \ \ ,\ \ \ {\cal X}=\fc{\mbb T^6}{\IZ_N\times \IZ_M}\ ,}
with the orbifold groups $\Gamma=\IZ_N$ and $\Gamma=\IZ_N\times \IZ_M$. 
To define the orbifold compactification ${\cal X}$, we must specify the six--torus $\mbb T^6$ and the
discrete point group $\Gamma$.  We will restrict ourselves to orbifolds with
Abelian point group without discrete torsion. The point group element $\theta$
can then be written as 
$\theta=\exp[2\pi i(v^1M^{12}+v^2M^{34}+v^3M^{56})],$ where the $M^{ij}$ are
the generators of the Cartan sub-algebra and $0\leq|v^i|<1,\ i=1,2,3$.
The group generator $\theta\in\Gamma$ acts as follows on the three
complex coordinates of $\mbb T^6$:
\zbe\label{oaction}
\theta:\ (z^1,z^2,z^3)\longrightarrow (e^{2\pi i\, {v^1}}\ z^1\ ,\ 
e^{2\pi i \, {v^2}}\ z^2\ ,\ e^{2\pi i \, {v^3}}\ z^3)\ .
\ee
To obtain ${\cal N}=2$ supersymmetry in $D=4$, the point group $\Gamma$
must be a subgroup of $SU(3)$, to furnish  $SU(3)$ holonomy ($\theta\ \Omega_3=\Omega_3$).
This requires $\pm v^1\pm v^2\pm v^3=0$ \cite{Dixon:1985jw,Dixon:1986jc}.
This condition together with the requirement that $\Gamma$ must act
crystallographically on the lattice specified by $\mbb T^6$ leads to $\Gamma$ being
either $\IZ_N$ with $N=3,4,6,7,8,12\,$ or $\IZ_M\times \IZ_N$ with $N$ a multiple of
$M$ and $N=2,3,4$. $\IZ_6,\ \IZ_8$ and $\IZ_{12}$ have two inequivalent embeddings
in $SO(6)$. In the following two tables we display the possible point groups
for the toroidal orbifolds \req{orbi}.
\begin{table}[h]
\centering
\begin{tabular}{|c|c|}
\hline
Point group\ $\theta$ & $v^i=\frac{1}{N}(n_1,n_2,n_3)$ \\ [2pt]
\hline
$\ \IZ_{3}  $      & $\frac{1}{3}\,(1,1,-2)$\\ [2pt]
$\ \IZ_{4}  $      & $\frac{1}{4}\,(1,1,-2)$\\ [2pt]
$\ \IZ_{6-I}  $      & $\frac{1}{6}\,(1,1,-2)$\\ [2pt]
$\ \IZ_{6-II}  $      & $\frac{1}{6}\,(1,2,-3)$\\ [2pt]
$\ \IZ_{7}  $      & $\frac{1}{7}\,(1,2,-3)$\\ [2pt]
$\ \IZ_{8-I}  $      & $\frac{1}{8}\,(1,2,-3)$\\ [2pt]
$\ \IZ_{8-II}  $      & $\frac{1}{8}\,(1,3,-4)$\\ [2pt]
$\ \IZ_{12-I}  $      & $\frac{1}{12}\,(1,4,-5)$\\ [2pt]
$\ \IZ_{12-II}  $      & $\frac{1}{12}\,(1,5,-6)$\\ [2pt]
\hline
\end{tabular}
\caption{Point group $\th$ of $\IZ_N$--orbifolds.\label{tab4}}
\end{table}
\begin{table}[h]
\centering
\begin{tabular}{|c|c|c|}
\hline
Point group\ $\theta,\omega$&$v^i=\frac{1}{N}(n_1,n_2,n_3)$ &$w^i=\frac{1}{M}(m_1,m_2,m_3)$\\ [2pt]
\hline
$\ \IZ_{2} \times \IZ_{2}$     & $\frac{1}{2}\,(1,0,-1)$   & $\frac{1}{2}\,(0,1,-1)$\\ [2pt]
$\ \IZ_{2} \times \IZ_{4} $      & $\frac{1}{2}\,(1,0,-1)$ & $\frac{1}{4}\,(0,1,-1)$\\ [2pt]
$\ \IZ_{2} \times \IZ_{6} $      & $\frac{1}{2}\,(1,0,-1)$ & $\frac{1}{6}\,(0,1,-1)$\\ [2pt]
$\ \IZ_{2} \times \IZ_{6'} $      & $\frac{1}{2}\,(1,0,-1)$ & $\frac{1}{6}\,(1,1,-2)$\\ [2pt]
$\ \IZ_{3} \times \IZ_{3} $      & $\frac{1}{3}\,(1,0,-1)$ & $\frac{1}{3}\,(0,1,-1)$\\ [2pt]
$\ \IZ_{3} \times \IZ_{6} $      & $\frac{1}{3}\,(1,0,-1)$ & $\frac{1}{6}\,(0,1,-1)$\\ [2pt]
$\ \IZ_{4} \times \IZ_{4} $      & $\frac{1}{4}\,(1,0,-1)$ & $\frac{1}{4}\,(0,1,-1)$\\ [2pt]
$\ \IZ_{6} \times \IZ_{6} $      & $\frac{1}{6}\,(1,0,-1)$ & $\frac{1}{6}\,(0,1,-1)$\\ [2pt]
\hline
\end{tabular}
\caption{Point groups $\th,\omega$ of $\IZ_N\times \IZ_M$--orbifolds.\label{tab5}}
\end{table}

\noindent
A six--torus $\mbb T^6$ has $15$ K\"ahler moduli associated to $(1,1)$--forms of $H^{1,1}(\mbb T^6)$
and six complex structure moduli related to the cohomology $H^{3}(\mbb T^6)$.
The orbifold group $\Gamma$ projects out some of these forms, resulting in
the untwisted Hodge numbers $h^{1,1}_{\rm untw.}({\cal X})$ and $h^{2,1}_{\rm untw.}({\cal X})$.
The twist elements $\theta,\ldots,\theta^{N-1}$ produce conical singularities
at the fixpoints $f_\alpha^{(n)}$. A fixpoint under $\theta^n$ is defined by 
$\theta^n\ f_\alpha^{(n)}=f_\alpha^{(n)}+\lambda$, with some lattice vector $\lambda\in\Lambda$.
In a small neighborhood around them, the space locally looks like 
$\ICC^3/\Gamma$ (isolated singularity) 
or $\ICC^2/\Gamma^{(2)}\times \ICC$  (non--isolated singularity).
In  \cite{blowup}  
these singularities are resolved using the methods of toric geometry resulting in
a smooth Calabi--Yau space ${\cal X}$. Each singularity is resolved locally in a patch and then
patches are put together according to the fixed set configuration.
The twisted Hodge numbers 
$h^{1,1}_{\rm twist.}({\cal X})$ and $h^{2,1}_{\rm twist.}({\cal X})$ count the number of exceptional
divisors necessary to obtain a smooth CY manifold.
In Table \ref{tab6} we give a list of possible $\IZ_N$ orbifolds, 
together with their torus lattices $\mbb T^6$ they live on and their Hodge numbers.
\begin{table}[h]
\centering
{\vbox{\ninepoint{$$
\vbox{\offinterlineskip\tabskip=0pt
\halign{\strut\vrule#
&~$#$~\hfil 
&\vrule$#$
&~$#$~\hfil 
&\vrule$#$ 
&~$#$~\hfil 
&\vrule$#$
&~$#$~\hfil 
&\vrule$#$ 
&~$#$~\hfil 
&\vrule$#$
&~$#$~\hfil 
&\vrule$#$&\vrule$#$\cr
\noalign{\hrule}
&\ \IZ_N &&{\rm lattice}\ \mbb T^6 && h_{\rm untw.}^{1,1} &&  h_{\rm untw.}^{2,1} &&
h^{1,1}_{\rm twist.}&& h_{\rm twist.}^{2,1} &\cr
\noalign{\hrule}\noalign{\hrule}
& \ \IZ_3      &&\   SU(3)^3            &&9 &&0&&27 &&0&\cr
& \  \IZ_4      &&\    SU(4)^2         &&5 &&  1&&20&&0&\cr
& \  \IZ_4      &&\     SU(2)\times SU(4)\times SO(5)  &&5&&  1&&22&&2&\cr
& \    \IZ_4       &&\     SU(2)^2\times SO(5)^2   &&5  &&  1&&26&&6&\cr
& \  \IZ_{6-I}     &&  (G_2\times SU(3)^{2})^{\flat}   &&5 &&0&&20&&1&\cr
& \  \IZ_{6-I}     &&    SU(3)\times G_2^2  &&5 && 0&&24&&5&\cr
& \  \IZ_{6-II}    && SU(2)\times SU(6)     && 3  && 1&&22&&0&\cr
& \  \IZ_{6-II}    && SU(3)\times SO(8)  &&3 &&1&& 26&&4&\cr
& \  \IZ_{6-II}    && (SU(2)^2\times SU(3)\times SU(3))^{\sharp}   &&3 &&1&&28&&6&\cr
& \  \IZ_{6-II}    &&  SU(2)^2\times SU(3)\times G_2   &&3 &&1&&32&&10&\cr
& \  \IZ_7         &&  SU(7)                               &&3 &&0&&21&&0&\cr
& \    \IZ_{8-I}     &&    (SU(4)\times SU(4))^*    &&3 && 0&&21&&0&\cr
& \    \IZ_{8-I}     &&   SO(5)\times SO(9)          &&3& &  0 &&24&&3&\cr
& \    \IZ_{8-II}    &&    SU(2)\times SO(10)        &&3 &&  1&&24&&2&\cr
& \    \IZ_{8-II}      &&    SO(4)\times SO(9)    &&3&&  1&&28&&6&\cr
& \  \IZ_{12-I}    &&  E_6    &&3&&0&&22&&1&\cr
& \  \IZ_{12-I}    &&  SU(3)\times F_4   &&3 &&0&&26&&5&\cr
& \    \IZ_{12-II}    &&   SO(4)\times F_4   &&3&&  1&&28&&6&\cr
\noalign{\hrule}}} $$}}}
\vskip-6pt
\caption{Orbifold groups, lattices and Hodge numbers for $\IZ_N$ orbifolds.\label{tab6}}
\end{table}
The lattices marked with $\flat$, $\sharp$, and $*$ are realized as generalized Coxeter twists, 
the automorphism being in the first and second case
$S_1S_2S_3S_4P_{36}P_{45}$ and in the third  
$S_1S_2S_3P_{16}P_{25}P_{34}$.
The Hodge numbers $h^{1,1}_{\rm twist.}, h^{2,1}_{\rm twist.}$ depend both on the orbifold
group $\Gamma$ and the underlying torus lattice $\mbb T^6$ \cite{Klemm,IMNQ}. 
In Table \ref{tab7} we give a list of possible $\IZ_N\times\IZ_M$ orbifolds, 
together with their torus lattices $\mbb T^6$ they live on and their Hodge numbers \cite{FIQ}.
\begin{table}[h]
\centering
{\vbox{\ninepoint{$$
\vbox{\offinterlineskip\tabskip=0pt
\halign{\strut\vrule#
&~$#$~\hfil 
&\vrule$#$
&~$#$~\hfil 
&\vrule$#$ 
&~$#$~\hfil 
&\vrule$#$
&~$#$~\hfil 
&\vrule$#$ 
&~$#$~\hfil 
&\vrule$#$
&~$#$~\hfil 
&\vrule$#$&\vrule$#$\cr
\noalign{\hrule}
&\ \IZ_N \times\IZ_M&&{\rm lattice}\ \mbb T^6 && h_{\rm untw.}^{1,1} &&  h_{\rm untw.}^{2,1} &&
h^{1,1}_{\rm twist.}&& h_{\rm twist.}^{2,1} &\cr
\noalign{\hrule}\noalign{\hrule}
& \    \IZ_2 \times\IZ_2      &&   SU(2)^6   &&3 && 3&&48&&0&\cr
& \    \IZ_2 \times\IZ_4     &&   SU(2)^2\times SO(5)^2  &&3&&  1&&58&&0&\cr
& \    \IZ_2 \times\IZ_6      &&    SU(2)^2\times SU(3)\times G_2 &&3&&  1&&48&&2&\cr
& \    \IZ_2 \times\IZ_{6'} &&    SU(3)\times G_2^2 &&3 &&  0&&33&&0&\cr
& \  \IZ_3 \times\IZ_3      && SU(3)^3    &&3&&0&&81&&0&\cr
& \  \IZ_3 \times\IZ_6      && SU(3)\times G_2^2  &&3 &&0&&70&&1&\cr
& \    \IZ_4 \times\IZ_4      &&   SO(5)^3   &&3&&  0&&87&&0&\cr
& \    \IZ_6 \times\IZ_6      &&   G_2^3  &&3&&  0&&81&&0&\cr
\noalign{\hrule}}} $$}}}
\caption{Orbifold groups, lattices and Hodge numbers for $\IZ_N\times\IZ_M$ orbifolds.\label{tab7}}
\end{table}
We shall now introduce an orientifold projection $\Oc$ and determine the closed string moduli 
space of the resulting orientifold ${\cal X}$.
Let us first discuss the orbifold case, \ie the $h_{\rm twist.}^{1,1}({\cal X})$ twisted
K\"ahler and $h^{2,1}_{\rm twist.}({\cal X})$ twisted complex structure moduli are fixed.
For those orientifolds an exact CFT description is at hand (\cff section 2).
According to subsection \ref{IIORI} the two orientifold projections $\Oc$, given
in {\it Eqs.}\req{Action} or \req{Action2}, are possible.
The orbifold group $\Gamma$ mixes with the orientifold group $\Om I_n$.
As a result, if the group $\Gamma$ contains $\IZ_2$--elements $\th$, 
which leave one complex plane fixed, we obtain in the case $(i)$ additional 
O$(9-|n-4|)$--planes 
or in the case $(ii)$ additional O$(3+|n-2|)$--planes from the element $\Om I_n \th$.
From the action of the reflection $\sigma^\ast$ on the untwisted
cohomology $H^{p,q}({\cal X})$
of the orbifolds \req{orbi}, one deduces that for the case $(i)$, \ie
$\sigma^\ast=I_6$, we obtain:
\bea
h^{1,1}_{+,{\rm untw.}}({\cal X})=h^{1,1}_{\rm untw.}({\cal X}) &,&
h^{1,1}_{-,{\rm untw.}}({\cal X})=0\ ,\nnn \\
h^{2,1}_{+,{\rm untw.}}({\cal X})=0 &,&
h^{2,1}_{-,{\rm untw.}}({\cal X})=h^{2,1}_{\rm untw.}({\cal X})\ ,\label{Hodgeuntw}
\eea
while the case $(ii)$, \ie $\sigma^\ast=I_0$ yields:
\bea
h^{1,1}_{+,{\rm untw.}}({\cal X})=h^{1,1}_{\rm untw.}({\cal X}) &,&
h^{1,1}_{-,{\rm untw.}}({\cal X})=0\ ,\nnn\\
h^{2,1}_{+,{\rm untw.}}({\cal X})=h^{2,1}_{\rm untw.}({\cal X})&,&
h^{2,1}_{-,{\rm untw.}}({\cal X})=0\ .\label{Hodgeuntww}
\eea
If the orbifold group $\Gamma$
contains $\IZ_2$--elements $\th$, the generators $\Oc\th$, which
produces an O7-- or O5--plane, respectively, does not put further
restrictions on the twist invariant forms.
Nonetheless, for both cases the geometry of the resulting orientifold ${\cal X}$ is described by 
$h^{1,1}_{\rm untw.}({\cal X})$ K\"ahler moduli $T^i$ and $h^{2,1}_{\rm untw.}({\cal X})$
structure moduli $U^\lambda$ as in the case without orientifold
projection. Hence, the orientifold group does not reduce the number of
untwisted moduli, and their number is determined by the orbifold
action. Besides we do not encounter the additional moduli fields $G^a,G_a$ 
(\cff Table \ref{tab2} and \ref{tab3}).

Without D--brane moduli locally 
the closed string moduli space $\Mc$ is a direct product of the dilaton $\Mc_S$, 
the K\"ahler $\Mc_K$ and complex structure moduli space \cite{lmrs04,seealso}\ 
(see also subsection \ref{SCATT}):
\eqn{DIRECT}{
\Mc=\Mc_S\otimes \Mc_{KM}\otimes \Mc_{CS}\ .}
Here, the manifold $\Mc_S$ is given in \eqq \req{Scoset}. The spaces 
$\Mc_{KM}$ and $\Mc_{CS}$ with dimensions $h_{untw.}^{1,1}({\cal X})$ and 
$h_{untw.}^{2,1}({\cal X})$, respectively are discussed in the following.
Depending on the specific numbers $h^{1,1}_{\rm untw.}, h^{2,1}_{\rm untw.}$
of untwisted K\"ahler $t^i$ and complex structure moduli $u^j$ 
the generic (untwisted) moduli spaces appearing in the toroidal orbifold 
compactifications ${\cal X}$ are described by the following six different cosets \cite{FKP1,FKP2,Ferrara:1990uu,IbanezHC}:
\begin{eqnarray}
h^{1,1}_{\rm untw.}=3\ \ ,\ \ h^{2,1}_{\rm untw.}=0,1,3&&:\ \   
\Mc_{KM}=\lf(\fc{SU(1,1)}{U(1)}\ri)^3\ \ ,\ \ 
\Mc_{CS}=\lf(\fc{SU(1,1)}{U(1)}\ri)^{h^{2,1}_{\rm untw.}}\ ,\nnn \\
h^{1,1}_{\rm untw.}=5\ \ ,\ \ h^{2,1}_{\rm untw.}=0,1&&:\ \  
\Mc_{KM}=\fc{SU(2,2)}{SU(2)\times SU(2)\times U(1)}\otimes
\lf(\fc{SU(1,1)}{U(1)}\ri)\ ,\nnn \\
&&\hskip0.6cm\Mc_{CS}=\lf(\fc{SU(1,1)}{U(1)}\ri)^{h^{2,1}_{\rm untw.}}\ ,\nnn \\
h^{1,1}_{\rm untw.}=9\ \ ,\ \ h^{2,1}_{\rm untw.}=0&&:\ \   
\Mc_{KM}=\fc{SU(3,3)}{SU(3)\times SU(3)\times U(1)}\ .
\label{cosets}
\end{eqnarray}
The corresponding K\"ahler potentials for these spaces \req{cosets} are known from heterotic 
string compactifications \cite{FKP1,FKP2,Ferrara:1990uu,IbanezHC}:
\begin{eqnarray}
h^{1,1}_{\rm untw.}=3\ \ ,\ \ h^{2,1}_{\rm untw.}=0,1,3&:&
\kappa_4^2\ K_{KM}=-\sum_{i=1}^3\ln(t^i-\ov t^i)\ ,\nnn\\
&&\kappa_4^2\ K_{CS}=-\sum_{j=1}^{h^{2,1}_{\rm untw.}}\ln(u^j-\ov u^j)\ ,\nnn\\
h^{1,1}_{\rm untw.}=5\ \ ,\ \ h^{2,1}_{\rm untw.}=0,1&:&\ \  
\kappa_4^2\ K=-\ln\det(t^{ij}-\ov t^{ij})-\ln(t^5-\ov t^5)\ ,\nnn\\
&&\ \ \kappa_4^2\ K_{CS}=-\sum_{j=1}^{h^{2,1}_{\rm untw.}}\ln(u^j-\ov u^j)\ ,\nnn\\
h^{1,1}_{\rm untw.}=9\ \ ,\ \ h^{2,1}_{\rm untw.}=0&:&\ \   
\kappa_4^2\ K_{KM}=-\ln\det(t^{ij}-\ov t^{ij})\ .
\label{Kcosets}
\end{eqnarray}
What is less known is the parameterization of the moduli fields $t^i,u^i$ 
in terms of the data of the torus $\mbb T^6$,
\ie the real metric $g$ and the discrete symmetries of the underlying effective
field theory. This has been worked out in great detail in \cite{LRSSi}.

The cosets \req{cosets} with their K\"ahler potentials \req{Kcosets} are the relevant
moduli spaces for the orbifold compactifications ${\cal X}$.
Since we are interested in the moduli for the \tb orientifold ${\cal X}$, we have to replace the
K\"ahler moduli $t^j$ by the proper good K\"ahler coordinates $T^j$.
Concretely, for the orientifold projection \req{Action} 
we have to determine the holomorphic K\"ahler moduli \req{defmod}, while for the projection
\req{Action2} we have to use the K\"ahler moduli \req{defmodd}.
At any rate, we only have to know the intersection numbers \req{2int} of the coset spaces
\req{cosets} to determine the holomorphic K\"ahler moduli \req{defmod} or \req{defmodd}
for the \tb orientifolds compactified on the orbifolds \req{orbi}.

Let us present an example. We briefly discuss the case $h^{1,1}_{\rm untw.}({\cal X})=3$.
The K\"ahler moduli space $\Mc_{KM}=\lf(\fc{SU(1,1)}{U(1)}\ri)^3$ 
is realized \eg in the $\IZ_2\times\IZ_2$ \tb orientifold, with the twists $Q_1,Q_2$
\zbe\label{QQtwist}
\begin{array}{cl}
&Q_1\ :\ e_{1,2}\lra-e_{1,2}\ \ \ ,\ \ \ e_{3,4}\lra-e_{4,5}\ \ \ ,\ \ \ e_{5,6}\lra e_{5,6}\ ,\\
&Q_2\ :\ e_{1,2}\lra-e_{1,2}\ \ \ ,\ \ \ e_{3,4}\lra e_{4,5}\ \ \ ,\ \ \ e_{5,6}\lra -e_{5,6}
\end{array}
\ee
acting on the integral basis $\{e_i\}$ of the torus $\mbb T^6$. 
The twists $Q_1,Q_2$ only allow for a factorizable lattice, \ie $\mbb
T^6$ being a direct product of three two--tori, \ie
$\mbb \mbb T^6=\otimes_{j=1}^3 \mbb T^2_j$, with the metrics $g^j$.
Each individual two--torus $\mbb T^2_j$ has one K\"ahler modulus $t^j$
and  one complex structure modulus $u^j$ describing the real parameters of the metric $g^j$. 
The complex structure moduli are given by (\cff subsection \ref{secinttor}): 
\eqn{CS}{
u^j=\fc{1}{g^j_{11}}\ (\ g^j_{12}+i\sqrt{\det g^j}\ )\ ,\ \ \ {\rm with:}\ \ \ 
g^j=\pmatrix{g_{11}^j&g_{12}^j\cr g_{12}^j&g_{22}^j}\ .}
Before introducing the orientifold projection, we would have ${\cal
  N}=2$ in $D=4$ and the full K\"ahler potential would be given by \req{KpotZ22}.
In \tb orientifolds the complex structure coordinates $u^j$ are already good K\"ahler coordinates,
\ie $U_2^j=u_1^j,\ U_1^j=u_2^j$.
The imaginary part of the K\"ahler modulus $t^j$ describes the size $\sqrt{\det g^j}$ 
of the subtorus $\mbb T^2_j$, \ie $\im(t^j)=\sqrt{\det g^j}$.
According to \req{Kcosets} the K\"ahler potential \req{KpotK} $K_{KM}$ for the K\"ahler 
moduli $t^i$ reads
$$\kappa_4^2\ 
K_{KM}=-\ln\lf[\ \fc{1}{8}\ (t^1-\ov t^1)\ (t^2-\ov t^2)\ (t^3-\ov t^3)\ \ri]\ ,$$
from which we determine the only non--vanishing intersection form (\cff \eqq \req{2int}):
$$\Kc_{123}=1\ .$$ 
With this information, from \req{defmod} we determine $\re(T^i)=\fc{3}{2}\ \im(t_j)\ \im(t_k)$.
The twist--invariant $4$--form $C_4$ is given by
\eqn{ffourform}{
C_4=\rho^1\ dx^2\wedge dy^2\wedge dx^3\wedge dy^3+\rho^2\ dx^1\wedge dy^1\wedge dx^3\wedge dy^3
+\rho^3\ dx^1\wedge dy^1\wedge dx^2\wedge dy^2\ .}
According to \req{defmod}, the K\"ahler modulus $T^j$ is complexified with the internal part of the
Ramond $4$--form $C_4$, \ie
$\im(T^j)=\fc{3}{2}\ \int\limits_{T^{2,k}\times T^{2,l}} C_4=\fc{3}{2}\ \rho^j$.
To this end, the three holomorphic K\"ahler moduli $T^i$ become:
\eqn{OBTAIN}{
T^j=\fc{3}{2}\ \im(t^k)\ \im(t^l)+\fc{3}{2}\ i\ c^j\ \ \ ,\ \ 
(j,k,l)=\overline{(1,2,3)}\ .}
In terms of these holomorphic coordinates $T^j$ the full K\"ahler potential \req{fullK} reads:
\eqn{Simply}{
\kappa_4^2\ K=-\ln(S+\ov S)-\sum_{j=1}^3\ln\left[\fc{1}{3}(T^j+\ov T^j)\right]
-\sum_{i=1}^3\ln(U^i+\ov U^i)\ .}

On the other hand, for the orientifold action \req{Action2}, following \req{defmodd}
we determine the three K\"ahler moduli (in the Einstein frame):
\zbe
T^j=e^{-\Phi}\ \im(t^j)-i\ c^j\ \ \ ,\ \ \ j=1,2,3
\ee
with the components $c^j$ of the RR two--form:
\zbe
C_2=c^1\ dx^1\wedge dy^1+c^2\ dx^2\wedge dy^2+c^3\ dx^3\wedge dy^3\ .
\ee
Furthermore, the dilaton field $S$ becomes:
\zbe
S=e^{-\Phi}\ \im(t^1)\im(t^2)\im(t^3)+i\ c_6\ ,
\ee
with the integrated $6$--form $c_6=\int_{{\cal X}} C_6$.
In terms of these holomorphic coordinates $S,T^j$ the full K\"ahler potential takes the same 
form as \req{Simply}.

A detailed account of the previous presentation with many more examples and details 
may be found in  \cite{LRSSi}.

Let us now move on to \tb orientifolds of the resolved orbifolds \req{orbi}.
We shall discuss orientifold actions, which allow for O3/O7--planes.
After resolving the orbifolds \req{orbi} resulting in a smooth Calabi--Yau space ${\cal X}$
a consistent orientifold action \req{Action} is 
introduced, resulting in the Calabi--Yau orientifold ${\cal X}$.
The involution $\sigma$ acting on the complex coordinates $z^i$ may also 
involve the new coordinates $y^j$ associated to the exceptional divisors \cite{blowup}.
After resolving the orbifold, three kinds of divisors $\Dc$ appear, 
namely $E_\alpha$, $D_{i\alpha}$, and $R_i$. 
The divisors $E_\alpha$ are the exceptional divisors arising from
the resolution of an orbifold singularity $f_\alpha$ (or an orbit under the orbifold group), 
while the divisors $D_{i\alpha}$ denote hyperplanes passing through fixed points: 
$D_{i\alpha}=\{z^i=z^i_{fixed,\alpha}\}$. The divisors $R_i=\{z^i=c\}$
for $c\not =z^i_{fixed,\alpha}$ are hyperplanes not passing through a
fixed point \cite{blowup}. As opposed to the $D_{i\alpha}$ the latter are allowed to move.
Furthermore, the divisors $R_i$ are directly related to the unresolved orbifold
and are independent on the resolution as they do not feel the blow up procedure.

In the case of $h^{1,1}_-({\cal X})\neq 0$ some divisors $E$ (or divisor orbits
under the orbifold group on the $\mbb T^6$) in the geometry of the covering space ${\cal X}$
may not be invariant under the orientifold action $\sigma$, but
are mapped to other divisors $\tilde E$ (or orbits under the orbifold group), \ie:
\eqn{happen}{
\sigma E =\tilde E\ .}
In this case, a pair of divisors $(E_i,E_a)$, which are eigenstates (with eigenvalues 
$\pm 1$) under $\sigma$ may be constructed:
\eqn{invDIV}{
E_{i}:=\h\ (E+\tilde E)\ \ \ ,\ \ \ E_{a}:=\h\ (E-\tilde E)\ }
such that: 
\eqn{eigenstates}{
\sigma E_i=E_i\ \ \ ,\ \ \ \sigma E_a=-E_a\ .}
Then we have $\omega_i\in H^{1,1}_+({\cal X})$ and $\omega_a\in H^{1,1}_-({\cal X})$ for their 
corresponding Hodge dual two--forms.
This way, the $h^{1,1}_-({\cal X})$ odd forms $\omega_a$ are paired with $h^{1,1}_-({\cal X})$ even forms $\omega_i$.
To this end, the original number of divisors $h^{1,1}({\cal X})$ is 
split into $h_{+}^{1,1}({\cal X})$ even  and $h_{-}^{1,1}({\cal X})$ odd  divisors.
These numbers are determined for the orientifolds of the resolved orbifolds 
\req{orbi} in  \cite{blowup} and are displayed in Table \ref{tab8}.

\begin{table}[h]
\centering
\vbox{\ninepoint{$$
\vbox{\offinterlineskip\tabskip=0pt
\halign{\strut\vrule#
&~$#$~\hfil 
&\vrule$#$
&~$#$~\hfil 
&\vrule$#$ 
&~$#$~\hfil 
&~$#$~\hfil 
&\vrule$#$&\vrule$#$
&~$#$~\hfil 
&\vrule$#$ 
&~$#$~\hfil 
&\vrule$#$&\vrule$#$\cr
\noalign{\hrule}
&\ \Gamma && h_{+}^{1,1} && h^{1,1}_{-}&&&\ \Gamma &&\ h_{+}^{1,1} 
&&\ h^{1,1}_{-}\ &\cr
\noalign{\hrule}\noalign{\hrule}
& \ \IZ_3             &&23 &&13&&& \    \IZ_{8-II}      &&\ 23 && \  4\ &\cr
& \  \IZ_4            &&19 &&  6&&& \    \IZ_{8-II}       &&\ 31 &&\   0\ &\cr
& \  \IZ_4       &&23&&  4&&& \  \IZ_{12-I}    &&\ 19&&\ 6\ &\cr
& \    \IZ_4        &&31  &&  0&&& \  \IZ_{12-I}       &&\ 23 &&\ 6\ &\cr
& \  \IZ_{6-I}       &&19 &&6&&& \    \IZ_{12-II}       &&\ 31&& \  0\ &\cr
& \  \IZ_{6-I}   && 23 && 6&&& \    \IZ_2 \times\IZ_2         &&\ 51 && \  0\ &\cr
& \  \IZ_{6-II}       && 19  && 6&&& \    \IZ_2 \times\IZ_4      &&\ 61&&\   0\ &\cr
& \  \IZ_{6-II}   &&23 &&6&&& \    \IZ_2 \times\IZ_6      &&\ 51&& \  0\ &\cr
& \  \IZ_{6-II}    &&23 &&8&&& \    \IZ_2 \times\IZ_{6'}  &&\ 36 && \  0\ &\cr
& \  \IZ_{6-II}     &&27 &&8&&& \  \IZ_3 \times\IZ_3      &&\ 47 &&\ 37\ &\cr
& \  \IZ_7            &&15 &&9&&& \  \IZ_3 \times\IZ_6      &&\ 51 &&\ 22\ &\cr
& \    \IZ_{8-I}         &&19 &&  5&&& \    \IZ_4 \times\IZ_4       &&\ 90&& \  0\ &\cr
& \    \IZ_{8-I}          &&27& &  0&&& \    \IZ_6 \times\IZ_6      &&\ 84&&\   0\ &\cr
\noalign{\hrule}}} $$}}
\caption{Hodge numbers $h^{1,1}({\cal X})$ after the orientifold action.\label{tab8}}
\end{table}
In the orbifold limit, discussed before, the involution
$\sigma: z^i\longrightarrow -z^i$ introduces 64 O3--planes.
Some of them may be identified under orbifold group $\theta$ or may be grouped
into orbits under the orbifold group. In the smooth case the orientifold action has to 
be chosen on the local patches in terms of the local coordinates~$(z^j,y^i)$.
Those of the $64$ O3--planes on the cover, which are located away from the 
resolved patches (resulting from the global involution)
remain the same in the resolved orientifold. 
The O3--plane solutions, which coincide with the orbifold
fixed sets are replaced 
by the solutions of the corresponding patch. This way an orientifold of a resolved orbifold
may have less than $64$ O3--planes.

The total D3--brane charge is calculated as follows.
The contribution from the O3--planes is (in the orientifold quotient ${\cal X}/\IZ_2$ of ${\cal X}$)
\eqn{tot3}{Q_3(O3)=-{1\over 4}\times n_{O3}\ ,}
where $n_{O3}$ denotes the number of O3--planes. 
The stacks $a$ of D7--branes contribute to the D3--tadpole (in the orientifold quotient ${\cal X}/\IZ_2$)
\eqn{tot7}{Q_3(D7)=-\h\ \sum_a\,{n_{D7,a}\,\chi(\Dc_a)\over 24}\ ,}
where $n_{D7,a}$ denotes the number of D7--branes in the stack located on the 
divisor $\Dc_a$. As we have seen, the $\Dc_a$ can be local D--divisors as well as 
exceptional divisors $E$.
The last contribution to the D3--brane tadpole comes from the O7--planes
(in the orientifold quotient ${\cal X}/\IZ_2$):
\eqn{tot7O}{Q_3(O7)=-\h\ \sum_a\,{\chi(\Dc_a)\over 6}\ .}
So the total D3--brane charge that must be canceled is
\eqn{totaltadpole}{Q_{3,tot}=-{n_{O3}\over4}-\h\ \sum_a\,{(n_{D7,a}+4)\,\chi(\Dc_a)\over 24}\ .}
These are the values for the orientifold quotient ${\cal X}/\IZ_2$, 
in the double cover ${\cal X}$ this value must be multiplied by two.
If we would like to avoid mobile D3--branes, this tadpole may be saturated by 
turning on 3--form flux $G_3$.

The formula \req{totaltadpole} for the total D3--brane charge $Q_{3,tot}$ differs from the 
known  tadpole equation for the singular orbifold case by the second term. 
The latter is induced by the curvature of the D7--branes which is absent in the singular 
case. 
In that case, the number of orientifold O3--planes is always $64$, \ie $n_{O3}=64$,
and \req{totaltadpole} boils down to $Q_{3,tot}=-16$ \cite{lrs04}.
In the CFT description, this tadpole originates from the total leading divergent contribution 
of the Klein bottle amplitude $Z_\Kc(1,1)$ of the untwisted orbifold sector.
However, there are additional tadpole contributions from other orbifold sectors 
to be canceled.
More precisely, the tadpole arising from the Klein bottle amplitude $Z_\Kc(1,\th^k)$ and in 
addition for even $N$ the $\IZ_2$--twisted tadpole related to 
 $Z_\Kc(\th^{N/2},\th^k)$ have to be canceled
($k=0,\ldots,N-1$).
The tadpoles from the sector $(1,1)$ and for even $N$ also from the sector $(1,\th^{N/2})$
may be canceled by introducing the right amount of D3--brane (or/and three--form flux) 
and D7--branes, respectively. On the other hand,
the divergences of the Klein bottle amplitude  $Z_\Kc(1,\th^k),\ k\neq 0$ 
or for even $N$ from the combination 
$Z_\Kc(1,\th^k)+Z_\Kc(\th^{N/2},\th^k),\ k\neq0,N/2$ 
can only be canceled against any of the annulus and M\"obius strip contributions 
in the case that the orbifold group $\Gamma$ is 
$\IZ_3,\,\IZ_{6-I},\,\IZ_{6-II},\,\IZ_7$ or $\IZ_{12-I}$ \cite{Aldazabal:1998mr} or 
$\IZ_2\times\IZ_2,
\IZ_3\times\IZ_3, \IZ_6\times\IZ_6, \IZ_2\times\IZ_3,\IZ_2\times \IZ_6, \IZ_2\times \IZ_6'$ 
\cite{Zwart:1997aj}. 
Hence singular orbifolds have much more constraining tadpole equations, which
are non--trivial to fulfill for all $\IZ_N$-- and $\IZ_N\times \IZ_M$ orbifolds.
However, if one introduces discrete torsion or vector structure tadpoles from all orbifold
sectors may be completely canceled in all singular orbifold cases 
\cite{Klein:2000qw,Klein:2000hf,Rabadan}.

Nevertheless, the orientifolds ${\cal X}$ constructed geometrically in  \cite{blowup}
in the large radius regime from resolved orbifolds ${\cal X}$ 
need not have a CFT counterpart in their orbifold limit, 
since D--branes (in particular stacks of D7 and O7--branes) wrapping cycles  
which vanish in the orbifold limit, give rise to  extra non--perturbative 
states in the orbifold limit.

As an example let us discuss the resolved $\IZ_2\times\IZ_2$ orientifold\footnote{As 
already pointed out in subsection \ref{secZ22} there are two different $\IZ_2\times \IZ_2$
type IIB orientifolds, c.f. table \ref{tabZ22models}. 
In the following we resolve the orbifold with the Hodge numbers
Hodge numbers $\npp=51$ and
$h_-^{2,1}({\cal X})=3$. On the other hand, in subsection \ref{FLUXORBI}
we shall discuss the first  orbifold, with $\npp=3$ and
$h_-^{2,1}({\cal X})=51$.} 
with the orientifold action \req{Action} allowing for O3 and O7--planes.
This orbifold allows for two different resolutions, one symmetric and one asymmetric.
In the following we shall report the symmetric resolution \cite{DenefMM}.
The orbifold has $48$ non--isolated singularities $f_{i\alpha,j\beta}$ with 
the local topology $\ICC^2/\IZ_2\times \ICC$ located at $z^i=f_\al,\ z^j=f_\bet$, 
with $i<j,\ i,j=1,2,3,\ \al,\bet=1,\ldots,4$. The points $f_\al$ denote the four fixed points
of $\mbb T^2/\IZ_2$. 
Each fixed line  is resolved locally by introducing one exceptional divisor $E_{i\alpha,j\beta}$,
with the topology of $\IP^1\times\IP^1$, blown up in four points.
The latter correspond to the four intersections of a fixed line with the fixed planes
in $\mbb T^6/\IZ_2\times\IZ_2$. In addition, there are $12$ divisors 
$D_{i\al},\ \al=1,\ldots4,\ i=1,2,3$, which
are hyperplanes passing through the fixed points $f_\al$, \ie $D_{i\al}=\{z^i=f_\al\}$.
The topology of the divisors $D_{i\al}$ is $\IP^1\times\IP^1$.
Note that both divisors $E_{i\alpha,j\beta}$ and $D_{i,\alpha}$ have 
the Hodge numbers $h^{0,1}=0=h^{0,2}$.
The orientifold action \req{Action} may be introduced such, that it is compatible with
the local patches, such providing $64$ O3--planes. 
Furthermore on each divisor $D_{i\al}$ an orientifold action producing an O7--plane may
be introduced. The tadpole from the O7--planes may be locally canceled by placing 
a stack of eight D7--branes on top of each O7--plane.
Thus, with $n_{O3}=64, n_{D7,a}=8$ and $\chi(\IP^1\times\IP^1)=4$, from \req{totaltadpole}
we obtain a total D3--brane charge of $Q_{3,tot}=-28$.
A basis of divisors is given by the three untwisted $R_i\equiv dz^i\wedge d\ov z^i$ divisors and 
the $48$ exceptional divisors $E_{i\alpha,j\beta}$. The divisors $D_{i\al}$ may be
expressed as linear combination of those divisors:
\bea
D_{1\al}&=&R_1-\sum_{\beta=1}^4 E_{1\al,2\beta}-\sum_{\gamma=1}^4
E_{1\al,3\gamma}\ ,\nnn\\
D_{2\al}&=&R_2-\sum_{\beta=1}^4 E_{1\beta,2\al}-\sum_{\gamma=1}^4
E_{2\al,3\gamma}\ ,\\
D_{3\al}&=&R_3-\sum_{\beta=1}^4 E_{1\beta,3\al}-\sum_{\gamma=1}^4
E_{2\gamma,3\al}\ .\nnn
\eea
Note, that these equations give rise to relations between the gauge
coupling ${\rm Vol}(D_{i\al})$ on a stack 
of D7--branes, which is wrapped on the divisor $D_{i\al}$, and the
divisor volumina ${\rm Vol}(R_i)$ and ${\rm Vol}(E_{i\al,j\beta})$, \eg
\eqn{gaugedivrelation}{
{\rm Vol}(D_{1\al})=r_1-\sum_{\beta=1}^4 t_{1\al,2\beta}-\sum_{\gamma=1}^4 t_{1\al,3\gamma}\ .}
The K\"ahler form $J_2$ is expanded with respect to  this basis:
\eqn{KEJ}{J_2=\sum_{i=1}^3 r_i\ R_i-\sum_{\al,\bet=1}^4
  t_{1\al,2\bet}\ E_{1\al,2\bet}-
\sum_{\bet,\gamma=1}^4 t_{2\bet,3\gamma}\ E_{2\bet,3\gamma}
-\sum_{\al,\gamma=1}^4t_{1\al,3\gamma}\ E_{1\al,3\gamma}\ .}
The intersection ring may be determined after gluing together the local patches:
\bea
{\rm Vol}({\cal X})&=&\fc{1}{6}\int_{{\cal X}}J_2\wedge J_2\wedge
J_2=\\
&=&r_1r_2r_3-\h\ \sum_{\beta,\gamma=1}^4\lf(r_1\  t_{2\beta,3\gamma}^2+r_2\
 t_{1\beta,3\gamma}^2+r_3\ 
  t_{1\beta,2\gamma}^2\ \ri)\nnn  \\
&-&\fc{1}{3}\
\sum_{\al,\bet=1}^4\lf(\ t_{1\al,2\bet}^3+t_{2\al,3\bet}^3+t_{1\al,3\bet}^3\
\ri)-\fc{1}{2}\ \sum_{\al,\bet,\gamma=1}^4
t_{1\al,2\bet}\ t_{2\bet,3\gamma}\ t_{1\alpha,3\gamma}\nnn\\
&+&\fc{1}{4}\sum_{\al,\bet,\gamma=1}^4\lf(\ t_{1\al,2\bet}\
t_{2\bet,3\gamma}^2+t_{1\al,2\bet}\ t_{1\alpha,3\gamma}^2+\ldots\ri)\ .
\eea
Hence, the K\"ahler potential for the $51$ K\"ahler moduli  of the
orientifold ${\cal X}$ becomes $\kappa_4^2\ K_{KM}=-2\ln{\rm Vol}({\cal X})$.

In  \cite{blowup} all other orbifolds \req{orbi} have been resolved 
and a consistent orientifold action \req{Action} introduced.
See also  \cite{Susanne} for a detailed pedagogical work on this.
The phenomenology of these orbifolds in view of moduli stabilization has been discussed in 
 \cite{LRSSii}.

\subsection{Open string moduli space}\label{OPEN}

The fluctuations of a D--brane are described by a set of massless open string states.
On the D--brane world volume there are the massless gauge fields $A_\mu^a$, 
which are represented by massless open string modes with Neumann boundary conditions. 
Furthermore, the transverse excitations of the D--brane are described by massless open 
string modes with Dirichlet boundary conditions. If the D--brane is wrapped around
a cycle, additional Wilson line moduli may appear, provided the wrapped part of the D--brane
contains non--trivial one--cycles. 
For a D$p$--brane wrapped around a $p-3$--cycle $C_j$, the spectrum of open string moduli
is determined by the cohomology of the cycle $C_j$.
We have ${\rm dim}\, H_+^{0}(C_j, NC_j)$
D--brane position moduli $C_k$ (matter) 
and ${\rm dim} H_-^{0,1}(C_j)$ Wilson line moduli $A_i$ \cite{jl04,Jockers:2005zy,Jockers:2005pn}.
Here $NC_j$ denotes the normal bundle of the $p-3$--cycle in ${\cal X}$.
The open string moduli are\footnote{E.g. for a D7--brane wrapping the four--cycle $C_j$
this dictionary reduces to \cite{jl04,Jockers:2005zy,Jockers:2005pn}:
\bea
{\rm D-brane\ positions:}&C_k&\ds{k=1,\ldots,
{\rm dim}\, H_-^{2,0}(C_j)}\ ,
\nnn\\
{\rm Wilson\ lines:}&A_i&\ds{i=1,\ldots,{\rm dim}\, H_-^{0,1}(C_j)}\ .\label{oopenM}
\eea}:
\bea
{\rm D-brane\ positions:}&C_k&\ds{k=1,\ldots,
{\rm dim}\, H_+^{0}(C_j, NC_j)}\ ,
\nnn\\
{\rm Wilson\ lines:}&A_i&\ds{i=1,\ldots,{\rm dim}\, H_-^{0,1}(C_j)}\ .\label{openM}
\eea
Both classes of moduli fields are in the adjoint representation of the  gauge group.
From point of view of the $D=4$ effective field theory open string moduli lead to an enlargement
of the moduli space discussed in the previous subsection \ref{GEOMETRY}.
Further open string moduli may arise from a non--trivial gauge bundle on the D--brane
world volume.
Finally, additional charged matter fields $C_{aa'}$ or $C_{ab}$ 
arise from self--intersections of one stack $a$ of D$p$--branes with its orientifold mirror
$a'$ or from intersections of different stacks $a,b$ of D$p$ and D$p'$--branes, respectively:
\bea
{\rm Matter\ fields:}&C_{aa'}&{\rm Open\ strings\ between\ stack}\ a\ {\rm and\ its\ orientifold\ 
mirror}\ a'\ ,\nnn\\
{\rm Matter\ fields:}&C_{ab}&{\rm Open\ strings\ between\ stack}\ a\ {\rm and\ stack}\ b\ .
\label{openMM}
\eea
To conclude, our set of open string moduli $\Cc$, introduced in subsection \ref{LOW}, 
is comprised by  the sets \req{openM} and \req{openMM}.
In this subsection we shall report on the dependence of the effective action on the 
open string moduli fields \req{openM}.
The dynamics of those fields may be derived
by a dimensional reduction of the Born--Infeld and CS world volume action.
This has been accomplished for a stack of space--time filling
D3--branes in 
\cite{ggjl03,ciu03} 
and for a single D7--brane in \cite{ciu04,jl04,Jockers:2005zy,Jockers:2005pn}.
Furthermore, for D5, D9--branes a similar analysis has been performed in \cite{kn03}.
More general work on the interplay between the open and closed string moduli space for
D--branes on Calabi--Yau spaces has been pioneered in 
\cite{Lerche:2001cw,Lerche:2002yw,Lerche:2002ck,Lerche:2003hs}.
On the other hand, the dependence of the effective action on the second type of moduli 
\req{openMM} is more complicated due to the lack of a local
action description of the intersection of two D--branes.
However studying scattering of matter fields from D--branes proves to be 
a powerful tool to determine those open string couplings (\cff subsection \ref{SCATT}).

In the following we show, how to obtain the kinetic energy terms for
Wilson line moduli 
$A_i$ and D-brane position moduli $C_k$ by performing a dimensional reduction of the Born--Infeld
action \req{dbi}. To understand better the structure of the interplay between the open and
closed string moduli sectors we shall focus on type IIB orientifolds of toroidal orbifolds.
The latter have been introduced in subsection \ref{ORBIFOLDS}.

\subsubsection{Open string moduli from D$p$--branes in toroidal orientifolds}

Let us start with the Born--Infeld action \req{dbi} for a stack of $n_p$ D$p$--branes
and dimensionally reduce this action on a $C_{p-3}$--cycle.
For this we assume the orthogonal splitting of the ten--dimensional metric $G$ into 
tangential and orthogonal components with respect to  the D$p$--brane world volume
\eqn{splittingG}{
G=\sum_{\mu,\nu=0}^3\eta_{\mu\nu}\ dx^\mu\ dx^\nu+
\sum_{i,j=4}^p g_{ij}\ dy^i\ dy^j + \sum_{i,j=p+1}^9\tilde g_{ij}\ dy^i\ dy^j}
Similarly, we make the following decomposition of the gauge field
strength ${\cal F}$ in \req{dbi}:
\def\Sc{{\cal S}}\def\Fc{{\cal F}}
\eqn{splitting}{
\Fc^a_{MN}=\lf(\ba{cccc}
0&0&0\\
0&\Fc_{ij}^a&-D_\nu a_i^a\\
0&D_\mu a_j^a & F_{\mu\nu}^a\ea\ri)\ .}
Here, the matrix $2\pi\ap\ \Fc^a_{ij}=B_{ij}+2\pi\ap F_{ij}^a,\ i,j=4,\ldots,p$  combines 
the anti--symmetric tensor $B$ with the constant background fluxes
$F_{ij}^a$. The latter are in the Cartan subalgebra 
of the gauge group $G_a$. Furthermore, the $a_i^a$ are Wilson lines with respect to  the $i$--th internal 
direction.
In the Born--Infeld action \req{dbi} the $9-p$ transverse position
moduli $\phi^i,\ i=p+1,\ldots,9$
are incorporated  by the pull--back, \ie by replacing in \req{dbi} the space--time part of
metric $G$ by \cite{Myers:1999ps}: 
\eqn{pullback}{P[G_{\mu\nu}]=\eta_{\mu\nu}+(2\pi\ap)^2\ \tilde g_{ij}\ 
D_\mu\phi^i\ D_\nu\phi^j\ .}
With the backgrounds \req{splittingG} and \req{splitting} and after 
replacing \req{pullback} into \req{dbi} at the leading order in $D=4$
we obtain the following terms from the Born--Infeld action for the D$p$--brane\footnote{
Note, that this is the type II result, while the type I result receives an additional factor
of $1/\sqrt{2}$.}:
\eqn{asdewq}{\ba{lcl}
\ds{\Sc_{Dp}}&=&\ds{-\fc{1}{2\pi}\ \int d^4x\ e^{-\Phi}\ \sqrt{-g_4}\ \sqrt{\det(g+2\pi\ap\Fc)}\ 
\lf\{\ \sum_{a=1}^{n_p} \lf[\ \fc{1}{(2\pi\ap)^{2}}+\fc{1}{4}\ F^a_{\mu\nu}F^{a\mu\nu}\ri.\ri.}\\
&+&\ds{\lf.\lf.\h\ \eta^{\mu\nu}\ D_\mu \phi_a^i\ 
\tilde g_{ij}\ D_\nu \phi^j+\h\ \eta^{\mu\nu}\ D_\mu a^a_i\ (g+2\pi\ap\Fc)^{ij}_{symm.}\ 
D_\nu a^a_j\ \ri]\ri\}+\ldots\ .}
\ea}
Here, the index symm. means symmetrization of the corresponding matrix.
Since in $D=4$ the gauge kinetic term is invariant under a Weyl rescaling,
we immediately extract the real part of the gauge kinetic function:
\eqn{immedial}{\re(f_{Dp_a})=(2\pi)^{-1}\ e^{-\Phi}\
  \sqrt{\det(g+2\pi\ap\ \Fc)}\ .}
Due to the calibration condition this result agrees with the results from the previous 
Subsections \ref{IIB37} and \ref{IIB59}.
The Weyl rescaling $g_{\mu\nu}\ra e^{2\Phi}(\det G)^{-1/2}g_{\mu\nu}$, which transforms
the Einstein term of the closed string sector into its canonical form,  yields 
the following kinetic energy term for the $9-p$ D$p$--brane position moduli $\phi^j_a$ and 
the $p-3$ Wilson lines $a^a_i$:
\bea\label{einstein}
&-&\fc{1}{4\pi}\ \int d^4x\ e^{\Phi}\ \sqrt{-g_4}\ 
\fc{\sqrt{\det(g+2\pi\ap\Fc)}}{\sqrt{\det G}}\\  
&&\times\lf\{\ \sum_{a=1}^{n_p}\eta^{\mu\nu}\ D_\mu \phi_a^i\ \tilde g_{ij}\ D_\nu \phi^j+
\eta^{\mu\nu}\ D_\mu a^a_i\ (g+2\pi\ap\Fc)^{ij}_{symm}\ D_\nu a^a_j\ \ri\}\ .\nnn
\eea

Additional kinetic energy terms mixing the Wilson line moduli $A_i$ with axions of the 
closed string moduli derive in type I from the coupling (\cff \eqq \req{mod3form})
\eqn{typeIG2}{
\int d^{10}x\ \sqrt{-g_{10}}\ e^{-2\Phi}\ 
\lf(dC_2-\fc{\kappa_{10}^2}{g_{10}^2}\ \omega_3\ \ri)^2\ ,}
with the CS three--form $\omega_3=A\wedge dA+\fc{2}{3}\ A\wedge A\wedge A$.
In type IIB orientifolds
with D9-- and D5--branes those couplings appear from the world--volume 
CS--coupling  $\int C_6\wedge F_2\wedge F_2$ of the D9--branes
and the CS--coupling $\int C_2\wedge F_2\wedge F_2$ of the D5--branes.
On the other hand, in  type IIB orientifolds
with D3-- and D7--branes these couplings appear from the world--volume 
CS--coupling  $\int C_4\wedge F_2\wedge F_2$ of the D7--branes.
Furthermore, kinetic energy terms mixing the position moduli $\phi^i$ with axions of the 
closed string moduli derive from the pullback of the Ramond forms $C_p$ \cite{jl04}
\bea
P[C_q]&=&\lf(\fc{1}{q!}\ C_{\mu\nu i_1\ldots i_{q-2}}-
\fc{1}{(q-1)!}\ D_{\mu} \phi^i\ C_{i\nu i_1\ldots i_{q-2}}
+\fc{1}{2(q-2)!}\  D_\mu \phi^i\ D_\nu \phi^j\ C_{ij i_1\ldots i_{q-2}}\ri)\nnn\\[4mm]
&\times& dx^{\mu}\wedge dx^{\nu}\wedge dx^{i_1}\ldots\wedge dx^{i_{q-2}}\label{PULLC}
\eea
applied on the CS--action of the D$p$--branes.

As an application let us discuss the case, that the internal manifold ${\cal X}$ is a direct product
\req{torfact} of three two--tori $\mbb T^2$, with the metric \req{tormetric}. 
Furthermore, we assume
$\Fc=0$. In that case we may simplify the integrand of \req{einstein} to the form:
\bea\label{simpli}
&&-\fc{1}{4\pi}\ \int d^4x\ e^{\Phi}\ \sqrt{-g_4}\ \fc{1}{\sqrt{\det \tilde g}}\\
&&\hskip-0.75cm\times\lf\{\sum_{a=1}^{n_p}
\sum_{i=1}^{(p-3)/2}\fc{1}{t_2^iu_2^i}\ | u^i\ D_\mu a^a_{2i+2}-D_\mu a^a_{2i+3}|^2
+\sum_{j=(p-3)/2+1}^3\fc{t_2^j}{u_2^j}\ |D_\mu \phi_a^{2j+2}+u^j\ D_\mu\phi_a^{2j+3}|^2\ri\}\ .\nnn
\eea
The (real) fields $A_i$ and $\phi^j$ do not yet correspond to complex scalars of 
supersymmetry multiplets and the proper complex fields $A$ and $C$ have to 
be defined by introducing the complex structures $u_1^j= U_2^j,\ u_2^j=U_1^j$:
\eqn{aC}{
\ba{lcl}
\ds{A^a_l=\ov U^l\ a^a_{2l+2}+i\ a^a_{2l+3}}&,&\ds{l=1,\ldots,\fc{p-3}{2}\ ,}\\
\ds{C^a_j=-i\ \phi_a^{2j+2}+\ov U^j\ D_\mu\phi_a^{2j+3}}&,&\ds{j=\fc{p-3}{2}+1,\ldots,3\ .}
\ea}
In the following we shall construct the D$p$--brane moduli spaces by combining
the actions \req{simpli} and \req{typeIG2}.

\subsubsection{D9-- and D5--branes}\label{DD99DD55}

\underline{D9--branes:}

For $p=9$ the kinetic energy terms \req{simpli} may be written
\eqn{simpliD9}{
-\fc{1}{4\pi}\ \int d^4x\ \sqrt{-g_4}\ \sum_{a=1}^{n_9}\ 
\sum_{l=1}^{3}\fc{1}{T_1^lU_1^l}\ |U^l\ D_\mu a^a_{2l+2}-i\ D_\mu a^a_{2l+3}|^2\ \ ,}
with the holomorphic moduli $T^l$, introduced in \req{defmodd}.
In $D=4$ the type I coupling \req{typeIG2}  gives rise to the terms:
\eqn{termsD9}{
\sum_{i=1}^3\ \fc{1}{(T^i+\ov T^i)^2}\ 
[\ \partial T_2^i+\h \sum\limits_{a=1}^{n_{D9}}(\partial a^a_{2i+2}\ a^a_{2i+3}+
a^a_{2i+2}\ \partial a^a_{2i+3})\ ]^2\ .}
On the other hand, in type IIB this coupling arises from from the world--volume 
CS--coupling  $\int C_6\wedge F_2\wedge F_2$ of the D9--branes.
Together with the closed string moduli metrics (for the fields $S,T^i$ and $U^j$), given in 
\eqq \req{Simply}, the metrics \req{simpliD9} and \req{termsD9} conspire together and 
may be derived from the  K\"ahler potential $K$ \cite{seealso}
\eqn{finalD9}{
\kappa_4^2\ 
K=-\ln(S+\ov S)-\sum_{i=1}^3\ln\lf[(T^i+\ov T^i)(U^i+\ov U^i)-\h\ \sum_{a=1}^{n_{D9}}
(A^a_i+\ov A^a_i)^2\ \ri]\ ,}
with the redefined K\"ahler moduli $T^i$:
\eqn{redT}{
T^i=\lf.T^i\ \ri|_{A=0}+\h\ \sum_{a=1}^{n_{D9}} a^a_{2i+2}\ A_i^a\ \ \ ,\ \ \ i=1,2,3\ .}
The K\"ahler potential \req{finalD9}
unifies the geometry of open and closed sting moduli in the coset space:
\eqn{geometryD9}{
\lf(\fc{SU(1,1)}{U(1)}\ri)_S\otimes \prod_{i=1}^3\lf(\fc{SO(2,2+n_{D9})}
{SO(2)\times SO(2+n_{D9})}\ \ri)_{T^i,U^i,A^a_i}\ .}
For toroidal orbifolds, some of the moduli are projected out by the orbifold group
and only a part of \req{geometryD9} survives.

\ \\
\underline{D5--branes:}

For $p=5$ the kinetic energy terms \req{simpli} may be written
\bea\label{simpliD5}
&&-\fc{1}{4\pi}\ \int d^4x\ \sqrt{-g_4}\ \\
&\times& \lf\{\sum_{a=1}^{n_5} 
\fc{1}{S_1U_1^1}\ |U^1\ D_\mu a^a_4-i\ D_\mu a^a_5|^2
+\sum_{j,k=2,3}\fc{1-\delta_{jk}}{U_1^jT_1^k}\ |i\ D_\mu \phi_a^{2j+2}+
U^j\ D_\mu\phi_a^{2j+3}|^2\ri\}\ ,\nnn
\eea
with the holomorphic moduli $S,T^i$, introduced in \req{defmodd}.
Again, additional kinetic energy terms, mixing the open and closed 
string moduli derive from type I coupling \req{typeIG2}.

For toroidal orbifolds, some of the moduli are projected out by the orbifold group
and only a part of \req{simpliD5} survives.
The Wilson line moduli $A^a_1=\ov U^1a_4^a+ia_5^a$ conspire 
with the closed string moduli $S,U^1$ into a single
K\"ahler potential. The type I coupling \req{typeIG2}  gives rise to the terms:
\eqn{termsD5}{
\fc{1}{(S+\ov S)^2}\ [\ \partial S_1+\h \sum\limits_{a=1}^{n_{D5}}
(\partial a^a_4\ a^a_5+a^a_4\ \partial a^a_5)\ ]^2\ .}
On the other hand, in type IIB this coupling arises from from the world--volume 
CS--coupling  $\int C_2\wedge F_2\wedge F_2$ of the D5--branes.
Together with the closed string moduli metrics (for the fields $S,T^1$ and $U^1$), given in 
\eqq \req{Simply}, the couplings \req{simpliD5} and \req{termsD5} 
may be derived from the  K\"ahler potential $K$ \cite{seealso}:
\eqn{finalD5}{
\kappa_4^2\ 
K=-\ln(T^1+\ov T^1)-\ln\lf[(S+\ov S)(U^1+\ov U^1)-\h\ \sum_{a=1}^{n_{D5}}(A_1^a+\ov A_1^a)^2
\ \ri]\ ,}
with the redefined dilaton $S$ field:
\eqn{redS}{
S=\lf.S\ \ri|_{A=0}+\h\ \sum_{a=1}^{n_{D5}} a^a_4\ A_1^a\ .}
The K\"ahler potential \req{finalD5} unifies the geometry of open and closed sting 
moduli in the coset space:
\eqn{geometryD5}{
\lf(\fc{SU(1,1)}{U(1)}\ri)_{T^1}\otimes \lf(\fc{SO(2,2+n_{D5})}{SO(2)\times SO(2+n_{D5})}\ 
\ri)_{S,U^1,A_1^a}\ .}

\ \\
\underline{D9/D5--branes:}

In the case of both D9-- and D5--branes the mixing between open and closed string
moduli becomes more complicated. 
We consider $n_{D9}$ D9--branes, which are wrapped around the full internal space ${\cal X}$  
and $n_{D5}$ D5--branes
wrapping the first torus. In addition, Wilson lines $a_4^a, a_5^a$ and $\tilde a_4^a, \tilde a_5^a$
with respect to  this torus are turned on. Here the first set of Wilson lines refers to the D9, while
the second group refers to the D5 gauge group. According to \req{aC} they are grouped
into the complex scalars $A^a=\ov U^1a_4^a +i\ a_5^a$ and 
$\tilde A^a=\ov U^1\tilde a_4^a +i\ \tilde a_5^a$, respectively.
For the case $\tilde A^a=0$ the moduli space and 
the corresponding K\"ahler  potential would be given by \req{finalD9} and \req{geometryD9} 
(with $A^a_2,A^a_3=0$), while
for $A^a=0$ the moduli space and the K\"ahler 
potential would be given by \req{finalD5} and \req{geometryD5}, \ie:
\bea
\lf(\fc{SU(1,1)}{U(1)}\ri)_{S}\otimes \lf(\fc{SO(2,2+n_{D9})}{SO(2)\times SO(2+n_{D9})}\ 
\ri)_{T^1,U^1,A^a}&,&\tilde A^a=0\ ,\label{mani1}  \\
\lf(\fc{SU(1,1)}{U(1)}\ri)_{T^1}\otimes \lf(\fc{SO(2,2+n_{D5})}{SO(2)\times SO(2+n_{D5})}\ 
\ri)_{S,U^1,\tilde A^a}&,&A^a=0\ . \label{mani2}
\eea
In the case $A,\tilde A\neq 0$ the full K\"ahler manifold of complex
dimension $3+n_5+n_9$ is no longer 
a symmetric space and the K\"ahler potential becomes \cite{seealso}
\bea
\kappa_4^2\ K&=&
-\ln\lf[\ (T^1+\ov T^1)(U^1+\ov U^1)-\h\ \sum_{a=1}^{n_{D9}}(A^a+\ov A^a)^2\ \ri]\ ,\nnn\\
&-&\ln\lf[\ (S+\ov S)(U^1+\ov U^1)-\h\ \sum_{a=1}^{n_{D5}} (\tilde A^a+\ov{\tilde A}^a)^2\ \ri]
+\ln (U^1+\ov U^1)\ ,\label{KAETOM}
\eea
with the redefined fields:
\bea
S&=&\lf.S\ \ri|_{\tilde A^a=0}+\h\ \sum_{a=1}^{n_{D5}} \tilde a^a_4\ \tilde A^a\ ,\nnn\\
T^1&=&\lf.T^1\ \ri|_{A^a=0}+\h\ \sum_{a=1}^{n_{D9}}  a^a_4\ A^a\ .
\eea
Instead of the K\"ahler potential \req{KAETOM} one may also use the following K\"ahler potential:
\bea
\kappa_4^2\ K&=&
-\ln\lf[(S+\ov S)(T^1+\ov T^1)(U^1+\ov U^1)-\h\ (S+\ov S)\ 
\sum_{a=1}^{n_{D9}}(A^a+\ov A^a)^2\ri.\nnn\\
&&\lf.-\h\ (T^1+\ov T^1)\ \sum_{a=1}^{n_{D5}} (\tilde A^a+\ov{\tilde A}^a)^2\ \ri]\ .
\eea
The corresponding scalar manifold is not anymore of special type, though it is K\"ahler.
Actually, it corresponds to the homogeneous not symmetric space $L(0,n_5,n_9)$ 
\cite{deWit:1991nm}.
This geometry describes the vector multiplet moduli space 
of $K3\times \mbb T^2$ or $\mbb T^4/\IZ_N\times \mbb T^2$ 
orientifolds with the action \req{Action2} and D9/D5--branes \cite{seealso}.

\subsubsection{D3-- and D7--branes}\label{DD33DD77}

\underline{D3--branes:}

Since the case with $n_{D3}$ space--time filling D3--branes is $T$--dual to the D9--brane case,
we may keep the discussion short.
From \eqq \req{simpli} for a D3--brane we derive the following kinetic energy terms
\eqn{trans3}{
-\fc{1}{4\pi}\ \int d^4x\ \sqrt{-g_4}\ 
 \sum_{a=1}^{n_{D3}} \sum_{j=1}^3\fc{1}{T_1^jU_1^j}\ |i\ D_\mu \phi_a^{2j+2}+U^j\ 
D_\mu\phi_a^{2j+3}|^2\ }
for the six transverse coordinates $\phi_a^i,\ i=1,\ldots,6$.
The latter are arranged into the three complex fields \req{aC}.
In addition, the pullback \req{PULLC} of the Ramond $4$--form 
yields the coupling on the D3--world volume
$$\fc{1}{4}\ D_\mu\phi^i\ D_\nu \phi^j\ C_{ij\rho\sigma}\ 
dx^\mu\wedge dx^\nu\wedge dx^\rho\wedge dx^\sigma\ ,$$
which mixes the kinetic energy terms of the brane positions $\phi^i$ with the axion of
the K\"ahler moduli $T^i$.
To this end, the closed string moduli $S,T^i,U^i$ conspire together with the three
complex fields $C^a_i$, defined in \eqq \req{aC}, into the coset space: 
\eqn{geometryD3}{
\lf(\fc{SU(1,1)}{U(1)}\ri)_S\otimes \prod_{i=1}^3\lf(\fc{SO(2,2+n_{D3})}
{SO(2)\times SO(2+n_{D3})}\ \ri)_{T^i,U^i,C^a_i}\ .}
The K\"ahler potential \req{Simply} is upgraded to:
\eqn{finalD3}{
\kappa_4^2\ K
=-\ln(S+\ov S)-\sum_{i=1}^3\ln\lf[(T^i+\ov T^i)(U^i+\ov U^i)-\h\ \sum_{a=1}^{n_{D3}}
(C^a_i+\ov C^a_i)^2\ \ri]\ ,}
with the redefined K\"ahler moduli $T^i$:
\eqn{redT3}{
T^i=\lf.T^i\ \ri|_{C=0}+\h\ \sum_{a=1}^{n_{D3}} \phi_a^{2i+3}\ C_i^a\ \ \ ,\ \ \ i=1,2,3\ .}

\ \\
\underline{D7--branes:}

For $n_{D7}$ D7--branes {\it Eq.} \req{simpli} yields:
\bea\label{trans7}
&-&\fc{1}{4\pi}\ \int d^4x\ \sqrt{-g_4}\ \\
&&\times\lf\{\sum_{a=1}^{n_{D7}} \sum_{j,k=1,2}^{3}\fc{1-\delta_{jk}}{T_1^kU_1^j}\ 
|U^j\ D_\mu a^a_{2j+2}-i\ D_\mu a^a_{2j+3}|^2+
\fc{1}{S_1U_1^3}\ |i\ D_\mu \phi_a^{8}+U^3\ D_\mu\phi_a^{9}|^2   \ri\}\ .\nnn
\eea
The definitions for complex fields $A_1,A_2$ and $C_3$ may be found in \eqq \req{aC}.
In addition, the pullback \req{PULLC} of the Ramond $8$--form 
yields the coupling on the D7--world volume
$$\fc{1}{2\cdot 6!}\ D_\mu\phi^i\ D_\nu \phi^j\ C_{ij\rho\sigma k_1\ldots k_4}\ 
dx^\mu\wedge dx^\nu\wedge dx^\rho\wedge dx^\sigma \wedge dx^{k_1}\wedge \ldots \wedge dx^{k_4}
\ ,$$
which mixes the kinetic energy terms of the brane positions $\phi^i$ with the axion of
the dilaton $S$.
To this end, the closed string moduli $S,T^i,U^i$ and the open string modulus $C_3$ 
give rise to the coset:
\eqn{geometryD7}{
\lf(\fc{SU(1,1)}{U(1)}\ri)_{T^3}\otimes \prod_{i=1}^3\lf(\fc{SO(2,2+n_{D7})}
{SO(2)\times SO(2+n_{D7})}\ \ri)_{S,U^3,C^a_3}\ .}
The corresponding K\"ahler potential is:
\eqn{finalD7}{
\kappa_4^2\ K=-\ln(T^3+\ov T^3)-\ln\lf[(S+\ov S)(U^3+\ov U^3)-\h\ \sum_{a=1}^{n_{D7}}
(C^a_3+\ov C^a_3)^2\ \ri]\ ,}
with the redefined dilaton $S$ field:
\eqn{redS7}{
S=\lf.S\ \ri|_{C=0}+\h\ \sum_{a=1}^{n_{D7}} \phi^9_a\ C_3^a\ .}

\ \\
\underline{D3/D7--branes:}

In the case of both D3-- and D7--branes the mixing between open and closed string
moduli becomes more complicated. In fact, this case is $T$--dual to
the D9/D5--case, discussed
before and a similar structure is exhibited. 
We consider $n_{D3}$  (space--time filling) D3--branes and $n_{D7}$ D7--branes, which are
transversal to the third torus.
In the following we consider the D3--brane positions $\phi^8,\phi^9$ 
and D7--brane positions $\tilde \phi^8_a,\tilde \phi^9_a$ along the third torus.
According to \req{aC} these positions are grouped
into the complex scalars $C^a=-i\phi^8_a+\ov U^3\ \phi^9_a$ and 
$\tilde C^a=-i\tilde\phi^8_a+\ov U^3\ \tilde\phi^9_a$, respectively.
For the case $\tilde C^a=0$ the moduli space and 
the corresponding K\"ahler  potential would be given by \req{finalD3} and \req{geometryD3} 
(with $C^a_1,C^a_2=0$), while
for $C^a=0$ the moduli space and the K\"ahler 
potential would be given by \req{finalD7} and \req{geometryD7}, \ie:
\bea
\lf(\fc{SU(1,1)}{U(1)}\ri)_{S}\otimes \lf(\fc{SO(2,2+n_{D3})}{SO(2)\times SO(2+n_{D3})}\ 
\ri)_{T^3,U^3,C^a}&,&\tilde C^a=0\ ,\label{Mani1}  \\
\lf(\fc{SU(1,1)}{U(1)}\ri)_{T^3}\otimes \lf(\fc{SO(2,2+n_{D7})}{SO(2)\times SO(2+n_{D7})}\ 
\ri)_{S,U^3,\tilde C^a}&,&C^a=0\ . \label{Mani2}
\eea
In the case $C,\tilde C\neq 0$ the full K\"ahler manifold of complex
dimension $3+n_3+n_7$ and the K\"ahler potential becomes:
\bea
\kappa_4^2\ K&=&-\ln\lf[(S+\ov S)(T^3+\ov T^3)(U^3+\ov U^3)-\h\ (S+\ov S)\ 
\sum_{a=1}^{n_{D3}}(C^a+\ov C^a)^2\ri.\nnn\\
&&\lf.-\h\ (T^3+\ov T^3)\ \sum_{a=1}^{n_{D7}} (\tilde C^a+\ov{\tilde C}^a)^2\ \ri]\ .
\label{KAETOMM}
\eea
with the redefined fields:
\bea
S&=&\lf.S\ \ri|_{\tilde C^a=0}+\h\ \sum_{a=1}^{n_{D7}} \tilde \phi_a^9\ \tilde C^a\ ,\nnn\\
T^3&=&\lf.T^3\ \ri|_{C^a=0}+\h\ \sum_{a=1}^{n_{D3}} \phi^9_a\ C^a\ .
\eea
This geometry describes the vector multiplet moduli space 
of $K3\times \mbb T^2$ or $\mbb T^4/\IZ_N\times\mbb T^2$ orientifolds with the action \req{Action} 
and D3/D7--branes  \cite{ADFT}.
The K\"ahler potential \req{KAETOM} may be derived from the following ${\cal N}=2$ prepotential 
\cite{seealso,ADFT}:
\eqn{prepp}{
F(S,T^3,U^3,C,\tilde C)=ST^3U^3-
\h\ S\ \sum_{b=1}^{n_3} (C^b)^2-\h\ T^3\ \sum_{a=1}^{n_7} (\tilde C^a)^2\ .}

\ \\
\underline{D3/D7--branes with two--form fluxes:}

On the internal D7--brane world volume the non--trivial (magnetic) two--form gauge flux 
\eqn{twoflux}{
\Fc^a_{NP}=\Fc^a_{45}+\Fc^a_{67}=2\pi\ap\ \lf(\ F^a_{45}\ dx^1\wedge dy^1+F^a_{67}\ 
dx^2\wedge dy^2\ \ri)} 
may be turned on. The latter obey the quantization rule $F^a_{ij}=2\pi\fc{n_{ij}^a}{m_{ij}^a}$, 
\ie: 
\eqn{quant}{
f^a_{ij}=\fc{1}{(2\pi)^2}\int_{C_{ij}} \Fc^a_{ij}=\ap\ \fc{n^a_{ij}}{m^a_{ij}}\ \ \ ,\ \ \ 
(i,j)=(4,5)\ ,\ (6,7)\ .}
The dependence of the moduli and matter field metrics on these two--form fluxes 
has been derived in \cite{lrs04,lmrs04}:
\bea
G_{T^3\ov T^3}&=&\fc{1}{(T^3+\ov T^3)^2}\ \ \ ,\ \ \ G_{S\ov S}=\fc{1}{(S+\ov S)^2}\ \ \ ,\ \ \ 
G_{U^3\ov U^3}=\fc{1}{(U^3+\ov U^3)^2}\ ,\nnn\\
G_{\tilde C^a\ov{\tilde C^a}}&=&\fc{1}{(S+\ov S)(U^3+\ov U^3)}-
\fc{\ap^{-2}\ f^a_{45} f^a_{67}}{(T^3+\ov T^3)(U^3+\ov U^3)}\ ,\nnn\\
G_{C^b\ov C^b}&=&\fc{1}{(T+\ov T)(U^3+\ov U^3)}\ .
\label{results}
\eea
In fact, up to second order in the open string moduli  $C_a$ and $\tilde C_a$,
we may summarize these results in the K\"ahler potential \cite{lmrs05}
\bea
\kappa_4^2\ K&=&-\ln\lf[(S+\ov S)(T^3+\ov T^3)(U^3+\ov U^3)-
\h\ \sum_{b=1}^{n_3} (S+\ov S)\ (C^b+\ov C^b)^2\ri.\nnn\\
&&-\lf.\h\ \sum_{a=1}^{n_7}\ [\ (T^3+\ov T^3)-(S+\ov S)\ f_{45}^af_{67}^a\ \ap^{-2}\ ]\ 
(\tilde C^a+\ov{\tilde C^a})^2 \ri]\ ,\label{kaehlerrr}
\eea
which is derived from the following ${\cal N}=2$ prepotential \cite{lmrs05}:
\eqn{prep}{
F(S,T^3,U^3,C,\tilde C)=ST^3U^3-\h\ S\ \sum_{b=1}^{n_3} (C^b)^2
-\h\ \sum_{a=1}^{n_7} (T^3-\ap^{-2}\ S\ f_{45}^af_{67}^a)\ (\tilde C^a)^2\ .}
Hence, \eqq \req{kaehlerrr} is the generalization of \req{KAETOMM} to the case of non--vanishing
two--form fluxes on the D7--brane world--volume.
It may be interesting to note, that a similar instanton--number dependent prepotential
arises in heterotic $K3\times \mbb T^2$ compactifications \cite{stieberg}.
Finally, the moduli space described by the K\"ahler potential \req{kaehlerrr} corresponds
to an orientifold limit of the underlying F--theory compactification on $K3\times K3$.
The latter has been investigated thoroughly in \cite{lmrs05}.

\subsection{Scattering of moduli and  matter fields on the disk}\label{SCATT}

\subsubsection{Disk amplitudes and  open/closed string moduli space}

The low--energy effective string action \req{eff4} may be reconstructed from string 
scattering amplitudes.
Up to the order (two--derivative level), displayed in \req{mostW} and \req{mostK}, 
the function $K(\Mc,\ov \Mc)$, the matter field metric $G_{C\ov C}$ and the trilinear couplings
$W_{ijk}$ may be derived 
by computing four--point scattering amplitudes involving moduli and matter fields.
This program has been pioneered in \cite{DKL} for heterotic CY compactifications
by considering tree--level four--point scattering on the sphere.
In \cite{lmrs04,progress} this idea is generalized to type II orientifolds
with D--branes.
Though partial results may be also obtained by considering two-- or three--point amplitudes
on the disk \cite{lmrs04,russo}, only four--point amplitudes can give the {\it full} answer 
\cite{lmrs04,progress}.
These amplitudes give non--trivial information on the Riemann curvature of the
geometric moduli space and matter field couplings.

In the following we shall discuss tree--level disk scattering from D--branes.
The world--sheet diagram of a string $S$--matrix describing the interaction of 
open and closed strings at (open string) tree--level can be conformally mapped to a surface with 
one boundary. The latter may be described by a disk, which is conformally 
equivalent to the upper (complex) half--plane ${\cal H}_+=\{z\in \ICC\ |\ \im(z)>0\}$. 
The string states, which correspond to asymptotic states in the string $S$--matrix formulation,
are created through vertex operators. The open string vertex operators are inserted 
at the boundary of the disk. On the other hand, the closed string vertex operators are 
inserted inside the disk. In theories with D--branes massless fields 
like gauge fields, Wilson line moduli or matter fields originate from open string excitations 
living on the D--brane world--volume.
Hence the boundary of the disk diagram is attached to the D--brane world--volume.
On the other hand, the closed strings, representing \eg the graviton, dilaton $S$ 
and metric moduli $M$ live in the bulk and are inserted in the bulk of the disk.
The disk, represented as upper half plane ${\cal H}_+$, may be obtained from the full complex plane
representing the sphere, through a $\IZ_2$ projection $z\mapsto \ov z$. 
It is convenient to perform the computations in the double cover, \ie
in the  complex plane $\ICC$, by taking into account the interaction
between left--moving and right--moving closed string fields (doubling trick).  
Disk scattering of gauge, matter and moduli fields in the presence of D--branes has been
pursued in \cite{bain,lmrs04}.

For each type II compactification on the CY manifold ${\cal X}$ on the 
closed string world--sheet a two--dimensional $(2,2)$ superconformal field theory (SCFT)
with stress tensors $(T,\ov T)$, supercurrents $(T_F,\ov T_F)$ and the  Abelian $U(1)$ 
currents $(J,\ov J)$ is furnished.
From the SCFT the string spectrum and all couplings from the closed string sector may be derived.
From the viewpoint of this CFT the geometric moduli 
fields $M$ describe marginal deformations of the
underlying $\sigma$--model. The vertex operators in the $(-1,-1)$ picture for NSNS moduli
are given by
\zbe
\begin{array}{lcl}
\ds{V^{(-1,-1)}_{t^K}(z,\ov z,k)}&=&\ds{c_{t^K}\ 
e^{-\tilde\phi(\ov z)}\ e^{-\phi(z)}\ \Pi^K(z,\ov z)
\ e^{ik_\rho X^\rho(z)}\ \ \ ,\ \ \ K=1,\ldots,h^{1,1}({\cal X})\ ,}\nnn\\[4mm]
\ds{V^{(-1,-1)}_{u^\Si}(z,\ov z,k)}&=&\ds{c_{u^\Si}\ 
e^{-\tilde\phi(\ov z)}\ e^{-\phi(z)}\ \Xi^\Si(z,\ov z)
\ e^{ik_\rho X^\rho(z)}\ \ \ ,\ \ \ \Si=1,\ldots,h^{2,1}({\cal X})\ ,}\nnn
\end{array}
\label{Vtu}
\ee
and their complex conjugates. The fields $\phi,\tilde \phi$ 
are related to the $U(1)$ currents of the superconformal system and $k$ is the space--time 
momentum. The latter obeys the massless on--shell constraints $k^2=0$.
The normalizations $c_{t^K},\ c_{u^\Si}$ may be fixed by considering four--point correlators. 
The conformal fields $\Pi^K(z,\ov z)\ ,\ \Xi^\Si(z,\ov z)$ have conformal dimensions
$(h,\ov h)=(\h,\h)$ with respect to  the internal left/right $(2,2)$ SCFT of type II
superstring compactifications on ${\cal X}$. 
The field $\Pi^K(z,\ov z)$ has $U(1)$ charges $(1,-1)$ in type IIA and
$(1,1)$ in type IIB, while the field $\Xi^\Si(z,\ov z)$ has the charges 
charges $(1,1)$ in type IIA and $(1,-1)$ in type IIB with respect to  the
internal $U(1)$ currents $J,\ov J$~\cite{Seiberg,Cecotti}. 
The vertex operators for the other sectors may be obtained by spectral flow.
Finally the vertex operator for the bosonic massless NSNS 
closed string modes describing a graviton, dilaton or anti--symmetric tensor in the 
$(-1,-1)$ ghost picture is given by:
\eqn{dilaton}{
V_G^{(-1,-1)}(\ov z,z,q)=\epsilon_{\mu\nu}\ e^{-\tilde\phi(\ov z)}\ 
e^{-\phi(z)}\ \tilde\psi^\mu(\ov z)\ \psi^\nu(z)\ e^{iq_\nu X^\nu(\ov z,z)}\ .}
The polarization tensor $\eps_{\mu\nu}$ is subject to the on--shell conditions 
$\eps_{\mu\nu} q^\mu=0=\eps_{\mu\nu} q^\nu$ and $q^2=0$.
For further details, see also \cite{MSD}.
Apart from these constraints we shall perform the calculation for arbitrary polarization 
$\epsilon_{\mu\nu}$, thus allowing to also extract the gauge and matter field couplings to 
the graviton and anti--symmetric tensor.
The polarization $\epsilon_{\mu\nu}$ in \req{dilaton} determines the relevant closed string 
state:
\bea\label{polarization}
\epsilon_{\mu\nu}=\epsilon_{\nu\mu}\ \ \ ,\ \ \ && {\rm Graviton\ ,}\cr
\epsilon_{\mu\nu}=-\epsilon_{\nu\mu}\ \ \ ,\ \ \ &&{\rm Kalb-Ramond\ ,}\cr
\epsilon_{\mu\nu}=\fc{1}{\sqrt 2}\ (\eta_{\mu\nu}-q_\mu \ov q_\nu-q_\nu \ov q_\mu)\ \ \ ,\ \ \ &&
\ov q^2=0\ \ ,\ \ q_\mu\ov q^\mu=1\ ,\ \ \ {\rm Dilaton\ .}
\eea
After introducing the orientifold action, the vertex operators \req{Vtu} do not correspond 
to moduli fields representing complex scalars of chiral multiplets in ${\cal N}=1$ in $D=4$.
We have to define new vertex operators related to the holomorphic
moduli fields $T^I,U^\Sigma$ through a linear combination.
As a result, \eg in \tb orientifolds with D3/D7--planes the vertex
operator \req{Vtu} for the K\"ahler moduli $T^I$ is replaced by (\cff \eqq \req{defmod}) 
\cite{progress}
\zbe\label{VT}
V^{(-1,-1)}_{T^I}(z,\ov z,k)=\fc{1}{6} \ \Kc^{IJ}\ V^{(-1,-1)}_{t^J}(z,\ov z,k)\ .
\ee 
with the inverse intersection form $\Kc^{IJ}$.

The uncharged moduli fields $M\in \{T^I,U^\Si\}$ interact with each other through their 
$\sigma$--model couplings or interact gravitationally. In addition there are couplings
due to a moduli dependent scalar potential $\Vc$ in \req{eff4}. At tree--level 
the string $S$--matrix $\Ac_{\rm Disk}(M^I,\ov M^{\ov J},M^K,\ov M^{\ov L})$ 
of four moduli fields $M$ decomposes into a power series in $\ap$, starting at $\ap^{-1}$. 
Each power in $\ap$ accounts for a class of irreducible and reducible Feynman diagrams with 
four external legs. 
The reducible diagrams are built from the above described interactions with 
intermediate moduli or graviton exchanges. The first order in $\ap$ gives:
\zbe\label{MMMM}
\lf.\Ac_{\rm Disk}(M^I,\ov M^{\ov J},M^K,\ov M^{\ov L})\ \ri|_{\ap^1}=t\ R_{I\ov J K\ov L}+
\kappa_4^2\ \fc{st}{u}\ G_{I\ov L}\ G_{K\ov J}+\kappa_4^2\ \fc{tu}{s}\ G_{I\ov J}\ G_{K\ov L}\ ,
\ee
with the metric $G_{I\ov J}$ and  the Riemann tensor $R_{I\ov J K\ov L}$ of the closed string 
moduli space $M$. Furthermore, we have introduced the kinematic invariants 
$s=\ap(k_1+k_2)^2,\ t=\ap(k_1+k_3)^2,\ u=\ap(k_1+k_4)^2$ with the four external space--time momenta
$k_i$ of the four incoming and outgoing moduli fields.
Subleading contributions $\alpha^{-1},\ \alpha^0$ originate for a non--vanishing potential $\Vc$
in the action \req{eff4}.
After extracting the first $\ap$ order of the string amplitude 
$\Ac_{\rm Disk}(M^I,\ov M^{\ov J},M^K,\ov M^{\ov L})$, the equation \req{MMMM} becomes a 
non--trivial differential equation for the closed string K\"ahler potential $K(M,\ov M)$, which
allows to fix uniquely the closed string moduli metrics \req{metric}.

For type II compactifications on ${\cal X}$
a similar relation \req{MMMM} holds for the geometric moduli
$m^I\in\{t^I,u^K\}$ of ${\cal X}$
\cite{DKL}:
\zbe\label{mmmm}
\lf.\Ac_{\rm Sphere}(m^I,\ov m^{\ov J},m^K,\ov m^{\ov L})\
\ri|_{\ap^1}=t\ \tilde R_{I\ov J K\ov L}+
\kappa_4^2\ \fc{st}{u}\ g_{I\ov L}\ g_{K\ov J}+\kappa_4^2\ \fc{tu}{s}\ g_{I\ov J}\ g_{K\ov L}\ ,
\ee
with the metric $g_{I\ov J}$ and Riemann tensor $\tilde R_{I\ov J K\ov L}$ of the closed 
string moduli space $m$.
The four moduli string $S$--matrix $\Ac_{\rm Sphere}(m^I,\ov m^{\ov J},m^K,\ov m^{\ov L})$
is computed on the sphere with the vertex operators \req{Vtu}. The set of equations \req{mmmm}
allow to probe the ${\cal N}=2$ results of subsection \ref{GEOMETRY}. 
In that case not any subleading corrections from a moduli--dependent scalar potential exist.
We may apply the relation between the ${\cal N}=2$ and ${\cal N}=1$ metrics
\zbe\label{interesting}
g_{I\ov J}\ \Kc^{IK}\ \Kc^{JL}=-\fc{9}{4}\ G_{K\ov L}+\fc{9}{4}\ K_{K}\ K_{\ov L}\ ,
\ee
to translate the r.h.s. of \req{mmmm} into ${\cal N}=1$ language.
However due to the second term in \req{interesting} the r.h.s. of \req{mmmm}
does not map to the r.h.s. of \req{MMMM} as a result of additional interactions between
left-- and right moving closed string fields on the disk.

Now we consider the scattering of two closed and two open string moduli.
Open string moduli may account for  Wilson lines $A_i$, 
D--brane positions $C_k$ (\cff \req{openM}) and charged matter
fields $C_{aa'},\ C_{ab}$ (\cff \req{openMM}).
The vertex operator for such an open string state is given by
\eqn{Vc}{
V_{C^\alpha}(z,k)=c_\alpha\ e^{-\phi(z)}\  \Lambda^\alpha(z)\ e^{i\ k_\rho X^\rho(z)}\ ,}
to be inserted at the boundary $z=\ov z$ of the disk ${\cal H}_+$.
Here, the field $C^\al$ collectively accounts for the open string fields $A_i,C_k,C_{aa'}$ 
and $C_{ab}$, introduced in \req{openM} and \req{openMM}.
On the boundary the world--sheet supercurrent is given by
$T_F(z,\ov z)=\h[\ T_F(z)+T_F(\ov z)\ ]$.
The conformal fields $\Lambda^\alpha(z)$ have conformal dimension $h=\h$.
Again, the normalization $c_\alpha$ may be fixed by considering scattering of \eg 
four open string moduli $C^\alpha$.
The linear $\ap$--order of a disk scattering of two closed string moduli
$M^I,\ov M^{\ov J}$ and two open string moduli $C^\alpha,\ov C^{\ov \beta}$ gives \cite{DKL}:
\eqn{MMCC}{
\lf.\Ac_{\rm Disk}(M^I,\ov M^{\ov J},C^\alpha,\ov C^{\ov \beta})\ \ri|_{\ap^1}=t\ 
R_{I\ov J \alpha\ov\beta}+\kappa_4^2\  \fc{tu}{s}\ G_{I\ov J}\ G_{\alpha\ov \beta}\ ,}
with mixed  components of the Riemann tensor for the whole moduli space:
\eqn{mixed}{
R_{I\ov J \alpha\ov\beta}=K_{I\ov J\alpha\ov\beta}-K_{I\alpha\ov\gamma}\ G^{\ov\gamma\delta}\ 
K_{\delta\ov\beta\ov J}\ .}
Again, subleading contributions $\alpha^{-1},\ \alpha^0$ originate for a non--vanishing 
potential $V$.
Finally we discuss the disk scattering of four open string moduli, which
gives non--trivial information on three--point functions $W_{ijk}$ of \req{mostW}.
\eqn{CCCC}{
\lf.\Ac_{\rm Disk}(C^\alpha,\ov C^{\ov \beta}, C^\gamma,\ov C^{\ov\delta})\ \ri|_{\ap^0}
\hskip-0.3cm=
\fc{t}{u}\ G_{\alpha\ov \delta}\ G_{\gamma\ov \beta}+\fc{t}{s}\ 
G_{\alpha\ov \beta}\ G_{\gamma\ov \delta}
+e^{\kappa_4^2 K(M,\ov M)}\ W_{\alpha\gamma\lambda}\ G^{\lambda\ov\mu}\ 
\ov W_{\ov\mu\ov\beta\ov\delta}\ ,}

As an application of the program outlined above we shall now determine the effective action 
\req{eff4} for orientifolds of toroidal orbifold compactifications, \ie we 
determine their closed and open sting moduli couplings \cite{lmrs04,progress}.
The advantage of these construction is, that for the orbifolds
\req{orbi}  an exact conformal field theory
description is accessible \cite{dfms87}.

\subsubsection{Scattering of moduli and matter fields in toroidal orientifolds}

In the following we shall concentrate on type II orbifolds \req{orbi}
endowed with the orientifold actions \req{Action2} or \req{Action}.
We refer the reader to subsection \ref{ORBIFOLDS} for more details on these constructions.

The vertex operators \req{Vtu} for the geometric untwisted moduli $t^i,u^j$, whose numbers are 
given in \eqq \req{Hodgeuntw}, are derived from the closed string
$\sigma$--model 
\req{sigmaaction} coupled to the orbifold background $g_{IJ}$ metric: 
\zbe
\ba{lcl}
\ds{V_{t^K}(z,\ov z,k)}&=&\ds{\h\ c_{t^K}\ e^{-\tilde\phi(\ov z)}\ e^{-\phi(z)}\ 
\fc{\partial}{\partial t^K}\ g_{IJ}\ \tilde \psi^I(\ov z)\ \psi^J(z) \
e^{ik_\rho X^\rho(z)}\ ,\ K=1,\ldots,h^{1,1}_{\rm untw.}\ ,}\\[4mm]
\ds{V_{u^\Si}(z,\ov z,k)}&=&\ds{\h\ c_{u^\Si}\ e^{-\tilde\phi(\ov z)}\ e^{-\phi(z)}\ 
\fc{\partial}{\partial u^\Si}\ g_{IJ}\ \tilde \psi^I(\ov z)\ \psi^J(z) \
e^{ik_\rho X^\rho(z)}\ ,\ \Si=1,\ldots,h^{2,1}_{\rm untw.}\ .}
\ea\label{VERTEXtu}
\ee
Hence, in \eqq \req{Vtu}  the $(h,\ov h)=(\h,\h)$ conformal fields $\Pi^K$ and $\Xi^\Si$ are
given by
\zbe
\Pi^I(z,\ov z)=\h\ \fc{\partial}{\partial t^K}\ g_{IJ}\ \tilde \psi^I(\ov z)\ \psi^J(z)\ \ \ ,\ \ \ 
\Xi^\Si(z,\ov z)=\h\ \fc{\partial}{\partial u^\Si}\ g_{IJ}\ \tilde \psi^I(\ov z)\ \psi^J(z) \ ,
\ee
with the internal left-- and right--moving fermion fields $\psi^I(z)$ and $\tilde\psi^J(\ov z)$,
respectively. 
On the other hand, for blown up moduli related to an isolated fixpoint $f_\al$
under the orbifold group element $\th$ the internal conformal fields are:
\zbe\label{fieldXC}
\Pi^{\theta,\al}(z,\ov z)=\prod_{i=1}^3 \tilde\sigma^i_{\theta_i,f^i_\al}(\ov
z)\ \tilde {\rm s}^i_{\theta_i}(\ov z)\ \sigma^i_{\theta_i,f^i_\al}(z)\ {\rm
  s}^i_{\theta_i}(z)\ .
\ee
We have introduced the bosonic 
twist fields $\si^j_{\theta_j}(z)$ and fermionic spin fields ${\rm s}^j_{\theta_j}(z)$.
The fields $\si, s$ generate branchings on the (internal) fields
$\partial X^J(z),\ov\partial X^J(\ov z)$ and $\psi^I(z), \tilde \psi^I(\ov z)$,
respectively. At a fixpoint $f_\al$ the local behavior
of those fields in the presence of twist fields is given by the operator 
products \cite{dfms87}. The twist field $\sigma_{\theta_i}^i$ has conformal dimension
$\h\theta_i(1-\theta_i)$, while the spin field ${\rm s}^i_{\theta_i}$ has dimension 
$\h(\theta_i)^2$.
Hence, the field \req{fieldXC} has conformal weights $(h,\ov h)=(\h,\h)$, subject
to the supersymmetry relation $\theta_1+\theta_2+\theta_3=1$ of the
orbifold twist \req{oaction}.
Furthermore, the internal conformal fields $\Lambda^\al(z)$ of the  open string 
moduli vertex operator are
\bea
\Lambda^i(z)&=&\lambda\  \psi^i(z)\ \ \ ,\ \ \ i=1,\ldots,6\ ,\label{twistedMAT}\\
\Lambda^\varphi(z)&=&\lambda\  \prod_{i=1}^3 \sigma^i_{\varphi_i}(z)\
s^i_{\varphi_i}(z)\ \ \ ,\ \ \ 1/4\ {\rm BPS\ matter}\ ,\label{twistedMAT1/2}\\
\Lambda^\varphi(z)&=&\lambda\  \prod_{i=1}^2 \sigma^i_{\varphi_i}(z)\  
s^i_{\varphi_i}(z)\ \ \ ,\ \ \ 1/2\ {\rm BPS\ matter}\ ,\label{twistedMAT1/4}
\eea
with $\lambda$ the Chan--Paton gauge degrees of freedom and $\psi^i(z)$ the internal 
fermionic open string field. While the fields $\Lambda^i$
describe untwisted matter fields, the fields $\Lambda^\varphi$
correspond to twisted matter fields. The latter originate in type IIA from open
strings stretched between two intersecting D6--branes with the relative
intersection angles $\varphi_i$ with respect to  the three internal complex planes. 
In the $T$--dual type IIB picture these fields describe
matter from open strings stretched between D$p$ and D$p'$--branes.
The latter may also carry non--trivial world volume two--form flux
$f^{j}$, to be specified in a moment.
In both cases the map to the intersecting D6--brane picture is
convenient. In \req{twistedMAT1/4} and \req{twistedMAT1/2}, 
the spin field $s_{\varphi_i}^i$ has conformal dimension
 $\h(1-\varphi_i/\pi)^2$. With the supersymmetry relation \req{anglesusy}
 the fields $\Lambda^\varphi(z)$ have conformal dimension $1/2$.

In the following  we shall restrict to the case of a factorizable six--torus
\req{torfact}. We refer the reader to subsection \ref{secinttor} for the definition of
the geometric moduli in \req{cplstr} and \req{Kahmod}. 
In addition, we turn on an internal constant magnetic background flux $2\pi\ap\
\Fc_{ij}^a=B_{ij}+2\pi\ap\ F_{ij}^a$. The latter is assumed to be
block--diagonal
with respect to  the three tori $\mbb T^2_j$. Hence we may write:
\eqn{introflux}{
\Fc_{ij}^a:=diag\lf(\ \Fc^a_1\ ,\ \Fc^a_2\ ,\ \Fc^a_3\ \ri)\ \ \
,\ \ \ \Fc_j^a=\lf(\ba{cc}  0 & f_a^{j}\\
                            -f_a^{j}&0  \ea\ri)\ .}
In the case of a factorizable six--torus \req{torfact}, with the
geometric moduli \req{cplstr}  and \req{Kahmod}, the vertex operators \req{VERTEXtu} reduce to:
\zbe
\ba{lcl}
\ds{V_{t^K}^{(-1,-1)}(\ov z,z,k)}&=&\ds{
\fc{1}{t^K-\ov t^K}\ e^{-\tilde\phi(\ov z)}\ e^{-\phi(z)}\ \tilde\Psi^j(\ov z)\ 
\ov{\Psi^j}(z)\ e^{ik_\rho X^\rho(z,\ov z)}\ ,\ K=1,\ldots,h^{1,1}_{\rm untw.}\ ,}\\[4mm]
\ds{V_{u^\Si}^{(-1,-1)}(\ov z,z,k)}&=&\ds{\fc{-1}{u^\Si-\ov u^\Si}\ 
e^{-\tilde\phi(\ov z)}\ e^{-\phi(z)}\ 
\tilde\Psi^j(\ov z)\ \Psi^j(z)\ e^{ik_\rho X^\rho(z,\ov z)}\ ,\ 
\Si=1,\ldots,h^{2,1}_{\rm untw.}\ .}
\ea\label{VERTEXTU}
\ee
More precisely the vertex operator for the imaginary part $t_2$ of $t$ 
is given by $V_{t_2}=i\ (V_t-V_{\ov t})$, which
amounts to symmetrizing the vertex operator $V_t$ with respect to  the left- and 
right-movers. The vertex operator for the real part of $t$ can be obtained
from space-time supersymmetry.
In the following we compute the amplitudes for the imaginary part 
$t_2$ by using the above operator $V_t$ and its conjugate 
while summing over the two resulting amplitudes at the end of the computation.
Furthermore, above we have introduced the complex bosonic and fermionic fields \cite{lmrs04}
($j=1,2,3$):
\bea\label{complexify}
\ov Z^j&=&\sqrt{\fc{t^j_2}{2 u^j_2}}\   (X^{2j-1}+u^j\ X^{2j})\ \ \ ,\ \ \ 
Z^j=\sqrt{\fc{t^j_2}{2 u^j_2}}\ \      (X^{2j-1}+\ov u^j\ X^{2j})\ ,\cr
\ov\Psi^j&=&\sqrt{\fc{t^j_2}{2 u^j_2}}\ (\psi^{2j-1}+u^j\ \psi^{2j})\ \ \ ,\ \ \ 
\Psi^j=\sqrt{\fc{t^j_2}{2 u^j_2}}\ (\psi^{2j-1}+\ov u^j\ \psi^{2j})\ .
\eea
In this writing, the  Green's functions for the internal bosonic fields
$\partial Z$ and fermions $\Psi$ take the simple form:
\bea\label{simplegreen}
&&\vev{\partial Z^j(z_1)\ \partial \ov Z^j(z_2)}=-\fc{1}{(z_1-z_2)^2}\ \ \ ,\ \ \ 
\vev{\partial Z^j(z_1)\ \partial Z^j(z_2)}=0\cr
&&\vev{\Psi^j(z_1)\ \ov \Psi^j(z_2)}=\fc{1}{z_1-z_2}\ \ \ ,\ \ \ 
\vev{\Psi^j(z_1)\ \Psi^j(z_2)}=0\ .
\eea
The interaction between (internal) left-- and right--moving closed string fields
on the double cover of the disk is described through the correlators \cite{lmrs04}
\bea\label{simplegreeni}
&&\vev{\partial Z^j(z_1)\ \ov\partial \ov Z^j(\ov z_2)}=-\fc{D^j}{(z_1-\ov z_2)^2}\ \ \ ,\ \ \ 
\vev{\partial Z^j(z_1)\ \ov\partial Z^j(\ov z_2)}=0\ ,\cr
&&\vev{\Psi^j(z_1)\ \ov{\tilde \Psi}^j(\ov z_2)}=\fc{D^j}{z_1-\ov z_2}\ \ \ ,\ \ \ 
\vev{\Psi^j(z_1)\ \tilde \Psi^j(\ov z_2)}=0\ ,
\eea
with the matrix
\eqn{DN}{
D^j=\cases{1\ , & {\rm Neumann\ ,}\cr
          -1\ , & {\rm Dirichlet\ ,}}}
for Neumann or Dirichlet boundary conditions in both directions of the
two--torus $\mbb T^2_j$, respectively. 
In the case of mixed boundary conditions
originating from an (internal) world--volume two--form flux $f^j$, the
matrix \req{DN} is generalized to \cite{lmrs04}:
\eqn{complexD}{
D^j=\fc{t^j-\ov t^j+2\ f^{j}}{t^j-\ov t^j-2\ f^{j}}=\fc{t_2^j-i\
  f^{j}}{t_2^j+i\ f^{j}}\ \ \ ,\ \ \ {\rm  Mixed\ D/N\ .}}
Obviously, we have $D^j \ov D^j=1$.
Note, that the two cases in \req{DN} follow from \req{complexD} 
in the limit $f^{j}\ra 0$ and $f^{j}\ra\infty$, respectively.
Moreover, in the dual type IIA case, we have $D^j=e^{-2i\varphi^j}$.
For our special background, a six--torus $\mbb T^6$ being the direct product of three single
two--tori $\mbb T^2_j$, correlators involving fields from different 
internal complex planes vanish. This is due to internal $U(1)$ charge conservation.

The open string vertex operator corresponding to the complex open
string moduli $C^i$, introduced in \eqq \req{aC}, becomes:
\eqn{zeromatter}{
V_{C^i}^{(-1)}(z,k)=\lambda\ e^{-\phi(z)}\ \Psi^i(z)\ e^{ik_\nu X^\nu(z)}\ \ \ ,\ \ \
i=1,2,3\ .}

Orbifolds with a factorized six--torus \req{torfact}
have the closed string moduli space given in the first
line of \req{cosets}, with $h^{2,1}_{\rm untw.}=0,1,3$. One example is the 
$\IZ_2\times \IZ_2$--orbifold, with $h^{2,1}_{\rm untw.}=3$. 
We shall perform our string S--matrix computations for this orbifold,
since the results may be easily truncated to simpler cases $h^{2,1}_{\rm untw.}\leq 2$, 
if we just fix the relevant complex structure moduli in the
corresponding expressions.
According to \req{MMMM} from a disk scattering of four closed string moduli we obtain
information on the Riemann tensor of the closed string moduli space.
This information is enough to deduce the closed string moduli  K\"ahlerpotential. 
The result is the first
line of \req{Kcosets} and given by the second last terms of \req{KpotZ22}:
\eqn{KAE}{
\kappa_4^2\ K=-\sum_{i=1}^3\ln(t^i+\ov t^i)-
\sum_{\Si=1}^3\ln(u^\Si+\ov u^{\ov \Si})\ .}
From scattering four closed string moduli fields on the 
sphere one ends up with the same result.
It has been already pointed out in \cite{DKL},
that $D=4$ heterotic ${\cal N}=1$ or ${\cal N}=2$ type II compactification on the same manifold 
have the same metric moduli spaces up to second order in the momenta. We have checked this
for the kind of models we discuss in this article.

Let us now move on to the metrics of the open string moduli~$\Cc$.
The power of the string computation shows up for deriving the
metrics for the fields \req{modss}, \ie
(twisted) matter fields originating from open strings stretched between
a D9--brane and a D5--brane or a D3-- and D7--brane.
A detailed analysis of the amplitudes \req{MMCC} and \req{CCCC} with the vertex operator
\req{twistedMAT1/4} gives
the results for the open string moduli metrics involving $1/4$ BPS matter $C_\varphi$ \cite{lmrs04}
\eqn{twisted1/4}{
\kappa_4^{2}\, G_{C_\varphi\ov C_\varphi}=e^{\Phi_4}\ \prod_{j=1}^3 
(u^j-\ov u^j)^{-\fc{\varphi^j}{\pi}}
\ \sqrt{\fc{\Gamma(\varphi^j/\pi)}{\Gamma(1-\varphi^j/\pi)}}\ \prod_{k=1}^3 
(t^k-\ov t^k)^{-\gamma\fc{\varphi^k}{\pi}-\beta}\ ,}
with $\varphi^1+\varphi^2+\varphi^3=2\pi$.
Here, the parameters $\beta,\gamma$ are rational numbers, to be explained
and fixed in \cite{progress}.
On the other hand, for the open string moduli metrics involving $1/2$ BPS matter $C_\varphi$ 
($\varphi^3=0$), with the vertex operator \req{twistedMAT1/2},  we obtain \cite{lrs04}
\eqn{twisted1/2}{
\kappa_4^{2}\,  G_{C_\varphi\ov C_\varphi}=e^{\Phi_4}\ (t^3-\ov t^3)^{\fc{\sigma}{2}}\ 
\prod_{j=1}^2 (u^j-\ov u^j)^{-\fc{\varphi^j}{\pi}}\ 
\sqrt\fc{\Gamma(\varphi^j/\pi)}{\Gamma(1-\varphi^j/\pi)}\ ,}
with $\sigma=\pm 1$, depending, whether the open strings in the third
plane have Neumann or Dirichlet boundary conditions, respectively.
The angles $\varphi^j$ measure the (relative) magnetic flux $f^{j}$ in the $j$--th subtorus 
$\mbb T^2_j$ felt by the open string, which is stretched between a D$p$--brane 
stack $a$ and a D$p'$--stack $b$. More precisely
\eqn{fluxyuk}{
\varphi^j\equiv\varphi^{j}_{ab}=\arctan\lf(\fc{f_b^{j}}{t_2^r}\ri)-
\arctan\lf(\fc{f_a^{j}}{t_2^r}\ri)\ ,}
with the two--form fluxes $f^{j}_a,\ f^{j}_b $ on stack $a$ and $b$, respectively. 
In the $T$--dual picture the above matter field metrics correspond to 
the metrics of matter fields of open strings stretched between two D6--branes
intersecting with the angles $\varphi^j$ with respect to  the $j$--th complex plane.

\paragraph{D9-- and D5--branes}
\ \\
\ \\
In the following we shall first discuss \tb orbifolds \req{orbi}
with the orientifold action \req{Action2}. The latter
allows for D9-- and D5--branes. Depending on which two--torus they are wrapped
there may be three kinds of D5--brane stacks. In the following $D5_j$ denotes 
a stack of D5--branes wrapped around the $j$--th subtorus $\mbb T^2_j$.
There are the two classes of open string moduli \req{openM} and \req{openMM}.
The first class \req{openM} comprises the complex fields (\cff \eqq \req{aC} for their
definitions):
\eqn{mods}{
\begin{array}{lcl}
{\rm Wilson\ line\ moduli\ of}\ D9     & A^9_i\ ,& i=1,2,3\ ,\\[2mm]
{\rm Wilson\ lines\ of\ D5}            & A_i^{5_i}\ ,& i=1,2,3\ ,\\[2mm]
{\rm Positions\ of\ D5}_i              & C_j^{5_i}\ ,& i,j=1,2,3\ ,\ i\neq j\ .
\end{array}}
The second class \req{openMM} is composed by the complex matter fields:
\eqn{modss}{
\begin{array}{lcr}
{\rm D9D5-matter}     & C^{95_i}\ ,  & i=1,2,3 \ ,\\[2mm]
{\rm D5D5-matter}     & C^{5_i5_j}\ , & i,j=1,2,3,\ i\neq j\ ,\\[2mm]
{\rm D9D5-matter\ with\ 2-flux}     & C^{95_i}\ ,  & i=1,2,3 \ ,\\[2mm]
\end{array}}
The open string fields $C^{95_i}$ and $C^{5_i5_j}$ correspond to $1/2$
BPS matter, while in the case, that there is a world--volume two--form
flux $f^j$ on the D9--brane, the fields $C^{95_i}$ correspond to $1/4$
BPS matter. A detailed analysis of the amplitudes \req{MMCC} and \req{CCCC} gives
the following results for the open string moduli metrics 
(in terms of the holomorphic closed string moduli) \cite{lmrs04}:
\eqn{metricsfield}{\ba{lll}
\ds{\kappa_4^{2}\ G_{A_i^9\ov A_i^9}=}&\ds{\fc{1}{(U^i+\ov U^i)\
    (T^i+\ov T^i)}\ \sqrt{\fc{[1+(\tf^k)^2]\
      [1+(\tf^l)^2]}{1+(\tf^i)^2}}\ ,} &
\ds{i,j,k=1,2,3\ ,\ j\neq i\neq k\ ,}\\[6mm]
\ds{\kappa_4^{2}\ G_{A_i^{5_i}\ov A^{5_i}_i}=}& \ds{\fc{1}{U^i+\ov
    U^i}\ \fc{1}{S+\ov S}\ ,} &
\ds{i=1,2,3\ ,}\\[6mm]
\ds{\kappa_4^{2}\ G_{C^{5_i}_j\ov C^{5_i}_j}=}& \ds{\fc{1}{U^j+\ov
    U^j}\ \fc{1}{T^k+\ov T^k}\ \ \ ,\ \ \ i,j,k=1,2,3\ ,\ } &
\ds{j\neq i\neq k\ ,}
\ea}
with $\tilde f^i=f^i/t_2^i$.
The non--diagonal matrix elements vanish for $\mbb T^6$ a direct
product of three two--tori:
\eqn{vanish}{G_{A_i\ov A_j}=0\ \ \ ,\ \ \ G_{C_i\ov C_j}=0\ \ \ ,\ \ \
  \i\neq j\ .}
The above metrics agree with the results from a dimensional reduction
of the Born--Infeld action (\cff subsection \req{DD99DD55}).
Furthermore, for vanishing two--form flux these metrics have been also derived in 
\cite{Rigolin}.

Let us now turn to the matter field metrics of the fields \req{modss}.
Expressed by the holomorphic closed string moduli \req{defmodd}
the metric \req{twisted1/4} for $1/4$ BPS matter becomes:
\eqn{Twisted1/4}{
\kappa_4^{2}\, G_{C_\varphi\ov C_\varphi}= (S+\ov
S)^{-\fc{1}{4}-\fc{3\beta}{2}-\gamma}\ \prod_{j=1}^3 (T^j+\ov
T^j)^{-\fc{1}{4}+\fc{\beta}{2}+\gamma(1+\fc{\varphi^j}{\pi})}\  
(U^j+\ov U^j)^{-\fc{\varphi^j}{\pi}}
\sqrt{\fc{\Gamma(\varphi^j/\pi)}{\Gamma(1-\varphi^j/\pi)}}\ .}
On the other hand, the metric \req{twisted1/2} for $1/2$ BPS matter becomes:
\bea
\kappa_4^{2}\, G_{C_\varphi\ov C_\varphi}&=&(S+\ov
S)^{\fc{\sigma-1}{4}}\ 
(T^3+\ov T^3)^{\fc{\sigma-1}{4}}\nnn\\
&\times&\prod_{j=1}^2 (T^j+\ov T^j)^{-\fc{\sigma+1}{4}}\  (U^j+\ov U^j)^{-\fc{\varphi^j}{\pi}}
\sqrt{\fc{\Gamma(\varphi^j/\pi)}{\Gamma(1-\varphi^j/\pi)}}\ \label{Twisted1/2}.
\eea
Equppied with these results we may derive the matter field metrics for the
fields \req{modss}. 
First, we consider a D5--brane, wrapped around the third two--torus $\mbb T^2_3$ and a 
D9--brane wrapped around the full six--torus $\mbb T^6$. This system, which is $1/2$ BPS, 
preserves ${\cal N}=2$ space--time supersymmetry and we shall apply
\req{Twisted1/2}. With respect to  the third torus, open strings
have Neumann boundary conditions, \ie $\sigma=1$. On the other hand, in the dual \ta picture, 
the two branes intersect at the angles $\pi/2$ within the other two internal planes, 
\ie $\varphi_{ab}^j=\h\ ,\ j=1,2$ and we obtain:
\eqn{ninefive}{
\kappa_4^{2}\ G_{C^{95_3}\ov C^{95_3}}=\fc{1}{(T^1+\ov T^1)^{1/2}(T^2+\ov T^2)^{1/2}}\ 
\fc{1}{(U^1+\ov U^1)^{1/2}\ (U^2+\ov U^2)^{1/2}}\ .}
Furthermore, for two D5--branes, with one wrapping the torus $\mbb T^{2}_1$ and the other
one wrapping the torus $\mbb T^2_2$ the open string coordinates have
pure Dirichlet boundary conditions w.r.t the third plane $\mbb
T^2_3$, \ie $\sigma=-1$.
Again, in the dual \ta picture, 
the two branes intersect at the angles $\pi/2$ within the other two internal planes, 
\ie $\varphi_{ab}^j=\h\ ,\ j=1,2$. Hence, from \eqq \req{Twisted1/2} we deduce:
\eqn{fivefive}{
\kappa_4^{2}\ G_{C^{5_15_2}\ov C^{5_15_2}}=\fc{1}{(S+\ov S)^{1/2}(T^3+\ov T^3)^{1/2}}\ 
\fc{1}{(U^1+\ov U^1)^{1/2}\ (U^2+\ov U^2)^{1/2}}\ .}

\paragraph{D3-- and D7--branes}
\ \\
\ \\
In the following we shall first discuss \tb orbifolds \req{orbi}
with the orientifold action \req{Action}. The latter
allows for D3-- and D7--branes. Depending on which two--torus they are transversal to
there may be three kinds of D7--brane stacks. In the following $D7_j$ denotes 
a stack of D7--branes transversal to the $j$--th subtorus $\mbb T^2_j$.
There are the two classes of open string moduli \req{openM} and \req{openMM}:
The first class \req{openM} comprises the complex fields (\cff \eqq \req{aC} for their
definitions):
\eqn{mods1}{
\begin{array}{lcl}
{\rm positions\ of}\ D3                & C^3_i\ ,& i=1,2,3\ ,\\[2mm]
{\rm positions\ of\ D7}_i              & C^{7_i}_i\ ,& i=1,2,3\ ,\\[2mm]
{\rm Wilson\ lines\ of\ D7}_i           & A_j^{7_i}\ ,& i,j=1,2,3\ ,\ j\neq i\ .
\end{array}}
The second class \req{openMM} is composed by the complex fields:
\eqn{modss1}{
\begin{array}{lcr}
{\rm D3D7-matter}     & C^{37_i}\ ,  & i=1,2,3 \ ,\\[2mm]
{\rm D7D7-matter}     & C^{7_i7_j}\ , & i,j=1,2,3,\ i\neq j\ ,\\[2mm]
{\rm D3D7-matter\ with\ 2-flux}     & C^{37_i}\ ,  & i=1,2,3\ ,\\[2mm]
{\rm D7D7-matter\ with\ 2-flux}     & C^{7_i7_j}\ ,& i,j=1,2,3\ ,\ i\neq j\ .
\end{array}}
The open string fields $C^{35_i}$ and $C^{7_i7_j}$ correspond to $1/2$
BPS matter, while in the case, that there is a world--volume two--form
flux $f^j$ on the D7--branes, the fields $C^{7_i7_j}$ correspond to $1/4$
BPS matter.
A detailed analysis of the amplitudes \req{MMCC} and \req{CCCC} gives
the following results for the open string moduli metrics (
in terms of the holomorphic closed string moduli) \cite{lrs04}:
\eqn{metricsfield1}{\ba{lcl}
\ds{\kappa_4^{2}\ G_{C^3_i\ov C^3_i}=}&\ds{\fc{1}{(U^i+\ov U^i)\ (T^i+\ov T^i)}\ ,}
&\ds{i=1,2,3\ ,}\\[6mm]
\ds{\kappa_4^{2}\ G_{C^{7_i}_i\ov C^{7_i}_i}=}& \ds{\fc{1}{U^i+\ov
    U^i}\ \fc{1}{S+\ov S}\ |1-\tilde{f}^j\tilde{f}^k|\ ,} &
\ds{i,j,k=1,2,3\ ,\ j\neq i\neq k\ ,}\\[6mm]
\ds{\kappa_4^{2}\ G_{A^{7_i}_j\ov A^{7_i}_j}=}& \ds{\fc{1}{U^j+\ov
    U^j}\ \fc{1}{T^k+\ov T^k}\ ,} &
\ds{i,j,k=1,2,3\ ,\ j\neq i\neq k\ .}
\ea}
The non--diagonal matrix elements vanish for $\mbb T^6$ a direct
product of three two--tori:
\eqn{vanish1}{G_{A_i\ov A_j}=0\ \ \ ,\ \ \ G_{C_i\ov C_j}=0\ \ \ ,\ \ \
  \i\neq j\ .}
The above metrics agree with the results from a dimensional reduction
of the Born--Infeld action (\cff subsection \req{DD33DD77}).
Furthermore, for vanishing two--form flux these metrics have been also derived in 
\cite{Rigolin}.

Let us now turn to the matter field metrics for the fields \req{modss1}, \ie
matter fields originating from open strings stretched between
a D3--brane and a D7--brane. 
Expressed by the holomorphic closed string moduli \req{defmod}
the metric \req{twisted1/4} becomes
\eqn{Twisted1/4b}{
\kappa_4^{2}\, G_{C_\varphi\ov C_\varphi}= (S+\ov
S)^{-\fc{1}{4}+\fc{3\beta}{2}+\gamma}\ \prod_{j=1}^3 
(T^j+\ov
T^j)^{-\fc{1}{4}-\fc{\beta}{2}-\gamma(1+\fc{\varphi^j}{\pi})}\  
(U^j+\ov U^j)^{-\fc{\varphi^j}{\pi}}
\sqrt{\fc{\Gamma(\varphi^j/\pi)}{\Gamma(1-\varphi^j/\pi)}}\ ,}
while the metric \req{twisted1/2} takes the form:
\bea
\kappa_4^{2}\, G_{C_\varphi\ov C_\varphi}&=&(S+\ov
S)^{-\fc{\sigma+1}{4}}\ 
(T^3+\ov T^3)^{-\fc{\sigma+1}{4}}\nnn\\
&\times&\prod_{j=1}^2 (T^j+\ov
T^j)^{\fc{\sigma-1}{4}}\  (U^j+\ov U^j)^{-\fc{\varphi^j}{\pi}}
\sqrt{\fc{\Gamma(\varphi^j/\pi)}{\Gamma(1-\varphi^j/\pi)}}\ \label{Twisted1/2b}.
\eea
Equipped with these results we may derive the matter field metrics for the
fields \req{modss1}.
First, we consider a D7--brane, transversal to the third two--torus $\mbb T^2_3$ and a 
(space--time filling) D3--brane. This system, which is $1/2$ BPS, 
preserves ${\cal N}=2$ space--time supersymmetry and we shall apply
\req{Twisted1/2b}. With respect to  the third torus, open strings
have Dirichlet boundary conditions, \ie $\sigma=-1$. On the other hand, in the dual \ta picture, 
the two branes intersect at the angles $\pi/2$ within the other two internal planes, 
\ie $\varphi_{ab}^j=\h\ ,\ j=1,2$ and we obtain:
\eqn{threeseven}{
\kappa_4^{2}\ G_{C^{37_3}\ov C^{37_3}}=\fc{1}{(T^1+\ov T^1)^{1/2}(T^2+\ov T^2)^{1/2}}\ 
\fc{1}{(U^1+\ov U^1)^{1/2}\ (U^2+\ov U^2)^{1/2}}\ .}
Furthermore, for two D7--branes, with one transversal to the torus $\mbb T^{2}_1$ and the other
transversal to the torus $\mbb T^2_2$ the open string coordinates have
pure Neumann boundary conditions w.r.t the third plane $\mbb T^2_3$, \ie $\sigma=1$.
Again, in the dual \ta picture, 
the two branes intersect at the angles $\pi/2$ within the other two internal planes, 
\ie $\varphi_{ab}^j=\h\ ,\ j=1,2$. Hence, from \eqq \req{Twisted1/2b} we deduce:
\eqn{sevenseven}{
\kappa_4^{2}\ G_{C^{7_17_2}\ov C^{7_17_2}}=\fc{1}{(S+\ov S)^{1/2}(T^3+\ov T^3)^{1/2}}\  
\fc{1}{(U^1+\ov U^1)^{1/2}\ (U^2+\ov U^2)^{1/2}}\ .}
Finally, the generalization of \req{threeseven} to the case, that
there is internal two--form flux (\cff \eqq \req{fluxyuk})
\eqn{fluxangle}{
\varphi^j=\arctan\lf(\fc{f^j}{t_2^j}\ri)\ \ \ ,\ \ \ j=1,2} 
on the D7--brane world--volume, may be obtained from \req{Twisted1/2b}:
\eqn{threeseven}{
\kappa_4^{2}\ G_{C^{37_3}\ov C^{37_3}}=\fc{1}{(T^1+\ov T^1)^{1/2}(T^2+\ov T^2)^{1/2}}\ 
\fc{1}{(U^1+\ov U^1)^{\varphi^1}\ (U^2+\ov U^2)^{\varphi^2}}\ .}

To conclude this subsection we refer the reader to
\cite{progress} for a more detailed and complete account on matter
field metrics.

\subsection{Yukawa couplings in type II orientifolds}\label{YUK}

The structure of Yukawa couplings in realistic string models is of particular interest, since 
it gives rise to non--trivial predictions on the couplings between the Higgs field 
and the SM fermions of the MSSM. In particular, these couplings have to correctly 
describe the observed masses and mixings of quarks and leptons. Hence a hierarchical structure
of different Yukawa couplings for the different generations has to emerge.
These properties are natural in heterotic orbifold compactifications with 
chiral matter arising from different fixpoints 
\cite{dfms87,Hamidi:1986vh,Erler:1992gt,Stieberger:1992bj,Casas:1991ac,
Stieberger:1992vb,Kobayashi:2003vi}.
In those compactifications an exponentially suppressed hierarchical structure 
emerges from the different locations of  fix--points, where chiral matter fields
from the twisted sectors are located.

In this subsection we shall review Yukawa couplings of chiral matter in type II
orientifolds \cite{Gava:1997jt,Antoniadis:2000jv,cim03,cp03,ao03,cim04}.
In the following we consider type IIB orientifolds with magnetized D9-- or D7--branes 
and IIA orientifolds with intersecting D6--branes. The chiral matter fields in the bifundamental
representation arise from open strings
stretched between different stacks of D--branes. Each of them generically comprises an $U(N)$ gauge
group. Non--trivial Yukawa couplings appear from a set of three different stacks of D--branes
allowing for three different types of bifundamental matter fields interacting.
In type IIA the Yukawa couplings $Y_{ijk}$ describe the interaction
of three chiral matter fields labeled by $i,j$ and $k$. 
The latter  originate from open strings stretched between 
two intersecting D6--branes. The coupling of three matter fields
$i,j$ and $k$ has been computed in \cite{cim03,cp03,ao03}:
\bea
Y^{\rm IIA}_{ijk}&=&e^{\Phi_4/2}\ \prod_{r=1}^3\sigma^{(r)}_{abc}\  (t_2^r)^{1/4}\ \lf(\ 
\fc{\Gamma(1-\fc{1}{\pi}\ \varphi_{ab}^{r})\ 
\Gamma(1-\fc{1}{\pi}\ \varphi_{ca}^{r})\ \Gamma(1-\fc{1}{\pi}\ \varphi_{bc}^{r})}{(2\pi)^3\ 
\Gamma(\fc{1}{\pi}\ \varphi_{ab}^{r})
\ \Gamma(\fc{1}{\pi}\ \varphi_{ca}^{r})\ \Gamma(\fc{1}{\pi}\ \varphi_{bc}^{r})}\ \ri)^{1/4}
\nnn\\
&\times&\vartheta\lf[{\delta_{ijk}^{r}\atop 0}\ri]
\lf(0;t^{r}\ I_{ab}^{r}I_{bc}^{r}I_{ca}^{r}\ri)\ ,\label{YUKIIA}
\eea
with $\varphi_{bc}^{r}=\pi-\varphi_{ab}^{r}-\varphi_{ca}^{r}$.
Here, the angles $\varphi_{ab}^{r}$ with respect to  the $r$--th subtorus $\mbb T^{2}_r$
are the relative angles of two intersecting stacks $a,b$.
The latter comprise one (twisted) matter field contributing in the Yukawa coupling.
The intersection numbers $I_{ab}^{r}$ with respect to  the $r$--th subtorus $\mbb T^{2}_r$ have been
introduced in \req{int1}. The latter may be expressed in terms of the relative angles 
$\varphi_{ab}^{r}$.
Furthermore, we have introduced the quantities:
\eqn{flvY}{\sigma^{(r)}_{abc}={\rm sign}(I^{r}_{ab}I^{r}_{bc}I^{r}_{ca})\ \ \ ,\ \ \ 
\delta_{ijk}^{r}=\fc{i^{r}}{I_{ab}^{r}}+\fc{j^{r}}{I_{ca}^{r}}+
\fc{k^{r}}{I_{bc}^{r}}\ .}
Strictly speaking, the expression \req{YUKIIA} is valid for $\sigma^{(r)}_{abc}=1$.
In the case $\sigma^{(r)}_{abc}=-1$, the K\"ahler modulus $t^r$ has to be replaced by $\ov t^r$.
The only flavor dependence of the Yukawa couplings \req{YUKIIA} 
enters through $\delta_{ijk}^{r}$. The Jacobi function $\vartheta$ 
accounts for world--sheet disk instantons originating
from maps of the open string disk world-sheet into the target space.
In \eqq \req{YUKIIA}
the product of $\Gamma$ functions is the quantum part of the three point function of three twisted matter fields.
It arises form a correlator of three twist fields \req{twistedMAT}
inserted at the boundary of the disk. 
This factor has been computed in \cite{cp03}. Since all vertex operators
are inserted at the boundary of the disk, the quantum part is just the square root of
an identical closed string computation performed in heterotic orbifold compactifications 
\cite{Erler:1992gt,Stieberger:1992bj}.

On the other hand, in type IIB with 
magnetized D9--branes, which are wrapped around the three 
subtori $\mbb T^{2}_r\ ,\ r=1,\ldots,3$, with the geometric moduli, 
introduced in \req{cplstr} and \req{Kahmod}, one finds\footnote{\label{footn}The product of $\Gamma$ functions 
in \req{YUKIIB} has the following power series expansion:
$\fc{\Gamma(1-\fc{1}{\pi}\ \varphi_{ab}^{r})\ 
\Gamma(1-\fc{1}{\pi}\ \varphi_{ca}^{r})\ \Gamma(\fc{1}{\pi}\ \varphi_{ab}^{r}+\fc{1}{\pi}\ \varphi_{ca}^{r})}
{\Gamma(\fc{1}{\pi}\ \varphi_{ab}^{r})
\ \Gamma(\fc{1}{\pi}\ \varphi_{ca}^{r})\ \Gamma(1-\fc{1}{\pi}\ \varphi_{ab}^{r}-\fc{1}{\pi}\ \varphi_{ca}^{r})}
=\fc{1}{\pi}\ 
\fc{\varphi_{ab}^r\varphi_{ca}^r}{\varphi_{ab}^r+\varphi_{ca}^r}-\fc{2}{\pi^4}\ \zeta(3)\ 
(\varphi_{ab}^r\varphi_{ca}^r)^2+\Oc(\varphi^6)$. In fact, in \cite{cim04} only the first term of this expansion
has been derived. In this reference the Yukawa couplings are computed in field theory
as overlap wave function integrals of two fermions and one complex scalar over the compact  dimensions.}
the Yukawa coupling \cite{Gava:1997jt,cim04}:
\bea
Y^{\rm IIB}_{ijk}&=&e^{\Phi_4/2}\ \prod_{r=1}^3\sigma^{(r)}_{abc}\  (u_2^r)^{1/4}\ \lf(\ 
\fc{\Gamma(1-\fc{1}{\pi}\ \varphi_{ab}^{r})\ 
\Gamma(1-\fc{1}{\pi}\ \varphi_{ca}^{r})\ \Gamma(1-\fc{1}{\pi}\ \varphi_{bc}^{r})}{(2\pi)^3\ 
\Gamma(\fc{1}{\pi}\ \varphi_{ab}^{r})
\ \Gamma(\fc{1}{\pi}\ \varphi_{ca}^{r})\ \Gamma(\fc{1}{\pi}\ \varphi_{bc}^{r})}\ \ri)^{1/4}
\nnn\\
&\times&\vartheta\lf[{\delta_{ijk}^{r}\atop 0}\ri]
\lf(0;u^{r}\ I_{ab}^{r}I_{bc}^{r}I_{ca}^{r}\ri)\ .\label{YUKIIB}
\eea
The intersection numbers $I_{ab}^{r}$ with respect to  the torus $\mbb T^{2}_r$
have been introduced in \req{int1}. They 
measure the (relative) magnetic flux $f^{r}$ in the $r$--th subtorus 
felt by the open string, which is stretched between stack $a$ and $b$. More precisely
\eqn{fluxyuk1}{
\ba{lcl}
\ds{I_{ab}^{r}}&=&\ds{p^{r}_a\ q^{r}_b-q^{r}_a\ p^{r}_b=\fc{1}{\ap}\ 
p^{r}_a\ p^{r}_b\ (\ f_b^{r}-f_a^{r}\ )\ ,}\\[3mm]
\ds{\varphi^{r}_{ab}}&=&\ds{\arctan\lf(\fc{f_b^{r}}{t_2^r}\ri)-
\arctan\lf(\fc{f_a^{r}}{t_2^r}\ri)\ ,}
\ea}
with the two--form fluxes $f^{r}_a,\ f^{r}_b $ on stack $a$ and $b$, respectively. 
The latter are quantized, \ie $f^{r}=\ap\fc{q_a}{p_a}$.
The two expression \req{YUKIIA} and \req{YUKIIB} are related by the $T$--duality \req{TonUT}.

According to \req{CCCC} the 
holomorphic trilinear couplings $W_{ijk}$, which appear in the superpotential \req{mostW}, 
give rise to the physical Yukawa couplings:
\eqn{physYUK}{
Y_{ijk}=(G_{ab}\ G_{bc}\ G_{ca})^{-1/2}\ e^{\kappa_4^2K_0/2}\ W_{ijk}\ .}
With the matter field metrics \req{Twisted1/4} and \req{Twisted1/4b}, we derive:
\bea
W^{\rm IIA}_{ijk}&=&\prod_{r=1}^3\ \vartheta\lf[{\delta_{ijk}^{r}\atop 0}\ri]
\lf(0;t^{r}\ I_{ab}^{r}I_{bc}^{r}I_{ca}^{r}\ri)\ ,\nnn\\
W^{\rm IIB}_{ijk}&=&\prod_{r=1}^3\ \vartheta\lf[{\delta_{ijk}^{r}\atop 0}\ri]
\lf(0;u^{r}\ I_{ab}^{r}I_{bc}^{r}I_{ca}^{r}\ri)\ .\label{trilinear}
\eea
The Yukawas and trilinear couplings between untwisted 
matter fields may be found in \cite{lmrs04}.

To conclude this subsection we want to  add a few remarks.
The only flavor dependence of the Yukawa couplings \req{YUKIIA} and \req{YUKIIB} 
enters through the parameters $\delta_{ijk}^{r}$, defined in \req{flvY}.
The flavor structure of Yukawa couplings in type II orientifolds allowing for a 
hierarchical structure
of different Yukawa couplings for the different generations has been investigated in 
\cite{Abel:2003yh,Kitazawa:2004nf,Higaki:2005ie}.
The Yukawa couplings $Y_{ijk}$ allow to study the proton decay in type II orientifolds.
This has been investigated in \cite{kw03,Cvetic:2006iz}. With respect to dimension six operators
the string proton decay rate is enhanced by a factor of $\alpha_{GUT}^{1/3}$ relative to 
the corresponding result in four--dimensional GUTs.
Finally, the computation of Yukawa couplings from intersecting D--branes in CY spaces
is pioneered by matrix factorization
in  \cite{Herbst:2006nn,Herbst:2004jp,Brunner:2004mt}.
Other work on disk scattering of open strings in type II orientifolds with D--branes
includes the scattering  of several matter fields \cite{ao03a}
and multi--gluon scattering \cite{Oprisa:2005wu,Stieberger:2006te}.

\subsection{One--loop gauge corrections in type II orientifolds}
\subsubsection{General aspects of one--loop gauge threshold corrections}

In this subsection we turn to the computation of  one--loop gauge
threshold corrections in Type II orientifolds with D--branes.
Unlike what happens \eg in perturbative heterotic string vacua, the tree--level gauge 
couplings for the various gauge groups, arising from different stacks
of D--branes, are generically
not  the same\footnote{See however the special construction in  \cite{bls03}.} 
at the string scale.
For example  in Type IIA they follow from dimensional reducing the Born--Infeld action of 
a D6--brane on a   three--cycle of the internal manifold 
${\cal X}$ and are given in \eqq \req{gcpl}.
Hence a priori there is no unification of gauge couplings at the string scale 
(at string tree--level) as for each gauge factor we may have a
different three--cycle
with different volume. Similar conclusions are drawn in type IIB. For example  in type IIB
with D3/D7--branes the gauge couplings
are given by the volumina of different four--cycles \cff \req{gaugef}.
One--loop gauge threshold corrections $\Delta_a$ (to the gauge group $G_a$), which take 
into account Kaluza--Klein and winding states from the internal dimensions and 
the heavy string modes, may change this picture  \cite{KaplunovskyRP}. 

\begin{figure}[h!]
  \begin{center}
    \includegraphics[width=80mm]{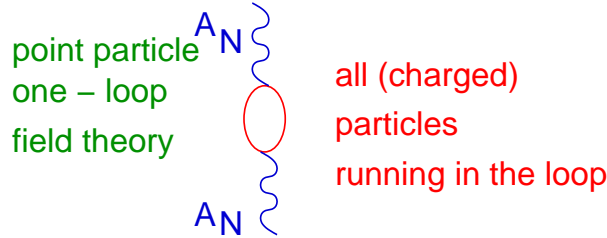}
    \caption{One--loop gauge threshold effect in field theory}
    \label{oneloopFT}
  \end{center}
\end{figure}

Generically the latter depend on the moduli fields $\phi$ and for certain regions in the moduli 
space these corrections may become huge and thus may have a substantial 
impact\footnote{This effect has been thoroughly investigated for heterotic ${\cal N}=1$ string vacua
in  \cite{IbanezHC,HPN,NEST}.} 
on the unification scale.
One--loop gauge corrections  are very important quantities to probe the 
low--energy physics below the string scale as they change the running of the gauge couplings 
for scales $\mu$ below the string scale
according to the Georgi, Quinn and Weinberg evolution equations of ordinary field theories:
\eqn{weinberg}{
\fc{1}{g^2_a(\mu)}=\fc{1}{g_{a,\ \rm tree}^2}+\fc{b_a}{16\pi^2}\ 
\ln\fc{M_{\rm string}^2}{\mu^2}+\fc{1}{16\pi^2}\Delta_a\ .}
Here $g_{a, {\rm tree}}$ is the bare tree--level gauge coupling of the gauge group $G_a$.
In string theory this coupling is related to the string coupling $g_{\rm string}$.
In field theory with charged point particles the quantity $\Delta_a$ is determined by
computing the respective Feynman loop-diagram, depicted in Figure \ref{oneloopFT}.  

\begin{figure}[h!]
  \begin{center}
    \includegraphics[width=160mm]{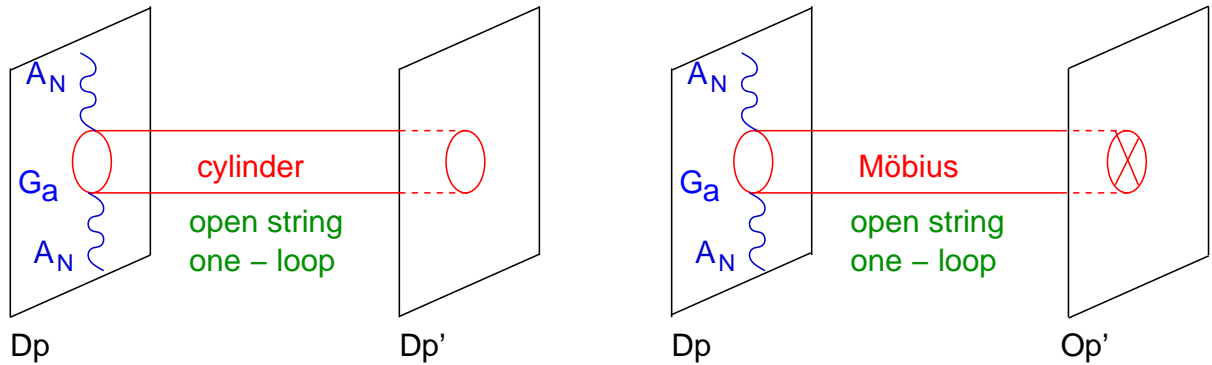}
    \caption{One--loop gauge threshold effects in string theories with D--branes.}
    \label{oneloopSTTH}
  \end{center}
\end{figure}

On the other hand, in superstring theory with D--branes the corresponding one--loop
gauge threshold effect $\Delta_a$ arises from a cylinder diagram with its
two boundaries attached to D$p$-- and D$p'$--branes. On one of them the two gauge vectors 
$A_M$ of the gauge group $G_a$ under consideration couple to the closed loop 
(for anomaly free gauge group), \cff Figure \ref{oneloopSTTH}. 
Hence one of the D--branes must carry the gauge group $G_a$ under consideration.
In addition, there is a M\"obius diagram starting and ending
at the same D--brane. In other words, the M\"obius diagram may be understood as a cylinder
with one boundary attached to the D$p$--brane with gauge group $G_a$, 
where the two gauge fields couple and
the second boundary representing a crosscap attached to an orientifold O--plane
(\cff Figure \ref{oneloopSTTH}).
On the other hand, since neither the Klein bottle nor the 
torus worldsheet have boundaries, where gauge fields from the branes could couple, these
diagrams do not contribute to one--loop gauge threshold corrections.

In a generic string background  with ${\cal N}=2$ SUSY in $D=4$ the gauge kinetic terms
are 1/2 BPS protected, which means that only 1/2 BPS states contribute in the one--loop diagram.
Furthermore, in that case the gauge kinetic function does not receive any further corrections
beyond one--loop. Though this is a special case, since only a subset of the charged states of 
the spectrum contributes in the loop, it is a more generic property, which also holds
in theories with ${\cal N}=1$ SUSY in $D=4$.

The gauge threshold corrections $\Delta_a$ for a gauge group $G_a$ 
are given by a (weighted) sum over the masses of
charged particles running in the loop. 
This effect may be written as a Schwinger type integral
\eqn{Schwinger}{\Delta_a=\int_\epsilon^\infty\ \fc{dt}{t}\ \Tr'\ (-1)^F\ F\ e^{-t\ M^2}\ ,}
with some UV--regularization $\epsilon$.
The masses $M$ depend on the geometry of the underlying
compactification manifold ${\cal X}$. Indeed, in a more geometric language, 
the gauge threshold corrections $\Delta_a$ are determined by a topological  invariant of
the compactification manifold ${\cal X}$, namely the analytic Ray--Singer 
torsion of the  manifold ${\cal X}$ \cite{RaySinger}.
The latter is  computed explicitly based on topological arguments, see
 \cite{BCOV,FKLZ,MS1} for examples in heterotic string theory or 
 \cite{fw02} in heterotic M--theory.
As an example, let us discuss type I superstring theory compactified 
on $K3\times \mbb T^2$ with D9--branes comprising the gauge group $G_a$
(in addition for consistency, there is a set of D5--branes). 
In that case the gauge kinetic term is a 1/2 BPS protected quantity
depending on the vectormultiplet moduli only (\cff subsection \ref{DD99DD55} for a detailed
account on that moduli space).
For the background $K3\times \mbb T^2$ we shall evaluate the integral \req{Schwinger},
which boils down to a sum over all Kaluza--Klein masses with respect to  the torus $\mbb T^2$. 
The latter has the K\"ahler modulus $t$, the complex structure $u$
and non--trivial Wilson lines  $a_1,a_2$. In terms of the latter the  
Kaluza--Klein BPS--masses of the open strings are given by:  
$$\ap\ M_{KK}^2=\fc{1}{t_2 u_2}\ |m_1+a_1+u\ (m_2+a_2)|^2\ ,$$ 
with the integers $m_1,m_2$. Therefore, \req{Schwinger} becomes:
\bea
\Delta_a &=&\int^\infty_0\ \fc{dt}{t}\ \lf\{\ \sum_{(m_1,m_2)}\ 
e^{-\fc{\pi t}{t_2 u_2}|m_1+a_1+u(m_2+a_2)|^2}
-\fc{t_2}{t}\ri\}\nnn\\
&=&t_2\ \int^\infty_0\ \fc{dt}{t^2}\ \sum_{(n_1,n_2)\neq (0,0)}\  
e^{-\fc{\pi t_2}{tu_2}|n_1+un_2|^2}\ 
e^{2\pi i (a_1n_2+a_2n_1)}\label{joint}\ .
\eea
The second term in the first line of \req{joint} is a regularization to remove
the state $(m_1,m_2)=(0,0)$.
The above integral may be evaluated following the techniques presented in \cite{kristin}, 
with the result \cite{JOINT}:
\eqn{totalRS}{
\Delta_a=-\ln\lf|e^{\pi i a_2^2 u}\ \fc{\th\lf[1/2\atop1/2\ri](a_1-u\ a_2,u)}{\eta(u)}\ri|^2\ .}
The $t$--dependence
has dropped out as a result of scale invariance of the integrand \req{joint}.
Indeed, the result \req{totalRS} is the analytic torsion of the genus one torus 
$\mbb T^2$ endowed with the non--trivial background $a_1,a_2$ \cite{RaySinger}.

In heterotic ${\cal N}=1$ string compactifications, the one--loop threshold correction $\Delta_a$ 
(more precisely the difference of two corrections $\Delta_a$ and $\Delta_{a'}$ for two
different gauge groups $G_a$ and $G_{a'}$) is related to the closed string topological partition 
function $F_1$ at genus one \cite{BCOV1}. 
Similarly, in type II with D--branes this correction is
given by the open string topological partition function $F_{0,2}$, \ie 
the open string topological partition function on an annulus \cite{BCOV}.
Hence \req{totalRS} is related to the open string topological partition function $F_{0,2}$ 
\cite{JOINT}.

\subsubsection{Gauge threshold corrections in type II orientifolds with D--branes}

Let us now turn to the concrete results for one--loop threshold corrections in
type II orientifolds with D--branes.
One--loop threshold corrections $\Delta_a$ for type IIB orientifolds with parallel
D--branes have been computed in \cite{fabre,seealso,APT,ABD},  while for
type IIA orientifolds with D--branes at angles or
for type IIB orientifolds with world volume two--form flux  these
corrections have been pioneered in \cite{ls03}.
The latter results have been generalized for non--diagonal two--form flux in 
\cite{Bianchi:2005sa}.

The relevant amplitudes to compute at string one--loop are the cylinder 
and the M\"obius amplitude (Figure \ref{oneloopSTTH}).
In type II orientifolds with ${\cal N}=1$ spacetime SUSY in $D=4$ the open strings
between a D$p$-- and D$p'$ or between a  D$p$-- and one O--plane may preserve
$1,1/2$ or $1/4$ of the original 16 supercharges of a type II orientifold in $D=10$, thus 
giving rise to a ${\cal N}=4,2$ or ${\cal N}=1$ SUSY sector contributing to the gauge threshold corrections,
respectively.
Since for ${\cal N}=4$ in $D=4$ gauge couplings are not renormalized at one--loop only the 
$1/2$ and $1/4$ BPS sectors give rise to a non--vanishing contribution 
to the one--loop gauge threshold corrections in $D=4$. 

In IIA orientifold with intersecting D6--branes and O6--planes, the  BPS sectors are 
classified according to the three relative angles $\varphi^j_{ba}=\varphi^j_b-\varphi^j_a,
\ j=1,2,3$ between two stacks of D6--branes $a$ and $b$ or between one
stack of a  D6--brane and an 
O6--plane. Here $j$ is the $j$--th internal complex direction
(\cff subsection \ref{sec23} for more details). For the case $\varphi_{ba}^j=0\ ,j=1,2,3$, 
\ie two parallel branes or one brane is parallel to an O6--plane all 16 supercharges are 
preserved. In that case the corresponding cylinder or M\"obius diagram does not give rise to
any one--loop threshold effect (${\cal N}=4$ sector).

On the other hand,  the ${\cal N}=1$ case $\varphi^1_{ba}+\varphi^2_{ba}+\varphi^3_{ba}=0\ {\rm mod}\ 2\pi$
gives a $1/4$ BPS contribution to the threshold corrections. It has been computed in
\cite{ls03} and gives:
\eqn{altertiinew}{
\Delta^{{\rm IIA},{\cal N}=1}_{ab}=-b_{ab}^{N=1}\  \ln \fc{\Gamma(1-\fc{1}{\pi}\ \varphi_{ba}^1)\ 
\Gamma(1-\fc{1}{\pi}\ \varphi_{ba}^2)\ \Gamma(1+\fc{1}{\pi}\ \varphi_{ba}^1+
\fc{1}{\pi}\ \varphi_{ba}^2)}{\Gamma(1+\fc{1}{\pi}\ \varphi_{ba}^1)\ 
\Gamma(1+\fc{1}{\pi}\ \varphi_{ba}^2)\ \Gamma(1-\fc{1}{\pi}\ \varphi_{ba}^1-
\fc{1}{\pi}\ \varphi_{ba}^2)}  \ .}
Here, the index $a$ refers to the D6--brane stack of the gauge group $G_a$ under consideration and 
the index $b$ accounts for any other stack of D6--branes or orientifold planes, with
the relative angles obeying $\varphi^1_{ba}+\varphi^2_{ba}+\varphi^3_{ba}=0\ {\rm mod}\ 2\pi$.
To obtain all $1/4$ contributions to the gauge threshold correction $\Delta_a$, one
has to sum over all those stacks $b$ under question.
The factor $b_{ab}$ is the corresponding beta--function coefficient accounting for
all charged open strings stretched between stack $a$ and $b$ and contributing in the 
corresponding cylinder or M\"obius diagram.
Note, that through the angles $\varphi_{ba}^j$ (\cff \eqq \req{tanphi}), the whole 
expression\footnote{It would be interesting to see, whether the expression \req{altertiinew} 
may be also anticipated from the Ray--Singer torsion of the underlying orientifold 
compactification ${\cal X}$, as we have demonstrated for the 
${\cal X}=K3\times \mbb T^2$ orientifold in the previous subsection. 
Interestingly in the corresponding M--theory setup the gauge 
threshold corrections derive from the Ray--Singer torsion \cite{fw02}.  Furthermore, 
it would be interesting, to see whether
\req{altertiinew} may be related to an open string topological partition function $F_{0,2}$.} 
\req{altertiinew} depends on the complex structure moduli.

Furthermore, the ${\cal N}=2$ case occurs, if $\varphi^k_{ba}=0$ for one $k$
and $\varphi^i_{ba}+\varphi_{ba}^j=0\ {\rm mod}\ 2\pi$ for the remaining two relative angles. That 
case gives rise to a 1/2 BPS contribution to the one--loop  threshold corrections.  
This contribution may be phrased as \req{joint}, however with a modified mass
contributing to the one--loop correction \req{Schwinger}.
More precisely, with respect to  the complex plane with $\varphi^k_{ba}=0$, 
the open strings stretched between the two stacks $a$ and $b$, which are parallel in this
subplane and each of them carries the wrapping numbers $(p_a^k,q_a^k)$, 
have non--vanishing Kaluza--Klein  momenta and windings. 
Their mass is given by the mass formula of open strings 
stretched between two parallel D1--branes, which are wrapped around the torus $T_2^k$
with wrapping numbers $(p_a^k,q_a^k)$, \cff \eqq \req{2192}:
\eqn{D1mass}{
\ap\ M_{KK}^2=\fc{u_2^k}{t_2^k}\ \fc{|n_k+m_kt^k|^2}{|p_a^k+q_a^ku^k|^2}\ .}
This mass enters the formula \req{Schwinger} and we compute \cite{ls03}:
\bea
\Delta_{ab}^{{\rm IIA},\rm N=2}&=&b^{\rm N=2}_{ab}\ \fc{|p_a^k+q_a^ku^k|^2}{u_2^k}\ 
\int_0^\infty\  \fc{dt}{t^2}\ 
\sum_{(n_k,m_k)\neq (0,0)}e^{-\fc{\pi}{t}\ 
\fc{|p_a^k+q_a^ku^k|^2}{t_2^ku_2^k}\ |m_k+n_kt^k|^2}\nnn\\
&=&-b_{ab}^{N=2}\ \lf[\ln t^k_2|\eta(t^k)|^4+\ln \fc{|p_a^k+q_a^ku^k|^2}{u_2^k}-\kappa\ri]\ ,
\label{thresholds}\eea
with the IR regularization constant $\kappa=\gamma_E-\ln(4\pi)$. The last two terms
of \req{thresholds} are effects of an IR--regularization \cite{kristin}.
Again to obtain all $1/2$ contributions to the gauge threshold correction $\Delta_a$, one
has to sum over all those stacks $b$, which are parallel to stack $a$ with respect to  one complex
plane. In the following, let us discuss $T$--duality on the result \req{thresholds}.
A $T$--duality along the $\vec {\bf e}_1$--axis of all two--tori $\mbb T^{2}_i$
results in the interchange $t^i\leftrightarrow u^i$ 
and the set of O6--planes is converted to a set of a O3/O7--planes.
The one--loop correction \req{thresholds} becomes:
\eqn{examp3}{
\Delta^{\rm IIB,\rm N=2}_{ab}=-b_{ab}^{N=2}\ \lf[\ln u^k_2|\eta(u^k)|^4+\ln \fc{|p_a^k+q_a^kt^k|^2}{t_2^k}
-\kappa\ri]\ .}
On the other hand, a $T$--duality along $\vec {\bf e}_2$--axis of all two--tori $\mbb T^{2}_i$
results in the interchange $t^i\leftrightarrow -1/u^i$ and $u^i\leftrightarrow -1/t^i$. 
Thereby the set of O6--planes is converted to a set of a O9/O5--planes
and \req{thresholds} becomes:
\eqn{examp2}{
\Delta^{\rm IIB,\rm N=2}_{ab}=-b_{ab}^{N=2}\ \lf[\ln u^k_2|\eta(u^k)|^4+\ln \fc{|p_a^kt^k-q_a^k|^2}{t_2^k}
-\kappa\ri]\ .}
Hence \eqq \req{examp3} is the type IIB analog of \eqq 
\req{thresholds} for the case of  D3/D7-- or D7/D7--branes with two--form fluxes
and \req{examp2} corresponds to the 
D9/D5 case. This case has been studied in great detail in \cite{fabre,ABD}.
Recall that in the  equations \req{examp3} and \req{examp2}  the real part of $t^j$ 
is given in terms of the NS
background field $b^j$, which gives rise to world--volume two--form flux in type IIB,
\cff subsection \ref{secinttor}.

In type IIA with intersecting D6-branes the threshold
correction \req{altertiinew} from the ${\cal N}=1$ sector depend on the
homology classes of the three--cycles 
(open string parameters) and also on the closed string geometrical moduli. 
In particular, these functions depend on the complex  structure moduli, 
while in IIB they are functions on the K\"ahler moduli.
The result \req{altertiinew} may be easily translated to type IIB orientifolds 
with two--form fluxes by using the relation \req{Fphi}.
In the equivalent $T$-dual picture the threshold corrections will be a function of the 
open string magnetic gauge fluxes and of the geometrical moduli of the
dual compact space, \ie they depend on the K\"ahler moduli.

The type IIA result \req{thresholds} and the type IIB results \req{examp3} and \req{examp2} 
are reminiscent of the heterotic threshold results in
toroidal orbifold compactification, where only ${\cal N}=2$ orbifold
sectors give rise to a moduli
dependent contribution \cite{DKL2}. On the other hand, in heterotic orbifold compactifications
the contributions from ${\cal N}=1$ sectors give rise to a non--moduli dependent constant.

Since the gauge fluxes are directly related to the non-commutativity parameters
of the internal torus, we obtain in this way some interesting, new informations
for one--loop threshold corrections on non-commutative tori
in string theory, a discussion which extends recent results
on one--loop corrections on compact non-commutative spaces in the literature.

To conclude this subsection, let us mention, that 
other research on one--loop string corrections in D--brane models 
investigates the K\"ahler potential and the superpotential, see 
\cite{Berg:2004ek,Berg:2005ja} and \cite{Abel:2004ue,Abel:2005qn}
and Fayet--Iliopoulos  terms \cite{Lawrence:2004sm,Anastasopoulos:2006cz}.
Furthermore in  \cite{Bianchi:2006nf}, 
four--gluon scattering at one--loop has been investigated. 
Finally, results for string amplitudes involving a world--sheet with one handle and one boundary
have been pioneered in \cite{Antoniadis:2005sd,Antoniadis:2004qn,Antoniadis:2005xa}.

\clearpage
\setcounter{equation}{0}


\section{FLUX COMPACTIFICATIONS}
\label{secfluxvacua}

All the string compactifications discussed so far, though showing
some of the salient features of the SM physics, suffer from two
major shortcomings. Their common origin is the existence of, in
general, a large number, of the order of $100$,  massless scalar fields.
These so-called string moduli arise from the possible marginal
deformations of the string background,
variations of the size of the total internal manifold or submanifolds thereof do not cost
any energy. These massless fields would give rise to long distance
interactions, which are not observed in nature. As a second issue,
the parameters in the effective four-dimensional
action depend on the expectation values of these
massless fields, i.e.\ without knowing these string theory cannot be
very predictive.

It was demonstrated that in particular non-perturbative effects like
gaugino condensation or world-sheet and space-time instanton
corrections could in principle lift the deformations and
in four-dimensional language generate a potential for the moduli. This would
freeze the moduli and potentially give rise to a small number of
supersymmetric or non-supersymmetric minima.

More recently, it was realized that there exist solutions to the
tree-level string equations of motions in which not only the
ten-dimensional metric varies non-trivially along the compact
directions, like in Calabi-Yau compactifications,
but also some of the other ten-dimensional fields. These solutions involve in particular
fluxes for the $p$-form tensor fields which have the power to
obstruct flat deformations of Calabi-Yau backgrounds and thus generate a potential. This leads to
the freezing of (at least part of) the moduli. The
induced potential is believed to be under sufficiently good control
to make precise statements about the structure of the set of string
vacua that follows (also called the string landscape).
As we will discuss in section \ref{secstatistic}, contrary to
earlier expectations, the fluxes give rise to a
plethora of stable vacua. By extrapolating this result our
universe seems to be by far less unique than initially hoped for.

The subject of flux compactifications has been developed into a
very broad subject and there exist two   recent review
articles on the subject \cite{grana,Douglas:2006es}.
Flux compactifications started with the heterotic string compactifications
with non-vanishing three-form $H$-flux \cite{Strominger:1986uh}
(see also \cite{deWit:1986xg}).
In many cases the effects of non-vanishing background fluxes can be formulated in terms
of an effective superpotential \cite{gvw99,tv99,Peter,cklt00,gkp01,kst02}.
in the effective
${\cal N}=1$ supergravity action.
Moreover, very often background fluxes in string theory can be explained in supergravity
by gauging isometries of the underlying scalar moduli spaces
respectively in terms of twisted tori compactifications
\cite{FERRARA1,glmw02,kstt02,deWit:2002vt,FERRARA,Angelantonj:2003up,Andrianopoli:2003jf,deWit:2004yr,aft03,Andrianopoli:2003sa,deWit:2003hq,Derendinger:2004kf,Andrianopoli:2004im,lmrs05,Dall'Agata:2005ff,Dall'Agata:2005mj}
along the lines of Scherk -- Schwarz compactifications    \cite{Scherk:1978ta}.

Making contact to the other parts of this review, our goal is to
provide the framework of how fluxes and D-branes can be
consistently coupled together in a single compactification ansatz.
This is necessary to obtain supersymmetric models in which moduli are
stabilized, making the low energy Lagrangian in principle predictable,
and which are still rich enough to contain a realistic open string
particle spectrum of non-abelian gauge symmetries and charged chiral
matter.
Besides preserving space-time supersymmetry, background fluxes can also
be part of breaking supersymmetry in the closed string bulk.
Coupling fluxes to the D-branes, one can calculate effects of
supersymmetry breaking via fluxes on the open string states. In this way soft
supersymmetry breaking terms are generated on the world volumes of
the D-branes. This will be discussed in subsection \ref{SOFT}.
In subsections \ref{FLUXORBI} and \ref{MODSTAB} we discuss how closed string
complex structure
moduli can be stabilized by background fluxes in type IIB orientifolds, where we also
include the effects of a non-perturbative superpotential, which depends on the geometrical
K\"ahler moduli fields. However we do not investigate how open string moduli, related
to the embedding of D-branes into the closed string geometry, can be stabilized
(see e.g. \cite{ADFT,lmrs05,Gomis:2005wc}).

We will restrict ourselves to
some central ideas mostly along the lines of the example of three-form fluxes
in type IIB string theory as discussed in \cite{Michelson:1996pn,gkp01,Dall'Agata:2001zh,kst02,Dall'Agata:2001zh}.
This is the situation that is under best control and has been
developed primarily in view of applications in string model
building. One of the reasons is that the backreaction of the 3-form fluxes is such
that the six-dimensional space is still a warped Calabi-Yau space.
However, type IIA or heterotic flux compactifications are more difficult to analyze.
Here the backreaction of the fluxes is so strong that the underlying geometry is
not any longer Calabi-Yau. We will discuss some aspects of type II and heterotic
flux compactifications in subsections \ref{sseciiaflux} and \ref{HETFLUX}.
Furthermore it is interesting to see how flux compactifications behave under various
geometrical string duality symmetries. For example,
concerning  mirror symmetry part of this discussion can be found
in \cite{Louis:2002ny,glmw02},
and for the heterotic/type IIA string duality in \cite{Curio:2001ae,Louis:2006kb}.
Finally, we will completely omit the subject of non-geometrical fluxes,
which recently received a lot of attention
\cite{Shelton:2005cf,Hull:2006va,Shelton:2006fd}.


\subsection{Fluxes in string compactifications}

In generality, we still only allow a background ansatz which preserves
four-dimensional Poincar\'e invariance or the corresponding anti-de
Sitter (AdS) or de Sitter (dS) symmetries. Fluxes are non-trivial
backgrounds for the anti-symmetric tensor fields strengths, the RR forms in type
II and the generic $H_3$ present in any ten-dimensional
theory. To preserve the symmetries, they can only be along the internal
space, or have to fill out the four-dimensional space-time (for which
the tensor field needs to be at least a four-form).

For a $p$-form potential $C_p$ with field strength $F_{p+1}=dC_p$ the
Bianchi-identity is $dF_{p+1}=0$. Following a argument similar to
Dirac's derivation of charge quantization, one arrives at the flux
quantization conditions. Upon integrating the field strength over
a $(p+1)$-dimensional manifold $\Sigma_{p+1}$ without boundary the
charge has to be integer,
\be\lab{Dirac}
\frac{1}{\ell_s^p} \int_{\Sigma_{p+1}} F_{p+1} \in \mathbb{Z}\  .
\ee
Due to the Bianchi identity, the integral only depends on the
homology class of $\Sigma_{p+1}$, and since $\S_{p+1}$ has no
boundaries, only the cohomology class of $F_{p+1}$ is relevant. Thus,
one speaks of a $(p+1)-$form flux through the $(p+1)-$cycle
$[\S_{p+1}]$, the homological class of $\S_{p+1}$.

This can be applied to the three-form $H_3$ assuming $dH_3=0$.
Choosing the compactification space $\cal{X}$
to be a six-dimensional compact Calabi-Yau manifold, one can
introduce the symplectic, integral basis $(A^\Lambda, B_{\Sigma})$
of the homology group $H_3({\cal X},\mbb Z)$ for the
three-cycles. These three-cycles are simply Poincar\'e-dual to the
basis of integer harmonic three-forms $(\a_\L, \bet^\S)$ that appear in
\reef{3int}. One can expand the generic
three-form $H_3$ in terms of flux quanta $(n^\L,e_\S)$ via
\bea
\label{hfluxq}
\frac{1}{\ell_s^2} \int_{A^\Lambda} H_{3}= m^{\Lambda}\ ,\quad
\frac{1}{\ell_s^2} \int_{B_\Sigma} H_{3}= e_{\Sigma}\ ,\quad\quad
\Lambda,\Sigma= 0, \dots,h^{2,1}\ .
\eea
Hence, $H_3$ in terms of the basis
$(\alpha_\L,\beta^\S)$ reads
\be
\frac{1}{\ell_s^2}  H_{3}=m^\L \alpha_\Lambda -e_\Sigma \beta^{\Sigma}\ .
\ee
Any closed field strength can be written as a linear combination of
harmonic forms of proper degree.

The major issue that we turn to in the following is the question under
which conditions the string equations of motion and the Killing
spinor equations, the supersymmetry conditions, allow solutions when
fluxes are present in the background. As mentioned already, the only
case where a trustworthy solution to the ten-dimensional equations of
motion is known is the case of type IIB orientifolds with three-form
flux. We will mostly concentrate on that.
To add fluxes into other string backgrounds very often a perturbative 
approach is used, starting with a Calabi-Yau solution and adding
fluxes as small perturbations while ignoring the backreaction of the
metric. This approach is justified in the large radius regime where
the fluxes become diluted. 


\subsection{Type IIB with three-form fluxes}

As a prototype example, where at least at the supergravity level
everything is quite well understood and under reasonable technical
control, let us discuss compactifications of the type IIB
superstring with background three-form fluxes. As explained in section
\ref{secIaII}, both type IIA and IIB string theories contain the
two-form field $B_{2}$ together with its field strength
$H_{3}=dB_{2}$. In addition type IIB has the RR potentials $C_{0}$,
$C_{2}$ and the four-form $C_{4}$. The complexified
axio-dilaton $\t$ and the complex three-form $G_3$ are often defined as
in \reef{dilG3}. Most of the following discussion is based in the effective
ten-dimensional type IIB supergravity action in Einstein frame and
with manifest $SL(2,\mbb Z)$ invariance written in \reef{IIBSL2}.

\subsubsection{Three--form fluxes in type IIB orientifolds}
\label{ANFANG}

Let us, before going through the details, first summarize some general
aspects of type IIB orientifold compactifications with
three-form fluxes. The relevant three-form background fields in IIB are $H_3$ and $F_3$
of section \ref{secIaII}. Note that the Bianchi identity of $F_3$ is
\bea
d F_3 = - dC_0 \wedge dB_2 = - F_1\wedge H_3\ .
\eea
Only after setting $F_1=0$ we can expand $F_3$ in the very same manner in
the cohomological basis used in \reef{hfluxq},
\bea
\label{ffluxq}
\frac{1}{\ell_s^2} \int_{A^\Lambda} F_{3}= \tilde m^{\Lambda}\ ,\quad
\frac{1}{\ell_s^2} \int_{B_\Sigma} F_{3}= \tilde e_{\Sigma}\ ,\quad\quad
\Lambda,\Sigma= 0, \dots, h^{2,1}\ .
\eea
This expansion is however only valid in IIB before passing to the
orientifold.

The orientifold projections \req{Action} and \req{Action2} put further
restrictions on the internal components that can appear.
For the orientifold projection \req{Action}, both $F_3$ and $H_3$ are
elements of $H_-^{3}(\cx)$, while for \req{Action2} $F_3$ is in
$H_+^{3}(\cx)$, and $H_3$ in $H_-^{3}(\cx)$ (\cff subsections \ref{IIB37} and
\ref{IIB59}). This leads to a modification of the expansion in
harmonics compared to \reef{hfluxq} or \reef{ffluxq}.
Let us first discuss the orientifold projection \req{Action}.
The two three-forms $F_3,\, H_3$ can be organized in terms of $G_3$
from \reef{dilG3} which also involves the dilaton $\t$.
The flux $G_3$ is now expanded in a symplectic integral basis of
$H^{3}_-(\cx,\mbb Z)$
\beqn\lab{G3exp}
{G_3\over \ell_s^2}=\sum_{\lambda=0}^{h^{2,1}_-}
m^\lambda \alpha_\lambda-e_\lambda \beta^\lambda\ ,
\eeqn
with the $2h^{2,1}_-(\cx)+2$ complex flux parameters $(m^\lambda,e_\lambda)$.
On the other hand, for the orientifold action \req{Action2} we have:
\zbe
{F_3\over \ell_s^2}=\sum_{\gamma=0}^{h^{2,1}_+(\cx)}
m^\gamma \alpha_\gamma-e_\gamma \beta^\gamma\ \ \ ,\ \ \
{H_3\over \ell_s^2}=\sum_{\lambda=1}^{h^{2,1}_-(\cx)}
m^\lambda \alpha_\lambda-e_\lambda \beta^\lambda\ ,
\ee
with $2(h^{2,1}_+(\cx)+2)$ real flux parameters $(m^\gamma,e_\gamma)$ for
$F_3$ and  in the same way $(m^\lambda,e_\lambda)$ for $H_3$.
These are the three-form fluxes that can exist in a IIB orientifold
background as a matter of principle.
The existence of an actual solution to the (classical) equations
of motion with a compact internal space then justifies their appearance in
orientifold compactifications.

Let us start with a list of qualitative consequences of these
fluxes in string compactifications. Afterwards we will elaborate
on each of the points in much more detail.

\begin{itemize}

\item{
As local conditions for supersymmetry the ten-dimensional Killing spinor equations, the variations of the
gravitinos and dilatino $\delta\psi_M=0$ and $\delta\lambda=0$
are modified in the presence of three-form fluxes.
As a result the metric is not even of the form
\reef{IIcomp}, i.e.\ four-dimensional Minkowski space times an
internal Ricci-flat space, instead a non-trivial warp factor appears in the metric
like in \reef{metricansatz}.
}

\item{
In the classical supergravity equations of motion
the background fluxes $H_{3}$ and $F_{3}$ provide extra source
terms, such as in the Einstein equations with a general
warped metric ansatz in \reef{Estequ}.
Thus, it also follows from the equations of motion, not even regarding
supersymmetry, that the internal metric is no longer
Ricci-flat. This implies that the deformations of the former
Calabi-Yau metric are at least partially obstructed by the fluxes.
}

\item{
Via CS terms in the ten-dimensional action \reef{IIBSL2}
the three-form fluxes act as sources for the four-form potential
$C_4$. They modify the topological charge cancellation condition
analogous to
\reef{loctad}, as can be read from the modified Bianchi identity
of the five-form field strength
\beqn \lab{5BI}
dF_5 =  - dC_2 \wedge dB_2 = H_3 \wedge F_3 \ .
\eeqn
Thus, the three-form fluxes effectively carry the charge of a number
of D3-branes as determined by the integral of \reef{5BI}.
The supersymmetry conditions that we will study in a
moment imply the positivity of the effective charge.
}

\item{
Dimensional reduction of the ten-dimensional action \reef{IIBSL2} in
the presence of three-form fluxes (plus other fields needed to
complete the solution) leads to a four-dimensional action with a
scalar potential. The relevant terms with three-form flux
contributions is the kinetic terms for $G_3$ and the CS
term. Roughly speaking, the potential depends on the moduli
controlling the size of the cycles $\S_{3}$ with non-vanishing
flux and the dilaton $\t$.\footnote{More precisely, one needs to pass
to the properly defined four-dimensional K\"ahler coordinates which
involve these ten-dimensional quantities, the complex structure moduli
and the four-dimensional dilaton.}
Minimizing this scalar potential can potentially fix these moduli
fields reflecting the obstruction of metric deformation via fluxes.
}

\item{
The fluxes also impose new consistency conditions for the world volume
theory of D-branes, if present. For instance, the Bianchi identity
for the diagonal $U(1)$ gauge field on a D-brane
\bea
2\pi\a' d{\cal F} = dB_2 + 2\pi\alpha'dF = 0
\eea
leads to the Freed-Witten anomaly cancellation condition.
The flux $H_3$ through a three-cycle wrapped by a D-brane has
to vanish.
}

\end{itemize}

In the following we discuss these points in more detail.


\subsubsection{Supersymmetry conditions with fluxes in IIB}
\label{subsubsecsusy}

Following the recent review article \cite{grana}, let us discuss the general supersymmetry conditions
resulting from the vanishing of the supersymmetry variations of
the gravitino and dilatino fields.
These are local conditions in ten dimensions that guarantee the
existence of Killing spinors in the given background. They lead to a
classification of supersymmetric geometries with fluxes, but a compact
internal space very often imposes additional constraints that are not
covered. Neither is always known if compact solutions of the described type
actually exist. Again, the simplest scenario of the type IIB
Calabi-Yau orientifold with three-form flux is an exception that can
be constructed on top of any Calabi-Yau space.

For the ten-dimensional metric one can use the general warped ansatz
\reef{metricansatz}.
The four-dimensional metric $g_{\mu\nu}$ describes either a
Minkowski, de Sitter (dS$_4$), or anti-de Sitter (AdS$_4$) space.
In general other bosonic fields are also allowed to acquire
non-trivial profiles and vacuum expectation values in the
background, but all fermion fields vanish.

In ten space-time dimensions the type II supersymmetry variations
for the two gravitinos $\psi_M^A$ ($A=1,2)$ and the two dilatinos
take the following form (in string frame)
\bea
\delta \psi_M &=& \nabla_M \epsilon
+ \frac{1}{4}\slash\!\!\!\!{H_M} \mathcal{P}\epsilon +
\frac{1}{16}e^{\Phi}
\sum_n \slash\!\!\! \!{F_{n}}\Gamma_M \mathcal{P}_n \epsilon\ , \nonumber\\
\delta \lambda &=&
\slash\!\!\!{\partial}\Phi\epsilon
+ \frac{1}{2}\slash\!\!\!\!{H} \mathcal{P}\epsilon +
\frac{1}{8}e^{\Phi}
\sum_n (-1)^n(5-n)
\slash\!\!\! \!{F_{n}}\mathcal{P}_n \epsilon\ .
\eea
Here the spinors $\psi_M$, $\l$ and $\e$ always combine two spinors
but we suppress the label $A$.
The $\mathcal{P}$ and $\mathcal{P}_n$ are $2\ti 2$ projection matrices,
whose form we do not need explicitly. For more details on the notation
we always refer to \cite{grana}.
The vanishing of these variations is required for supersymmetry. The
number of Killing spinors $\epsilon$ determines
the number of supercharges that are preserved.
It is now evident that without fluxes and with constant dilaton the solutions are just the
covariantly constant spinors of the Calabi-Yau, while
fluxes and dilaton profiles turn the conditions into much more
complicated looking differential equations.

To examine the constraints, we can decompose the ten-dimensional
spinors into tensor products of four- and
six-dimensional spinors,
\bea\lab{IIBeps}
\epsilon^A_{\rm IIB}(x^M) = \xi^A_{+}(x^\mu) \otimes
\eta_{+}(x^i) + \xi^A_{-}(x^\mu) \otimes \eta_{-}(x^i)\ .
\eea

Supersymmetry in the absence of fluxes thus leads to the constraint
\reef{Killspin} implying reduced holonomy for the internal space,
from $SO(6)\simeq SU(4)$ to $SU(3)$, so that ${\cal X}$ is a Calabi-Yau manifold.
The two covariantly constant spinors of type II theories
of course lead to ${\cal N}=2$ supersymmetry in four dimensions.
With non-vanishing fluxes $\eta_\pm$ can be viewed as covariantly constant with respect
to a new connection $\nabla'$ different from the Levi-Civita connection.
The internal manifold will no longer have $SU(3)$ holonomy (with
respect to the Levi-Civita connection).
Instead, the  requirement of having $SU(3)$ holonomy
with respect to the new connection means that the six-dimensional
internal manifold has a so-called $SU(3)$ group structure, i.e.\ the
transition functions of the frame bundle take values in an
$SU(3)\subset SO(6)$ subgroup.

This allows to decompose spinors and forms of the internal
six-dimensional manifold with respect to their transformation
properties under $SU(3)$.
One finds that there exist a non-vanishing real two-form
$J_2$ and a complex three-form $\Omega_3$ like on a Calabi-Yau.
Moreover, these forms can be related to spinors $\eta_\pm$ in the
following way (we have set $\ell_s=1$), 
\bea
\label{kahlom}
(J_2)_{ij}&=& \mp 2i \eta^{\dagger}_{\pm}\G_{ij} \eta_{\pm} \ , \non
(\Omega_3)_{ijk}&=& - 2i \eta_{-}^{\dagger}\G_{ijk}\eta_{+} \ .
\eea
While these forms are closed on a Calabi-Yau, this is now no longer the
case. Their deviation from being closed is measured by what is called
torsion.\footnote{The torsion of a connection is defined as
$[\nabla'_i,\nabla'_j]\, V_k =-R_{ijk}^l\, V_l -2 T^l_{mn}\,
\nabla'_l\, V_k$, where $R$ is the Riemann tensor.} The
classification of the different classes of torsion under $SU(3)$
helps in understanding the properties of the underlying geometry \cite{Chiossi:2002tw}.
The exterior derivative of $J_2$ and $\Omega_3$ can be expressed
using these torsion classes
\bea\lab{dJdOm}
dJ_2 &=& \frac{3}{2} \textrm{Im}(\overline{\cal W}_1 \Omega_3)+ {\cal W}_4
\wedge J_2 + {\cal W}_3 \ , \non
d\Omega_3&=& {\cal W}_1 J_2 \wedge J_2 + {\cal W}_2 \wedge J_2 + \ov{\cal W}_5 \wedge
\Omega_3 \ .
\eea
Here ${\cal W}_1$ is a complex
scalar, ${\cal W}_2$ a complex primitive $(1,1)$-form, ${\cal W}_3$ a real
primitive three-form built out of $(2,1)+(1,2)$-forms,
${\cal W}_{4}$ a real vector, and ${\cal W}_5$ is a complex
$(1,0)$-form.\footnote{Primitivity
means that the contraction with $J_2$ vanishes.}

A manifold with $SU(3)$ structure is  complex
if ${\cal W}_1={\cal W}_2=0$ and symplectic if ${\cal W}_1={\cal W}_3={\cal W}_4=0$. A
K\"ahler manifold is at the same time complex and symplectic, and
therefore the only non-zero torsion class can be ${\cal
W}_5$.\footnote{For a K\"ahler manifold the Levi-Civita connection has $U(3)$ holonomy.}
Finally for a Calabi-Yau manifold with $SU(3)$ holonomy all five
torsion classes are zero.

Since the connection and the torsion depend on the fluxes, the
above conditions on the torsion classes can be translated into
conditions on fluxes. A detailed analysis for ${\cal N}=1$
solutions in type IIB with both RR and NSNS three-form fluxes $H_3$ and $F_3$, yields three
different classes
\beqn
\label{susytyp}
\left.
\begin{array}{lc}
{\rm Type\ A}: & dJ_2\pm iH_3 \\
{\rm Type\ B}: & F_3\mp ie^{-\Phi}H_3  \\
{\rm Type\ C}: & d(e^{-\Phi}J_2)\pm iF_3
\end{array}
\right\}
{\rm is~(2,1)~and~primitive}\ .
\eea
On a Calabi-Yau the $(2,1)$-forms are automatically primitive, so
there is no extra condition. For extended supersymmetry, such as
for instance in case of a flat torus, there are additional
$(2,1)$-forms such as $dz\wedge J_2$ which are not primitive.

In the following we will be mainly interested in the type B class
solutions. They can be
reformulated in terms of the complex three-form \reef{dilG3}, saying
that $G_3$ is imaginary self dual (ISD) under internal Hodge duality
and has no $(0,3)$-component,\footnote{As noted in appendix
\ref{conventions} we use $*$ to denote the ten-dimensional Hodge duality and
$\star$ for the Hodge duality along the internal six-dimensional
space.}
\be\lab{G3ISD}
{\rm Type\ B}:\quad \star G_3 = i G_3\  , \quad G_3^{0,3}=0\ .
\ee
One can now immediately see that this condition \reef{G3ISD} can
fix geometrical moduli, since it depends on the internal metric
through the Hodge star operation. However, the Hodge duality in
six dimensions is invariant under an overall rescaling of the
metric, such that the overall volume of the compactification space
is not constrained by \reef{G3ISD}.

In the following we will discuss how the supersymmetry condition are
related to the condition for a solution to the equations of motion and
to the conditions that follow from extremizing an effective
four-dimensional potential after dimensional reduction. It is striking
to note at this point that only for the type B case of solutions
to the Killing spinor equations actual
classes of solutions to the equations of motions are known and an
effective four-dimensional approach can be defined. We will later also
comment on type IIA and heterotic flux vacua, but we basically leave
all other solutions aside.


\subsubsection{Compactification with fluxes: equations of motion}
\lab{seceomsol}

In order to start a program of flux compactifications one has to make
sure that solutions to the equations of motion exist with an internal
space that is actually compact. To preserve supersymmetry, the Killing
spinor equations of the previous section also have to be satisfied, in
our case the conditions \reef{G3ISD}. The corresponding solution was
found by Giddings, Kachru and Polchinski in \cite{gkp01}. We will not go
through all details but only explain the logic of the equations and
list the result.

To start with, recall the modification of the Bianchi identity (or
equation of motion) of the self-dual five-form $F_5$ in
\reef{5BI}. The supersymmetry condition on $G_3$ in \reef{G3ISD} now
immediately implies the positivity of the right-hand-side. Integration
over the compact internal space leads to a contradiction
such that other sources are needed. Anti-branes would be an option at
this stage, but in order to preserve $\cn=1$ supersymmetry one has to
pass to an orientifold with negative background D3-brane charge. The
natural framework is that of type IIB orientifolds of the type with
O3/O7-planes and potential D3/D7-branes. Including localized D3-brane
and O3-plane charges
the Bianchi identity gets modified to
\beqn \lab{mod5BI}
dF_5 =  H_3 \wedge F_3 + 2\k_{10}^2 \mu_3 \sum_a \pi_6^a + 2\k_{10}^2 Q_3 \mu_3 \pi_6^{\rm Opl}
\ ,
\eeqn
where the six-forms $\pi_6^a$ and $\pi_6^{\rm Opl}$ have support at the
locations of the branes (labeled again by $a$) and the
O3-planes. Integration schematically leads to
\beqn
N_{\rm flux} + N_{\rm D3} + Q_3 N_{\rm O3} = 0 \ .
\eeqn
where in particular the effective charge induced by fluxes is defined
\begin{equation}
N_{\rm flux}={1\over \ell_s^4} \int_{\cal X} H_3\wedge F_3\ .
\label{tadpoleb}
\end{equation}
Thus the equation for the five-form imposes the topological charge
cancellation condition analogous to \reef{loctad} for the D3-brane
charge.

The Einstein equation for the general metric ansatz
\reef{metricansatz} is of the form given in
\reef{Estequ}. On the right-hand-side the contribution of the
three-form fluxes, the five-form and the dilaton appear. Integration
of the component with four-dimensional indices
again gives a constraint since the left-hand-side is a total
derivative. A solution is found in the following way, for the
five-form an ansatz is made
\bea \lab{5form}
F_5= (1+*) d\a \wedge dx^0\wedge dx^1\wedge dx^2\wedge dx^3 \ ,
\eea
and the dilaton is taken to be constant (which is possible in the
presence of only D3-branes without D7-branes).\footnote{If D7-branes
are present one has to pass to an F-theory description, which is not
very well known how to handle in detail.} One can then combine
the Einstein equation with \reef{mod5BI} and finds a solution under
the following restrictions on the closed string background fields,
\beqn \lab{eomcond}
\star G_3 = i G_3 \ , \quad
\D^{-2} = \a \ , \quad
\Phi = {\rm const}\ .
\eeqn
In other words, the warp factor is given by the five-form and the
three-form flux has to be ISD. The remaining Einstein equation for the
internal components becomes equivalent to the equation of the five
form if the internal metric without the warp factor is Ricci-flat. The
internal space is therefore a conformal Calabi-Yau manifold, i.e. a Calabi-Yau
manifold only up to the warp factor. This is very good news, since we
know such manifolds do exist, we know their moduli spaces,
and also how to perform dimensional reduction on such a background.

It is important to note that \reef{eomcond} are precisely the supersymmetry
conditions of the type B solution in type IIB, except for the missing
constraint that $G^{0,3}=0$ from \reef{G3ISD}. Thus, ISD three-form
flux of Hodge type $(0,3)$ can break supersymmetry without destroying
the solution.


\subsubsection{The effective four-dimensional potential}
\label{subsecscalpot}

Let us now investigate the type IIB orientifold flux compactifications
based on the solution of the previous section from a four-dimensional
perspective. We mainly concentrate on the derivation of the scalar
potential.

A complication in such an approach is the presence of the non-trivial
warp factor in the metric \reef{metricansatz}. In principle, one
expects that terms from the ten-dimensional action with internal
derivatives in the reduction contribute, i.e.\ that $\a'$ corrections
are getting important due to the dependence of the warp factor on
the internal coordinates. This effect is usually treated as a small
effect and neglected which is justified by the scaling of  the warp
factor with the inverse total volume of the internal space. At
sufficiently large volume effects of warping are believed to be
controllably small, at least away from sources and regions of strong
warping, so-called throats. We will also follow this philosophy and
implicitly set $\D$ to a constant.

The easiest way to obtain a potential from fluxes is to just
naively insert a reduction ansatz into the ten-dimensional kinetic action of $G_3$.
It is given by the following expression
\be
\cv_{\rm kin} =
{1\over 4\kappa_{10}^2 {\Im}\, \t  } \int_{\cal X}
     G_3\wedge \star \bar G_3\ .
\ee
Here $G_3$ only refers to the background flux and no fluctuations of
the two-forms are included. In type IIB these would be present but in
the O3/O7 orientifold both $C_2$ and $B_2$ are projected out of the
spectrum.
Using that $\star^2 G_3 =- G_3$ one can now split the three-form flux into an
imaginary self-dual piece $G_3^+$ and an imaginary anti-self-dual
piece $G_3^-$, defined by
\bea
\star G_3^\pm =\pm i G_3^\pm\ .
\eea
After some algebra one finds that
\be
\label{potent}
     \cv_{\rm kin} =
{1\over 2\kappa_{10}^2 {\Im}\, \t} \int_{\cal X}
       G^-_3\wedge \star \bar G^-_3 -
      {i\over 4\kappa_{10}^2 {\Im}\, \t}  \int_{\cal X}
G_3\wedge  \bar G_3\ ,
\ee
The second term is equal to $\mu_3 N_{\rm flux}$. When the vacuum
energy of the localized sources, the tension of the D3-branes and
O3-planes is added to the full potential, one can make use of the
charge cancellation condition \reef{mod5BI} to cancel these against
the second term of \reef{potent}. It turns out that these
contributions can be interpreted as D-terms in the effective theory as
we have already seen at various instances for the tension of brane
systems. Their cancellation is thus a D-flatness condition.
Thus the full potential energy is
then given as
\beqn \lab{fluxpot}
\cv = \cv_{\rm kin} + \cv_{\rm DBI} = {1\over 2\kappa_{10}^2 {\Im}\, \t} \int_{\cal X}
       G^-_3\wedge \star \bar G^-_3 \ .
\eeqn
It is positive definite and can be written as an F-term in the
effective theory. Obviously, the potential has global minima whenever
$G_3^-=0$, which is the condition that also followed from the
ten-dimensional equations of motion in \reef{eomcond}.

One can now reproduce the structure of a standard supergravity
F-term scalar potential as in \reef{Fterm}.
The superpotential is of the Gukov-Taylor-Vafa-Witten type
\cite{gvw99,tv99,Peter}\footnote{Our conventions for dimensions are
that the components of the potentials inside $G_3$ are dimensionless such that $G_3$ has
dimension two, $\O_3$ dimension three and $W$ is bound to have
dimension minus three.}
\begin{equation}
\label{gvw}
W = \frac1{\k_{10}^2} \int_{\cal X} G_3\wedge \Omega_3\ ,
\end{equation}
which depends only on the complex structure moduli encoded in
$\O_3$, and on the dilaton $\t$ inside $G_3$. The K\"ahler potential
of the background Calabi-Yau was given in \reef{KpotOm}, \reef{KpotK}
and \reef{Kpotdil}. The use of this K\"ahler potential is of course
not fully accurate since there exist corrections already at the
classical level through the fluxes. We still use it in a heuristic
fashion, following \cite{tv99}.

For an orientifold with O3/O7--planes $\Om_3\in H^{3}_-(\cx)$ and
$G_3\in H^{3}_-(\cx)$. The expansion of $\O_3$ in the Calabi-Yau
itself from \reef{expJOm} then needs to be truncated to odd harmonic
three-forms. The analogous expansion of $G_3$ was given in \reef{G3exp}.
Since $W$ does not depend on the K\"ahler moduli, the remaining sum over these in
the term $G^{\a\bbe} \ck_\a \ck_\bbe |W|^2$ in
(\ref{Fterm}) cancels against the $-3|W|^2$. This condition,
\beqn \lab{no-scale}
G^{A\bar B} \ck_A \ck_{\bar B} = 3 \ ,
\eeqn
$A,\bar B$ labeling K\"ahler moduli only, is the so-called
no-scale structure of the classical K\"ahler potential on a
Calabi-Yau. It was shown in \cite{Becker:2002nn} that $\alpha'$ corrections
to the K\"ahler potential for the K\"ahler moduli break the no-scale
structure (see also \cite{Balasubramanian:2004uy,conlon1}). 
Using labels $I,
\bar J$ collectively for the remaining complex structure moduli and the
dilaton the F-term
potential is
\bea
\label{FFG}
\cv = \kappa^2_4\,
e^{{\k^2_4 \cal K}} G^{I\bar J}
               D_I W D_{\bar J} \bar {W}  \ .
\eea
This expression is positive definite and agrees with the first term in 
(\ref{fluxpot}).

Unbroken ${\cal N}=1$ supersymmetry now requires the vanishing of the F-terms.
Writing $F_\t,\, F_{T^A}$ and $F_{U^I}$ for the F-terms of the
dilaton, the K\"ahler and the complex structure moduli, the
constraints are
\bea
\begin{array}{ccc}
D_{T^A}W=(\partial_{T^A} {\cal K})\int_{\cal X}G_3\wedge\Omega_3 =0 &
\Leftrightarrow &
G_3^{0,3}=0 \ , \\[.4cm]
D_{\t}W= {1\over \t-\bar \t}\int_{\cal X}\bar G_3\wedge\Omega_3 =0 &
\Leftrightarrow & G_3^{3,0}=0 \ , \\[.4cm]
D_{U^I}\, W =\int_{\cal X} G_3\wedge \chi_I =0 & \Leftrightarrow &
G_3^{1,2}=0\ ,
\end{array}
\eea
where the $\chi_I$ denote a basis of odd $(2,1)$-forms.
The only component which survives is $G_3^{2,1}$.
These are absolute minima of the scalar potential with vanishing
vacuum energy and
supersymmetric solutions to the ten-dimensional equations of motion
(if suitably completed with solution for $F_5$, a constant dilaton and
the warped metric). Only up to the warp factor, the groundstate is always a
four-dimensional Minkowski space-time.

Simple counting shows
that to set all F-terms to zero one has $2h_-^{2,1}+4$ real equations
compared to $2h_-^{2,1}+2 $ real parameters in ($\t,U^I$) to solve
for. The system of equations is heavily over-constrained and for
general fluxes there are no solutions.

We have so far ignored the D-term constraint for the cancellation
of the tension of the branes against the topological contribution
of the fluxes, the second term in \reef{potent}. In case of
D3-branes only, there is no extra condition, since their tension
(in four-dimensional Einstein frame) only depends on the overall
volume of the internal space which is unconstrained. This goes
along with the fact that $G_3$ is automatically primitive on a
Calabi-Yau. Instead, for $H^1({\cal X})\ne 0$ the additional
supersymmetry condition $J_2\wedge G_3=0$ imposes a non-trivial
constraint on the K\"ahler moduli. This can be demonstrated in toy
model orientifold flux compactifications on $\mbb T^6$ for example
\cite{FERRARA1,FERRARA}.

Coming back to \reef{susytyp} the type IIB supersymmetry conditions
can be also satisfied by three-form fluxes, for which $G_3$ is not ISD
(type A and type C class solutions). Here the back-reaction on the internal
geometry is qualitatively more severe in the sense that ${\cal X}$ is
not anymore a warped CY space. Furthermore, no explicit solution to
the equations of motion along the lines outlined in section
\ref{seceomsol} is known. Nevertheless there are proposals for
effective superpotentials also for these cases. They involve the
piece $dJ_2$ which is non-vanishing due to torsion but no NSNS
three-form flux,
\bea
W = \frac1{\k_{10}^2} \int_\cx  (e^{-\Phi}dJ_2+F_3)\wedge\Omega_3\ .
\eea
This type of effective superpotential is believed to be relevant for orientifolds
with O5/O9-planes instead of O3/O7-planes.
The idea is that the F-flatness conditions of this superpotential
reproduce the supersymmetry constraints of the ten-dimensional Killing
spinor equations. But the difficulty is to identify the proper moduli
with respect to which one should perform a variation and to justify
the procedure by an explicit compact solution.

\subsection{Three--form fluxes in type IIB $\mbb Z_N$ and $ {\mbb
Z_N\times\mbb Z_M}$
orientifolds}\label{FLUXORBI}

After going through the general foundations of type IIB orientifold
flux compactifications, we now provide a detailed treatment of a class
of examples in which the process of moduli stabilization can be
studied particularly well.

In this subsection we discuss the flux space for the toroidal $\mbb Z_N$ and
$\mbb Z_N\times\mbb Z_M$ orbifolds \req{orbi}.
On a six--torus $\mbb T^6$ we have $h^{2,1}(\mbb T^6)=9$, thus ${\rm dim} H^{3}(\mbb T^6)=20$.
We introduce the six real periodic coordinates $x^i,y^i$ on the torus $\mbb T^6$, \ie
$x^i\cong x^i+1$ and $y^i\cong y^i+1$,\ $i=1,2,3$.
Furthermore we define the following  basis
\bea\label{realbase}
\ell_s^3\, \alpha_0&=&dx^1 \wedge dx^2  \wedge dx^3\ \ \ ,\ \ \
\ell_s^3\, \beta^0=dy^1 \wedge dy^2 \wedge dy^3\ ,\cr
\ell_s^3\,\alpha_1&=&dy^1 \wedge dx^2  \wedge dx^3\ \ \ ,\ \ \ \ell_s^3\,\beta^1=-dx^1 \wedge dy^2\wedge dy^3\ ,\cr
\ell_s^3\,\alpha_2&=&dx^1 \wedge dy^2  \wedge dx^3\ \ \ ,\ \ \ \ell_s^3\,\beta^2=-dy^1 \wedge dx^2 \wedge dy^3\ ,\cr
\ell_s^3\,\alpha_3&=&dx^1 \wedge dx^2  \wedge dy^3\ \ \ ,\ \ \ \ell_s^3\,\beta^3=-dy^1 \wedge dy^2 \wedge dx^3\ ,\cr
\ell_s^3\,\gamma_1&=&dx^1 \wedge dy^1  \wedge dx^2\ \ \ ,\ \ \ \ell_s^3\,\delta^1=-dy^2 \wedge dx^3 \wedge dy^3\ ,\cr
\ell_s^3\,\gamma_2&=&dx^1 \wedge dy^1  \wedge dx^3\ \ \ ,\ \ \ \ell_s^3\,\delta^2=-dx^2 \wedge dy^2 \wedge dy^3\ ,\cr
\ell_s^3\,\gamma_3&=&dx^1 \wedge dx^2  \wedge dy^2\ \ \ ,\ \ \ \ell_s^3\,\delta^3=-dy^1 \wedge dx^3\wedge dy^3\ ,\cr
\ell_s^3\,\gamma_4&=&dx^2 \wedge dy^2  \wedge dx^3\ \ \ ,\ \ \
\ell_s^3\,\delta^4=-dx^1 \wedge dy^1 \wedge dy^3\ ,\cr
\ell_s^3\,\gamma_5&=&dx^1 \wedge dx^3  \wedge dy^3\ \ \ ,\ \ \ \ell_s^3\,\delta^5=-dy^1 \wedge dx^2 \wedge dy^2\ ,\cr
\ell_s^3\,\gamma_6&=&dx^2 \wedge dx^3  \wedge dy^3\ \ \ ,\ \ \ \ell_s^3\,\delta^6=-dx^1 \wedge dy^1\wedge dy^2\ ,
\eea
for the integer cohomology class $H^{3}(\mbb T^6,\mbb Z)$ with respect to ambient space $\mbb T^6$.
This basis enjoys the following intersection properties (with the choice of orientation 
$\int_{\cal X} dx^1\wedge dx^2\wedge dx^3\wedge dy^1\wedge dy^2\wedge dy^3=l_s^6$):
\zbe\label{prop}
\int_{\mbb T^6} \alpha_i \wedge \beta^j=\delta^j_i\ \ \ ,\ \ \ \int_{\mbb T^6} \gamma_i
\wedge \delta^j=\delta^j_i\ .
\ee
Before the orbifold projection, there are $20+20$ independent real components
for $H_{ijk}$ and $F_{ijk}$.  Hence the general three-form
flux may be expanded  with respect to that basis \req{realbase}:
\eqn{GR}{ {1\over \ell_s^2}{G_3}=\sum_{i=0}^{3} \lf[(a^i+iS\
c^i)\alpha_i+(b_i+iS\ d_i)\beta^i\ri]+\sum_{j=1}^{6} \lf[(e^j+iS\
g^j)\gamma_j+(f_j+iS\ h_j)\delta^j\ri]\ .}
The coefficients $a^i,\ b_i,\ e^i,\ f_i$ refer to the Ramond part of $G_3$, while the
coefficients $c^i,\ d_i,\ g^i,\ h_i$ refer to its  Neveu-Schwarz part.

Under the involution $\sigma^\ast=I_6$ (\eqq \req{Action}) all $20$ untwisted
three-forms \req{realbase} pick up a minus sign, \ie belong to the
cohomology  $H^{3}_-(\mbb T^6/\si,\mbb Z)$. Hence in that case all $20$ forms
\req{realbase} may give rise to flux components for both $F_3$ and $H_3$.
On the other hand, under the second involution $\sigma^\ast=1$ (\eqq \req{Action2})
all $20$ untwisted three-forms \req{realbase} are inert, \ie belong to the
cohomology  $H^{3}_+(\mbb T^6/\si,\mbb Z)$. Hence, $H_3=0$ and only $F_3$ may
receive contributions from all $20$ elements \req{realbase} \cite{aft03}.

Let us now discuss the effect of the orbifold action \req{oaction}.
Only a portion of the forms \req{realbase} is invariant under the orbifold group $\Gamma$.
First of all, the three-forms  $\gamma_i$ and $\delta^j$ are never invariant under
the orbifold twist $\Gamma$ as a result, that those forms have two legs from the same complex
plane. However they play an important role in constructing a twist invariant basis
of three-forms.
The number of untwisted flux components of $F_3,H_3$ subject to the orbifold action
\req{oaction} and the orientifold involutions \req{Action} or
\req{Action2} is given by the Hodge
number $h_{\rm untwist.}^{2,1}({\cal X})$ of the original orbifold ${\cal X}$. For
details, see \eqq \req{Hodgeuntw} for the involution $\sigma^\ast=I_6$ and \eqq
\req{Hodgeuntww} for the involution $\sigma^\ast=1$, respectively.
Hence, for the orientifold action \req{Action} the flux $G_3$ has
$2h_{\rm untwist.}^{2,1}({\cal X})+2$ untwisted complex components,
while the action \req{Action2} allows only for components of $F_3$ and $H_3=0$.

The untwisted three-forms, invariant under the orbifold action
\req{oaction} may be most conveniently found in the complex basis, in which
the orbifold group $\Gamma$ acts diagonally by the representation \req{oaction}
on the complex basis $\{z^i\}$. To pass from the real basis \req{realbase} to a complex basis,
one introduces complex structures, \ie defines the complex coordinates:
\eqn{achieved}{
dz^j=\sum_{i=1}^3\ \rho^j_i\ dx^i+ \tau^j_i\ dy^i\ \ \ ,\ \ \ j=1,2,3\ .}
Most of the parameters $\rho^j_i$ and $\tau^j_i$ are fixed through the
orbifold twist $\Gamma$
\cite{LRSSi}, with  only those remaining undetermined, which correspond to the
$\mbb Z_2$--elements of $\Gamma$.
Eventually the latter are  fixed through the flux quantization condition.
As we shall see in a moment, the  specific values of the constants $\rho^j_i$ and $\tau^j_i$
are relevant for finding flux solutions.
With respect to  to the complex coordinates, defined in \req{achieved},  we may
introduce the complex three-form cohomology
$H^3=H^{3,0}\oplus H^{2,1}\oplus H^{1,2}\oplus H^{0,3}$ of the
six--torus $\mbb T^6$:
\bea\label{cplxz}
{\ell_s^3}\, \om_{A_0}&=&d z^1\wedge dz^2\wedge d z^3\ \ \ ,\ \ \  {\ell_s^3}\, \om_{B_0}=d\ov
z^1\wedge d \ov z^2\wedge d \ov z^3\cr
{\ell_s^3}\,\om_{A_1}&=&d\ov z^1\wedge
dz^2\wedge dz^3\ \ \ ,\ \ \ {\ell_s^3}\,\om_{B_1}= dz^1\wedge d\ov z^2\wedge d\ov z^3\cr
{\ell_s^3}\,\om_{A_2}&=&dz^1\wedge
d\ov z^2\wedge dz^3\ \ \ ,\ \ \ {\ell_s^3}\,\om_{B_2}=d\ov z^1\wedge dz^2\wedge d\ov z^3\cr
{\ell_s^3}\,\om_{A_3}&=&dz^1\wedge dz^2\wedge d\ov z^3\ \ \ ,\ \ \ {\ell_s^3}\,\om_{B_3}= d\ov
z^1\wedge d\ov z^2\wedge dz^3\cr
{\ell_s^3}\,\om_{C_1}&=&dz^1\wedge d\ov z^1\wedge dz^2\ \ \ ,\ \ \
{\ell_s^3}\,\om_{D_1}=dz^1\wedge d \ov z^1\wedge d \ov z^2\cr
 {\ell_s^3}\,\om_{C_2}&=&dz^1\wedge
d\ov z^1\wedge dz^3\ \ \ ,\ \ \ {\ell_s^3}\,\om_{D_2}= dz^1\wedge d\ov z^1\wedge d\ov z^3\cr
{\ell_s^3}\,\om_{C_3}&=&dz^1\wedge
d z^2\wedge d\ov z^2\ \ \ ,\ \ \ \ell_s^3\, \om_{D_3}=d\ov z^1\wedge dz^2\wedge d\ov z^2\cr
{\ell_s^3}\,\om_{C_4}&=&dz^2\wedge d\ov z^2\wedge d z^3\ \ \ ,\ \ \ {\ell_s^3}\,\om_{D_4}=
d z^2\wedge d\ov z^2\wedge d\ov z^3\cr
{\ell_s^3}\,\om_{C_5}&=&dz^1\wedge
dz^3\wedge d\ov z^3\ \ \ ,\ \ \ \ell_s^3\, \om_{D_5}=d\ov z^1\wedge dz^3\wedge d\ov z^3\cr
{\ell_s^3}\,\om_{C_6}&=&dz^2\wedge dz^3\wedge d\ov z^3\ \ \ ,\ \ \ {\ell_s^3}\,\om_{D_6}=
d\ov z^2\wedge d z^3\wedge d\ov z^3\ .
\eea
In the following two
tables we list for all toroidal orbifolds \req{orbi} those three-forms, which are left
invariant under the orbifold action $\Gamma$, defined in \req{oaction}.
\begin{table}[h]
\centering
{\vbox{\ninepoint{$$
\vbox{\offinterlineskip\tabskip=0pt
\halign{\strut\vrule#
&~$#$~\hfil
&\vrule$#$
&~$#$~\hfil
&~$#$~\hfil
&~$#$~\hfil
&~$#$~\hfil
&~$#$~\hfil
&~$#$~\hfil
&~$#$~\hfil
&~$#$~\hfil
&~$#$~\hfil
&\vrule$#$\cr
\noalign{\hrule}
&\ \  G_3&&\ \mbb Z_3
&\mbb Z_4&\mbb Z_{6-I}&\mbb Z_{6-II}&\mbb Z_7&\mbb Z_{8-I}&\mbb Z_{8-II}&\mbb Z_{12-I}&\mbb Z_{12-II}&\cr
\noalign{\hrule}\noalign{\hrule}
&dz^1\wedge dz^2\wedge dz^3&&\ +&+&+&+&+&+&+&+&+&\cr
&d\ov z^1\wedge dz^2\wedge dz^3&&\ -&-&-&-&-&-&-&-&-&\cr
&dz^1\wedge d\ov z^2\wedge dz^3&&\ -&-&-&-&-&-&-&-&-&\cr
&dz^1\wedge dz^2\wedge d\ov z^3&&\ -&+&-&+&-&-&+&-&+&\cr
&d z^1\wedge d\ov z^2\wedge d\ov z^3&&\ -&-&-&-&-&-&-&-&-&\cr
&d\ov z^1\wedge d z^2\wedge d\ov z^3&&\ -&-&-&-&-&-&-&-&-&\cr
&d\ov z^1\wedge d\ov z^2\wedge d z^3&&\ -&+&-&+&-&-&+&-&+&\cr
&d\ov z^1\wedge d\ov z^2\wedge d\ov z^3&&\ +&+&+&+&+&+&+&+&+&\cr
\noalign{\hrule}}}$$}}}
\label{tab9}
\caption{Allowed three-form fluxes for point group $\mbb Z_N$.}
\end{table}

\begin{table}[h]
\centering
\hskip-1.5cm
{\vbox{\ninepoint{$$
\vbox{\offinterlineskip\tabskip=0pt
\halign{\strut\vrule#
&~$#$~\hfil
&\vrule$#$
&~$#$~\hfil
&~$#$~\hfil
&~$#$~\hfil
&~$#$~\hfil
&~$#$~\hfil
&~$#$~\hfil
&~$#$~\hfil
&~$#$~\hfil
&\vrule$#$\cr
\noalign{\hrule}
&\ \ G_3&&\mbb Z_2\times\mbb Z_2&\mbb Z_3\times \mbb Z_3&\mbb Z_2\times \mbb Z_4
&\mbb Z_4\times \mbb Z_4&\mbb Z_2\times \mbb Z_{6-I}&\mbb Z_2 \times \mbb Z_{6-II}&\mbb Z_3\times \mbb Z_6
&\mbb Z_6\times \mbb Z_6&\cr
\noalign{\hrule}\noalign{\hrule}
&dz^1\wedge dz^2\wedge dz^3&&\ +&+&+&+&+&+&+&+&\cr
&d\ov z^1\wedge dz^2\wedge dz^3&&\ +&-&+&-&+&-&-&-&\cr
&dz^1\wedge d\ov z^2\wedge dz^3&&\ +&-&-&-&-&-&-&-&\cr
&dz^1\wedge dz^2\wedge d\ov z^3&&\ +&-&-&-&-&-&-&-&\cr
&d z^1\wedge d\ov z^2\wedge d\ov z^3&&\ +&-&+&-&+&-&-&-&\cr
&d\ov z^1\wedge d z^2\wedge d\ov z^3&&\ +&-&-&-&-&-&-&-&\cr
&d\ov z^1\wedge d\ov z^2\wedge d z^3&&\ +&-&-&-&-&-&-&-&\cr
&d\ov z^1\wedge d\ov z^2\wedge d\ov z^3&&\ +&+&+&+&+&+&+&+&\cr
\noalign{\hrule}}}$$}}}
\label{tab10}
\caption{Allowed three-form fluxes for point group $\mbb Z_M\times \mbb Z_N$.}
\end{table}

The remaining 12 fluxes of the form $dz^a\wedge d\ov z^a\wedge dz^b$ and
$dz^a\wedge d\ov z^a\wedge d \ov z^b$, respectively are always projected out
and therefore do not appear in the tables. They are related to the real basis elements $\gamma_i$
and $\delta^j$.
In terms of the complex basis \req{cplxz} the most general three-form flux $G_3$
on $\mbb T^6$ has 20 complex components and assumes the following expansion
\eqn{GC}{{1\over \ell_s^2}\ G_3=\sum_{i=0}^3 (\ A^i\ \omega_{A_i}+B^i\ \omega_{B_i}\ )+
\sum_{j=1}^6(\ C^j\ \omega_{C_j}+D^j\ \omega_{D_j}\ )\ ,}
with the complex coefficients $A^i,B^i,C^j,D^j\in \ICC$.
The three-forms $\om_{A_i}$,  $\om_{B_i}$ correspond to flux components
with all one--forms coming from different planes,
while the forms $\om_{C_i}$, $\om_{D_i}$ are flux components
with two one--forms coming from the same plane.
The latter we have just written down for completeness, as they are projected out in
all orbifolds.
With \req{achieved} we may switch from \req{GC} to \req{GR} and express
all $20$ complex coefficients $A^i,B^i,C^j,D^j$ in terms of the $40$ real coefficients
$a^i,\ b_i,\ e^i,\ f_i, c^i,\ d_i,\ g^i,\ h_i$ or vice versa.

In the $\mbb Z_2\times \mbb Z_2$ orientifold/orbifold, which allows for the largest number of
(untwisted) fluxes components (\cff \cite{lrs04} and Table 10), all $\om_{A_i}$ and $\om_{B_i}$ remain,
whereas in most other orbifolds only the $(3,0)$ and $(0,3)$ components, given by
$\om_{A_0}$ and $\om_{B_0}$ survive.
In these cases, no
supersymmetric $(2,1)$--form flux can be turned on.
That the $(0,3)$ and $(3,0)$-flux always
survive is obvious, as the $(3,0)$-flux corresponds to the Calabi-Yau
3-form $\Om_3$, which is always present, and the $(0,3)$-flux to its
conjugate.

While in the form \req{GC}, the cohomology structure of $G_3$ is manifest,
in order to impose the flux quantization \req{hfluxq} on $G_3$, \ie
\eqn{fluxqu}{
\fc{1}{\ell_s^2}\int_{C_3} F_3 \in N_{min}\ \mbb Z\ \ \ ,\ \ \
\fc{1}{\ell_s^2}\int_{C_3} H_3 \in N_{min}\ \mbb Z\ ,}
with some integer $N_{min}$ (depending on the orbifold group $\Gamma$)
to be specified in a moment, one has to transform the forms \req{cplxz}
into a real basis \req{realbase}.
Let us briefly comment on the integers $N_{min}$, introduced in the flux quantization conditions
\req{fluxqu}.
It has been pointed out in  \cite{FP},
that there are subtleties for toroidal orientifolds
due to additional three-cycles, which are not present in the covering space $\mbb T^6$.
If some integers are odd, additional discrete flux has to be
turned on in order to meet the quantization rule for those three-cycles.
We may bypass these problems in the $\mbb Z_N$ ($\mbb Z_N\times \mbb Z_M$)--orientifolds,
if we choose the quantization numbers to be multiples of $N_{min}=2N$ ($N_{min}=2NM$)
and do not allow for discrete flux at the orientifold planes \cite{blt03,cu03,AF04}.
Note, that for $h^{2,1}_{\rm twist.}\neq 0$,
in addition to the untwisted flux components $H_{ijk}$ and $F_{ijk}$ there may
be also NSNS-- and RR--flux components from the twisted sector. We do not
consider them here. It is assumed, that their quantization rules freeze the
twisted complex structure moduli.

Generically one starts with a specific flux choice \req{GC}, \ie some choice for
the complex numbers $A^i,B^i,C^j,D^j$ allowing \eg  only some flux directions.
Then one imposes the flux
quantization rule \req{fluxqu} for the fluxes $F_3,H_3$ with respect  to an
integral basis.  This is achieved by transforming  the flux \req{GC} with the help
of \req{achieved} into the integral basis \req{GR}.
However, then one realizes that the flux quantization conditions \req{fluxqu} may only
be fulfilled for specific values of the dilaton $S$ and complex structure moduli $U$.
This way flux quantization is related to discrete values for the dilaton and
complex structure moduli.
This may be also seen dynamically through the superpotential \req{gvw} (\cff subsection \ref{DYNFIX}).

One way one can think of turning on flux on a D$p$-brane
is via the generalized Scherk-Schwarz Ansatz:
\eqn{SchSch}{B_{mn}=H_{mnp}\, x^p.}
The index $p$ may run only over the coordinates transversal to the brane, so in the case
of a D7-brane, which fills the directions $x_0,\ldots,x_7$ (wrapping the
tori $\mbb T_1^2\times \mbb T_2^2$), we have
\eqn{SSseven}{B_{mn}^7=H_{mn8}\, x^8+H_{mn9}\, x^9.}
As $H_{ijk}$ must always have one index equal to either 8 or 9, not all of the
20 possible components are allowed in our case. Not allowed are the fluxes
\bea
&& dx^1 \wedge dy^1  \wedge dx^2 \qquad dx^1 \wedge dy^1 \wedge dy^2 \nnn \\
&& dx^1 \wedge dx^2  \wedge dy^2 \qquad dy^1 \wedge dx^2
 \wedge dy^2\ .
\label{notallowed}
\eea
Expressed in complex notation, this would correspond to the fluxes
$H_{1\ov{1}2}$, $H_{1\ov{2}2}$, $H_{1\ov{1}\ov{2}}$, $H_{2\ov{1}\ov{2}}$.

For D7-branes wrapping the tori $\mbb T_1^2\times \mbb T_3^2$ or $\mbb T_2^2\times \mbb T_3^2$,
we find similar results. Having a setup of three stacks of D7-branes, one
stack not wrapping $\mbb T_3^2$, one not wrapping  $\mbb T_2^2$, and one not wrapping
$\mbb T_1^2$, we lose 12 of the twenty flux components and are left with fluxes,
which have one index on each of the tori.

\subsubsection{Example: Three--form fluxes on the $\mbb T^6/ \mathbb{Z}_2\times \mathbb{Z}_2$ orientifold}
\label{sssecexam}

As an example we consider the   $\mbb T^6 / \mathbb{Z}_2\times
\mathbb{Z}_2$ orientifold with the 
Hodge numbers $h^{1,1}=3$ and $h^{2,1}=51$, \cff Table \ref{tabZ22models}.
Let us discuss the three-form flux $G_3$ for
the $\mbb Z_2\times\mbb Z_2$ orbifold with the 
orientifold action \req{Action} \cite{lrs04}.
According to Table \ref{tab6} 
we have $h_{\rm untw.}^{2,1}({\cal X})=3$. Hence the (untwisted) three-form flux $G_3$ has
$2h_{\rm untw.}^{2,1}+2h^{3,0}=2\cdot 3+2=8$ 
complex flux components. The latter correspond to the forms
$\omega_{A_i}$ and $\omega_{B_i}$, with $i=0,\ldots,3$. These eight flux
components are related to a linear combination
of the primitive elements of the (untwisted) cohomology $H^3({\cal X},\ICC)$.
After choosing the complex structure
\eqn{chCPL}{dz^i=dx^i+i\, U^i\ dy^i\ \ \ ,\ \ \ i=1,2,3}
we may go back to the flux in the real representation \req{GR}.
With respect to the basis \req{realbase} the flux $G_3$ is expressible as a linear
combination of the eight real forms
\eqn{above3}{\alpha_i\ \ ,\ \ \ \beta^i\ \ \ ,\ \ \ i=0,\ldots,3\ ,}
given in \eqq \req{realbase}.
The expansion coefficients correspond to the eight allowed real components
\eqn{fluxsurvive}{
H_{135}\ , \ H_{136}\ ,\ H_{145}\ ,\ H_{146}\ ,\ H_{235}\ ,\ H_{236}\ ,\ H_{245}\ ,\ H_{246}\ ,}
of the NSNS three-form $H_3$ and to the eight possible real components
\eqn{fluxsurvivei}{
F_{135}\ , \ F_{136}\ ,\ F_{145}\ ,\ F_{146}\ ,\ F_{235}\ ,\ F_{236}\
,\ F_{245}\ ,\ F_{246}}
of the RR three-form $F_3$, each.
The above three-forms \req{above3}, which are invariant under the $\mathbb{Z}_2\times
\mathbb{Z}_2$ orbifold symmetry and form a symplectic basis,  are Poincar\'e-dual to the obvious invariant
three-cycles on $\mbb T^6$.

There are the three moduli, $U^I$, $I=1,2,3$, which
define the complex structure \req{chCPL} on the orbifold cover $\mbb T^6$. The holomorphic
three-form
\bea
{1 \over \ell_s^3}\, \Omega_3=dz^1\wedge dz^2\wedge
dz^3 = X^{\Lambda}\alpha_{\Lambda}- F_{\Lambda}\beta^{\Lambda}
\eea
defines the homogeneous coordinates $X^{\Lambda}$ and the
derivatives $F_{\Lambda}=\partial_\Lambda F$ of the prepotential
\begin{eqnarray}
X^0&=&1\ ,  \quad\quad
F_0=iU^1\, U^2\, U^3\ , \nonumber\\
                       X^1&=&iU^1,\quad\quad F_1= -U^2\, U^3\ , \nonumber\\
                       X^2&=&iU^2,\quad\quad F_2= -U^1\, U^3\ , \nonumber\\
                       X^3&=&iU^3,\quad\quad F_3= -U^1\, U^2\ .
\end{eqnarray}
Therefore the prepotential is given by
\bea
F={X^1X^2X^3\over
X^0}=-i\, U^1U^2U^3\ .
\eea
The K\"ahler potential is given by
\begin{eqnarray*}
\kappa_4^2\, {K}= -\ln \prod_{I=1}^3 (U^I+\bar{U}^I)\ ,
\end{eqnarray*}
and finally the superpotential takes the form
\bea
{\ell_s^{3}\over 4\pi}\, W&=& (b_0+i S\, d_0) + (b_1+i S\, d_1)\, iU_1+ (b_2+i S\, d_2)\, iU_2+ (b_3+i S\, d_3)\, iU_3
                \non
&&     +(a_1+i S\, c_1)\, U_2\, U_3  +(a_2+i S\, c_2)\, U_1\, U_3 +(a_3+i S\, c_3)\, U_1\, U_2\non
&&
         - (a_0+i S\, c_0)\, iU_1\, U_2\, U_3  .\label{withh}
\eea


\subsubsection{Complex structure and dilaton stabilization through three-form flux}\label{DYNFIX}

In this subsection we demonstrate how the stabilization of the dilaton and complex structure
moduli is achieved dynamically with the superpotential \req{gvw}
for a general  three-form flux $G_3$.
To keep the expressions short we shall discuss
orbifolds ${\cal X}$ with $h_{-}^{2,1}({\cal X})= 1$.
The K\"ahler potential ${K}_0$ for the dilaton and complex structure modulus 
($U\equiv U^3$) is:
\eqn{BECOMES}{
\kappa_4^2\, K_0=-\ln(S+\bar S)-\ln(U+\bar U)\ .}
For $h_{-}^{2,1}({\cal X})= 1$  there are four
complex parameters to parameterize the most general three-form flux $G_3$. The contribution of
the latter to the tree--level superpotential \req{gvw} may be written
\eqn{WandK}{
{\ell_s^{3}\over 4\pi}\, W_0=A+B\ S+U\ (C+D\ S) \ ,}
with four complex numbers $A,B,C,D \in \ICC$, to specified later.
With the resulting $F$--terms
\bea\label{FSU}
{\ell_s^{3}\over 4\pi}\, \ov F^{\ov S}&=&\kappa_4^2\ \lf(\fc{S+\ov S}{U+\ov U}\ri)^{1/2}\ \lf[\
-A+B\ \ov S-U\ (C-D\ \ov S)\ \ri]\ ,\cr
{\ell_s^{3}\over 4\pi}\, \ov F^{\ov U}&=&\kappa_4^2\ \lf(\fc{U+\ov U}{S+\ov S}\ri)^{1/2}\ \lf[\
-A-B\ S+\ov U\ (C+D\ S)\ \ri]\ ,
\eea
we may cast the scalar potential
\bea
\Vc=g_{S\ov S}\ F^S\ov F^{\ov S}+g_{U\ov U}\ F^U\ov F^{\ov U}-
 3\ \kappa_4^2\ e^{\kappa_4^2\, {\cal K}_0}\ |W_0|^2
\eea
into the form:
\bea\label{superpotential}
\Vc&=& \fc{1}{(S+\bar S)(U+\bar U)}\ \lf[\ |\ A-B\ \ov S+U\ (C-D\ \ov S)\ |^2
+|\ A+B\ S-\ov U\ (C+D\ S)\ |^2\ri.\cr
&&\hskip3cm\lf.-3\ |\ A+B\ S+ U\ (C+D\ S)\ |^2\ \ri]\ \kappa_4^2\fc{(4\pi)^2}{l_s^6}.
\eea
The extremal points in the moduli space $(S,U)$ are determined
by the solutions of the equations $F^S,F^U=0$:
\eqn{AXION}{
s_2=\fc{i}{2}\ \fc{\ov B\ C-B\ \ov C-\ov A\ D+A\ \ov D}{\ov B\ D+B\ \ov D}\ \ \ ,
\ \ \
u_2=\fc{i}{2}\ \fc{-\ov B\ C+B\ \ov C-\ov A\ D+A\ \ov D}{\ov C\ D+C\ \ov D}\ ,}
and similarly for the real parts $s_1,u_1$.

The three-form flux \req{GR} entering \req{gvw} is given as a linear combination  with respect to
the integer cohomology basis $\{\al_i,\beta^i\}_{i=0,\ldots,3}$
and $\{\gamma_j,\delta^j\}_{j=1,\ldots,6}$, given in \eqq \req{realbase}.
This gives rise to 20 real flux components to be constrained by the
respective orbifold group $\mbb Z_N$ (\cff subsection \ref{FLUXORBI}).
This allows to express the complex parameters
$A,B,C,D$ through the eight integers $a^0,a^1,b_0,b_1,c^0,c^1,d_0,d_1$. For more details
\cff \cite{LRSSi}. The F--flatness conditions $F^S,F^{U}=0$ force
the complex structure to align such, that the flux $G_3$
\bea\label{FLUXG}
\hskip-0.5cm\ds{l_s\ G_3}&=&\ds{\fc{i}{2\ \re(U)}\ \lf\{\
\lf[\ \ov A-\ov B\ S+\ov U\ (\ov C-\ov D\ S)\ \ri]\ dz^1\wedge dz^2\wedge dz^3\ri.}\nnn\\[2mm]
&-&\ds{\lf[\ A+B\ S+U\ (C+D\ S)\ \ri]\ d\ov z^1\wedge d\ov z^2\wedge d\ov z^3}\nnn\\[2mm]
&+&\ds{\lf[\ A+B\ S-\ov U\ (C+D\ S)\ \ri]\ d\ov z^1\wedge d\ov z^2\wedge d z^3}\nnn\\[2mm]
&-&\ds{\lf.\lf[\ \ov A-\ov B\ S-U\ (\ov C-\ov D\ S)\ \ri]\ d z^1\wedge d z^2\wedge d\ov z^3\
\ri\}}
\eea
corresponding to the choice \req{WandK} becomes ISD, \ie it has only
$(2,1)$ and $(0,3)$--components at the extremum.
The flux $G_3$ induces the contribution of
\eqn{NFLUX}{
N_{\rm flux}={1\over \ell_s^4 }\ \int_{\cal X} F_3\wedge H_3}
to the total D3-brane charge \req{totaltadpole}. Generically, this integral is calculated
in the orientifold cover ${\cal X}$. Therefore the number $N_{flux}$ has to be twice
the negative value of the total D3-brane charge \req{totaltadpole}, \ie
\eqn{idFLUX}{N_{\rm flux}=-2\ Q_{3,tot}}
to cancel the latter by flux only.

As an example, let us discuss the
$\mbb Z_{6-II}$--orbifold  with the torus lattice $SU(2)\times SU(6)$ \cite{LRSSii}.
The $\mbb Z_{6-II}$--orbifold has the action $(v^1,\,v^2,\,v^3)=({1\over6},\,{1\over3},\,-{1\over2})$.
The three-form flux \req{GR} constrained by the $\mbb Z_{6-II}$--orbifold group becomes:
\bea\label{GdreiZsechs}
\ds{\fc{1}{\ell_s^2}\ G_3}&=&\ds{\fc{1}{3}\ (a_0+iS c_0)\
(3\ \alpha_0+2\ \beta_3+\gamma_1-2\gamma_2-
2\ \gamma_3+\gamma_4-\delta_5)}\nnn\\[2mm]
&+&\ds{(b_0+iSd_0)\ (-\alpha_3+\beta_0+\gamma_5-\gamma_6)} \nnn\\[2mm]
&+&\ds{\fc{1}{2}\ (b_1+iSd_1)\ (2\ \beta_1+\beta_2+\delta_1-\delta_2-2\ \delta_3-\delta_4)}
\nnn\\[2mm]
&+&\ds{(a_1+iSc_1)\ (\alpha_1+\alpha_2+\beta_3-\gamma_2-\gamma_3-\delta_6)\ .}
\eea
This flux corresponds to the flux number:
\eqn{NflussZ}{
N_{\rm {flux}}=2\ b_0\ c_0+b_1\ (c_0+3\ c_1)-2\ a_0\ d_0-d_1\ (a_0+3\ a_1)\ .}
For the $\mbb Z_{6-II}$ orbifold with $SU(2)\times SU(6)$ lattice the coefficients
$A,B,C,D$ entering \req{superpotential} become:
\bea\label{Zsechs}
A&=&-\fc{\sqrt 3}{2}\ b_1+i b_0 \ \ \ ,\ \ \ B=-d_0-\fc{\sqrt 3\,i}{2}d_1 \ , \cr
C&=&a_0+i \lf(\fc{a_0}{\sqrt 3}+\sqrt 3 a_1\ri) \ \ \ ,\ \ \
D=-\lf(\fc{c_0}{\sqrt 3}+\sqrt 3 c_1\ri)+ic_0\ .
\eea
With this information, \eqq \req{WandK} yields the superpotential:
\bea\label{WflussZ}
{\ell_s^{3}\over 4\pi}\, W_0&=&-\fc{\sqrt{3}}{2}\ b_1+i\ b_0-S\ \lf(d_0+\fc{\sqrt{3}i}{2}\
d_1\ri) \cr
&+&U\ \lf[\ a_0+i\lf(\fc{a_0}{\sqrt{3}}+\sqrt{3}\ a_1\ri)\ \ri]-
S\ U\ \lf(\fc{c_0}{\sqrt{3}}+\sqrt{3}\ c_1-i\ c_0\ri)\ .
\eea
Since the total D3-brane charge
in the CY orientifold is $Q_{3,tot}=-22$ (see  \cite{LRSSii}), we look for  fluxes
\req{GdreiZsechs}
with $N_{flux}=44$ in the covering space. Furthermore, the fields
$S=s_1+is_2$ and $U=u_1+iu_2$ should be fixed (\cff \req{AXION}) to realistic values.
A reasonable  value for $\re S$ is $s_1\sim 3.6$, which corresponds to a string
coupling constant $g_{\rm string}\sim 0.27$ at the string scale. The complex structure
modulus $U$ is expected to be around the $\rho$--point in the fundamental 
region with $\rho=\h+\fc{i}{2}\sqrt 3$.
After a systematic scan in the flux space $(a^0,a^1,b_0,b_1,c^0,c^1,d_0,d_1)\in \mbb Z^8$
we find hundreds of vacua, which meet these criteria.
A set of equivalent vacua, differing only in the discrete flux parameters
$(a^0,a^1,b_0,b_1,c^0,c^1,d_0,d_1)$, is given in the following Table
\ref{tab11}.

\begin{table}[h]\hskip-0.6cm
{\vbox{\ninepoint{$$
\vbox{\offinterlineskip\tabskip=0pt
\halign{\strut\vrule#
&~$#$~\hfil 
&\vrule$#$&\vrule$#$
&~$#$~\hfil 
&\vrule$#$ 
&~$#$~\hfil 
&\vrule$#$
&~$#$~\hfil 
&\vrule$#$ 
&~$#$~\hfil 
&\vrule$#$
&~$#$~\hfil 
&\vrule$#$
&~$#$~\hfil 
&\vrule$#$&\vrule$#$\cr
\noalign{\hrule}
&\ (a^0,\ b_0,\ c^0,\ d_0,\ a^1,\ b_1,\ c^1,\ d_1)\  &&&  s_1 && s_2   && u_1&& u_2 && m_S &&  m_U &\cr
\noalign{\hrule}\noalign{\hrule}
&(-5 ,\  12 ,\  0 ,\  2 ,\  -4 ,\  -8 ,\  -1 ,\  0)\  &&& 3.15788 && 5.83333 && 1.26315 && \
0.0666667 &&   2.18 && 13.68 &\cr
 &(-5 ,\  10 ,\  0 ,\  2 ,\  -3 ,\  -8 ,\  -1 ,\  0)\  &&& 3.15788 && 4.83333 && 1.26315 && \
0.0666667 &&   2.18 && 13.68 &\cr
&(-5 ,\  6 ,\  0 ,\  2 ,\  -1 ,\  -8 ,\  -1 ,\  0)\  &&& 3.15788 && 2.83333 && 1.26315 && \
0.0666667 &&   2.18 && 13.68 &\cr
&(-5 ,\ 0 ,\  0 ,\  2 ,\  2 ,\  -8 ,\  -1 ,\  0)\  &&& 3.15788 && -0.166667 && 1.26315 && \
0.0666667 &&   2.18 && 13.68 &\cr
&(-5 ,\  -4 ,\  0 ,\  2 ,\  4 ,\  -8 ,\  -1 ,\  0)\   &&& 3.15788 && -2.16667 && 1.26315 && \
0.0666667 &&   2.18 && 13.68 &\cr
&(-5 ,\  -8 ,\  0 ,\  2 ,\  6 ,\  -8 ,\  -1 ,\  0)\  &&& 3.15788 && -4.16667 && 1.26315 && \
0.0666667 &&   2.18 && 13.68 &\cr
&(-5 ,\ -12 ,\  0 ,\  2 ,\  8 ,\  -8 ,\  -1 ,\  0)\  &&& 3.15788 && -6.16667 && 1.26315 &&\ 
0.0666667 &&   2.18 && 13.68 &\cr
&(5 ,\  10 ,\  0 ,\  -2 ,\  -7 ,\  8 ,\  1 ,\  0)\  &&& 3.15788 && -5.16667 && 1.26315 && \
0.0666667 && 2.18 && 13.68 &\cr
&(5 ,\  8 ,\  0 ,\  -2 ,\  -6 ,\  8 ,\  1 ,\  0)\  &&& 3.15788 && -4.16667 && 1.26315 && \
0.0666667 &&   2.18 && 13.68 &\cr
&(5 ,\  6 ,\  0 ,\  -2 ,\  -5 ,\  8 ,\  1 ,\  0)\   &&& 3.15788 && -3.16667 && 1.26315 && \
0.0666667 &&    2.18 && 13.68 &\cr
&(5 ,\  2 ,\  0 ,\  -2 ,\  -3 ,\  8 ,\  1 ,\  0)\   &&& 3.15788 && -1.16667 && 1.26315 && \
0.0666667 &&   2.18 && 13.68 &\cr
\noalign{\hrule}}}$$}}}
\vskip-6pt
\caption{Discrete landscape of supersymmetric AdS minima.\label{tab11}}
\end{table}
Clearly, the axionic vacuum expectation value 
$s_2$ may be shifted back into the fundamental region 
$s_2\equiv-0.166667$,
while the flux number $N_{\rm flux}$ in \req{NFLUX} and $K_0,W_0$ are 
preserved \cite{KachruA}.
Indeed with these solutions, the three-form flux \req{FLUXG} becomes an ISD--flux.


\subsection{Type IIB fluxes and D-branes}

As we have seen, fluxes provide at  least on the supergravity
level a mechanism to stabilize closed string moduli. With such a
mechanism at our disposal we now need to combine it with a
realistic D-brane sector. Such a combined construction of string
vacua should enable us at least at the string scale to compute concrete values not only for the
topological  data like gauge group and matter but also for the
``continuous'' parameters like coupling constants. 
This would bring us one step closer to a predictive
framework for a single concrete string compactification. Let us
describe such D-brane models with fluxes for our example of
three-form fluxes in type IIB orientifold models (for a collection
of D-brane models with fluxes see
\cite{blt03,cu03,cu03a,ms04,cl04,ms04a,cll05}).


\subsubsection{Freed-Witten anomalies}

Three-form flux induces a tadpole for the RR four form $C_4$. In
this case we were forced to also introduce orientifold O3-planes,
which arise in $\Omega \sigma (-1)^{F_L}$ orientifolds. Such
orientifolds can in addition contain O7-planes. One  may now
introduce magnetized D9-branes (labeled by $a,b,...$) with field
strength $F_a$ to cancel these tadpoles. Under the orientifold
projection such D9-branes are mapped to magnetized anti-D9-branes
labeled $a'$ with opposite field strength $F_{a'}=-F_a$.

However, one first has to make sure that fluxes and D-branes do
not interfere. It is, for instance, known that on D-branes
the Bianchi-identity for the diagonal $U(1)$ gauge field strength $F$ gets
modified in the presence of $H_3$ form flux to
\bea
              dF=-H_3.
\eea
Integrating this equation over a three-cycle inside the brane
world-volume $\cw$ , i.e.\ $\S_3\in H_3(\cw,\mbb Z)$, one finds
that the $H_3$ form flux through any such cycle has to vanish, i.e
\bea
   \int_{\S_3} H_3=0\ .
\eea
This is the vanishing of the Freed-Witten anomaly \cite{Freed:1999vc}.
Another manifestation of this condition arises from the gauge
invariance of the flux-brane supergravity action.
Following \cite{Villadoro:2006ia} let us discuss
this for the magnetized D9-brane above, for which the Freed-Witten
anomaly would simply mean that such space-filling D9-branes cannot
be introduced if there is non-vanishing $H_3$ flux in the internal
directions. On the world-volume of these D9-branes, there exists
the GS coupling
\bea
           \int_{\mbb R^{1,3}\times \cx} C_8\wedge F
         \simeq \int_{\mbb R^{1,3}} C_2 \wedge f\,
\eea
where we have expanded $C_8=C_2 \wedge \ell_s^6 d{\rm vol}_{\cx}$
and $f$
denotes the field strength of the four-dimensional $U(1)$ gauge
potential. Therefore, on the D9-brane there exists an axion-gauge
boson coupling, which is needed for canceling the abelian gauge
anomalies
In other words, the
axion $C_2$ transforms like in \reef{axshift} under gauge
transformations. However, $C_2$ is in four dimensions related via
Hodge-duality of its field strength to $C_0$
(just like $F_9$ and $F_1$ are related in ten dimensions), where $C_0$
is the axionic scalar in the complex dilaton $\t$.
Therefore, the term
\bea
      W=\int_{\cx} \Omega_3\wedge \t H_3
\eea
in the superpotential is not gauge invariant under the shift of
$C_0$ unless $H_3$ vanishes. Thus, we conclude that for
non-vanishing $H_3$ no space-filling D9-branes can be
introduced.
For more general fluxes and D-branes, this
rule can be generalized to:
\begin{center}
{\it
The F-terms induced by fluxes and the D-terms on the D-branes\\
    should depend on and restrict different moduli fields}
\end{center}

As was pointed out in \cite{Marchesano:2006ns},
there seems to be an intriguing  relation between fluxes
and chirality in the sense that:
\begin{center}
{\it
If a flux contributes to a Dp-brane tadpole, then a generalization
of the Freed-Witten anomaly condition forbids the branes leading
to chiral matter with these Dp-branes.}
\end{center}
In our case, the fluxes contribute to the D3-brane tadpole and the
relatively supersymmetric magnetized D9-branes are forbidden.
Indeed, these magnetized D9-branes are the only branes which give
rise to chiral bifundamental matter from open strings stretched
between them and the D3-branes. It is not clear yet how general
such a statement is, but its consequences are clearly very
important for realistic chiral string model building in the
presence of fluxes.


\subsubsection{Example: MSSM-like model on $\mbb T^6/\mbb{Z}_2\times \mbb{Z}_2$}\label{MSSMlike}

As in section \ref{sssecexam} let us consider the orbifold $M=
\mbb T^6/\mbb{Z}_2\times \mbb{Z}_2$ with Hodge numbers
$(h^{2,1},h^{1,1})=(51,3)$
(the
T-dual of the type IIA orientifold studied in
\cite{csu01,csu01a}). In addition one performs the orientifold
projection $\Omega I_6 (-1)^{F_L}$, where $I_6$ reflects all six
internal directions. Flux compactifications on the mirror
symmetric Calabi-Yau given by the orbifold with discrete torsion
were also considered \cite{Blumenhagen:2005tn}.

Turning on three-form fluxes $H_3$ and $F_3$, the Chern-Simons
term in the type IIB effective action induces a four-form tadpole given by
\reef{tadpoleb}.
The field strengths obey the Bianchi identities $dH_3=dF_3=0$,
i.e. obey the flux quantization rules \req{fluxqu}
for any three-cycle $\S_3$. Here $N_{min}$ is an integer
guaranteeing  that in orbifold models only untwisted three-form
fluxes are turned on, for which we can trust the supergravity
approximation. Taking also the orientifold projection into account
for the $\mbb T^6/\mbb{Z}_2\times \mbb{Z}_2$ orbifold, one gets
$N_{\rm min}=8$.

Neglecting the Freed-Witten anomalies for a moment, in order to
cancel the resulting tadpoles, one introduces in the usual way
magnetized D9-branes. We will come back to the Freed-Witten
anomaly at the end of this section. Such a magnetized brane is
characterized by three pairs of integers $(p^I_a,q^I_a)$ which
satisfy
\bea
{q^I_a\over 2\pi}\int_{\mbb T^2_I}   F_a^I = p^I_a\ ,
\eea
where the $q^I_a$ denote the wrapping number of the D9-brane
around the torus $\mbb T^2_I$ and $p^I_a$ is the magnetic flux.
The orientifold projection acts as follows on these quantum
numbers $\Omega I_6 (-1)^{F_L}:(p^I_a,q^I_a)\to (p^I_a,-q^I_a)$.
Since $h^{1,1}=3$ one gets in the orientifold four tadpole
cancellation conditions
\bea
\label{tadflux}
\sum_a N_a\, p^1_a\, p^2_a\, p^3_a &=& 8-{N_{\rm flux}\over 4}\ ,  \\
\sum_a N_a\, p^I_a\, q^J_a\, q^K_a &=& -8\quad {\rm for}\ I\ne J\ne
K\ne I\  .
\eea
In order for each brane to preserve the same supersymmetry as the
orientifold planes, they have to satisfy
\bea
\sum_I \arctan\left(  {q^I_a {t_2}^I\over p^I_a} \right) =0\ ,
\eea
where ${t_2}^I$ denotes the volume of $\mbb T_I^2$ in units of
$\alpha'$. The number of chiral fermions between two different
magnetized branes is given by the index \reef{inter}
and as usual can lead to matter in bifundamental, symmetric or anti-symmetric
representations of the gauge group.

Taking the flux quantization with $N_{\rm min}=8$ into account,
the contribution of the flux to the D3-brane tadpole is given by
$N_{\rm flux}/4\in 16\, \mbb{Z}$. Therefore, for non-trivial flux
the right hand side of the D3-brane tadpole cancellation condition
(\ref{tadflux}) is always negative. One might conclude that
therefore no supersymmetric solutions to the tadpole cancellation
conditions  do exist. However this is too naive, namely in
\cite{ms04,ms04a} it was shown that there exist supersymmetric branes
which give the "wrong" sign in one of the four tadpole
cancellation conditions. Still neglecting the Freed-Witten anomaly,
consider for instance the magnetized brane
$(p_a^I,q_a^I)=((-2,1)(-3,1)(-4,1))$,
which contributes as $(-24,-4,-2,-3)$ to the four tadpole
conditions. Precisely branes of this type were used in
\cite{ms04}
to construct  supersymmetric, chiral, MSSM like flux compactifications.
For illustrative purposes, let us present here only one
of their examples.

Choosing the three-form flux as
\bea
{\ell_s}\, G_3={8\over \sqrt 3}\, e^{-{\pi i\over 3}}\, ( d \overline{z} _1 dz_2 dz_3 +
        dz_1 d \overline{z} _2 dz_3 +   dz_1 dz_2 d \overline{z} _3 ),
\eea
yields a contribution $N_{\rm flux}/4= 48$ to the tadpole
condition and freezes the moduli at $\tau = e^{2\pi
i/3}$. Introducing the supersymmetric
branes shown in Table \ref{susyfluxmodel} cancels all the tadpoles
and gives rise to a one-generation MSSM-like model with gauge
group
\bea
\cg = SU(3)\times SU(2)\times SU(2)\times U(1)_{B-L}\times
    [ U(1)'\times USp(8) ].
\eea
 For more technical and phenomenological details of such models
please consult  the original literature.
\begin{table}
\centering
\vspace{3mm}
\label{twrap}
\begin{tabular}{|c||c|c|c|}
\hline
$N_a$ & $(p^1_a,q^1_a)$ & $(p^2_a,q^2_a)$ & $(p^3_a,q^3_a)$  \\
\hline\hline
$N_a=3$ &  $(1,0)$ & $(1,1)$ & $(1,-1)$ \\
\hline
$N_b=1$ &  $(0,1)$ & $(1,0)$ & $(0,-1)$ \\
\hline
$N_c=1$ &  $(0,1)$ & $(0,-1)$ & $(1,0)$ \\
\hline
$N_d=1$ &  $(1,0)$ & $(1,1)$ & $(1,-1)$ \\
\hline
\hline
$N_{h_1}=1$ &  $(-2,1)$ & $(-3,1)$ & $(-4,1)$ \\
\hline
$N_{h_2}=1$ &  $(-2,1)$ & $(-4,1)$ & $(-3,1)$ \\
\hline
$N_f=4$ &  $(1,0)$ & $(1,0)$ & $(1,0)$ \\
\hline
\end{tabular}
\caption{Wrapping numbers for semi-realistic model.\label{susyfluxmodel}}
\end{table}
Note that the branes $b,c$ can be placed directly on top
the corresponding O7-planes yielding a gauge group
$SU(2)\times SU(2)$.

Most of the branes in the model are actually magnetized D7-branes
and one can show that there is no $H_3$ form flux through these
branes. However, the branes $h_{1,2}$ are truly magnetized
D9-branes giving rise to a non-vanishing Freed-Witten anomaly. One
way to reconcile this is to let the brane $h_i$ and its  $\Omega
I_6 (-1)^{F_L}$ mirror anti-branes recombine (assuming that
this flat direction exists). This new object only carries D7- and
D3-brane charges and can be understood as a D7-brane wrapping a
four cycle on $\mbb T^6$ endowed with a vector bundle. For this
D7-brane the Freed-Witten anomaly vanishes.

This example shows that it is indeed possible to construct
supersymmetric semi-realistic string models with fluxes and partly
frozen moduli. This is an encouraging observation, but of course
much more work is needed to really establish an entire class of
such models (see \cite{cl04,AF04,cll05,Kumar:2005hf,Chen:2005mj,Chen:2005cf} 
for some work in this direction).

\subsection{Flux--induced soft--supersymmetry breaking terms}\label{SOFT}

Whether the minimal supersymmetric standard model
(MSSM) or some of its ramifications will be experimentally discovered at the
LHC is of burning
interest also for theoretical particle physics. In the MSSM, supersymmetry
breaking
is usually parameterized by a set of soft supersymmetry breaking parameters,
like gaugino, squark and slepton masses,
which have the virtue that they
do not spoil the good renormalization behavior of supersymmetric
field theories. But the MSSM does not offer any deeper microscopic explanation
of the origin of the soft supersymmetry breaking parameters.
Nevertheless there are some phenomenological constraints on
the structure of the soft terms, e.g. the absence of flavor
changing neutral currents strongly favors squark masses,
which are universal for all squark flavors.

A controllable way to obtain the soft supersymmetry breaking terms of the MSSM
is provided by coupling the matter sector of the MSSM to local
${\cal N}=1$ supergravity. Then
spontaneous supersymmetry breaking
by non-vanishing F-- or D--terms induces soft supersymmetry breaking terms in the
matter field action.
Superstring theory offers a concrete, microscopic realization
of soft supersymmetry breaking in ${\cal N}=1$ supergravity:
the effective low energy action of supersymmetric string compactifications
to four space-time dimensions is given by the
${\cal N}=1$ supergravity action \req{eff4}.
Furthermore, spontaneous supersymmetry breaking is due to F--terms
of the gauge singlet scalar fields $\Mc$,
namely the dilaton $S$ or the geometric moduli $M$, whose F--terms
are called $F^S$ and $F^{M}$, respectively.
 Then supersymmetry breaking is transmitted
from the gauge neutral sector to the charged sector of the MSSM
by gravitational interactions. This scenario already allows
for a fairly model independent analysis of the soft terms, which are all
proportional to certain combinations of  $F^S$ or $F^{M}$.
In particular, the dilaton dominated scenario
with $F^S\neq 0$, $F^{M}=0$ possesses the feature
of flavor universal soft scalar masses, which is usually spoiled
by non-vanishing vevs for $F^{M}$.
In more generic
scenarios, in which both  $F^S\neq 0$ and $F^{M}\neq 0$, the
soft supersymmetry breaking terms
can be nicely parameterized
by a so-called goldstino angle $\tan\theta_g\sim F^S/F^T$, where
$T$ is the overall volume modulus of the internal space.

The final step for a complete understanding of the soft-terms
is undertaken by knowing
$(i)$ how the matter sector of the MSSM is microscopically built in string
theory, and $(ii)$ how the supersymmetry breaking auxiliary fields $F^S$,
$F^{M}$
are induced, i.e. by knowing how a non-trivial effective superpotential
for the fields $S$ and $M$ is generated.
In this subsection we review generic aspects of soft--supersymmetry
terms in type IIB orientifolds with D3 and D7-branes and three--form  flux.

In order to derive the soft supersymmetry breaking parameters, one has
to compute the couplings between the open string matter fields on the
D3/D7-branes and the closed string
three-form field strengths $G_3$.
The soft supersymmetry breaking terms can be derived
either by studying the Born-Infeld action
on the D-brane world volumes coupled to the flux $G_3$ as accomplished
for D3-branes in \cite{ciu03,ggjl03}, and for D7--branes in
\cite{ciu04}, or by coupling the effective action
from open/closed string scattering amplitudes to the effective closed
string action with three-form fluxes turned on, as it was performed
for D3- and D7-branes in \cite{lrs04,lrs04a}.
Note that  in  the last references also the open string
two--form $f$--flux on the D7-branes has been taken into account, which is crucial
for realistic model building with chiral fermions.
See also  \cite{fi04,recently,lmrs05}.
In any case, the results of the two different approaches \cite{lrs04}
and \cite{ciu03,ggjl03,ciu04} are completely consistent with each other
and lead to identical results for vanishing $f$--flux.
Results for type IIA orientifolds may be found in \cite{kn03,kklw04,msw04}.

We start with the effective action \req{eff4} and the expansions
\req{mostK}.
The superpotential is modified by potential supersymmetry breaking
terms:
\bea\label{mostWW}
W(\Mc,\Cc)&=&W_0(\Mc)+\sum_{\al}\tilde a_\al(\Mc)\ \Cc_\al+\h
\sum_{\al,\bet} \tilde\mu_{\al\bet}(\Mc)\ \Cc_\al\ \Cc_\bet\\
&+&\fc{1}{3}\ \sum_{\al,\bet,\gamma} \tilde
Y_{\al\bet\gamma}(\Mc)\ \Cc_\al\ \Cc_\bet\ \Cc_\gamma+\ldots\ .\nnn
\eea
The first term encodes the three-form flux dependence \req{gvw}, while the second
term may appear \eg in the presence of (primitive) world--volume two-form fluxes on D7-branes
\cite{lmrs05,Jockers:2005zy,Jockers:2005pn}.
Furthermore, the third term gives rise to the supersymmetric mass term
\zbe\label{susymass}
m_{\al\ov\bet,Susy}^2=\mu_{\al\gamma}\ G^{\gamma\ov\delta}\ \ov\mu_{\ov\delta\ov\bet}
\ee
for the chiral fields $\Cc$, with:
\eqn{Susymass}{\mu_{\al\bet}=
e^{\kappa_4^2K_0/2}\ \tilde\mu_{\al\bet}+m_{3/2}\ H_{\al\bet}-\ov F^{\ov I}\
\ov\partial_{\ov I}\ H_{\al\bet}\ ,}
with the (complex) gravitino mass $m_{3/2}=\kappa_4^2\ e^{\kappa_4^2K_0/2}W_0$.
We shall see in this subsection that such a mass term is generated
by ISD $(2,1)$--form fluxes $G_3$ \cite{ciu04,lmrs05}.
The second term in \req{mostWW}, which originates from primitive world--volume two-form
fluxes \cite{lmrs05},  gives rise to non--vanishing $F_\Cc$--terms
and a non--vanishing scalar potential even at $\Cc=0$. In the following we consider the case
$\tilde a_\al=0$.

Before computing the scalar potential ${\cal V}(\phi,\ov \phi)$ it is convenient to introduce
the effective superpotential $W_{\rm eff}$ encoding all supersymmetry
preserving terms of \req{mostWW}:
\eqn{Weff}{W_{\rm eff}=\h
\sum_{\al,\bet} \mu_{\al\bet}(\Mc)\ \Cc_\al\ \Cc_\bet+
\fc{1}{3}\ \sum_{\al,\bet,\gamma} Y_{\al\bet\gamma}(\Mc)\ \Cc_\al\ \Cc_\bet\ \Cc_\gamma\ ,}
with
\eqn{}{
Y_{\al\bet\gamma}=e^{\kappa_4^2K_0/2}\ \tilde Y_{\al\bet\gamma}\ .}
The full scalar potential is a real function of all moduli fields $\phi,\ov\phi$
and may be determined with the \eqq \req{SGpot}.
Up to second order in the matter fields $\Cc$ it assumes the form
\be\label{effPot}
\ba{lcl}
\ds{\Vc(\Mc,\ov \Mc,\Cc,\ov \Cc)}&=&\ds{ \Vc_{\rm F}(\Mc,\ov \Mc)+\Vc_{\rm D}(\Mc,\ov \Mc)+
\Vc_{\rm D}(\Cc,\ov \Cc)
+\partial_\al W_{eff}\ G^{\al\ov\bet}\ \ov\partial_{\ov \bet} \ov W_{eff} }\\[3mm]
&+&\ds{m_{\al\ov\bet}^2\ \Cc^\al\ \ov \Cc^{\ov\bet}+\lf(\ \fc{1}{3}\ A_{\al\bet\gamma}\
\Cc^\al\ \Cc^\bet\ \Cc^\gamma+\h\ B_{\al\bet}\ \Cc^\al\ \Cc^\bet+hc.\ \ri)+\ldots\ ,}\nnn
\ea
\ee
with the D--term potentials $\Vc_{\rm D}$ given in \req{Dterms} and the scalar potential
\zbe\label{with}
 \Vc_F(\Mc,\ov \Mc)=e^{\kappa_4^2K_0}\ \lf(\ K_0^{I\ov J}\  D_IW_0\ D_{\ov
  I} \ov W_0-3\kappa_4^2\ |W_0|^2\ \ri)\ ,
\ee
derived from the lowest order K\"ahler potential $K_0$ and
superpotential $W_0$ (closed string sector only).
In the following, in a sum
capital roman letters denote the closed string moduli, while greek letters
run over the matter fields. Both types of moduli fields are summarized in small roman letters.
The second line of \req{effPot} gives rise to a
series of bosonic scalar soft--supersymmetry breaking terms \cite{Hall}.
The latter split the masses of the scalars $\Cc$ of the chiral
multiplets while guaranteeing the absence of quadratic divergences.
For $\tilde a_\al=0$ the parameters $m^2,A$ and $B$ have been computed for vanishing
F--term scalar
potential, \ie $\Vc_F(\Mc,\ov \Mc)=0$, in  \cite{kaplu,IbanezHC} and for
$\Vc_F(\Mc,\ov \Mc)\neq 0$ in \cite{Spain1,ggjl03}.
In that case these parameters take the form:
\bea
m_{\al\ov\bet}^2&=&\lf[\ |m_{3/2}|^2+\Vc_F(\Mc,\ov \Mc)\ \ri]\ G_{\al\ov\bet}-
F^I\ \ov F^{\ov j}\ R_{I\ov J\al\ov\bet}\ ,\nnn\\[3mm]
A_{\al\bet\gamma}&=&F^I\ \lf[\ \partial_I Y_{\al\bet\gamma}+\h\ \partial_IK_0\ Y_{\al\bet\gamma}-
\Gamma_{I(\al}^\delta\ Y_{\bet\gamma)\delta}\ \ri]\ ,\label{parameter}\nnn\\[3mm]
B_{\al\bet}&=&\lf[\ 2\ |m_{3/2}|^2+\Vc_F(\Mc,\ov \Mc)\ \ri]\ H_{\al\bet}-\ov
m_{3/2}\ \ov F^{\ov I}\ \ov\partial_{\ov I} H_{\al\bet}\\[3mm]
&+&m_{3/2}\ F^I\ \lf(\ \partial_IH_{\al\bet}-\Gamma_{I\alpha}^\delta\
H_{\delta\beta}-\Gamma_{I\beta}^\delta\ H_{\alpha\delta}\ \ri)\nnn\\[3mm]
&-&F^I\ov F^{\ov J}\ \lf(\ \partial_I\ov\partial_{\ov J}
H_{\al\bet}-\Gamma_{I\alpha}^\delta\
\ov \partial_{\ov J}H_{\delta\beta}-\Gamma_{I\beta}^\delta\
\ov \partial_{\ov J}H_{\alpha\delta}\ \ri)\nnn\\[3mm]
&-&e^{K_0/2}\ \tilde\mu_{\al\bet}\ \ov m_{3/2}+e^{K_0/2}\ F^I\
\lf(\ \partial_I\tilde\mu_{\al\bet}+\partial_IK_0\ \tilde \mu_{\al\bet}
-\Gamma_{I\alpha}^\delta\ \tilde\mu_{\delta\beta}
-\Gamma_{I\beta}^\delta\  \tilde\mu_{\alpha\delta}\ \ri)\ ,\nnn
\eea
with the F--terms:
\eqn{Fterms}{
\ov F^{\ov N}=e^{\kappa_4^2K_0/2}\ K_0^{\ov NL}\ (\ \partial_LW_0+\kappa_4^2\ W\ \partial_L K_0\ )\ .}
The connection for Hermitean manifolds is $\Gamma^\al_{I\bet}=G^{\al\ov j}\ \partial_I
G_{\beta\ov j}$ and the curvature tensor $R_{I\ov
  J\al\ov\bet}=\partial_I\partial_{\ov J} G_{\al\ov
  \beta}-\Gamma^i_{I\al}G_{i\ov j}\Gamma^{\ov j}_{\ov J\ov\beta}$ has been expressed
in \eqq \req{mixed} in terms of the K\"ahler potential $K(\phi,\ov\phi)$.
The gaugino masses $m_a$ are determined by the formula
\eqn{gauginomasses}{
m_a=\h\ F^I\ \partial_I \ln \re(f_a)\ ,}
with $f_a$ the gauge kinetic function of the gauge group $G_a$.

In the following we shall discuss the soft--supersymmetry breaking
terms \req{parameter} induced by three- and two-form
background fluxes in \tb orientifolds.
All the relevant properties of fluxed--induced soft--supersymmetry
breaking terms in \tb orientifolds with matter fields originating from
D--branes may be seen at one particular example, namely
the \tb orientifold of $\mbb T^4/\mbb Z_2\times \mbb T^2$ (\cff subsection \ref{DD33DD77} for more details).
For this model we have four stacks of eight D7-branes, which are wrapped around the orbifold
$\mbb T^4/\mbb Z_2$ and  placed at the four orientifold O7-planes.
The latter are located at the four $\mbb Z_2$ fixpoints of the $\mbb T^2$.
Within this framework we calculate the flux--induced
soft supersymmetry breaking terms \req{parameter} stemming from the effective four--dimensional
D7-brane action.
From the complex three-forms \req{cplxz} the following basis elements represent
$$\omega_{A_i}\ ,\ \omega_{B_j}\ \ ,\ \ \omega_{C_2}\ ,\ \omega_{D_2}\ ,\ \omega_{C_4}\ ,\
\omega_{D_4}\ \ \ ,\ \ \ i,j=0,\ldots,3\ $$
a basis of $H^{3}(\mbb T^4/\mbb Z_2\times \mbb T^2)$, \ie they are
invariant under the orbifold group.
Only the primitive forms $\omega_{A_i}, \omega_{B_j}$ will contribute to the scalar potential
\req{with} and give rise to soft--terms.
The (untwisted) three-form flux may be expanded as in \req{GR}.

In the model under consideration there are two kinds of matter fields
$\Cc$. They originate from open strings  stretched between different
stacks of D7-branes or they describe the position $C_{a}$ of one stack $a$
of D7-branes along the $\mbb T^2$. In the following we shall focus on the latter.
The K\"ahler potential for the closed and open string moduli has been given in \req{KAETOMM}.
According to \req{mostK} it assumes the expansion
\eqn{mostKK}{
K(\Mc,\ov\Mc,C,\ov C)= K_0(\Mc,\ov\Mc)+
G_{C_{a}\ov C_{a}}\ C_{a}\ \ov C_{a}+\lf(\ \h\ H_{C_{a}C_{a}}\ C_{a}\ C_{a}+
h.c.\ \ri)+\ldots\ ,}
with ($U\equiv U^3)$:
\eqn{withGH}{
\kappa_4^{2}\ G_{C_{a}\ov C_{a}}=\fc{1}{(S+\ov S)(U+\ov U)}\ \ \ ,\ \ \
\kappa_4^{2}\ H_{C_{a}C_{a}}=\fc{1}{(S+\ov S)(U+\ov U)}\ .}
The expansion of the superpotential $W$ takes the form \req{mostWW}
\eqn{mostWWWW}{
W(\Mc,C)=W_0(\Mc)+{1\over 2}\ \tilde\mu_{C_{a}C_{a}}\ C_{a}\ C_{a}+{1\over 3}\
\tilde Y_{ABC}\ C_{A}\ C_{B}\ C_{C}+\ldots\ ,}
with the superpotential term $W_0$, given in \req{withh} with $U^3\equiv U$, 
the three-form flux dependent $\mu$--term
\eqn{muu}{
\tilde \mu_{C_{7_a}C_{7_a}}=-d_3 - c_2\ U^1 - c_1\ U^2 + c_0\ U^1\ U^2\ ,}
and the holomorphic three--point couplings $\tilde Y_{ABC}$.
The above superpotential for the orientifold $\mbb T^4/\mbb Z_2\times \mbb T^2$ may be derived
from gauging some PQ--symmetries of axions of ${\cal N}=2$ hypermultiplet or
from F--theory \cite{lmrs05}.
According to \req{Weff} the $\mu$--term \req{muu} and the Yukawa
couplings $\tilde Y_{ABC}$ give rise to the effective superpotential
\eqn{Wefff}{
W_{\rm eff}(C)=\h\ \mu_{C_{a}C_{a}}\ (C_{a})^2+\fc{1}{3}\ Y_{ABC}\ C_AC_BC_C+\ldots\ ,}
with the effective $\mu$--term \req{Susymass}:
\eqn{effmu}{
\mu_{C_{7_a}C_{7_a}}=e^{\kappa_4^2K_0/2}\ \tilde\mu_{C_{7_a}C_{7_a}}+m_{3/2}\ H_{C_{7_a}C_{7_a}}-
\ov F^{\ov M}\ \ov \partial_{\ov M}\ H_{C_{7_a}C_{7_a}}=-Y^{-1/2}\ {G_{C_{7_a}\ov C_{7_a}}}
\ \int \ov G_3\wedge \omega_{A_3}\ ,}
with $Y=e^{-\kappa_4^2K_0}$.
From \req{parameter} we determine the following soft--masses $m^2_{C_{7_a}C_{7_a}}$
for the D7-brane position moduli $C_{7_a}$:
\eqn{scalarmasss}{
m^2_{C_{7_a}C_{7_a}}=|Y|^{-1}\ \lf\{\ \lf|\!\int G_3\wedge \Omega\,\ri|^2+\lf|\!\int G_3\wedge
\omega_{A_1}\ri|^2+\lf|\!\int G_3\wedge \omega_{A_2}\ri|^2\ \ri\}\ G_{C_{7_a}\ov C_{7_a}}\ .}
Furthermore, according to \req{susymass}
the supersymmetric mass becomes:
\eqn{SUSYMASS}{
m^2_{C_{7_a}\ov C_{7_a},Susy}=G^{C_{7_a}\ov C_{7_a}}\ |\mu_{C_{7_a}C_{7_a}}|^2=
|Y|^{-1}\ \lf|\ \int \ov G_3\wedge \omega_{A_3}\ \ri|^2\ G_{C_{7_a}\ov C_{7_a}}\ .}
From the two expressions \req{scalarmasss} we conclude, that both  ISD
$(0,3)$-- and $(1,2)$--form fluxes contribute to the soft--mass
\req{scalarmasss}.
On the other hand, a $(2,1)$--flux preserving supersymmetry
gives rise to the effective $\mu$--term \req{SUSYMASS}.
The fact, that an ISD--flux gives rise to non--vanishing scalar masses
for D7-brane position moduli \cite{lrs04,ciu04} is to be contrasted to scalar masses
of  D3-brane position moduli, which only receive contributions from
IASD--fluxes \cite{ciu03,ggjl03}.

Let us now discuss the gaugino masses. The gauge kinetic function for
$SO(8)$ gauge theory on one stack of D7-branes is given in
\req{gaugef}.
With this we may determine the gaugino masses \req{gauginomasses}
\eqn{Gauginomasses}{
m_a=\fc{1}{T+\ov T}\ F^T\ .}
Hence the latter is sensitive to ISD $(0,3)$--form fluxes only.
In Table \ref{Pattern} we show the
flux components contributing to soft--supersymmetry breaking terms
\req{parameter} on D7--branes.
\begin{table}[h]
\centering
\vspace{3mm}
\begin{tabular}{|c||c|c|c|c|}
\hline
\ & $(2,1)$& $(0,3)$&$(3,0)$& $(1,2)$\\
\hline\hline
$m_{\al\al}$& $-$&$+$&$-$&$+$\\
$\mu_{\al\bet}$& $+$&$-$&$-$&$-$\\
$m_a$&$-$&$+$&$-$&$-$\\
\hline
\end{tabular}
\caption{Fluxes contributing to soft--supersymmetry breaking terms on D7-branes.\label{Pattern}}
\end{table}

If in addition to the three-form flux $G_3$ also (non--primitive) two-form fluxes $f$
on the D7-brane world--volume  are
turned on, the results \req{scalarmasss}, \req{SUSYMASS} and
\req{Gauginomasses} change. Note, that non--primitive two-form fluxes contribute to
the D--term potential, while $\tilde a_\al=0$ in \req{mostWW}.
The expressions become rather long and can  be found  in  \cite{lmrs05}.
In that case all flux components contribute to the soft--mass
\req{scalarmasss}. Furthermore, the gaugino mass receives also a contribution
from an IASD $(3,0)$--flux. From those results the soft--masses for a stack
of (space--time filling) D3-branes may be easily anticipated by sending the two-form
flux on the D7-branes to infinity.
In Table \ref{pattern} we show the
flux components contributing to soft--supersymmetry breaking terms
\req{parameter} on  D3-branes. In particular, no soft--masses are generated from ISD--fluxes.
This fact is a manifestation of the no--scale structure of the effective action of
a D3-brane (at leading order in $\ap$).
\begin{table}[h]
\centering
\vspace{3mm}
\begin{tabular}{|c||c|c|c|c|}
\hline
\ & $(2,1)$& $(0,3)$&$(3,0)$& $(1,2)$\\
\hline\hline
$m_{\al\al}$& $-$&$-$&$+$&$-$\\
$\mu_{\al\bet}$& $-$&$-$&$-$&$+$\\
$m_a$&$-$&$-$&$+$&$-$\\
\hline
\end{tabular}
\caption{Fluxes contributing to soft--supersymmetry breaking terms on D3-branes.\label{pattern}}
\end{table}

In the presence of world--volume two-form flux on the D7-brane, \ie
with mixed $D/N$--boundary conditions, the pattern of both tables \ref{Pattern} and \ref{pattern}
are combined  and give a rich structure of soft--terms \cite{lrs04,lrs04a}.
Most importantly, chiral scalar fields, which correspond to twisted
open string sectors, \ie open strings which stretch
between two D7-branes with
different type of $f$--flux boundary conditions, get also masses from $(3,0)$-
as well as from $(0,3)$-fluxes.
In particular, these effects appear in realistic models with
three chiral generations as discussed in subsection \ref{MSSMlike}.
We refer the reader to the following literature \cite{lrs04,lrs04a,fi04,recently}
for a detailed account. In addition, some phenomenological research on flux--induced
soft supersymmetry terms
may be found in \cite{IBANEZ04,Allanach:2005yq}.

The soft-supersymmetry breaking pattern, discussed above, 
holds also quite generically for other orientifold compactifications,
see recent calculations on a CY manifold in the large radius approximation
\cite{conlon2,conlon3,conlon4,conlon5}. In this non--toroidal case
the K\"ahler potential for the charged matter fields is the hardest part  of
the computation. However, due to non--renormalization theorems the
scaling behavior (encoded in modular weights) of the K\"ahler potential with respect
to certain K\"ahler moduli could
indirectly be obtained from the physical Yukawa couplings in type IIB
orientifolds, thereby neglecting the dependence on the complex structure moduli \cite{conlon4}.
It should however be stressed, that such an attempt cannot provide non--trivial functions\footnote{Confer 
also footnote \ref{footn} for a comparison of
a field--theory vs. string theory computation of the Yukawa couplings.} encoding stringy effects
as they appear in the matter metrics \req{twisted1/4}.
The latter allow for a non--trivial and rich structure in the
soft--supersymmetry breaking terms \cite{lrs04a}.
Nonetheless, the computation of the resulting 
soft supersymmetry breaking terms in \cite{conlon5} showed
unexpected cancellations and in leading order led to
flavor universal soft terms as in the cases described above.
It would clearly be of some interest to
perform such computations of soft supersymmetry breaking terms for
a larger set of models. Given these terms, available software
packages allow to run these string derived couplings down to the
TeV scale and compare with LHC results \cite{Kane:2006yi}
(of course assuming that  supersymmetry 
at the TeV scale is found at LHC).

\subsection{Moduli stabilization in \tb orientifolds}
\label{MODSTAB}

For \tb orientifolds the work of KKLT \cite{kklt03} proposes a mechanism to stabilize
all moduli at a small positive cosmological constant. Let us review this propoasal.

In the following a flux compactification of a \tb CY orientifold
with $\m=0$ and the general K\"ahler potential \req{FULL} is assumed.
Furthermore, we consider the racetrack superpotential:
\eqn{SUP}{
W=W_0(S,U)+\sum_{j=1}^\npp\beta_j(\Cc)\ \gamma_j(S,U)\ e^{a_j\ T^j}\ .}
The first term $W_0$ of \req{SUP} represents the tree--level flux superpotential \req{gvw}.
On the other hand, the sum of exponentials accounts for D3--brane instantons and gaugino
condensation on stacks of D7--branes.
Here the function $\gamma_j(S,U)$ includes possible further contributions from the dilaton and
the complex structure moduli, whereas $\beta_j(\Cc)$ includes
possible contributions of 
charged matter fields
$\Cc$ (see below).
The D3--instantons come from wrapping (Euclidean)  D3--branes on internal
four--cycles $C_j$ of the CY orientifold ${\cal X}$. The latter have the volume $\re(T^j)$
(\cff \eqq \req{NICE})
and lead to the instanton effect $e^{-2\pi T^j}$ in the superpotential, \ie $a_j=-2\pi$.
In order for such a superpotential being generated, one needs precisely
two fermionic zero modes being present on the world volume of
the Euclidean D3--branes.
In F/M--theory, where one considers
instead of the D3-branes Euclidean M5--branes, wrapped around 6-dimensional
divisors of a CY fourfold, a necessary
condition for having two fermionic zero
modes is that the arithmetic genus $\chi$ of the divisor
is equal to one \cite{Witten:1996bn}:
\begin{equation}\label{countD}
\chi({\rm wrapped~divisor})=h^{(0,0)}-h^{(0,1)}+h^{(0,2)}-h^{(0,3)}=1\ .
\end{equation}
However the number of zero modes may change in the presence of
background fluxes \cite{Kallosh:2005yu,Saulina:2005ve,Kallosh:2005gs,Bergshoeff:2005yp,Park:2005hj,Gomis:2005wc,Lust:2005cu, LRSSii}. 
In type IIB a similar condition than \req{countD} can be formulated under certain circumstances. 
In fact in \cite{Bergshoeff:2005yp,LRSSii} it is shown directly in type IIB, 
how the zero modes counting is changed in the presence of background 
three--form fluxes $G_3$ and orientifolds. 
Additionally, it was shown in \cite{Lust:2005cu} that only the $(2,1)$--component of the $G_3$--flux may  
lift zero modes. The advantage of the latter counting procedure is that it is not 
necessary to do an F-theory lift, the calculations can be done directly in the type IIB picture. 
Although there exist criteria for the existence of D3--instantons contributing 
in \req{SUP}, the coefficient $\gamma_j(S,U)$, which  represents the one--loop 
determinant of the instanton solution, is hard to compute directly. 
So far  this has been achieved (in)directly in F-- or M--theory \cite{Witten:1996bn} or by some duality arguments 
\cite{Berglund:2005dm,Aspinwall:2005ad}.

On the other hand, gaugino condensation on a D7--brane, which is wrapped on the four--cycle $C_j$,
can only occur under certain instances. The gauge coupling on a D7--brane
is given by $\re(T^j)$, \cff \eqq \req{gaugef}.
Hence, gaugino condensation on this D7--brane yields the effect
$e^{-T^j/b_a}$ in the superpotential, with $b_a$ the $\beta$--function
coefficient of the effective super Yang--Mills gauge theory, which lives on the world volume of
the wrapped D7--branes.
Gaugino condensation can only arise if the gauge theory is asymptotically free,
\ie if $b_a>0$. This puts some constraints on the possible matter
spectrum on the D7--branes.
The simplest possibility, which always leads to a gaugino
condensate, is that the gauge theory on the D7--branes is a pure
${\cal N}=1$ Yang-Mills theory without any massless
fundamental or adjoint chiral matter
fields. For example,  for the gauge group $SU(N_c)$ we have
$b_{SU(N_c)}=\fc{N_c}{2\pi}$, \ie $a_j=-\fc{2\pi}{N_c}$ and $\beta_j(\Cc)=1$.

Gaugino condensation can also occur if there are $N_f$ charged matter fields in the fundamental
plus antifundamental representations of the confining gauge group $SU(N_c)$.
Specifically, for the case $N_f<N_c$, one obtains the following
Affleck-Dine-Seiberg (ADS) non-perturbative
superpotential \cite{Affleck:1983mk}
\begin{equation} \label{wads}
 W_{ADS}=
\gamma_j(S,U) \ \left( \frac{e^{- 8 \pi^2 T^{j}}}{\det(M)} \right)^{\frac{1}{N_c-N_f}}\ ,
\end{equation}
which now contains   an
 additional factor involving the meson determinant
 \begin{equation}
 \det(M^{i}_{j})\equiv \det (\tilde{\Cc}^{ia}\Cc_{ja}),
 \end{equation}
 where the $\Cc^{ia}$ are charged matter fields, and $i,j$ denotes
 the flavor and $a,b$ the 
color index.

 In order to realize this superpotential in type IIB orientifold
 compactifications, one considers two different stacks of D7-branes
 one denoted
by D7$_c$ and  the other one by D7$_f$  \cite{Haack:2006cy}.
On the first stack, which consists of $N_c$ D7$_c$-branes wrapped around
the four-cycle $C_c$,  we assume a gauge group
$G_{c}$ that can undergo gaugino condensation.
As generalizations of the ADS superpotential
in the presence of (anti-)symmetric tensor representations
  are not very well
understood,  we assume that
these representations are absent. The simplest situation in which this is the case is when
 the D7$_{c}$ stack does not intersect the O-planes.
 The gauge group   $G_{c}$  will therefore be
  of the form $U(N_{c})\cong SU(N_c)\times U(1)_{c}$.

The other stack of $\tilde N_f$ D7$_{f}$-branes being wrapped around
$C_f$,  has a gauge group
 $G_{f}$ that contains  at least one  $U(1)$ factor
denoted by $U(1)_{f}$.
In the generic case, when  the D7$_{f}$ stack does not lie on top of the
O7-planes, one has the usual unitary gauge group including an Abelian $U(1)_f$
factor.  On the other hand,   in the  case when  the D7$_{f}$ stack coincides
with  an O7-plane,  $G_{f}$ becomes enhanced to a symplectic or orthogonal
gauge group. This group   can be broken to a unitary group with Abelian
factors by switching on appropriate
  world volume fluxes, so this might a priori also be a valid option.
However,  for simplicity we want to assume
that
 the D7$_{f}$ branes are not on top of
O7-planes, and hence $G_f=U(1)_f\times SU(\tilde N_f)$.
The bifundamental matter fields $\Cc$ are coming from
open strings stretched between the $f$- and $c$-stacks. They each transform
under $(SU(\tilde N_f),SU(N_c))_{U(1)_f}$ in the representation
\begin{equation} \label{representation}
(\tilde N_f, N_c \oplus \bar N_c)_{q_f=1}\ ,
\end{equation}
where the subscript denotes their charge under the  $U(1)_f$. The fields
in the $(\tilde N_f, \bar N_c)$-representation originate from
strings stretched from D7$_{c}$ to
D7$_{f}$, whereas the ones transforming in the $(\tilde N_f, N_c)$-representation
arise from open strings stretched from the orientifold image of D7$_{c}$ to
D7$_{f}$.
Obviously, with only this particle content $SU(\tilde N_{f})$ would be
anomalous. Thus, additional fields charged under $SU(\tilde N_{f})$, for instance from
other  brane stacks that intersect D7$_{f}$, have to be present in a globally
consistent model. Alternatively if $\tilde N_f=1$ there is only an $U(1)_f$ anomaly
that can be canceled by the Green-Schwarz mechanism.

In order to have a non-zero flavor number $N_f$ one has to turn on open string
F-flux on the D7$_f$-branes.
Then the number of
chiral bifundamentals is given by the index of the Dirac operator on the intersection
of the two D7-branes and in the background of the world volume flux along this intersection.
We assume that there is no world-volume flux on the D7$_{c}$-stack,
otherwise the difference
of the fluxes on the two stacks along their intersection locus
would enter the index. This would lead to different numbers of
$(\tilde N_f, \bar N_c)$- and $(\tilde N_f, N_c)$-representations.
Under this assumption, the number of $(\tilde N_f, \bar N_c)$-representations
is given by
\begin{equation} \label{index}
n={\rm index}(\nabla)  = \alpha^{\prime -1}
\int_{C_f \cap C_c} \frac{F}{2 \pi} \ ,
\end{equation}
where we introduced the factors $\alpha^{\prime -1}$ for dimensional reasons.
The number of $(\tilde N_f, N_c)$-representations is given by (\ref{index}) as well.
Therefore, the total number of bifundamental fields $\Cc$ is given by $N_f=n\tilde N_f$.

The ADS superpotential $W_{ADS}$ is also important if D-term potentials arise
from world-volume F-flux on the D7$_f$-branes.  In this case the non-trivial shift
of the K\"ahler modulus $T^j$ under  the anomalous $U(1)_f$ gauge symmetry is compensated
by the transformation of the matter fields $\Cc$ under this group, such that the superpotential
$W_{ADS}$ is completely invariant \cite{Achucarro:2006zf,Haack:2006cy}.
This observation becomes relevant when performing
the uplift from an AdS-vacuum to a dS-vacuum by the D-term potential 
instead of the inclusion of anti--$D3$--branes (see below).

Hence the closed string moduli stabilization procedure is generically influenced by
local brane properties and open string moduli. As a result one should stabilize
both open and closed string moduli at once. However in practice one may
first solve for $F_\Cc=0$ and substitute the resulting solution for the open string
modulus $\Cc$ into the remaining closed string system. This reduces the
problem to a KKLT scenario. This way it has been shown in \cite{leb0},
that it is safe to ignore the effects of anomalous $U(1)$'s in the KKLT setup.
Therefore, in the following we shall assume the absence of charged 
matter fields $\Cc$ in the  superpotential and vanishing D--terms.

In \eqq \req{SUP} we assume $\beta_j(\Cc)=1$, $W_0\in \ICC$, $\gamma_j\in \ICC$, and $a_j \in \IR_{-}$.
We do not consider a possible open string moduli dependence of the superpotential
\cite{gktt04,lmrs05}.
On the $D7$--brane,  $\gamma_j(S,U)$ may comprise one--loop effects and further instanton
effects from $D(-1)$--branes: One loop corrections to the gauge coupling
give rise to \cite{ls03}
\eqn{ext}{\gamma_j\sim \eta(U)^{-2/b_a}\ ,}
while additional instantons in the $D7$--gauge theory amount to:
\eqn{exti}{\gamma_j\sim e^{-S/b_a\int_{C^j_4} F\wedge F}\ .}

Stabilization of all moduli and breaking supersymmetry at a positive vacuum energy
is accomplished through three steps.\footnote{The possibility to stabilize all moduli in Minkowski vacua in type
IIB orientifolds via racetrack superpotentials was discussed in
\cite{Blanco-Pillado:2005fn,Krefl:2006vu}.}
One first dynamically fixes the dilaton $S$ and the complex structure moduli $U^\lambda$
through the tree--level piece $W_0$ (given in \eqq \req{gvw}) of the superpotential.
This is achieved with a generic three--form flux $G_3$ with both ISD-- and IASD--flux
components. At the minimum of the scalar potential
in the complex structure and dilaton directions,
the flux becomes ISD and the potential assumes the value $\Vc_0(S,U)=-3e^K |W_0|^2$.
The soft masses $m_S,m_U$ (\cff subsection \ref{SOFT})
for the dilaton and complex structure scalars are generically
of the order $\ap/R^3$
\cite{lrs04a}. In the large radius approximation $\re(T)\gg 1$,
the non--perturbative terms in \req{SUP} only amount to
a small exponentially suppressed additional contribution to $m_S,m_{U}$. According to
\cite{HPNii} the latter is negligible.
The second step is the addition of the non--perturbative
piece to the superpotential \req{SUP}, which allows the stabilization of the K\"ahler moduli
$T^j$ at a supersymmetric AdS minimum. The soft masses for the K\"ahler moduli
are much smaller than soft masses $m_S$ and $m_{U}$. This property allows us to
separate the first and second step, \ie to effectively first integrate out the dilaton
and complex structure moduli.
Nonetheless, strictly speaking these two steps should be treated at the same time.

The dynamics of the effective ${\cal N}=1$ supergravity theory with the superpotential \req{SUP}
is determined by the associated scalar potential \req{SGpot}
\eqn{Scalarpot}{
\Vc_{AdS}=e^{\kappa_4^2K}\ \lf(|D_S W|^2+\sum_{i=1}^{h_+^{1,1}({\cal X})}|D_{T^i}W|^2+
\sum_{j=1}^{h_-^{2,1}({\cal X})}|D_{U^j}W|^2-3\ \kappa_4^2\ |W|^2\ \ri)\ ,}
with the K\"ahler potential for the fields $S,T^j,U^j$, given in \req{FULL}.
Supersymmetric vacuum solutions are found  by finding the zeros of the F--terms:
\zbe\label{FTERMS}
\ov F^{\ov M}=K^{\ov M J}\ (\partial_J W+\kappa_4^2\ W\ K_J)\ .
\ee
Solutions to the equations $F^M=0$ give rise to extremal points of the scalar potential.
In addition, it has to be verified whether those zeros lead to a stable minimum.
Since the matrix $K^{\ov M J}$ is positive definite, the zeros
in the moduli space are determined by the $1+h_-^{2,1}({\cal X})+\npp$ equations
\bea
\ds{\partial_{S} W+\kappa_4^2\ W\ K_{S}}&=&0\ ,\nnn\\
\ds{\partial_{U^\lambda} W+\kappa_4^2\ W\ K_{U^\lambda}}&=&\ds{0\ \ \ ,\ \ \ \lambda=1,\ldots,
h_-^{2,1}({\cal X})\ ,}\label{FEQ}\\
\ds{\partial_{T^j} W+\kappa_4^2\ W\ K_{T^j}}&=&\ds{0\ \ \ ,\ \ \ j=1,\ldots,\npp}\nnn
\eea
following from the requirement of vanishing F--terms \req{FTERMS}.
These equations turn into the $1+\npp+h_{-}^{2,1}({\cal X})$ equations, whose solution
allow  to fix the vev for dilaton $S$, the $h_{-}^{2,1}({\cal X})$ complex
structure $U^\lambda$ and $\npp$ K\"ahler moduli $T^j$.
At these values, the scalar potential assumes the negative value
\eqn{issl}{\Vc_{AdS}=-3\ \kappa_4^2\ e^{\kappa_4^2K}\ |W|^2}
and supersymmetry is restored at the AdS minimum.
Typically the sizes of the four--cycles are fixed at reasonable large values.
To summarize, with the first two steps of the KKLT mechanism
stabilization of all moduli is achieved.
This has been demonstrated in much detail for some selected \tb orientifold examples
in \cite{Race,DenefMM,Linde,LRSSii}.

The third step in the KKLT proposal is the inclusion of one anti--D3--brane.
The effect of the latter is an additional positive energy amount (with some constant $D$)
\zbe\label{posD3}
\Vc_{\ov{D3}}=\fc{D}{{\rm Vol}({\cal X})^2}\ ,
\ee
to be added to the scalar potential \req{Scalarpot}:
\zbe\label{Vtotal}
\Vc_{total}=\Vc_{AdS}+ \Vc_{\ov{D3}}\ .
\ee
With this contribution the full scalar potential becomes positive and its
minimum may be adjusted to a small positive value.

\begin{figure}[h!]
  \begin{center}
    \includegraphics[width=110mm]{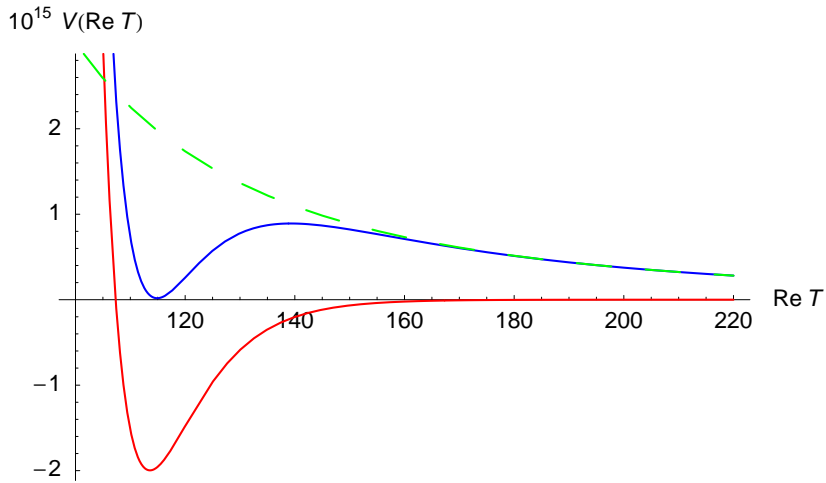}
    \caption{Scalar potentials in a KKLT scenario: $V_{AdS}$ in red, $\Vc_{\ov{D3}}$ in green,
and $\Vc_{total}$ in blue.}
    \label{kklt1}
  \end{center}
\end{figure}

In Figure \ref{kklt1} we have depicted the situation for one K\"ahler modulus
(overall radius $\re T$ of the CY manifold ${\cal X}$).
With the first and second step we obtain the  scalar potential \req{Scalarpot}, shown in red.
The additional contribution \req{posD3} of the anti--D3 is shown in green.
The resulting final potential \req{Vtotal} is shown in blue.

The minimum is lifted from AdS to dS, however the values for the moduli derived from
the vacuum equations \req{FEQ} do not significantly change during the uplift.
Furthermore, the supersymmetry breaking parameters like the gravitino mass or
soft--masses are essentially determined by the depth of the AdS--minimum
and most of the soft--supersymmetry breaking pattern (exhibited in subsection \ref{SOFT})
does apply.

The stability of AdS vacua in gravity coupled to scalar fields has been
investigated in  \cite{BF}. 
Stability is guaranteed, if all scalar masses fulfill the Breitenlohner--Freedman
(BF) bound \cite{BF}, \ie their mass eigenvalues do not fall below a certain minimal bound.
The latter is a negative number related to the scalar potential at the minimum \req{issl}.
It can be shown in a completely model independent way that
all scalars have masses above this bound at any ${\cal N}=1$ supersymmetric AdS minimum in
supergravity theories.
However, the third and final step in the KKLT scenario
consists in the addition of one anti D3--brane,
\ie a positive contribution to the scalar
potential, which lifts the AdS minimum to a dS minimum.
The masses for the moduli fields do not change significantly during this process.
However stable dS vacua require positive mass eigenvalues. Hence, any negative mass
eigenvalue before the uplift is unacceptable since the effect of the anti D3--branes
on the mass eigenvalues is too small to change a  negative mass to positive.

The question under which conditions one may achieve a stable uplift
from AdS to dS has been asked in \cite{ChoiSX,LRSSi} (see also 
\cite{dealwis,leb2,Japan1,Japan2,Japan3})
and further investigated in full generality in \cite{LRSSii}.
One important conclusion of the work \cite{LRSSii} is, that compactifications
without complex structure moduli, \ie $h^{2,1}_-({\cal X})=0$ generically do
not allow for a stable uplift. More precisely, there are certain
constraints on the form of the K\"ahler potential for the K\"ahler moduli in order
to yield positive mass eigenvalues.
In particular, all $\mbb Z_N$--and $\mbb Z_N\times \mbb
Z_M$--orientifolds ${\cal X}$
with $h^{2,1}_-({\cal X})=0$ are excluded for a KKLT scenario. This concerns
$\mbb Z_3, \mbb Z_7, \mbb Z_3\times \mbb Z_3, \mbb Z_4\times \mbb Z_4,
\mbb Z_6\times \mbb Z_6$
and $\mbb Z_2\times \mbb Z_{6'}$ --both at the orbifold point and away from it.

Finally we want to mention, that other scenarios for moduli stabilization
and supersymmetry breaking in type IIB orientifolds 
have been developed in \cite{Burgess:2003ic,conlon1,leb2,leb1}.
Instead of the potential \req{posD3} the uplift from an AdS- to a
dS-vacuum may be also achieved by a D-term potential (D--term uplift) 
\cite{Burgess:2003ic}. However, it has been pointed out in \cite{HPNii}
that a supersymmetric AdS minimum \req{FEQ} 
cannot be uplifted by non--vanishing D--terms.
On the other hand, in \cite{leb2,leb1} an appealing alternative has been proposed
to break supersymmetry by F--terms $F_\Cc$ of open string moduli or matter fields 
(F--term uplift). In the matter dominated supersymmetry breaking 
scenario $F_\Mc\sim 0,\ F_\Cc\sim m_{3/2}$ supersymmetry is broken by
non--vanishing F--terms of matter fields while essentially 
saving the stabilization results \req{FEQ} of the closed string moduli 
furnished at the AdS minimum,  \cff \cite{leb2,leb1} for more details.


\subsection{Type IIA flux compactifications}
\label{sseciiaflux}

In order to explain the idea of flux compactifications we have
focused on the prototype example of three-form fluxes in type IIB
string theory. The main reason was that the backreaction of the
fluxes on the geometry is under rather good control insofar as one
deals with a conformal Calabi-Yau manifold as internal space.
Neglecting the warp factor, one can employ the standard Calabi-Yau
techniques and rely on the mathematical proofs for their
existence.

Compared to type IIB, matters are more complicated in type IIA
\cite{glmw02,LopesCardoso:2002hd,bc03,Dall'Agata:2003ir,bc04,bc04a,Derendinger:2004jn,Kachru:2004jr,DeWolfe:2005uu,Camara:2005dc,Aldazabal:2006up,Villadoro:2005cu,Villadoro:2006ia,Acharya:2006ne}.
In type IIA a similar
analysis of the Killing spinor equations as sketched in section
\ref{subsubsecsusy} leads to the conclusions that IIA solutions with
non-vanishing NSNS or RR fluxes are either not symplectic or not
complex, or neither.
More concretely, a specific class of solutions of the type IIA
Killing spinor equations is given by almost-K\"ahler, half-flat manifolds with the following
torsion class\footnote{These are six-dimensional manifolds, which allow for
an uplift to a seven-dimensional M-theory manifold with $G_2$ group structure
or even $G_2$ holonomy, using the fibration structure introduced by
Hitchin \cite{Hitchin:2001rw}.}

\beqn\lab{IIatorsion}
\tau\in {\cal W}_2^+\ .
\eeqn
It means that the holomorphic 3-form $\Omega_3$ is non-closed
(cfr. eq.\reef{dJdOm}):
\beqn\lab{nonclosedOmega}
d\Omega_3={\cal W}_2\wedge J_2\ .
\eeqn
This geometrical quantity can the be related to the non-vanishing
RR fluxes in the following way:
\beqn\lab{f2f4w}
{\cal W}_2^+=e^\phi (F_2+F_4)\ .
\eeqn

A similar analysis is actually true for flux
compactifications with a four-dimensional AdS$_4$ space instead of
Minkowski space-time \cite{Lust:2004ig}.  Here the torsion is constrained to fall
into the following classes:
\beqn\lab{IIatorsionmasse}
\tau\in {\cal W}_1^++ {\cal W}_2^+\ .
\eeqn
Therefore the six-dimensional space is a half-flat manifold.
In addition to \ref{IIatorsionmasse}, the Bianchi identities require
the exterior derivate of ${\cal W}_2^+$ to be proportional to the real
part of the (3,0)-form on the six-dimensional space:
\beqn\lab{IIatorsionbianci}
d{\cal W}_2^+\sim {\rm Re}(\Omega)\ .
\eeqn

About the type IIA non-Calabi-Yau  solutions to the supersymmetry
constraints much less is known and examples are rare. To address
the moduli problem one should first know the moduli space, which
is not the case for these solutions. Nevertheless, one can try a
perturbative approach and treat the fluxes and the backreaction of
the metric and other fields as small effects. Effectively, one can
then use the Calabi-Yau background and its moduli space. Let us
sketch how one can proceed in type IIA in this way.

Type IIA theory contains the RR
one-form $C_{1}$ and three-form $C_{3}$ as well as their Hodge dual forms $C_7$ and
$C_5$. In addition,
in type IIA Calabi-Yau orientifolds we have to deal with
orientifold O6-planes that couple to the RR seven-form $C_7$ via
\reef{Oplact}, explicitly
\beqn\lab{O6cpl}
Q_6 \mu_6 \int_{\cw} C_7\ .
\eeqn
Regarding the effects of fluxes on the 6-brane charge one has to
take care of a subtlety which was mentioned in the course of
introducing the democratic type II action in section
\ref{secIaII}, namely the potential presence of RR field strengths
without dynamical RR potentials. This includes the RR zero-form
$F_0$ which is conventionally referred to as the mass parameter
$m$ of IIA. Including this zero-form, the RR two-form field
strength is defined
\beqn
F_2 = dC_1 + F_0 B_2 \ .
\eeqn
By the democratic formulation, the equation of motion for $C_7$ is
the Bianchi identity of $F_2$. Thus, the coupling \reef{O6cpl}
leads to a modification of this Bianchi identity in the presence
of O6-planes in the form
\beqn
dF_2 = F_0 H_3 - Q_6 \mu_6 \pi_3 \ .
\eeqn
Here $\pi_3$ is the Poincar\'e-dual three-form of the six-cycle
wrapped by the O6-plane. Integrating this equation over any
three-cycle produces a cancellation condition among the
combination $F_0 H_3$ of RR zero-form flux and NSNS three-form
flux and the background O6-plane charge. Of course, adding
D6-branes in addition would also contribute. This is the analogue
of the effective three-brane charge
\reef{tadpoleb} of three-form fluxes in IIB. Note that in IIA
there are other RR fluxes which remain unconstrained.

The potential energy induced by the fluxes is obtained again by
inserting a general ansatz for the fields into the action. All
potentials are expanded into harmonic forms of proper degree on
the Calabi-Yau. Fluxes are included for their field strengths. The
dilaton is considered to be constant, and the metric a direct
product of Minkowski times the Calabi-Yau.\footnote{As remarked
above, this is neither supersymmetric nor a solution to the
equations of motion, but valid only in a perturbative sense.} The
kinetic action of the RR forms $F_{2p}$, $p=0,2,4$, is of course
positive, and from the tension of the O6-planes there comes a
negative contribution. This allows a stabilization with negative
vacuum energy in a potentially supersymmetric AdS$_4$ groundstate.

The explicit dimensional reduction was performed in
\cite{Grimm:2004ua}, where it was shown how to capture the effects
of the fluxes in terms of a superpotential that has two pieces,
\beqn\lab{supoIIA}
W_{\rm IIA-flux} = {1\over \kappa_{10}^2}\, \int_\cx \Big[ \Om_3 \wedge H_3 +
\bigoplus_{p=0}^3 F_{2p} \wedge e^{B+iJ_2}\Big] \ ,
\eeqn
which combines the general NSNS flux $H_3$ with the even RR fluxes
$F_{2p}$.
In addition, it is very plausible that the effective superpotential
for type IIA compactifications on non-Calabi-Yau spaces contains
a geometrical piece, which is determined by the non-vanishing torsion,
i.e. the non-closure of $\Omega_3$, of the internal manifold (see also
the similar expression for heterotic flux compactifications):
\beqn\lab{supoIIAgeom}
W_{\rm IIA-geometry} = {1\over \kappa_{10}^2}\, \int_\cx  (d\Om_3 \wedge J_2)  \ .
\eeqn
The important difference compared to the case of IIB is the fact
that \reef{supoIIA} depends on both types of moduli, and thus
complete moduli stabilization may in principle be achieved in IIA
compactifications with fluxes
\cite{Derendinger:2004jn,Kachru:2004jr,Villadoro:2005cu,Derendinger:2005ph,DeWolfe:2005uu,Camara:2005dc,Aldazabal:2006up,Villadoro:2006ia}.

Now, the superpotential $W_{\rm IIA-flux}+W_{\rm IIA-geometry}$ is
supposed to produce the correct conditions for preserving
supersymmetry from its F-flatness. That means  vanishing F-terms
should impose the same constraints that follow from the Killing
spinor equations. However this is quite difficult to be explicitly
shown, since the moduli spaces of the non-Calabi Yau manifolds are
not very well understood.

An alternative effective description was suggested basically based
on experience with toroidal models and T-duality. The idea is that
the deviation of the metric from being Calabi-Yau, i.e.\ the
geometric torsion, could possibly be described by adding a
background value for the spin connection $\omega_{ij}^k$, which
can also be viewed as a three-index object. This has lead to the
term ``geometric flux''. Such geometric fluxes arise naturally
from mirror symmetry to Type IIB flux compactifications and
contribute both to the RR tadpole cancellation conditions as well
as to the superpotential \cite{glmw02,kstt02}. They also give rise
to new consistency conditions among fluxes and branes
\cite{Camara:2005dc,Villadoro:2006ia}.
The geometric fluxes deform the torus to a twisted torus. The
resulting consistency conditions can be understood from their
geometry and topology. In particular, it was pointed out that on
the twisted torus the cohomology groups $H^3(\mbb T^6_\omega,\mbb
Z)$ contain $\mbb Z_N$ torsion pieces which give rise to extra
tadpole cancellation conditions \cite{Marchesano:2006ns}. First
attempts with concrete models with D6-branes and fluxes in IIA
have been reported in
\cite{Camara:2005dc,Chen:2006gd,Chen:2006ip,Ihl:2006pp,Floratos:2006hs}. From T-duality it
was also conjectured that even more fluxes (or better flux-like
deformations) could exist in so-called non-geometric
compactifications \cite{Shelton:2005cf}.

Since in general type IIA flux vacua and other more general
deformations of Calabi-Yau compactifications are not very well
understood, we stop here.

\subsection{Heterotic flux compactifications}
\label{HETFLUX}

Heterotic string compactifications on an internal space with
non--trivial warp factor, dilaton and H--field background were first
discussed in \cite{Strominger:1986uh} (see also
\cite{deWit:1986xg}. We will follow essentially the discussion
in \cite{LopesCardoso:2002hd,LopesCardoso:2003af}.
More and also subsequent work on that subject can be found in
\cite{Ivanov:2000fg,Louis:2001uy,Gutowski:2002bc,Gauntlett:2002sc,Becker:2002sx,Becker:2002jj,Goldstein:2002pg,Becker:2003yv,Gauntlett:2003cy,Becker:2003gq,Curio:2003ur,LopesCardoso:2003sp,Becker:2003sh,Becker:2003dz,Becker:2004qh,Becker:2004ii,Curio:2005ew,Kimura:2006af,Curio:2006dc,Kim:2006qs}.

In the following, we will be interested in ${\cal N}=1$ supersymmetric
compactifications of heterotic string theory on spaces with metric
given by
\begin{equation}
ds^2 = g^0_{MN} \, dx^M \otimes dx^N =
{\rm e}^{2\Delta(y)} \left( dx^\mu \otimes dx^\nu \,  g_{\mu\nu}(x)
+ dy^m \otimes dy^n \, \hat g_{mn}(y)\right)\,.
\label{metric}
\end{equation}
Here ${ g}_{\mu\nu}(x)$ denotes the metric of a four--dimensional
maximally symmetric spacetime and
$\Delta$ denotes a warp factor which we take to only depend on the
internal coordinates $y^m$.
The
ten--dimensional supersymmetry equations (in the absence of gaugino
condensates) can be cast into the following form
\begin{eqnarray}
\label{dilatinor}
\delta \psi_{M} &=& {\cal D}_M \epsilon  \equiv
\nabla_{M}\epsilon - \frac14 H_{M} \epsilon \,,
\label{gravitinor}\\
\delta \chi &=& - \frac14 \,\Gamma^{MN}\epsilon
\,F_{MN}\,,
\label{gauginor}\\
\delta \lambda &=&  -{1\over 4}\nabla\!\!\!\!\slash \phi + \frac{1}{24} \,
H\, \epsilon  \,,
\label{dilatinor}
\end{eqnarray}

\noindent
where $H \equiv \Gamma^{MNP} H_{MNP}$, $H_{M}
\equiv H_{MNP} \,\Gamma^{NP}$,
and where the covariant derivative $\nabla$ is
constructed from the rescaled metric $g_{MN} = {\rm e}^{-2 \phi} g_{MN}^0$.
Necessary and sufficient conditions for
${\cal N}=1$ spacetime supersymmetry in four dimensions were
derived in \cite{Strominger:1986uh} and are given by:
\begin{enumerate}
\item the four--dimensional spacetime has to be Minkowski, i.e. $
{g}_{\mu \nu} = \eta_{\mu \nu}$;
\item the internal six--dimensional manifold has to be complex, i.e.
the Nijenhuis tensor
$N_{mnp} $ has to vanish;
\item up to a constant factor,
there is exactly one holomorphic $(3,0)$--form $\omega$,
whose norm is related to
the complex structure $J$ by
\begin{equation}
\star d \star
J =
i  \left( \bar \partial - \partial\right)\, \log ||\omega|| \,;
\label{ddagger}
\end{equation}
\item the Yang--Mills background field strength must be a $(1,1)$-- form
and must satisfy
\begin{equation}
\hbox{tr} F \wedge F = \hbox{tr} \, \tilde R \wedge
\tilde R  - i \, \partial \bar \partial J \,
\label{ddJ}
\end{equation}
as well as
\begin{equation}
\label{FJ}
F_{mn}J^{mn} = 0\,;
\end{equation}
\item the warp factor $\Delta$ and the dilaton $\phi$ are determined by
\begin{eqnarray}
\Delta (y) &=& -{1\over 4}\phi (y) + {\rm constant} \;,\nonumber\\
\phi ( y) & =& -\frac{1}{2} \log ||\omega|| + {\rm constant} \;;
\end{eqnarray}
\item
the background three--form $H$ is determined in terms of $J$ by
\begin{equation}
H = \frac{i}{2} \,\left(\bar\partial - \partial\right) J\,,
\label{HJ}
\end{equation}
where $i (\partial - { \bar \partial} ) = dx^n \, J_n\,^m \,
\partial_m$.

\end{enumerate}
Inspection of (\ref{ddJ}) shows that if
${\rm tr}
\,{\tilde R} \wedge {\tilde R}$ is non--vanishing, then it has to be a
$(2,2)$--form for consistency.

Next we will reformulate the conditions just mentioned
in terms of torsional constraints, using the language of section
\ref{subsubsecsusy}.
Since the internal manifold is taken to be complex,
it immediately follows from  that
\begin{equation}
{\cal W}_1 = {\cal W}_2 = 0\,.
\label{cW}
\end{equation}
The torsion is therefore left in
\begin{equation}
\tau \in {\cal W}_3 \oplus {\cal W}_4 \oplus {\cal W}_5 \,,
\label{torsion1}
\end{equation}
but it cannot be completely generic, because there is one
further geometric constraint to be satisfied, namely (\ref{ddagger}).
This equation relates the dual of the complex structure to the
holomorphic (3,0)--form and therefore can be interpreted as a relation
among the ${\cal W}_4$ and ${\cal W}_5$ classes.
The ${\cal W}_4$ class is determined by $J \wedge d J$ which, using
the duality relation $\star J = \frac12 J \wedge J$,
can be interpreted as $d \star J$.
This implies that information about this class is encoded in the
left--hand side of equation (\ref{ddagger}), as this is given by the
one--form $\star d \star J$.
Moreover, from the definition of ${\cal W}_4$, it follows that it
must be described by a one--form, so it is interesting to establish the
precise relation among the two quantities.
We can rephrase the Hodge star dual and show that
\begin{equation}
{\cal W}_4 =\frac12 \, J \, \lrcorner \, dJ =
\frac12\, J \cdot \left( \star d \star J\right)\,.
\label{W4Stro}
\end{equation}
The proof follows directly from the definition of the contraction operator
\begin{equation}
 J \, \lrcorner \, dJ = \frac32 \;
J^{sn} \; dx^p \;\nabla_{[s} J_{np]} = dx^n\; {J_s}^p\, \nabla_p
 {J_n}^s \,,
\label{proof1}
\end{equation}
and of the Hodge dual:
\begin{equation}
 J \cdot \left( \star d \star J\right) = - dx^n\; {J_n}^s\; \nabla_p
 {J_s}^p =  dx^n\; {J_s}^p\; \nabla_p
 {J_n}^s \,.
\label{proof2}
\end{equation}
Going back to (\ref{ddagger}), and in order to determine the precise relation
between ${\cal W}_4$ and ${\cal W}_5$,
we better consider multiplying (\ref{ddagger})
with $J$.
In this way the equation gets simplified to
\begin{equation}
{\cal W}_4 =-\frac12\, d \log ||\omega||\,,
\label{deq}
\end{equation}
which gives a further constraint on ${\cal W}_4$, namely that it
is an exact real
1--form.
On the right hand side of this equation we find the norm of the holomorphic
form, which is related to ${\cal W}_5$.
Our classification of the torsion relies on the definition of a
unit norm (3,0)--form, which in this case is simply
\begin{equation}
\Psi = \frac{\omega}{||\omega||} \,.
\label{Psidef}
\end{equation}
This form is not holomorphic anymore for a generic dilaton profile
(which is then related to the $\omega$ norm) and that implies
${\cal W}_5 \neq 0$.
From the definition of the unit--norm (3,0)--form (\ref{Psidef}) it
follows that
\begin{equation}
d \psi_+ = \frac12 \, \left( d \Psi + d \bar \Psi\right) = - d \log ||\omega||
\wedge \psi_{+}\,.
\label{dpsi}
\end{equation}
The contraction with $\psi_+$ will therefore lead to
\begin{equation}
{\cal W}_5 = \frac12 \psi_+ \, \lrcorner \, d \psi_+ = d \log ||\omega||\,,
\label{W5Stro}
\end{equation}
and this finally translates into
\begin{equation}
2 \,{\cal W}_4 + {\cal W}_5 =0\,.
\label{Wrelation}
\end{equation}

With respect to concrete solutions of the above equations, it is possible to find manifolds that satisfy all
these geometrical constraints. For instance  the Iwasawa manifold
is an explicit example for a solution of the geometrical heterotic
supersymmetry conditions..
However there are additional constraints that are related to the
heterotic Bianchi identities and to the equations in the heterotic gauge
bundle sector.
For  non-Calabi Yau spaces with H-flux explicit constructions
and solutions for heterotic gauge bundles are very hard to find.
Hence we omit this part of the discussion and refer the reader to
some of the relevant work in this direction  \cite{Goldstein:2002pg,Kimura:2006af}.

Now we want to discuss the
effective action of heterotic flux compactifications. The aim is, like for
type II flux compactifications, to find an effective superpotential $W_{het}$, which
implements via $DW_{het}=0$ the same supersymmetry conditions that we have derived in the
previous paragraph.
The discussion will be still somewhat qualitative, since the moduli spaces of
non-Calabi Yau manifolds are largely unknown.
We start with
the bosonic part of the Lagrangean up to second order in
$\alpha^\prime$ is given by \cite{Bergshoeff:1989de}
\begin{eqnarray}
S &=& \int d^{10}x \, \sqrt{-g}\, e^{-2\phi}\left[\frac14 \, R -\frac{1}{12}
H_{MNP}H^{MNP} +  (\partial_M \phi)^2 \right. \nonumber\\
&&\qquad \qquad \qquad \;\;  \left. - \frac14  \alpha^\prime
\left(F_{MN}^{I} F^{I\,MN}-  R^+_{MNPQ} R^{+\,MNPQ}  \right)
\,\right]\,.
\label{action}
\end{eqnarray}
This action is written in the string frame (we have set $\kappa_{10}^2=2$) and its fermionic
completion makes it supersymmetric using the three--form Bianchi
identity given by
\begin{equation}
dH = \alpha^\prime \left( \hbox{tr} \,  R^{+} \wedge
R^{+}  -\hbox{tr} F \wedge F\right)\,,
\label{HBI0}
\end{equation}
where the curvature $R^{+}$ is the generalized Riemann curvature built
from the generalized connection $\nabla^{+}$.

To simplify the discussion we limit ourselves to the case with
dilaton and  warp factor identified, i.e. $\phi = -4\Delta$, but the
generalization of the following results is straightforward.
After some manipulations, the action (\ref{action}) can be written as
\begin{eqnarray}
S  = && \int d^4 x \, \sqrt{-g_{4}}\, \left\{- \frac{1}{2} \int_{{\cal
X}} \, {\rm e}^{-2\phi}\left(-2d\phi + \theta \right)\wedge \star \left(-2 d\phi + \theta \right)
+\frac18 \int_{{\cal X}} {\rm e}^{-2\phi}\, J \wedge
J \wedge {\hat R }^{ab}J_{ab} \right.\nonumber \\
& -&\frac14 \int_{{\cal X}} d^{6}y \; \sqrt{g_6} \,{\rm e}^{-2\phi}\,
N_{mn}{}^p \,g^{mq}g^{nr}g_{ps}\,N_{qr}{}^s
\,
\nonumber \\
&+&  \frac12 \int_{{\cal X}} {\rm e}^{-2\phi}\,
\left(H + \frac12 \star {\rm e}^{2\phi}\, d({\rm e}^{-2\phi}\, J)\right) \wedge \star \left(H + \frac12 \star
{\rm e}^{-2\phi}\, d({\rm e}^{-2\phi}\, J)\right) \,\nonumber\\
&-&  \frac{\alpha^\prime}{2}\int_{{\cal X}} d^{6}y\; \sqrt{g_6}\,
 {\rm e}^{-2\phi}\, \left[\hbox{tr} (F^{(2,0)})^{2} +
{\rm tr}(F^{(0,2)})^{2} + \frac14\,\hbox{tr} (J^{mn}F_{mn})^{2}\right] \nonumber \\
 &+& \left. \frac{\alpha^\prime}{2} \int_{{\cal X}} d^{6}y\;\sqrt{g_6}
\, {\rm e}^{-2\phi}\, \left[\hbox{tr} (R^{+\,(2,0)})^{2} +
\,\hbox{tr} (R^{+\,(0,2)})^{2} + \frac14 \,\hbox{tr} (J^{mn}
 R^{+}_{mn})^{2}\right]\right\}\,.
\label{finalaction}
\end{eqnarray}
In this expression the traces are taken with respect to the fiber
indices $a,b,\ldots$, whereas the Hodge type refers to the base
indices $m,n,\ldots$ of the curvatures.
The other geometrical objects appearing in the above
expression are the Lee--form
\begin{equation}
\theta \equiv J \lrcorner dJ= \frac{3}{2} J^{mn} \,
\partial_{[m}J_{np]} \, dx^p\,,
\label{eqLee}
\end{equation}
the Nijenhuis tensor
\begin{equation}
N_{mn}{}^p = {J_m}^q \partial_{[q}J_{n]}{}^p -  {J_n}^q
\partial_{[q}J_{m]}{}^p\,,
\label{eq:Nijenhuis}
\end{equation}
and the generalized curvature $\hat R$, which is constructed using the
Bismut connection built from the standard Levi--Civita connection and a
totally antisymmetric torsion $T^B$ proportional to the complex structure,
\begin{equation}
T^B_{mnp} = \frac32 \,{J_m}^{q} {J_n}^{r}
{J_p}^{s}\partial_{[q}J_{rs]} = -\frac32 J_{[m}{}^{q} \nabla_{|q|} J_{np]}\,.
\label{eq:tor}
\end{equation}
The action (\ref{finalaction}) will now be used to find the conditions
determining the background geometry.
Namely,  (\ref{finalaction}) can be used as the expression
for the scalar potential of the effective four--dimensional theory,
\begin{equation}
S = - \int d^4 x \,\sqrt{-g_4} \,{\cal V}\,.
\label{eq:Vdef}
\end{equation}

The action (\ref{finalaction}) consists of a sum of squares as well as
of one linear term.
In order to have a solution of the equations of motion one sets to
zero all the squares and proves that the linear term does not
contribute to the equations of motion.
First we would
like to exhibit the correspondence between the supersymmetry
conditions and the terms which are squared in the action.
The geometrical conditions resulting from the vanishing of the
BPS--like squares are the vanishing of the Nijenhuis tensor
$$N^{m}{}_{np} = 0$$ and of some components of the generalized Riemann
curvature constructed from the $\nabla^{+}$ connection,
$$R^{+\,(2,0)}=R^{+\,(0,2)}=J^{mn}R^{+}_{mn} = 0.$$
The vanishing of the Nijenhuis tensor states that the internal
manifold is complex (which means ${\cal W}_{1} = {\cal W}_{2} = 0$ in
the torsion classes language).
The conditions on the $R^{+}$ curvature can be translated into the
integrability constraints following from the vanishing of the
gravitino supersymmetry transformation (\ref{gravitinor}), which leads
to the requirement of $SU(3)$ holonomy for the $\nabla^{-}$
connection.
The proof requires the identity
\begin{equation}
R^{+}_{ab\,cd} = R^{-}_{cd\,ab} - (dH)_{abcd}\,,
\label{eq:rpm}
\end{equation}
which relates the $R^{+}$ and $R^{-}$ curvatures with the base and
fiber indices swapped.
Using this identity and the fact that $dH$ gives higher order terms
in $\alpha^\prime$ the conditions on the base indices of $R^{+}$ become
conditions on the $R^{-}$ fiber indices, to lowest order in $\alpha^\prime$,
\begin{equation}
R^{-\,(2,0)} = R^{-\,(0,2)}=J^{ab}R^{-}_{ab} = 0\,.
\label{eq:rm}
\end{equation}
These conditions precisely state that the generalized curvature $R^-$
is in the adjoint representation of $SU(3) \subset SO(6)$ and
therefore its holonomy group is contained in $SU(3)$.
We also obtain the relation between the ${\cal W}_{4}$ and ${\cal
W}_{5}$ torsion classes, expressed by the identification of the
differential of the dilaton with the Lee--form,
\begin{equation}
d\phi -\frac12 \, \theta =0\, \,.
\label{dfth}
\end{equation}
The final BPS--like square precisely yields the locking condition of
the three--form $H$ onto the almost complex structure $J$ which was
derived from the vanishing of the supersymmetry transformations.
Indeed, on a complex manifold, and using (\ref{dfth}), the following
identity holds,
\begin{eqnarray}
H = -  \frac12 \star {\rm e}^{2\phi}\,d({\rm e}^{-2\phi}\,J) =
\frac12 i ( \partial- {\bar \partial }) J \;. \label{locking}
\end{eqnarray}

Since the potential essentially consists of a sum of BPS--like squares, and is hence
positive definite, we can derive it as a F-term potential from a holomorphic superpotential, plus
various D--terms.
The necessary requirements for having an ${\cal N} = 1$ vacuum, i.e.
$W = 0$ and $\partial W =0$, then impose certain conditions leading to
moduli stabilization.
A rigorous derivation of the superpotential for flux compactifications
requires a detailed knowledge of the moduli space of the
compactification manifolds, which is not available at present.
However an educated guess for the superpotential is given by \cite{Becker:2003gq,LopesCardoso:2003af}
\begin{equation}
W_{het}= {1\over 2} \int \Omega_3 \wedge \left(H + \frac{i}{2} dJ\right) \, .
\label{supo}
\end{equation}
Notice that for generic flux compactifications the internal space is
not complex, i.e. $J$ is not integrable, and therefore $dJ \wedge
\Omega \neq 0$.
Since the above superpotential explicitly depends on the three--form
flux $H$, its extremisation should give rise to the torsional
constraints leading to supersymmetric configurations.
Again, a rigorous derivation requires an explicit
knowledge of the metric moduli, but it is plausible that under
certain assumptions the expected torsional constraints do follow.
More precisely, the superpotential $W_{het}$ has to lead to a determination
of $H$ in terms of the deviation of the internal space from being a
Calabi--Yau manifold.
$W_{het}$ must therefore also include pieces which are purely geometrical
and which measure the non--Calabi--Yau--ness of the internal space.
This is captured by the additional piece proportional to $dJ$ in
(\ref{supo}).

Let us
briefly comment on the limiting case with constant dilaton $\phi= const.$ and
vanishing flux $H = 0$.
The locking condition (\ref{locking}) simply becomes the
requirement for the internal manifold to be K\"ahler, imposing $dJ =
0$.
In addition, the square involving the dilaton becomes a condition imposing
the vanishing of the Lee form, $d\star J = 0$.
Moreover, now $\nabla^{\pm}=\nabla$ and $R^+ = R^- = R$.
Therefore the conditions on the holonomy of $\nabla^{-}$ become
conditions on the Levi--Civita connection.
The solution is obviously given by Calabi--Yau manifolds,
which are  K\"ahler and have
vanishing first Chern class.

\clearpage
\setcounter{equation}{0}


\section{STATISTICS OF FLUX AND  D-BRANE MODELS}
\label{secstatistic}

So far we have reviewed the various classes of string compactifications
with ${\cal N}=1$ space-time supersymmetry in four dimensions.
Both  heterotic as well as D-brane constructions allow to realize
many of the features for the SM, like gauge symmetry, chiral
matter particles, gauge symmetry breaking, family replication
etc. These discrete parameters are mostly of topological origin
in string theory and are relatively well understood.
When it comes to the continuous parameters like
gauge or Yukawa couplings, string theory is far less under control,
for first these terms in the effective action are only possible to  be computed
for simple toroidal orbifold models so far and second they  depend
sensitively  on the moduli notoriously present in string compactifications.

However, as a big step forward it was realized that flux backgrounds
(in addition to non-perturbative corrections) in general freezes
these moduli either in  supersymmetric or non-supersymmetric minima.
However, with a method really determining (the classical or naked)
energy density in a string vacuum, the so far neglected
cosmological constant problem arises.
In general one expects  that quantum corrections generate a cosmological
constant of the order of the supersymmetry breaking scale $M^4_{susy}=(1$TeV)$^4$. In order
to obtain the observed value of $\Lambda=(0.003$eV)$^4$ a fine-tuning
of the classical value is needed for which so far no physical reason has been
identified.

As we will review in the next subsection, an estimate
of the number of  supersymmetric flux vacua  in Type IIB orientifolds \cite{Ashok:2003gk}
comes to the conclusion that for general Calabi-Yau manifolds
there exist of the order of $10^{500}$. This number dramatically exceeds
most of the estimates for the number of string vacua made before. 
For instance the number of toric Calabi threefolds is 184026 \cite{Skarke:1996hq}.
An exception is the early estimate of the number of self-dual lattices in \cite{Lerche:1986cx}(see also \cite{Schellekens:2006xz}).
Taking this number seriously,  i.e. assuming that the number of vacua
is not drastically reduced by taking non-perturbative corrections to the
flux superpotential into account,
led R. Bousso and J. Polchinski \cite{Bousso:2000xa} to the proposal
that the cosmological constant might not be fixed by a dynamical
principle to such a small value but that this fine tuning
is "solved" by merely  the enormous vacuum degeneracy
of string vacua.
Along this same line of reasoning,
M.R. Douglas  proposed that,  complementary to a model by model
search, one should  follow
a statistical approach to the string vacuum problem \cite{Douglas:2003um},
as this large degeneracy seemed to make it very unlikely
that we will easily identify models, which come close to
our universe.

In \cite{Ashok:2003gk,Denef:2004ze} very powerful statistical methods were developed
to determine to distribution of Type IIB flux vacua over the complex
structure moduli space. The generalization to M-theory flux vacua
was performed in \cite{Acharya:2005ez}.
Since in this review we restrict ourselves to  deal mainly with
D-brane constructions, we do not review the entire story about
the statistics in the flux sector.
For short reviews on this
subject we refer the reader to \cite{Douglas:2004zg,Kumar:2006tn}
and for a more elaborate review to \cite{Douglas:2006es}. 
Here we mainly review
the methods developed so far to describe the statistics in the
D-brane sector. This should only be regarded as a first approach
to the problem, as  in a completely realistic setting the statistics
of the D-brane sector has to be  combined with the statistics of  the flux sector.

\subsection{Counting supersymmetric Type IIB flux vacua}

In section \ref{secfluxvacua} we have provided the formalism to describe flux vacua in
Type IIB string theory. Just taking the tree level induced potential over
the complex structure moduli space of a chosen Calabi-Yau manifold, one
could ask the question
\vskip 0.2cm
\centerline{{\it  How many different flux vacua are there?}}
\vskip 0.2cm
\noindent
Following \cite{Denef:2004ze}, let us make an estimate. Consider Type IIB compactified on a Calabi-Yau manifold
with $b_3$ three-forms $\alpha^i$.
Now we turn on general G-flux through these three-cycles
\bea
    {1\over \ell_s^2}  H_3=\sum_i N^i_{NS}\,\alpha_i  , \quad\quad
    {1\over \ell_s^2}   F_3=\sum_i N^i_{R}\,\alpha_i ,
\eea
so that
\bea
      N_{\rm flux}=    {1\over \ell_s^4 } \int F_3\wedge H_3 = \eta_{ij} N^i_{NS}\, N^j_{R}
          >0
\eea
with $\eta_{ij}=\int \alpha_i\wedge \alpha_j$.
The tadpole cancellation condition reads
\bea
            {N_{\rm flux}\over 2} + N_{D-branes} = L_*
\eea
where $L_*$ denotes the contribution of the orientifold planes.
Now, we want to count the number of solutions with $0\le L\le L_*$ with
$L=N_{\rm flux}/2$.
This number is given by
\bea
     {\cal N}_{\rm flux}(L\le L_*)&=&\sum_{{\rm susy\ vac}} \theta(L_*-L) \nonumber \\
                              &=&\sum_{vac} {1\over 2\pi i} \int_C {d\alpha\over \alpha}
                             e^{\alpha\, (L_* -L)}  \\
                             &=&  {1\over 2\pi i} \int_C {d\alpha\over \alpha}
                e^{\alpha\, L_*}\left(\sum_{vac} e^{-{\alpha\over 2} {\bf N}\,
                    \eta\,  {\bf N}}  \right)
\nonumber
\eea
where the path $C$ in the complex plane runs parallel to the $y$ axis with small positive
$x$. After approximating the discrete sum over the flux quanta
by an integral,   the sum over all vacua can be written
as
\bea
\label{ahok}
 {\cal N}(\alpha)&=&\sum_{vac} e^{-{\alpha\over 2} {\bf N}\, \eta\, {\bf N}} \nonumber \\
 &=&\int_{M} d^{2m}z \int d^{4m}N  e^{-{\alpha\over 2} {\bf N}\, \eta\,  {\bf N}} \,
         \delta^{2m}(DW)\, |\det D^2 W |
\eea
with $m=b_3/2$ and $M$ denoting a fundamental region in
the complex structure/dilaton moduli space.
However, the scaling of the number of vacua with $L_*$ can be estimated
without evaluating this integral. Let us rescale $N\to N/\sqrt\alpha$, which
implies ${\cal N}(\alpha)\to \alpha^{-2m} {\cal N}(1)$ so that gets
\bea
         N_{\rm flux}(L\le L_*)&=&{1\over 2\pi i} \int_C {d\alpha\over \alpha^{2m+1}}
                e^{\alpha\, L_*}\, {\cal N}(1) \nonumber \\
              &=& \theta(L_*)\, { L_*^{2m} \over (2m)!}\, {\cal N}(1) .
\eea
This is good approximation as long as the radius of the sphere in $N$-space is
large enough, i.e. $L_*\gg 2m$.
For typical numbers such as $L_*\simeq 1000$ (as they appear in F-theory)
and $m\simeq 200$ one gets
${\cal N}_{\rm flux}\simeq 10^{250}$.
This is an amazingly large number, which is may orders of magnitudes larger
than the ensembles of string vacua people have dealt with before,
like for instance toric varieties \cite{Kreuzer:2000xy}, where numbers like $4\cdot 10^{9}$ occurred
\footnote{The only exception seems to be an earlier estimate of the number
of covariant lattices \cite{Lerche:1986cx}.}.

In \cite{Ashok:2003gk} the integral (\ref{ahok}) was evaluated further (with $|\det D^2 W|$
$\to$ $\det D^2 W$ ) leading eventually to the formula
\bea
   N_{\rm flux}(L\le L_*)&=& { 2\pi L_*^{2m} \over \pi^n\, (2m)!}\,
         \int_{{\cal F}\times {\cal H}}  \det\left( -R-\omega\right)
\eea
where ${\cal F}$ denotes the fundamental region of $SL(2, \mbb{Z})$,
${\cal H}$ the  fundamental region of  the complex structure
moduli space and $R$ and $\omega$ the curvature and K\"ahler two forms.

In view of this huge vacuum degeneracy, the so-called string landscape,
M.R. Douglas has proposed a statistical approach to the string vacuum
problem \cite{Douglas:2003um}.
In its pragmatic version it says that complementary to a (necessary)
model by model search one should study the statistical
distribution of various physical quantities in the ensemble of string vacua.
Such an approach might be helpful to
\footnote{More philosophically, one can also combine the landscape picture with
the weak anthropic principle, saying that in some meta-world
all string theory vacua are realized and from the many possibilities,
we of course happen to live in  a  (meta-)stable one where the physical
parameters of course must have the right values to bring about almost intelligent
life forms. This might  explain why some anthropically essential
physical quantities like the cosmological constant have the
 "fine tuned" value we observe \cite{Susskind:2003kw} (see also \cite{Schellekens:2006xz} for
earlier discussion of the anthropic idea in string theory). }
\begin{itemize}
\item{ estimate the frequency with which standard-like  string
models arise}
\item{ get an idea in which regions of the landscape to look
    for realistic models}
\item{ find statistical evidence that standard-like properties
      are extremely rare, i.e. almost excluded $\to$ falsification
    of string theory \footnote{In the terminology of \cite{Vafa:2005ui}, this means
    that the SM would lie in the swampland of string theory.}}
\item{ argue for a uniform distribution of certain physical quantities
like for instance the cosmological constant, shedding a new light
on so-called fine tuning problems}
\end{itemize}

Our viewpoint is rather pragmatic stating that  no final word has been spoken in this matter and one
should carry on investigating the set of string vacua with all
possible means. It might also be that without a completely understanding
of M-theory we are missing some essential (non-perturbative) consistency
conditions, which eventually will reduces the number
of vacua.

\subsection{Statistical approach for D-brane models}

The statistics due to the closed string fluxes provide estimates for
the frequency of cosmological parameters like the cosmological
constant. Of course, for making contact with elementary
particle physics and the SM, we have to also include
the statistics of the open string sector in Type II orientifolds.

The statistics of D3-brane gauge groups in flux vacua has been
investigated in \cite{Kumar:2004pv}.
A more general study of D-brane statistics was initiated in 
\cite{Blumenhagen:2004xx}, where
for the ensemble of intersecting branes
on certain toroidal orientifolds, the statistical distribution
of various gauge theoretic quantities was studied, like the rank of the gauge group,
the number of models with an $SU(M)$ gauge factor and the number
of generations. In particular,  the examples
of supersymmetric intersecting branes on the $T^2$, $T^4/\mbb{Z}_2$ and
$T^6/\mbb{Z}_2\times \mbb{Z}_2$
orientifolds were discussed in detail.
These results were supplemented and confirmed by concrete results from
a long term brute force computer search as reported in \cite{Gmeiner:2005vz, Gmeiner:2005nh,Gmeiner:2006vb,Gmeiner:2006qw}.
In \cite{Douglas:2006xy} the statistics of supersymmetric intersecting D-brane models
on $T^6/\mbb{Z}_2\times \mbb{Z}_2$ orbifolds was also discussed.
In contrast to \cite{Gmeiner:2005vz, Gmeiner:2005nh,Gmeiner:2006vb,Gmeiner:2006qw}  here
the distributions were discussed for fixed numbers of D-branes i.e. by
neglecting the exponential degeneracy
of the hidden sector branes.  In particular
it was proven that the number of tadpole canceling configurations
is finite.
Another example of models where a brute force statistical analysis was performed
are the aforementioned Gepner model orientifolds \cite{dhs04a,Anastasopoulos:2006da}.

The models we are interested in are still supersymmetric intersecting D-brane models
with the main constraints given by supersymmetry and the tadpole cancellation
condition.
The first step is to determine all or at least a large,
preferably  representative subset of supersymmetric branes.
After solving the supersymmetry constraints, in all the  examples
discussed in \cite{Blumenhagen:2004xx},
this was given by a subset $S$ of the naively allowed wrapping numbers $X_I$.
As a constraint one faces the various tadpole cancellation
conditions
\begin{equation}
\label{tadpoles}
       \sum_{a=1}^k  N_a\,  X_{a,I} = L_I
\end{equation}
with $I=1,\ldots,b_3/2$  and $L_I$ denoting the contribution
from the orientifold planes and the three-form fluxes.
A gauge theoretic quantity is often a function of the
wrapping numbers  and the number of D-branes.
Therefore it is necessary to have methods to compute statistical
distributions in the unconstrained set of solutions to
(\ref{tadpoles}).

\subsubsection{Counting tadpole solutions}

Let us discuss such a method at a very simple example, which
however shows already the general idea.
Say we would first  count the solutions of
the single tadpole condition
\bea
\label{tadea}
       \sum_{a=1}^k  N_a\,  X_a = L.
\eea
By writing the Kronecker delta function as
\bea
    \delta_{n,0}={1\over 2\pi i} \oint dq\,  q^{n-1},
\eea
we can express this number as
\bea
{\cal N}(L)
                     &\simeq&{1\over 2\pi i} \oint dq {1\over q^{L+1} }
              \sum_{k=1}^\infty {1\over k!}\,
                \sum_{N_1=1}^\infty\sum_{X_1=1}^L \ldots
                  \sum_{N_k=1}^\infty\sum_{X_k=1}^L
                           q^{\sum_a  N_a X_a} \nonumber \\
           &=&{1\over 2\pi i} \oint dq {1\over q^{L+1} }
                     \sum_{k=1}^\infty {1\over k!}\,
                  \left(\sum_{X=1}^L {q^X\over 1-q^X} \right)^k
\eea

A common method to evaluate the asymptotic expansion of such integrals
is the saddle point approximation.
First we write
\bea
{\cal N}(L)={1\over 2\pi i} \oint {dq} \,
            e^{f(q)} .
\eea
To evaluate this integral one assumes that the main contribution
comes from the saddle points $q_0$, which are determined by the
condition  $df/dq|_{q_0}=0$. Here we assume that there exists only
one saddle point. Otherwise one has to sum over all
saddle  points.
The leading order saddle point approximation is then simply given by
\bea
 {\cal N}^{(0)}(L)=  e^{f(q_{0})}.
\eea
One can compute the next to leading order approximation
\bea
 {\cal N}^{(2)}(L)=  {1\over \sqrt{2\pi}}\, { e^{f(q_{0})} \over
    \sqrt{{\partial^2 f \over  \partial q^2}|_{q_0} }} .
\eea
In our case  the saddle point function $f$ reads
\bea
       f(q)= \sum_{X=1}^L {q^X\over 1-q^X } - (L+1)\, \log q.
\eea
Close to $q\simeq 1$ we find the analytical expression
\bea
  f(q)\simeq {1\over 1-q}\sum_{X=1}^L {1\over X} -L\,
                \log q   \simeq {\log L\over 1-q} -L\, \log q .
\eea
In this approximation the saddle point is at $q_{0}=1-\sqrt{\log L\over L}$
yielding the leading order approximation
\bea
\label{tadee}
   {\cal N}(L)\simeq e^{2\sqrt{ L\log L}} .
\eea
A rough intuitive understanding  of this result can be gained  as follows:
In order to solve (\ref{tadea}) one first divides $L$ into its partitions and
then one writes each term as a product of two positive integer numbers.
We know already that the number of partitions scales like $e^{2\sqrt{ L}}$.
On the other hand it is known in number theory that the function $\sigma_0(n)$
of divisors of an integer $n$ has the property
\bea
   {1\over L} \sum_{n=1}^L \sigma_0(n)\simeq \log L + (2\gamma_E -1)
\eea
for $L>>1$, where $\gamma_E$ denotes the Euler-Mascheroni constant.

For the more general case of $b_3/2$ tadpole cancellation conditions,
the number of such solutions  is given by the expression
\bea
\label{numb}
  {\cal N}(\vec L )
                     &\simeq& {1\over (2\pi i)^{b_3\over 2}}
    \oint \prod_I {dq_I \over q_I^{L_I+1} }\,
                \exp\left(\sum_{X_I\in S} {\prod_I q_I^{X_I}
                \over 1-\prod_I q_I^{X_I}} \right),
\eea
which can be  evaluated at leading order by a saddle point
approximation with
\begin{equation}
       f(\vec q)= \sum_{X_I\in S} {\prod_I q_I^{X_I}
                \over 1-\prod_I q_I^{X_I}} - \sum_I (L_I+1)\, \log q_I.
\end{equation}
The saddle point is determined by the condition
$\nabla f(\vec q)|_{\vec q_0}=0$,
and the second order saddle point approximation reads
\begin{equation}
{\cal N}^{(2)}(\vec L)=  {1\over  \sqrt{2\pi}^{b_3\over 2}}\, { e^{f(\vec q_{0})}
         \over
    \sqrt{ \det\left[ \left( {\partial^2 f\over \partial q_i\partial q_j}\right)
        \right]_{q_0}}}.
\end{equation}

\subsubsection{Distribution of physical quantities}

\begin{figure}[h!]
  \begin{center}
    \includegraphics[width=80mm]{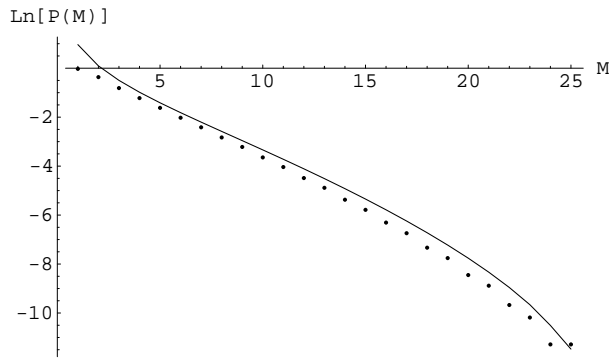}
    \caption{Frequency  of at least one $SU(M)$ factor for L=25. The dots show the exact results and the curve the saddle point approximation.}
    \label{frea}
  \end{center}
\end{figure}

We can also ask, what the percentage of models  with
at least one $SU(M)$ gauge factor is. Using the same methods as above this
can be written
\bea
   P(M) &\simeq& {1\over 2\pi i\, {\cal N(L)} } \oint dq\,
               {1\over q^{L+1} }
               \sum_{k=1}^\infty {1\over (k-1)!}\,
                  \left(\sum_{X=1}^L {q^X\over 1-q^X} \right)^{k-1}\,
            \sum_{X=1}^L \sum_{N=1}^\infty q^{NX} \delta_{N,M} \nonumber \\
               &=& {1\over 2\pi i\, {\cal N(L)} } \oint dq\,
               {1\over q^{L+1} }\,
               \exp\left( \sum_{X=1}^L  {q^X\over 1-q^X} \right)\, q^M
                   \left({1-q^{ML}\over 1- q^M}\right)
\eea
The corresponding saddle point function reads
\bea
\label{probb}
       f(q)= \sum_{X=1}^L {q^X\over 1-q^X } + \log\left[q^M
                   \left({1-q^{ML}\over 1- q^M}\right) \right]- (L+1)\, \log
                   q.
\eea
We can either numerically search for saddle points of this function or
we can observe that for $M\ll L$ the second, $M$ dependent, term in
(\ref{probb}) is just a small perturbation. In this case we expect
that in leading order the saddle point does not change, so that we have only
to evaluate (\ref{probb}) at $q_{0}=1-\sqrt{ L\log L}$.
Doing this for large $L$ one expects that the probability
to find an $SU(M)$ gauge factor scales like
\bea
\label{probc}
      P(M)\simeq \exp\left( -\sqrt{\log L \over L} M \right).
\eea
In Figure \ref{frea}  we have shown the frequency  for at least
one $SU(M)$ factor for L=25.

We see that the exact and saddle point approximation
nicely agree in the regime $M\ll L$ and that for
$M\simeq L$ we get deviations from the simple (\ref{probc})
behavior.

\begin{figure}[h!]
  \begin{center}
    \includegraphics[width=80mm]{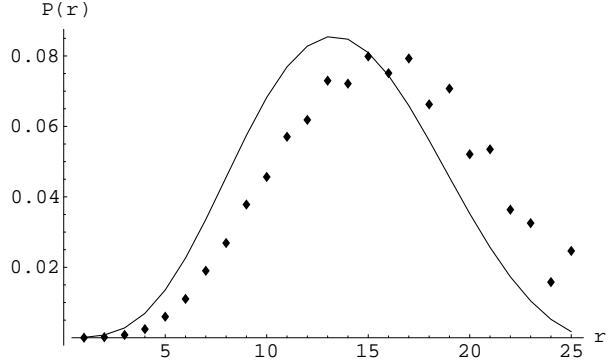}
    \caption{The rank distribution for $L=16$. The dots show the exact results and the curve the saddle point approximation.}
    \label{rankpr}
  \end{center}
\end{figure}

Next we investigate what the frequency  is to get
a gauge group of total rank $r$. This means that we also have to implement
the constraint
\bea
\label{defrank}
\sum_{a}  N_a =r,
\eea
which we again do by writing the Kronecker delta function as a contour integral
\bea
  P(r)
                     &\simeq& {1\over 2\pi i\, {\cal N}(L)}
        \oint dq {1\over q^{L+1} }  \oint dz {1\over z^{r+1} }
              \sum_{k=1}^\infty {1\over k!}\,
                \sum_{N_1=1}^\infty\sum_{X_1=1}^L \ldots
                  \sum_{N_k=1}^\infty\sum_{X_k=1}^L
                           q^{\sum_a  N_a X_a}\, z^{\sum_a  N_a} \nonumber \\
           &=&{1\over 2\pi i{\cal N}(L)} \oint dq {1\over q^{L+1} }
           \oint dz {1\over z^{r+1} }
             \exp\left( \sum_{X=1}^L  {z\, q^X\over 1-z\, q^X} \right).
\eea
Then the saddle point function reads
\bea
       f(q,z)= \sum_{X=1}^L {z\, q^X\over 1-z\, q^X } - (L+1)\, \log q-(r+1)\, \log z.
\eea
Numerically determining   the saddle point now in the two variables, $q$ and
$z$ we find the Gaussian like distribution shown in Figure \ref{rankpr}.

After having checked in \cite{Blumenhagen:2004xx} that the number of solutions to the tadpole
cancellation conditions for fixed complex structures for
the 8D, 6D and 4D examples are finite,
various gauge theoretic distributions were computed and compared to
a brute force computer classification.
Cutting a long story short, the following qualitative results
we obtained
\begin{itemize}
\item{The frequency to find an $SU(M)$ gauge factor scales
like
\bea
P(M)\simeq \exp\left( -\sqrt{\log L\over L} M \right).
\eea
For a product gauge group $\prod_{i=1}^k SU(M_i)$ with $\sum_{i=1}^k M_i\ll L$
it satisfies mutual independence, i.e.
$P(M_1\, \ldots M_k)=\prod_i P(M_i)$.
}
\item{The rank distribution yields approximately a Gauss curve
with the maximum depending on the complex structure moduli.
}
\item{Defining a measure for the chirality of a solution by
    $\chi=\langle \Pi'\circ \Pi\rangle$,  in the 6D case a dependence
like
\bea
P(\chi)\simeq \exp\left( -\kappa\,  \sqrt{\chi}  \right)
\eea
was found
with $\kappa$ denoting some constant depending presumably on the $L_I$.
}
\end{itemize}

\subsubsection{Statistical correlations}

An important question is whether one can see any statistical correlations
of physical quantities.
Here one can distinguish two kinds of correlations:
\begin{itemize}
\item{Correlations which are directly the effect of the stringy consistency
conditions like for instance the tadpole cancellation conditions. Though rather
obvious from the stringy point of view, from a pure field theoretical  or
mathematical point of view, these correlations may nevertheless be quite surprising.
A good example of such a correlation is for instance mirror symmetry, which is
rather trivial from the world-sheet  point of view, but rather surprising
from the target space point of view.}
\item{Correlations which are not obvious at all and cannot easily be traced
back to the defining string equations.}
\end{itemize}

\begin{figure}[h!]
  \begin{center}
    \includegraphics[width=\textwidth]{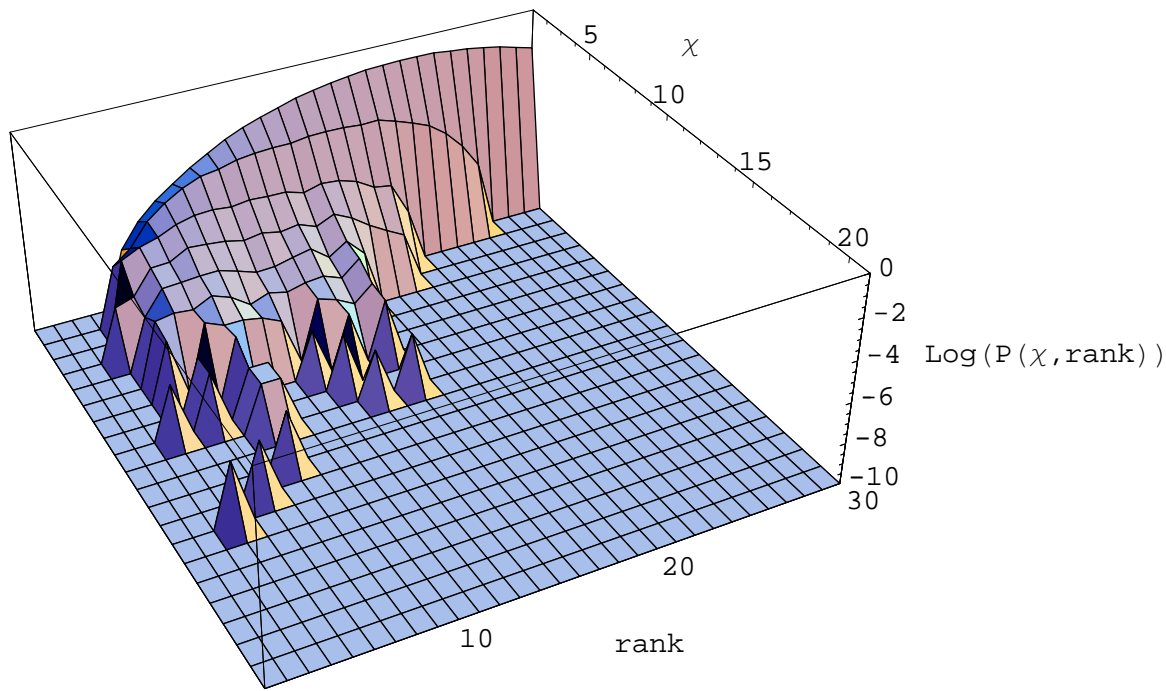}
    \caption{The frequency rank distribution of models of specific rank and
chirality. $L_0=L_1=L_2=L_3=8$ and complex structures $U_I=1$.}
    \label{correl}
  \end{center}
\end{figure}

All correlations known so far are of the first type. In the
ensemble of intersecting D-brane models one finds for instance a
correlation between the rank of the gauge group (\ref{defrank})
and the mean chirality of the model (see figure \ref{correl}),
which we have defined as
\bea\label{EqChiDef}
  \chi=<\Pi_a'\circ \Pi_b-\Pi_a\circ \Pi_b>.
\eea

This correlation was anticipated in \cite{Blumenhagen:2004xx} employing  the saddle point
approximation and confirmed in \cite{Gmeiner:2005vz} by a direct computer search.
It roughly speaking says that the higher the rank of the total
gauge symmetry is the smaller is the number of generations in the model.

\subsubsection{Statistical distributions of gauge theoretic quantities}

In view of phenomenological applications of intersecting D-brane models
it is interesting to learn something about the
frequency of MSSM like models in this framework.
In \cite{Gmeiner:2005vz} of the order of $10^{8}$ intersecting D-brane models
have been constructed, for which various statistical distributions were
plotted. The main emphasis was on the statistics of semi-realistic
models containing a subset of D-branes realizing the Standard Model  quiver
or a variation thereof. It was exemplified that most of the
Standard Model features can be considered as being
statistically independent,
which allowed one to make an estimate
for the frequency of MSSM-like models in this set-up.
The individual suppression factors are listed in Table \ref{tab_corr}.
\begin{table}[htb]
\begin{center}
\begin{tabular}{|l|r|}\hline
Restriction                    & Factor\\\hline
gauge factor $U(3)$            & $8.16\cdot10^{-2}$\\
gauge factor $U(2)/SP(2)$            & $9.92\cdot10^{-1}$\\
No symmetric representations   & $8.39\cdot10^{-1}$\\
Massless $U(1)$                & $4.23\cdot10^{-1}$\\
Three generations of quarks    & $2.92\cdot10^{-5}$\\
Three generations of leptons   & $1.62\cdot10^{-3}$\\\hline
\emph{Total}                   & $1.3\cdot10^{-9}$\\\hline
\end{tabular}
\caption{Suppression factors for various constraints of Standard Model
properties.}
\label{tab_corr}
\end{center}
\end{table}
Combining all these factors leads to an overall suppression factor of
$\approx\,1.3\cdot10^{-9}$.
For plots of other  statistical distributions we refer the
reader to the original paper \cite{Gmeiner:2005vz}.

\subsection{Outlook on statistics}

Clearly, we are just beginning to approach the problem of unraveling
the statistics on the landscape of string theory. The final
aim would be to perform the statistics over as many parameters
as possible to really get a realistic picture of what overall statistical
averages can tell us about the distribution of various
physical quantities.
The methods shown above might play an important role whenever one
encounters string theoretic constraints similar
to the tadpole cancellation conditions.
More modestly, as a next step it would be interesting to
study the distributions of heterotic string vacua and to see
whether, as expected from string dualities,
they feature  similar patterns as the orientifolds \footnote{The statistics
of a certain set of non-supersymmetric heterotic string vacua was investigated
in \cite{Dienes:2006ut} and free fermion constructions were scaned
in \cite{Faraggi:2006bc}.}.
Gepner model orientifolds \cite{Dijkstra:2004ym,Anastasopoulos:2006da}
might also provide a nice testing ground for comparing
and possibly refining the technical statistical tools.

In principle, having agreed upon a good statistical ensemble one would
like to address questions concerned directly with the
Standard Model, like:
\begin{itemize}
\item{What is the percentage of models having the right gauge group,
       matter and number of families?}
\item{How drastically is this number reduced by requiring more
     detailed constraints, like the correct  gauge and Yukawa couplings,
     the right Higgs couplings, absence of exotic matter?}
\item{Having installed all phenomenological constraints, how
     does the distribution of the supersymmetry  breaking scale and
    the cosmological constant look like?}
\end{itemize}
The answers to these questions will  strongly depend on possible
statistical correlations among the various quantities.

Finally, let us mention that the idea of a stringy landscape has also influenced the way
people think about other fine tuning problems, like for instance
the gauge hierarchy problem. This has led to the idea
of split supersymmetry \cite{Arkani-Hamed:2004fb}, which still
gives rise to gauge coupling unification at the GUT scale but
does not employ low energy supersymmetry for stabilizing the
weak scale.

\section{CONCLUSIONS}
\label{secconclusions}

Let us close with a summary of the status of the attempts we have
covered here to find a realistic string vacuum. We find it fair to
say that despite the enormous effort and the unquestionable
successes in understanding the structure of string models so far
there is no fully realistic candidate. We have discussed models
with essential features of the Standard Model, but all promising
candidates fail to be realistic at a certain step. In the end, a
successful string compactification would need to solve the moduli
problem explicitly and make all parameters computable which is
still beyond our capabilities. Before getting desperate about this
situation one should keep in mind, though, that we are only
looking in very special corners of the overall configuration
space, namely those which are technically accessible and under
good control. Toroidal orbifolds are the most computable examples,
Gepner models or geometric compactification at large radius
defined by vector bundles on Calabi-Yau manifolds are
alternatives. For most other string compactifications we cannot
even answer the most elementary questions.

Therefore, a scan of all possible string vacua is far beyond our
understanding and computational abilities. As argued in
\cite{Denef:2006ad} the complexity of the task to identify
the physically relevant vacua among the set of all vacua may fall
in the category of so-called NP-hard problems. This may be a
reason why we have not found the completely successful model yet.
So far, we can only see the tip of the iceberg. The search for a
realistic vacuum is unquestionably one of the most essential tasks
in string theory, arguably the most ambitious computational
problem ever encountered in theoretical physics. The reputation of
string theorists will have to be measured by the answer delivered
to this question.

In this review article we have collected a number of technical
tools usable for building models describing various classes of
${\cal N}=1$ supersymmetric four-dimensional string
compactifications. There are techniques applicable in the
geometric large radius regime and conformal field theory methods
which can be used for certain special points in the parameter
space. A better understanding of the physics in the intermediate
regime would be highly desirable. It would be of great impact to
obtain concrete information on superpotentials, both classical
contributions and those generated by world-sheet or space-time
instantons. The way these can lift part of the moduli space is
still not well understood. In order to really quantitatively
relate string compactifications to low energy physics higher order
corrections in perturbation theory are equally important. This
refers in particular to the K\"ahler potential. A numerical
approach along the line proposed in \cite{Douglas:2006hz} might
turn out to be promising.

The problem of moduli stabilization needs to find a solution
before any string compactification can lead to fully a determined
effective Lagrangian. It is also at the heart of the naturalness
problems of standard field theory. In this sense, without a fully
satisfactory moduli stabilization, there are no completely
realistic string compactifications anyways. We have here also
partly reviewed the mechanism of moduli stabilization through
fluxes or similar deformations of Calabi-Yau compactifications in
type II string theories. It provides a simple and, at least in
cases, controllable way to stabilize part of the moduli scalars.
This subject is currently still debated very actively and final
conclusions are hard to identify yet. The general set of fluxes
permissable in a given model, the consistency conditions that have
to be satisfied when D-branes are present as well, and a number of
conceptual issues such as the proper treatment of backreaction are
examples for open questions that bear some impact on the models
one can construct. It is not finally excluded that quantum physics
of fluxes in string theory could spoil some of the results
obtained in the classical approach mostly used at present.

On the other hand, taking the known formulas for potentials
induced by fluxes seriously leads unavoidably to the enormous
proliferation in the number of string vacua. This would change the
nature of the vacuum problem significantly. It would provide at
least a logical rationale for a statistical solution of the
naturalness problems, the gauge hierarchy of the Standard Model
and the cosmological constant problem, but at the prize of
sacrificing the strict predictability of the theory. Statistical
methods might then become really mandatory to get any insight into
the distributions of physical quantities. This would, of course,
leave many problems. It would be hard or impossible to decide
which parameters have statistical or environmental explanations
and which ones are uniquely determined and predictable by the
dynamics of the theory. A simple example where both possibilities
could in principle apply is the question why the space we live in
has four macroscopic space-time dimensions.

Regarding this issue, we do not expect to be any close to an
answer. We know far too little about the structure of string
theory and its space of solutions. Within the available
experimental and theoretical understanding of nature none of the
two possibilities is excluded, there may exist a multiverse and
part of the physics in our own universe is explained
environmentally, or maybe a better understanding of full quantum
M-theory could tell us the dynamical selection principle for four
dimensions. Research remains a historical process, and time will
tell.

\vskip 1cm
 {\noindent  {\Large \bf Acknowledgements}}
 \vskip 0.5cm
R.B. and D.L. would like to thank the KITP Santa Barbara and
St.St.  CERN for hospitality. R.B.
also thanks Rutgers University, where part of this review was written.

We would like to thank all our collaborators for working with us during the last six years on the material presented in this article:
C. Angelantonj, K. Behrndt,  M. Berg, V. Braun, J. Conlon, M. Cveti{\v c}, G. Curio, G. Dall'Agata, F. Epple, M. Gaberdiel, F. Gmeiner, L. G\"orlich, M. Haack, R. Helling, G. Honecker, A. Klemm, A. Krause, D. Krefl,  P. Langacker,  G. Lopes Cardoso, P. Manousselis, F. Marchesano, P. Mayr, S. Moster, P. Nath, T. Ott, E. Plauschinn, S. Reffert, R. Richter, E. Scheidegger, W. Schulgin, G. Shiu, K. Suruliz, T.R. Taylor, S. Theisen, P.K. Tripathy, D. Tsimpis, A. Van Proeyen, T. Weigand, M. Zagermann and G. Zoupanos.

Moreover, over the years we have learnt very much  from interesting discussions with
B. Acharya, N. Akerblom, G. Aldazabal, M. Alim, S. de Alwis, I. Antoniadis,
R. Apreda, C. Bachas, K. Becker, M. Becker, M. Bianchi, D. Cremades, F. Denef,
J.-P. Derendinger, K. Dienes, M. Douglas, E. Dudas, G. Dvali, J. Erdmenger,
M. Faux, A. Faraggi, S. Ferrara, B. Florea, S. F\"orste, A. Font, M. Gra{\~
  n}a, T. Grimm, Y.H. He, A. Hebecker, L. Ib{\' a}{\~ n}ez, C. Jeschek,
H. Jockers, S. Kachru, R. Kallosh, E. Kiritsis, S. K\"ors,  
C. Kounnas, W. Lerche, A. Linde, J. Louis, A. Lukas, A. Miemiec, H.P. Nilles, B. Ovrut, F. Quevedo, G. Pradisi, R. Rabadan,  M. Ratz, R. Reinbacher, A. Sagnotti, B. Schellekens, M. Schmidt-Sommerfeld, 
C. Sieg, E. Silverstein, M. Trigiante,  A. Uranga and F. Zwirner.

This work is supported in part by the European Community's Human
Potential Programme under contract MRTN-CT-2004-005104 `Constituents,
fundamental forces and symmetries of the universe'.

\clearpage

\begin{appendix}

\setcounter{equation}{0}


\section{Some basics of CFT}
\setcounter{equation}{0}
\lab{cftbasics}

We collect some basic formulas for the world sheet CFT of the
superstring, sticking to the conventions of \cite{JP98,JP98a}.

\subsection{Free closed string CFT}

The world sheet sigma-model for a ten-dimensional flat target space with
free scalars and fermions in super conformal gauge is
\beqn \label{flatsigma}
\cs = \frac{1}{2\pi\alpha'} \int_\Sigma d^2z\, \Big[ \partial X^M
\bar\partial X_M + \frac{\alpha'}{2} \Big( \psi^M \bar\partial \psi_M
+ \tilde\psi^M \partial \tilde\psi_M \Big) \Big] \ .
\eeqn
It is obtained from the general closed string sigma-model
(\ref{sigmaaction}) by specifying to a target space with
$G_{MN}=\eta_{MN},\, B_{MN}=0,$ and constant $e^\Phi=g_s$. The
gauge field at the boundary is set to zero, $A_M=0$. The complex
world sheet coordinates are defined
\beqn \lab{cplws}
z=\exp(-i\sigma_1 + \sigma_2)\ .
\eeqn
The equations of motion are
\beqn \label{eom}
\partial\bar\partial X^M (z,\bar z) = 0\ , \quad
\partial\tilde\psi^M(z,\bar z) = \bar\partial \psi^M(z,\bar z) = 0 \ .
\eeqn
Periodicity condition are imposed by $\sigma_1 \equiv \s_1+2\pi$,
while $-\infty < \sigma_2 < \infty$. They are
\beqn \label{period}
&&
X^M(\sigma_1,\sigma_2) ~=~ X^M(\sigma_1+2\pi, \sigma_2) \ , \\
&&
\psi^M(\sigma_1,\sigma_2) ~=~ e^{2\pi i\nu}  \psi^M(\sigma_1+2\pi, \sigma_2)\ , \quad
\tilde\psi^M(\sigma_1,\sigma_2) ~=~ e^{2\pi i\tilde\nu}  \tilde\psi^M(\sigma_1+2\pi, \sigma_2)\ ,
\nonumber
\eeqn
the signs referring to the periodic Ramond ($\nu,\tilde\nu=0$) or the anti-periodic
Neveu-Schwarz ($\nu,\tilde\nu=\frac12$) sectors. Periodicities for left- and right-moving
fermions can be chosen independently, giving rise to four different
sectors, RR, NSNS, RNS, and NSR.
One can now split the fields into holomorphic left- and anti-holomorphic
right-movers, $X^M = X^M_L(z) + X^M_R(\bar z)$,
$\psi^M =  \psi^M(z)$, $\tilde\psi^M = \tilde\psi^M(\bar z)$,
with
\beqn \label{modes}
X^M_L (z) &=& \frac{x^M}{2} - i \frac{\alpha'}{2} p_L^M \ln(z) +
i \sqrt{\frac{\alpha'}{2}} \sum_{n\neq 0} \frac{\alpha_n^M}{n}
z^{-n} \ , \non
X^M_R (\bar z) &=& \frac{x^M}{2} - i \frac{\alpha'}{2} p_R^M
\ln(\bar z) +
i \sqrt{\frac{\alpha'}{2}} \sum_{n\neq 0} \frac{\tilde\alpha_n^M}{n}
\bar z^{-n} \ , \non
\psi^M (z) &=& \sum_{r \in \mbb Z+\nu} \psi_r^M z^{-r-1/2} \ , \non
\tilde\psi^M (\bar z) &=& \sum_{r \in \mbb Z+\tilde \nu}
 \tilde\psi_r^M \bar z^{-r-1/2} \ .
\eeqn
The total momentum and winding is defined
\beqn \lab{KKpw}
p^M = \frac12 \Big( p^M_L+p^M_R\Big) \ , \quad
w^M = \frac12 \Big( p^M_L-p^M_R\Big) \  .
\eeqn
The algebra of raising and lowering operators is
\beqn \label{wsalgebra}
[ \alpha^M_n, \alpha^N_m] &=&
[ \tilde\alpha^M_n,\tilde\alpha^N_m] ~=~ n \delta_{n,-m} \eta^{MN}\ , \non
\{ \psi^M_r,\psi^N_s\} &=&
\{ \tilde\psi^M_r,\tilde\psi^N_s\} ~=~ \delta_{r,-s}  \eta^{MN}\ , \non
\big[ x^M , p^N \big] & = &
i \eta^{MN}\ .
\eeqn
The closed string Hamiltonian is given by
\beqn \lab{clHam}
{\cal H}_{\rm cl} = L_0 + \bar L_0 \ ,
\eeqn
with
\beqn \label{L0cl}
L_0 = \frac{\alpha'}{4} p_L^2 + \frac12 \sum_{n\neq 0}
\alpha^M_{-n}\alpha_{M n}  + \frac12 \sum_{r\in \mbb Z+\nu} r
\psi^M_{-r} \psi_{M r}  + a\ ,
\eeqn
and similarly $\bar L_0$ for the right-movers. The normal ordering
constant $a$ takes values $\frac1{24}$ for each real periodic
fermion, and $-\frac1{48}$ for antiperiodic fermions, and opposite
signs for bosons. Thus, in flat Minkowski space-time it is $a=0$
in a R and $a=-\frac12$ in NS sectors. The world sheet fermion
number operators are
\beqn \label{numop}
F = \sum_{r\in \mbb Z+\nu} \psi^M_{-r} \psi_{M r} \ , \quad
\tilde F = \sum_{r\in \mbb Z+\tilde\nu} \tilde\psi^M_{-r} \tilde\psi_{M
  r} \ .
\eeqn
%


\subsection{Free closed strings with boundaries}

One may now impose boundary conditions e.g.\ at $\s_2=0$, i.e.\ on the
unit circle $z=1/\bar z$.
For the bosons Neumann boundary conditions are $\pa_{\s_2} X^M =0$, Dirichlet conditions
$\pa_{\s_1} X^M =0$. For the bosonic modes this translates into
\beqn \label{NDbos}
{\rm N:} && \quad p^M = 0\ , \quad \alpha^M_n +
\tilde\alpha_{-n}^M = 0 \ , \non
{\rm D:} && \quad w^M  = 0\ , \quad \alpha^M_n -
\tilde\alpha_{-n}^M = 0 \ .
\eeqn
The mode expansion that satisfy the boundary conditions are
\beqn \label{NDbosmodes}
{\rm N:} && \quad X^M(z,\bar z) = x^M - i\frac{\alpha'}{2} w^M \ln( z{\bar z} ) +
i\sqrt{\frac{\alpha'}{2}}
\sum_{n\neq 0} \frac{\alpha_n^M}{n} \bigg[ \frac1{z^{n}} + \frac1{\bar z^{n}}
\bigg]\ ,
\non
{\rm D:} && \quad X^M(z,\bar z) = x^M - i\frac{\alpha'}{2}  p^M \ln( z\bar z) +
i\sqrt{\frac{\alpha'}{2}}
\sum_{n\neq 0} \frac{\alpha_n^M}{n} \bigg[ \frac1{z^{n}} - \frac1{\bar z^{n}}
\Big]\ .
\eeqn
By forming linear combinations more general boundary conditions
(\ref{mixedbc}) can be imposed for constant $G_{MN}$ and ${\cal
F}_{MN}$. The mode expansions are also just superpositions of the
free fields.

The boundary conditions for the fermionic coordinates leave a
freedom of choice for the spin structure at the boundary
\cite{Seiberg:1986by}. This is is reflected in the value of $\eta =
\pm 1$ in
\beqn \label{NDferm}
{\rm N:}  \quad \psi^M + i\eta \tilde\psi^M = 0 \Big|_{\pa\S} \ , \quad
{\rm D:}  \quad \psi^M - i\eta \tilde\psi^M = 0 \Big|_{\pa\S} \ .
\eeqn
Mode expansions are easily constructed from (\ref{modes}).


\subsection{Free open string CFT}
\lab{openCFT}

The open string is defined by supplementing (\ref{flatsigma}) and
(\ref{eom}) with boundary conditions at $\s_1=0,\pi$, the real axis $z=\bar z$, while
$0\leq \s_1\leq \pi$ and $-\infty<\s_2<\infty$. Neumann conditions
read $\partial_{\s_1} X^M =0$ and Dirichlet $\partial_{\s_2} X^M
=0$, opposite from the condition for the closed string imposed at
fixed $\s_2$. The mode expansion of the open string reads
\beqn \label{NDopenmodes}
{\rm N:} && \quad X^M(z,\bar z) = x^M - i\alpha' p^M \ln( z\bar z) +
i\sqrt{\frac{\alpha'}{2}}
\sum_{n\neq 0} \frac{\alpha_n^M}{n} \Big[ \frac1{z^n} - \frac1{\bar z^n}
\Big]\ ,
\non
{\rm D:} && \quad X^M(z,\bar z) = x^M - i\alpha' w^M \ln( z\bar z) +
i\sqrt{\frac{\alpha'}{2}}
\sum_{n\neq 0} \frac{\alpha_n^M}{n} \Big[ \frac1{z^n} + \frac1{\bar z^n}
\Big]\ .
\eeqn
The normalization of the space-time momentum $p^M$ and the winding
$w^M$ differs by a factor 2 compared to (\ref{NDbosmodes}). At the
boundary there is again a choice of spin structure for the
fermions.
By the ``doubling trick'' one defines $\psi^M(\s_1,\s_2)$ for
$\s_1\in[0,2\pi]$ via $\psi^M(\s_1,\s_2) = \tilde\p(2\pi -
\s_1,\s_2)$ for $\s_1\in[\pi,2\pi]$. With this construction the open
string fermion behaves in all respect like a single holomorphic
fermion with either periodic or anti-periodic
periodicity, i.e.\ with a R and an NS sector.
The periodicity of the fields also depends ont he types of boundary
conditions at both ends, NN or DD leading to periodic R and
antiperiodic NS fields, and vice vers for ND or DN boundary
conditions.

The algebra of the open string mode operators is then also just the
holomorphic part of (\ref{wsalgebra}) and the open string Hamiltonian
is
\beqn \label{L0op}
{\cal H}_{\rm op} = L_0 = \a' p^2 + \frac12 \sum_{n\neq 0} \a^M_{-n}
\a_{M n} + \frac12 \sum_{r\in\mbb Z+\nu} r \p^M_{-r}\p_{M r} + a \ .
\eeqn
The normal ordering constant $a$ depends on the periodicity of
the fields just as for the closed string sectors.


\subsection{Free CFT of intersecting branes}
\lab{appinters}

It is straightforward to generalize the above to the situation of
closed and open strings that end on D-branes that are rotated and
intersecting. Consider for simplicity the boundary conditions
(\ref{Thbc}) in a two-dimensional plane or on a two-dimensional
torus. It is useful to introduce complex coordinates
\beqn \label{complplane}
Z = X^1 + U X^2 \ , \quad \bar Z = X^1 + \bar U X^2 \ , \quad 
\P = \p^1 + U \p^2 \ , \quad \bar\P = \p^1 + \bar U \p^2\ ,
\eeqn
where $U$ can be the complex structure of the torus.
Let us simplify by setting $U=i$. The mode operators are similarly
complexified, e.g.\ $\a_n = \a^1_n + i \a^2_n$, $\bar\a_n = \a^1_n
- i \a^2_n$, $\tilde\a_n = \tilde\a^1_n + i \tilde\a^2_n$, $\tilde{\bar\a}_n = \tilde{\bar\a}^1_n + i
\tilde{\bar\a}^2_n$, and so on.

Boundary conditions for a one-dimensionally extended D-brane along the
plane, rotated by an angle $\vf$, can be written
\beqn
\pa_{\s_1} {\rm Re} ( e^{i\vf} Z ) = 0 \Big|_{\pa\S} \ , \quad
\pa_{\s_2} {\rm Im} ( e^{i\vf} Z ) = 0 \Big|_{\pa\S} \ .
\eeqn
It is again straightforward to write the linear combinations of
mode expansions that satisfy these boundary conditions. An
alternative formulation is the rotated boundary state that
satisfies the closed string boundary conditions above. Starting
from
\reef{oscbst} one only has to apply a rotation to a state with ND
boundary conditions. One gets
\beqn \label{complbst}
|B, \vf, \eta \>_{\rm osc} &=& \exp\Big( -\frac12 \sum_{n>0} \frac1n ( e^{2i\vf}
\a_{-n}\tilde\a_{-n} + e^{-2i\vf} \bar\a_{-n} \tilde{\bar\a}_{-n} ) \\
&& \hspace{2cm}
- \frac i2 \eta \sum_{r>0} (e^{2i\vf} \p_{-r} \tilde\p_{-r}
+ e^{-2i\vf} \bar\p_{-r} \tilde{\bar\p}_{-r}) \Big) | B, \eta,
0\>_{\rm osc} \ .
\nonumber
\eeqn
Note that this describes only the oscillator part of the full
boundary state, its zero mode pieces will also change under the
rotation. From here one can read off that in the annulus tree
channel amplitude, the overlap of two such states with potentially
different rotation angles, a phase factor appears.

Open strings with rotated boundary conditions can best be described by their mode expansions. As
long as the boundary conditions at the two ends $\s_1=0,\pi$
are identical (the two rotation angles equal) the solutions are again obtained by simple
rotation from \reef{NDopenmodes}. When the open strings stretch between two sets of branes
that are rotated at different relative angles, they have to satisfy
\beqn
&&
\pa_{\s_2} {\rm Re} ( e^{i\vf_a} Z ) = 0 \Big|_{\s_1=0} \ , \quad
\pa_{\s_1} {\rm Im} ( e^{i\vf_a} Z ) = 0 \Big|_{\s_1=0} \ , \non
&&
\pa_{\s_2} {\rm Re} ( e^{i\vf_b} Z ) = 0 \Big|_{\s_1=\pi} \ , \quad
\pa_{\s_1} {\rm Im} ( e^{i\vf_b} Z ) = 0 \Big|_{\s_1=\pi} \ .
\eeqn
Fields are no longer
periodic or anti-periodic, and modes are not integer or
half-integer. Instead, the mode levels are shifted in the form $n \in
\mbb Z\pm\d_{ab}$ and $r\in\mbb Z+\nu\pm\d_{ab}$ by
\beqn \label{del}
\delta_{ab} = \frac{\varphi_a -\varphi_b}{\pi} \ .
\eeqn
There are no linear modes (linear in $\s_1$ and $\s_2$) that can satisfy the boundary conditions,
and thus no zero modes exist, neither fermionic nor bosonic. The only
degeneracy of the ground state (besides zero modes along transverse
directions) comes from the center of mass coordinate which is confined
to the intersection locus of the two branes. If the branes have
multiple intersection, there is a tower of open string states at each
intersection point.

One can continue to use most formulas of the case for parallel branes,
such as the open string Hamiltonian \reef{L0op},
keeping the shift of modings in mind. The general zero-point energy
for a complex bosonic field with modings in $\mbb Z+\d$ is
\beqn\lab{vacen}
a(\d) = \frac{1}{24} - \frac18 ( 2\delta -1)^2\ .
\eeqn
Fermions come with opposite sign.


\subsection{Modular functions and useful identities}

The eta- and theta-functions we use are
\beqn
\eta(\tau) &=& q^{1/24} \prod_{n=1}^\infty \left( 1-q^n \right) \ , \non
\tht\zba{\vec \alpha}{\vec\beta}(\vec \nu,G) &=& \sum_{\vec n\in \mathbb{Z}^N}
 e^{i\pi(\vec n+\vec \alpha)^{T} G (\vec n+\vec \alpha)}
e^{2\pi i(\vec \nu+\vec \beta)^{T}(\vec n+\vec\alpha)}  \ ,
\label{thetamatrix}
\eeqn
where $G$ can be any $N\times N$ matrix with ${\rm Im}(G)>0$, and
$q=e^{2\pi i \tau}$ as always. The case $N=1$ is the usual set of
genus one theta-functions. Their product expansion is
\beqn \lab{thetaprod}
\frac{\thba\a\beta(\nu,\t)}{\eta(\t)} &=& e^{2\pi i\a(\nu+\beta)} q^{\a^2/2-1/24}\\
&& \times
 \prod_{n=1}^\infty \Bigg[ \Big(1+q^{n-1/2+\a}  e^{2\pi i (\nu+\beta)} \Big)
                          \Big(1+q^{n-1/2-\a} e^{-2\pi i (\nu+\beta)} \Big)
\Bigg] \nonumber \ .
\eeqn
We sometimes use the notation
\beqn
\tht \zba{0}{0}(\nu,\tau) = \tht_3(\nu,\tau)\ , \quad
\tht \zba{1/2}{0}(\nu,\tau) = \tht_2(\nu,\tau)\ , \non
\tht \zba{0}{1/2}(\nu,\tau) = \tht_4(\nu,\tau)\ , \quad
\tht \zba{1/2}{1/2}(\nu,\tau) = -\tht_1(\nu,\tau)\ .
\eeqn
The $S$ and $T$ modular transformation of the eta-function are
\beqn \lab{Seta}
\eta(\t) &=& (-i\t)^{-1/2} \eta(-1/\t)\ , \non
\eta(\t) &=& e^{-i\pi/12} \eta(\t+1)
\ .
\eeqn
Those of the theta-functions are
\beqn
\thba{\alpha}{\beta}(\nu,\tau)
&=& e^{i \pi \alpha(\alpha+1)}
\thba{\alpha}{\beta-\alpha-1/2}(\nu,\tau+1) \label{T} \ , \non
\thba{\alpha}{\beta}(\nu,\tau)
&=& (-i \tau)^{-1/2} e^{2\pi i \alpha \beta - i \pi \nu^2/\tau}
\thba{-\beta}{\alpha}(\nu/\tau,-1/\tau) \label{S} \ .
\eeqn
The $S$ transformation of the general
theta-function (\ref{thetamatrix}) relevant for transforming KK sums is
\beqn \lab{poisson}
\tht\zba{\vec \alpha}{\vec 0}(0,it G^{-1}) &=& \sqrt{{\rm det}(G)} \, t^{-N/2}\,
 \tht\zba{\vec 0}{\vec 0 }(\vec \alpha,it^{-1} G) \ .
\eeqn
For the M\"obius strip the following sequence $P=TST^2S$ of
modular transformations is useful
\[
\tau ~\rightarrow~ -{1 \over \tau} ~\rightarrow~ -{1 \over \tau}
+2 ~\rightarrow~ -{1 \over -{1 \over \tau}+2}~\rightarrow~ -{1 \over -{1 \over \tau}+2} + 1 \ ,
\]
which maps $\frac12+it$ to $\frac12+\frac i{4t}$. It gives
\beqn\lab{SMoeb}
\eta(\t) &=& e^{-i\pi/4} (1-2\t)^{-1/2} \eta\Big( \frac{1-\t}{1-2\t}\Big) \ , \\
\thba{\alpha}{\beta}( 0,\tau)
&=& (1-2\tau)^{-1/2}
e^{2\pi i(\a^2/2 + \beta^2 + 2\a\beta -\a/2 -2\beta)}
\thba{\alpha+2\beta}{1/2-\a-\beta}\Big(0, \frac{1-\tau}{1-2\tau}\Big) \;
\nonumber .
\eeqn
The basic quartic Riemann identity is
\beqn\lab{jacobi}
{1 \over 2}\sum_{\alpha,\beta} \eta_{\alpha\beta}
\prod_{i=1}^4 \thba{\alpha}{\beta}(g_i,\tau)&=&
-\prod_{i=1}^4 \thba{1/2}{1/2}(g_i',\tau)
\eeqn
with
\beqn \label{gs}
g_1' &=& {1 \over 2}(g_1+g_2+g_3+g_4) \; , \qquad
g_2' = {1 \over 2}(g_1+g_2-g_3-g_4) \; , \non
g_3' &=& {1 \over 2}(g_1-g_2+g_3-g_4) \; , \qquad
g_4' = {1 \over 2}(g_1-g_2-g_3+g_4) \ .
\eeqn
%


\section{Differential forms and characteristic classes}
\label{conventions}

We use the following standard conventions for differential forms.
For a real or complex valued $p$-form $\Omega_p$ in ten dimensions write
\beqn \label{defforms}
\Omega_p &=& \frac{1}{p!} \Omega_{M_1\, ... \, M_p}
 dx^{M_1} \wedge\, \cdots\, \wedge dx^{M_p}\ , \non
|\Omega_p|^2 &=& \frac{1}{p!} \Omega_{M_1\, ... \, M_p}
   \bar\Omega^{M_1\, ... \, M_p} \ ,
\eeqn
Ten-dimensional Hodge duality is defined by
\beqn
* \Omega_p = \frac{1}{p!(10-p)!} {\epsilon^{M_1\, ...\, M_{p}}}_{M_{p+1}\, ...\, M_{10}}
\Omega_{M_{1}\, ...\, M_{p}}
dx^{M_{p+1}} \wedge\, \cdots\, \wedge dx^{M_{10}}\ ,
\eeqn
We sometimes also perform a Hodge duality of only the internal
six-dimensional space. It is denoted by $\star \O_p$ and given by
\beqn \label{hodgesix}
(\star \Omega_p)_{\mu_1 \ldots \mu_n j_1 \ldots j_{6-p+n}}
= \frac{1}{(p-n)!} {\epsilon^{i_1 \ldots i_{p-n}}}_{j_1 \ldots j_{6-p+n}}
\Omega_{\mu_1\, ...\, \mu_n i_{1}\, ...\, i_{p-n}}\ .
\eeqn
For the totally antisymmetric tensor $\epsilon$ in $D$ dimensions
we use the convention
\beqn
\epsilon_{1\, ... \, D} = \pm \sqrt{|g_D|}\ ,\quad
\epsilon^{1\, ... \, D} = \frac{1}{\sqrt{|g_D|}}\ ,
\eeqn
with $g_D$ the determinant of the metric
and the sign depends on the signature of the metric.

The characteristic classes are polynomials in the curvature form
of a vector bundle given by a matrix $x$ with eigenvalues $x_i$.
The Chern class of a complex bundle is
\beqn \lab{Cherncl}
c(x) = {\rm det}(1+x) = \sum_{n=0}^\infty c_n(x) = 1 + \tr\, x -\frac12 ( \tr\, x^2 - (\tr\, x)^2 )
 + \ \cdots\ .
\eeqn
The $c_n(x)$ are of degree $n$ in $x$.
The Chern character is
\beqn \lab{Chernch}
{\rm ch} (x) = \tr\, \exp(x) = \sum_{n=0}^\infty {\rm ch}_n(x) \ .
\eeqn
It has the important properties
\beqn
{\rm ch}(x\oplus y) = {\rm ch}(x) + {\rm ch}(y) \ , \quad
{\rm ch}(x\otimes y) = {\rm ch}(x) \wedge {\rm ch}(y) \ .
\eeqn
The Pontryagin class of a real bundle is
\beqn \lab{Pontryagin}
p(x)={\rm det}(1-x) = \sum_{n=0}^\infty p_n (x) \ .
\eeqn
The individual terms $p_n(x)$ are of degree $2n$ in $x$, since
$x^T=-x$. This is also true for the following real classes. The
A-roof genus is
\beqn \lab{Aroof}
\hat A(x) = \prod_i \frac{\frac12 x_i}{\sinh(\frac12 x_i)} = 1 -\frac1{24} p_1
 + \frac1{5760} ( 7p_1^2-4p_2) + \ \cdots\ .
\eeqn
The Hirzebruch $L$-polynomial is
\beqn \lab{Hirzebruch}
L(x) = \prod_i \frac{x_i}{\tanh(x_i)} =1+{1\over 3} p_1 +{1\over 45}(-p_1^2 +7 p_2) + \ \cdots\ .
\eeqn
To evaluate traces in $SU(N)$ explicitly we note some relation for
traces in different representations,
\beqn\lab{anomtr}
\tr_{\rm adj}\, F^2 &=& 2N \tr_{N}\, F^2\ , \non
\tr_{\rm S}\, F^2 &=& (N+4) \tr_{N}\, F^2\ , \non
\tr_{\rm A}\, F^2 &=& (N-4) \tr_{N}\, F^2\ ,
\eeqn
where representations are denoted by adj, $N$, A, and S for the
adjoint, fundamental, anti-symmetric and symmetric.


\end{appendix}

\clearpage
\nocite{*}
\bibliography{rev}

\providecommand{\href}[2]{#2}\begingroup\raggedright\begin{thebibliography}{10%
0}

\bibitem{JP95}
J.~Polchinski, {\em Dirichlet-Branes and Ramond-Ramond Charges,} Phys. Rev.
  Lett. {\bf 75} (1995) 4724--4727,
\href{http://www.arXiv.org/abs/hep-th/9510017}{{\tt hep-th/9510017}}.

\bibitem{pcj96}
J.~Polchinski, S.~Chaudhuri, and C.~V. Johnson, {\em Notes on D-Branes,}
\href{http://www.arXiv.org/abs/hep-th/9602052}{{\tt hep-th/9602052}}.

\bibitem{JP96}
J.~Polchinski, {\em Lectures on D-branes,}
\href{http://www.arXiv.org/abs/hep-th/9611050}{{\tt hep-th/9611050}}.

\bibitem{Alvarez-Gaume:1983ig}
L.~Alvarez-Gaume and E.~Witten, {\em Gravitational anomalies,} Nucl. Phys. {\bf
  B234} (1984)
269.

\bibitem{Green:1984sg}
M.~B. Green and J.~H. Schwarz, {\em Anomaly Cancellation in Supersymmetric D=10
  Gauge Theory and Superstring Theory,} Phys. Lett. {\bf B149} (1984)
117--122.

\bibitem{GrossDD}
D.~J. Gross, J.~A. Harvey, E.~J. Martinec, and R.~Rohm, {\em The Heterotic
  String,} Phys. Rev. Lett. {\bf 54} (1985)
502--505.

\bibitem{Candelas:1985en}
P.~Candelas, G.~T. Horowitz, A.~Strominger, and E.~Witten, {\em Vacuum
  configurations for superstrings,} Nucl. Phys. {\bf B258} (1985)
46--74.

\bibitem{Dixon:1985jw}
L.~J. Dixon, J.~A. Harvey, C.~Vafa, and E.~Witten, {\em Strings on orbifolds,}
  Nucl. Phys. {\bf B261} (1985)
678--686.

\bibitem{Dixon:1986jc}
L.~J. Dixon, J.~A. Harvey, C.~Vafa, and E.~Witten, {\em Strings on orbifolds.
  2,} Nucl. Phys. {\bf B274} (1986)
285--314.

\bibitem{Ibanez:1986tp}
L.~E. Ib{\'a}{\~n}ez, H.~P. Nilles, and F.~Quevedo, {\em Orbifolds and Wilson
  lines,} Phys. Lett. {\bf B187} (1987)
25--32.

\bibitem{Kawai:1986va}
H.~Kawai, D.~C. Lewellen, and S.~H.~H. Tye, {\em Construction of
  four-dimensional fermionic string models,} Phys. Rev. Lett. {\bf 57} (1986)
1832.

\bibitem{Antoniadis:1986rn}
I.~Antoniadis, C.~P. Bachas, and C.~Kounnas, {\em Four-dimensional
  superstrings,} Nucl. Phys. {\bf B289} (1987)
87.

\bibitem{Lerche:1986cx}
W.~Lerche, D.~L{\"u}st, and A.~N. Schellekens, {\em Chiral four-dimensional
  heterotic strings from selfdual lattices,} Nucl. Phys. {\bf B287} (1987)
477.

\bibitem{Sagnotti:1987tw}
A.~Sagnotti, {\em Open strings and their symmetry groups,}
\href{http://www.arXiv.org/abs/hep-th/0208020}{{\tt hep-th/0208020}}.

\bibitem{Polchinski:1987tu}
J.~Polchinski and Y.~Cai, {\em Consistency of open superstring theories,} Nucl.
  Phys. {\bf B296} (1988)
91.

\bibitem{Pradisi:1988xd}
G.~Pradisi and A.~Sagnotti, {\em Open String orbifolds,} Phys. Lett. {\bf B216}
  (1989)
59.

\bibitem{Horava:1989vt}
P.~Horava, {\em Strings on world sheet orbifolds,} Nucl. Phys. {\bf B327}
  (1989)
461.

\bibitem{Bianchi:1990yu}
M.~Bianchi and A.~Sagnotti, {\em On the systematics of open string theories,}
  Phys. Lett. {\bf B247} (1990)
517--524.

\bibitem{Bianchi:1991eu}
M.~Bianchi, G.~Pradisi, and A.~Sagnotti, {\em Toroidal compactification and
  symmetry breaking in open string theories,} Nucl. Phys. {\bf B376} (1992)
365--386.

\bibitem{as02}
C.~Angelantonj and A.~Sagnotti, {\em Open strings,} Phys. Rept. {\bf 371}
  (2002) 1--150,
\href{http://www.arXiv.org/abs/hep-th/0204089}{{\tt hep-th/0204089}}.

\bibitem{Gimon:1996rq}
E.~G. Gimon and J.~Polchinski, {\em Consistency Conditions for Orientifolds and
  D-Manifolds,} Phys. Rev. {\bf D54} (1996) 1667--1676,
\href{http://www.arXiv.org/abs/hep-th/9601038}{{\tt hep-th/9601038}}.

\bibitem{Gimon:1996ay}
E.~G. Gimon and C.~V. Johnson, {\em K3 Orientifolds,} Nucl. Phys. {\bf B477}
  (1996) 715--745,
\href{http://www.arXiv.org/abs/hep-th/9604129}{{\tt hep-th/9604129}}.

\bibitem{Berkooz:1996dw}
M.~Berkooz and R.~G. Leigh, {\em A D = 4 N = 1 orbifold of type I strings,}
  Nucl. Phys. {\bf B483} (1997) 187--208,
\href{http://www.arXiv.org/abs/hep-th/9605049}{{\tt hep-th/9605049}}.

\bibitem{Angelantonj:1996uy}
C.~Angelantonj, M.~Bianchi, G.~Pradisi, A.~Sagnotti, and Y.~S. Stanev, {\em
  Chiral asymmetry in four-dimensional open- string vacua,} Phys. Lett. {\bf
  B385} (1996) 96--102,
\href{http://www.arXiv.org/abs/hep-th/9606169}{{\tt hep-th/9606169}}.

\bibitem{Kakushadze:1997wx}
Z.~Kakushadze, {\em Aspects of N = 1 type I-heterotic duality in four
  dimensions,} Nucl. Phys. {\bf B512} (1998) 221--236,
\href{http://www.arXiv.org/abs/hep-th/9704059}{{\tt hep-th/9704059}}.

\bibitem{Aldazabal:1998mr}
G.~Aldazabal, A.~Font, L.~E. Ib{\'a}{\~n}ez, and G.~Violero, {\em D = 4, N = 1,
  type IIB orientifolds,} Nucl. Phys. {\bf B536} (1998) 29--68,
\href{http://www.arXiv.org/abs/hep-th/9804026}{{\tt hep-th/9804026}}.

\bibitem{bdl96}
M.~Berkooz, M.~R. Douglas, and R.~G. Leigh, {\em Branes intersecting at
  angles,} Nucl. Phys. {\bf B480} (1996) 265--278,
\href{http://www.arXiv.org/abs/hep-th/9606139}{{\tt hep-th/9606139}}.

\bibitem{CB95}
C.~Bachas, {\em A Way to break supersymmetry,}
\href{http://www.arXiv.org/abs/hep-th/9503030}{{\tt hep-th/9503030}}.

\bibitem{bgk99}
R.~Blumenhagen, L.~G{\"o}rlich, and B.~K{\"o}rs, {\em Supersymmetric
  orientifolds in 6D with D-branes at angles,} Nucl. Phys. {\bf B569} (2000)
  209--228,
\href{http://www.arXiv.org/abs/hep-th/9908130}{{\tt hep-th/9908130}}.

\bibitem{GP99}
G.~Pradisi, {\em Type I vacua from diagonal Z(3)-orbifolds,} Nucl. Phys. {\bf
  B575} (2000) 134--150,
\href{http://www.arXiv.org/abs/hep-th/9912218}{{\tt hep-th/9912218}}.

\bibitem{bgk99a}
R.~Blumenhagen, L.~G{\"o}rlich, and B.~K{\"o}rs, {\em Supersymmetric 4D
  orientifolds of type IIA with D6-branes at angles,} JHEP {\bf 01} (2000) 040,
\href{http://www.arXiv.org/abs/hep-th/9912204}{{\tt hep-th/9912204}}.

\bibitem{bgkl00a}
R.~Blumenhagen, L.~G{\"o}rlich, B.~K{\"o}rs, and D.~L{\"u}st, {\em
  Noncommutative compactifications of type I strings on tori with magnetic
  background flux,} JHEP {\bf 10} (2000) 006,
\href{http://www.arXiv.org/abs/hep-th/0007024}{{\tt hep-th/0007024}}.

\bibitem{aads00}
C.~Angelantonj, I.~Antoniadis, E.~Dudas, and A.~Sagnotti, {\em Type-I strings
  on magnetised orbifolds and brane transmutation,} Phys. Lett. {\bf B489}
  (2000) 223--232,
\href{http://www.arXiv.org/abs/hep-th/0007090}{{\tt hep-th/0007090}}.

\bibitem{fhs00}
S.~F{\"o}rste, G.~Honecker, and R.~Schreyer, {\em Supersymmetric Z(N) x Z(M)
  orientifolds in 4D with D-branes at angles,} Nucl. Phys. {\bf B593} (2001)
  127--154,
\href{http://www.arXiv.org/abs/hep-th/0008250}{{\tt hep-th/0008250}}.

\bibitem{bkl00}
R.~Blumenhagen, B.~K{\"o}rs, and D.~L{\"u}st, {\em Type I strings with F- and
  B-flux,} JHEP {\bf 02} (2001) 030,
\href{http://www.arXiv.org/abs/hep-th/0012156}{{\tt hep-th/0012156}}.

\bibitem{afiru00}
G.~Aldazabal, S.~Franco, L.~E. Ib{\'a}{\~n}ez, R.~Rabad{\' a}n, and A.~M.
  Uranga, {\em Intersecting brane worlds,} JHEP {\bf 02} (2001) 047,
\href{http://www.arXiv.org/abs/hep-ph/0011132}{{\tt hep-ph/0011132}}.

\bibitem{afiru00a}
G.~Aldazabal, S.~Franco, L.~E. Ib{\'a}{\~n}ez, R.~Rabad{\' a}n, and A.~M.
  Uranga, {\em D = 4 chiral string compactifications from intersecting branes,}
  J. Math. Phys. {\bf 42} (2001) 3103--3126,
\href{http://www.arXiv.org/abs/hep-th/0011073}{{\tt hep-th/0011073}}.

\bibitem{imr01}
L.~E. Ib{\'a}{\~n}ez, F.~Marchesano, and R.~Rabad{\' a}n, {\em Getting just the
  standard model at intersecting branes,} JHEP {\bf 11} (2001) 002,
\href{http://www.arXiv.org/abs/hep-th/0105155}{{\tt hep-th/0105155}}.

\bibitem{bklo01}
R.~Blumenhagen, B.~K{\"o}rs, D.~L{\"u}st, and T.~Ott, {\em The standard model
  from stable intersecting brane world orbifolds,} Nucl. Phys. {\bf B616}
  (2001) 3--33,
\href{http://www.arXiv.org/abs/hep-th/0107138}{{\tt hep-th/0107138}}.

\bibitem{csu01}
M.~Cveti{\v c}, G.~Shiu, and A.~M. Uranga, {\em Three-family supersymmetric
  standard like models from intersecting brane worlds,} Phys. Rev. Lett. {\bf
  87} (2001) 201801,
\href{http://www.arXiv.org/abs/hep-th/0107143}{{\tt hep-th/0107143}}.

\bibitem{csu01a}
M.~Cveti{\v c}, G.~Shiu, and A.~M. Uranga, {\em Chiral four-dimensional N = 1
  supersymmetric type IIA orientifolds from intersecting D6-branes,} Nucl.
  Phys. {\bf B615} (2001) 3--32,
\href{http://www.arXiv.org/abs/hep-th/0107166}{{\tt hep-th/0107166}}.

\bibitem{Strominger:1986uh}
A.~Strominger, {\em Superstrings with Torsion,} Nucl. Phys. {\bf B274} (1986)
253.

\bibitem{drs99}
K.~Dasgupta, G.~Rajesh, and S.~Sethi, {\em M theory, orientifolds and G-flux,}
  JHEP {\bf 08} (1999) 023,
\href{http://www.arXiv.org/abs/hep-th/9908088}{{\tt hep-th/9908088}}.

\bibitem{gvw99}
S.~Gukov, C.~Vafa, and E.~Witten, {\em CFT's from Calabi-Yau four-folds,} Nucl.
  Phys. {\bf B584} (2000) 69--108,
\href{http://www.arXiv.org/abs/hep-th/9906070}{{\tt hep-th/9906070}}.

\bibitem{tv99}
T.~R. Taylor and C.~Vafa, {\em RR flux on Calabi-Yau and partial supersymmetry
  breaking,} Phys. Lett. {\bf B474} (2000) 130--137,
\href{http://www.arXiv.org/abs/hep-th/9912152}{{\tt hep-th/9912152}}.

\bibitem{Peter}
P.~Mayr, {\em On supersymmetry breaking in string theory and its realization in
  brane worlds,} Nucl. Phys. {\bf B593} (2001) 99--126,
\href{http://www.arXiv.org/abs/hep-th/0003198}{{\tt hep-th/0003198}}.

\bibitem{gss00}
B.~R. Greene, K.~Schalm, and G.~Shiu, {\em Warped compactifications in M and F
  theory,} Nucl. Phys. {\bf B584} (2000) 480--508,
\href{http://www.arXiv.org/abs/hep-th/0004103}{{\tt hep-th/0004103}}.

\bibitem{cklt00}
G.~Curio, A.~Klemm, D.~L{\" u}st, and S.~Theisen, {\em On the vacuum structure
  of type II string compactifications on Calabi-Yau spaces with H-fluxes,}
  Nucl. Phys. {\bf B609} (2001) 3--45,
\href{http://www.arXiv.org/abs/hep-th/0012213}{{\tt hep-th/0012213}}.

\bibitem{gkp01}
S.~B. Giddings, S.~Kachru, and J.~Polchinski, {\em Hierarchies from fluxes in
  string compactifications,} Phys. Rev. {\bf D66} (2002) 106006,
\href{http://www.arXiv.org/abs/hep-th/0105097}{{\tt hep-th/0105097}}.

\bibitem{kst02}
S.~Kachru, M.~B. Schulz, and S.~P. Trivedi, {\em Moduli stabilization from
  fluxes in a simple IIB orientifold,} JHEP {\bf 10} (2003) 007,
\href{http://www.arXiv.org/abs/hep-th/0201028}{{\tt hep-th/0201028}}.

\bibitem{bcls05}
R.~Blumenhagen, M.~Cveti{\v c}, P.~Langacker, and G.~Shiu, {\em Toward
  realistic intersecting D-brane models,} Ann. Rev. Nucl. Part. Sci. {\bf 55}
  (2005) 71--139,
\href{http://www.arXiv.org/abs/hep-th/0502005}{{\tt hep-th/0502005}}.

\bibitem{kw03}
I.~R. Klebanov and E.~Witten, {\em Proton decay in intersecting D-brane
  models,} Nucl. Phys. {\bf B664} (2003) 3--20,
\href{http://www.arXiv.org/abs/hep-th/0304079}{{\tt hep-th/0304079}}.

\bibitem{Cvetic:2006iz}
M.~Cveti{\v c} and R.~Richter, {\em Proton decay via dimension-six operators in
  intersecting D6-brane models,}
\href{http://www.arXiv.org/abs/hep-th/0606001}{{\tt hep-th/0606001}}.

\bibitem{Conlon:2006tq}
J.~P. Conlon, {\em The QCD axion and moduli stabilisation,} JHEP {\bf 05}
  (2006) 078,
\href{http://www.arXiv.org/abs/hep-th/0602233}{{\tt hep-th/0602233}}.

\bibitem{Svrcek:2006yi}
P.~Svrcek and E.~Witten, {\em Axions in string theory,} JHEP {\bf 06} (2006)
  051,
\href{http://www.arXiv.org/abs/hep-th/0605206}{{\tt hep-th/0605206}}.

\bibitem{Svrcek:2006hf}
P.~Svrcek, {\em Cosmological Constant and Axions in String Theory,}
\href{http://www.arXiv.org/abs/hep-th/0607086}{{\tt hep-th/0607086}}.

\bibitem{Conlon:2006ur}
J.~P. Conlon, {\em Seeing the invisible axion in the sparticle spectrum,}
\href{http://www.arXiv.org/abs/hep-ph/0607138}{{\tt hep-ph/0607138}}.

\bibitem{gsw87}
M.~B. Green, J.~H. Schwarz, and E.~Witten, {\em Superstring theory, Vol. 1:
  Introduction,}. Cambridge, Uk: Univ. Pr. ( 1987) 469 P. ( Cambridge
  Monographs On Mathematical Physics).

\bibitem{gsw87a}
M.~B. Green, J.~H. Schwarz, and E.~Witten, {\em Superstring theory, Vol. 2:
  Loop amplitudes, anomalies and phenomenology,}. Cambridge, Uk: Univ. Pr. (
  1987) 596 P. ( Cambridge Monographs On Mathematical Physics).

\bibitem{Lust:1989tj}
D.~L{\"u}st and S.~Theisen, {\em Lectures on string theory,} Lect. Notes Phys.
  {\bf 346} (1989)
1--346.

\bibitem{JP98}
J.~Polchinski, {\em String theory. Vol. 1: An introduction to the bosonic
  string,}. Cambridge, UK: Univ. Pr. (1998) 402 p.

\bibitem{JP98a}
J.~Polchinski, {\em String theory. Vol. 2: Superstring theory and beyond,}.
  Cambridge, UK: Univ. Pr. (1998) 531 p.

\bibitem{CJ03}
C.~V. Johnson, {\em D-branes,}. Cambridge, USA: Univ. Pr. (2003) 548 p.

\bibitem{BZ04}
B.~Zwiebach, {\em A first course in string theory,}. Cambridge, UK: Univ. Pr.
  (2004) 558 p.

\bibitem{AD98}
A.~Dabholkar, {\em Lectures on orientifolds and duality,}
\href{http://www.arXiv.org/abs/hep-th/9804208}{{\tt hep-th/9804208}}.

\bibitem{Quevedo}
F.~Quevedo, {\em Lectures on string / brane cosmology,} Class. Quant. Grav.
  {\bf 19} (2002) 5721--5779,
\href{http://www.arXiv.org/abs/hep-th/0210292}{{\tt hep-th/0210292}}.

\bibitem{AU03}
A.~M. Uranga, {\em Chiral four-dimensional string compactifications with
  intersecting D-branes,} Class. Quant. Grav. {\bf 20} (2003) S373--S394,
\href{http://www.arXiv.org/abs/hep-th/0301032}{{\tt hep-th/0301032}}.

\bibitem{EK03}
E.~Kiritsis, {\em D-branes in standard model building, gravity and cosmology,}
  Fortsch. Phys. {\bf 52} (2004) 200--263,
\href{http://www.arXiv.org/abs/hep-th/0310001}{{\tt hep-th/0310001}}.

\bibitem{DL04}
D.~L{\"u}st, {\em Intersecting brane worlds: A path to the standard model?,}
  Class. Quant. Grav. {\bf 21} (2004) S1399--1424,
\href{http://www.arXiv.org/abs/hep-th/0401156}{{\tt hep-th/0401156}}.

\bibitem{RB04}
R.~Blumenhagen, {\em Recent progress in Intersecting D-brane models,}
\href{http://www.arXiv.org/abs/hep-th/0412025}{{\tt hep-th/0412025}}.

\bibitem{FM03}
F.~G. Marchesano, {\em Intersecting D-brane models,}
\href{http://www.arXiv.org/abs/hep-th/0307252}{{\tt hep-th/0307252}}.

\bibitem{OTT03}
T.~Ott, {\em Aspects of stability and phenomenology in type IIA orientifolds
  with intersecting D6-branes,} Fortsch. Phys. {\bf 52} (2004) 28--137,
\href{http://www.arXiv.org/abs/hep-th/0309107}{{\tt hep-th/0309107}}.

\bibitem{LG04}
L.~G{\"o}rlich, {\em N=1 and non-supersymmetric open string theories in six and
  four space-time dimensions,}
\href{http://www.arXiv.org/abs/hep-th/0401040}{{\tt hep-th/0401040}}.

\bibitem{grana}
M.~Grana, {\em Flux compactifications in string theory: A comprehensive
  review,} Phys. Rept. {\bf 423} (2006) 91--158,
\href{http://www.arXiv.org/abs/hep-th/0509003}{{\tt hep-th/0509003}}.

\bibitem{Douglas:2006es}
M.~R. Douglas and S.~Kachru, {\em Flux compactification,}
\href{http://www.arXiv.org/abs/hep-th/0610102}{{\tt hep-th/0610102}}.

\bibitem{Frey:2003tf}
A.~R. Frey, {\em Warped strings: Self-dual flux and contemporary
  compactifications,}
\href{http://www.arXiv.org/abs/hep-th/0308156}{{\tt hep-th/0308156}}.

\bibitem{Susanne}
S.~Reffert, {\em Toroidal orbifolds: Resolutions, orientifolds and applications
  in string phenomenology,}
\href{http://www.arXiv.org/abs/hep-th/0609040}{{\tt hep-th/0609040}}.

\bibitem{Kumar:2006tn}
J.~Kumar, {\em A review of distributions on the string landscape,}
\href{http://www.arXiv.org/abs/hep-th/0601053}{{\tt hep-th/0601053}}.

\bibitem{Fradkin:1985qd}
E.~S. Fradkin and A.~A. Tseytlin, {\em Nonlinear electrodynamics from quantized
  strings,} Phys. Lett. {\bf B163} (1985)
123.

\bibitem{Callan:1985ia}
J.~Callan, Curtis~G., E.~J. Martinec, M.~J. Perry, and D.~Friedan, {\em Strings
  in background fields,} Nucl. Phys. {\bf B262} (1985)
593.

\bibitem{Abouelsaood:1986gd}
A.~Abouelsaood, J.~Callan, Curtis~G., C.~R. Nappi, and S.~A. Yost, {\em Open
  strings in background gauge fields,} Nucl. Phys. {\bf B280} (1987)
599.

\bibitem{Dorn:1986jf}
H.~Dorn and H.~J. Otto, {\em Open bosonic strings in general backgrounds,} Z.
  Phys. {\bf C32} (1986)
599.

\bibitem{Leigh:1989jq}
R.~G. Leigh, {\em Dirac-Born-Infeld action from Dirichlet sigma models,} Mod.
  Phys. Lett. {\bf A4} (1989)
2767.

\bibitem{Witten:1995im}
E.~Witten, {\em Bound states of strings and p-branes,} Nucl. Phys. {\bf B460}
  (1996) 335--350,
\href{http://www.arXiv.org/abs/hep-th/9510135}{{\tt hep-th/9510135}}.

\bibitem{Callan:1988wz}
J.~Callan, Curtis~G., C.~Lovelace, C.~R. Nappi, and S.~A. Yost, {\em Loop
  corrections to superstring equations of motion,} Nucl. Phys. {\bf B308}
  (1988)
221.

\bibitem{Cardy:1989ir}
J.~L. Cardy, {\em Boundary conditions, fusion rules and the verlinde formula,}
  Nucl. Phys. {\bf B324} (1989)
581.

\bibitem{Li:1995pq}
M.~Li, {\em Boundary States of D-Branes and Dy-Strings,} Nucl. Phys. {\bf B460}
  (1996) 351--361,
\href{http://www.arXiv.org/abs/hep-th/9510161}{{\tt hep-th/9510161}}.

\bibitem{Green:1996um}
M.~B. Green and M.~Gutperle, {\em Light-cone supersymmetry and D-branes,} Nucl.
  Phys. {\bf B476} (1996) 484--514,
\href{http://www.arXiv.org/abs/hep-th/9604091}{{\tt hep-th/9604091}}.

\bibitem{Gaberdiel:2000jr}
M.~R. Gaberdiel, {\em Lectures on non-BPS Dirichlet branes,} Class. Quant.
  Grav. {\bf 17} (2000) 3483--3520,
\href{http://www.arXiv.org/abs/hep-th/0005029}{{\tt hep-th/0005029}}.

\bibitem{Dai:1989ua}
J.~Dai, R.~G. Leigh, and J.~Polchinski, {\em New connections between string
  theories,} Mod. Phys. Lett. {\bf A4} (1989)
2073--2083.

\bibitem{Seiberg:1986by}
N.~Seiberg and E.~Witten, {\em Spin structures in string theory,} Nucl. Phys.
  {\bf B276} (1986)
272.

\bibitem{Witten:1981me}
E.~Witten, {\em Search for a realistic Kaluza-Klein theory,} Nucl. Phys. {\bf
  B186} (1981)
412.

\bibitem{Paton:1969je}
J.~E. Paton and H.-M. Chan, {\em Generalized veneziano model with isospin,}
  Nucl. Phys. {\bf B10} (1969)
516--520.

\bibitem{Schwarz:1982md}
J.~H. Schwarz, {\em Gauge groups for Type I superstrings,}. Presented at 6th
  Johns Hopkins Workshop on Current Problems in High-Energy Particle Theory,
  Florence, Italy, Jun 2-4, 1982.

\bibitem{Marcus:1982fr}
N.~Marcus and A.~Sagnotti, {\em Tree level constraints on gauge groups for Type
  I superstrings,} Phys. Lett. {\bf B119} (1982)
97.

\bibitem{Douglas:1996sw}
M.~R. Douglas and G.~W. Moore, {\em D-branes, Quivers, and ALE Instantons,}
\href{http://www.arXiv.org/abs/hep-th/9603167}{{\tt hep-th/9603167}}.

\bibitem{Klein:2000hf}
M.~Klein and R.~Rabadan, {\em D = 4, N = 1 orientifolds with vector structure,}
  Nucl. Phys. {\bf B596} (2001) 197--230,
\href{http://www.arXiv.org/abs/hep-th/0007087}{{\tt hep-th/0007087}}.

\bibitem{Douglas:1995bn}
M.~R. Douglas, {\em Branes within branes,}
\href{http://www.arXiv.org/abs/hep-th/9512077}{{\tt hep-th/9512077}}.

\bibitem{Green:1996dd}
M.~B. Green, J.~A. Harvey, and G.~W. Moore, {\em I-brane inflow and anomalous
  couplings on D-branes,} Class. Quant. Grav. {\bf 14} (1997) 47--52,
\href{http://www.arXiv.org/abs/hep-th/9605033}{{\tt hep-th/9605033}}.

\bibitem{Cheung:1997az}
Y.-K.~E. Cheung and Z.~Yin, {\em Anomalies, branes, and currents,} Nucl. Phys.
  {\bf B517} (1998) 69--91,
\href{http://www.arXiv.org/abs/hep-th/9710206}{{\tt hep-th/9710206}}.

\bibitem{Morales:1998ux}
J.~F. Morales, C.~A. Scrucca, and M.~Serone, {\em Anomalous couplings for
  D-branes and O-planes,} Nucl. Phys. {\bf B552} (1999) 291--315,
\href{http://www.arXiv.org/abs/hep-th/9812071}{{\tt hep-th/9812071}}.

\bibitem{Stefanski:1998yx}
J.~Stefanski, Bogdan, {\em Gravitational couplings of D-branes and O-planes,}
  Nucl. Phys. {\bf B548} (1999) 275--290,
\href{http://www.arXiv.org/abs/hep-th/9812088}{{\tt hep-th/9812088}}.

\bibitem{Scrucca:1999uz}
C.~A. Scrucca and M.~Serone, {\em Anomalies and inflow on D-branes and
  O-planes,} Nucl. Phys. {\bf B556} (1999) 197--221,
\href{http://www.arXiv.org/abs/hep-th/9903145}{{\tt hep-th/9903145}}.

\bibitem{Scrucca:1999jq}
C.~A. Scrucca and M.~Serone, {\em Anomaly inflow and RR anomalous couplings,}
\href{http://www.arXiv.org/abs/hep-th/9911223}{{\tt hep-th/9911223}}.

\bibitem{Myers:1999ps}
R.~C. Myers, {\em Dielectric-branes,} JHEP {\bf 12} (1999) 022,
\href{http://www.arXiv.org/abs/hep-th/9910053}{{\tt hep-th/9910053}}.

\bibitem{Tseytlin:1997cs}
A.~A. Tseytlin, {\em On non-abelian generalisation of the Born-Infeld action in
  string theory,} Nucl. Phys. {\bf B501} (1997) 41--52,
\href{http://www.arXiv.org/abs/hep-th/9701125}{{\tt hep-th/9701125}}.

\bibitem{Hashimoto:1997gm}
A.~Hashimoto and I.~Taylor, Washington, {\em Fluctuation spectra of tilted and
  intersecting D-branes from the Born-Infeld action,} Nucl. Phys. {\bf B503}
  (1997) 193--219,
\href{http://www.arXiv.org/abs/hep-th/9703217}{{\tt hep-th/9703217}}.

\bibitem{Denef:2000rj}
F.~Denef, A.~Sevrin, and J.~Troost, {\em Non-Abelian Born-Infeld versus string
  theory,} Nucl. Phys. {\bf B581} (2000) 135--155,
\href{http://www.arXiv.org/abs/hep-th/0002180}{{\tt hep-th/0002180}}.

\bibitem{Bergshoeff:2000ik}
E.~A. Bergshoeff, M.~de~Roo, and A.~Sevrin, {\em Non-Abelian Born-Infeld and
  kappa-symmetry,} J. Math. Phys. {\bf 42} (2001) 2872--2888,
\href{http://www.arXiv.org/abs/hep-th/0011018}{{\tt hep-th/0011018}}.

\bibitem{Stieberger:2002fh}
S.~Stieberger and T.~R. Taylor, {\em Non-Abelian Born-Infeld action and type I
  - heterotic duality. I: Heterotic F**6 terms at two loops,} Nucl. Phys. {\bf
  B647} (2002) 49--68,
\href{http://www.arXiv.org/abs/hep-th/0207026}{{\tt hep-th/0207026}}.

\bibitem{Stieberger:2002wk}
S.~Stieberger and T.~R. Taylor, {\em Non-Abelian Born-Infeld action and type I
  - heterotic duality. II: Nonrenormalization theorems,} Nucl. Phys. {\bf B648}
  (2003) 3--34,
\href{http://www.arXiv.org/abs/hep-th/0209064}{{\tt hep-th/0209064}}.

\bibitem{Coletti:2003ai}
E.~Coletti, I.~Sigalov, and W.~Taylor, {\em Abelian and nonabelian vector field
  effective actions from string field theory,} JHEP {\bf 09} (2003) 050,
\href{http://www.arXiv.org/abs/hep-th/0306041}{{\tt hep-th/0306041}}.

\bibitem{Cederwall:1996pv}
M.~Cederwall, A.~von Gussich, B.~E.~W. Nilsson, and A.~Westerberg, {\em The
  Dirichlet super-three-brane in ten-dimensional type IIB supergravity,} Nucl.
  Phys. {\bf B490} (1997) 163--178,
\href{http://www.arXiv.org/abs/hep-th/9610148}{{\tt hep-th/9610148}}.

\bibitem{Aganagic:1996pe}
M.~Aganagic, C.~Popescu, and J.~H. Schwarz, {\em D-brane actions with local
  kappa symmetry,} Phys. Lett. {\bf B393} (1997) 311--315,
\href{http://www.arXiv.org/abs/hep-th/9610249}{{\tt hep-th/9610249}}.

\bibitem{Cederwall:1996ri}
M.~Cederwall, A.~von Gussich, B.~E.~W. Nilsson, P.~Sundell, and A.~Westerberg,
  {\em The Dirichlet super-p-branes in ten-dimensional type IIA and IIB
  supergravity,} Nucl. Phys. {\bf B490} (1997) 179--201,
\href{http://www.arXiv.org/abs/hep-th/9611159}{{\tt hep-th/9611159}}.

\bibitem{Bergshoeff:1996tu}
E.~Bergshoeff and P.~K. Townsend, {\em Super D-branes,} Nucl. Phys. {\bf B490}
  (1997) 145--162,
\href{http://www.arXiv.org/abs/hep-th/9611173}{{\tt hep-th/9611173}}.

\bibitem{Aganagic:1996nn}
M.~Aganagic, C.~Popescu, and J.~H. Schwarz, {\em Gauge-invariant and
  gauge-fixed D-brane actions,} Nucl. Phys. {\bf B495} (1997) 99--126,
\href{http://www.arXiv.org/abs/hep-th/9612080}{{\tt hep-th/9612080}}.

\bibitem{Grana:2002tu}
M.~Grana, {\em D3-brane action in a supergravity background: The fermionic
  story,} Phys. Rev. {\bf D66} (2002) 045014,
\href{http://www.arXiv.org/abs/hep-th/0202118}{{\tt hep-th/0202118}}.

\bibitem{deAzcarraga:1989gm}
J.~A. de~Azcarraga, J.~P. Gauntlett, J.~M. Izquierdo, and P.~K. Townsend, {\em
  Topological extensions of the supersymmetry algebra for extended objects,}
  Phys. Rev. Lett. {\bf 63} (1989)
2443.

\bibitem{Dabholkar:1990yf}
A.~Dabholkar, G.~W. Gibbons, J.~A. Harvey, and F.~Ruiz~Ruiz, {\em Superstrings
  and solitons,} Nucl. Phys. {\bf B340} (1990)
33--55.

\bibitem{Horowitz:1991cd}
G.~T. Horowitz and A.~Strominger, {\em Black strings and P-branes,} Nucl. Phys.
  {\bf B360} (1991)
197--209.

\bibitem{Callan:1991at}
J.~Callan, Curtis~G., J.~A. Harvey, and A.~Strominger, {\em Supersymmetric
  string solitons,}
\href{http://www.arXiv.org/abs/hep-th/9112030}{{\tt hep-th/9112030}}.

\bibitem{Stelle:1998xg}
K.~S. Stelle, {\em BPS branes in supergravity,}
\href{http://www.arXiv.org/abs/hep-th/9803116}{{\tt hep-th/9803116}}.

\bibitem{Dabholkar:1997zd}
A.~Dabholkar, {\em Lectures on orientifolds and duality,}
\href{http://www.arXiv.org/abs/hep-th/9804208}{{\tt hep-th/9804208}}.

\bibitem{Zwart:1997aj}
G.~Zwart, {\em Four-dimensional N = 1 Z(N) x Z(M) orientifolds,} Nucl. Phys.
  {\bf B526} (1998) 378--392,
\href{http://www.arXiv.org/abs/hep-th/9708040}{{\tt hep-th/9708040}}.

\bibitem{Klein:2000qw}
M.~Klein and R.~Rabadan, {\em Z(N) x Z(M) orientifolds with and without
  discrete torsion,} JHEP {\bf 10} (2000) 049,
\href{http://www.arXiv.org/abs/hep-th/0008173}{{\tt hep-th/0008173}}.

\bibitem{Frau:1999qs}
M.~Frau, L.~Gallot, A.~Lerda, and P.~Strigazzi, {\em Stable non-BPS D-branes in
  type I string theory,} Nucl. Phys. {\bf B564} (2000) 60--85,
\href{http://www.arXiv.org/abs/hep-th/9903123}{{\tt hep-th/9903123}}.

\bibitem{Lerda:1999um}
A.~Lerda and R.~Russo, {\em Stable non-BPS states in string theory: A
  pedagogical review,} Int. J. Mod. Phys. {\bf A15} (2000) 771--820,
\href{http://www.arXiv.org/abs/hep-th/9905006}{{\tt hep-th/9905006}}.

\bibitem{Fukuma:1999jt}
M.~Fukuma, T.~Oota, and H.~Tanaka, {\em Comments on T-dualities of
  Ramond-Ramond potentials on tori,} Prog. Theor. Phys. {\bf 103} (2000)
  425--446,
\href{http://www.arXiv.org/abs/hep-th/9907132}{{\tt hep-th/9907132}}.

\bibitem{Bergshoeff:2001pv}
E.~Bergshoeff, R.~Kallosh, T.~Ortin, D.~Roest, and A.~Van~Proeyen, {\em New
  formulations of D = 10 supersymmetry and D8 - O8 domain walls,} Class. Quant.
  Grav. {\bf 18} (2001) 3359--3382,
\href{http://www.arXiv.org/abs/hep-th/0103233}{{\tt hep-th/0103233}}.

\bibitem{Hassan:2000zk}
S.~F. Hassan and R.~Minasian, {\em D-brane couplings, RR fields and Clifford
  multiplication,}
\href{http://www.arXiv.org/abs/hep-th/0008149}{{\tt hep-th/0008149}}.

\bibitem{Berg:2003ri}
M.~Berg, M.~Haack, and B.~K{\"o}rs, {\em An orientifold with fluxes and branes
  via T-duality,} Nucl. Phys. {\bf B669} (2003) 3--56,
\href{http://www.arXiv.org/abs/hep-th/0305183}{{\tt hep-th/0305183}}.

\bibitem{Jockers:2005pn}
H.~Jockers, {\em The effective action of D-branes in Calabi-Yau orientifold
  compactifications,} Fortsch. Phys. {\bf 53} (2005) 1087--1175,
\href{http://www.arXiv.org/abs/hep-th/0507042}{{\tt hep-th/0507042}}.

\bibitem{Rabadan:2002wy}
R.~Rabadan and F.~Zamora, {\em Dilaton tadpoles and D-brane interactions in
  compact spaces,} JHEP {\bf 12} (2002) 052,
\href{http://www.arXiv.org/abs/hep-th/0207178}{{\tt hep-th/0207178}}.

\bibitem{Dudas:2004nd}
E.~Dudas, G.~Pradisi, M.~Nicolosi, and A.~Sagnotti, {\em On tadpoles and vacuum
  redefinitions in string theory,} Nucl. Phys. {\bf B708} (2005) 3--44,
\href{http://www.arXiv.org/abs/hep-th/0410101}{{\tt hep-th/0410101}}.

\bibitem{Fischler:1986tb}
W.~Fischler and L.~Susskind, {\em Dilaton tadpoles, string condensates and
  scale invariance. 2,} Phys. Lett. {\bf B173} (1986)
262.

\bibitem{Fischler:1986ci}
W.~Fischler and L.~Susskind, {\em Dilaton tadpoles, string condensates and
  scale invariance,} Phys. Lett. {\bf B171} (1986)
383.

\bibitem{Dudas:2000ff}
E.~Dudas and J.~Mourad, {\em Brane solutions in strings with broken
  supersymmetry and dilaton tadpoles,} Phys. Lett. {\bf B486} (2000) 172--178,
\href{http://www.arXiv.org/abs/hep-th/0004165}{{\tt hep-th/0004165}}.

\bibitem{Blumenhagen:2000dc}
R.~Blumenhagen and A.~Font, {\em Dilaton tadpoles, warped geometries and large
  extra dimensions for non-supersymmetric strings,} Nucl. Phys. {\bf B599}
  (2001) 241--254,
\href{http://www.arXiv.org/abs/hep-th/0011269}{{\tt hep-th/0011269}}.

\bibitem{deWit:1986xg}
B.~de~Wit, D.~J. Smit, and N.~D. Hari~Dass, {\em Residual supersymmetry of
  compactified d=10 supergravity,} Nucl. Phys. {\bf B283} (1987)
165.

\bibitem{Maldacena:2000mw}
J.~M. Maldacena and C.~Nunez, {\em Supergravity description of field theories
  on curved manifolds and a no go theorem,} Int. J. Mod. Phys. {\bf A16} (2001)
  822--855,
\href{http://www.arXiv.org/abs/hep-th/0007018}{{\tt hep-th/0007018}}.

\bibitem{Ivanov:2000fg}
S.~Ivanov and G.~Papadopoulos, {\em A no-go theorem for string warped
  compactifications,} Phys. Lett. {\bf B497} (2001) 309--316,
\href{http://www.arXiv.org/abs/hep-th/0008232}{{\tt hep-th/0008232}}.

\bibitem{Dine:1987xk}
M.~Dine, N.~Seiberg, and E.~Witten, {\em Fayet-Iliopoulos terms in string
  theory,} Nucl. Phys. {\bf B289} (1987)
589.

\bibitem{Witten:1981nf}
E.~Witten, {\em Dynamical Breaking of Supersymmetry,} Nucl. Phys. {\bf B188}
  (1981)
513.

\bibitem{Fischler:1981zk}
W.~Fischler, H.~P. Nilles, J.~Polchinski, S.~Raby, and L.~Susskind, {\em
  Vanishing renormalization of the D-term in supersymmetric U(1) theories,}
  Phys. Rev. Lett. {\bf 47} (1981)
757.

\bibitem{Kachru:2002he}
S.~Kachru, M.~B. Schulz, and S.~Trivedi, {\em Moduli stabilization from fluxes
  in a simple IIB orientifold,} JHEP {\bf 10} (2003) 007,
\href{http://www.arXiv.org/abs/hep-th/0201028}{{\tt hep-th/0201028}}.

\bibitem{gl04}
T.~W. Grimm and J.~Louis, {\em The effective action of N = 1 Calabi-Yau
  orientifolds,} Nucl. Phys. {\bf B699} (2004) 387--426,
\href{http://www.arXiv.org/abs/hep-th/0403067}{{\tt hep-th/0403067}}.

\bibitem{Grimm:2004ua}
T.~W. Grimm and J.~Louis, {\em The effective action of type IIA Calabi-Yau
  orientifolds,} Nucl. Phys. {\bf B718} (2005) 153--202,
\href{http://www.arXiv.org/abs/hep-th/0412277}{{\tt hep-th/0412277}}.

\bibitem{bgk00}
R.~Blumenhagen, L.~G{\"o}rlich, and B.~K{\"o}rs, {\em A new class of
  supersymmetric orientifolds with D-branes at angles,}
\href{http://www.arXiv.org/abs/hep-th/0002146}{{\tt hep-th/0002146}}.

\bibitem{bcs04}
R.~Blumenhagen, J.~P. Conlon, and K.~Suruliz, {\em Type IIA Orientifolds on
  General Supersymmetric Z(N) Orbifolds,} JHEP {\bf 07} (2004) 022,
\href{http://www.arXiv.org/abs/hep-th/0404254}{{\tt hep-th/0404254}}.

\bibitem{Bergshoeff:1981um}
E.~Bergshoeff, M.~de~Roo, B.~de~Wit, and P.~van Nieuwenhuizen, {\em
  Ten-dimensional Maxwell-Einstein supergravity, its currents, and the issue of
  its auxiliary fields,} Nucl. Phys. {\bf B195} (1982)
97--136.

\bibitem{Sugimoto:1999tx}
S.~Sugimoto, {\em Anomaly cancellations in type I D9-D9-bar system and the
  USp(32) string theory,} Prog. Theor. Phys. {\bf 102} (1999) 685--699,
\href{http://www.arXiv.org/abs/hep-th/9905159}{{\tt hep-th/9905159}}.

\bibitem{jl04}
H.~Jockers and J.~Louis, {\em The effective action of D7-branes in N = 1
  Calabi-Yau orientifolds,} Nucl. Phys. {\bf B705} (2005) 167--211,
\href{http://www.arXiv.org/abs/hep-th/0409098}{{\tt hep-th/0409098}}.

\bibitem{Burgess:1986ah}
C.~P. Burgess and T.~R. Morris, {\em Open and unoriented strings a la
  Polyakov,} Nucl. Phys. {\bf B291} (1987)
256.

\bibitem{Burgess:1986wt}
C.~P. Burgess and T.~R. Morris, {\em Open superstrings a la Polyakov,} Nucl.
  Phys. {\bf B291} (1987)
285.

\bibitem{Kakushadze:1998bw}
Z.~Kakushadze, G.~Shiu, and S.~H.~H. Tye, {\em Type IIB orientifolds with NS-NS
  antisymmetric tensor backgrounds,} Phys. Rev. {\bf D58} (1998) 086001,
\href{http://www.arXiv.org/abs/hep-th/9803141}{{\tt hep-th/9803141}}.

\bibitem{Angelantonj:1999jh}
C.~Angelantonj, {\em Comments on open-string orbifolds with a non-vanishing
  B(ab),} Nucl. Phys. {\bf B566} (2000) 126--150,
\href{http://www.arXiv.org/abs/hep-th/9908064}{{\tt hep-th/9908064}}.

\bibitem{Kakushadze:1999if}
Z.~Kakushadze, {\em Non-perturbative K3 orientifolds with NS-NS B-flux,} Phys.
  Lett. {\bf B459} (1999) 497--506,
\href{http://www.arXiv.org/abs/hep-th/9905033}{{\tt hep-th/9905033}}.

\bibitem{ab99}
C.~Angelantonj and R.~Blumenhagen, {\em Discrete deformations in type I vacua,}
  Phys. Lett. {\bf B473} (2000) 86--93,
\href{http://www.arXiv.org/abs/hep-th/9911190}{{\tt hep-th/9911190}}.

\bibitem{Kakushadze:2000hm}
Z.~Kakushadze, {\em Geometry of orientifolds with NS-NS B-flux,} Int. J. Mod.
  Phys. {\bf A15} (2000) 3113--3196,
\href{http://www.arXiv.org/abs/hep-th/0001212}{{\tt hep-th/0001212}}.

\bibitem{seealso}
I.~Antoniadis, C.~Bachas, C.~Fabre, H.~Partouche, and T.~R. Taylor, {\em
  Aspects of type I - type II - heterotic triality in four dimensions,} Nucl.
  Phys. {\bf B489} (1997) 160--178,
\href{http://www.arXiv.org/abs/hep-th/9608012}{{\tt hep-th/9608012}}.

\bibitem{Vafa:1994rv}
C.~Vafa and E.~Witten, {\em On orbifolds with discrete torsion,} J. Geom. Phys.
  {\bf 15} (1995) 189--214,
\href{http://www.arXiv.org/abs/hep-th/9409188}{{\tt hep-th/9409188}}.

\bibitem{bgkl00}
R.~Blumenhagen, L.~G{\"o}rlich, B.~K{\"o}rs, and D.~L{\"u}st, {\em Asymmetric
  orbifolds, noncommutative geometry and type I string vacua,} Nucl. Phys. {\bf
  B582} (2000) 44--64,
\href{http://www.arXiv.org/abs/hep-th/0003024}{{\tt hep-th/0003024}}.

\bibitem{Bergshoeff:1997kr}
E.~Bergshoeff, R.~Kallosh, T.~Ortin, and G.~Papadopoulos, {\em Kappa-symmetry,
  supersymmetry and intersecting branes,} Nucl. Phys. {\bf B502} (1997)
  149--169,
\href{http://www.arXiv.org/abs/hep-th/9705040}{{\tt hep-th/9705040}}.

\bibitem{Font:2006na}
A.~Font, L.~E. Ibanez, and F.~Marchesano, {\em Coisotropic D8-branes and
  model-building,} JHEP {\bf 09} (2006) 080,
\href{http://www.arXiv.org/abs/hep-th/0607219}{{\tt hep-th/0607219}}.

\bibitem{Marino:1999af}
M.~Marino, R.~Minasian, G.~W. Moore, and A.~Strominger, {\em Nonlinear
  instantons from supersymmetric p-branes,} JHEP {\bf 01} (2000) 005,
\href{http://www.arXiv.org/abs/hep-th/9911206}{{\tt hep-th/9911206}}.

\bibitem{RR01}
R.~Rabad{\' a}n, {\em Branes at angles, torons, stability and supersymmetry,}
  Nucl. Phys. {\bf B620} (2002) 152--180,
\href{http://www.arXiv.org/abs/hep-th/0107036}{{\tt hep-th/0107036}}.

\bibitem{el03}
F.~Epple and D.~L{\"u}st, {\em Tachyon condensation for intersecting branes at
  small and large angles,} Fortsch. Phys. {\bf 52} (2004) 367--387,
\href{http://www.arXiv.org/abs/hep-th/0311182}{{\tt hep-th/0311182}}.

\bibitem{cim02a}
D.~Cremades, L.~E. Ib{\'a}{\~n}ez, and F.~Marchesano, {\em Intersecting brane
  models of particle physics and the Higgs mechanism,} JHEP {\bf 07} (2002)
  022,
\href{http://www.arXiv.org/abs/hep-th/0203160}{{\tt hep-th/0203160}}.

\bibitem{bgo02}
R.~Blumenhagen, L.~G{\"o}rlich, and T.~Ott, {\em Supersymmetric intersecting
  branes on the type IIA T**6/Z(4) orientifold,} JHEP {\bf 01} (2003) 021,
\href{http://www.arXiv.org/abs/hep-th/0211059}{{\tt hep-th/0211059}}.

\bibitem{ho04}
G.~Honecker and T.~Ott, {\em Getting just the Supersymmetric Standard Model at
  Intersecting Branes on the Z6-orientifold,} Phys. Rev. {\bf D70} (2004)
  126010,
\href{http://www.arXiv.org/abs/hep-th/0404055}{{\tt hep-th/0404055}}.

\bibitem{Sagnotti:1992qw}
A.~Sagnotti, {\em A Note on the Green-Schwarz mechanism in open string
  theories,} Phys. Lett. {\bf B294} (1992) 196--203,
\href{http://www.arXiv.org/abs/hep-th/9210127}{{\tt hep-th/9210127}}.

\bibitem{Ibanez:1998qp}
L.~E. Ib{\'a}{\~n}ez, R.~Rabadan, and A.~M. Uranga, {\em Anomalous U(1)'s in
  type I and type IIB D = 4, N = 1 string vacua,} Nucl. Phys. {\bf B542} (1999)
  112--138,
\href{http://www.arXiv.org/abs/hep-th/9808139}{{\tt hep-th/9808139}}.

\bibitem{Ibanez:1999pw}
L.~E. Ib{\'a}{\~n}ez, R.~Rabadan, and A.~M. Uranga, {\em Sigma-model anomalies
  in compact D = 4, N = 1 type IIB orientifolds and Fayet-Iliopoulos terms,}
  Nucl. Phys. {\bf B576} (2000) 285--312,
\href{http://www.arXiv.org/abs/hep-th/9905098}{{\tt hep-th/9905098}}.

\bibitem{Scrucca:1999zh}
C.~A. Scrucca and M.~Serone, {\em Gauge and gravitational anomalies in D = 4 N
  = 1 orientifolds,} JHEP {\bf 12} (1999) 024,
\href{http://www.arXiv.org/abs/hep-th/9912108}{{\tt hep-th/9912108}}.

\bibitem{Antoniadis:2002cs}
I.~Antoniadis, E.~Kiritsis, and J.~Rizos, {\em Anomalous U(1)s in type I
  superstring vacua,} Nucl. Phys. {\bf B637} (2002) 92--118,
\href{http://www.arXiv.org/abs/hep-th/0204153}{{\tt hep-th/0204153}}.

\bibitem{Derendinger:1991kr}
J.-P. Derendinger, S.~Ferrara, C.~Kounnas, and F.~Zwirner, {\em All loop gauge
  couplings from anomaly cancellation in string effective theories,} Phys.
  Lett. {\bf B271} (1991)
307--313.

\bibitem{Louis:1996ya}
J.~Louis and K.~F{\"o}rger, {\em Holomorphic couplings in string theory,} Nucl.
  Phys. Proc. Suppl. {\bf 55B} (1997) 33--64,
\href{http://www.arXiv.org/abs/hep-th/9611184}{{\tt hep-th/9611184}}.

\bibitem{Klein:1999im}
M.~Klein, {\em Anomaly cancellation in D = 4, N = 1 orientifolds and
  linear/chiral multiplet duality,} Nucl. Phys. {\bf B569} (2000) 362--390,
\href{http://www.arXiv.org/abs/hep-th/9910143}{{\tt hep-th/9910143}}.

\bibitem{Coriano':2005js}
C.~Coriano', N.~Irges, and E.~Kiritsis, {\em On the effective theory of low
  scale orientifold string vacua,} Nucl. Phys. {\bf B746} (2006) 77--135,
\href{http://www.arXiv.org/abs/hep-ph/0510332}{{\tt hep-ph/0510332}}.

\bibitem{Anastasopoulos:2006cz}
P.~Anastasopoulos, M.~Bianchi, E.~Dudas, and E.~Kiritsis, {\em Anomalies,
  anomalous U(1)'s and generalized Chern-Simons terms,}
\href{http://www.arXiv.org/abs/hep-th/0605225}{{\tt hep-th/0605225}}.

\bibitem{Antoniadis:2000en}
I.~Antoniadis, E.~Kiritsis, and T.~N. Tomaras, {\em A D-brane alternative to
  unification,} Phys. Lett. {\bf B486} (2000) 186--193,
\href{http://www.arXiv.org/abs/hep-ph/0004214}{{\tt hep-ph/0004214}}.

\bibitem{Krause:2000gp}
A.~Krause, {\em Critical vacuum energy, warped geometry and grand unification,}
  Nucl. Phys. {\bf B748} (2006) 98--125,
\href{http://www.arXiv.org/abs/hep-th/0006226}{{\tt hep-th/0006226}}.

\bibitem{Antoniadis:2002qm}
I.~Antoniadis, E.~Kiritsis, J.~Rizos, and T.~N. Tomaras, {\em D-branes and the
  standard model,} Nucl. Phys. {\bf B660} (2003) 81--115,
\href{http://www.arXiv.org/abs/hep-th/0210263}{{\tt hep-th/0210263}}.

\bibitem{Blumenhagen:2005mu}
R.~Blumenhagen, M.~Cveti{\v c}, P.~Langacker, and G.~Shiu, {\em Toward
  realistic intersecting D-brane models,} Ann. Rev. Nucl. Part. Sci. {\bf 55}
  (2005) 71--139,
\href{http://www.arXiv.org/abs/hep-th/0502005}{{\tt hep-th/0502005}}.

\bibitem{bbkl02a}
R.~Blumenhagen, V.~Braun, B.~K{\"o}rs, and D.~L{\"u}st, {\em The standard model
  on the quintic,}
\href{http://www.arXiv.org/abs/hep-th/0210083}{{\tt hep-th/0210083}}.

\bibitem{AU02}
A.~M. Uranga, {\em Local models for intersecting brane worlds,} JHEP {\bf 12}
  (2002) 058,
\href{http://www.arXiv.org/abs/hep-th/0208014}{{\tt hep-th/0208014}}.

\bibitem{AU03a}
A.~M. Uranga, {\em Local intersecting brane worlds,} Fortsch. Phys. {\bf 51}
  (2003)
879--884.

\bibitem{cgqu03}
J.~F.~G. Cascales, M.~P. Garcia~del Moral, F.~Quevedo, and A.~M. Uranga, {\em
  Realistic D-brane models on warped throats: Fluxes, hierarchies and moduli
  stabilization,} JHEP {\bf 02} (2004) 031,
\href{http://www.arXiv.org/abs/hep-th/0312051}{{\tt hep-th/0312051}}.

\bibitem{Verlinde:2005jr}
H.~Verlinde and M.~Wijnholt, {\em Building the standard model on a D3-brane,}
\href{http://www.arXiv.org/abs/hep-th/0508089}{{\tt hep-th/0508089}}.

\bibitem{Garcia-Etxebarria:2006rw}
I.~Garcia-Etxebarria, F.~Saad, and A.~M. Uranga, {\em Local models of gauge
  mediated supersymmetry breaking in string theory,} JHEP {\bf 08} (2006) 069,
\href{http://www.arXiv.org/abs/hep-th/0605166}{{\tt hep-th/0605166}}.

\bibitem{Buican:2006sn}
M.~Buican, D.~Malyshev, D.~R. Morrison, M.~Wijnholt, and H.~Verlinde, {\em
  D-branes at singularities, compactification, and hypercharge,}
\href{http://www.arXiv.org/abs/hep-th/0610007}{{\tt hep-th/0610007}}.

\bibitem{bbkl02}
R.~Blumenhagen, V.~Braun, B.~K{\"o}rs, and D.~L{\"u}st, {\em Orientifolds of K3
  and Calabi-Yau manifolds with intersecting D-branes,} JHEP {\bf 07} (2002)
  026,
\href{http://www.arXiv.org/abs/hep-th/0206038}{{\tt hep-th/0206038}}.

\bibitem{Minasian:1997mm}
R.~Minasian and G.~W. Moore, {\em K-theory and Ramond-Ramond charge,} JHEP {\bf
  11} (1997) 002,
\href{http://www.arXiv.org/abs/hep-th/9710230}{{\tt hep-th/9710230}}.

\bibitem{Witten:1998cd}
E.~Witten, {\em D-branes and K-theory,} JHEP {\bf 12} (1998) 019,
\href{http://www.arXiv.org/abs/hep-th/9810188}{{\tt hep-th/9810188}}.

\bibitem{Uranga:2000xp}
A.~M. Uranga, {\em D-brane probes, RR tadpole cancellation and K-theory
  charge,} Nucl. Phys. {\bf B598} (2001) 225--246,
\href{http://www.arXiv.org/abs/hep-th/0011048}{{\tt hep-th/0011048}}.

\bibitem{Witten:1982fp}
E.~Witten, {\em An SU(2) anomaly,} Phys. Lett. {\bf B117} (1982)
324--328.

\bibitem{Maiden:2006qe}
J.~Maiden, G.~Shiu, and J.~Stefanski, Bogdan, {\em D-brane spectrum and
  K-theory constraints of D = 4, N = 1 orientifolds,} JHEP {\bf 04} (2006) 052,
\href{http://www.arXiv.org/abs/hep-th/0602038}{{\tt hep-th/0602038}}.

\bibitem{Garcia-Etxebarria:2005qc}
I.~Garcia-Etxebarria and A.~M. Uranga, {\em From F/M-theory to K-theory and
  back,} JHEP {\bf 02} (2006) 008,
\href{http://www.arXiv.org/abs/hep-th/0510073}{{\tt hep-th/0510073}}.

\bibitem{bbs95}
K.~Becker, M.~Becker, and A.~Strominger, {\em Five-branes, membranes and
  nonperturbative string theory,} Nucl. Phys. {\bf B456} (1995) 130--152,
\href{http://www.arXiv.org/abs/hep-th/9507158}{{\tt hep-th/9507158}}.

\bibitem{Kachru:1999vj}
S.~Kachru and J.~McGreevy, {\em Supersymmetric three-cycles and (super)symmetry
  breaking,} Phys. Rev. {\bf D61} (2000) 026001,
\href{http://www.arXiv.org/abs/hep-th/9908135}{{\tt hep-th/9908135}}.

\bibitem{cu04}
J.~F.~G. Cascales and A.~M. Uranga, {\em Branes on generalized calibrated
  submanifolds,} JHEP {\bf 11} (2004) 083,
\href{http://www.arXiv.org/abs/hep-th/0407132}{{\tt hep-th/0407132}}.

\bibitem{cim02}
D.~Cremades, L.~E. Ib{\'a}{\~n}ez, and F.~Marchesano, {\em SUSY quivers,
  intersecting branes and the modest hierarchy problem,} JHEP {\bf 07} (2002)
  009,
\href{http://www.arXiv.org/abs/hep-th/0201205}{{\tt hep-th/0201205}}.

\bibitem{bls03}
R.~Blumenhagen, D.~L{\"u}st, and S.~Stieberger, {\em Gauge unification in
  supersymmetric intersecting brane worlds,} JHEP {\bf 07} (2003) 036,
\href{http://www.arXiv.org/abs/hep-th/0305146}{{\tt hep-th/0305146}}.

\bibitem{Brunner:1999jq}
I.~Brunner, M.~R. Douglas, A.~E. Lawrence, and C.~Romelsberger, {\em D-branes
  on the quintic,} JHEP {\bf 08} (2000) 015,
\href{http://www.arXiv.org/abs/hep-th/9906200}{{\tt hep-th/9906200}}.

\bibitem{Kachru:2000ih}
S.~Kachru, S.~Katz, A.~E. Lawrence, and J.~McGreevy, {\em Open string
  instantons and superpotentials,} Phys. Rev. {\bf D62} (2000) 026001,
\href{http://www.arXiv.org/abs/hep-th/9912151}{{\tt hep-th/9912151}}.

\bibitem{Villadoro:2006ia}
G.~Villadoro and F.~Zwirner, {\em D terms from D-branes, gauge invariance and
  moduli stabilization in flux compactifications,} JHEP {\bf 03} (2006) 087,
\href{http://www.arXiv.org/abs/hep-th/0602120}{{\tt hep-th/0602120}}.

\bibitem{Diaconescu:2006nk}
D.-E. Diaconescu, A.~Garcia-Raboso, and K.~Sinha, {\em A D-brane landscape on
  Calabi-Yau manifolds,} JHEP {\bf 06} (2006) 058,
\href{http://www.arXiv.org/abs/hep-th/0602138}{{\tt hep-th/0602138}}.

\bibitem{Dine:1986zy}
M.~Dine, N.~Seiberg, X.~G. Wen, and E.~Witten, {\em Nonperturbative effects on
  the string world sheet,} Nucl. Phys. {\bf B278} (1986)
769.

\bibitem{Dine:1987bq}
M.~Dine, N.~Seiberg, X.~G. Wen, and E.~Witten, {\em Nonperturbative effects on
  the string world sheet. 2,} Nucl. Phys. {\bf B289} (1987)
319.

\bibitem{Kachru:2000an}
S.~Kachru, S.~Katz, A.~E. Lawrence, and J.~McGreevy, {\em Mirror symmetry for
  open strings,} Phys. Rev. {\bf D62} (2000) 126005,
\href{http://www.arXiv.org/abs/hep-th/0006047}{{\tt hep-th/0006047}}.

\bibitem{Blumenhagen:2006xt}
R.~Blumenhagen, M.~Cveti{\v c}, and T.~Weigand, {\em Spacetime instanton
  corrections in 4D string vacua - the seesaw mechanism for D-brane models,}
\href{http://www.arXiv.org/abs/hep-th/0609191}{{\tt hep-th/0609191}}.

\bibitem{Florea:2006si}
B.~Florea, S.~Kachru, J.~McGreevy, and N.~Saulina, {\em Stringy Instantons and
  Quiver Gauge Theories,}
\href{http://www.arXiv.org/abs/hep-th/0610003}{{\tt hep-th/0610003}}.

\bibitem{Ooguri:1999bv}
H.~Ooguri and C.~Vafa, {\em Knot invariants and topological strings,} Nucl.
  Phys. {\bf B577} (2000) 419--438,
\href{http://www.arXiv.org/abs/hep-th/9912123}{{\tt hep-th/9912123}}.

\bibitem{Aganagic:2000gs}
M.~Aganagic and C.~Vafa, {\em Mirror symmetry, D-branes and counting
  holomorphic discs,}
\href{http://www.arXiv.org/abs/hep-th/0012041}{{\tt hep-th/0012041}}.

\bibitem{Aganagic:2001nx}
M.~Aganagic, A.~Klemm, and C.~Vafa, {\em Disk instantons, mirror symmetry and
  the duality web,} Z. Naturforsch. {\bf A57} (2002) 1--28,
  \href{http://www.arXiv.org/abs/hep-th/0105045}{{\tt hep-th/0105045}}.

\bibitem{Lerche:2001cw}
W.~Lerche and P.~Mayr, {\em On N = 1 mirror symmetry for open type II strings,}
\href{http://www.arXiv.org/abs/hep-th/0111113}{{\tt hep-th/0111113}}.

\bibitem{Ibanez:2006da}
L.~E. Ib{\'a}{\~n}ez and A.~M. Uranga, {\em Neutrino Majorana Masses from
  String Theory Instanton Effects,}
\href{http://www.arXiv.org/abs/hep-th/0609213}{{\tt hep-th/0609213}}.

\bibitem{ekn02}
J.~R. Ellis, P.~Kanti, and D.~V. Nanopoulos, {\em Intersecting branes flip
  SU(5),} Nucl. Phys. {\bf B647} (2002) 235--251,
\href{http://www.arXiv.org/abs/hep-th/0206087}{{\tt hep-th/0206087}}.

\bibitem{Axenides:2003hs}
M.~Axenides, E.~Floratos, and C.~Kokorelis, {\em SU(5) unified theories from
  intersecting branes,} JHEP {\bf 10} (2003) 006,
\href{http://www.arXiv.org/abs/hep-th/0307255}{{\tt hep-th/0307255}}.

\bibitem{CK02a}
C.~Kokorelis, {\em New standard model vacua from intersecting branes,} JHEP
  {\bf 09} (2002) 029,
\href{http://www.arXiv.org/abs/hep-th/0205147}{{\tt hep-th/0205147}}.

\bibitem{CK02b}
C.~Kokorelis, {\em Exact standard model compactifications from intersecting
  branes,} JHEP {\bf 08} (2002) 036,
\href{http://www.arXiv.org/abs/hep-th/0206108}{{\tt hep-th/0206108}}.

\bibitem{CK02}
C.~Kokorelis, {\em GUT model hierarchies from intersecting branes,} JHEP {\bf
  08} (2002) 018,
\href{http://www.arXiv.org/abs/hep-th/0203187}{{\tt hep-th/0203187}}.

\bibitem{CK02d}
C.~Kokorelis, {\em Deformed intersecting D6-brane GUTs. I,} JHEP {\bf 11}
  (2002) 027,
\href{http://www.arXiv.org/abs/hep-th/0209202}{{\tt hep-th/0209202}}.

\bibitem{fhs01}
S.~F{\"o}rste, G.~Honecker, and R.~Schreyer, {\em Orientifolds with branes at
  angles,} JHEP {\bf 06} (2001) 004,
\href{http://www.arXiv.org/abs/hep-th/0105208}{{\tt hep-th/0105208}}.

\bibitem{bkl01}
D.~Bailin, G.~V. Kraniotis, and A.~Love, {\em Standard-like models from
  intersecting D4-branes,} Phys. Lett. {\bf B530} (2002) 202--209,
\href{http://www.arXiv.org/abs/hep-th/0108131}{{\tt hep-th/0108131}}.

\bibitem{GH01}
G.~Honecker, {\em Non-supersymmetric orientifolds with D-branes at angles,}
  Fortsch. Phys. {\bf 50} (2002) 896--902,
\href{http://www.arXiv.org/abs/hep-th/0112174}{{\tt hep-th/0112174}}.

\bibitem{GH02}
G.~Honecker, {\em Intersecting brane world models from D8-branes on (T(2) x
  T(4)/Z(3))/Omega R(1) type IIA orientifolds,} JHEP {\bf 01} (2002) 025,
\href{http://www.arXiv.org/abs/hep-th/0201037}{{\tt hep-th/0201037}}.

\bibitem{cim02b}
D.~Cremades, L.~E. Ib{\'a}{\~n}ez, and F.~Marchesano, {\em Standard model at
  intersecting D5-branes: Lowering the string scale,} Nucl. Phys. {\bf B643}
  (2002) 93--130,
\href{http://www.arXiv.org/abs/hep-th/0205074}{{\tt hep-th/0205074}}.

\bibitem{CK02c}
C.~Kokorelis, {\em Exact standard model structures from intersecting D5-
  branes,} Nucl. Phys. {\bf B677} (2004) 115--163,
\href{http://www.arXiv.org/abs/hep-th/0207234}{{\tt hep-th/0207234}}.

\bibitem{bkl02a}
D.~Bailin, G.~V. Kraniotis, and A.~Love, {\em New standard-like models from
  intersecting D4-branes,} Phys. Lett. {\bf B547} (2002) 43--50,
\href{http://www.arXiv.org/abs/hep-th/0208103}{{\tt hep-th/0208103}}.

\bibitem{bkl02b}
D.~Bailin, G.~V. Kraniotis, and A.~Love, {\em Standard-like models from
  intersecting D5-branes,} Phys. Lett. {\bf B553} (2003) 79--86,
\href{http://www.arXiv.org/abs/hep-th/0210219}{{\tt hep-th/0210219}}.

\bibitem{DB02}
D.~Bailin, {\em Standard-like models from D-branes,}
\href{http://www.arXiv.org/abs/hep-th/0210227}{{\tt hep-th/0210227}}.

\bibitem{bkl02c}
D.~Bailin, G.~V. Kraniotis, and A.~Love, {\em Intersecting D5-brane models with
  massive vector-like leptons,} JHEP {\bf 02} (2003) 052,
\href{http://www.arXiv.org/abs/hep-th/0212112}{{\tt hep-th/0212112}}.

\bibitem{bkl02}
R.~Blumenhagen, B.~K{\"o}rs, and D.~L{\"u}st, {\em Moduli stabilization for
  intersecting brane worlds in type 0' string theory,} Phys. Lett. {\bf B532}
  (2002) 141--151,
\href{http://www.arXiv.org/abs/hep-th/0202024}{{\tt hep-th/0202024}}.

\bibitem{cim02c}
D.~Cremades, L.~E. Ib{\'a}{\~ n}ez, and F.~Marchesano, {\em More about the
  standard model at intersecting branes,}
\href{http://www.arXiv.org/abs/hep-ph/0212048}{{\tt hep-ph/0212048}}.

\bibitem{Angelantonj:2005hs}
C.~Angelantonj, M.~Cardella, and N.~Irges, {\em Scherk-Schwarz breaking and
  intersecting branes,} Nucl. Phys. {\bf B725} (2005) 115--154,
\href{http://www.arXiv.org/abs/hep-th/0503179}{{\tt hep-th/0503179}}.

\bibitem{Arkani-Hamed:2004fb}
N.~Arkani-Hamed and S.~Dimopoulos, {\em Supersymmetric unification without low
  energy supersymmetry and signatures for fine-tuning at the LHC,} JHEP {\bf
  06} (2005) 073,
\href{http://www.arXiv.org/abs/hep-th/0405159}{{\tt hep-th/0405159}}.

\bibitem{Antoniadis:2004dt}
I.~Antoniadis and S.~Dimopoulos, {\em Splitting supersymmetry in string
  theory,} Nucl. Phys. {\bf B715} (2005) 120--140,
\href{http://www.arXiv.org/abs/hep-th/0411032}{{\tt hep-th/0411032}}.

\bibitem{Antoniadis:2006eb}
I.~Antoniadis, K.~Benakli, A.~Delgado, M.~Quiros, and M.~Tuckmantel, {\em Split
  extended supersymmetry from intersecting branes,} Nucl. Phys. {\bf B744}
  (2006) 156--179,
\href{http://www.arXiv.org/abs/hep-th/0601003}{{\tt hep-th/0601003}}.

\bibitem{Ott:2005bf}
T.~Ott, {\em A new picture for 4-dimensional 'spacetime' from intersecting
  D-branes on the T**9,} Phys. Rev. {\bf D73} (2006) 046003,
\href{http://www.arXiv.org/abs/hep-th/0509108}{{\tt hep-th/0509108}}.

\bibitem{Blumenhagen:2005tn}
R.~Blumenhagen, M.~Cveti{\v c}, F.~Marchesano, and G.~Shiu, {\em Chiral D-brane
  models with frozen open string moduli,} JHEP {\bf 03} (2005) 050,
\href{http://www.arXiv.org/abs/hep-th/0502095}{{\tt hep-th/0502095}}.

\bibitem{Dudas:2005jx}
E.~Dudas and C.~Timirgaziu, {\em Internal magnetic fields and supersymmetry in
  orientifolds,} Nucl. Phys. {\bf B716} (2005) 65--87,
\href{http://www.arXiv.org/abs/hep-th/0502085}{{\tt hep-th/0502085}}.

\bibitem{Blumenhagen:2006ab}
R.~Blumenhagen and E.~Plauschinn, {\em Intersecting D-branes on shift Z(2) x
  Z(2) orientifolds,} JHEP {\bf 08} (2006) 031,
\href{http://www.arXiv.org/abs/hep-th/0604033}{{\tt hep-th/0604033}}.

\bibitem{Antoniadis:1999ux}
I.~Antoniadis, G.~D'Appollonio, E.~Dudas, and A.~Sagnotti, {\em Open
  descendants of Z(2) x Z(2) freely-acting orbifolds,} Nucl. Phys. {\bf B565}
  (2000) 123--156,
\href{http://www.arXiv.org/abs/hep-th/9907184}{{\tt hep-th/9907184}}.

\bibitem{GP02}
G.~Pradisi, {\em Magnetized (shift-)orientifolds,}
\href{http://www.arXiv.org/abs/hep-th/0210088}{{\tt hep-th/0210088}}.

\bibitem{LG03}
M.~Larosa and G.~Pradisi, {\em Magnetized four-dimensional Z(2) x Z(2)
  orientifolds,} Nucl. Phys. {\bf B667} (2003) 261--309,
\href{http://www.arXiv.org/abs/hep-th/0305224}{{\tt hep-th/0305224}}.

\bibitem{cps02}
M.~Cveti{\v c}, I.~Papadimitriou, and G.~Shiu, {\em Supersymmetric three family
  SU(5) grand unified models from type IIA orientifolds with intersecting
  D6-branes,} Nucl. Phys. {\bf B659} (2003) 193--223,
\href{http://www.arXiv.org/abs/hep-th/0212177}{{\tt hep-th/0212177}}.

\bibitem{Cvetic:2002qa}
M.~Cveti{\v c}, P.~Langacker, and G.~Shiu, {\em Phenomenology of a three-family
  standard-like string model,} Phys. Rev. {\bf D66} (2002) 066004,
\href{http://www.arXiv.org/abs/hep-ph/0205252}{{\tt hep-ph/0205252}}.

\bibitem{Cvetic:2002wh}
M.~Cveti{\v c}, P.~Langacker, and G.~Shiu, {\em A three-family standard-like
  orientifold model: Yukawa couplings and hierarchy,} Nucl. Phys. {\bf B642}
  (2002) 139--156,
\href{http://www.arXiv.org/abs/hep-th/0206115}{{\tt hep-th/0206115}}.

\bibitem{cp03a}
M.~Cveti{\v c} and I.~Papadimitriou, {\em More supersymmetric standard-like
  models from intersecting D6-branes on type IIA orientifolds,} Phys. Rev. {\bf
  D67} (2003) 126006,
\href{http://www.arXiv.org/abs/hep-th/0303197}{{\tt hep-th/0303197}}.

\bibitem{clll04}
M.~Cvetic, P.~Langacker, T.-j. Li, and T.~Liu, {\em D6-brane splitting on type
  IIA orientifolds,} Nucl. Phys. {\bf B709} (2005) 241--266,
\href{http://www.arXiv.org/abs/hep-th/0407178}{{\tt hep-th/0407178}}.

\bibitem{cll04}
M.~Cveti{\v c}, T.~Li, and T.~Liu, {\em Supersymmetric Pati-Salam models from
  intersecting D6- branes: A road to the standard model,} Nucl. Phys. {\bf
  B698} (2004) 163--201,
\href{http://www.arXiv.org/abs/hep-th/0403061}{{\tt hep-th/0403061}}.

\bibitem{Chen:2005ab}
C.~M. Chen, G.~V. Kraniotis, V.~E. Mayes, D.~V. Nanopoulos, and J.~W. Walker,
  {\em A supersymmetric flipped SU(5) intersecting brane world,} Phys. Lett.
  {\bf B611} (2005) 156--166,
\href{http://www.arXiv.org/abs/hep-th/0501182}{{\tt hep-th/0501182}}.

\bibitem{Chen:2005mm}
C.~M. Chen, G.~V. Kraniotis, V.~E. Mayes, D.~V. Nanopoulos, and J.~W. Walker,
  {\em A K-theory anomaly free supersymmetric flipped SU(5) model from
  intersecting branes,} Phys. Lett. {\bf B625} (2005) 96--105,
\href{http://www.arXiv.org/abs/hep-th/0507232}{{\tt hep-th/0507232}}.

\bibitem{GP03}
G.~Pradisi, {\em Magnetic fluxes, NS-NS B field and shifts in four- dimensional
  orientifolds,}
\href{http://www.arXiv.org/abs/hep-th/0310154}{{\tt hep-th/0310154}}.

\bibitem{GH03}
G.~Honecker, {\em Chiral supersymmetric models on an orientifold of Z(4) x Z(2)
  with intersecting D6-branes,} Nucl. Phys. {\bf B666} (2003) 175--196,
\href{http://www.arXiv.org/abs/hep-th/0303015}{{\tt hep-th/0303015}}.

\bibitem{GH03a}
G.~Honecker, {\em Supersymmetric intersecting D6-branes and chiral models on
  the T(6)/(Z(4) x Z(2)) orbifold,}
\href{http://www.arXiv.org/abs/hep-th/0309158}{{\tt hep-th/0309158}}.

\bibitem{GH04}
G.~Honecker, {\em Chiral N=1 4D orientifolds with D-branes at angles,} Mod.
  Phys. Lett. {\bf A19} (2004) 1863--1879,
\href{http://www.arXiv.org/abs/hep-th/0407181}{{\tt hep-th/0407181}}.

\bibitem{Bailin:2006zf}
D.~Bailin and A.~Love, {\em Towards the supersymmetric standard model from
  intersecting D6-branes on the Z'(6) orientifold,}
\href{http://www.arXiv.org/abs/hep-th/0603172}{{\tt hep-th/0603172}}.

\bibitem{Antoniadis:2004pp}
I.~Antoniadis and T.~Maillard, {\em Moduli stabilization from magnetic fluxes
  in type I string theory,} Nucl. Phys. {\bf B716} (2005) 3--32,
\href{http://www.arXiv.org/abs/hep-th/0412008}{{\tt hep-th/0412008}}.

\bibitem{Bianchi:2005yz}
M.~Bianchi and E.~Trevigne, {\em The open story of the magnetic fluxes,} JHEP
  {\bf 08} (2005) 034,
\href{http://www.arXiv.org/abs/hep-th/0502147}{{\tt hep-th/0502147}}.

\bibitem{Aspinwall:2004jr}
P.~S. Aspinwall, {\em D-branes on Calabi-Yau manifolds,}
\href{http://www.arXiv.org/abs/hep-th/0403166}{{\tt hep-th/0403166}}.

\bibitem{Douglas:2000ah}
M.~R. Douglas, B.~Fiol, and C.~Romelsberger, {\em Stability and BPS branes,}
\href{http://www.arXiv.org/abs/hep-th/0002037}{{\tt hep-th/0002037}}.

\bibitem{Blumenhagen:2005zh}
R.~Blumenhagen, G.~Honecker, and T.~Weigand, {\em Non-abelian brane worlds: The
  open string story,}
\href{http://www.arXiv.org/abs/hep-th/0510050}{{\tt hep-th/0510050}}.

\bibitem{Blumenhagen:2005zg}
R.~Blumenhagen, G.~Honecker, and T.~Weigand, {\em Non-abelian brane worlds: The
  heterotic string story,} JHEP {\bf 10} (2005) 086,
\href{http://www.arXiv.org/abs/hep-th/0510049}{{\tt hep-th/0510049}}.

\bibitem{Witten:1985mj}
E.~Witten, {\em Global Anomalies In String Theory,}
  \href{http://www.arXiv.org/abs/Print-85-0620 (PRINCETON)}{{\tt Print-85-0620
  (PRINCETON)}}.

\bibitem{Freed:1986zx}
D.~S. Freed, {\em Determinants, Torsion And Strings,} Commun. Math. Phys. {\bf
  107}
(1986).

\bibitem{Witten:1984dg}
E.~Witten, {\em Some properties of O(32) superstrings,} Phys. Lett. {\bf B149}
  (1984)
351--356.

\bibitem{Blumenhagen:2005pm}
R.~Blumenhagen, G.~Honecker, and T.~Weigand, {\em Supersymmetric (non-)abelian
  bundles in the type I and SO(32) heterotic string,} JHEP {\bf 08} (2005) 009,
\href{http://www.arXiv.org/abs/hep-th/0507041}{{\tt hep-th/0507041}}.

\bibitem{Gepner:1987qi}
D.~Gepner, {\em Space-time supersymmetry in compactified string theory and
  superconformal models,} Nucl. Phys. {\bf B296} (1988)
757.

\bibitem{Gepner:1987vz}
D.~Gepner, {\em Exactly solvable string compactifications on manifolds of SU(N)
  holonomy,} Phys. Lett. {\bf B199} (1987)
380--388.

\bibitem{abpss96}
C.~Angelantonj, M.~Bianchi, G.~Pradisi, A.~Sagnotti, and Y.~S. Stanev, {\em
  Comments on Gepner models and type I vacua in string theory,} Phys. Lett.
  {\bf B387} (1996) 743--749,
\href{http://www.arXiv.org/abs/hep-th/9607226}{{\tt hep-th/9607226}}.

\bibitem{bw98}
R.~Blumenhagen and A.~Wisskirchen, {\em Spectra of 4D, N=1 type I string vacua
  on non-toroidal CY threefolds,} Phys. Lett. {\bf B438} (1998) 52--60,
\href{http://www.arXiv.org/abs/hep-th/9806131}{{\tt hep-th/9806131}}.

\bibitem{Govindarajan:2003vp}
S.~Govindarajan and J.~Majumder, {\em Crosscaps in Gepner models and type IIA
  orientifolds,} JHEP {\bf 02} (2004) 026,
\href{http://www.arXiv.org/abs/hep-th/0306257}{{\tt hep-th/0306257}}.

\bibitem{aaln03}
G.~Aldazabal, E.~C. Andres, M.~Leston, and C.~Nunez, {\em Type IIB orientifolds
  on Gepner points,} JHEP {\bf 09} (2003) 067,
\href{http://www.arXiv.org/abs/hep-th/0307183}{{\tt hep-th/0307183}}.

\bibitem{RB03a}
R.~Blumenhagen, {\em Supersymmetric orientifolds of Gepner models,} JHEP {\bf
  11} (2003) 055,
\href{http://www.arXiv.org/abs/hep-th/0310244}{{\tt hep-th/0310244}}.

\bibitem{bhhw04}
I.~Brunner, K.~Hori, K.~Hosomichi, and J.~Walcher, {\em Orientifolds of Gepner
  models,}
\href{http://www.arXiv.org/abs/hep-th/0401137}{{\tt hep-th/0401137}}.

\bibitem{bw04}
R.~Blumenhagen and T.~Weigand, {\em Chiral supersymmetric Gepner model
  orientifolds,} JHEP {\bf 02} (2004) 041,
\href{http://www.arXiv.org/abs/hep-th/0401148}{{\tt hep-th/0401148}}.

\bibitem{dhs04}
T.~P.~T. Dijkstra, L.~R. Huiszoon, and A.~N. Schellekens, {\em Chiral
  supersymmetric standard model spectra from orientifolds of Gepner models,}
  Phys. Lett. {\bf B609} (2005) 408--417,
\href{http://www.arXiv.org/abs/hep-th/0403196}{{\tt hep-th/0403196}}.

\bibitem{aaj04}
G.~Aldazabal, E.~C. Andres, and J.~E. Juknevich, {\em Particle models from
  orientifolds at Gepner-orbifold points,} JHEP {\bf 05} (2004) 054,
\href{http://www.arXiv.org/abs/hep-th/0403262}{{\tt hep-th/0403262}}.

\bibitem{bw04a}
R.~Blumenhagen and T.~Weigand, {\em A note on partition functions of Gepner
  model orientifolds,} Phys. Lett. {\bf B591} (2004) 161--169,
\href{http://www.arXiv.org/abs/hep-th/0403299}{{\tt hep-th/0403299}}.

\bibitem{dhs04a}
T.~P.~T. Dijkstra, L.~R. Huiszoon, and A.~N. Schellekens, {\em Supersymmetric
  standard model spectra from RCFT orientifolds,} Nucl. Phys. {\bf B710} (2005)
  3--57, \href{http://www.arXiv.org/abs/hep-th/0411129}{{\tt hep-th/0411129}}.

\bibitem{Dijkstra:2004ym}
T.~P.~T. Dijkstra, L.~R. Huiszoon, and A.~N. Schellekens, {\em Chiral
  supersymmetric standard model spectra from orientifolds of Gepner models,}
  Phys. Lett. {\bf B609} (2005) 408--417,
  \href{http://www.arXiv.org/abs/hep-th/0403196}{{\tt hep-th/0403196}}.

\bibitem{Gato-Rivera:2005qd}
B.~Gato-Rivera and A.~N. Schellekens, {\em Remarks on global anomalies in RCFT
  orientifolds,} Phys. Lett. {\bf B632} (2006) 728--732,
\href{http://www.arXiv.org/abs/hep-th/0510074}{{\tt hep-th/0510074}}.

\bibitem{Aldazabal:2006nz}
G.~Aldazabal, E.~Andres, and J.~E. Juknevich, {\em On SUSY standard-like models
  from orbifolds of D = 6 Gepner orientifolds,} JHEP {\bf 07} (2006) 039,
\href{http://www.arXiv.org/abs/hep-th/0603217}{{\tt hep-th/0603217}}.

\bibitem{Anastasopoulos:2006da}
P.~Anastasopoulos, T.~P.~T. Dijkstra, E.~Kiritsis, and A.~N. Schellekens, {\em
  Orientifolds, hypercharge embeddings and the standard model,}
\href{http://www.arXiv.org/abs/hep-th/0605226}{{\tt hep-th/0605226}}.

\bibitem{Fuchs:2000cm}
J.~Fuchs, L.~R. Huiszoon, A.~N. Schellekens, C.~Schweigert, and J.~Walcher,
  {\em Boundaries, crosscaps and simple currents,} Phys. Lett. {\bf B495}
  (2000) 427--434,
\href{http://www.arXiv.org/abs/hep-th/0007174}{{\tt hep-th/0007174}}.

\bibitem{Sagnotti:1996eb}
A.~Sagnotti and Y.~S. Stanev, {\em Open descendants in conformal field theory,}
  Fortsch. Phys. {\bf 44} (1996) 585--596,
\href{http://www.arXiv.org/abs/hep-th/9605042}{{\tt hep-th/9605042}}.

\bibitem{Recknagel:1997sb}
A.~Recknagel and V.~Schomerus, {\em D-branes in Gepner models,} Nucl. Phys.
  {\bf B531} (1998) 185--225,
\href{http://www.arXiv.org/abs/hep-th/9712186}{{\tt hep-th/9712186}}.

\bibitem{Brunner:2000nk}
I.~Brunner and V.~Schomerus, {\em D-branes at singular curves of Calabi-Yau
  compactifications,} JHEP {\bf 04} (2000) 020,
\href{http://www.arXiv.org/abs/hep-th/0001132}{{\tt hep-th/0001132}}.

\bibitem{Fuchs:2000fd}
J.~Fuchs {\em et al.}, {\em Boundary fixed points, enhanced gauge symmetry and
  singular bundles on K3,} Nucl. Phys. {\bf B598} (2001) 57--72,
\href{http://www.arXiv.org/abs/hep-th/0007145}{{\tt hep-th/0007145}}.

\bibitem{Braun:2005eg}
V.~Braun and S.~Schafer-Nameki, {\em D-brane charges in Gepner models,}
\href{http://www.arXiv.org/abs/hep-th/0511100}{{\tt hep-th/0511100}}.

\bibitem{Brunner:2000wx}
I.~Brunner and V.~Schomerus, {\em On superpotentials for D-branes in Gepner
  models,} JHEP {\bf 10} (2000) 016,
\href{http://www.arXiv.org/abs/hep-th/0008194}{{\tt hep-th/0008194}}.

\bibitem{Polchinski:1995df}
J.~Polchinski and E.~Witten, {\em Evidence for Heterotic - Type I String
  Duality,} Nucl. Phys. {\bf B460} (1996) 525--540,
\href{http://www.arXiv.org/abs/hep-th/9510169}{{\tt hep-th/9510169}}.

\bibitem{Blumenhagen:2005ga}
R.~Blumenhagen, G.~Honecker, and T.~Weigand, {\em Loop-corrected
  compactifications of the heterotic string with line bundles,} JHEP {\bf 06}
  (2005) 020,
\href{http://www.arXiv.org/abs/hep-th/0504232}{{\tt hep-th/0504232}}.

\bibitem{Blumenhagen:2006ux}
R.~Blumenhagen, S.~Moster, and T.~Weigand, {\em Heterotic GUT and standard
  model vacua from simply connected Calabi-Yau manifolds,} Nucl. Phys. {\bf
  B751} (2006) 186--221,
\href{http://www.arXiv.org/abs/hep-th/0603015}{{\tt hep-th/0603015}}.

\bibitem{Honecker:2006dt}
G.~Honecker, {\em Massive U(1)s and heterotic five-branes on K3,} Nucl. Phys.
  {\bf B748} (2006) 126--148,
\href{http://www.arXiv.org/abs/hep-th/0602101}{{\tt hep-th/0602101}}.

\bibitem{Andreas:2004ja}
B.~Andreas and D.~Hernandez~Ruiperez, {\em U(n) vector bundles on Calabi-Yau
  threefolds for string theory compactifications,} Adv. Theor. Math. Phys. {\bf
  9} (2005) 253--284,
\href{http://www.arXiv.org/abs/hep-th/0410170}{{\tt hep-th/0410170}}.

\bibitem{Andreas:2006dm}
B.~Andreas and G.~Curio, {\em Standard models from heterotic string theory,}
\href{http://www.arXiv.org/abs/hep-th/0602247}{{\tt hep-th/0602247}}.

\bibitem{Tatar:2006dc}
R.~Tatar and T.~Watari, {\em Proton decay, Yukawa couplings and underlying
  gauge symmetry in string theory,} Nucl. Phys. {\bf B747} (2006) 212--265,
\href{http://www.arXiv.org/abs/hep-th/0602238}{{\tt hep-th/0602238}}.

\bibitem{Silverstein:1995re}
E.~Silverstein and E.~Witten, {\em Criteria for conformal invariance of (0,2)
  models,} Nucl. Phys. {\bf B444} (1995) 161--190,
\href{http://www.arXiv.org/abs/hep-th/9503212}{{\tt hep-th/9503212}}.

\bibitem{Basu:2003bq}
A.~Basu and S.~Sethi, {\em World-sheet stability of (0,2) linear sigma models,}
  Phys. Rev. {\bf D68} (2003) 025003,
\href{http://www.arXiv.org/abs/hep-th/0303066}{{\tt hep-th/0303066}}.

\bibitem{Beasley:2003fx}
C.~Beasley and E.~Witten, {\em Residues and world-sheet instantons,} JHEP {\bf
  10} (2003) 065,
\href{http://www.arXiv.org/abs/hep-th/0304115}{{\tt hep-th/0304115}}.

\bibitem{Blumenhagen:1995tt}
R.~Blumenhagen and A.~Wisskirchen, {\em Exactly solvable (0,2) supersymmetric
  string vacua with GUT gauge groups,} Nucl. Phys. {\bf B454} (1995) 561--586,
\href{http://www.arXiv.org/abs/hep-th/9506104}{{\tt hep-th/9506104}}.

\bibitem{Blumenhagen:1995ew}
R.~Blumenhagen, R.~Schimmrigk, and A.~Wisskirchen, {\em The (0,2) Exactly
  Solvable Structure of Chiral Rings, Landau-Ginzburg Theories, and Calabi-Yau
  Manifolds,} Nucl. Phys. {\bf B461} (1996) 460--492,
\href{http://www.arXiv.org/abs/hep-th/9510055}{{\tt hep-th/9510055}}.

\bibitem{Friedman:1997yq}
R.~Friedman, J.~Morgan, and E.~Witten, {\em Vector bundles and F theory,}
  Commun. Math. Phys. {\bf 187} (1997) 679--743,
\href{http://www.arXiv.org/abs/hep-th/9701162}{{\tt hep-th/9701162}}.

\bibitem{Curio:1998vu}
G.~Curio, {\em Chiral matter and transitions in heterotic string models,} Phys.
  Lett. {\bf B435} (1998) 39--48,
\href{http://www.arXiv.org/abs/hep-th/9803224}{{\tt hep-th/9803224}}.

\bibitem{Ovrut:1999xu}
B.~A. Ovrut, {\em N = 1 supersymmetric vacua in heterotic M-theory,}
\href{http://www.arXiv.org/abs/hep-th/9905115}{{\tt hep-th/9905115}}.

\bibitem{Donagi:1999fa}
R.~Donagi, B.~A. Ovrut, T.~Pantev, and D.~Waldram, {\em Standard model vacua in
  heterotic M-theory,}
\href{http://www.arXiv.org/abs/hep-th/0001101}{{\tt hep-th/0001101}}.

\bibitem{Donagi:2000zf}
R.~Donagi, B.~A. Ovrut, T.~Pantev, and D.~Waldram, {\em Standard-model bundles
  on non-simply connected Calabi-Yau threefolds,} JHEP {\bf 08} (2001) 053,
\href{http://www.arXiv.org/abs/hep-th/0008008}{{\tt hep-th/0008008}}.

\bibitem{Donagi:2000zs}
R.~Donagi, B.~A. Ovrut, T.~Pantev, and D.~Waldram, {\em Standard-model
  bundles,} Adv. Theor. Math. Phys. {\bf 5} (2002) 563--615,
\href{http://www.arXiv.org/abs/math.ag/0008010}{{\tt math.ag/0008010}}.

\bibitem{Ovrut:2002hi}
B.~A. Ovrut, {\em Lectures on heterotic M-theory,}
\href{http://www.arXiv.org/abs/hep-th/0201032}{{\tt hep-th/0201032}}.

\bibitem{Donagi:2004qk}
R.~Donagi, Y.-H. He, B.~A. Ovrut, and R.~Reinbacher, {\em Moduli dependent
  spectra of heterotic compactifications,} Phys. Lett. {\bf B598} (2004)
  279--284,
\href{http://www.arXiv.org/abs/hep-th/0403291}{{\tt hep-th/0403291}}.

\bibitem{Donagi:2004ia}
R.~Donagi, Y.-H. He, B.~A. Ovrut, and R.~Reinbacher, {\em The particle spectrum
  of heterotic compactifications,} JHEP {\bf 12} (2004) 054,
\href{http://www.arXiv.org/abs/hep-th/0405014}{{\tt hep-th/0405014}}.

\bibitem{Donagi:2004ub}
R.~Donagi, Y.-H. He, B.~A. Ovrut, and R.~Reinbacher, {\em The spectra of
  heterotic standard model vacua,} JHEP {\bf 06} (2005) 070,
\href{http://www.arXiv.org/abs/hep-th/0411156}{{\tt hep-th/0411156}}.

\bibitem{Kobayashi:2004ya}
T.~Kobayashi, S.~Raby, and R.-J. Zhang, {\em Searching for realistic 4d string
  models with a Pati-Salam symmetry: Orbifold grand unified theories from
  heterotic string compactification on a Z(6) orbifold,} Nucl. Phys. {\bf B704}
  (2005) 3--55,
\href{http://www.arXiv.org/abs/hep-ph/0409098}{{\tt hep-ph/0409098}}.

\bibitem{Buchmuller:2005jr}
W.~Buchmuller, K.~Hamaguchi, O.~Lebedev, and M.~Ratz, {\em The supersymmetric
  standard model from the heterotic string,} Phys. Rev. Lett. {\bf 96} (2006)
  121602,
\href{http://www.arXiv.org/abs/hep-ph/0511035}{{\tt hep-ph/0511035}}.

\bibitem{Buchmuller:2006ik}
W.~Buchmuller, K.~Hamaguchi, O.~Lebedev, and M.~Ratz, {\em Supersymmetric
  standard model from the heterotic string. II,}
\href{http://www.arXiv.org/abs/hep-th/0606187}{{\tt hep-th/0606187}}.

\bibitem{Nilles:2006np}
H.~P. Nilles, S.~Ramos-Sanchez, P.~K.~S. Vaudrevange, and A.~Wingerter, {\em
  Exploring the SO(32) heterotic string,} JHEP {\bf 04} (2006) 050,
\href{http://www.arXiv.org/abs/hep-th/0603086}{{\tt hep-th/0603086}}.

\bibitem{Bouchard:2005ag}
V.~Bouchard and R.~Donagi, {\em An SU(5) heterotic standard model,} Phys. Lett.
  {\bf B633} (2006) 783--791,
\href{http://www.arXiv.org/abs/hep-th/0512149}{{\tt hep-th/0512149}}.

\bibitem{Braun:2005ux}
V.~Braun, Y.-H. He, B.~A. Ovrut, and T.~Pantev, {\em A heterotic standard
  model,} Phys. Lett. {\bf B618} (2005) 252--258,
\href{http://www.arXiv.org/abs/hep-th/0501070}{{\tt hep-th/0501070}}.

\bibitem{Braun:2005bw}
V.~Braun, Y.-H. He, B.~A. Ovrut, and T.~Pantev, {\em A standard model from the
  E(8) x E(8) heterotic superstring,} JHEP {\bf 06} (2005) 039,
\href{http://www.arXiv.org/abs/hep-th/0502155}{{\tt hep-th/0502155}}.

\bibitem{Braun:2005zv}
V.~Braun, Y.-H. He, B.~A. Ovrut, and T.~Pantev, {\em Vector bundle extensions,
  sheaf cohomology, and the heterotic standard model,}
\href{http://www.arXiv.org/abs/hep-th/0505041}{{\tt hep-th/0505041}}.

\bibitem{Bouchard:2006dn}
V.~Bouchard, M.~Cveti{\v c}, and R.~Donagi, {\em Tri-linear couplings in an
  heterotic minimal supersymmetric standard model,} Nucl. Phys. {\bf B745}
  (2006) 62--83,
\href{http://www.arXiv.org/abs/hep-th/0602096}{{\tt hep-th/0602096}}.

\bibitem{LauerKH}
D.~L{\"u}st, S.~Theisen, and G.~Zoupanos, {\em Four-Dimensional Heterotic
  Strings And Conformal Field Theory,} Nucl. Phys. {\bf B296} (1988)
800.

\bibitem{Lauer:1987kh}
J.~Lauer, D.~L{\"u}st, and S.~Theisen, {\em Four-Dimensional Supergravity From
  Four-Dimensional Strings,} Nucl. Phys. {\bf B304} (1988)
236--268.

\bibitem{DKL}
L.~J. Dixon, V.~Kaplunovsky, and J.~Louis, {\em On Effective Field Theories
  Describing (2,2) Vacua Of The Heterotic String,} Nucl. Phys. {\bf B329}
  (1990)
27--82.

\bibitem{IbanezHC}
L.~E. Ib{\'a}{\~n}ez and D.~L{\"u}st, {\em Duality anomaly cancellation,
  minimal string unification and the effective low-energy Lagrangian of 4-D
  strings,} Nucl. Phys. {\bf B382} (1992) 305--364,
\href{http://www.arXiv.org/abs/hep-th/9202046}{{\tt hep-th/9202046}}.

\bibitem{HPN}
H.~P. Nilles and S.~Stieberger, {\em String unification, universal one-loop
  corrections and strongly coupled heterotic string theory,} Nucl. Phys. {\bf
  B499} (1997) 3--28,
\href{http://www.arXiv.org/abs/hep-th/9702110}{{\tt hep-th/9702110}}.

\bibitem{CremmerEN}
E.~Cremmer, S.~Ferrara, L.~Girardello, and A.~Van~Proeyen, {\em Yang-Mills
  Theories With Local Supersymmetry: Lagrangian, Transformation Laws And
  Superhiggs Effect,} Nucl. Phys. {\bf B212} (1983)
413.

\bibitem{Cand}
P.~Candelas and X.~de~la Ossa, {\em Moduli Space Of Calabi-Yau Manifolds,}
  Nucl. Phys. {\bf B355} (1991)
455--481.

\bibitem{Acharya:2002ag}
B.~Acharya, M.~Aganagic, K.~Hori, and C.~Vafa, {\em Orientifolds, mirror
  symmetry and superpotentials,}
\href{http://www.arXiv.org/abs/hep-th/0202208}{{\tt hep-th/0202208}}.

\bibitem{Brunner:2003zm}
I.~Brunner and K.~Hori, {\em Orientifolds and mirror symmetry,} JHEP {\bf 11}
  (2004) 005,
\href{http://www.arXiv.org/abs/hep-th/0303135}{{\tt hep-th/0303135}}.

\bibitem{blowup}
D.~L{\"u}st, S.~Reffert, E.~Scheidegger, and S.~Stieberger, {\em Resolved
  toroidal orbifolds and their orientifolds,}
\href{http://www.arXiv.org/abs/hep-th/0609014}{{\tt hep-th/0609014}}.

\bibitem{Jockers:2005zy}
H.~Jockers and J.~Louis, {\em D-terms and F-terms from D7-brane fluxes,} Nucl.
  Phys. {\bf B718} (2005) 203--246,
\href{http://www.arXiv.org/abs/hep-th/0502059}{{\tt hep-th/0502059}}.

\bibitem{GL}
T.~W. Grimm, {\em The effective action of type II Calabi-Yau orientifolds,}
  Fortsch. Phys. {\bf 53} (2005) 1179--1271,
\href{http://www.arXiv.org/abs/hep-th/0507153}{{\tt hep-th/0507153}}.

\bibitem{lmrs04}
D.~L{\"u}st, P.~Mayr, R.~Richter, and S.~Stieberger, {\em Scattering of Gauge,
  Matter, and Moduli Fields from Intersecting Branes,} Nucl. Phys. {\bf B696}
  (2004) 205--250,
\href{http://www.arXiv.org/abs/hep-th/0404134}{{\tt hep-th/0404134}}.

\bibitem{lrs04}
D.~L{\"u}st, S.~Reffert, and S.~Stieberger, {\em Flux-induced soft
  supersymmetry breaking in chiral type IIb orientifolds with D3/D7-branes,}
  Nucl. Phys. {\bf B706} (2005) 3--52,
\href{http://www.arXiv.org/abs/hep-th/0406092}{{\tt hep-th/0406092}}.

\bibitem{LRSSii}
D.~Lust, S.~Reffert, E.~Scheidegger, W.~Schulgin, and S.~Stieberger, {\em
  Moduli stabilization in type IIB orientifolds. II,} Nucl. Phys. {\bf B766}
  (2007) 178--231,
\href{http://www.arXiv.org/abs/hep-th/0609013}{{\tt hep-th/0609013}}.

\bibitem{Klemm}
J.~Erler and A.~Klemm, {\em Comment on the generation number in orbifold
  compactifications,} Commun. Math. Phys. {\bf 153} (1993) 579--604,
\href{http://www.arXiv.org/abs/hep-th/9207111}{{\tt hep-th/9207111}}.

\bibitem{IMNQ}
L.~E. Ib{\'a}{\~n}ez, J.~Mas, H.-P. Nilles, and F.~Quevedo, {\em Heterotic
  Strings In Symmetric And Asymmetric Orbifold Backgrounds,} Nucl. Phys. {\bf
  B301} (1988)
157.

\bibitem{FIQ}
A.~Font, L.~E. Ib{\'a}{\~n}ez, and F.~Quevedo, {\em $\IZ_N\times \IZ_M$
  orbifolds and discrete torsion,} Phys. Lett. {\bf B217} (1989)
272.

\bibitem{FKP1}
S.~Ferrara, C.~Kounnas, and M.~Porrati, {\em General Dimensional Reduction Of
  Ten-Dimensional Supergravity And Superstring,} Phys. Lett. {\bf B181} (1986)
263.

\bibitem{FKP2}
M.~Cveti{\v c}, J.~Louis, and B.~A. Ovrut, {\em A String Calculation Of The
  Kahler Potentials For Moduli Of Z(N) Orbifolds,} Phys. Lett. {\bf B206}
  (1988)
227.

\bibitem{Ferrara:1990uu}
S.~Ferrara and S.~Theisen, {\em Moduli Spaces, Effective Actions And Duality
  Symmetry In String Compactifications,}. Based on lectures given at 3rd
  Hellenic Summer School, Corfu, Greece, Sep 13-23, 1989.

\bibitem{LRSSi}
D.~Lust, S.~Reffert, W.~Schulgin, and S.~Stieberger, {\em Moduli stabilization
  in type IIB orientifolds. I: Orbifold limits,} Nucl. Phys. {\bf B766} (2007)
  68--149,
\href{http://www.arXiv.org/abs/hep-th/0506090}{{\tt hep-th/0506090}}.

\bibitem{Rabadan}
M.~Klein and R.~Rabadan, {\em Orientifolds with discrete torsion,} JHEP {\bf
  07} (2000) 040,
\href{http://www.arXiv.org/abs/hep-th/0002103}{{\tt hep-th/0002103}}.

\bibitem{DenefMM}
F.~Denef, M.~R. Douglas, B.~Florea, A.~Grassi, and S.~Kachru, {\em Fixing all
  moduli in a simple F-theory compactification,}
\href{http://www.arXiv.org/abs/hep-th/0503124}{{\tt hep-th/0503124}}.

\bibitem{ggjl03}
M.~Gra{\~ n}a, T.~W. Grimm, H.~Jockers, and J.~Louis, {\em Soft supersymmetry
  breaking in Calabi-Yau orientifolds with D-branes and fluxes,} Nucl. Phys.
  {\bf B690} (2004) 21--61,
\href{http://www.arXiv.org/abs/hep-th/0312232}{{\tt hep-th/0312232}}.

\bibitem{ciu03}
P.~G. C{\' a}mara, L.~E. Ib{\' a}{\~ n}ez, and A.~M. Uranga, {\em Flux-induced
  SUSY-breaking soft terms,} Nucl. Phys. {\bf B689} (2004) 195--242,
\href{http://www.arXiv.org/abs/hep-th/0311241}{{\tt hep-th/0311241}}.

\bibitem{ciu04}
P.~G. Camara, L.~E. Ib{\'a}{\~n}ez, and A.~M. Uranga, {\em Flux-induced
  SUSY-breaking soft terms on D7-D3 brane systems,} Nucl. Phys. {\bf B708}
  (2005) 268--316,
\href{http://www.arXiv.org/abs/hep-th/0408036}{{\tt hep-th/0408036}}.

\bibitem{kn03}
B.~K{\"o}rs and P.~Nath, {\em Effective action and soft supersymmetry breaking
  for intersecting D-brane models,} Nucl. Phys. {\bf B681} (2004) 77--119,
\href{http://www.arXiv.org/abs/hep-th/0309167}{{\tt hep-th/0309167}}.

\bibitem{Lerche:2002yw}
W.~Lerche, P.~Mayr, and N.~Warner, {\em N = 1 special geometry, mixed Hodge
  variations and toric geometry,}
\href{http://www.arXiv.org/abs/hep-th/0208039}{{\tt hep-th/0208039}}.

\bibitem{Lerche:2002ck}
W.~Lerche, P.~Mayr, and N.~Warner, {\em Holomorphic N = 1 special geometry of
  open-closed type II strings,}
\href{http://www.arXiv.org/abs/hep-th/0207259}{{\tt hep-th/0207259}}.

\bibitem{Lerche:2003hs}
W.~Lerche, {\em Special geometry and mirror symmetry for open string
  backgrounds with N = 1 supersymmetry,}
\href{http://www.arXiv.org/abs/hep-th/0312326}{{\tt hep-th/0312326}}.

\bibitem{deWit:1991nm}
B.~de~Wit and A.~Van~Proeyen, {\em Special geometry, cubic polynomials and
  homogeneous quaternionic spaces,} Commun. Math. Phys. {\bf 149} (1992)
  307--334,
\href{http://www.arXiv.org/abs/hep-th/9112027}{{\tt hep-th/9112027}}.

\bibitem{ADFT}
C.~Angelantonj, R.~D'Auria, S.~Ferrara, and M.~Trigiante, {\em $K3 \times
  T^2/\IZ_2$ orientifolds with fluxes, open string moduli and critical points,}
  Phys. Lett. {\bf B583} (2004) 331--337,
\href{http://www.arXiv.org/abs/hep-th/0312019}{{\tt hep-th/0312019}}.

\bibitem{lmrs05}
D.~L{\"u}st, P.~Mayr, S.~Reffert, and S.~Stieberger, {\em F-theory flux,
  destabilization of orientifolds and soft terms on D7-branes,} Nucl. Phys.
  {\bf B732} (2006) 243--290,
\href{http://www.arXiv.org/abs/hep-th/0501139}{{\tt hep-th/0501139}}.

\bibitem{stieberg}
S.~Stieberger, {\em (0,2) heterotic gauge couplings and their M-theory origin,}
  Nucl. Phys. {\bf B541} (1999) 109--144,
\href{http://www.arXiv.org/abs/hep-th/9807124}{{\tt hep-th/9807124}}.

\bibitem{progress}
D. L{\"u}st and S. Stieberger, work to appear.

\bibitem{russo}
M.~Bertolini, M.~Billo, A.~Lerda, J.~F. Morales, and R.~Russo, {\em Brane world
  effective actions for D-branes with fluxes,} Nucl. Phys. {\bf B743} (2006)
  1--40,
\href{http://www.arXiv.org/abs/hep-th/0512067}{{\tt hep-th/0512067}}.

\bibitem{bain}
P.~Bain and M.~Berg, {\em Effective action of matter fields in four-dimensional
  string orientifolds,} JHEP {\bf 04} (2000) 013,
\href{http://www.arXiv.org/abs/hep-th/0003185}{{\tt hep-th/0003185}}.

\bibitem{Seiberg}
N.~Seiberg, {\em Observations On The Moduli Space Of Superconformal Field
  Theories,} Nucl. Phys. {\bf B303} (1988)
286.

\bibitem{Cecotti}
S.~Cecotti, S.~Ferrara, and L.~Girardello, {\em Geometry Of Type II
  Superstrings And The Moduli Of Superconformal Field Theories,} Int. J. Mod.
  Phys. {\bf A4} (1989)
2475.

\bibitem{MSD}
P.~Mayr and S.~Stieberger, {\em Dilaton, antisymmetric tensor and gauge fields
  in string effective theories at the one loop level,} Nucl. Phys. {\bf B412}
  (1994) 502--522,
\href{http://www.arXiv.org/abs/hep-th/9304055}{{\tt hep-th/9304055}}.

\bibitem{dfms87}
L.~J. Dixon, D.~Friedan, E.~J. Martinec, and S.~H. Shenker, {\em The Conformal
  Field Theory Of Orbifolds,} Nucl. Phys. {\bf B282} (1987)
13--73.

\bibitem{Rigolin}
L.~E. Ib{\'a}{\~n}ez, C.~Munoz, and S.~Rigolin, {\em Aspects of type I string
  phenomenology,} Nucl. Phys. {\bf B553} (1999) 43--80,
\href{http://www.arXiv.org/abs/hep-ph/9812397}{{\tt hep-ph/9812397}}.

\bibitem{Hamidi:1986vh}
S.~Hamidi and C.~Vafa, {\em Interactions on orbifolds,} Nucl. Phys. {\bf B279}
  (1987)
465.

\bibitem{Erler:1992gt}
J.~Erler, D.~Jungnickel, M.~Spalinski, and S.~Stieberger, {\em Higher twisted
  sector couplings of Z(N) orbifolds,} Nucl. Phys. {\bf B397} (1993) 379--416,
\href{http://www.arXiv.org/abs/hep-th/9207049}{{\tt hep-th/9207049}}.

\bibitem{Stieberger:1992bj}
S.~Stieberger, D.~Jungnickel, J.~Lauer, and M.~Spalinski, {\em Yukawa couplings
  for bosonic Z(N) orbifolds: Their moduli and twisted sector dependence,} Mod.
  Phys. Lett. {\bf A7} (1992) 3059--3070,
\href{http://www.arXiv.org/abs/hep-th/9204037}{{\tt hep-th/9204037}}.

\bibitem{Casas:1991ac}
J.~A. Casas, F.~Gomez, and C.~Munoz, {\em Complete structure of Z(n) Yukawa
  couplings,} Int. J. Mod. Phys. {\bf A8} (1993) 455--506,
\href{http://www.arXiv.org/abs/hep-th/9110060}{{\tt hep-th/9110060}}.

\bibitem{Stieberger:1992vb}
S.~Stieberger, {\em Moduli and twisted sector dependence on Z(N) x Z(M)
  orbifold couplings,} Phys. Lett. {\bf B300} (1993) 347--353,
\href{http://www.arXiv.org/abs/hep-th/9211027}{{\tt hep-th/9211027}}.

\bibitem{Kobayashi:2003vi}
T.~Kobayashi and O.~Lebedev, {\em Heterotic Yukawa couplings and continuous
  Wilson lines,} Phys. Lett. {\bf B566} (2003) 164--170,
\href{http://www.arXiv.org/abs/hep-th/0303009}{{\tt hep-th/0303009}}.

\bibitem{Gava:1997jt}
E.~Gava, K.~S. Narain, and M.~H. Sarmadi, {\em On the bound states of p- and
  (p+2)-branes,} Nucl. Phys. {\bf B504} (1997) 214--238,
\href{http://www.arXiv.org/abs/hep-th/9704006}{{\tt hep-th/9704006}}.

\bibitem{Antoniadis:2000jv}
I.~Antoniadis, K.~Benakli, and A.~Laugier, {\em Contact interactions in D-brane
  models,} JHEP {\bf 05} (2001) 044,
\href{http://www.arXiv.org/abs/hep-th/0011281}{{\tt hep-th/0011281}}.

\bibitem{cim03}
D.~Cremades, L.~E. Ib{\'a}{\~n}ez, and F.~Marchesano, {\em Yukawa couplings in
  intersecting D-brane models,} JHEP {\bf 07} (2003) 038,
\href{http://www.arXiv.org/abs/hep-th/0302105}{{\tt hep-th/0302105}}.

\bibitem{cp03}
M.~Cveti{\v c} and I.~Papadimitriou, {\em Conformal field theory couplings for
  intersecting D-branes on orientifolds,} Phys. Rev. {\bf D68} (2003) 046001,
\href{http://www.arXiv.org/abs/hep-th/0303083}{{\tt hep-th/0303083}}.

\bibitem{ao03}
S.~A. Abel and A.~W. Owen, {\em Interactions in intersecting brane models,}
  Nucl. Phys. {\bf B663} (2003) 197--214,
\href{http://www.arXiv.org/abs/hep-th/0303124}{{\tt hep-th/0303124}}.

\bibitem{cim04}
D.~Cremades, L.~E. Ib{\'a}{\~n}ez, and F.~Marchesano, {\em Computing Yukawa
  couplings from magnetized extra dimensions,} JHEP {\bf 05} (2004) 079,
\href{http://www.arXiv.org/abs/hep-th/0404229}{{\tt hep-th/0404229}}.

\bibitem{Abel:2003yh}
S.~A. Abel, O.~Lebedev, and J.~Santiago, {\em Flavour in intersecting brane
  models and bounds on the string scale,} Nucl. Phys. {\bf B696} (2004)
  141--173,
\href{http://www.arXiv.org/abs/hep-ph/0312157}{{\tt hep-ph/0312157}}.

\bibitem{Kitazawa:2004nf}
N.~Kitazawa, T.~Kobayashi, N.~Maru, and N.~Okada, {\em Yukawa coupling
  structure in intersecting D-brane models,} Eur. Phys. J. {\bf C40} (2005)
  579--587,
\href{http://www.arXiv.org/abs/hep-th/0406115}{{\tt hep-th/0406115}}.

\bibitem{Higaki:2005ie}
T.~Higaki, N.~Kitazawa, T.~Kobayashi, and K.-j. Takahashi, {\em Flavor
  structure and coupling selection rule from intersecting D-branes,} Phys. Rev.
  {\bf D72} (2005) 086003,
\href{http://www.arXiv.org/abs/hep-th/0504019}{{\tt hep-th/0504019}}.

\bibitem{Herbst:2006nn}
M.~Herbst, W.~Lerche, and D.~Nemeschansky, {\em Instanton geometry and quantum
  A(infinity) structure on the elliptic curve,}
\href{http://www.arXiv.org/abs/hep-th/0603085}{{\tt hep-th/0603085}}.

\bibitem{Herbst:2004jp}
M.~Herbst, C.-I. Lazaroiu, and W.~Lerche, {\em Superpotentials, A(infinity)
  relations and WDVV equations for open topological strings,} JHEP {\bf 02}
  (2005) 071,
\href{http://www.arXiv.org/abs/hep-th/0402110}{{\tt hep-th/0402110}}.

\bibitem{Brunner:2004mt}
I.~Brunner, M.~Herbst, W.~Lerche, and J.~Walcher, {\em Matrix factorizations
  and mirror symmetry: The cubic curve,}
\href{http://www.arXiv.org/abs/hep-th/0408243}{{\tt hep-th/0408243}}.

\bibitem{ao03a}
S.~A. Abel and A.~W. Owen, {\em N-point amplitudes in intersecting brane
  models,} Nucl. Phys. {\bf B682} (2004) 183--216,
\href{http://www.arXiv.org/abs/hep-th/0310257}{{\tt hep-th/0310257}}.

\bibitem{Oprisa:2005wu}
D.~Oprisa and S.~Stieberger, {\em Six gluon open superstring disk amplitude,
  multiple hypergeometric series and Euler-Zagier sums,}
\href{http://www.arXiv.org/abs/hep-th/0509042}{{\tt hep-th/0509042}}.

\bibitem{Stieberger:2006te}
S.~Stieberger and T.~R. Taylor, {\em Multi-gluon scattering in open superstring
  theory,} Phys. Rev. {\bf D74} (2006) 126007,
\href{http://www.arXiv.org/abs/hep-th/0609175}{{\tt hep-th/0609175}}.

\bibitem{KaplunovskyRP}
V.~S. Kaplunovsky, {\em One loop threshold effects in string unification,}
  Nucl. Phys. {\bf B307} (1988) 145,
\href{http://www.arXiv.org/abs/hep-th/9205068}{{\tt hep-th/9205068}}.

\bibitem{NEST}
H.~P. Nilles and S.~Stieberger, {\em How to Reach the Correct $\sin^2\theta_W$
  and $\alpha_S$ in String Theory,} Phys. Lett. {\bf B367} (1996) 126--133,
\href{http://www.arXiv.org/abs/hep-th/9510009}{{\tt hep-th/9510009}}.

\bibitem{RaySinger}
D.~B. Ray and I.~M. Singer, {\em Analytic torsion for complex manifolds,}
  Annals Math. {\bf 98} (1973)
154--177.

\bibitem{BCOV}
M.~Bershadsky, S.~Cecotti, H.~Ooguri, and C.~Vafa, {\em Kodaira-Spencer theory
  of gravity and exact results for quantum string amplitudes,} Commun. Math.
  Phys. {\bf 165} (1994) 311--428,
\href{http://www.arXiv.org/abs/hep-th/9309140}{{\tt hep-th/9309140}}.

\bibitem{FKLZ}
S.~Ferrara, C.~Kounnas, D.~L{\"u}st, and F.~Zwirner, {\em Duality invariant
  partition functions and automorphic superpotentials for (2,2) string
  compactifications,} Nucl. Phys. {\bf B365} (1991)
431--466.

\bibitem{MS1}
P.~Mayr and S.~Stieberger, {\em Threshold corrections to gauge couplings in
  orbifold compactifications,} Nucl. Phys. {\bf B407} (1993) 725--748,
\href{http://www.arXiv.org/abs/hep-th/9303017}{{\tt hep-th/9303017}}.

\bibitem{fw02}
T.~Friedmann and E.~Witten, {\em Unification scale, proton decay, and manifolds
  of G(2) holonomy,} Adv. Theor. Math. Phys. {\bf 7} (2003) 577--617,
\href{http://www.arXiv.org/abs/hep-th/0211269}{{\tt hep-th/0211269}}.

\bibitem{kristin}
K.~F{\"o}rger and S.~Stieberger, {\em Higher derivative couplings and
  heterotic-type I duality in eight dimensions,} Nucl. Phys. {\bf B559} (1999)
  277--300,
\href{http://www.arXiv.org/abs/hep-th/9901020}{{\tt hep-th/9901020}}.

\bibitem{JOINT}
P. Mayr and S. Stieberger, unpublished notes.

\bibitem{BCOV1}
M.~Bershadsky, S.~Cecotti, H.~Ooguri, and C.~Vafa, {\em Holomorphic anomalies
  in topological field theories,} Nucl. Phys. {\bf B405} (1993) 279--304,
\href{http://www.arXiv.org/abs/hep-th/9302103}{{\tt hep-th/9302103}}.

\bibitem{fabre}
C.~Bachas and C.~Fabre, {\em Threshold Effects in Open-String Theory,} Nucl.
  Phys. {\bf B476} (1996) 418--436,
\href{http://www.arXiv.org/abs/hep-th/9605028}{{\tt hep-th/9605028}}.

\bibitem{APT}
I.~Antoniadis, H.~Partouche, and T.~R. Taylor, {\em Duality of N = 2
  heterotic-type I compactifications in four dimensions,} Nucl. Phys. {\bf
  B499} (1997) 29--44,
\href{http://www.arXiv.org/abs/hep-th/9703076}{{\tt hep-th/9703076}}.

\bibitem{ABD}
I.~Antoniadis, C.~Bachas, and E.~Dudas, {\em Gauge couplings in
  four-dimensional type I string orbifolds,} Nucl. Phys. {\bf B560} (1999)
  93--134,
\href{http://www.arXiv.org/abs/hep-th/9906039}{{\tt hep-th/9906039}}.

\bibitem{ls03}
D.~L{\"u}st and S.~Stieberger, {\em Gauge threshold corrections in intersecting
  brane world models,}
\href{http://www.arXiv.org/abs/hep-th/0302221}{{\tt hep-th/0302221}}.

\bibitem{Bianchi:2005sa}
M.~Bianchi and E.~Trevigne, {\em Gauge thresholds in the presence of oblique
  magnetic fluxes,} JHEP {\bf 01} (2006) 092,
\href{http://www.arXiv.org/abs/hep-th/0506080}{{\tt hep-th/0506080}}.

\bibitem{DKL2}
L.~J. Dixon, V.~Kaplunovsky, and J.~Louis, {\em Moduli dependence of string
  loop corrections to gauge coupling constants,} Nucl. Phys. {\bf B355} (1991)
649--688.

\bibitem{Berg:2004ek}
M.~Berg, M.~Haack, and B.~K{\"o}rs, {\em Loop corrections to volume moduli and
  inflation in string theory,} Phys. Rev. {\bf D71} (2005) 026005,
\href{http://www.arXiv.org/abs/hep-th/0404087}{{\tt hep-th/0404087}}.

\bibitem{Berg:2005ja}
M.~Berg, M.~Haack, and B.~K{\"o}rs, {\em String loop corrections to Kaehler
  potentials in orientifolds,} JHEP {\bf 11} (2005) 030,
\href{http://www.arXiv.org/abs/hep-th/0508043}{{\tt hep-th/0508043}}.

\bibitem{Abel:2004ue}
S.~A. Abel and B.~W. Schofield, {\em One-loop Yukawas on intersecting branes,}
  JHEP {\bf 06} (2005) 072,
\href{http://www.arXiv.org/abs/hep-th/0412206}{{\tt hep-th/0412206}}.

\bibitem{Abel:2005qn}
S.~A. Abel and M.~D. Goodsell, {\em Intersecting brane worlds at one loop,}
  JHEP {\bf 02} (2006) 049,
\href{http://www.arXiv.org/abs/hep-th/0512072}{{\tt hep-th/0512072}}.

\bibitem{Lawrence:2004sm}
A.~Lawrence and J.~McGreevy, {\em D-terms and D-strings in open string models,}
  JHEP {\bf 10} (2004) 056,
\href{http://www.arXiv.org/abs/hep-th/0409284}{{\tt hep-th/0409284}}.

\bibitem{Bianchi:2006nf}
M.~Bianchi and A.~V. Santini, {\em String predictions for near future colliders
  from one-loop scattering amplitudes around D-brane worlds,}
\href{http://www.arXiv.org/abs/hep-th/0607224}{{\tt hep-th/0607224}}.

\bibitem{Antoniadis:2005sd}
I.~Antoniadis, K.~S. Narain, and T.~R. Taylor, {\em Open string topological
  amplitudes and gaugino masses,} Nucl. Phys. {\bf B729} (2005) 235--277,
\href{http://www.arXiv.org/abs/hep-th/0507244}{{\tt hep-th/0507244}}.

\bibitem{Antoniadis:2004qn}
I.~Antoniadis and T.~R. Taylor, {\em Topological masses from broken
  supersymmetry,} Nucl. Phys. {\bf B695} (2004) 103--131,
\href{http://www.arXiv.org/abs/hep-th/0403293}{{\tt hep-th/0403293}}.

\bibitem{Antoniadis:2005xa}
I.~Antoniadis and T.~R. Taylor, {\em Note on mediation of supersymmetry
  breaking from closed to open strings,} Nucl. Phys. {\bf B731} (2005)
  164--170,
\href{http://www.arXiv.org/abs/hep-th/0509048}{{\tt hep-th/0509048}}.

\bibitem{FERRARA1}
R.~D'Auria, S.~Ferrara, and S.~Vaula, {\em N = 4 gauged supergravity and a IIB
  orientifold with fluxes,} New J. Phys. {\bf 4} (2002) 71,
\href{http://www.arXiv.org/abs/hep-th/0206241}{{\tt hep-th/0206241}}.

\bibitem{glmw02}
S.~Gurrieri, J.~Louis, A.~Micu, and D.~Waldram, {\em Mirror symmetry in
  generalized Calabi-Yau compactifications,} Nucl. Phys. {\bf B654} (2003)
  61--113,
\href{http://www.arXiv.org/abs/hep-th/0211102}{{\tt hep-th/0211102}}.

\bibitem{kstt02}
S.~Kachru, M.~B. Schulz, P.~K. Tripathy, and S.~P. Trivedi, {\em New
  supersymmetric string compactifications,} JHEP {\bf 03} (2003) 061,
\href{http://www.arXiv.org/abs/hep-th/0211182}{{\tt hep-th/0211182}}.

\bibitem{deWit:2002vt}
B.~de~Wit, H.~Samtleben, and M.~Trigiante, {\em On Lagrangians and gaugings of
  maximal supergravities,} Nucl. Phys. {\bf B655} (2003) 93--126,
\href{http://www.arXiv.org/abs/hep-th/0212239}{{\tt hep-th/0212239}}.

\bibitem{FERRARA}
R.~D'Auria, S.~Ferrara, F.~Gargiulo, M.~Trigiante, and S.~Vaula, {\em N = 4
  supergravity Lagrangian for type IIB on T**6/Z(2) in presence of fluxes and
  D3-branes,} JHEP {\bf 06} (2003) 045,
\href{http://www.arXiv.org/abs/hep-th/0303049}{{\tt hep-th/0303049}}.

\bibitem{Angelantonj:2003up}
C.~Angelantonj, S.~Ferrara, and M.~Trigiante, {\em Unusual gauged
  supergravities from type IIA and type IIB orientifolds,} Phys. Lett. {\bf
  B582} (2004) 263--269,
\href{http://www.arXiv.org/abs/hep-th/0310136}{{\tt hep-th/0310136}}.

\bibitem{Andrianopoli:2003jf}
L.~Andrianopoli, R.~D'Auria, S.~Ferrara, and M.~A. Lledo, {\em 4-D gauged
  supergravity analysis of type IIB vacua on K3 x T**2/Z(2),} JHEP {\bf 03}
  (2003) 044,
\href{http://www.arXiv.org/abs/hep-th/0302174}{{\tt hep-th/0302174}}.

\bibitem{deWit:2004yr}
B.~de~Wit, H.~Nicolai, and H.~Samtleben, {\em Gauged supergravities in three
  dimensions: A panoramic overview,}
\href{http://www.arXiv.org/abs/hep-th/0403014}{{\tt hep-th/0403014}}.

\bibitem{aft03}
C.~Angelantonj, S.~Ferrara, and M.~Trigiante, {\em New D = 4 gauged
  supergravities from N = 4 orientifolds with fluxes,} JHEP {\bf 10} (2003)
  015,
\href{http://www.arXiv.org/abs/hep-th/0306185}{{\tt hep-th/0306185}}.

\bibitem{Andrianopoli:2003sa}
L.~Andrianopoli, S.~Ferrara, and M.~Trigiante, {\em Fluxes, supersymmetry
  breaking and gauged supergravity,}
\href{http://www.arXiv.org/abs/hep-th/0307139}{{\tt hep-th/0307139}}.

\bibitem{deWit:2003hq}
B.~de~Wit, H.~Samtleben, and M.~Trigiante, {\em Maximal supergravity from IIB
  flux compactifications,} Phys. Lett. {\bf B583} (2004) 338--346,
\href{http://www.arXiv.org/abs/hep-th/0311224}{{\tt hep-th/0311224}}.

\bibitem{Derendinger:2004kf}
J.-P. Derendinger, C.~Kounnas, and F.~Zwirner, {\em Potentials and
  superpotentials in the effective N = 1 supergravities from higher
  dimensions,} Nucl. Phys. {\bf B691} (2004) 233--248,
\href{http://www.arXiv.org/abs/hep-th/0403043}{{\tt hep-th/0403043}}.

\bibitem{Andrianopoli:2004im}
L.~Andrianopoli, S.~Ferrara, and M.~A. Lledo, {\em Scherk-Schwarz reduction of
  D = 5 special and quaternionic geometry,} Class. Quant. Grav. {\bf 21} (2004)
  4677--4696,
\href{http://www.arXiv.org/abs/hep-th/0405164}{{\tt hep-th/0405164}}.

\bibitem{Dall'Agata:2005ff}
G.~Dall'Agata and S.~Ferrara, {\em Gauged supergravity algebras from twisted
  tori compactifications with fluxes,} Nucl. Phys. {\bf B717} (2005) 223--245,
\href{http://www.arXiv.org/abs/hep-th/0502066}{{\tt hep-th/0502066}}.

\bibitem{Dall'Agata:2005mj}
G.~Dall'Agata, R.~D'Auria, and S.~Ferrara, {\em Compactifications on twisted
  tori with fluxes and free differential algebras,} Phys. Lett. {\bf B619}
  (2005) 149--154,
\href{http://www.arXiv.org/abs/hep-th/0503122}{{\tt hep-th/0503122}}.

\bibitem{Scherk:1978ta}
J.~Scherk and J.~H. Schwarz, {\em Spontaneous breaking of supersymmetry through
  dimensional reduction,} Phys. Lett. {\bf B82} (1979)
60.

\bibitem{Gomis:2005wc}
J.~Gomis, F.~Marchesano, and D.~Mateos, {\em An open string landscape,} JHEP
  {\bf 11} (2005) 021,
\href{http://www.arXiv.org/abs/hep-th/0506179}{{\tt hep-th/0506179}}.

\bibitem{Michelson:1996pn}
J.~Michelson, {\em Compactifications of type IIB strings to four dimensions
  with non-trivial classical potential,} Nucl. Phys. {\bf B495} (1997)
  127--148,
\href{http://www.arXiv.org/abs/hep-th/9610151}{{\tt hep-th/9610151}}.

\bibitem{Dall'Agata:2001zh}
G.~Dall'Agata, {\em Type IIB supergravity compactified on a Calabi-Yau manifold
  with H-fluxes,} JHEP {\bf 11} (2001) 005,
\href{http://www.arXiv.org/abs/hep-th/0107264}{{\tt hep-th/0107264}}.

\bibitem{Louis:2002ny}
J.~Louis and A.~Micu, {\em Type II theories compactified on Calabi-Yau
  threefolds in the presence of background fluxes,} Nucl. Phys. {\bf B635}
  (2002) 395--431,
\href{http://www.arXiv.org/abs/hep-th/0202168}{{\tt hep-th/0202168}}.

\bibitem{Curio:2001ae}
G.~Curio, A.~Klemm, B.~K{\"o}rs, and D.~L{\"u}st, {\em Fluxes in heterotic and
  type II string compactifications,} Nucl. Phys. {\bf B620} (2002) 237--258,
\href{http://www.arXiv.org/abs/hep-th/0106155}{{\tt hep-th/0106155}}.

\bibitem{Louis:2006kb}
J.~Louis and A.~Micu, {\em Heterotic-type IIA duality with fluxes,}
\href{http://www.arXiv.org/abs/hep-th/0608171}{{\tt hep-th/0608171}}.

\bibitem{Shelton:2005cf}
J.~Shelton, W.~Taylor, and B.~Wecht, {\em Nongeometric flux compactifications,}
  JHEP {\bf 10} (2005) 085,
\href{http://www.arXiv.org/abs/hep-th/0508133}{{\tt hep-th/0508133}}.

\bibitem{Hull:2006va}
C.~M. Hull, {\em Doubled geometry and T-folds,}
\href{http://www.arXiv.org/abs/hep-th/0605149}{{\tt hep-th/0605149}}.

\bibitem{Shelton:2006fd}
J.~Shelton, W.~Taylor, and B.~Wecht, {\em Generalized flux vacua,}
\href{http://www.arXiv.org/abs/hep-th/0607015}{{\tt hep-th/0607015}}.

\bibitem{Chiossi:2002tw}
S.~Chiossi and S.~Salamon, {\em The intrinsic torsion of SU(3) and $G_2$
  structures,}
\href{http://www.arXiv.org/abs/math.dg/0202282}{{\tt math.dg/0202282}}.

\bibitem{Becker:2002nn}
K.~Becker, M.~Becker, M.~Haack, and J.~Louis, {\em Supersymmetry breaking and
  alpha'-corrections to flux induced potentials,} JHEP {\bf 06} (2002) 060,
\href{http://www.arXiv.org/abs/hep-th/0204254}{{\tt hep-th/0204254}}.

\bibitem{Balasubramanian:2004uy}
V.~Balasubramanian and P.~Berglund, {\em Stringy corrections to Kaehler
  potentials, SUSY breaking, and the cosmological constant problem,} JHEP {\bf
  11} (2004) 085,
\href{http://www.arXiv.org/abs/hep-th/0408054}{{\tt hep-th/0408054}}.

\bibitem{conlon1}
V.~Balasubramanian, P.~Berglund, J.~P. Conlon, and F.~Quevedo, {\em Systematics
  of moduli stabilisation in Calabi-Yau flux compactifications,} JHEP {\bf 03}
  (2005) 007,
\href{http://www.arXiv.org/abs/hep-th/0502058}{{\tt hep-th/0502058}}.

\bibitem{FP}
A.~R. Frey and J.~Polchinski, {\em N = 3 warped compactifications,} Phys. Rev.
  {\bf D65} (2002) 126009,
\href{http://www.arXiv.org/abs/hep-th/0201029}{{\tt hep-th/0201029}}.

\bibitem{blt03}
R.~Blumenhagen, D.~L{\"u}st, and T.~R. Taylor, {\em Moduli stabilization in
  chiral type IIB orientifold models with fluxes,} Nucl. Phys. {\bf B663}
  (2003) 319--342,
\href{http://www.arXiv.org/abs/hep-th/0303016}{{\tt hep-th/0303016}}.

\bibitem{cu03}
J.~F.~G. Cascales and A.~M. Uranga, {\em Chiral 4d N = 1 string vacua with
  D-branes and NSNS and RR fluxes,} JHEP {\bf 05} (2003) 011,
\href{http://www.arXiv.org/abs/hep-th/0303024}{{\tt hep-th/0303024}}.

\bibitem{AF04}
A.~Font, {\em Z(N) orientifolds with flux,} JHEP {\bf 11} (2004) 077,
\href{http://www.arXiv.org/abs/hep-th/0410206}{{\tt hep-th/0410206}}.

\bibitem{KachruA}
O.~DeWolfe, A.~Giryavets, S.~Kachru, and W.~Taylor, {\em Enumerating flux vacua
  with enhanced symmetries,} JHEP {\bf 02} (2005) 037,
\href{http://www.arXiv.org/abs/hep-th/0411061}{{\tt hep-th/0411061}}.

\bibitem{cu03a}
J.~F.~G. Cascales and A.~M. Uranga, {\em Chiral 4d string vacua with D-branes
  and moduli stabilization,}
\href{http://www.arXiv.org/abs/hep-th/0311250}{{\tt hep-th/0311250}}.

\bibitem{ms04}
F.~Marchesano and G.~Shiu, {\em MSSM vacua from flux compactifications,} Phys.
  Rev. {\bf D71} (2005) 011701,
\href{http://www.arXiv.org/abs/hep-th/0408059}{{\tt hep-th/0408059}}.

\bibitem{cl04}
M.~Cvetic and T.~Liu, {\em Three-family supersymmetric standard models, flux
  compactification and moduli stabilization,} Phys. Lett. {\bf B610} (2005)
  122--128,
\href{http://www.arXiv.org/abs/hep-th/0409032}{{\tt hep-th/0409032}}.

\bibitem{ms04a}
F.~Marchesano and G.~Shiu, {\em Building MSSM flux vacua,} JHEP {\bf 11} (2004)
  041,
\href{http://www.arXiv.org/abs/hep-th/0409132}{{\tt hep-th/0409132}}.

\bibitem{cll05}
M.~Cvetic, T.~Li, and T.~Liu, {\em Standard-like models as type IIB flux
  vacua,} Phys. Rev. {\bf D71} (2005) 106008,
\href{http://www.arXiv.org/abs/hep-th/0501041}{{\tt hep-th/0501041}}.

\bibitem{Freed:1999vc}
D.~S. Freed and E.~Witten, {\em Anomalies in string theory with D-branes,}
\href{http://www.arXiv.org/abs/hep-th/9907189}{{\tt hep-th/9907189}}.

\bibitem{Marchesano:2006ns}
F.~Marchesano, {\em D6-branes and torsion,} JHEP {\bf 05} (2006) 019,
\href{http://www.arXiv.org/abs/hep-th/0603210}{{\tt hep-th/0603210}}.

\bibitem{Kumar:2005hf}
J.~Kumar and J.~D. Wells, {\em Surveying standard model flux vacua on T**6/Z(2)
  x Z(2),} JHEP {\bf 09} (2005) 067,
\href{http://www.arXiv.org/abs/hep-th/0506252}{{\tt hep-th/0506252}}.

\bibitem{Chen:2005mj}
C.-M. Chen, T.~Li, and D.~V. Nanopoulos, {\em Standard-like model building on
  type II orientifolds,} Nucl. Phys. {\bf B732} (2006) 224--242,
\href{http://www.arXiv.org/abs/hep-th/0509059}{{\tt hep-th/0509059}}.

\bibitem{Chen:2005cf}
C.-M. Chen, V.~E. Mayes, and D.~V. Nanopoulos, {\em Flipped SU(5) from D-branes
  with type IIB fluxes,} Phys. Lett. {\bf B633} (2006) 618--626,
\href{http://www.arXiv.org/abs/hep-th/0511135}{{\tt hep-th/0511135}}.

\bibitem{lrs04a}
D.~L{\"u}st, S.~Reffert, and S.~Stieberger, {\em MSSM with soft SUSY breaking
  terms from D7-branes with fluxes,} Nucl. Phys. {\bf B727} (2005) 264--300,
\href{http://www.arXiv.org/abs/hep-th/0410074}{{\tt hep-th/0410074}}.

\bibitem{fi04}
A.~Font and L.~E. Ib{\'a}{\~n}ez, {\em SUSY-breaking soft terms in a MSSM
  magnetized D7-brane model,} JHEP {\bf 03} (2005) 040,
\href{http://www.arXiv.org/abs/hep-th/0412150}{{\tt hep-th/0412150}}.

\bibitem{recently}
M.~P. Garcia~del Moral, {\em A new mechanism of Kahler moduli stabilization in
  type IIB theory,} JHEP {\bf 04} (2006) 022,
\href{http://www.arXiv.org/abs/hep-th/0506116}{{\tt hep-th/0506116}}.

\bibitem{kklw04}
G.~L. Kane, P.~Kumar, J.~D. Lykken, and T.~T. Wang, {\em Some phenomenology of
  intersecting D-brane models,} Phys. Rev. {\bf D71} (2005) 115017,
\href{http://www.arXiv.org/abs/hep-ph/0411125}{{\tt hep-ph/0411125}}.

\bibitem{msw04}
F.~Marchesano, G.~Shiu, and L.-T. Wang, {\em Model building and phenomenology
  of flux-induced supersymmetry breaking on D3-branes,} Nucl. Phys. {\bf B712}
  (2005) 20--58,
\href{http://www.arXiv.org/abs/hep-th/0411080}{{\tt hep-th/0411080}}.

\bibitem{Hall}
L.~J. Hall, J.~D. Lykken, and S.~Weinberg, {\em Supergravity as the messenger
  of supersymmetry breaking,} Phys. Rev. {\bf D27} (1983)
2359--2378.

\bibitem{kaplu}
V.~S. Kaplunovsky and J.~Louis, {\em Model independent analysis of soft terms
  in effective supergravity and in string theory,} Phys. Lett. {\bf B306}
  (1993) 269--275,
\href{http://www.arXiv.org/abs/hep-th/9303040}{{\tt hep-th/9303040}}.

\bibitem{Spain1}
A.~Brignole, L.~E. Ib{\'a}{\~n}ez, and C.~Munoz, {\em Soft
  supersymmetry-breaking terms from supergravity and superstring models,}
\href{http://www.arXiv.org/abs/hep-ph/9707209}{{\tt hep-ph/9707209}}.

\bibitem{IBANEZ04}
L.~E. Ib{\'a}{\~n}ez, {\em The fluxed MSSM,} Phys. Rev. {\bf D71} (2005)
  055005,
\href{http://www.arXiv.org/abs/hep-ph/0408064}{{\tt hep-ph/0408064}}.

\bibitem{Allanach:2005yq}
B.~C. Allanach, A.~Brignole, and L.~E. Ib{\'a}{\~n}ez, {\em Phenomenology of a
  fluxed MSSM,} JHEP {\bf 05} (2005) 030,
\href{http://www.arXiv.org/abs/hep-ph/0502151}{{\tt hep-ph/0502151}}.

\bibitem{conlon2}
J.~P. Conlon, F.~Quevedo, and K.~Suruliz, {\em Large-volume flux
  compactifications: Moduli spectrum and D3/D7 soft supersymmetry breaking,}
  JHEP {\bf 08} (2005) 007,
\href{http://www.arXiv.org/abs/hep-th/0505076}{{\tt hep-th/0505076}}.

\bibitem{conlon3}
J.~P. Conlon and F.~Quevedo, {\em Gaugino and scalar masses in the landscape,}
  JHEP {\bf 06} (2006) 029,
\href{http://www.arXiv.org/abs/hep-th/0605141}{{\tt hep-th/0605141}}.

\bibitem{conlon4}
J.~P. Conlon, D.~Cremades, and F.~Quevedo, {\em Kaehler potentials of chiral
  matter fields for Calabi-Yau string compactifications,}
\href{http://www.arXiv.org/abs/hep-th/0609180}{{\tt hep-th/0609180}}.

\bibitem{conlon5}
J.~P. Conlon, S.~S. Abdussalam, F.~Quevedo, and K.~Suruliz, {\em Soft SUSY
  breaking terms for chiral matter in IIB string compactifications,}
\href{http://www.arXiv.org/abs/hep-th/0610129}{{\tt hep-th/0610129}}.

\bibitem{Kane:2006yi}
G.~L. Kane, P.~Kumar, and J.~Shao, {\em LHC string phenomenology,}
\href{http://www.arXiv.org/abs/hep-ph/0610038}{{\tt hep-ph/0610038}}.

\bibitem{kklt03}
S.~Kachru, R.~Kallosh, A.~Linde, and S.~P. Trivedi, {\em De Sitter vacua in
  string theory,} Phys. Rev. {\bf D68} (2003) 046005,
\href{http://www.arXiv.org/abs/hep-th/0301240}{{\tt hep-th/0301240}}.

\bibitem{Witten:1996bn}
E.~Witten, {\em Non-Perturbative Superpotentials In String Theory,} Nucl. Phys.
  {\bf B474} (1996) 343--360,
\href{http://www.arXiv.org/abs/hep-th/9604030}{{\tt hep-th/9604030}}.

\bibitem{Kallosh:2005yu}
R.~Kallosh and D.~Sorokin, {\em Dirac action on M5 and M2 branes with bulk
  fluxes,} JHEP {\bf 05} (2005) 005,
\href{http://www.arXiv.org/abs/hep-th/0501081}{{\tt hep-th/0501081}}.

\bibitem{Saulina:2005ve}
N.~Saulina, {\em Topological constraints on stabilized flux vacua,} Nucl. Phys.
  {\bf B720} (2005) 203--210,
\href{http://www.arXiv.org/abs/hep-th/0503125}{{\tt hep-th/0503125}}.

\bibitem{Kallosh:2005gs}
R.~Kallosh, A.-K. Kashani-Poor, and A.~Tomasiello, {\em Counting fermionic zero
  modes on M5 with fluxes,} JHEP {\bf 06} (2005) 069,
\href{http://www.arXiv.org/abs/hep-th/0503138}{{\tt hep-th/0503138}}.

\bibitem{Bergshoeff:2005yp}
E.~Bergshoeff, R.~Kallosh, A.-K. Kashani-Poor, D.~Sorokin, and A.~Tomasiello,
  {\em An index for the Dirac operator on D3 branes with background fluxes,}
  JHEP {\bf 10} (2005) 102,
\href{http://www.arXiv.org/abs/hep-th/0507069}{{\tt hep-th/0507069}}.

\bibitem{Park:2005hj}
J.~Park, {\em D3 instantons in Calabi-Yau orientifolds with(out) fluxes,}
\href{http://www.arXiv.org/abs/hep-th/0507091}{{\tt hep-th/0507091}}.

\bibitem{Lust:2005cu}
D.~L{\"u}st, S.~Reffert, W.~Schulgin, and P.~K. Tripathy, {\em Fermion zero
  modes in the presence of fluxes and a non- perturbative superpotential,} JHEP
  {\bf 08} (2006) 071,
\href{http://www.arXiv.org/abs/hep-th/0509082}{{\tt hep-th/0509082}}.

\bibitem{Berglund:2005dm}
P.~Berglund and P.~Mayr, {\em Non-perturbative superpotentials in F-theory and
  string duality,}
\href{http://www.arXiv.org/abs/hep-th/0504058}{{\tt hep-th/0504058}}.

\bibitem{Aspinwall:2005ad}
P.~S. Aspinwall and R.~Kallosh, {\em Fixing all moduli for M-theory on K3 x
  K3,} JHEP {\bf 10} (2005) 001,
\href{http://www.arXiv.org/abs/hep-th/0506014}{{\tt hep-th/0506014}}.

\bibitem{Affleck:1983mk}
I.~Affleck, M.~Dine, and N.~Seiberg, {\em Dynamical supersymmetry breaking in
  supersymmetric QCD,} Nucl. Phys. {\bf B241} (1984)
493--534.

\bibitem{Haack:2006cy}
M.~Haack, D.~Krefl, D.~L{\"u}st, A.~Van~Proeyen, and M.~Zagermann, {\em Gaugino
  Condensates and D-terms from D7-branes,}
\href{http://www.arXiv.org/abs/hep-th/0609211}{{\tt hep-th/0609211}}.

\bibitem{Achucarro:2006zf}
A.~Achucarro, B.~de~Carlos, J.~A. Casas, and L.~Doplicher, {\em de Sitter vacua
  from uplifting D-terms in effective supergravities from realistic strings,}
  JHEP {\bf 06} (2006) 014,
\href{http://www.arXiv.org/abs/hep-th/0601190}{{\tt hep-th/0601190}}.

\bibitem{leb0}
D.~Cremades, M.~P. Garcia~del Moral, F.~Quevedo, and K.~Suruliz, {\em Moduli
  stabilisation and de Sitter string vacua from magnetised D7 branes,}
\href{http://www.arXiv.org/abs/hep-th/0701154}{{\tt hep-th/0701154}}.

\bibitem{gktt04}
L.~G{\"o}rlich, S.~Kachru, P.~K. Tripathy, and S.~P. Trivedi, {\em Gaugino
  condensation and nonperturbative superpotentials in flux compactifications,}
  JHEP {\bf 12} (2004) 074,
\href{http://www.arXiv.org/abs/hep-th/0407130}{{\tt hep-th/0407130}}.

\bibitem{Blanco-Pillado:2005fn}
J.~J. Blanco-Pillado, R.~Kallosh, and A.~Linde, {\em Supersymmetry and
  stability of flux vacua,} JHEP {\bf 05} (2006) 053,
\href{http://www.arXiv.org/abs/hep-th/0511042}{{\tt hep-th/0511042}}.

\bibitem{Krefl:2006vu}
D.~Krefl and D.~L{\"u}st, {\em On supersymmetric Minkowski vacua in IIB
  orientifolds,} JHEP {\bf 06} (2006) 023,
\href{http://www.arXiv.org/abs/hep-th/0603166}{{\tt hep-th/0603166}}.

\bibitem{HPNii}
K.~Choi, A.~Falkowski, H.~P. Nilles, and M.~Olechowski, {\em Soft supersymmetry
  breaking in KKLT flux compactification,} Nucl. Phys. {\bf B718} (2005)
  113--133,
\href{http://www.arXiv.org/abs/hep-th/0503216}{{\tt hep-th/0503216}}.

\bibitem{Race}
F.~Denef, M.~R. Douglas, and B.~Florea, {\em Building a better racetrack,} JHEP
  {\bf 06} (2004) 034,
\href{http://www.arXiv.org/abs/hep-th/0404257}{{\tt hep-th/0404257}}.

\bibitem{Linde}
J.~J. Blanco-Pillado {\em et al.}, {\em Inflating in a better racetrack,} JHEP
  {\bf 09} (2006) 002,
\href{http://www.arXiv.org/abs/hep-th/0603129}{{\tt hep-th/0603129}}.

\bibitem{BF}
P.~Breitenlohner and D.~Z. Freedman, {\em Stability In Gauged Extended
  Supergravity,} Ann. Phys. {\bf 144} (1982)
249.

\bibitem{ChoiSX}
K.~Choi, A.~Falkowski, H.~P. Nilles, M.~Olechowski, and S.~Pokorski, {\em
  Stability of flux compactifications and the pattern of supersymmetry
  breaking,} JHEP {\bf 11} (2004) 076,
\href{http://www.arXiv.org/abs/hep-th/0411066}{{\tt hep-th/0411066}}.

\bibitem{dealwis}
S.~P. de~Alwis, {\em Effective potentials for light moduli,} Phys. Lett. {\bf
  B626} (2005) 223--229,
\href{http://www.arXiv.org/abs/hep-th/0506266}{{\tt hep-th/0506266}}.

\bibitem{leb2}
O.~Lebedev, H.~P. Nilles, and M.~Ratz, {\em de Sitter vacua from matter
  superpotentials,} Phys. Lett. {\bf B636} (2006) 126,
\href{http://www.arXiv.org/abs/hep-th/0603047}{{\tt hep-th/0603047}}.

\bibitem{Japan1}
H.~Abe, T.~Higaki, and T.~Kobayashi, {\em Remark on integrating out heavy
  moduli in flux compactification,} Phys. Rev. {\bf D74} (2006) 045012,
\href{http://www.arXiv.org/abs/hep-th/0606095}{{\tt hep-th/0606095}}.

\bibitem{Japan2}
H.-X. Yang, {\em On moduli stabilization scheme in type IIB flux
  compactifications,}
\href{http://www.arXiv.org/abs/hep-th/0608155}{{\tt hep-th/0608155}}.

\bibitem{Japan3}
H.-X. Yang, {\em Moduli stabilization in type IIB flux compactifications,}
  Phys. Rev. {\bf D73} (2006) 066006,
\href{http://www.arXiv.org/abs/hep-th/0511030}{{\tt hep-th/0511030}}.

\bibitem{Burgess:2003ic}
C.~P. Burgess, R.~Kallosh, and F.~Quevedo, {\em de Sitter string vacua from
  supersymmetric D-terms,} JHEP {\bf 10} (2003) 056,
\href{http://www.arXiv.org/abs/hep-th/0309187}{{\tt hep-th/0309187}}.

\bibitem{leb1}
O.~Lebedev, V.~Lowen, Y.~Mambrini, H.~P. Nilles, and M.~Ratz, {\em Metastable
  vacua in flux compactifications and their phenomenology,} JHEP {\bf 02}
  (2007) 063,
\href{http://www.arXiv.org/abs/hep-ph/0612035}{{\tt hep-ph/0612035}}.

\bibitem{LopesCardoso:2002hd}
G.~Lopes~Cardoso, G.~Curio, G.~Dall'Agata, D.~L{\"u}st, P.~Manousselis, and
  G.~Zoupanos, {\em Non-Kaehler string backgrounds and their five torsion
  classes,} Nucl. Phys. {\bf B652} (2003) 5--34,
\href{http://www.arXiv.org/abs/hep-th/0211118}{{\tt hep-th/0211118}}.

\bibitem{bc03}
K.~Behrndt and M.~Cveti{\v c}, {\em Supersymmetric intersecting D6-branes and
  fluxes in massive type IIA string theory,} Nucl. Phys. {\bf B676} (2004)
  149--171,
\href{http://www.arXiv.org/abs/hep-th/0308045}{{\tt hep-th/0308045}}.

\bibitem{Dall'Agata:2003ir}
G.~Dall'Agata and N.~Prezas, {\em N = 1 geometries for M-theory and type IIA
  strings with fluxes,} Phys. Rev. {\bf D69} (2004) 066004,
\href{http://www.arXiv.org/abs/hep-th/0311146}{{\tt hep-th/0311146}}.

\bibitem{bc04}
K.~Behrndt and M.~Cveti{\v c}, {\em General N = 1 supersymmetric flux vacua of
  (massive) type IIA string theory,} Phys. Rev. Lett. {\bf 95} (2005) 021601,
\href{http://www.arXiv.org/abs/hep-th/0403049}{{\tt hep-th/0403049}}.

\bibitem{bc04a}
K.~Behrndt and M.~Cveti{\v c}, {\em General N = 1 supersymmetric fluxes in
  massive type IIA string theory,} Nucl. Phys. {\bf B708} (2005) 45--71,
\href{http://www.arXiv.org/abs/hep-th/0407263}{{\tt hep-th/0407263}}.

\bibitem{Derendinger:2004jn}
J.-P. Derendinger, C.~Kounnas, P.~M. Petropoulos, and F.~Zwirner, {\em
  Superpotentials in IIA compactifications with general fluxes,} Nucl. Phys.
  {\bf B715} (2005) 211--233,
\href{http://www.arXiv.org/abs/hep-th/0411276}{{\tt hep-th/0411276}}.

\bibitem{Kachru:2004jr}
S.~Kachru and A.-K. Kashani-Poor, {\em Moduli potentials in type IIA
  compactifications with RR and NS flux,} JHEP {\bf 03} (2005) 066,
\href{http://www.arXiv.org/abs/hep-th/0411279}{{\tt hep-th/0411279}}.

\bibitem{DeWolfe:2005uu}
O.~DeWolfe, A.~Giryavets, S.~Kachru, and W.~Taylor, {\em Type IIA moduli
  stabilization,} JHEP {\bf 07} (2005) 066,
\href{http://www.arXiv.org/abs/hep-th/0505160}{{\tt hep-th/0505160}}.

\bibitem{Camara:2005dc}
P.~G. Camara, A.~Font, and L.~E. Ib{\'a}{\~n}ez, {\em Fluxes, moduli fixing and
  MSSM-like vacua in a simple IIA orientifold,} JHEP {\bf 09} (2005) 013,
\href{http://www.arXiv.org/abs/hep-th/0506066}{{\tt hep-th/0506066}}.

\bibitem{Aldazabal:2006up}
G.~Aldazabal, P.~G. Camara, A.~Font, and L.~E. Ib{\'a}{\~n}ez, {\em More dual
  fluxes and moduli fixing,} JHEP {\bf 05} (2006) 070,
\href{http://www.arXiv.org/abs/hep-th/0602089}{{\tt hep-th/0602089}}.

\bibitem{Villadoro:2005cu}
G.~Villadoro and F.~Zwirner, {\em N = 1 effective potential from dual type-IIA
  D6/O6 orientifolds with general fluxes,} JHEP {\bf 06} (2005) 047,
\href{http://www.arXiv.org/abs/hep-th/0503169}{{\tt hep-th/0503169}}.

\bibitem{Acharya:2006ne}
B.~S. Acharya, F.~Benini, and R.~Valandro, {\em Fixing moduli in exact type IIA
  flux vacua,}
\href{http://www.arXiv.org/abs/hep-th/0607223}{{\tt hep-th/0607223}}.

\bibitem{Hitchin:2001rw}
N.~J. Hitchin, {\em Stable forms and special metrics,}
\href{http://www.arXiv.org/abs/math.dg/0107101}{{\tt math.dg/0107101}}.

\bibitem{Lust:2004ig}
D.~L{\"u}st and D.~Tsimpis, {\em Supersymmetric AdS(4) compactifications of IIA
  supergravity,} JHEP {\bf 02} (2005) 027,
\href{http://www.arXiv.org/abs/hep-th/0412250}{{\tt hep-th/0412250}}.

\bibitem{Derendinger:2005ph}
J.~P. Derendinger, C.~Kounnas, P.~M. Petropoulos, and F.~Zwirner, {\em Fluxes
  and gaugings: N = 1 effective superpotentials,} Fortsch. Phys. {\bf 53}
  (2005) 926--935,
\href{http://www.arXiv.org/abs/hep-th/0503229}{{\tt hep-th/0503229}}.

\bibitem{Chen:2006gd}
C.-M. Chen, T.~Li, and D.~V. Nanopoulos, {\em Type IIA Pati-Salam flux vacua,}
  Nucl. Phys. {\bf B740} (2006) 79--104,
\href{http://www.arXiv.org/abs/hep-th/0601064}{{\tt hep-th/0601064}}.

\bibitem{Chen:2006ip}
C.-M. Chen, T.~Li, and D.~V. Nanopoulos, {\em Flipped and unflipped SU(5) as
  type IIA flux vacua,} Nucl. Phys. {\bf B751} (2006) 260--284,
\href{http://www.arXiv.org/abs/hep-th/0604107}{{\tt hep-th/0604107}}.

\bibitem{Ihl:2006pp}
M.~Ihl and T.~Wrase, {\em Towards a realistic type IIA T**6/Z(4) orientifold
  model with background fluxes. I: Moduli stabilization,} JHEP {\bf 07} (2006)
  027,
\href{http://www.arXiv.org/abs/hep-th/0604087}{{\tt hep-th/0604087}}.

\bibitem{Floratos:2006hs}
E.~Floratos and C.~Kokorelis, {\em MSSM GUT string vacua, split supersymmetry
  and fluxes,}
\href{http://www.arXiv.org/abs/hep-th/0607217}{{\tt hep-th/0607217}}.

\bibitem{LopesCardoso:2003af}
G.~Lopes~Cardoso, G.~Curio, G.~Dall'Agata, and D.~L{\"u}st, {\em BPS action and
  superpotential for heterotic string compactifications with fluxes,} JHEP {\bf
  10} (2003) 004,
\href{http://www.arXiv.org/abs/hep-th/0306088}{{\tt hep-th/0306088}}.

\bibitem{Louis:2001uy}
J.~Louis and A.~Micu, {\em Heterotic string theory with background fluxes,}
  Nucl. Phys. {\bf B626} (2002) 26--52,
\href{http://www.arXiv.org/abs/hep-th/0110187}{{\tt hep-th/0110187}}.

\bibitem{Gutowski:2002bc}
J.~Gutowski, S.~Ivanov, and G.~Papadopoulos, {\em Deformations of generalized
  calibrations and compact non- Kahler manifolds with vanishing first Chern
  class,}
\href{http://www.arXiv.org/abs/math.dg/0205012}{{\tt math.dg/0205012}}.

\bibitem{Gauntlett:2002sc}
J.~P. Gauntlett, D.~Martelli, S.~Pakis, and D.~Waldram, {\em G-structures and
  wrapped NS5-branes,} Commun. Math. Phys. {\bf 247} (2004) 421--445,
\href{http://www.arXiv.org/abs/hep-th/0205050}{{\tt hep-th/0205050}}.

\bibitem{Becker:2002sx}
K.~Becker and K.~Dasgupta, {\em Heterotic strings with torsion,} JHEP {\bf 11}
  (2002) 006,
\href{http://www.arXiv.org/abs/hep-th/0209077}{{\tt hep-th/0209077}}.

\bibitem{Becker:2002jj}
M.~Becker and D.~Constantin, {\em A note on flux induced superpotentials in
  string theory,} JHEP {\bf 08} (2003) 015,
\href{http://www.arXiv.org/abs/hep-th/0210131}{{\tt hep-th/0210131}}.

\bibitem{Goldstein:2002pg}
E.~Goldstein and S.~Prokushkin, {\em Geometric model for complex non-Kaehler
  manifolds with SU(3) structure,} Commun. Math. Phys. {\bf 251} (2004) 65--78,
\href{http://www.arXiv.org/abs/hep-th/0212307}{{\tt hep-th/0212307}}.

\bibitem{Becker:2003yv}
K.~Becker, M.~Becker, K.~Dasgupta, and P.~S. Green, {\em Compactifications of
  heterotic theory on non-Kaehler complex manifolds. I,} JHEP {\bf 04} (2003)
  007,
\href{http://www.arXiv.org/abs/hep-th/0301161}{{\tt hep-th/0301161}}.

\bibitem{Gauntlett:2003cy}
J.~P. Gauntlett, D.~Martelli, and D.~Waldram, {\em Superstrings with intrinsic
  torsion,} Phys. Rev. {\bf D69} (2004) 086002,
\href{http://www.arXiv.org/abs/hep-th/0302158}{{\tt hep-th/0302158}}.

\bibitem{Becker:2003gq}
K.~Becker, M.~Becker, K.~Dasgupta, and S.~Prokushkin, {\em Properties of
  heterotic vacua from superpotentials,} Nucl. Phys. {\bf B666} (2003)
  144--174,
\href{http://www.arXiv.org/abs/hep-th/0304001}{{\tt hep-th/0304001}}.

\bibitem{Curio:2003ur}
G.~Curio and A.~Krause, {\em Enlarging the parameter space of heterotic
  M-theory flux compactifications to phenomenological viability,} Nucl. Phys.
  {\bf B693} (2004) 195--222,
\href{http://www.arXiv.org/abs/hep-th/0308202}{{\tt hep-th/0308202}}.

\bibitem{LopesCardoso:2003sp}
G.~Lopes~Cardoso, G.~Curio, G.~Dall'Agata, and D.~L{\"u}st, {\em Heterotic
  string theory on non-Kaehler manifolds with H- flux and gaugino condensate,}
  Fortsch. Phys. {\bf 52} (2004) 483--488,
\href{http://www.arXiv.org/abs/hep-th/0310021}{{\tt hep-th/0310021}}.

\bibitem{Becker:2003sh}
K.~Becker, M.~Becker, P.~S. Green, K.~Dasgupta, and E.~Sharpe, {\em
  Compactifications of heterotic strings on non-Kaehler complex manifolds. II,}
  Nucl. Phys. {\bf B678} (2004) 19--100,
\href{http://www.arXiv.org/abs/hep-th/0310058}{{\tt hep-th/0310058}}.

\bibitem{Becker:2003dz}
M.~Becker and K.~Dasgupta, {\em Kaehler versus non-Kaehler compactifications,}
\href{http://www.arXiv.org/abs/hep-th/0312221}{{\tt hep-th/0312221}}.

\bibitem{Becker:2004qh}
M.~Becker, K.~Dasgupta, A.~Knauf, and R.~Tatar, {\em Geometric transitions,
  flops and non-Kaehler manifolds. I,} Nucl. Phys. {\bf B702} (2004) 207--268,
\href{http://www.arXiv.org/abs/hep-th/0403288}{{\tt hep-th/0403288}}.

\bibitem{Becker:2004ii}
K.~Becker, M.~Becker, K.~Dasgupta, and R.~Tatar, {\em Geometric transitions,
  non-Kaehler geometries and string vacua,} Int. J. Mod. Phys. {\bf A20} (2005)
  3442--3448,
\href{http://www.arXiv.org/abs/hep-th/0411039}{{\tt hep-th/0411039}}.

\bibitem{Curio:2005ew}
G.~Curio, A.~Krause, and D.~L{\"u}st, {\em Moduli stabilization in the
  heterotic / IIB discretuum,} Fortsch. Phys. {\bf 54} (2006) 225--245,
\href{http://www.arXiv.org/abs/hep-th/0502168}{{\tt hep-th/0502168}}.

\bibitem{Kimura:2006af}
T.~Kimura and P.~Yi, {\em Comments on heterotic flux compactifications,} JHEP
  {\bf 07} (2006) 030,
\href{http://www.arXiv.org/abs/hep-th/0605247}{{\tt hep-th/0605247}}.

\bibitem{Curio:2006dc}
G.~Curio and A.~Krause, {\em S-Track stabilization of heterotic de Sitter
  vacua,}
\href{http://www.arXiv.org/abs/hep-th/0606243}{{\tt hep-th/0606243}}.

\bibitem{Kim:2006qs}
S.~Kim and P.~Yi, {\em A heterotic flux background and calibrated five-branes,}
\href{http://www.arXiv.org/abs/hep-th/0607091}{{\tt hep-th/0607091}}.

\bibitem{Bergshoeff:1989de}
E.~A. Bergshoeff and M.~de~Roo, {\em The quartic effective action of the
  heterotic string and supersymmetry,} Nucl. Phys. {\bf B328} (1989)
439.

\bibitem{Ashok:2003gk}
S.~Ashok and M.~R. Douglas, {\em Counting flux vacua,} JHEP {\bf 01} (2004)
  060,
\href{http://www.arXiv.org/abs/hep-th/0307049}{{\tt hep-th/0307049}}.

\bibitem{Skarke:1996hq}
H.~Skarke, {\em Weight systems for toric Calabi-Yau varieties and reflexivity
  of Newton polyhedra,} Mod. Phys. Lett. {\bf A11} (1996) 1637--1652,
\href{http://www.arXiv.org/abs/alg-geom/9603007}{{\tt alg-geom/9603007}}.

\bibitem{Schellekens:2006xz}
A.~N. Schellekens, {\em The landscape 'avant la lettre',}
\href{http://www.arXiv.org/abs/physics/0604134}{{\tt physics/0604134}}.

\bibitem{Bousso:2000xa}
R.~Bousso and J.~Polchinski, {\em Quantization of four-form fluxes and
  dynamical neutralization of the cosmological constant,} JHEP {\bf 06} (2000)
  006,
\href{http://www.arXiv.org/abs/hep-th/0004134}{{\tt hep-th/0004134}}.

\bibitem{Douglas:2003um}
M.~R. Douglas, {\em The statistics of string / M theory vacua,} JHEP {\bf 05}
  (2003) 046,
\href{http://www.arXiv.org/abs/hep-th/0303194}{{\tt hep-th/0303194}}.

\bibitem{Denef:2004ze}
F.~Denef and M.~R. Douglas, {\em Distributions of flux vacua,} JHEP {\bf 05}
  (2004) 072,
\href{http://www.arXiv.org/abs/hep-th/0404116}{{\tt hep-th/0404116}}.

\bibitem{Acharya:2005ez}
B.~S. Acharya, F.~Denef, and R.~Valandro, {\em Statistics of M theory vacua,}
  JHEP {\bf 06} (2005) 056,
\href{http://www.arXiv.org/abs/hep-th/0502060}{{\tt hep-th/0502060}}.

\bibitem{Douglas:2004zg}
M.~R. Douglas, {\em Basic results in vacuum statistics,} Comptes Rendus
  Physique {\bf 5} (2004) 965--977,
\href{http://www.arXiv.org/abs/hep-th/0409207}{{\tt hep-th/0409207}}.

\bibitem{Kreuzer:2000xy}
M.~Kreuzer and H.~Skarke, {\em Complete classification of reflexive polyhedra
  in four dimensions,} Adv. Theor. Math. Phys. {\bf 4} (2002) 1209--1230,
\href{http://www.arXiv.org/abs/hep-th/0002240}{{\tt hep-th/0002240}}.

\bibitem{Susskind:2003kw}
L.~Susskind, {\em The anthropic landscape of string theory,}
\href{http://www.arXiv.org/abs/hep-th/0302219}{{\tt hep-th/0302219}}.

\bibitem{Vafa:2005ui}
C.~Vafa, {\em The string landscape and the swampland,}
\href{http://www.arXiv.org/abs/hep-th/0509212}{{\tt hep-th/0509212}}.

\bibitem{Kumar:2004pv}
J.~Kumar and J.~D. Wells, {\em Landscape cartography: A coarse survey of gauge
  group rank and stabilization of the proton,} Phys. Rev. {\bf D71} (2005)
  026009,
\href{http://www.arXiv.org/abs/hep-th/0409218}{{\tt hep-th/0409218}}.

\bibitem{Blumenhagen:2004xx}
R.~Blumenhagen, F.~Gmeiner, G.~Honecker, D.~L{\"u}st, and T.~Weigand, {\em The
  statistics of supersymmetric D-brane models,} Nucl. Phys. {\bf B713} (2005)
  83--135,
\href{http://www.arXiv.org/abs/hep-th/0411173}{{\tt hep-th/0411173}}.

\bibitem{Gmeiner:2005vz}
F.~Gmeiner, R.~Blumenhagen, G.~Honecker, D.~Lust, and T.~Weigand, {\em One in a
  billion: MSSM-like D-brane statistics,} JHEP {\bf 01} (2006) 004,
\href{http://www.arXiv.org/abs/hep-th/0510170}{{\tt hep-th/0510170}}.

\bibitem{Gmeiner:2005nh}
F.~Gmeiner, {\em Standard model statistics of a type II orientifold,} Fortsch.
  Phys. {\bf 54} (2006) 391--398,
\href{http://www.arXiv.org/abs/hep-th/0512190}{{\tt hep-th/0512190}}.

\bibitem{Gmeiner:2006vb}
F.~Gmeiner and M.~Stein, {\em Statistics of SU(5) D-brane models on a type II
  orientifold,} Phys. Rev. {\bf D73} (2006) 126008,
\href{http://www.arXiv.org/abs/hep-th/0603019}{{\tt hep-th/0603019}}.

\bibitem{Gmeiner:2006qw}
F.~Gmeiner, {\em Gauge sector statistics of intersecting D-brane models,}
\href{http://www.arXiv.org/abs/hep-th/0608227}{{\tt hep-th/0608227}}.

\bibitem{Douglas:2006xy}
M.~R. Douglas and W.~Taylor, {\em The landscape of intersecting brane models,}
\href{http://www.arXiv.org/abs/hep-th/0606109}{{\tt hep-th/0606109}}.

\bibitem{Dienes:2006ut}
K.~R. Dienes, {\em Statistics on the heterotic landscape: Gauge groups and
  cosmological constants of four-dimensional heterotic strings,} Phys. Rev.
  {\bf D73} (2006) 106010,
\href{http://www.arXiv.org/abs/hep-th/0602286}{{\tt hep-th/0602286}}.

\bibitem{Faraggi:2006bc}
A.~E. Faraggi, C.~Kounnas, and J.~Rizos, {\em Chiral family classification of
  fermionic Z(2) x Z(2) heterotic orbifold models,}
\href{http://www.arXiv.org/abs/hep-th/0606144}{{\tt hep-th/0606144}}.

\bibitem{Denef:2006ad}
F.~Denef and M.~R. Douglas, {\em Computational complexity of the landscape. I,}
\href{http://www.arXiv.org/abs/hep-th/0602072}{{\tt hep-th/0602072}}.

\bibitem{Douglas:2006hz}
M.~R. Douglas, R.~L. Karp, S.~Lukic, and R.~Reinbacher, {\em Numerical solution
  to the hermitian Yang-Mills equation on the Fermat quintic,}
\href{http://www.arXiv.org/abs/hep-th/0606261}{{\tt hep-th/0606261}}.

\end{thebibliography}\endgroup
\bibliographystyle{utphys}

\end{document}